\documentclass[final,3p,times]{elsarticle}

\usepackage{graphicx}
\usepackage{amssymb}
\usepackage{amsmath}
\usepackage{bm}
\usepackage{color}

\usepackage{cleveref}
\crefformat{appendix}{}

\biboptions{sort&compress}

\journal{Physics Reports}

\begin{document}

\begin{frontmatter}

\title{Quantum field theory in a magnetic field:\\
From quantum chromodynamics to graphene and Dirac semimetals}

\author[VAM]{Vladimir A. Miransky}
\ead{vmiransk@uwo.ca}
\author[IAS]{Igor A. Shovkovy\corref{CorrespondingAuthor}}
\ead{igor.shovkovy@asu.edu}
\ead[url]{http://shovkovy.faculty.asu.edu}

\address[VAM]{Department of Applied Mathematics, Western University, London, Ontario N6A 5B7, Canada}
\address[IAS]{College of Letters and Sciences, Arizona State University, Mesa, Arizona 85212, USA}

\cortext[CorrespondingAuthor]{Corresponding author}

\begin{abstract}
A range of quantum field theoretical phenomena driven by external magnetic fields and their 
applications in relativistic systems and quasirelativistic condensed matter ones, such as graphene 
and Dirac/Weyl semimetals, are reviewed. We start by introducing the underlying physics of 
the magnetic catalysis. The dimensional reduction of the low-energy dynamics of relativistic 
fermions in an external magnetic field is explained and its role in catalyzing spontaneous 
symmetry breaking is emphasized. The general theoretical consideration is supplemented 
by the analysis of the magnetic catalysis in quantum electrodynamics, chromodynamics and 
quasirelativistic models relevant for condensed matter physics. By generalizing the ideas of 
the magnetic catalysis to the case of nonzero density and temperature, we argue that other 
interesting phenomena take place. The chiral magnetic and chiral separation effects are 
perhaps the most interesting among them. In addition to the general discussion of the 
physics underlying chiral magnetic and separation effects, we also review their possible 
phenomenological implications in heavy-ion collisions and compact stars. We also discuss 
the application of the magnetic catalysis ideas for the description of the quantum Hall effect 
in monolayer and bilayer graphene, and conclude that the generalized magnetic catalysis, 
including both the magnetic catalysis condensates and the quantum Hall ferromagnetic ones, 
lies at the basis of this phenomenon. We also consider how an external magnetic field affects 
the underlying physics in a class of three-dimensional quasirelativistic condensed matter 
systems, Dirac semimetals. While at sufficiently low temperatures and zero density of 
charge carriers, such semimetals are expected to reveal the regime of the magnetic 
catalysis, the regime of Weyl semimetals with chiral asymmetry is realized at nonzero 
density. Finally, we discuss the interplay between relativistic quantum field theories 
(including quantum electrodynamics and quantum chromodynamics) in a magnetic 
field and noncommutative field theories, which leads to a new type of the latter, nonlocal 
noncommutative field theories.
\end{abstract}


\date{\today}

\end{frontmatter}

\tableofcontents



\section{Introduction}
\label{sec:Introduction}

The gravitational and electromagnetic forces are the only known long-range interactions 
in Nature. Because of this property, they lead to numerous easily observed 
macroscopic phenomena and become an integral part of our everyday life. This is also the 
main reason why these two forces were discovered long time ago. Systematic studies of 
the gravitational forces with the use of modern scientific methods started in the 16th-17th 
centuries (with the works of Galileo Galilei, Isaac Newton and others), and the studies 
of electromagnetism began in the 19th century (with the works of Alessandro Volta, 
Hans Christian Oersted, Andre-Marie Ampere, Michael Faraday, and James Clerk 
Maxwell, and many others). The other two known fundamental interactions of nature, 
responsible for the strong and weak forces, are short range and can be detected only 
under very special laboratory conditions. They were discovered much later, in the middle 
of the 20th century.

It is not surprising, therefore, that the first two classical field theories were the 
theory of electromagnetism (the Maxwell's theory) and the theory of gravity (the 
Newton's law of universal gravitation and later the Einstein's theory of general 
relativity). The Maxwell's theory led to remarkable discoveries, one of which was 
the understanding of the electromagnetic nature of light. 

The electromagnetic interactions became even more important in the quantum epoch, 
when the quantum nature of numerous phenomena was revealed. The paramagnetism 
of metals was described by the Pauli theory \cite{1927ZPhy...41...81P,1927ZPhy...43..601P}, the 
diamagnetism was described by the Landau theory \cite{1930ZPhy...64..629L}, and the 
ferromagnetism found the explanation in the Heisenberg theory \cite{1928ZPhy...49..619H}. 
New effects, such as the Meissner effect in superconductivity \cite{1933NW.....21..787M}, 
Shubnikov-de Haas effect (oscillations) \cite{Shubnikov-deHaas:1930a,Shubnikov-deHaas:1930b,
Shubnikov-deHaas:1930c} in metals and semimetals, quantum Hall effect \cite{1975JPSJ...39..279A,
PhysRevLett.45.494,PhysRevB.23.5632} in semiconductors, to name few, were discovered. 

With the development of quantum field theory, a whole range of other quantum 
phenomena was discovered and explained theoretically. Of particular importance among 
them was the Bardeen-Cooper-Schrieffer (BCS) theory of low-temperature superconductivity 
\cite{Bardeen:1957mv,Bardeen:1957kj}. In addition to being of great value on its own, 
it inspired numerous applications of spontaneous (dynamical) symmetry breaking 
mechanisms to various subfields of physics. In particle physics, for example, the 
mechanism of symmetry breaking helped to understand the puzzling low-energy 
spectrum of QCD, allowed a natural unification of the electromagnetic and weak 
interactions, and gave a real hope for the ultimate unification of all forces in Nature. 
The success of the corresponding ideas also shaped our understanding of cosmology.
In particular, they implied that symmetry breaking phase transitions should have 
occurred during the evolution of the early Universe. Considering the likely existence 
of very strong magnetic fields at that epoch \cite{Vilenkin:1980fu,Vachaspati:1991nm,Enqvist:1993kf,
Cheng:1994yr,Baym:1995fk,Grasso:2000wj}, it is natural to ask how the underlying 
dynamics and the phase transition were affected by such fields. Can one utilize the 
longtime experience from low-temperature superconductivity to answer the question? 
In general, the answer is ``No".

As is well known, in low-temperature superconductors, an external magnetic field has a tendency 
to break superconductivity by interfering with Cooper pairing. At the microscopic 
level, this can be understood as the result of interaction of the electrons' magnetic 
moments with the field. The seemingly favorable alignment of the magnetic 
moments with the direction of the magnetic field comes in a direct conflict with 
the antiparallel alignment inside Cooper pairs. However, this example alone does 
not help to shed light on the underlying dynamics in relativistic systems 
undergoing other types of phase transitions, associated with spontaneous (dynamical) 
symmetry breaking. In fact, as we discuss in this review, there exist (quasi-)relativistic 
systems, in which external magnetic fields not only enhance, but even induce 
symmetry breaking.
 
Here it may be appropriate to emphasize that magnetic fields indeed play an 
important role in the dynamics of many relativistic and quasirelativistic systems. 
These range from the already mentioned evolution of the early Universe to the fireballs 
of quark-gluon plasma created in heavy-ion collisions, from compact stars to numerous 
condensed matter systems. The latter, in particular, include a growing number of 
quasirelativistic systems, such as 2-dimensional graphene and a number of
3-dimensional Dirac (semi-)metals. It is also hoped that, by making use of certain 
topological properties of materials and crystal symmetries, other types of quasirelativistic 
materials will be possible to engineer in the future. One of such exotic possibilities is 
Weyl semimetals.

The origin of the magnetic field may differ from one system to another. In some cases, 
one may simply apply external (electro-)magnetic fields in order to either probe its physical 
properties, or to better understand the underlying physics. In other cases, the production 
of the magnetic field is the native property of the system. In heavy-ion collisions, for 
example, very strong magnetic fields are generated during the early stages of the 
collisions as the result of the electric currents from the colliding positively charges 
ions \cite{PhysRevLett.36.517,Kharzeev:2007jp,Skokov:2009qp,Bzdak:2011yy,
Voronyuk:2011jd,Deng:2012pc}. Because of the high electric conductivity of the 
medium, the corresponding fields may survive for as long as the lifetime of the 
quark-gluon plasma itself and, thus, have a profound effect on its dynamics. 
In the case of the early Universe, several competing mechanisms responsible
for the generation of very strong magnetic fields were proposed \cite{Vilenkin:1980fu,
Vachaspati:1991nm,Enqvist:1993kf,Cheng:1994yr,Baym:1995fk,Grasso:2000wj}. 
Despite the differences in the details, the consensus is that rather strong fields 
should have been generated. This is required by the present day observations 
of weak, but nonvanishing intergalactic magnetic fields. In the case of compact 
stars, the existence of strong magnetic fields is inferred from the observational 
data \cite{Kouveliotou:1998ze,Rea:2011fa,Olausen:2013bpa}, even though the exact 
nature of the underlying mechanism \cite{Thompson:1993hn,Duncan:1992hi} 
responsible for generation of such fields may still be debated. It is clear, however, 
that such fields exist and play a profound role in the stellar physics. 

With the discovery of quasirelativistic condensed matter systems such as graphene 
\cite{2004Sci...306..666N}, it became clear that a whole new realm of interesting quantum 
phenomena induced by a magnetic field can be realized in simple table-top experiments. 
In fact, the experimental proof that the low-energy excitations in graphene are Dirac 
quasiparticles was obtained by analyzing the observational features of the quantum 
Hall effect in a weak field \cite{2005Natur.438..197N,Zhang:2005zz}. In a strong 
magnetic field, on the other hand, a number of additional quantum Hall states were 
observed. It can be argued that such states are the consequence of a series of 
symmetry breaking quantum phase transitions, induced by the external field.

In this review, we will describe in detail the role of the magnetic field in a number of 
(quasi-)relativistic systems, using the results derived by analytic quantum field theoretical 
methods. In particular, we will try to place the emphasis on the underlying physics. 
When possible, the theoretical results will be compared with lattice simulations of 
the magnetic field induced dynamics in quantum chromodynamics and in graphene. 
In the case of graphene, we will also make a comparison of theoretical predictions
with the experimental data. 

We will start the review by introducing the underlying physics of the magnetic catalysis 
phenomenon \cite{Gusynin:1994re,Gusynin:1994va,Gusynin:1994xp}. We will discuss 
in detail the origin of the dimensional reduction in low-energy theories describing 
the interactions of charged relativistic fermions in an external magnetic field, the 
implication of such dimensional reduction for the particle-antiparticle pairing 
dynamics and the spontaneous (dynamical) symmetry breaking associated with it. 
Finally, we will make the case about the universality of the underlying mechanism. The 
general theoretical considerations will be followed by the discussion of applications of 
the magnetic catalysis in quantum electrodynamics (QED), quantum chromodynamics 
(QCD), and other relativistic models. We will also discuss the magnetic catalysis 
phenomenon in a class of quasirelativistic systems, such as highly oriented pyrolytic 
graphite, monolayer and bilayer graphene.

By generalizing the ideas of the magnetic catalysis to the case of nonzero 
density and temperature, we will show that other interesting phenomena are 
realized in various types of relativistic matter. The chiral magnetic effect 
\cite{Kharzeev:2007tn,Kharzeev:2007jp,Fukushima:2008xe} and the chiral 
separation effect \cite{Vilenkin:1980fu,Metlitski:2005pr,Newman:2005as} 
are the two important examples of such phenomena that take place at 
nonzero density. We will present recent theoretical ideas about these two 
effects and their possible phenomenological implications in heavy-ion
collisions and in compact stars. We will also discuss condensed matter 
analogues of these phenomena. In particular, we will argue about an 
interesting possibility of dynamical transformation of Dirac semimetals into 
Weyl ones, and suggest possible experimental signatures associated with 
such a transformation.

It is worth noting that the studies of relativistic quantum field theories in magnetic fields have
a long history starting from the classic papers by Heisenberg and Euler \cite{Heisenberg:1935qt}, 
and Schwinger \cite{Schwinger:1951nm} (for an excellent historical review of 
these papers, see Ref.~\cite{Dunne:2012vv}). A lot of theoretical progress in this area has been 
achieved since then, especially in connection to theories with spontaneous symmetry breaking, 
in which external fields may play a profound role in the underlying dynamics. Currently, however, 
there exist no comprehensive reviews on the topic. We hope that this review will fill the gap at 
least partly.

\section{Magnetic catalysis}
\label{sec:MagneticCatalysis}

The fact that an external magnetic field enhances the generation of a fermion mass in $3+1$ 
dimensions was first established in the framework of the Nambu--Jona-Lasinio (NJL) model 
in Refs.~\cite{Klevansky:1989vi,Suganuma:1990nn}. Also, in the 
context of the $(2+1)$-dimensional Gross-Neveu model, which was expected to
give a simple effective description for certain condensed matter planar systems, the 
authors of Refs.~\cite{Klimenko:1991he,Klimenko:1992ch,Krive:1992xh,Krive:1991uu} 
showed that the dynamical generation of a nonzero fermion mass takes place in a 
magnetic field as soon as there is an attractive interaction between fermions and 
antifermions. It will be fair to say, however, that the underlying mechanism in these 
early studies remained mysterious.

Most of the early model calculations did not address the underlying reason for the 
puzzling role of an external magnetic field as a trigger of symmetry breaking. To put this in 
perspective, one should recall that, at the time, the standard source of physics intuition for 
many nuclear and particle physics models with dynamical symmetry breaking was the BCS 
theory of superconductivity. By all accounts, however, the results of the NJL model revealed 
a drastically different role of the magnetic field. It helped to break, rather than restore symmetry. 
Also, there was no perfect diamagnetism and no Meissner effect associated with the broken phase.

Perhaps one exception from the general rule was the review of the NJL model by 
Klevansky \cite{Klevansky:1992qe}, which indeed offered a general qualitative 
explanation why an electric field tends to restore chiral symmetry and the magnetic field 
tends to break it. The reasoning went as follows: ``the electric field destroys the condensate 
by pulling the pairs apart, while the magnetic field aids in antialigning the helicities which 
are bound by the NJL interaction." Also, there it was clarified that the ``surprising" 
response to the magnetic field was so different from that in the theory of superconductivity 
because the pairing in the NJL model was between particles and antiparticles that carry 
opposite charges. This is in contrast to the usual Cooper pairing, which involves particles
of the same charge. 

While later it was found that there is much more to it, such an explanation 
was the first step in establishing the underlying physics of the magnetic catalysis. Also, as 
we will demonstrate in this section, despite the obvious differences pointed above, there 
are in fact profound, although more subtle, similarities between the dynamics of the magnetic catalysis and superconductivity.

Without a solid understanding of the underlying physics, it is hard to appreciate how 
important or how general is the mechanism, found in the models with local four-fermion 
interaction. Will it also work in more realistic gauge models with long-range interaction?
What are the similarities and differences between models in various space-time dimensions? 
Answering these and other related questions is the main goal of this section.

This mystery was resolved and the theoretical basis for the magnetic catalysis in both 
$2 + 1$ and $3 + 1$ dimensions was built in Refs.~\cite{Gusynin:1994re,Gusynin:1994va,
Gusynin:1994xp}. It was revealed that the origin of the magnetic catalysis is the 
dimensional reduction $D\to D-2$ (i.e., $2+1 \to 0+1$ and $3+1 \to 1+1$) in the 
infrared dynamics of the fermion pairing in a magnetic field. As is well known, in a 
lower dimension the infrared dynamics become stronger. A classical example of this 
phenomenon is provided by the Bardeen-Cooper-Schrieffer (BCS) theory of superconductivity 
\cite{Bardeen:1957kj,Bardeen:1957mv}. The physical reason of this reduction 
is the fact that the motion of charged particles is restricted in those directions that 
are perpendicular to the magnetic field.  This is in turn connected with the point 
that, at weak coupling between fermions, the fermion pairing, leading to the chiral 
(in general case, flavor) condensate, is mostly provided by fermions from the lowest 
Landau level (LLL) whose dynamics is $(D-2)$-dimensional. The effect was called 
the magnetic catalysis \cite{Gusynin:1994re}.

Thus, a constant magnetic field in $2+1$ and $3+1$ dimensions is a strong catalyst 
of dynamical chiral symmetry breaking, leading to the generation of a dynamical
fermion mass even at the weakest attractive interaction between
fermions and antifermions. It is crucial that the magnetic catalysis effect is 
universal and takes place for any fermion-antifermion attractive interaction.

The model-independent nature of magnetic catalysis was tested in numerous $(2+1)$- 
and $(3+1)$-dimensional models with local four-fermion interactions \cite{Gusynin:1994re,
Gusynin:1994va,Gusynin:1994xp,Babansky:1997zh,Klimenko:1998su,Ebert:1999ht,
Vdovichenko:2000sa,Zhukovsky:2000yd,Ishii:2001xg,Inagaki:2003yi,Ebert:2003yk,
Inagaki:2004ih,Ghosh:2005rf,Osipov:2007je,Hiller:2008eh,Klimenko:2008mg,
Menezes:2008qt,Menezes:2009uc,Boomsma:2009yk,Fayazbakhsh:2010bh,Chatterjee:2011yi,
Chatterjee:2011ry,Avancini:2012ee,Ferrari:2012yw,Allen:2013lda,Fayazbakhsh:2014mca}, 
as well in $d > 3$ spatial dimensions \cite{Gorbar:2000ku} and in models with additional 
gauge interactions \cite{Ishii:2001xg} and Polyakov loop effects \cite{Mizher:2010zb,
Gatto:2010qs,Gatto:2010pt,Gatto:2012sp,Ferreira:2013tba,Ferreira:2014kpa,
Andersen:2013swa,Andersen:2014oaa}, ${\cal N}=1$ supersymmetric models 
\cite{Elias:1996zu}, quark-meson models \cite{Andersen:2011ip,
Andersen:2012bq,Fraga:2013ova,Ruggieri:2013cya,Ruggieri:2014bqa,Fraga:2008qn}, models in curved space 
\cite{Geyer:1996np,Gitman:1996mk,Inagaki:1997nv,Gorbar:2007kd}. The realization of magnetic 
catalysis was investigated in chiral perturbation theory \cite{Shushpanov:1997sf,Agasian:1999sx,
Agasian:2000hw,Agasian:2001ym,Agasian:2001hv,Cohen:2007bt,Cohen:2008bk} 
and in models with the Yukawa interaction \cite{Elizalde:2002ca,Ferrer:1998xp,Ferrer:1998fd,
Ferrer:2000ed}. The mechanism of the magnetic catalysis is supported by the arguments of the 
renormalization group \cite{Hong:1996pv,Semenoff:1999xv,Fukushima:2012xw,Scherer:2012nn,
Andersen:2013swa,Kamikado:2013pya,Kojo:2013uua,Braun:2014fua}.

By now, the magnetic catalysis has been extensively studied not only in the NJL model 
but also in realistic theories, such as quantum electrodynamics (QED) and quantum 
chromodynamics (QCD) by using both analytic quantum-field theoretical approaches 
\cite{Gusynin:1995gt,Gusynin:1995nb,Parwani:1994an,Parwani:1995am,Leung:1995mh,
Gusynin:1997kj,Gusynin:1998nh,Gusynin:1998zq,Lee:1998sr,Gusynin:1999pq,
Farakos:1999qc,Gusynin:2000tv,Alexandre:2000yf,Alexandre:2001vu,Kabat:2002er,
Miransky:2002rp,Gusynin:2003dz,Leung:2005xz,Leung:2005yq,
Sadooghi:2007ys,Agasian:2008tb,Ayala:2009fv,Ayala:2010fm,Mueller:2014tea}, 
holographic models \cite{Filev:2007gb,Filev:2007qu,Erdmenger:2007bn,Zayakin:2008cy,
Argyres:2008sw,Filev:2009xp,Filev:2010pm,Filev:2011mt,Evans:2010iy,Evans:2010hi,
Evans:2010xs,Evans:2011mu,Evans:2011tk,Preis:2010cq,Preis:2011sp,Erdmenger:2011bw,
Preis:2012fh,Bolognesi:2011un,Bolognesi:2012pi,Alam:2012fw,Rebhan:2014rxa},
lattice simulations of $(2 + 1)$-dimensional QED \cite{Farakos:1998fe,Farakos:1998ds,
Alexandre:2001pa,Cea:2011hu,Cea:2012up} and QCD-like gauge theories 
\cite{Buividovich:2008wf,Buividovich:2009bh,Braguta:2010ej,2010PhRvD..82e1501D,2011PhRvD..83k4028D,Bali:2011qj,
Bali:2011uf,Bali:2012zg,2013LNP...871..181D,Ilgenfritz:2012fw,Ilgenfritz:2013ara,DElia:2013twa,
Bruckmann:2013oba,Bruckmann:2013ufa}.
It is also noticeable that the magnetic catalysis is a basic mechanism in the dynamics 
of the quantum Hall effect in graphene \cite{Gorbar:2007xh,Gorbar:2008hu,
Gorbar:2011kc,Gusynin:2006gn,Herbut:2008ui,Semenoff:2010zd,Semenoff:2011ya} 
(see Section~\ref{sec:QHEgraphene}). Ideas inspired by the magnetic 
catalysis were even extended to solid state systems such as high-temperature superconductors  
\cite{Ferrer:2001fz,Ferrer:2002gf,Liu:1998mg,Semenoff:1998bk,Zhukovsky:2003bs,Zhukovsky:2000jm}
and highly oriented pyrolytic graphite \cite{Gorbar:2002iw,Khveshchenko:2001zz,Khveshchenko:2001zza}.
Finally, the generalization of magnetic catalysis was also made to non-Abelian chromomagnetic fields  
\cite{Ebert:2001bb,Ebert:1997um,Gusynin:1997vh,Klimenko:1993ec,Shovkovy:1995td,
Vshivtsev:1994si,Zhukovsky:2001iq}, where the dynamics is 
dimensionally reduced by one unit of space, $D\to D-1$.

In this section, we will start from the description of the simplest realization of the 
magnetic catalysis phenomenon in the NJL models with a large number of fermion colors $N$. 
We will discuss models in $2 + 1$ and $3 + 1$ space-time dimensions, and emphasize 
the role of the number of dimensions. (The magnetic catalysis effect in gauge theories, 
including QED and QCD, will be considered in Section~\ref{sec:MagCatGauge}.) 
The rest of the section is organized as follows. In Section~\ref{sec:MagCat2+1General}, 
general remarks concerning the magnetic catalysis effect in $2 + 1$ dimensions are made. 
In Sections~\ref{sec:Free2+1MagField} and \ref{sec:SymBr2+1Free}, we consider the problem 
of free relativistic fermions in a magnetic field in $2+1$ dimensions. We show that the roots 
of the fact that a magnetic field is a strong catalyst of dynamical flavor symmetry breaking 
are already present in this problem. In Section~\ref{sec:NJL2+1General}, we study the NJL model 
in a magnetic field in $2+1$ dimensions. We derive the low-energy effective action and 
determine the spectrum of long wavelength collective excitations in the NJL model. We 
also study the thermodynamic properties of the NJL model in a magnetic field in $2 + 1$ 
dimensions. In particular, the phase transitions with respect to temperature and the fermion 
chemical potential are described. The discussion of the $(3+1)$-dimensional 
free fermions in a magnetic field is presented in Section~\ref{sec:MagCat3+1General}, 
and the analysis of the magnetic catalysis in the NJL model in $3 + 1$ dimensions 
is given in Section~\ref{sec:NJL3+1General}. In Section~\ref{sec:BS-NJL}, the Bethe-Salpeter
equations for the Nambu-Goldstone bosons in the NJL model in a magnetic field in both 
$2 + 1$ and $3 + 1$ dimensions are analyzed. The role on magnetic translations is
discussed in Section~\ref{sec:MagneticTranslations}. In Section~\ref{sec:MagCatSummary}, 
we summarize the main results in $2 + 1$ and $3 + 1$ dimensions.

\subsection{Opening remarks}
\label{sec:MagCat2+1General}

During the past three decades, there has been a considerable interest in studying relativistic 
field models in $2+1$ dimensions. Not only the rich and sophisticated dynamics in $2+1$ 
dimensions is interesting from a theoretical viewpoint, the corresponding models also serve as 
effective theories for the description of long wavelength excitations in a class of planar condensed 
matter systems. In particular, as will be discussed at length in Section~\ref{sec:QHEgraphene}, 
they play an important role for the description of the dynamics in graphene.

In this section, we will show that a constant magnetic field acts as a strong catalyst of 
dynamical flavor symmetry breaking leading to the generation of fermion masses in 
$2+1$ dimensions. We will in particular show that there is a striking similarity between 
the role of the magnetic field in $(2+1)$-dimensional models and the role of the Fermi 
surface in the Bardeen-Cooper-Schrieffer (BCS) theory of superconductivity 
\cite{Bardeen:1957kj,Bardeen:1957mv}: both of them enhance the interactions 
of fermions in the infrared region (at low energies) and, therefore, are responsible 
for a dynamical generation of a fermion mass (an energy gap in the spectrum) even 
at the weakest attractive interaction between fermions.

We note that, in absence of a magnetic field, supercritical dynamics (with an 
effective coupling constant $g$ being larger than a critical value $g_c$) is a common 
prerequisite for generating a dynamical fermion mass in $3+1$ and $2+1$ dimensions 
(for reviews, see Refs.~\cite{Fomin:1984tv,Miransky:1985aq,Miransky:1994vk}).
As will be shown below, in $2+1$ dimensions, a magnetic field reduces the 
value of the critical coupling to zero. In other words, in a magnetic field, the 
generation of a fermion mass happens even at the weakest attractive interaction 
between fermions and antifermions. The essence of this effect is that in a magnetic 
field, in $2+1$ dimensions, the dynamics of fermion pairing (relating essentially to
fermions in the lowest Landau level) is one-dimensional, see Section~\ref{sec:SymBr2+1Free}.

We stress that  this effect is universal (i.e., model independent) in $2+1$ dimensions. 
This point, in particular, will be important in connection with the use of the magnetic 
catalysis phenomenon for the description of the quantum Hall effect in graphene, 
see Section~\ref{sec:QHEgraphene}. As a soluble example, in this section we will 
consider the NJL model in a magnetic field, in the leading order in $1/N$ expansion, 
where $N$ is the number of fermion ``colors''. Another type of the $(2 + 1)$-dimensional 
model, which is more relevant for the description of condensed matter systems (in,
particular, graphene), will be considered in Section~\ref{sec:MagCatQEDreduced}. It is 
the reduced QED \cite{Gorbar:2001qt,Alexandre:2001ps,Gorbar:2002iw}, in which
fermions are restricted to a plane and the (electromagnetic) gauge fields propagate
in a three-dimensional bulk.

\subsection{Free fermions in a magnetic field in $2 + 1$ dimensions}
\label{sec:Free2+1MagField}

Before starting a detailed analysis of the dynamics underlying the phenomenon of 
magnetic catalysis, it is instructive to review the problem of free relativistic fermions 
in a magnetic field in $2+1$ dimensions. By first introducing a nonzero fermion mass
$m$ as an infrared regulator in such a theory, and then taking the limit $m\to 0$, we 
will show that the roots of magnetic catalyst of flavor symmetry breaking are already 
present in the free theory. Indeed, as we will show in Section~\ref{sec:SymBr2+1FreeReal}, 
the structure of the corresponding ground state bares some resemblance to superfluidity 
in an almost ideal Bose gas \cite{Bogolyubov:1947zz}.

\subsubsection{Two-component spinor representation} 

In $2+1$ dimensions, the Lagrangian density in the problem of a relativistic fermion 
in a constant magnetic field $B$ takes the following form:
\begin{equation}
{\cal L} = \frac{1}{2} \left[\bar{\Psi}(u), (i\tilde{\gamma}^\mu D_\mu-m)\Psi(u)\right],    
\label{eq:fre}
\end{equation}
where the space-time position four-vector is denoted by $u^{\mu} = (t,\mathbf{r})$ 
and $\mathbf{r}=(x,y)$. According to our convention, the Lorentz indices are denoted by 
Greek letters ($\mu$, $\nu$, $\lambda$, etc.) and run over $0$, $1$, $2$ or $t$, $x$, $y$. 
(In $3+1$ dimensions, another spatial direction, denoted by either $3$ or $z$, will be added.)
The spatial indices are denoted by Latin letters from the middle of the alphabet  
($i$, $j$, $k$, etc.) and run over $1$, $2$ or  $x$, $y$. The Minkowski metric 
tensor is given by $\eta^{\mu\nu}=\mbox{diag}(1,-1,-1)$, and the spatial vectors 
are identified with the {\it contravariant} components of four-vectors. The covariant 
derivative is defined as usual,
\begin{equation}
D_\mu =\partial_\mu+ ie A_\mu,
\label{eq:dmu}
\end{equation}
where $e$ is the charge of the fermion (e.g., in the case of the electron $e<0$). 
We will assume that the external magnetic field is described by the vector potential 
in the Landau gauge, 
\begin{equation}
A^k = -B y \delta^{k}_{1} \qquad \mbox{(Landau gauge)}
\label{eq:Landau_gauge}
\end{equation}
or in the vector notation, $\mathbf{A}=(-By,0)$. As is easy to check, the corresponding 
field strength tensor is given by $F_{\mu\nu} =\partial_{\mu}A_\nu - \partial_{\nu}A_\mu
=-\epsilon_{0\mu\nu}B$, where $\epsilon_{012} =1$. We note that the magnetic field $B$ 
is a pseudoscalar in $2+1$ dimensions.

The Dirac gamma matrices $\tilde{\gamma}^\mu$ obey the Clifford-Dirac algebra 
$C l_{1,2}(\mathbb{C})$ in $2+1$ dimensions, i.e., 
$\tilde{\gamma}^\mu\tilde{\gamma}^\nu+\tilde{\gamma}^\mu\tilde{\gamma}^\nu=2\eta^{\mu\nu}$. 
This algebra has two inequivalent matrix representations, which can be chosen, for example, 
as follows: 
\begin{eqnarray}
\mbox{Representation I:}&\quad &
\tilde{\gamma}^0=\sigma_3, \qquad
\tilde{\gamma}^1=i\sigma_1,\qquad
\tilde{\gamma}^2=i\sigma_2,
\label{eq:pauli}
\\
\mbox{Representation II:} &\quad &
\tilde{\gamma}^0=-\sigma_3, \qquad
\tilde{\gamma}^1=-i\sigma_1,\qquad
\tilde{\gamma}^2=-i\sigma_2 ,
\label{eq:pauli1}
\end{eqnarray}
where $\sigma_i$ are the Pauli matrices.

Let us begin by considering the representation in Eq.~(\ref{eq:pauli}) and determine the 
corresponding energy spectrum in the model described by Eq.~(\ref{eq:fre}). In the 
Landau gauge (\ref{eq:Landau_gauge}), it is convenient to write the general solution 
in the form $\Psi(u) = e^{-i E t+ik x} \psi (\xi)$, where $\xi = y/l+kl\,\mathrm{sign}(eB)$ is 
a new dimensionless coordinate replacing $y$, and $l=1/\sqrt{|eB|}$ is the magnetic length. 
Then, the equation of motion for the two-component spinor $\psi (\xi)$ takes the following 
form: 
\begin{equation}
\left(\begin{array}{cc}
E-m & \frac{i}{l}\left[\frac{d}{d\xi} - \xi \, \mathrm{sign}(eB)\right]
\\
-\frac{i}{l}\left[\frac{d}{d\xi} + \xi \, \mathrm{sign}(eB)\right]  & -E-m
\end{array}
\right)\psi (\xi) =0 .
\label{eq:Dirac2x2sign(eB)}
\end{equation}
For concreteness, let us first assume that $eB>0$ and $m\geq0$. Then, by solving 
the eigenvalue problem, we obtain the following energy spectrum \cite{Gusynin:1994va}:
\begin{eqnarray}
E_0 &=& \omega_0=m, \label{spectr-n=0} \\
E_n &=& \pm \omega_n=\pm\sqrt{m^2+2|eB|n},\qquad  n=1,2,\dots .
\label{spectr-n>0}
\end{eqnarray}
This energy spectrum describes the Landau levels labeled by a nonnegative 
integer index $n$. The corresponding eigenstates are given by \cite{Akhiezer:1965}
\begin{eqnarray}
u_{0k}(u) &=& \frac{1}{(lL_x)^{1/2}} 
    \left(\begin{array}{c} w_0   (\xi) \\ 0     \end{array}\right) e^{-i\omega_0t+ikx}, 
\label{u_0k} \\
u_{nk}(u) &=& \frac{1}{(lL_x)^{1/2}} \frac{1}{\sqrt{2\omega_n}}
        \left( \begin{array}{c} \sqrt{\omega_n+m} \, w_n(\xi) \\
               -i\sqrt{\omega_n-m} \,  w_{n-1}(\xi)  \end{array}\right)e^{-i\omega_nt+ikx} , 
\qquad
n\geq1,    
\label{u_nk}
\\
v_{nk}(u) & = & \frac{1}{(lL_x)^{1/2}} 
\frac{1}{\sqrt{2\omega_n}} \left(\begin{array}{c} \sqrt{\omega_n-m}\, w_n(\xi) \\ 
        i\sqrt{\omega_n+m} \,  w_{n-1}(\xi) \end{array}\right)e^{i\omega_nt+ikx},
\qquad
n\geq1,
\label{v_nk}
\end{eqnarray}
where we introduced the following harmonic oscillator wave functions:
\begin{equation}
w_n(\xi)=\frac{e^{-\frac{\xi^2}{2}}}{\sqrt{2^nn!\sqrt{\pi}}} H_n(\xi),
\end{equation}
which are given in terms of the Hermite polynomials $H_n(\xi)$ \cite{1980tisp.book.....G}. 
Note that functions $w_n(\xi)$ satisfy the following ``ladder" identities: 
\begin{eqnarray}
\hat{a}\, w_{n}(\xi)  & = &  \sqrt{n}\,w_{n-1}(\xi) ,\\
\hat{a}^{\dagger}w_{n}(\xi)  & = &  \sqrt{n+1}\,w_{n+1}(\xi), 
\end{eqnarray}
where 
\begin{eqnarray}
\hat{a} & = & \frac{1}{\sqrt{2}}\left(\xi + \frac{d}{d\xi} \right),\\
\hat{a}^{\dagger} & = & \frac{1}{\sqrt{2}} \left(\xi - \frac{d}{d\xi} \right) 
\end{eqnarray}
are the annihilation and creation operators, respectively. They satisfy the canonical 
commutation relation $[\hat{a},\hat{a}^{\dagger}]=1$. By making use of these operators, 
in fact, the spectral problem in Eq.~(\ref{eq:Dirac2x2sign(eB)}) can be solved 
in a symbolic form without the need to explicitly solve any differential equations.

From the above solutions, we see that the Landau levels are highly (infinitely) 
degenerate. Indeed, their energies depend only on the integer index $n$, but 
have no dependence on the quantum number $k$. In the Landau gauge used, 
the latter can be formally interpreted as the momentum associated with the 
spatial direction $x$. An alternative dual interpretation of $k$, as the center of 
the fermion orbit in coordinate space, follows from the fact that the corresponding 
wave functions are localized around $\xi =0$, or equivalently around 
$y_{\rm center}=-kl^2\mathrm{sign}(eB)$ in the spatial direction $y$. 

In order to determine the density of states in the Landau levels, we can use the 
following simple arguments. First, we assume that the system has a finite size, 
$L_x$ and $L_y$, in the spatial directions $x$ and $y$, respectively. By enforcing 
the periodic boundary conditions in the $x$ direction, we will find that the values of
the corresponding momentum are quantized: $k=2\pi p/L_x$ where $p=0,\pm1,\pm2,\ldots$. 
Then, by taking into account that the same momentum $k=2\pi p/L_x$ also determines the 
$y$ position around which the wave function is localized, we have to require that 
$-L_y/2 \lesssim y_{\rm center} \lesssim L_y/2$. In terms of the integer quantum 
number $p$, the constraint takes the following form: $ -L_x L_y |eB|/(4\pi^2) \lesssim 
p \lesssim L_x L_y |eB|/(4\pi^2)$. Therefore, the total number of degenerate 
states in each Landau level is $L_xL_y|eB|/(2\pi^2)$, and the density of states per 
unit area is $|eB|/(2\pi^2)$ \cite{Akhiezer:1965}.

The general solution for the spinor field is given by
\begin{equation}
\Psi(u)=\sum_{n=0}^{\infty} \sum_{p} a_{np} u_{np}(u) + \sum_{n=1}^{\infty} \sum_{p}  b^{+}_{np} v_{np}(u),
\label{eq:sol}
\end{equation}
where $a_{np}$ and $b^{+}_{np}$ are operator-valued coefficients, which can be
interpreted as particle annihilation and antiparticle creation operators, respectively.  

As is obvious from the general solution, the lowest Landau level with $n=0$ is special. 
It contains only positive energy states (particles) with $E_0=m$. This is in contrast to 
all higher Landau levels with $n\geq1$, which include solutions with both positive 
energies $E_n=\omega_n$, describing particle states, and negative energies 
$E_n=-\omega_n$, describing antiparticle states.

The spectral asymmetry of the lowest Landau level is sensitive to the sign of $eB$. 
To show this, let us write down the general solution for the fermion field in the case 
of $eB<0$. Instead of solving the spectral equation in Eq.~(\ref{eq:Dirac2x2sign(eB)}) 
directly, we could use the fact that the two sets of solutions (i.e., for $eB>0$ and $eB<0$) 
are related by the charge conjugation symmetry, $\Psi \to \Psi^C=\tilde{\gamma}_2\bar{\Psi}^T$. 
Therefore, the general solution for the spinor field at $eB<0$ can be written in the following 
form:
\begin{equation}
\Psi(u) = \sum_{n=1}^{\infty} \sum_{p} a_{np}v^C_{np}(u) + \sum_{n=0}^{\infty} \sum_{p} b^{+}_{np} u^C_{np} (u) ,
\label{eq:sol_eB<0}
\end{equation}
where the charge conjugate spinors are defined as follows: 
$v_{np}^C=\tilde{\gamma}_2\bar{v}_{np}^T$ and $u_{np}^C=\tilde{\gamma}_2\bar{u}_{np}^T$.
Note that $\bar{v}_{np}$ and $\bar{u}_{np}$ are the Dirac conjugate of the eigenfunctions
$v_{np}$ and $u_{np}$ in Eqs.~(\ref{u_0k}) -- (\ref{v_nk}). In this solution, the lowest 
Landau level with $n=0$ describes negative energy states (antiparticles) with $E_0=-m$.

It is straightforward to check that, if we use the second representation of the Dirac matrices, 
given in Eq.~(\ref{eq:pauli1}), the general solution will be given again by an expression 
like that in Eq.~(\ref{eq:sol}), but with the eigenstates $u_{np}(u)$ and $v_{np}(u)$ 
replaced by $(-1)^nv_{np}(-u)$ and $(-1)^nu_{np}(-u)$, respectively, i.e.,
\begin{equation}
\Psi(u)=\sum_{n=1}^{\infty} \sum_{p} (-1)^n c_{np} v_{np}(-u)
+\sum_{n=0}^{\infty} \sum_{p} (-1)^n d_{np}^\dagger u_{np}(-u) .
\label{eq:9}
\end{equation}
[The factor $(-1)^n$ is introduced here for convenience.] In this representation, the LLL 
solutions correspond to antiparticle states with $E=-m$ when $eB>0$ and particle states 
with $E=m$ when $eB<0$.

\subsubsection{Four-component spinor representation} 

We note that the mass term in the Lagrangian density (\ref{eq:fre}) violates
parity defined by the following transformation of the fermion field:
\begin{equation}
{\cal P}: \quad \Psi (t,x,y)\to\sigma_1\Psi(t,-x,y).
\end{equation}
In order to construct a $(2+1)$-dimensional theory of massive Dirac fermions that preserves
this discrete symmetry, it is convenient to use the language of four-component fermions 
\cite{Pisarski:1984dj,Appelquist:1986fd}, connected with a (reducible) four-dimensional 
representation of the Dirac matrices,
\begin{equation}
\gamma^0= \left(\begin{array}{cc} \sigma_3 & 0 \\ 0&
-\sigma_3\end{array}\right), 
\qquad
\gamma^1= \left(\begin{array}{cc}i\sigma_1&0\\0&-i\sigma_1\end{array}\right),
\qquad
\gamma^2= \left(\begin{array}{cc}i\sigma_2&0\\0&-i\sigma_2\end{array}\right).
\label{eq:11}
\end{equation}
Then, the mass term in the corresponding Lagrangian density
\begin{equation}
{\cal L} =\frac{1}{2} \left[\bar{\Psi}, \left(i\gamma^\mu D_\mu-m\right)\Psi\right]
\label{eq:free12}
\end{equation}
will indeed preserve parity defined by
\begin{equation}
{\cal P}:\quad
\Psi (t,x,y)\to-i\gamma^3\gamma^1\Psi (t,-x,y),
\end{equation}
where the additional Dirac matrix $\gamma^3$ is given by
\begin{equation}
\gamma^3=i\left(\begin{array}{cc}0&I\\ I&0\end{array}\right).
\label{eq:gamma3}
\end{equation}
In the massless limit ($m=0$), the Lagrangian density (\ref{eq:free12}) is invariant under 
the global (flavor) $\mathrm{U}(2)$ symmetry. The corresponding transformations are 
determined by the following generators:
\begin{equation}
T_0=I,\qquad
T_1=\gamma_5,\qquad
T_2=-i\gamma^3, \qquad
T_3=\gamma^3\gamma^5,
\label{T_i-generators}
\end{equation}
where
\begin{equation}
\gamma^5=i\gamma^0\gamma^1\gamma^2\gamma^3=i
\left(\begin{array}{cc}0&I\\ -I&0\end{array}\right).
\label{gamma_5_2+1}
\end{equation}
The mass term breaks this symmetry down to the $\mathrm{U}(1)\times \mathrm{U}(1)$ with the
generators $T_0$ and $T_3$.

In the case of the four-component fermions, the energy spectrum is given by
\begin{eqnarray}
E_0 &=& \pm \omega_0=\pm m, 
\label{4comp_E0}\\
E_n &=& \pm \omega_n =\pm \sqrt{m^2+2|eB|n}, \quad n\geq1.
\label{4comp_En}
\end{eqnarray}
[Compare with Eqs.~(\ref{spectr-n=0}) and (\ref{spectr-n>0}).] The density of the 
lowest Landau level ($n=0$) states with the energies $E_0=\pm m$ is $|eB|/2\pi$, 
and it is $|eB|/\pi$ for higher Landau level states ($n\geq1$).

We note that the four-component fermions naturally appear in low-energy effective 
actions describing planar condensed matter systems with two sublattices, see 
Section~\ref{sec:QHEgraphene}. Moreover, usually they appear 
in the actions without a mass term, and the important problem is to establish 
a criterion of dynamical flavor symmetry breaking, which may occur as a result 
of interaction between fermions \cite{Miransky:1994vk,Appelquist:1988sr,
Dagotto:1988id,Hands:1989mv,Pisarski:1991kg,Pennington:1990bx,Gusynin:1998kz}.
As was already indicated in Section~\ref{sec:SymBr2+1Free}, dynamical flavor
symmetry breaking in $2+1$ dimensions usually takes place only at a rather
strong effective coupling between fermions.

\subsection{Symmetry breaking for free fermions in a magnetic field in $2 + 1$ dimensions}
\label{sec:SymBr2+1Free}

Let us now show that at $m=0$ and $B\neq 0$, the dynamical breakdown of the
$\mathrm{U}(2)$ flavor symmetry takes place already in the theory (\ref{eq:free12}),
even without any additional interaction between fermions. In order to prove this, 
we will show that in the limit $m\to0$, there is a nonzero symmetry breaking flavor 
condensate, $\langle0|\bar{\Psi}\Psi|0\rangle=-|eB|/2\pi$.

\subsubsection{Nonzero flavor condensate}

The condensate $\langle0|\bar{\Psi}\Psi|0\rangle$ can be expressed in terms of the 
fermion propagator $S(u,u^\prime)=\langle0| T\Psi(u) \bar{\Psi} (u^\prime)| 0\rangle$ as follows:
\begin{equation}
\langle0|\bar{\Psi}\Psi|0\rangle=-\lim_{u\to u^\prime} \mathrm{tr}\left[S(u,u^\prime)\right].
\label{eq:tr-psi-bar-psi}
\end{equation}
The explicit form of the fermion propagator in an external magnetic field, $S(u,u^\prime)$, 
is given in Eq.~(\ref{eq:green}) in Appendix~\ref{App:FermionProp} in the 
Schwinger proper-time representation \cite{Schwinger:1951nm}. The result has 
the form of a product of the Schwinger phase, $e^{i\Phi(\mathbf{r},\mathbf{r}^\prime)}$, and a 
translation invariant function $\bar{S}(u-u^\prime)$. In Eq.~(\ref{eq:Fourier-transl-inv}) in the same 
Appendix, we also present the Fourier transform $\bar{S}(k) = \int d^3x e^{iku}\bar{S} (u)$ 
of the translation invariant part of the propagator. In Euclidean space $(k^0\to ik_3, s\to -is)$,
the latter takes the following form:
\begin{equation}
\bar{S}_E(k) =-i \int^\infty_0 ds \exp \left[-s\Bigg(m^2+k^2_3+
\mathbf{k}^2
\frac{\tanh(eBs)}{eBs}\Bigg)\right]\left(-k_\mu\gamma^{E}_\mu+m
+ i(k_2\gamma^{E}_1-k_1\gamma^{E}_2)
\tanh(eBs)\right) \left(1+ i\gamma^{E}_1\gamma^{E}_2\tanh(eBs)\right).
\label{eq:21} 
\end{equation}
Here $\gamma^{E}_3\equiv-i\gamma^0$, $\gamma^{E}_1\equiv\gamma^1$ and 
$\gamma^{E}_2\equiv\gamma^2$ are antihermitian matrices. [In Euclidean space, 
the metric tensor coincides with the unit matrix and, therefore, there is no difference between the covariant 
and contravariant components of vectors.]

In the limit $\mathbf{r}\to\mathbf{r}^\prime$, the Schwinger phase $\Phi(\mathbf{r},\mathbf{r}^\prime)$ 
vanishes and, thus, plays no role in the calculation of the condensate defined by Eq.~(\ref{eq:tr-psi-bar-psi}). 
Then, by making use of the translation invariant part of the propagator in Eq.~(\ref{eq:21}), 
we derive the following result:
\begin{equation}
\langle0|\bar{\Psi}\Psi|0\rangle_{(2+1)} = - \frac{4m}{(2\pi)^3} \int dk_3 d^2 \mathbf{k}
\int^\infty_{1/\Lambda^2} ds 
\exp \left[-s\Bigg(m^2+k^2_3+\mathbf{k}^2 \frac{\tanh(eBs)}{eBs}\Bigg)\right] ,
\end{equation}
where the Gaussian integration over momenta can be performed exactly. In the limit $m\to 0$, 
in particular, the result reads
\begin{eqnarray}
\langle0|\bar{\Psi}\Psi|0\rangle_{(2+1)} &=&-\lim_{\Lambda\to\infty}\lim_{m\to 0} \frac{m}{2\pi^{3/2}}
\int^\infty_{1/\Lambda^2} \frac{ds}{\sqrt{s}} e^{-sm^2} eB\coth(eBs) 
\nonumber\\
&=&-\lim_{\Lambda\to\infty}\lim_{m\to 0} \frac{m}{2\pi^{3/2}}
\left[2\Lambda +\sqrt{\pi}\frac{|eB|}{|m|}+O\left(m,\sqrt{|eB|}, \frac{m^2}{\Lambda}\right) \right]=-
\frac{|eB|}{2\pi} \mathrm{sign}(m),
\label{eq:condensate}
\end{eqnarray}
where an ultraviolet cutoff $\Lambda$ was used as a regulator at intermediate steps of the calculation.
Note that the limit $m\to 0$ is taken before the limit $\Lambda\to \infty$.

The result in Eq.~(\ref{eq:condensate}) indicates that, in the presence of a constant 
magnetic field, spontaneous breakdown of the flavor $\mathrm{U}(2)$ symmetry takes 
place even in the free massless theory! This is exclusively a $(2+1)$-dimensional 
phenomenon. Indeed, as we will see later, the corresponding result in $3+1$ dimensions
is very different: $\langle0|\bar{\Psi}\Psi|0\rangle_{(3+1)}\sim |eB|m\ln m$, which 
goes to zero as $m\to0$ [see Eq.~(\ref{9Cond3+1Perturb})].

What is the physical origin of this phenomenon? In order to answer this question, 
we note that the singular $1/m$ behavior of the integral in Eq.~(\ref{eq:condensate}) 
comes from the infrared region at large values of the proper-time coordinate, 
$s\to\infty$. Moreover, one can see from Eq.~(\ref{eq:condensate}) that the presence 
of a nonzero magnetic field is responsible for such a behavior. At large values of
the proper time, $s\to\infty$, the function $eB\coth(eBs)$ in the integrand approaches 
a constant value $|eB|$, which is in contrast to $eB\coth(eBs) \to 1/s$ in the case of 
$B\to 0$. From the physics viewpoint, this reflects the fact that the magnetic field 
removes the two perpendicular space dimensions in the low-energy dynamics,
making it effectively one-dimensional (in Euclidean space) and, thus, much more
singular in infrared. 

It can be also argued that the spontaneous breakdown of the flavor $\mathrm{U}(2)$ 
symmetry in $2+1$ dimensions is intimately connected with the form of the energy 
spectrum of fermions in a constant magnetic field. As we see from Eq.~(\ref{4comp_E0}), 
in the limit $m\to 0$, the energies of all lowest Landau states go to zero and the vacuum 
becomes infinitely degenerate. It is not coincidental that the value of the condensate 
(\ref{eq:condensate}) is determined by the density of the LLL states per unit area of the 
system. Also, the effective one-dimensional dynamics behind the flavor symmetry breaking 
mechanism, mentioned above, finds a simple explanation in the one-dimensional character 
of the LLL, which is described by a single continuous variable $k_3=-ik^0$.

This observation suggests that the effect of the magnetic field 
in the present problem is similar to that of the Fermi surface in the BCS theory of 
superconductivity \cite{Bardeen:1957kj,Bardeen:1957mv}. However, as we discuss in 
Section~\ref{sec:NJL2+1EffPot}, there are important differences between them. 
In particular, the magnetic catalysis effect in $2 + 1$ dimensions is much stronger than 
in the BCS theory. [At the same time, as we will see in Section~\ref{sec:MagCat3+1General}, 
there is indeed a close similarity between the dynamics in the BCS theory and the 
magnetic catalysis in $3 + 1$ dimensions.]

To make this statement more rigorous mathematically, let us rewrite the fermion propagator 
in a form of an expansion over Landau levels. The desired form of the propagator can be obtained from Eq.~(\ref{eq:21}) 
by making use of the following identity $\tanh(x)=1-2\exp(-2x)/[1+\exp(-2x)]$ and the relation 
\cite{1980tisp.book.....G}
\begin{equation}
(1-z)^{-(\alpha+1)}\exp\left(\frac{xz}{z-1}\right)=\sum_{n=0}^{\infty}
L_n^{\alpha}(x)z^n,
\label{sum-Laguerre}
\end{equation}
where $L_n^{\alpha}(x)$ are the generalized Laguerre polynomials. The resulting 
alternative representation of the propagator $\bar{S}_E (k)$ reads \cite{Chodos:1990vv}
\begin{equation}
\bar{S}_E (k)=-i\exp\left(-\frac{\mathbf{k}^2}{|eB|}\right)
\sum_{n=0}^{\infty}(-1)^n\frac{D_n(eB,k)}{k_3^2+m^2+2|eB|n},
\label{eq:poles}
\end{equation}
where
\begin{eqnarray}
D_n(eB,k)&=&(m-k_3\gamma^{E}_3)\left[\left(1 + i\gamma^{E}_1\gamma^{E}_2 \mathrm{sign}(eB)\right)
L_n\left(2\frac{\mathbf{k}^2}{|eB|}\right) 
- \left(1- i\gamma^{E}_1\gamma^{E}_2\mathrm{sign}(eB)\right)L_{n-1}\left(2\frac{\mathbf{k}^2}{|eB|}\right)\right]
\nonumber\\
&& +4\left(k_1\gamma^{E}_1+k_2\gamma^{E}_2\right)L_{n-1}^1\left(2\frac{\mathbf{k}^2}{|eB|}\right).
\end{eqnarray}
Here $L_n(x) \equiv L_n^0(x)$ and $L_{-1}^{\alpha}(x)=0$ by definition. 
The expansion over Landau levels in Eq.~(\ref{eq:poles}) makes it easy to see that, in the limit $m\to 0$, 
the chiral condensate (\ref{eq:condensate}) comes exclusively from the lowest Landau level:
\begin{equation}
\langle0|\bar{\Psi}\Psi|0\rangle_{(2+1)}
\simeq-\frac{m}{2\pi^3} \int dk_3 d^2\mathbf{k}
\frac{\exp\left(-\mathbf{k}^2/|eB|\right)}{k_3^2+m^2}=
-\frac{|eB|}{2\pi} \mathrm{sign}(m) .
\end{equation}
Last but not least, this chiral condensate is equal (up to an overall sign) to the density of 
the LLL states per unit area in the $xy$-plane. Due to the Atiyah-Singer index theorem \cite{Atiyah:1968mp},
the latter is a topological invariant. Beyond the zero mass limit, the chiral condensate will also receive
contributions from the higher Landau levels and, therefore, will not be topologically protected.

In the next subsection, we will discuss aspects of spontaneous flavor symmetry
breaking for ($2+1$)-dimensional fermions in a magnetic field in more detail.

\subsubsection{Haag criteria of spontaneous symmetry breaking}
\label{sec:SymBr2+1FreeReal}

As was shown in the preceding section, the flavor condensate $\langle 0|\bar{\Psi}\Psi|0\rangle$ 
is nonzero as the fermion mass $m$ goes to zero. Although usually such a nonvanishing 
condensate is considered as a firm signature of spontaneous flavor (chiral) symmetry 
breaking, the following questions may arise in this case:

\begin{itemize}
\item[(i)] Unlike the conventional spontaneous flavor (chiral) symmetry
breaking, the dynamical mass of fermions is vanishing in this problem.
Therefore, one may wonder whether there is {\it real} 
symmetry breaking in this case.

\item[(ii)] The vacuum $|0\rangle$ was defined as $\lim_{m\to 0}|0\rangle_m$ from
the vacuum $|0\rangle_m$ in the theory with $m\neq 0$. Such a vacuum
corresponds to a particular filling of the lowest Landau level. Indeed, in the vacuum 
$|0\rangle_m$ at $m\neq 0$, the particle states with $E_0=m>0$ are empty and the 
antiparticle states with $E_0=-m$ are filled. Thus, the vacuum 
$|0\rangle=\lim_{m\to 0}|0\rangle_m$ is annihilated by all the operators
$a_{0p}$, $d_{0p}$ and $a_{np}$, $b_{np}$, $c_{np}$, $d_{np}$ ($n\geq 1$).
On the other hand, at $m=0$, there is an infinite degeneracy of the vacua, 
connected with different fillings of the lowest Landau level. Why should one 
choose the filling leading to the vacuum $|0\rangle$? Also, is there a filling of 
the lowest Landau level leading to the ground state which is invariant under 
the flavor $\mathrm{U}(2)$? One might think that the latter possibility would 
imply that spontaneous flavor symmetry breaking can be avoided.
\end{itemize}

In this section we will show that there is a genuine realization of spontaneous 
breaking of the flavor symmetry in the present problem. More precisely, we will 
show that this phenomenon satisfies the Haag criteria of spontaneous symmetry 
breaking \cite{Haag:1962}. We will also discuss the status of the Nambu--Goldstone (NG) 
modes and the induced quantum numbers \cite{Niemi:1983rq,Kovner:1990pz} in the 
problem at hand.

Let us begin by constructing the charge operators 
$Q_i=1/2\int d^2\mathbf{r}\, [\Psi^{\dagger}(t,\mathbf{r}),T_i\Psi(t,\mathbf{r})]$ of the 
flavor $\mathrm{U}(2)$ group with the generators $T_i$ defined in Eq.~(\ref{T_i-generators}). 
By making use of the fermion fields in Eqs.~(\ref{eq:sol}) and (\ref{eq:9}), 
we derive
\begin{eqnarray}
Q_0 &=& \sum\limits_p\left(a_{0p}^\dagger a_{0p}-d_{0-p}^\dagger d_{0-p}\right)
+\sum\limits_{n=1}^\infty\sum\limits_p\left(a_{np}^\dagger a_{np}
-b_{np}^\dagger b_{np}+c_{np}^\dagger c_{np} -d_{np}^\dagger  d_{np}\right),
\\
Q_1 &=&i\sum\limits_p\left(a_{0p}^\dagger d_{0-p}^\dagger -d_{0-p}a_{0p}\right)
+i\sum\limits_{n=1}^\infty\sum\limits_p\left(a_{np}^\dagger c_{np}-c_{np}^
\dagger a_{np}+b_{np}^\dagger d_{np}-d_{np}^\dagger b_{np}
\right), \label{eq:charge1}
\\
Q_2 &=&
\sum\limits_p\left(a_{0p}^\dagger d_{0-p}^\dagger +d_{0-p}a_{0p}\right)
+\sum\limits_{n=1}^\infty\sum\limits_p\left(a_{np}^\dagger c_{np}+c_{np}^
\dagger a_{np}+b_{np}^\dagger d_{np}+d_{np}^\dagger b_{np}
\right),     \label{eq:charge2}
\\
Q_3 &=&\frac{|eB|}{2\pi}S+\sum\limits_p\left(a_{0p}^\dagger a_{0p}
+d_{0-p}^\dagger  d_{0-p}\right)+\sum\limits_{n=1}^\infty\sum\limits_p
\left(a_{np}^\dagger a_{np}-b_{np}^\dagger b_{np}-c_{np}^\dagger c_{np}+
d_{np}^\dagger d_{np}\right),
\label{eq:charge3}
\end{eqnarray}
where $a_{np}$, $c_{np}$, ($b_{np}$, $d_{np}$) are annihilation operators of
fermions (antifermions) from the $n$th Landau level and $S=L_xL_y$ is the
two-dimensional volume. Now we can construct an infinite set of degenerate
vacua
\begin{equation}
|\theta_1,\theta_2\rangle=\exp\left(iQ_1\theta_1+iQ_2\theta_2\right)|0\rangle
\end{equation}
where the vacuum $|0\rangle=\lim_{m\to 0}|0\rangle_m$ is annihilated by all the 
operators $a_{np}$, $b_{np}$, $c_{np}$, and $d_{np}$. As one can see from 
Eqs.~(\ref{eq:charge1}) and (\ref{eq:charge2}), the crucial point for the existence 
of the continuum of degenerate vacua is the first sum, over the states in the 
lowest Landau level, in the definitions of the charges $Q_1$ and $Q_2$.

The presence of such a set of degenerate vacua is a signal of the spontaneous 
breakdown of flavor symmetry, $\mathrm{U}(2)\to \mathrm{U}(1)\times \mathrm{U}(1)$. 
Note that the vacua $|\theta_1,\theta_2\rangle$ can be also constructed by replacing 
the mass term $m\bar{\Psi}\Psi$ with $m\bar{\Psi}_{\theta_1,\theta_2}\Psi_{\theta_1,\theta_2}$, 
where $\Psi_{\theta_1,\theta_2}=\exp\left(iQ_1\theta_1+iQ_2\theta_2\right)\Psi$, 
and then taking the limit $m\to 0$. Again, this is a standard way of constructing
degenerate vacua in the case of spontaneous symmetry breaking.

One can check that different vacua $|\theta_1,\theta_2\rangle$ become
orthogonal when the system size in the $x$-direction, $L_x$, goes to infinity. For
example,
\begin{equation}
\left|\langle 0,\theta_2|0,\theta_2^\prime\rangle\right|=\prod\limits_p
|\cos\theta|=\exp\left(L_x\int dk\ln|\cos\theta|\right),
\label{theta2-theta2-example}
\end{equation}
where $\theta=\theta_2^\prime-\theta_2$. This goes to zero as $L_x\to\infty$ for all
values of $\theta\neq 0$. Here we took into account that $|0,\theta_2+\pi\rangle=-|0,\theta_2\rangle$, 
implying that all nonequivalent vacua are spanned by the values of $\theta_2$ in the range from $0$ 
to $\pi$. The matrix element in Eq.~(\ref{theta2-theta2-example}) also goes to zero when the 
maximum momentum (i.e., the ultraviolet cutoff) $|k_{\rm max}|=\Lambda$ goes to infinity.
As usual, this point reflects the fact that spontaneous symmetry breaking occurs only in a 
system with an infinite number of degrees of freedom. One can check that in this case all 
states (and not just vacua) from different Fock spaces $F_{\{\theta_1\theta_2\}}$, defined 
by different vacua $|\theta_1,\theta_2\rangle$, are orthogonal. In other words, different 
vacua $|\theta_1,\theta_2\rangle$ define nonequivalent representations of canonical 
commutation relations.

On the other hand, taking the ground state
\begin{equation}
|\Omega\rangle=C\int d\mu(\theta_1,\theta_2,\theta_3)
|\theta_1,\theta_2\rangle,\label{eq:omega}
\end{equation}
where $d\mu$ is the invariant measure of $\mathrm{SU}(2)$ and $C$ is a normalization
constant, we are led to the vacuum $|\Omega\rangle$ which is a singlet
with respect to the flavor $\mathrm{U}(2)$. In fact, the set of the vacua
$\{|\theta_1,\theta_2\rangle\}$ can be decomposed into irreducible
representations of $\mathrm{SU}(2)$:
\begin{equation}
\{|\theta_1,\theta_2\rangle\}=\{|\Omega^{(i)}\rangle\}
\end{equation}
Why should we consider the vacua $|\theta_1,\theta_2\rangle $ instead of the
vacua $|\Omega^{(i)}\rangle$?

To answer to this question, we consider, following Haag \cite{Haag:1962}, the
clusterization property of Green's functions. To start with, let us consider a 
Green's function of the following type:
\begin{equation}
G^{(n+k)}=\langle 0|T\prod_{i=1}^{n}A_i(u_i)\prod_{j=1}^{k}B_j(u^\prime_j)|0\rangle,
\end{equation}
where $A_i(u_i)$, $B_j(u^\prime_j)$ are some local operators. The clusterization
property implies that, at $r^2_{ij}=(\mathbf{r}_i-\mathbf{r}^\prime_j)^2\to \infty$ 
for all $i$ and $j$, the Green's function factorizes as follows:
\begin{equation}
G^{(n+k)}\to \langle 0|T\prod_{i=1}^{n}A_i(u_i)|0\rangle
\langle 0|T\prod_{j=1}^{k}B_j(u^\prime_j)|0\rangle.
\end{equation}
The physical meaning of this property is clear: it reflects the absence of 
instantaneous long-range correlations in the system, so that the
dynamics in two distant spatially-separated regions are independent.

The clusterization property takes place for all vacua $|\theta_1,\theta_2\rangle $. 
The simplest way to show this is to note that the vacuum $|\theta_1,\theta_2\rangle $ 
appears in the limit $m\to 0$ from the vacuum in the system with the mass term 
$m\bar{\Psi}_{\theta_1\theta_2}\Psi_{\theta_1\theta_2}$. Since the vacuum is unique
at $m\neq 0$, the clusterization is valid for {\it every} value of $m\neq 0$. Therefore, 
it is also valid in the limit $m\to 0$, provided the Green's functions exist in this limit.
In connection with that, we would like to note that, in thermodynamic limit
$L_x$, $L_y\to \infty$, the vacuum $|\theta_1,\theta_2\rangle $ is the only
normalizable and translation invariant state in the Fock space
$F_{\theta_1\theta_2}$. To show this, let us introduce the
operators $a_n(k)=(L_x/2\pi)^{1/2}a_{np}$, $b_n(k)=(L_x/2\pi)^{1/2}b_{np}$,
$c_n(k)=(L_x/2\pi)^{1/2}c_{np}$, $d_n(k)=(L_x/2\pi)^{1/2}d_{np}$, where
$k=2\pi p/L_x$. They satisfy the conventional commutation relations $[a_{n}(k),
a_{n^\prime}^{\dagger}(k^{\prime})]=\delta_{nn^{\prime}}\delta(k-k^{\prime})$,
etc. Therefore, despite the fact that the states of the form 
$\prod_{i}a^{\dagger}_0(k_i)\prod_{j}d^{\dagger}_0(k_j)
|\theta_1,\theta_2\rangle $ have zero energy, they are not normalizable and,
at $\sum_{i}k_i+\sum_{j}k_j\neq 0$, are not translation invariant.

In contrast, the clusterization property is not valid for all Green's functions in 
the vacua $|\Omega^{(i)}\rangle$. As an example, let us consider the Green's 
function
\begin{equation}
G^{(4)}= \langle \Omega|T\left(\bar{\Psi}(u_1)\Psi(u_2)\right)
\left(\bar{\Psi}(u^\prime_1)\Psi(u^\prime_2)\right)|\Omega\rangle,
\end{equation}
where $|\Omega\rangle$ is the vacuum singlet (\ref{eq:omega}).
Since the bilocal operator $\bar{\Psi}(u_1)\Psi(u_2)$ is assigned to the
triplet of $\mathrm{SU}(2)$, the clusterization property would imply that
\begin{equation}
G^{(4)}\to \langle \Omega|T\left(\bar{\Psi}(u_1)\Psi(u_2)\right)|\Omega\rangle
\langle\Omega|\left(\bar{\Psi}(u^\prime_1)\Psi(u^\prime_2)\right)|\Omega\rangle\to 0
\end{equation}
when $r^2_{ij}=(\mathbf{r}_i-\mathbf{r}^\prime_j)^2\to \infty$. However,
since
\begin{eqnarray}
\langle\Omega|T\left(\bar{\Psi}(u_1)\Psi(u_2)\right)|\Omega^{(3)}\rangle &\neq& 0, \\
\langle\Omega^{(3)}|T\left(\bar{\Psi}(u^\prime_1)\Psi(u^\prime_2)\right)|\Omega\rangle &\neq& 0,
\end{eqnarray}
where $|\Omega^{(3)}\rangle$ is a state from the vacuum triplet, we see that $G^{(4)}$ 
does not vanish at $r_{ij}^2\to \infty$. Thus, the clusterization property does not take 
place for the $|\Omega^{(i)}\rangle$-vacua.

This is a common feature of the systems with spontaneous breaking of continuous symmetries 
\cite{Miransky:1994vk,Haag:1962}: an orthogonal set of vacua can either be labeled
by the continuous parameters $\{\theta_i\}$, connected with the generators $Q_i$ of 
the broken symmetry, or it can be decomposed in irreducible representations of the 
initial group. However, the latter vacua do not satisfy the clusterization property.

All Fock spaces $F_{\{\theta_1\theta_2\}}$ yield physically equivalent descriptions
of the dynamics: in the space $F_{\{\theta_1\theta_2\}}$, the $\mathrm{SU}(2)$
spontaneously breaks down to $\mathrm{U}_{\{\theta_1\theta_2\}}(1)$, where the
$\mathrm{U}_{\{\theta_1\theta_2\}}(1)$ symmetry is connected with the generator
$Q_3^{\{\theta_1\theta_2\}}=\exp(iQ_1\theta_1+iQ_2\theta_2)Q_3
\exp(-iQ_1\theta_1-iQ_2\theta_2)$. It is natural to ask whether there are any 
NG modes associated with the symmetry breaking.

To answer to the question about the NG modes, let us consider the thermodynamic 
limit $L_x$, $L_y\to \infty$. One can see that, in every Fock space $F_{\{\theta_1\theta_2\}}$ 
with the vacuum $|\theta_1, \theta_2\rangle$, there are a lot of ``excitations''
with nonzero momentum $k$ and zero energy $E$ created by the operators
$a_{0}^\dagger(k)$ and $d_{0}^\dagger(k)$. However, there are no genuine 
NG modes with a nontrivial dispersion law because the energies of all states 
in the lowest Landau level are vanishing, $E\equiv0$. Since the Lorentz
symmetry is broken by a magnetic field, this point does not contradict to
the Goldstone theorem \cite{Goldstone:1962es}. This of course does not 
imply that the existence of NG modes is incompatible with a magnetic field: 
the situation is more subtle. As will be shown in Section~\ref{sec:NJL2+1General},
even the weakest attractive interaction in the problem of $(2+1)$-dimensional 
fermions in a magnetic field is enough to ``resurrect'' the genuine
NG modes. The key point for their existence is that the flavor condensate
$\langle 0 |\bar{\Psi} \Psi | 0\rangle$ and the NG modes are neutral, and the
translation symmetry in neutral channels is not violated by a magnetic
field, see Section~\ref{sec:MagneticTranslations}.

We will also see that the excitations from the lowest Landau level (with quantum 
numbers of the NG modes) in the problem of free fermions in a magnetic field can 
be interpreted as ``remnants'' of the genuine NG modes in the limit when the interaction 
between fermions is switched off, see Section~\ref{sec:NJL2+1BosSpectr}. Moreover, 
as we will see in Section~\ref{sec:NJL2+1EffPot}, the vacua $|\theta_1,\theta_2\rangle$ 
constructed above yield a good approximation for the vacua of  systems with weakly 
interacting fermions in a magnetic field. In fact, it appears that the vacua 
$|\theta_1,\theta_2\rangle$ play the same role as the $\theta$-vacua of the ideal Bose 
gas in the description of an almost ideal Bose gas in the theory of superfluidity 
\cite{Bogolyubov:1947zz}.

Before concluding this section, let us also discuss the phenomenon of induced 
quantum numbers \cite{Niemi:1983rq,Kovner:1990pz} in the problem at hand. 
As follows from Eq.~(\ref{eq:charge3}), the vacuum $|\theta_1, \theta_2\rangle$ is an 
eigenstate of the density operator $\rho_3^{\{\theta_1 \theta_2\}}=\lim_{S\to \infty}
Q_3^{\{\theta_1 \theta_2\}}/S$ with a nonzero value
\begin{equation}
\rho_3^{\{\theta_1 \theta_2\}}|\theta_1, \theta_2\rangle=
\frac{|eB|}{2\pi}|\theta_1, \theta_2\rangle.
\end{equation}
The nonzero density is an induced quantum number in the 
$|\theta_1 \theta_2\rangle$-vacuum, which is intimately connected with the phenomenon 
of spontaneous flavor symmetry breaking in the presence of a magnetic field. Indeed, 
since $Q_3^{\{\theta_1 \theta_2\}}$ is one of the generators of the non-Abelian 
$\mathrm{SU}(2)$ symmetry, its vacuum eigenvalue would be zero if the symmetry 
were exact and the vacuum were assigned to the singlet (trivial) representation 
of $\mathrm{SU}(2)$. This is in contrast to the less restrictive case of Abelian 
$\mathrm{U}(1)$ symmetries because $\mathrm{U}(1)$ has an infinite number of 
one-dimensional representations and, as a result, the vacuum can be an eigenstate 
of the charge density $\rho=\lim_{S\to \infty}Q/S$ with an arbitrary eigenvalue.

Note that, since the $\mathrm{SU}(2)$ is spontaneously broken here, it is appropriate
to redefine the generator of the exact $\mathrm{U}_{\{\theta_1\theta_2\}}(1)$
symmetry as $\tilde{Q}_3^{\{\theta_1\theta_2\}}=Q_3^{\{\theta_1\theta_2\}}
-|eB|S/2\pi$.

\subsection{The NJL model in a magnetic field in $2+1$ dimensions}
\label{sec:NJL2+1General}

In this section, we will consider the NJL model in $2+1$ dimensions. 
This model gives a clear illustration of the general fact that a constant magnetic 
field is a strong catalyst of a dynamical mass generation for fermions in $2+1$ 
dimensions.

Let us consider the following NJL model invariant under the $\mathrm{U}(2)$ flavor 
transformations:
\begin{equation}
{\cal L} = \frac{1}{2} \left[\bar{\Psi}, i\gamma^\mu D_\mu\Psi\right]
+ \frac{G}{2} \left[\left(\bar{\Psi}\Psi\right)^2+\left(\bar{\Psi}i\gamma_5 \Psi\right)^2 
+ \left(\bar{\Psi}\gamma_3\Psi\right)^2\right],   \label{eq:lag1}
\end{equation}
where $D_\mu$ is the covariant derivative (\ref{eq:dmu}) and fermion fields
carry an additional, ``color'' index $\alpha=1,2,\dots, N$. By making use of the 
Hubbard-Stratonovich transformation \cite{1957SPhD....2..416S,1959PhRvL...3...77H},
this theory becomes equivalent to a theory with the Lagrangian density
\begin{equation}
{\cal L} = \frac{1}{2} \left[\bar{\Psi}, i\gamma^\mu D_\mu\Psi\right] -
\bar{\Psi} \left(\sigma+\gamma^3\tau+i\gamma^5\pi\right)
\Psi 
- \frac{1}{2G} \left(\sigma^2+\pi^2+\tau^2\right).\label{eq:lag2}
\end{equation}
The Euler-Lagrange equations for the auxiliary fields $\sigma$, $\tau$ and $\pi$
take the form of constraints:
\begin{equation}
\sigma=-G\left(\bar{\Psi}\Psi\right), \qquad 
\tau=-G\left(\bar{\Psi}\gamma^3\Psi\right),  \qquad 
\pi=-G \left(\bar{\Psi}i\gamma^5\Psi\right).  
\label{eq:cons}
\end{equation}
The Lagrangian density (\ref{eq:lag2}) reproduces Eq.~(\ref{eq:lag1})
upon application of the constraints in Eq.~(\ref{eq:cons}).

The effective action for the composite fields is expressed through the path
integral over the fermion fields, i.e.,
\begin{equation}
\Gamma(\sigma,\tau,\pi) =-\frac{1}{2G} \int d^3u \left(\sigma^2+\tau^2+\pi^2\right) 
+ \tilde{\Gamma}(\sigma,\tau,\pi), 
\label{eq:effact}
\end{equation}
where 
\begin{equation}
\exp\left(i\tilde{\Gamma}\right) = \int [d\Psi][d\bar{\Psi}] 
\exp \left\{ \frac{i}{2} \int d^3u \left[ 
\bar{\Psi}(u), \left( i\gamma^\mu D_\mu - (\sigma+\gamma^3\tau+i\gamma^5\pi) \right) \Psi(u)
\right]\right\}.
\label{eq:exp}
\end{equation}
After performing the Gaussian integration, we obtain
\begin{equation}
\tilde{\Gamma}(\sigma,\tau,\pi) =- i \, \mathrm{Tr} \, \mbox{Ln}  \left[i\gamma^\mu D_\mu-
(\sigma+\gamma^3\tau+i\gamma^5\pi)\right].\label{eq:tildefact}
\end{equation}
As $N\to\infty$, the path integral over the composite (auxiliary) fields is dominated by 
the stationary points of the action, determined by the classical equations of motion,
i.e., $\delta\Gamma/\delta\sigma =\delta\Gamma/\delta\tau =\delta\Gamma/\delta\pi=0$. 
We will analyze the dynamics in this limit by using the expansion of the action 
$\Gamma(\sigma,\tau,\pi)$ up to second order in powers of derivatives of the
composite fields.

It might be natural to ask whether the $1/N$ expansion could be reliable in this problem. 
Indeed, as was emphasized in Section~\ref{sec:MagneticCatalysis}, the external magnetic field 
reduces the dimension of the fermion pairing dynamics by two units. If such a reduction 
took place for the whole dynamics and not just for the fermion pairing, the $1/N$ 
perturbative expansion would be problematic. In particular, the contribution of the NG 
modes to the gap equation in the next-to-leading order in $1/N$ would lead to infrared 
divergences. As is well know, this is exactly what happens in the $(1+1)$-dimensional 
Gross-Neveu model with a continuous chiral symmetry ~\cite{Gross:1974jv,Witten:1978qu}. 
The corresponding infrared divergences reflect the well known fact that a spontaneous 
breakdown of continuous symmetries in space dimensions lower than two are forbidden by the 
Mermin-Wagner-Coleman theorem \cite{Mermin:1966fe,Coleman:1973ci}.

Fortunately, this is not the case in the present problem because of the neutrality 
of the condensate $\langle 0|\bar{\Psi}\Psi|0\rangle$ and the NG modes. Most importantly, 
the NG modes are not subject to the dimensional reduction in infrared. The argument is 
intimately connected with the status of the spatial translation symmetry in a constant 
magnetic field, see Section~\ref{sec:MagneticTranslations} for details. As we know, in the presence 
of a magnetic field, the usual translation symmetry is broken. This is seen, for example, 
from the fact that the fermion propagator is not translation invariant. This is also reflected 
in the fact that the transverse momenta are not good quantum numbers to characterize 
the fermionic states. Instead, one has a Landau level spectrum and the low-energy 
fermionic dynamics is dimensionally reduced.

The situation is drastically different for NG modes because they are neutral. 
For such states, the transverse components of the momentum of the center of mass, 
$k_x$ and $k_y$, are conserved quantum numbers \cite{Zak:1964zz,1978AnPhy.114..431A},
see also Section~\ref{sec:MagneticTranslations}. This is also confirmed by direct 
calculations in Section~\ref{sec:NJL2+1BosSpectr} and Appendix~\ref{App:CompProp},
where we find that the structure of the propagator of the NG modes has a genuine 
$(2+1)$-dimensional structure. As a result, the contribution of the NG modes to the 
dynamics does not lead to infrared divergences, and the $1/N$ expansion is reliable. 
[Note, however, that the dispersion relations of the NG modes may still be highly 
nonisotropic in the presence of an external magnetic field. This is the result of their 
composite nature: they are made of charged particles.]

\subsubsection{The effective potential and gap equation}
\label{sec:NJL2+1EffPot}

As mentioned earlier, we want to calculate the low-energy effective action $\Gamma(\sigma,\tau,\pi)$ 
by making use of the derivative expansion. We will start from calculating the effective 
potential $V(\sigma,\tau,\pi)$ in this subsection, and than proceed to the derivation of  
the kinetic part of the action in the next subsection. 

Since $V(\sigma,\tau,\pi)$ depends only on the $\mathrm{SU}(2)$-invariant 
$\rho^2=\sigma^2+\tau^2+\pi^2$, it is sufficient to consider a configuration
with $\tau=\pi=0$ and a nonzero $\sigma$-field independent of space-time 
coordinates. By also taking into account that $\Gamma(\sigma) \equiv -\int d^3u 
V(\sigma)$ and by using the definition in Eq.~(\ref{eq:effact}), we obtain
\begin{equation}
V(\sigma) =\frac{\sigma^2}{2G}
+\frac{i}{{\cal V}} \,  \mathrm{Tr} \, \mbox{Ln}\left(i\hat{D}-\sigma\right)
=\frac{\sigma^2}{2G}
+\frac{i}{{\cal V}}  \, \mbox{Ln}\, \mbox{Det}\left(i\hat{D}-\sigma\right),
\end{equation}
where $\hat{D}\equiv\gamma^\mu D_\mu$ and ${\cal V} \equiv \int d^3u$ is the 
space-time volume. By making use of the following identity:
\begin{equation}
\mbox{Det}\left(i\hat{D}-\sigma\right)
=\mbox{Det}\left(\gamma^5(i\hat{D}-\sigma)\gamma^5\right)
=\mbox{Det}\left(-i\hat{D} -\sigma\right),
\end{equation}
we see that the one-loop part of the effective potential, which we will denote as 
$\tilde{V}(\sigma)$, can be rewritten more conveniently as follows:
\begin{equation}
\tilde{V}(\sigma) 
= \frac{i}{2{\cal V}} \mathrm{Tr}\left[\mbox{Ln}(i\hat{D}-\sigma)+\mbox{Ln}(-i\hat{D}-\sigma)\right] 
= \frac{i}{2{\cal V}} \mathrm{Tr} \, \mbox{Ln}  \left(\hat{D}^2+\sigma^2\right).
\end{equation}
By making use of the proper-time representation, the effective potential $\tilde{V}(\sigma)$ 
can be expressed as follows:
\begin{equation}
\tilde{V}(\sigma) =-\frac{i}{2} \int^\infty_0\frac{ds}{s}\mathrm{tr}
\langle u |e^{-is(\hat{D}^2+\sigma^2)}| u \rangle ,
\label{eq:gamm}
\end{equation}
where
\begin{equation}
\hat{D}^2 = D_\mu D^\mu + \frac{ie}{2} \gamma^\mu\gamma^\nu F_{\mu\nu} 
=  D_\mu D^\mu - i\gamma^1\gamma^2 e B.  
\end{equation}
The matrix element $\langle u |e^{-is(\hat{D}^2+\sigma^2)}| u^\prime \rangle$ can be
calculated by using the Schwinger's proper-time method \cite{Schwinger:1951nm}. 
The corresponding result reads
\begin{eqnarray}
\langle u |e^{-is(\hat{D}^2+\sigma^2)}| u^\prime \rangle &=& e^{-is\sigma^2}
\langle u | e^{-isD_\mu D^\mu}| u^\prime \rangle \Bigg[ \cos (eBs)
- \gamma^1\gamma^2 \sin (eBs)\Bigg] \nonumber \\
&=& \frac{e^{-i\frac{\pi}{4}}}{8(\pi s)^{3/2}} e^{-i(s\sigma^2-S_{\rm cl})}
  \Bigg[eBs \cot(eBs) - \gamma^1\gamma^2 eBs\Bigg],
\label{eq:heatker}
\end{eqnarray}
where the ``classical" part of the action is given by
\begin{equation}
S_{\rm cl} = - e\int^{u}_{u^\prime} A _\lambda dz^\lambda - \frac{1}{4s} (u-u^\prime)_\nu
   \left( g^{\nu\mu}+\frac{(F^2)^{\nu\mu}}{B^2}  \left[1-eBs\cot(eBs)\right]\right) (u-u^\prime)_\mu.
\end{equation}
By definition, the line integral $\int^{u}_{u^\prime} A _\lambda dz^\lambda$ in the last expression 
is calculated along the straight line between the space-time points $(u^\prime)^{\mu}$ and $u^{\mu}$.

By substituting the matrix element (\ref{eq:heatker}) into Eq.~(\ref{eq:gamm}), we obtain
the following result for the one-loop part of the effective potential:
\begin{equation}
\tilde{V}(\sigma) = - \frac{iNe^{-i\frac{\pi}{4}}}{4\pi^{3/2}} 
\int^\infty_0 \frac{ds}{s^{3/2}} e^{-is\sigma^2} eB \cot (eBs).
\end{equation}
Therefore, the complete expression for the effective potential is given by
\begin{eqnarray}
V(\sigma) &=& \frac{\sigma^2}{2G}+\tilde{V}(\sigma)= \frac{\sigma^2}{2G}
+\frac{N}{4\pi^{3/2}} \int^\infty_{1/\Lambda^2} \frac{ds}{s^{3/2}}
e^{-s\sigma^2}eB \coth (eBs) \nonumber\\
&\simeq& \frac{\sigma^2}{2G}
+\frac{N}{4\pi^{3/2}} \int^\infty_{1/\Lambda^2} \frac{ds}{s^{5/2}} e^{-s\sigma^2}
+\frac{N}{4\pi^{3/2}} \int^\infty_{1/\Lambda^2} \frac{ds}{s^{3/2}}
e^{-s\sigma^2}\left[eB \coth (eBs)-\frac{1}{s}\right],
\label{eq:poten}
\end{eqnarray}
where we changed the integration variable to the imaginary proper time, $s\to e^{-i\pi/2} s$,
and introduced explicitly the ultraviolet cutoff $\Lambda$. Taking into account that the 
last integral is finite in the limit $\Lambda\to \infty$, it is justified to replace its lower 
limit with $0$. Then, by using the integral representation for the generalized Riemann 
zeta function $\zeta$ \cite{1980tisp.book.....G},
\begin{equation}
\int^\infty_0 ds s^{\mu-1} e^{-\beta s} \coth s  = \Gamma(\mu)
\left[2^{1-\mu}\zeta \left(\mu,\frac{\beta}{2}\right)-\beta^{-\mu}\right]
= \Gamma(\mu)
\left[2^{1-\mu}\zeta \left(\mu,1+\frac{\beta}{2}\right)+\beta^{-\mu}\right]
,\label{eq:zeta}
\end{equation}
which is valid for $\mu>1$, and analytically continuing this representation
to $\mu=-\frac{1}{2}$, we can rewrite Eq.~(\ref{eq:poten}) as
\begin{equation}
V(\sigma) = \frac{N}{\pi} \Bigg[\frac{\Lambda}{2}
\left(\frac{1}{g}-\frac{1}{\sqrt{\pi}}\right) \sigma^2-\frac{\sqrt{2}}{l^3}\zeta
\left(-\frac{1}{2}, \frac{(\sigma l)^2}{2}+1\right) -  \frac{\sigma}{2l^2}\Bigg]+O\left(\frac{1}{\Lambda}\right),     \label{eq:poten41}
\end{equation}
where we introduced the dimensionless coupling constant
\begin{equation}
g\equiv \frac{N G \Lambda }{\pi}.       \label{eq:gG}
\end{equation}
Let us now analyze the gap equation $dV/d\sigma=0$. In the integral form, it reads
\begin{equation}
\frac{\Lambda\sigma}{\pi g}=\frac{\sigma}{2\pi^{3/2}} \int^\infty_{1/\Lambda^2}
\frac{ds}{s^{3/2}} e^{-s\sigma^2}eBs\coth(eBs) .
\label{eq:gap1}
\end{equation}
This can be also rewritten as
\begin{eqnarray}
2\Lambda l\left(\frac{1}{g}-\frac{1}{\sqrt{\pi}}\right)\sigma &=& \frac{1}{l} +\sqrt{2} \sigma
\zeta\left(\frac{1}{2},1+\frac{\sigma^2l^2}{2}\right)+O\left(\frac{1}{\Lambda}\right).
\label{eq:gap2}
\end{eqnarray}
In the limit of the vanishing magnetic field, $B\to0$, we recover the 
known gap equation \cite{Rosenstein:1990nm}:
\begin{equation}
\sigma^2=\sigma\Lambda\left(\frac{1}{g_c}-\frac{1}{g}\right),
\label{eq:gap}
\end{equation}
where $g_c\equiv \sqrt{\pi}$ is the critical value of coupling. This equation admits a nontrivial 
solution $\sigma = m^{(0)}_{\rm dyn}\equiv (g-g_c)\Lambda/(g g_c)$ only if the coupling 
constant $g$ is supercritical, i.e., $g>g_c$. [As is clear from Eq.~(\ref{eq:lag2}), a solution 
to the gap equation, $\sigma=\bar{\sigma}$, coincides with the fermion dynamical mass,
$\bar{\sigma}=m_{\rm dyn}$.] Let us now show that the magnetic field changes 
the situation dramatically, allowing a nontrivial solution to exist at all $g>0$.
The reason for this is the enhancement of the interaction in the infrared region 
(or at large $s$), which is formally seen from the fact that, at $B\neq0$, the 
integral in Eq.~(\ref{eq:gap1}) becomes proportional to $1/\sigma$ as $\sigma\to0$.

Let us first consider the case of subcritical coupling, $g<g_c$, which
in turn can be subdivided into two cases: (i) $g\lesssim g_c$ (weak coupling) and 
(ii) $g\to g_c-0$ (nearcritical region). As we will see below, the solution in the weak 
coupling case will be such that $|\bar{\sigma}l|\ll1$. In the nearcritical region,
on the other hand, we will find that $|\bar{\sigma}l|\simeq 1$.

Assuming that $|\bar{\sigma}l|\ll1$ in the weak coupling region, 
from Eq.~(\ref{eq:gap2}) we find the following approximate solution:
\begin{equation}
m_{\rm dyn}\equiv\bar{\sigma}\simeq\frac{g g_c |eB|}{2(g_c-g)\Lambda},
\qquad\mbox{(weak coupling)} .
\label{eq:mdyn}
\end{equation}
The form of the solution implies that the condition $|\bar{\sigma}l|\ll1$
is satisfied when $g_c-g\gg \sqrt{|eB|}/ \Lambda$. Taking into account the 
natural model assumption that the magnetic field scale is much smaller than 
the ultraviolet cutoff $\Lambda$, we conclude that the relation (\ref{eq:mdyn}) 
is actually valid in a large window of subcritical couplings, excluding only the 
nearcritical region.

Several comment are in order about the solution (\ref{eq:mdyn}) in the 
weakly coupled regime. From the first constraint in Eq.~(\ref{eq:cons}), we
see that the dynamical mass is related to the condensate as follows: $m_{\rm dyn}=\langle 0|\sigma|0\rangle
=- \pi g\langle 0|\bar{\Psi}\Psi|0\rangle/(N\Lambda)$. By comparing this relation with the 
solution in Eq.~(\ref{eq:mdyn}), we conclude that the corresponding weak-coupling result 
for the condensate should read $\langle 0|\bar{\Psi}\Psi|0\rangle=-N|eB|/(2\pi)$. The latter 
coincides with the value of the condensate calculated in the problem of free fermions in a 
magnetic field, see Eq.~(\ref{eq:condensate}). The agreement between the two results is a 
clear indication that the $| \theta_1,\theta_2\rangle$-vacua constructed in Section~\ref{sec:SymBr2+1FreeReal} 
are good trial states for the vacua in the problem of weakly interacting fermions. In turn, this 
also explains why the dynamical mass $m_{\rm dyn}$ in Eq.~(\ref{eq:mdyn}) is an analytic 
function of the coupling at $g\to 0$. In essence, the dynamical mass is a result of a small 
perturbation in the theory of free fermions. This argument justifies our earlier claim that the 
magnetic catalysis effect in $2 + 1$ dimensions is much stronger than the Cooper pairing in 
the BCS theory, where the energy gap is exponentially suppressed at small coupling 
\cite{Bardeen:1957kj,Bardeen:1957mv}.

Let us now turn to the analysis of the gap equation in the nearcritical region, $g\to g_c-0$. 
In this case, is it convenient to introduce the following mass scale: $m^{*}=(g_c-g)\Lambda/(g g_c)$. 
Then, the gap equation (\ref{eq:gap2}) can be rewritten in an equivalent form, 
\begin{equation}
2m^{*}l=\frac{1}{|\sigma|l}+\sqrt{2}\zeta \left(\frac{1}{2}, \frac{(\sigma l)^2}{2} +1\right).
\end{equation}
In the nearcritical region $g\to g_c$, where the additional condition $m^{*}l\ll 1$ is satisfied, 
the dynamical mass is approximately given by 
\begin{equation}
m_{\rm dyn}=\bar{\sigma}\simeq 0.446 \sqrt{|eB|},
\qquad\mbox{(scaling region)} . 
\label{eq:mdynB}
\end{equation}
As we see, in the scaling region when $m^{*}l\to0$ (or equivalently $g_c-g\ll \sqrt{|eB|}/\Lambda$), 
the cutoff dependence disappears from the expression for the dynamical mass $m_{\rm dyn}$. 
This agrees with the well-known fact that the critical value $g_c=\sqrt{\pi}$ is an ultraviolet
stable fixed point at leading order in $1/N$ \cite{Rosenstein:1990nm}. The relation (\ref{eq:mdynB}) 
can be considered as a scaling law in the nearcritical region.

In the supercritical region, $g>g_c$, the analytic expression for $m_{\rm dyn}$
can be obtained in the limit of a weak magnetic field, satisfying the condition
$\sqrt{|eB|}\ll m^{(0)}_{\rm dyn}$ where $m^{(0)}_{\rm dyn}$ is the solution
of the gap equation (\ref{eq:gap}) at $B=0$. From Eq.~(\ref{eq:gap2}), we 
derive the following solution: 
\begin{equation}
m_{\rm dyn}=\bar{\sigma}\simeq m_{\rm dyn}^{(0)}\left(1+\frac{(eB)^2}{12\left(m^{(0)}_{\rm dyn}\right)^4}
\right),
\qquad\mbox{(strong coupling, weak field)} .  
\label{eq:md51}
\end{equation}
In the derivation of this result, we made use of the following asymptotic
expansion for the zeta-function~\cite{BatemanErdelyi_Vol1,NIST:DLMF,Olver:2010:NHMF}:
\begin{equation}
\zeta (s,q) \simeq \frac{1}{(s-1)q^{s-1}} +\frac{1}{2q^{s}}
+\sum_{n=1}^{\infty} \frac{B_{2n} \, \Gamma(s+2n-1)}{(2n)! \, \Gamma(s) \, q^{2n+s-1} },
\quad\mbox{for}\quad
q\to\infty,
\label{eq:asym}
\end{equation}
where $B_{2n}$ are the Bernoulli numbers. The numerical values of the first few 
nontrivial Bernoulli numbers are
$B_{0} = 1$,  $B_{1} = -\frac{1}{2}$, $B_{2} = \frac{1}{6}$,  $B_{4} = -\frac{1}{30}$,  
$B_{6} = \frac{1}{42}$,  $B_{8} = -\frac{1}{30}$,  $B_{10} = \frac{5}{66}$, etc.

\begin{figure}[t]
\begin{center}
\includegraphics[width=0.47\textwidth]{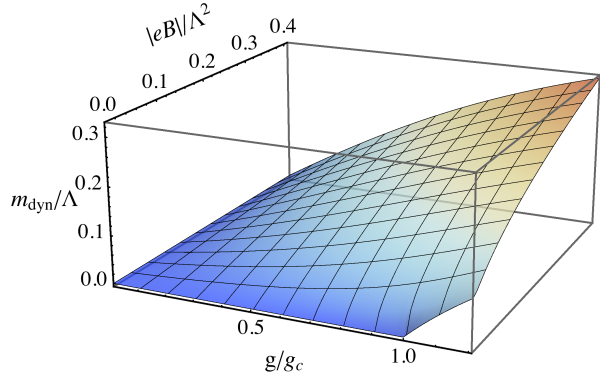}
\caption{(Color online) The dynamical mass as a function of the coupling constant
and magnetic field.}
\label{fig-DynMass}
\end{center}
\end{figure}

The weak-field expression (\ref{eq:md51}) shows that the dynamical mass increases 
with $B$. The same is also true beyond the weak-field approximation. In the general 
case, the numerical solution to the gap equation (\ref{eq:gap2}) is shown in 
Fig.~\ref{fig-DynMass}. The results indeed reveals that the dynamical mass 
increases with $B$ at all values of $g$ and $B$.

A striking property of the gap equation at $B\neq0$, which is not shared by the 
gap equation (\ref{eq:gap}) in the absence of the magnetic field, is that it does 
not have the trivial solution $\sigma=0$. Indeed, Eq.~(\ref{eq:poten}) implies that 
$dV/d\sigma|_{\sigma=0}=d\tilde{V}/d\sigma|_{\sigma=0}$, and
then we find from Eqs.~(\ref{eq:lag2}) and (\ref{eq:exp}) that
\begin{equation}
\left.\frac{d\tilde{V}}{d\sigma}\right|_{\sigma=0}
=\left.\langle0|\bar{\Psi}\Psi|0\rangle_{(2+1)} \right|_{g=0}
=-N\frac{|eB|}{2\pi}\neq 0
\end{equation}
[compare with Eq.~(\ref{eq:condensate})]. Thus, despite the spontaneous
breaking of the $\mathrm{U}(2)$ flavor symmetry, no trivial solution (stable 
or unstable) exists at any values of $g$ when the magnetic field is nonzero.

\subsubsection{The spectrum of the collective excitations}
\label{sec:NJL2+1BosSpectr}

Let us now consider the kinetic term ${\cal L}_k$ in the effective action (\ref{eq:effact}).
The $\mathrm{U}(2)$ symmetry implies that ${\cal L}_k$ should have the general form
\begin{eqnarray}
{\cal L}_k &=& N\frac{F_1^{\mu\nu}}{2} \left(\partial_\mu\rho_j\partial_\nu \rho_j\right)
+ N\frac{F^{\mu\nu}_2}{\rho^2}\left(\rho_j\partial_\mu\rho_j\right)  \left(\rho_i\partial_\nu\rho_i\right),
\label{eq:Lk}
\end{eqnarray}
where $\bm{\rho}=(\sigma,\tau,\pi)$ and $F^{\mu\nu}_1$, $F^{\mu\nu}_2$ are functions of 
$\rho^2=\sigma^2+\tau^2+\pi^2$. In the literature, there exist different methods to calculate 
the kinetic terms of the action. We will use the method of Refs.~\cite{Gusynin:1991ny,
Gusynin:1991wm,Miransky:1992bj}. The details of the derivation of functions $F^{\mu\nu}_1$
and $F^{\mu\nu}_2$ are given in Appendix~\ref{App:CompProp}. Here we will present only 
the final results.

The off-diagonal components (with $\mu\neq \nu$) of both tensor functions $F^{\mu\nu}_1$ 
and $F^{\mu\nu}_2$ vanish. The diagonal components are given by the following expressions:
\begin{eqnarray}
F^{00}_1 &=& \frac{l}{8\pi} \left[\frac{1}{\sqrt{2}}\zeta \left(\frac{3}{2},
  \frac{(\rho l)^2}{2}+1\right)+(\rho l)^{-3}\right],  \\
F^{11}_1 &=& F^{22}_1=\frac{1}{4\pi\rho},  \\
F^{00}_2 &=&- \frac{l}{16\pi} \left[\frac{(\rho l)^2}{2\sqrt{2}} \zeta
  \left(\frac{5}{2}, \frac{(\rho l)^2}{2}+1\right) + (\rho l)^{-3}\right],
  \label{eq:Fij} \\
F_2^{11} &=& F^{22}_2 = \frac{l}{8\pi} \left[\frac{(\rho l)^4}{\sqrt{2}} \zeta
   \left(\frac{3}{2}, \frac{(\rho l)^2}{2}+1\right) 
   +  \sqrt{2}(\rho l)^2 \zeta \left(\frac{1}{2},
   \frac{(\rho l)^2}{2}+1\right) +2\rho l-(\rho l)^{-1}\right] .
\end{eqnarray}
By making use of the effective potential and the kinetic term, now we can determine the
low-energy spectrum (dispersion relations) of the collective excitations $\tau$, $\pi$,
and $\sigma$. 

In order to calculate the mass (energy gap) $M_\sigma$ of the $\sigma$-mode, we
also need the explicit expression for the second derivative of the effective potential 
at its minimum, $\sigma=\bar{\sigma}$. The corresponding result is given by
\begin{eqnarray}
V^{\prime\prime}(\bar{\sigma}) &=& N
\frac{\bar{\sigma}^2l}{\pi^{3/2}} \int^\infty_0ds \sqrt{s}
e^{-(\bar{\sigma}l)^2s}
\coth s
=N\frac{\bar{\sigma}^2l}{2\pi} \left[\frac{1}{\sqrt{2}}
\zeta\left(\frac{3}{2},\frac{(\bar{\sigma}l)^2}{2}+1\right)
+(\bar{\sigma}l)^{-3}\right] .
\label{eq:d2V}
\end{eqnarray}
This can be easily derived either from Eq.~(\ref{eq:poten}) or from Eq.~(\ref{eq:poten41}) after 
taking into account that $\bar{\sigma}$ satisfies the gap equation, $V^{\prime}(\bar{\sigma}) =0$. 
By making use of the above expression for $V^{\prime\prime}(\bar{\sigma})$ together with the kinetic 
term ${\cal L}_k$ in Eq.~(\ref{eq:Lk}), we readily determine the long-wavelength dispersion relations 
for the two degenerate NG bosons ($\pi$ and $\tau$), as well as for the massive $\sigma$-mode. 
The corresponding relations take the form
\begin{eqnarray}
E_{\pi,\tau} &=& v_{\pi,\tau}\sqrt{\mathbf{k}^2}, \\
E_{\sigma} &=& \sqrt{M^2_\sigma+v^2_\sigma \mathbf{k}^2},
\end{eqnarray}
where the expressions for the two velocity parameters and the $\sigma$-mass are determined by
\begin{eqnarray}
v_{\pi,\tau}^2 &=&  \frac{F_1^{11}}{ F_1^{00}}  
=\frac{4 (\bar{\sigma} l)^2}{2+\sqrt{2}(\bar{\sigma} l)^3 \zeta \left(\frac{3}{2},\frac{(\bar{\sigma} l)^2}{2}+1\right)}, 
\label{v_pi_general}
\\
v_{\sigma} ^2&=&  \frac{F_1^{11}+2F_2^{11}}{ F_1^{00}+2F_2^{00}} 
=\frac{ 4 \bar{\sigma} l \left[ 2\sqrt{2}
+2\bar{\sigma} l \zeta \left(\frac{1}{2},\frac{(\bar{\sigma} l)^2}{2}+1\right)
+(\bar{\sigma} l)^3 \zeta \left(\frac{3}{2},\frac{(\bar{\sigma} l)^2}{2}+1\right)\right]}
{2\zeta \left(\frac{3}{2},\frac{(\bar{\sigma} l)^2}{2}+1\right)-(\bar{\sigma} l)^2 \zeta \left(\frac{5}{2},\frac{(\bar{\sigma} l)^2}{2}+1\right)}, 
\label{v_sigma_general}\\
M_\sigma^2 &=&   \frac{V^{\prime\prime}(\bar{\sigma})}{N(F_1^{00}+2F_2^{00})} 
=\frac{8\left[\sqrt{2} + (\bar{\sigma} l)^3 \zeta \left(\frac{3}{2},\frac{(\bar{\sigma} l)^2}{2}+1\right)\right]}
{\bar{\sigma} l \left[2\zeta \left(\frac{3}{2},\frac{(\bar{\sigma} l)^2}{2}+1\right)-(\bar{\sigma} l)^2 \zeta \left(\frac{5}{2},\frac{(\bar{\sigma} l)^2}{2}+1\right)\right]} .
\label{m_sigma_general}
\end{eqnarray}
We would like to emphasize that the propagator of the NG modes in leading order in $1/N$ 
has a genuine $(2+1)$-dimensional form, see Appendix~\ref{App:CompProp} and a brief 
discussion at the end of this section. This fact is also crucial for providing the reliability of the 
$1/N$ expansion in the problem at hand. 

At weak coupling, when $g_c-g\gg\sqrt{|eB|}/\Lambda$ is satisfied, the solution to the gap equation (\ref{eq:mdyn}) 
implies that $|\bar{\sigma}l|\ll1$. In this case, the approximate expressions for the parameters $v_{\pi,\tau}$, 
$v_{\sigma}$ and $M_\sigma$ read 
\begin{eqnarray}
v_{\pi,\tau}^2 &\simeq & 2 (\bar{\sigma}l)^2
= \frac{(gg_c)^2 |eB| }{2(g_c-g)^2\Lambda^2}, 
\label{v_pi_weak} \\
v_{\sigma}^2&\simeq &   \frac{4\sqrt{2} \bar{\sigma}l }{ \zeta\left(\frac{3}{2}\right) }
=\frac{2\sqrt{2} g g_c\sqrt{|eB|}}{\zeta\left(\frac{3}{2}\right)(g_c-g)\Lambda}, 
\label{v_sigma_weak}\\
M_\sigma^2 &\simeq &  \frac{4\sqrt{2} }{ \zeta\left(\frac{3}{2}\right) \bar{\sigma}l^3 }
=\frac{8\sqrt{2} (g_c-g)\Lambda\sqrt{|eB|}}{\zeta\left(\frac{3}{2}\right)g g_c} .
\label{m_sigma_weak}
\end{eqnarray}
We see that, as the interaction is switched off, $g\to0$, the velocity of the $\pi$ and $\tau$ 
NG (gapless) modes becomes zero. (The velocity parameter $v_{\sigma}$ also vanishes, but it 
is of little importance for the heavy $\sigma$-mode.) In addition, the mass of the $\sigma$-mode goes 
to infinity, implying that the corresponding massive collective excitation decouples from the low-energy 
spectrum. Therefore, by taking the limit $g\to0$, we return to the dynamics with spontaneous 
flavor symmetry breaking, but without genuine NG modes. This is in agreement with the earlier
discussion in Section~\ref{sec:SymBr2+1FreeReal}.

The above analysis of the dynamics in the problem of a relativistic fermion in an 
external magnetic field shows that, as expected, there are no propagating low-energy 
bosonic modes when the interaction between fermions is turned off. However, adding 
even an arbitrarily weak attractive $(g>0)$ interaction ``resurrects'' the NG modes 
and they acquire a small, but nonzero velocity $v$ proportional to the coupling 
constant $g$.

Let us now consider the nearcritical region with $g_c-g\ll\sqrt{|eB|}/\Lambda$. 
From the general expressions in Eqs.~(\ref{v_pi_general}),  (\ref{v_sigma_general}) 
and (\ref{m_sigma_general}), we find that
\begin{eqnarray}
v_{\pi,\tau} &\simeq &  0.587, 
\label{v_pi_scaling} \\
v_{\sigma} &\simeq & 0.791 , 
\label{v_sigma_scaling}\\
M_\sigma  &\simeq &  2.509 \sqrt{|eB|},
\label{m_sigma_scaling}
\end{eqnarray}
where we used the solution to the gap equation in the nearcritical (scaling) 
region, $\bar{\sigma}\simeq 0.446 |eB|^{1/2}$, see Eq.~(\ref{eq:mdynB}). 
As expected, the cutoff $\Lambda$ disappears from all the observables in 
this scaling region.

Let us now turn to the supercritical region with $g>g_c$. The analytic expressions
for $E_{\tau,\pi}$ and $M^2_\sigma$ can be obtained for small magnetic fields 
satisfying the condition $\sqrt{|eB|}\ll m^{(0)}_{\rm dyn}$, where $m^{(0)}_{\rm dyn}$ 
is the solution of the gap equation (\ref{eq:gap}) at $B=0$. In this case, from 
Eqs.~(\ref{v_pi_general}),  (\ref{v_sigma_general}) and (\ref{m_sigma_general}), 
we obtain:
\begin{eqnarray}
v_{\tau,\pi}^2 &=& \left( 1 - \frac{(eB)^2}{4\left(m^{(0)}_{\rm dyn}\right)^4} \right),
\label{v_pi_strong} \\
v_{\sigma}^2 &=& \left( 1 - \frac{(eB)^2}{4\left(m^{(0)}_{\rm dyn}\right)^4} \right),
\label{v_sigma_strong} \\
M_\sigma^2 &=& 6\left(m^{(0)}_{\rm dyn}\right)^2 
\left(1+\frac{2(eB)^2}{3\left(m^{(0)}_{\rm dyn}\right)^4}\right).  
\label{m_sigma_strong}
\end{eqnarray}
Here we used the solution for the gap equation in Eq.~(\ref{eq:md51}), as well as the
asymptotic expansion (\ref{eq:asym}) for the $\zeta$-functions at large values of the 
argument. The above results show that the velocity of the NG modes (which is 
always less than 1 in units of the speed of light) decreases and the mass of the 
$\sigma$-mode increases as functions of the magnetic field. [In the original paper 
\cite{Gusynin:1994va}, the results at strong coupling and weak magnetic field had 
several mistakes, which were corrected in Eqs.~(\ref{v_pi_strong}) -- (\ref{m_sigma_strong}).]

\begin{figure}[t]
\includegraphics[width=0.47\textwidth]{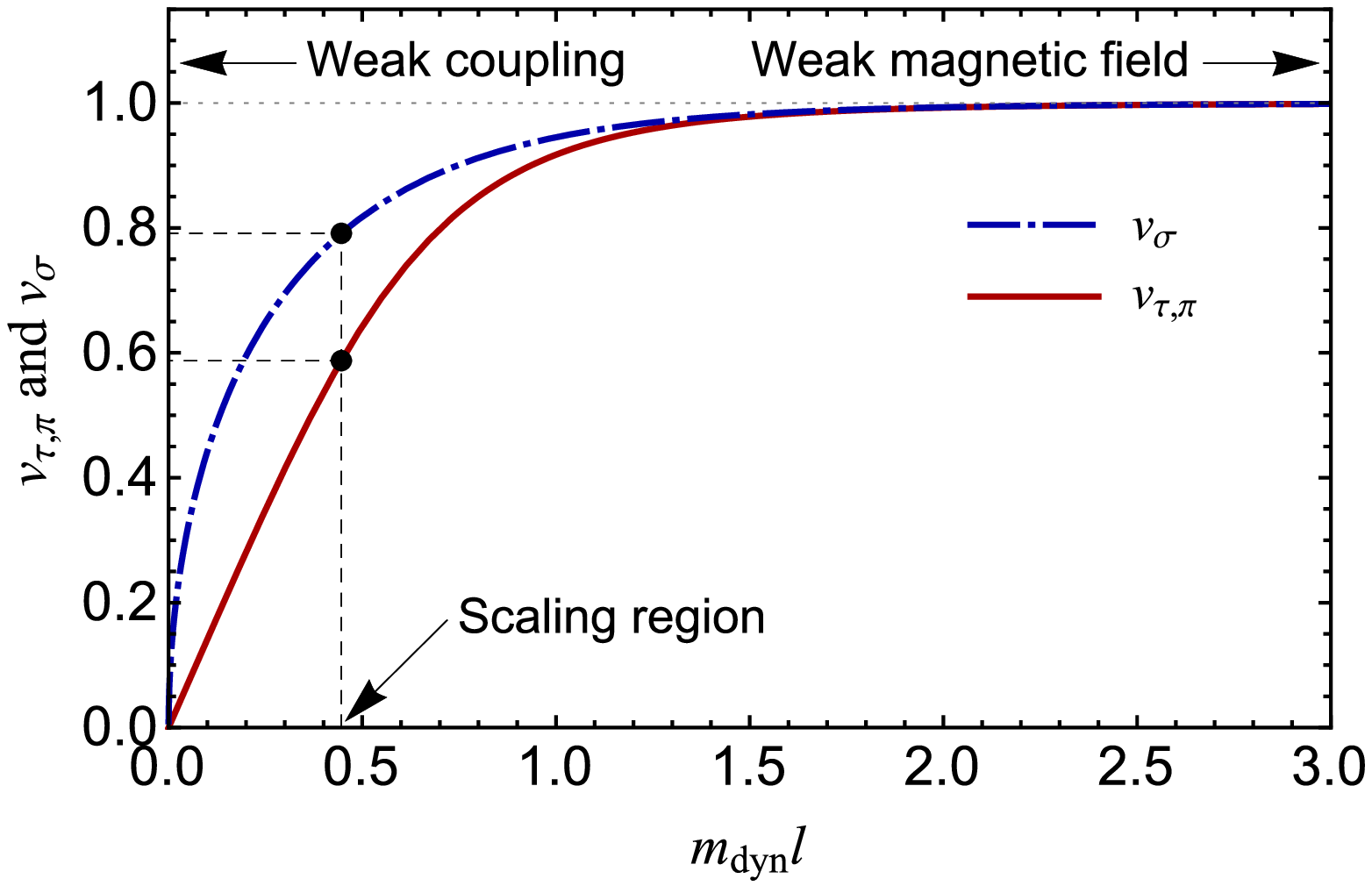}\hspace{0.05\textwidth}
\includegraphics[width=0.47\textwidth]{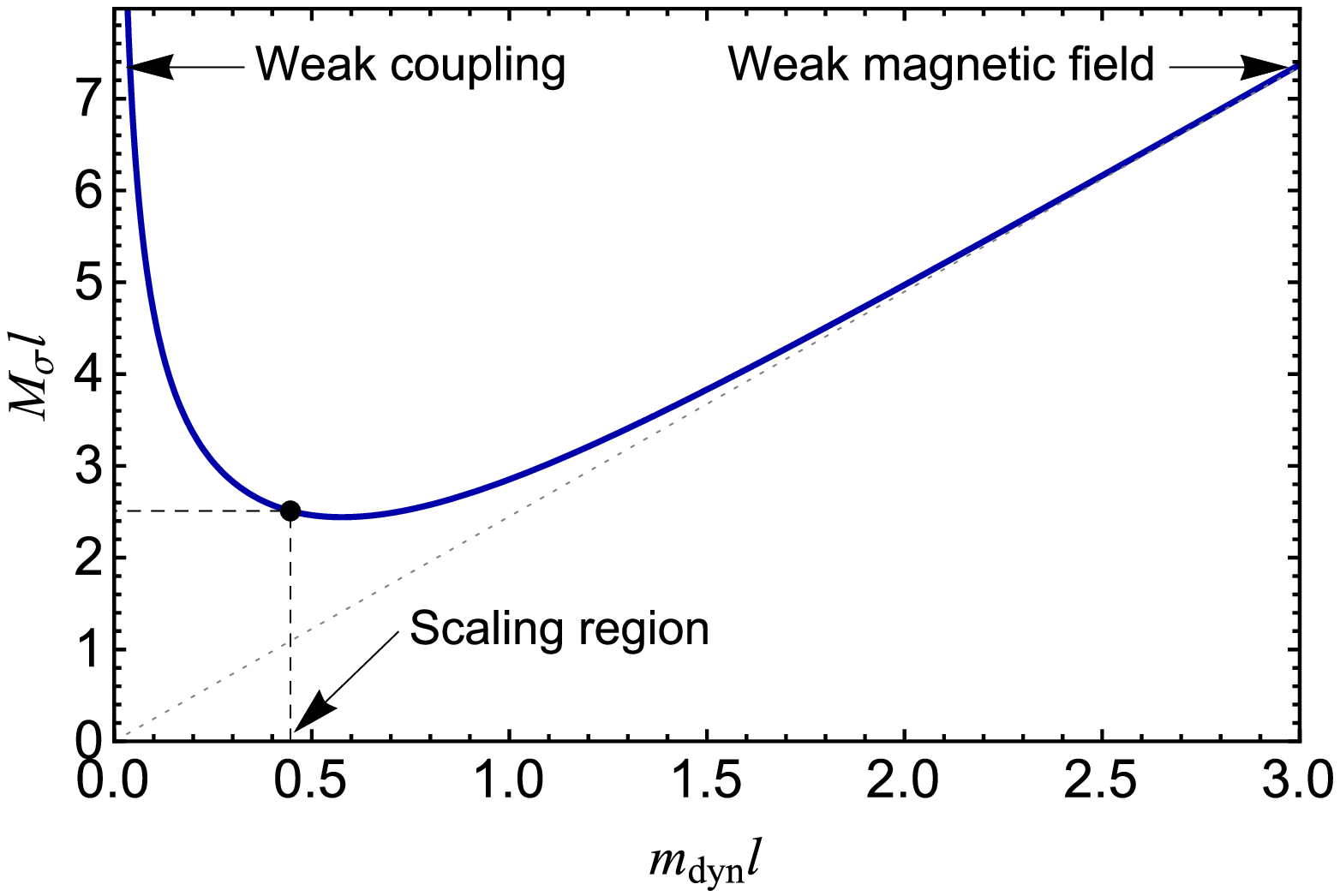}
\caption{(Color online) The velocities of the NG modes ($\pi$ and $\tau$) and the massive 
$\sigma$-mode (left panel), as well as the mass of the $\sigma$-mode (right panel) as 
functions of the $m_{\rm dyn}l$. The weak coupling (large 
$B$) limit corresponds to $m_{\rm dyn}l\ll1$ and the weak magnetic field (strong coupling) 
limit corresponds to $m_{\rm dyn}l\gg1$. In the scaling region, $m_{\rm dyn}l\approx 0.446$.}
\label{fig-Vpitau-Msigma}
\end{figure}

In the general case, the numerical results for the velocities $v_{\tau,\pi}$ and $v_{\sigma}$,
and the mass $M_\sigma$ of the $\sigma$-mode are shown in Fig.~\ref{fig-Vpitau-Msigma}. 
The results are plotted as functions of $m_{\rm dyn}l\equiv m_{\rm dyn}/\sqrt{|eB|}$, where 
$m_{\rm dyn}$ is the dynamical mass that satisfies the gap equation (\ref{eq:gap2}). 
In Fig.~\ref{fig-Vpitau-Msigma}, the weak coupling (large $B$) limit corresponds to 
$m_{\rm dyn}l\ll1$, and the weak magnetic field (strong coupling) limit corresponds 
to $m_{\rm dyn}l\gg1$. In the scaling region, $m_{\rm dyn}l\approx 0.446$, see 
Eq.~(\ref{eq:mdynB}).

We emphasize once again that, unlike the low-energy charged fermions, 
the neutral ($\pi$ and $\tau$) NG modes are genuinely $(2+1)$-dimensional. The 
$(2+1)$-dimensional character is clear from the fact that the velocity in Eq.~(\ref{v_pi_general}) 
is nonzero in general. Only in the extreme case when the interaction is completely switched 
off ($g\to 0$), does the velocity of the NG bosons vanish. This is evident from 
the weakly-coupling result in Eq.~(\ref{v_pi_weak}) where the velocity $v_{\pi,\tau}$ 
decreases with $\bar{\sigma}$ ($g$) and becomes zero when $\bar{\sigma}\to 0$ ($g\to 0$). 
The reason for this should be clear: since at $g=0$ the energy of the neutral system 
which is made up of a free fermion and a free antifermion from the lowest Landau
level is identically zero, its velocity should be zero too. Of course, this is also related to 
the fact that the motion of charged fermions is restricted in the $xy$ 
plane by the magnetic field. On the other hand, at $g>0$, there are genuine neutral 
NG bound states (with the binding energy $E_{b} \equiv -2m_{\rm dyn}$). Since the motion
of the center of the mass of {\it neutral} bound states is not restricted
by a magnetic field, their dynamics is $(2+1)$-dimensional.

Let us now discuss the continuum limit $\Lambda\to\infty$ in more detail. As is known, 
at $B=0$, in this model, an interacting continuum theory appears only at the critical 
value $g=g_c$ (the continuum theory is trivial at $g<g_c$) 
\cite{Miransky:1994vk,Rosenstein:1990nm,Kondo:1992sq,Hands:1992be}. Therefore, 
since at $g<g_c$, in the continuum limit, there is no attractive interaction between 
fermions, it is not surprising that at $g<g_c$, the dynamical mass
$m_{\rm dyn}\sim g|eB|/\Lambda$ disappears as $\Lambda\to\infty$.

Without the magnetic field ($B=0$), the continuum theory is in the symmetric phase 
at $g\to g_{c}-0$ and in the broken phase at $g\to g_{c}+0$. On the other hand, in a 
magnetic field, it is in the broken phase for both $g\to g_{c}-0$ and $g\to g_{c}+0$ 
(although the dispersion relations for the fermions and collective excitations $\pi$, $\tau$
and $\sigma$ are different at $g\to g_{c}-0$ and $g\to g_{c}+0$).

Up to now we have considered four-component fermions. In the case of
two-component fermions, the effective potential, $V_2$ is 
$V_2(\sigma)=V(\sigma)/2$ where $V(\sigma)$ is defined in Eqs.~(\ref{eq:poten}) and
(\ref{eq:poten41}). However, the essential new point is that there is no continuous 
$\mathrm{U}(2)$ symmetry (and therefore NG modes) in this case. As in the case of 
four-component fermions, in an external magnetic field, the dynamical fermion
mass (now breaking parity) is generated at any positive value of the coupling
constant $g$.

The above analysis of the NJL model illustrates the general phenomenon in $2+1$ 
dimensions: in the infrared region, an external magnetic field reduces the dynamics 
of fermion pairing to a $(0+1)$-dimensional dynamics (associated with the lowest 
Landau level) and, thus, catalyzes the generation of a dynamical mass for fermions.

\subsubsection{Thermodynamic properties}
\label{sec:NJL2+1Thermo}

In this section, we discuss the thermodynamic properties of the NJL model in
a magnetic field. In particular, we will show that there is a symmetry restoring 
phase transition at high temperature or large chemical potential.

The derivation of the thermodynamic potential at nonzero temperature and chemical 
potential to the leading order in $1/N$ is given in Appendix~\ref{App:ThermoPotNJL} 
(see also Refs.~\cite{Gusynin:1994va,Andersen:1995ia,Cangemi:1996tp,Elmfors:1993wj,
Elmfors:1993bm,Elmfors:1994mq,Persson:1994pz}). The method uses the imaginary 
time formalism \cite{Dolan:1973qd}. Here we present only the final result for the potential, 
\begin{eqnarray}
V_{\beta,\mu}(\sigma) &=& V(\sigma)+\tilde{V}_{\beta,\mu}(\sigma)
= \frac{N}{\pi} \left[\frac{\Lambda}{2} \left(\frac{1}{g}-\frac{1}{\sqrt{\pi}}\right)
\sigma^2-\frac{\sqrt{2}}{l^3}\zeta \Big(-\frac{1}{2}, \frac{(\sigma l)^2}{2}
+1\Big) -\frac{\sigma}{2l^2}\right]-\nonumber\\
&-& N \frac{|eB|}{2\pi\beta} \left[\ln (1+e^{-\beta(\sigma-\mu)})
+2\sum_{k=1}^\infty\ln\left(1+  e^{-\beta(\sqrt{\sigma^2+\frac{2k}{l^2}}-\mu)}\right)
+(\mu\to-\mu)\right] .
\label{eq:temperature}
\end{eqnarray}
where we also used the explicit form of the effective potential in Eq.~(\ref{eq:poten41}), 
obtained at $T=\mu=0$.

\begin{figure}[t]
\includegraphics[width=0.47\textwidth]{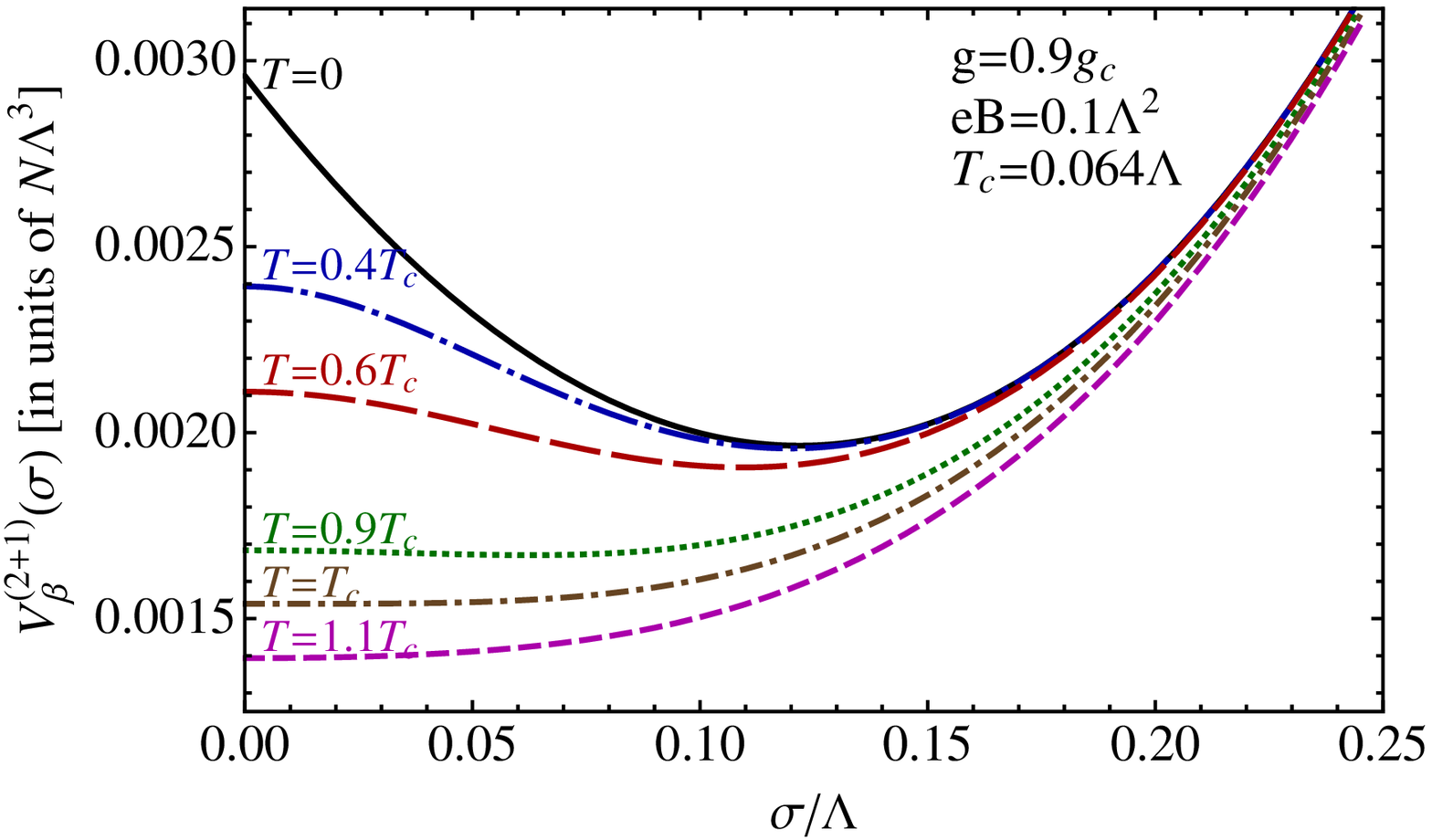}\hspace{0.05\textwidth}
\includegraphics[width=0.47\textwidth]{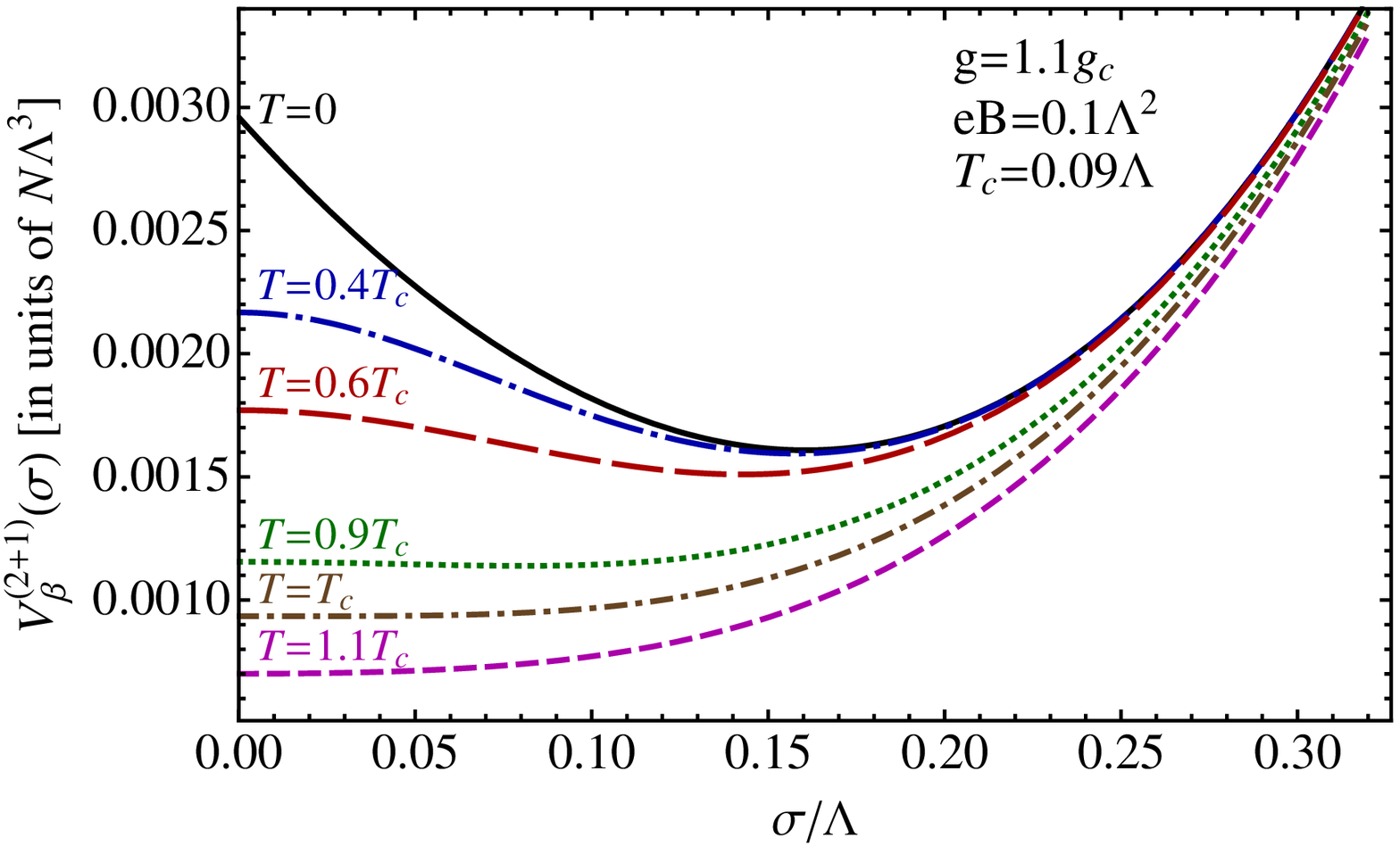}
\caption{(Color online) The thermodynamic potential $V_\beta(\sigma)$ in $2+1$ dimensions 
for several values of temperature and two fixed values of the coupling constant: 
$g=0.9g_c$ (left panel) and $g=1.1g_c$ (right panel).}
\label{fig-effVvsSigmaT}
\end{figure}

Let us first consider the case of nonzero temperature, but vanishing chemical 
potential. This models a system with equal densities of thermally excited fermions 
and antifermions, but with the vanishing net fermion number density. In this case, 
the thermodynamic potential $V_\beta(\sigma) \equiv V_{\beta,\mu=0}(\sigma)$ 
is given by
\begin{equation}
V_\beta(\sigma) = \frac{N}{\pi} \Bigg[\frac{\Lambda}{2} \left(\frac{1}{g}-\frac{1}{\sqrt{\pi}}\right) 
\sigma^2- \frac{\sqrt{2}}{l^3}\zeta \Big(-\frac{1}{2},
\frac{(\sigma l)^2}{2}+1\Big) - \frac{\sigma}{2l^2}\Bigg] 
- N \frac{|eB|}{\pi\beta}
\left[\ln(1+e^{-\beta\sigma})+   2 \sum^\infty_{k=1} \ln
\left(1+e^{-\beta\sqrt{\sigma^2+\frac{2k}{l^2}}}\right)\right].
\label{eq:Veff_T}
\end{equation}
At sufficiently low temperatures, this potential has a nontrivial minimum at $\bar{\sigma}\neq 0$, 
while at high temperatures the only minimum of the potential is at $\bar{\sigma}= 0$. This is 
illustrated in Fig.~\ref{fig-effVvsSigmaT} for several values of the temperature and
two fixed values of the coupling constant. The left panel shows the results for a 
coupling constant smaller than $g_c$ (i.e., the critical value of coupling in absence 
of the magnetic field), while the right panel shows the results for a coupling 
larger than $g_c$. We observe a similar qualitative behavior of the effective 
potentials in both cases. This should not be surprising because the magnetic 
field drastically changed the dynamics in the subcritical regime ($g<g_c$) by 
effectively turning it into a supercritical one.

By studying how the minimum of the effective potential evolves with the temperature, 
we determine the temperature dependence of the dynamically generated fermion mass, i.e., $m_{\rm dyn}(T) 
= \bar{\sigma}(T)$. Formally, the minimum of the potential can be located by solving the following 
gap equation:
\begin{equation}
\frac{dV_\beta(\sigma)}{d\sigma}=0.
\end{equation}
The numerical solution to this equation reveals that the dynamical mass gradually decreases with the 
temperature $T$ and turns to zero at $T\geq T_c$. This is in agreement with the qualitative 
behavior of the effective potentials in Fig.~\ref{fig-effVvsSigmaT}. Such a temperature dependence 
implies that a symmetry restoring phase transition occurs at $T=T_c$. Taking into account that the 
dynamical mass is a continuous function of the temperature at $T=T_c$, the corresponding transition
is a second order one.

\begin{figure}[t]
\begin{center}
\includegraphics[width=0.47\textwidth]{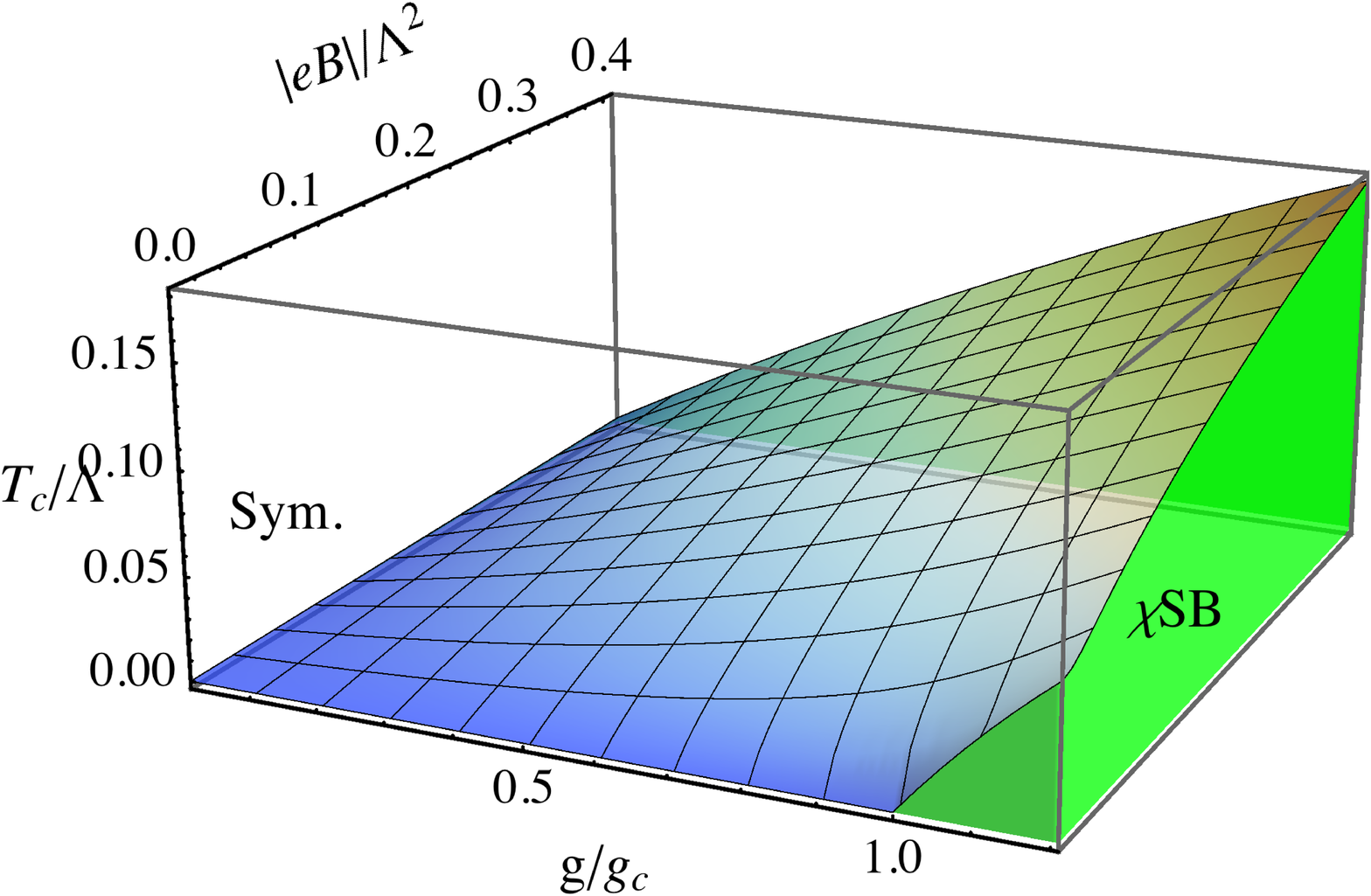}\hspace{0.05\textwidth}
\includegraphics[width=0.47\textwidth]{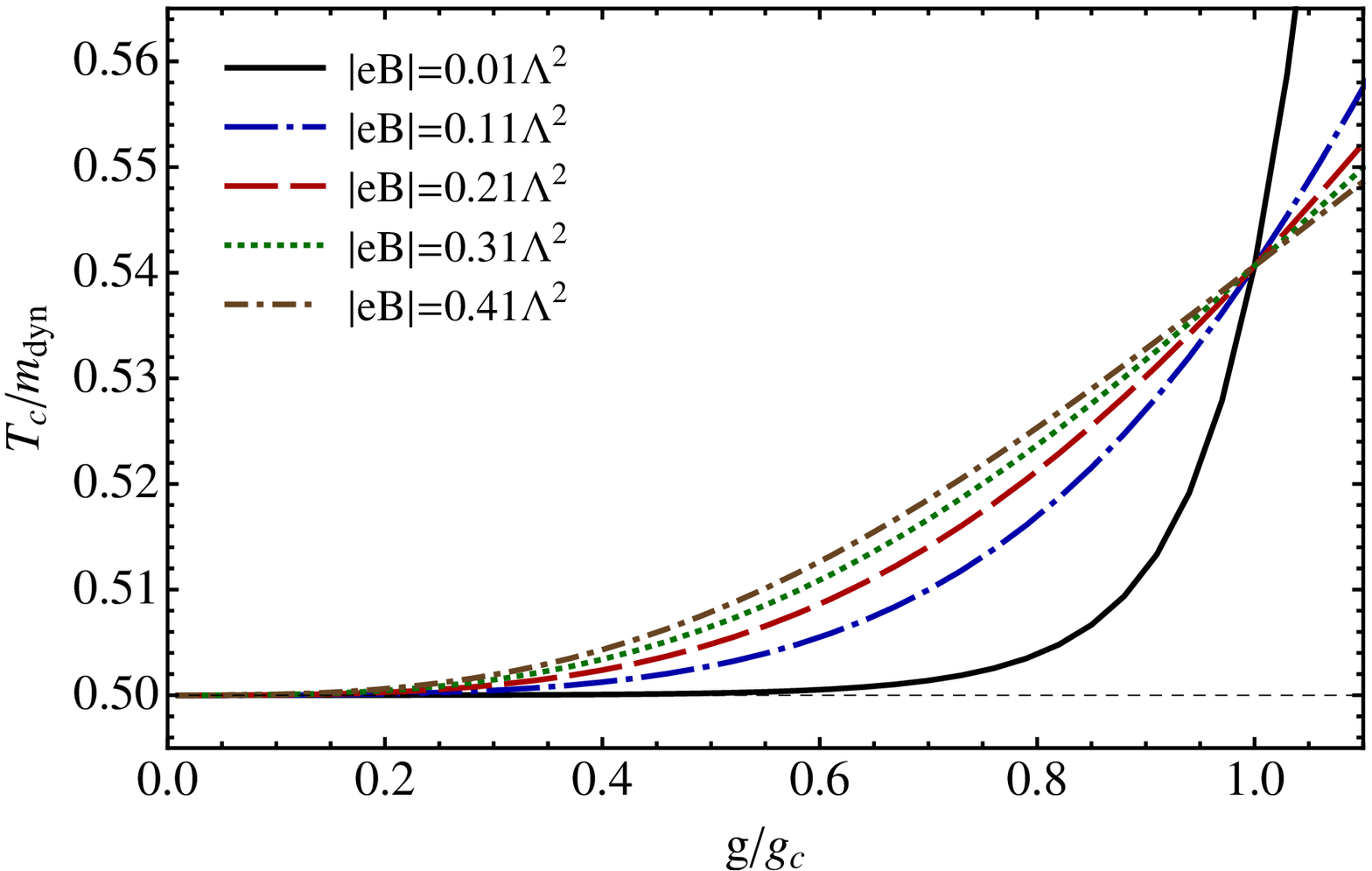}
\caption{(Color online) Left panel: Critical temperature as a function of the coupling constant $g/g_c$ and 
the magnetic field in the NJL model in $2+1$ dimensions (cf. with the dynamical mass in Fig.~\ref{fig-DynMass}). 
Right panel: The ratio of the critical temperature to the zero-temperature dynamical mass for several 
fixed values of the magnetic field.}
\label{fig-TcgB}
\end{center}
\end{figure}

The compilation of the numerical results for the critical temperature $T_c$ as a function of 
the coupling constant and magnetic field is shown in the left panel of Fig.~\ref{fig-TcgB}. It may be 
instructive to compare these results with those for the dynamical mass at zero temperature, shown 
in Fig.~\ref{fig-DynMass}. The qualitative resemblance between the two is obvious and can be 
easily understood. The larger is the value of the dynamical mass $m_{\rm dyn}$ at $T=0$, the 
stronger thermal fluctuations are needed to destroy it. A more careful comparison shows that the 
approximate relation between the critical temperature and the zero-temperature mass is 
$T_c\approx m_{\rm dyn}/2$, i.e., the ratio of the two is about $1/2$. This is supported by the numerical 
results in the right panel of Fig.~\ref{fig-TcgB}, which shows the ratio $T_c/m_{\rm dyn}$ as a function of the 
coupling constant for several fixed values of the magnetic field. As we see, the above approximate 
relation indeed holds at weak coupling. As the coupling increases towards $g_c$, however, some 
deviations become evident. At exactly $g=g_c$, the ratio of the critical temperature to the 
zero-temperature dynamical mass approaches $T_c/m_{\rm dyn} \approx 0.541$, which is 
about $8\%$ larger than the result at weak coupling ($g\to 0$). The ratio departs from $1/2$ even more 
when $g>g_c$. It is important to emphasize that the ratio $T_c/m_{\rm dyn} \approx 0.541$ at 
$g=g_c$ (scaling region) is independent of the magnetic field. In the corresponding critical 
(scaling) region, there is only one independent scale in the problem and it is associated 
with the external magnetic field. Therefore, at $g=g_c$, all dimensionfull dynamical 
parameters scale as appropriate powers of the magnetic scale, while all dimensionless 
quantities (e.g., such as $T_c/m_{\rm dyn}$) must be independent of the field. 

Here it may be appropriate to recall that, in accordance with the Mermin-Wagner-Coleman
theorem \cite{Mermin:1966fe,Coleman:1973ci}, a continuous symmetry cannot be 
spontaneously broken at nonzero temperature in $2+1$ dimensions. This is because 
the dynamics of zero modes in $(2+1)$-dimensional field theories is two dimensional 
at nonzero temperature. As a result, strong fluctuations of the would-be-NG modes 
destroy the order parameter, which means that the corresponding continuous symmetry 
is not broken. 

In the NJL model at finite temperature (in both cases with and without the magnetic field), the 
Mermin-Wagner-Coleman theorem manifests itself only beyond the leading order in $1/N$. 
One plausible possibility of what happens at $T\neq 0$ beyond the leading order in $1/N$ 
is the following. The dynamics of the zero mode in this model is essentially equivalent
to that of the $\mathrm{SU}(2)$ $\sigma$-model in two-dimensional Euclidean space. As is known, 
the $\mathrm{SU}(2)$ symmetry is exact in the latter model and, as a result, the would-be-NG bosons 
become massive excitations \cite{Polyakov:1983tt}. Therefore, it seems plausible that, in the
$(2+1)$-dimensional NJL model in a magnetic field, the $\mathrm{SU}(2)$ symmetry will be restored 
at any finite temperature, and the dynamically generated mass $m_{\rm dyn}$ of fermions 
will disappear. The question whether this, or another, scenario is realized at finite temperature 
in this model deserves further study. One of the possibilities has been recently discussed in 
Ref.~\cite{Cao:2014uva}, where the authors argued that the finite temperature phase transition 
in this model should be of the Berezinski-Kosterlitz-Thouless type 
\cite{1971JETP...32..493B,1973JPhC....6.1181K}.

\begin{figure}[t]
 \includegraphics[width=0.47\textwidth]{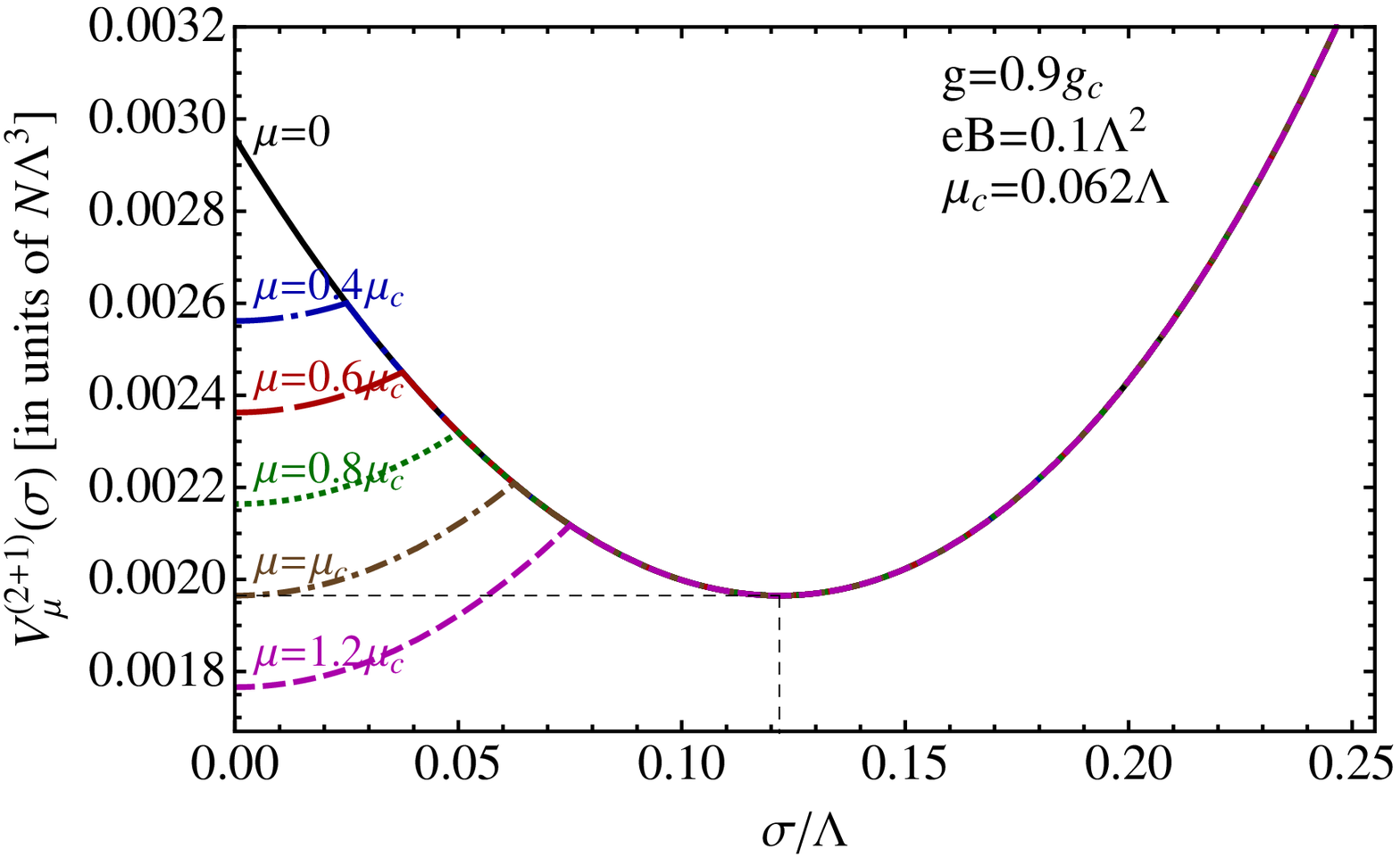}\hspace{0.05\textwidth}
\includegraphics[width=0.47\textwidth]{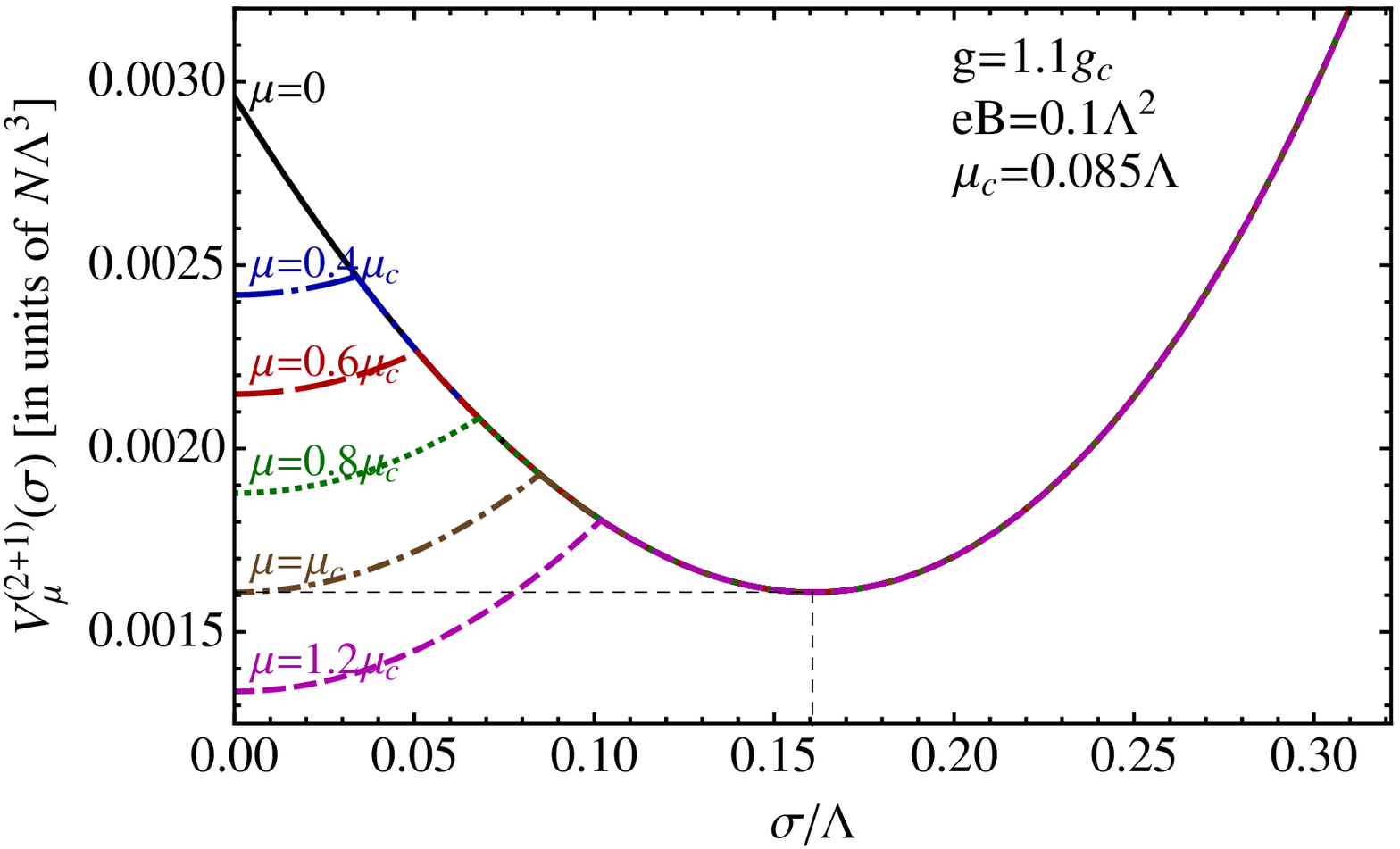}
\caption{(Color online) The thermodynamic potential $V_\mu(\sigma)$ in $2+1$ dimensions 
for several values of the chemical potential and two fixed values of the coupling constant: 
$g=0.9g_c$ (left panel) and 
$g=1.1g_c$ (right panel).}
\label{fig-effVvsSigmaMu}
\end{figure}

Before concluding this section, let us also address the effect of a nonzero chemical potential 
on the dynamical generation of fermion mass. The corresponding effective potential can be 
easily obtained from the general expression in Eq.~(\ref{eq:temperature}) by taking the limit 
$\beta\to \infty$ (i.e., $T\to 0$). The explicit form of the resulting potential $V_\mu(\sigma) \equiv 
V_{\beta=\infty,\mu}(\sigma)$ reads 
\begin{eqnarray}
V_{\mu}(\sigma) &=& \frac{N}{\pi} \left[\frac{\Lambda}{2\sqrt{\pi}} \left(\frac{\sqrt{\pi}}{g}-1\right)
\sigma^2-\frac{\sqrt{2}}{l^3}\zeta \Big(-\frac{1}{2}, \frac{(\sigma l)^2}{2}
+1\Big) -\frac{\sigma}{2l^2}\right]-\nonumber\\
&-& N \frac{|eB|}{2\pi } \left[(|\mu|-\sigma)\theta(|\mu|-\sigma)
+2\sum_{k=1}^\infty\left(|\mu|-\sqrt{\sigma^2+\frac{2k}{l^2}}\right)\theta\left(|\mu|-\sqrt{\sigma^2+\frac{2k}{l^2}}\right)\right].
\label{eq:Veff_Mu}
\end{eqnarray}
The graphical representation of this effective potential at a moderately strong magnetic field, $|eB|=0.1\Lambda^2$, 
is shown in Fig.~\ref{fig-effVvsSigmaMu} for several values of the chemical potential and two fixed values of 
the coupling constant. As we see, a nonzero chemical potential $\mu$ affects the behavior of the 
potential only at $\sigma<|\mu|$. By recalling that the variational parameter $\sigma$ plays the role of a 
dynamical fermion mass, this feature should not be surprising at all in the limit of zero temperature.
When $|\mu|$ is smaller than the fermion mass $\sigma$, the chemical potential cannot affect the 
corresponding variational state simply because it lies in the energy gap between the occupied (negative 
energy) states in the Dirac see and the empty states with positive energies.

\begin{figure}[t]
\begin{center}
\includegraphics[width=0.47\textwidth]{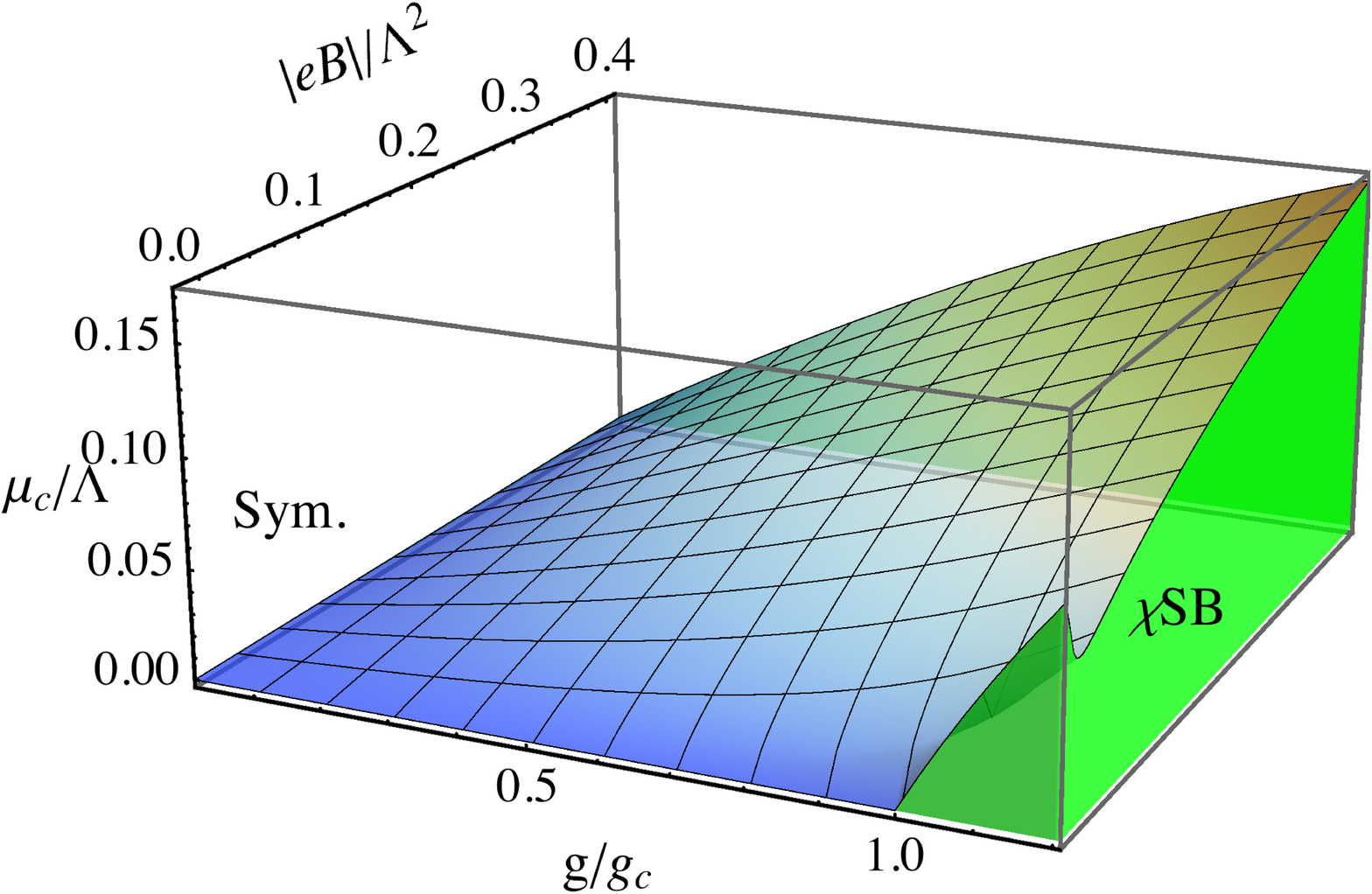}\hspace{0.05\textwidth}
\includegraphics[width=0.47\textwidth]{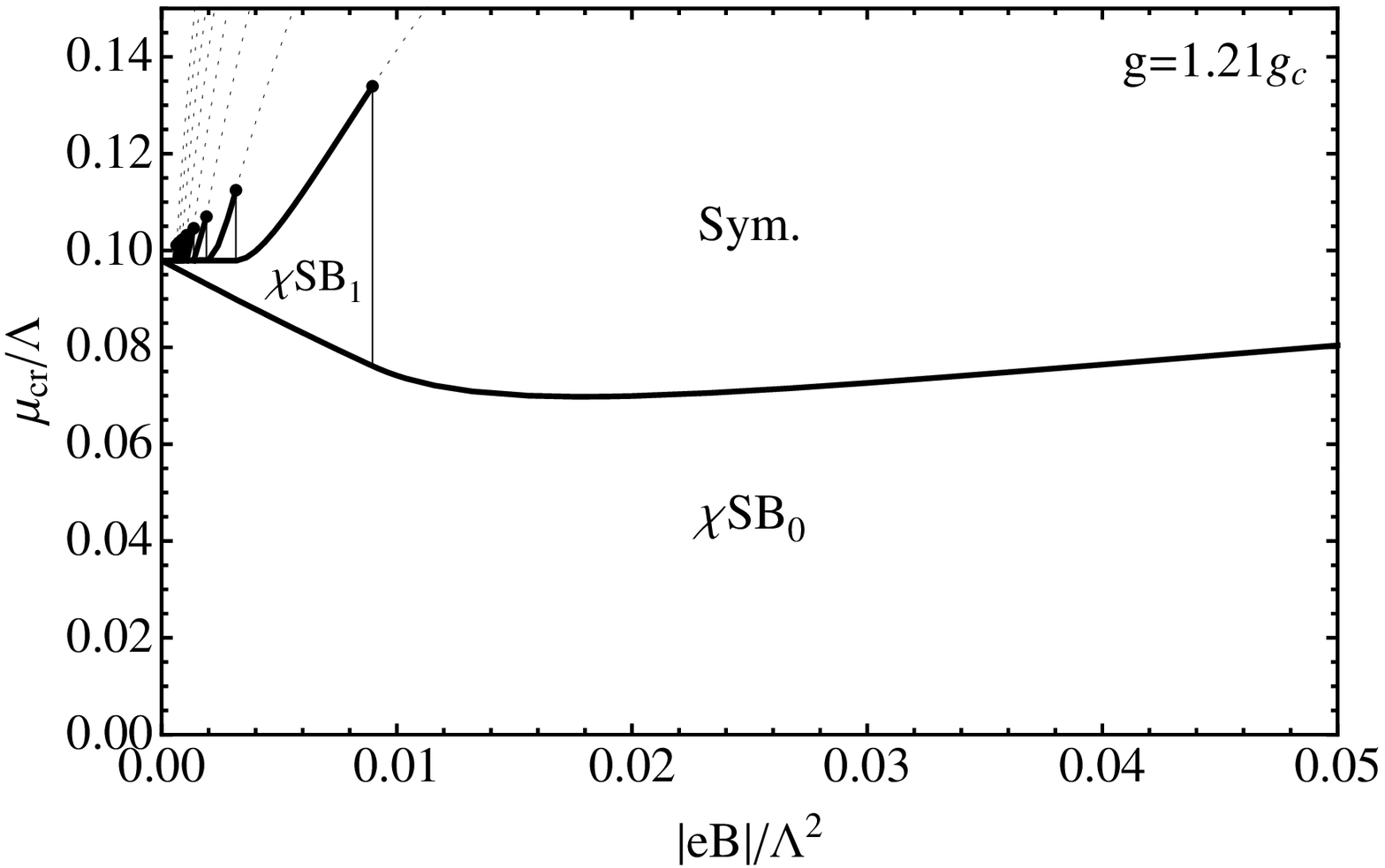}
 \caption{(Color online) Left panel: Critical chemical potential at $T=0$ as functions of the coupling constant $g/g_c$ and the 
magnetic field $|eB|/\Lambda^2$ in the NJL model in $2+1$ dimensions. Right panel: The rich structure of the 
phase diagram in the regime of weak magnetic field with the model assumption that the chiral condensate is 
the only possible order parameter.}
\label{fig-MucgB}
\end{center}
\end{figure}

The compilation of the numerical results for the critical chemical potential  $\mu_c$ as a function of 
the coupling constant and magnetic field is shown in the left panel of Fig.~\ref{fig-MucgB}. It may 
be instructive to compare these results with those for the dynamical mass at zero temperature  
in Fig.~\ref{fig-DynMass} and the critical temperature in Fig.~\ref{fig-TcgB}. In most regions 
of the parameter space, there is a clear qualitative similarity. One exception is the region of 
sufficiently small magnetic fields and $g>g_c$, where the behavior of the critical chemical potential 
appears to be counterintuitive: there $\mu_c$ decreases with the field. Moreover, upon a closer 
examination of the effective potential in the regime of weak fields, we find that the dynamics is even more 
complicated than the simplified 3D phase diagram in the left panel of Fig.~\ref{fig-MucgB} 
reveals. In fact, the corresponding phase diagram shows only one of the branches of the critical 
chemical potential.

Because of the highly oscillatory behavior of the effective potential, associated with a large 
number of nearly degenerate Landau levels, the model at hand formally experiences an infinite 
number of quantum phase transitions. A representative snapshot of the detailed phase diagram 
at a fixed value of coupling is shown in the right panel of Fig.~\ref{fig-MucgB}. The thick (thin) 
solid lines represent first (second) order phase transitions. The dotted lines separate the regions
of the symmetric phase with different number of Landau levels filled, but do not formally correspond 
to phase transitions in the conventional sense. While there is only one symmetric phase, there appear 
to be an infinite number of symmetry broken states ($\chi$SB$_{i}$ where $i=0,1,2,\ldots$). The 
latter differ by the value of the Dirac mass and the number of filled Landau levels. They are 
generically separated from one another by first order phase transitions. 

Before concluding this section, let us mention that the phase diagram in Fig.~\ref{fig-MucgB}
should not be taken too seriously. In the analysis that lead to this diagram, we made a very strong
assumption that the chiral condensate is the only possible order parameter. This may be justifiable
in the vacuum state at zero density, as well as at nonzero temperature. The inclusion of a nonzero 
chemical potential, on the other hand, may bring in new features that cannot be captured by the 
simplest variational ansatz with the Dirac mass alone. For example, a fractional filling of Landau 
levels may cause the dynamical generation of other order parameters, associated with the 
Mott-type metal-insulator phase transitions. In fact, we will discuss a very important example 
of such dynamics in more details in Section~\ref{sec:QHEgraphene} when studying the quantum 
Hall effect in graphene.

\subsection{Free fermions in a magnetic field in $3 + 1$ dimensions}
\label{sec:MagCat3+1General}

In this section, we will consider the phenomenon of magnetic catalysis in $3 + 1$ 
dimensions within the framework of the NJL model. Although its realization is similar to that in $2 + 1$ 
dimensions, we will show that there are important differences between the two cases. In particular, 
we will argue that the underlying dynamics of the magnetic catalysis in $3 + 1$ dimensions 
is much closer to the BCS dynamics \cite{Bardeen:1957kj,Bardeen:1957mv} than the magnetic 
catalysis dynamics in $2 + 1$ dimensions.

Let us start from the analysis of the spectrum of $(3 + 1)$-dimensional Dirac fermions 
in a constant magnetic field $B$. Without loss of generality, we assume that the field points 
in the $z$-direction. The Lagrangian density reads
\begin{equation}
{\cal L} = \frac{1}{2} \left[\bar{\Psi},(i\gamma^\mu D_\mu-m)\Psi\right],
\label{1}
\end{equation}
where the covariant derivative $D_\mu$ has the same form as in Eq.~(\ref{eq:dmu}). 
Also, the same Landau gauge (\ref{eq:Landau_gauge}) is used for the vector potential. 
In the vector notation, $\mathbf{A}=(-By,0,0)$. The energy spectrum is well known and 
is given by \cite{Akhiezer:1965}
\begin{equation}
E_n(k_z) = \pm\sqrt{m^2+2|eB|n+k_z^2},
\label{3}
\end{equation}
where index $n=0,1,2,\dots$ labels the Landau levels. As in $2+1$ dimensions, each 
Landau level is highly degenerate: for a fixed value of the longitudinal momentum 
$k_z$, the number of states is equal to $\frac{|eB|}{2\pi}S_{xy}$ at $n=0$ and 
$\frac{|eB|}{\pi}S_{xy}$ at $n>0$, where $S_{xy}$ is the area occupied by the 
system in the $xy$-plane. In the Landau gauge used, this degeneracy is connected 
with the momentum $k_x$, which also determines the $y$-coordinate of the center 
of a fermion orbit \cite{Akhiezer:1965}. This is very similar to the  $(2+1)$-dimensional 
case, discussed in detail in Section~\ref{sec:Free2+1MagField}.

As the fermion mass $m$ goes to zero, there is no energy gap between the vacuum 
and the lowest Landau level with $n=0$. In this case, the density of states at zero energy
is
\begin{equation}
\nu_0 =\left. \frac{1}{V}\frac{dN_0}{dE_0} \right|_{E_0=0}=\frac{1}{S_{xy}L_z}
\left. \frac{dN_0}{dE_0}\right|_{E_0=0}=\frac{|eB|}{4\pi^2},
\label{DOSatE0}
\end{equation}
where $E_0=|k_z|$ and $dN_0=S_{xy}L_z\frac{|eB|}{2\pi} \frac{dk_z}{2\pi}$ 
and $L_z$ is the system size in $z$-direction. We will see that the dynamics of the 
LLL plays the crucial role in catalyzing spontaneous chiral symmetry breaking. 
In particular, the density of states $\nu_0$ plays the same role here as the density of states 
on the Fermi surface $\nu_F$ in the theory of superconductivity \cite{Bardeen:1957kj,Bardeen:1957mv}.

The important feature of the dynamics in the analysis below will be associated with the 
$(1+1)$-dimensional nature of the LLL, which dominates at low energies. It is described 
by two continuous variables, i.e., the momentum $k_z$ and the energy $E$. Recall that 
the momentum $k_y$ is replaced by the discrete quantum number $n$ and the dynamics 
does not depend on the momentum $k_x$. Just like in $2+1$ dimensions, the LLL dynamics
is subject to the dimensional reduction $D \rightarrow D-2$.

In order to see the implications of the dimensional reduction in $3+1$ dimensions, let 
us calculate the chiral condensate $\langle 0|{\bar\Psi}\Psi|0 \rangle$ in this problem.
The condensate is expressed through the fermion propagator
$S(u,u^\prime)=\langle 0|T\Psi(u){\bar\Psi}(u^\prime)|0 \rangle$:
\begin{equation}
\langle0|\bar{\Psi}\Psi|0\rangle=-\lim_{u\to u^\prime} \mathrm{tr}\left[S(u,u^\prime) \right].
\label{5defCondensate}
\end{equation}
The propagator in a magnetic field was calculated by Schwinger \cite{Schwinger:1951nm}
and is quoted in Appendix~\ref{App:A-SchwingerProp}. It has the form of a product of the 
Schwinger phase factor $\exp \left( -  ie\int^u_{u^\prime} A_\lambda  dz^\lambda\right)$, 
where the line integral is calculated along the straight line, and a translation invariant 
function $\bar{S} (u-u^\prime)$. The Fourier transform of $\bar{S}(u)$ in Euclidean 
space (with $k^0\to ik_4$, $s\to -is$) is given by
\begin{eqnarray}
\bar{S}_E(k) &=&-i \int^\infty_0 ds 
\exp \left[-s\Bigg(m^2+k^2_4+k^2_3+\mathbf{k}^2_{\perp}\frac{\tanh(eBs)}{eBs}\Bigg)\right]
\nonumber\\
&\times &\left[m-k_\mu\gamma_\mu + i(k_2\gamma_1-k_1\gamma_2)\tanh(eBs)\right] 
\left[1 + i\gamma_1\gamma_2\tanh(eBs)\right],
\label{8SFourier}
\end{eqnarray}
where $\mathbf{k}_{\perp}=(k_1,k_2)$. Note that matrices $\gamma_4=-i\gamma^0$ and
$\gamma_i\equiv\gamma^i$ $(i=1,2,3)$ in the Euclidean space are antihermitian. From 
Eqs.~(\ref{5defCondensate}) and (\ref{8SFourier}) we obtain the following expression for the 
condensate:
\begin{eqnarray}
\langle0|\bar{\Psi}\Psi|0\rangle_{(3+1)} & = & 
- \frac{4m}{(2\pi)^4}\int d^4 k \int^\infty_{1/\Lambda^2}ds
\exp\left[-s\left(m^2+k_4^2+k_3^2+\mathbf{k}^2_{\perp} \frac{\tanh(eBs)}{eBs}
\right) \right]
\nonumber\\
&=& - \frac{m eB}{4\pi^{2}} \int^\infty_{1/\Lambda^2} \frac{ds}{s}
e^{-sm^2} \coth(eBs),
\label{Condensate3+1Free}
\end{eqnarray}
where $\Lambda$ is an ultraviolet cutoff. 
In the limit $m\to 0$, this gives
\begin{equation}
\langle0|\bar{\Psi}\Psi|0\rangle_{(3+1)} 
\simeq -\lim_{\Lambda\to\infty}\lim_{m\to 0} 
\frac{m}{4\pi^2} \left[\Lambda^2 +|eB|\ln\frac{|eB|}{\pi m^2} - m^2 \ln\frac{\Lambda^2}{2|eB|} 
+O\left(\frac{m^4}{|eB|}\right)\right]
=0,
\label{9Cond3+1Perturb}
\end{equation}
This result implies that there 
is no spontaneous chiral symmetry breaking for the system of {\it free} fermions in 
a magnetic field in $3+1$ dimensions. This is very different from the $(2+1)$-dimensional 
case.

We can trace the most singular contribution,
proportional to $m|eB|\ln(m)$ at $m\to 0$, in the condensate in Eq.~(\ref{9Cond3+1Perturb}) to 
the integration at large values of the proper-time coordinate, $s \rightarrow \infty$. The 
corresponding singularity is directly connected with the effective dimensional reduction in the infrared 
dynamics, $D\to D-2$, produced by the magnetic field. When the field is switched off ($B\to 0$), 
the integrand in Eq.~(\ref{Condensate3+1Free}) is proportional to $1/s^2$ and the proper-time 
integration is convergent in infrared ($s \rightarrow \infty$). In contrast, at nonzero $B$, the 
integrand behaves as $1/s$ and produces the mentioned logarithmic singularity. The 
corresponding $1/s$ asymptote of the proper-time function is characteristic for a $(1+1)$-dimensional 
model without external fields. From the physics viewpoint, the dimensional reduction 
$D\to D-2$ can be also interpreted as the restriction of the fermion motion in the two 
directions perpendicular to the magnetic field.

Let us show that the logarithmic singularity is due to the LLL dynamics. Instead of using the 
propagator in Eq.~(\ref{8SFourier}), for this purpose, it is more convenient to utilize the 
Landau level representation of the propagator. It can be obtained in the same way as in 
$2+1$ dimensions from Eq.~(\ref{8SFourier}) by making use of the identity (\ref{sum-Laguerre}).
Alternatively, it can be derived by using the method in Appendix~\ref{CSE:App:Landau-level-rep}. 
The final result reads
\begin{equation}
\bar{S}_E (k)=-i\exp\left(-\frac{\mathbf{k}^2_{\perp}}{|eB|}\right)
\sum_{n=0}^{\infty}(-1)^n\frac{D_n(eB,k)}{k_4^2+k_3^2+m^2+2eBn}
\end{equation}
where
\begin{eqnarray}
D_n(eB,k) &=& (m-k_4\gamma_4-k_3\gamma_3)\left\{\left[1+i\gamma_1\gamma_2\mathrm{sign}(eB)\right]
L_n\left(2\frac{\mathbf{k}^2_{\perp}}{|eB|}\right)-\left[1-i\gamma_1\gamma_2\mathrm{sign}(eB)\right]L_{n-1}
\left(2\frac{\mathbf{k}^2_{\perp}}{|eB|}\right)\right\} ,\nonumber\\
&&
+4(k_1\gamma_1+k_2\gamma_2)L_{n-1}^1\left(2\frac{\mathbf{k}^2_{\perp}}{|eB|}\right).
\end{eqnarray}
By definition, $L_n(x) \equiv L_n^0(x) $ and $L_{-1}^{\alpha}(x) =0$.

In the lowest Landau level approximation, the propagator $\bar{S}_E(k)$ becomes 
\begin{equation}
\bar{S}_{(3+1)} ^{0}(k)=-i\exp\left(-\frac{\mathbf{k}^2_{\perp}}{|eB|}\right)
\frac{m-k_4\gamma_4-k_3\gamma_3}{k_4^2+k_3^2+m^2}
\left[1+ i\gamma_1\gamma_2\mathrm{sign}(eB)\right],
\label{LLL-prop-in-3+1}
\end{equation}
where the projection operator $\left[1+i\gamma_1\gamma_2\mathrm{sign}(eB)\right]/2$ 
on the right hand side reflects the spin polarized nature of the LLL states. The orientation of the spin 
is parallel (antiparallel) to the magnetic field in the case of a positive (negative) electric charge 
of the fermions. The propagator in Eq.~(\ref{LLL-prop-in-3+1}) explicitly demonstrates the 
$(1+1)$-dimensional character of the LLL dynamics, which is described by two continuous 
variables $k_3$ and $k_4$. The corresponding approximate expression for the condensate
is given by
\begin{equation}
\langle0|\bar{\Psi}\Psi|0\rangle_{(3+1)} =
- \lim_{\Lambda\to\infty}\lim_{m\to 0}
\frac{4m}{(2\pi)^4}\int d^4 k \exp\left(-\frac{\mathbf{k}^2_{\perp}}{|eB|}\right)
\frac{1}{k_4^2+k_3^2+m^2} 
\simeq - \lim_{\Lambda\to\infty}\lim_{m\to 0}
\frac{m|eB|}{(2\pi)^2} \ln\frac{\Lambda^2}{m^2} 
= 0.
\end{equation}
This result reconfirms that the logarithmic singularity $m|eB|\ln(m)$ originates from the LLL. 

The analysis in both $2+1$ and $3+1$ dimensions suggests that there is a universal 
mechanism of the enhancement of the generation of fermion masses in a magnetic 
field. The fermion pairing is dominated by fermions in the LLL, which is subject to the 
dimensional reduction $D\to D-2$. Because of such a reduction, the pairing in the 
presence of a magnetic field is effectively much stronger than without the field. As
a consequences, the spontaneous chiral (flavor) symmetry breaking takes place 
even at the weakest attractive interaction between fermions. 

\subsection{The NJL model in a magnetic field in $3+1$ dimensions}
\label{sec:NJL3+1General}

Let us consider the Nambu--Jona-Lasinio (NJL) model with the $\mathrm{U_L(1)\times
U_R(1)}$ chiral symmetry:
\begin{equation}
{\cal L} = \frac{1}{2} \left[\bar{\Psi}, (i\gamma^\mu D_\mu)\Psi\right] +
\frac{G}{2} \left[ \left(\bar{\Psi}\Psi\right)^2+\left(\bar{\Psi}i\gamma^5\Psi\right)^2 \right],
\label{12L_NJL_U1xU1}
\end{equation}
where $D_\mu$ is the covariant derivative (\ref{eq:Landau_gauge}) and fermion fields carry an
additional, ``color'', index $\alpha=1,2,\dots,N$. The theory is equivalent
to the theory with the Lagrangian density
\begin{equation}
{\cal L}=\frac{1}{2} \left[\bar{\Psi}, \left( i\gamma^\mu D_\mu\right)
 \Psi\right] - \bar{\Psi}\left(\sigma+i\gamma^5\pi\right)\Psi
- \frac{1}{2G} \left(\sigma^2+\pi^2\right),
\label{13}
\end{equation}
where $\sigma=- G\bar{\Psi}\Psi$ and $\pi=-G\bar{\Psi}i\gamma^5\Psi$.
The effective action for the composite fields $\sigma$ and $\pi$ is
\begin{equation}
\Gamma(\sigma,\pi) = -\frac{1}{2G}\int d^4u\left(\sigma^2+\pi^2\right) +
\tilde{\Gamma}(\sigma,\pi),
\label{14}
\end{equation}
where
\begin{equation}
\tilde{\Gamma}(\sigma,\pi) =- i \mathrm{Tr}\, \mbox{Ln} \left[i\gamma^\mu D_\mu
- (\sigma+i\gamma^5\pi)\right].
\label{15}
\end{equation}
As $N\to\infty$, the path integral over the composite fields is dominated
by stationary points of their action:
$\delta\Gamma/\delta\sigma=\delta\Gamma/\delta\pi=0$. We
will analyze the dynamics in this limit by using the expansion of the action
$\Gamma$ in powers of derivatives of the composite fields.

\subsubsection{The effective potential and gap equation}
\label{NJL3+1:EffPot}

\begin{figure}[t]
\begin{center}
\includegraphics[width=0.47\textwidth]{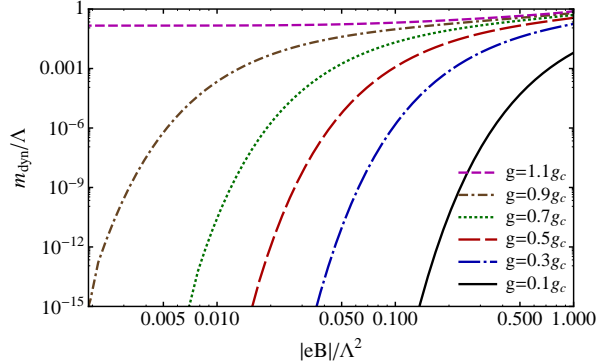}
\caption{(Color online) The dynamical mass as a function of $|eB|$ for several fixed values of the 
coupling constant.}
\label{fig-DynMass3+1}
\end{center}
\end{figure}

We begin by calculating the effective potential $V$. Since $V$ depends only on the 
$\mathrm{U_L(1)\times U_R(1)}$-invariant $\rho^2=\sigma^2+\pi^2$, it is sufficient to
consider a configuration with $\pi=0$ and $\sigma$ independent of the space-time coordinate 
$u$. Then, by rewriting the expression in Eq.~(\ref{15}) with the help of proper-time method 
\cite{Schwinger:1951nm}, we derive the following potential: 
\begin{eqnarray}
V(\rho)&=&\frac{\rho^2}{2G}+\tilde{V}(\rho)=\frac{\rho^2}{2G} +
\frac{N}{8\pi^2} \int^\infty_{1/\Lambda^2} \frac{ds}{s^2}
e^{-s\rho^2} \frac{1}{l^2}\coth\left(\frac{s}{l^2}\right)
\nonumber\\
&=&\frac{\rho^2}{2G}+\frac{N}{8\pi^2} \Bigg[\frac{\Lambda^4}{2}+
\frac{1}{3l^4}\ln(\Lambda l)^2 + \frac{1-\gamma-\ln2}{3l^4}-(\rho\Lambda)^2
+\frac{\rho^4}{2}\ln(\Lambda l)^2+\frac{\rho^4}{2}(1-\gamma-\ln2)
\nonumber\\
&+&\frac{\rho^2}{l^2}\ln\frac{\rho^2l^2}{2}-\frac{4}{l^4}\zeta'\left(-1,
\frac{\rho^2l^2}{2}+1\right)\Bigg],
\label{16}
\end{eqnarray}
where the magnetic length $l\equiv|eB|^{-1/2}$, $\zeta'(-1,x)=
\frac{d}{d\nu}\zeta(\nu,x)|_{\nu=-1}$, $\zeta(\nu,x)$ is the generalized
Riemann $\zeta$-function \cite{1980tisp.book.....G}, and $\gamma\approx0.577$ is the Euler
constant. The gap equation $dV/d\rho=0$ is
\begin{equation}
\Lambda^2\left(\frac{1}{g}-1\right)=-\rho^2\ln\frac{(\Lambda l)^2}{2}+
\gamma\rho^2+\frac{1}{l^{2}}\ln\frac{(\rho l)^2}{4\pi}+\frac{2}{l^{2}}\ln\Gamma\left(
\frac{\rho^2 l^2}{2}\right)+O\left(\frac{1}{\Lambda}\right),
\label{17}
\end{equation}
where the dimensionless coupling constant $g=NG\Lambda^2/(4\pi^2)$. In the
derivation of this equation, we used the relations \cite{1980tisp.book.....G}
\begin{eqnarray}
\frac{d}{dx}\zeta(\nu,x)&=&-\nu\zeta(\nu+1,x),
\label{18}
\\
\frac{d}{d\nu}\zeta(\nu,x)\Big|_{\nu=0}&=&\ln\Gamma(x)-\frac{1}{2}\ln2\pi,\\
\zeta(0,x)&=&\frac{1}{2}-x.
\label{19}
\end{eqnarray}
As $B\to0$ ($l\to\infty$), we recover the known gap equation in the NJL
model (for a review see the book \cite{Miransky:1994vk}):
\begin{equation}
\Lambda^2\left(\frac{1}{g}-1\right)=-\rho^2\ln\frac{\Lambda^2}{\rho^2}.
\label{20}
\end{equation}
This equation admits a nontrivial solution only if $g$ is supercritical,
$g>g_c=1$. As Eq.~(\ref{13}) implies, a solution to Eq.~(\ref{20}), $\rho=\bar{\sigma}$,
coincides with the fermion dynamical mass, $\bar{\sigma}=m_{\rm dyn}$, and the
dispersion relation for fermions is given by Eq.~(\ref{3}) with $m$ replaced by
$\bar{\sigma}$. We will show that the magnetic field changes the situation
dramatically: at $B\neq0$, a nontrivial solution exists at all $g>0$.
We will first consider the case of subcritical $g$, $g<g_c=1$, which
in turn can be divided into two subcases: (i) $g\ll g_c$ and
(ii) $g\rightarrow g_c-0$ (nearcritical $g$). Since at $g<g_c=1$
the left-hand
side in Eq.~(\ref{17}) is positive and the first term on the right-hand side in
this equation is negative, we conclude that a nontrivial solution to this
equation may exist only at $\rho^2 \ln\Lambda^2 l^2\ll l^{-2}
\ln(\rho l)^{-2}$. Then, at $g\ll 1$, we find the solution:
\begin{equation}
m^2_{\rm dyn}\equiv\bar{\sigma}^2\simeq 
\frac{1}{\pi l^2}\exp\left(-\frac{1-g}{g}\Lambda^2l^2\right) 
= \frac{eB}{\pi} \exp\left(\frac{\Lambda^2}{|eB|}\right) 
\exp\left( -\frac{4\pi^2}{|eB| N G}\right) .
\label{m_dyn_weak_3+1}
\end{equation}
The numerical solutions for the dynamical mass as a function of $|eB|$ for 
several fixed values of the coupling constant are shown in Fig.~\ref{fig-DynMass3+1}.
We checked that the numerical results agree quite well with the approximate analytical 
expression in Eq.~(\ref{m_dyn_weak_3+1}).

It is instructive to compare this result with that in $(2+1)$-dimensional NJL model 
in a magnetic field and with the BCS relation for the energy gap in the theory of 
superconductivity \cite{Bardeen:1957kj,Bardeen:1957mv}. While the expression
for $m_{\rm dyn}^2$ in Eq.~(\ref{m_dyn_weak_3+1}) has an essential singularity 
at $g=0$, in $(2+1)$-dimensional NJL model $m_{\rm dyn}^2$ is analytic at $g=0$: 
$m_{\rm dyn}^2\sim |eB|^{2}g^{2}/\Lambda^2$ [see Eq.~(\ref{eq:mdyn})]. The latter is 
connected with the fact that the condensate $\langle0|\bar{\Psi}\Psi|0\rangle_{(2+1)}$ 
in $2+1$ dimensions is nonzero even for free fermions in a 
magnetic field, see Section~\ref{sec:SymBr2+1Free}. As a result, the dynamical 
mass is analytic in $g$ in that case, as was shown in Section~\ref{sec:NJL2+1EffPot}. 
This is in turn connected with the point that, because of the dimensional reduction
$D\rightarrow D-2$ in a magnetic field, the LLL dynamics of the fermion pairing
in a magnetic field in $2+1$ dimensions is effectively $(0+1)$-dimensional. In fact, 
as will be shown in Section~\ref{sec:BS-NJL}, it can be described by the one-dimensional 
Schr\"odinger equation in quantum mechanics.

On the other hand, the dynamics of the fermion pairing in a magnetic field
in $3+1$ dimensions is $(1+1)$-dimensional. We recall that, because of the
Fermi surface, the dynamics of the electron pairing in BCS theory is also
$(1+1)$-dimensional. This analogy is rather deep. In particular, the expression 
(\ref{m_dyn_weak_3+1}) for $m_{\rm dyn}$ can be rewritten in a form similar to 
that for the energy gap $\Delta$ in the BCS theory: while $\Delta\sim\omega_D
\exp\left(-\mbox{const}/G_S\nu_F\right)$, where $\omega_D$ is the Debye
frequency, $G_S$ is a coupling constant and $\nu_F$ is the density of states
on the Fermi surface. Similarly, the expression for the mass $m_{\rm dyn}$ 
can be rewritten as $m_{\rm dyn}\sim \sqrt{eB}\exp\left(-1/2 N G \nu_0 \right)$, where 
$\nu_0$  is the density of LLL states on the energy surface $E=0$ given in
Eq.~(\ref{DOSatE0}). As we see, the energy surface $E=0$ plays the role analogous 
of the Fermi surface.

Let us now consider the nearcritical $g$ with $\Lambda^2(1-g)/g\rho^2\ll\ln\Lambda^2l^2$. 
Looking for a solution with $\rho^2l^2\ll 1$, we come to the following equation:
\begin{equation}
\frac{1}{\rho^2l^2}\ln\frac{1}{\pi\rho^2l^2}\simeq\ln\Lambda^2l^2,
\label{22}
\end{equation}
that gives the solution
\begin{equation}
m_{\rm dyn}^2=\bar{\sigma}^2\simeq|eB|\frac{\ln\left[(\ln\Lambda^2l^2)/\pi
\right]}{\ln\Lambda^2l^2}.
\label{23}
\end{equation}
In order to reveal the physical meaning of this solution, let us recall that, 
at $g=g_c$, the NJL model is equivalent to the (renormalizable) Yukawa model 
\cite{Miransky:1994vk}. To the leading order in $1/N$ expansion, the renormalized 
Yukawa coupling $\alpha_Y^{(l)}=(g_Y^{(l)})^2/(4\pi)$, related to the energy scale 
$\mu\simeq l^{-1}$, is given by $\alpha_Y^{(l)}\simeq\pi/\ln\Lambda^2l^2$ \cite{Miransky:1994vk}. 
Therefore, the dynamical mass (\ref{23}) can be equivalently rewritten as
\begin{equation}
m_{\rm dyn}^2\simeq|eB|\frac{\alpha_Y^{(l)}}{\pi}\ln\frac{1}{\alpha_Y^{(l)}}.
\label{24}
\end{equation}
Thus, as expected in a renormalizable theory, the ultraviolet cutoff $\Lambda$ is removed 
from the observable $m_{\rm dyn}$ through the renormalization of parameters (the coupling 
constant in the case at hand).

At $g>g_c$, an analytic expression for $m_{\rm dyn}$ can be obtained at a weak
magnetic field, satisfying the condition $|eB|^{1/2}/m^{(0)}_{\rm dyn} \ll1$, where 
$m^{(0)}_{\rm dyn}$ is the solution to the gap equation (\ref{20}) at $B=0$. 
Then, from Eq.~(\ref{17}), we find
\begin{equation}
m_{\rm dyn}^2\simeq\left(m^{(0)}_{\rm dyn}\right)^2\left(1+\frac{|eB|^2}
{3(m^{(0)}_{\rm dyn})^4\ln(\Lambda/m^{(0)}_{\rm dyn})^2}\right),
\label{25}
\end{equation}
 i.e., $m_{\rm dyn}$ increases with $|eB|$. In the nearcritical region, with $g-g_c\ll 1$, this
expression for $m_{\rm dyn}^2$ can be rewritten as
\begin{equation}
m_{\rm dyn}^2\simeq\left(m^{(0)}_{\rm dyn}\right)^2\left(1+\frac{1}{3\pi}
\alpha^{(m_{\rm dyn})}_Y \frac{|eB|^2}
{(m^{(0)}_{\rm dyn})^4}\right),
\label{26}
\end{equation}
where, in leading order in $1/N$, $\alpha_Y^{(m_{\rm dyn})}\simeq\pi/\ln(\Lambda/
m_{\rm dyn}^{(0)})^2$ is the renormalized Yukawa coupling related to the scale 
$\mu=m_{\rm dyn}^{(0)}$.

\subsubsection{The spectrum of the collective excitations}
\label{sec:NJL3+1KineticTerm}

Let us now turn to the kinetic term. The chiral $\mathrm{U_L(1)\times U_R(1)}$ symmetry 
implies that the general form of the kinetic term is
\begin{equation}
{\cal L}_k = \frac{F_1^{\mu\nu}}{2} (\partial_\mu\rho_j\partial_\nu \rho_j)
+ \frac{F^{\mu\nu}_2}{\rho^2}(\rho_j\partial_\mu\rho_j)(\rho_i
\partial_\nu\rho_i),
\label{27}
\end{equation}
where $\bm{\rho}=(\sigma,\pi)$ and $F^{\mu\nu}_1$, $F^{\mu\nu}_2$ are functions 
of $\rho^2$. These functions can be obtained by using the same method that was 
used in Section~\ref{sec:NJL2+1BosSpectr} in the study of the NJL model in $2 + 1$ dimensions
(see also Appendix~\ref{App:C1}). We find that the diagonal components of $F^{\mu\nu}_1$ 
and $F^{\mu\nu}_2$ are the only nonzero functions, i.e.,
\begin{eqnarray}
F^{00}_1 &=& F^{33}_1=\frac{N}{8\pi^2}\left[\ln\frac{\Lambda^2l^2}{2}-
\psi\left(\frac{\rho^2l^2}{2}+1\right)+\frac{1}{\rho^2l^2}-\gamma+
\frac{1}{3} \right], \\
F^{11}_1 &=& F^{22}_1=\frac{N}{8\pi^2}\left[\ln\frac{\Lambda^2}{\rho^2}-
\gamma+\frac{1}{3}\right], \\
F^{00}_2 &=&F^{33}_2=-\frac{N}{24\pi^2}\left[\frac{\rho^2 l^2}{2}
\zeta\left(2,\frac{\rho^2 l^2}{2}+1\right)+\frac{1}{\rho^2 l^2}\right], \\
F_2^{11} &=& F^{22}_2=\frac{N}{8\pi^2}\Bigg[\rho^4 l^4
\psi\left(\frac{\rho^2 l^2}{2}+1\right)-2\rho^2 l^2\ln\Gamma\left(
\frac{\rho^2 l^2}{2}\right)
- \rho^2 l^2 \ln\frac{\rho^2 l^2}{4\pi}-\rho^4 l^4-\rho^2 l^2+1\Bigg],
\label{28}
\end{eqnarray}
where $\psi(x)=d\left[\ln\Gamma(x)\right]/dx$. 

By making use of the kinetic term for the bosonic fields in the long-wavelength 
approximation (\ref{27}) and the explicit form of the effective potential (\ref{16}), 
we formally obtain the following spectrum for the $\sigma$ and $\pi$ excitations:
\begin{eqnarray}
E_{\pi} &=& \sqrt{v_{\pi,\perp}^2 \mathbf{k}_\perp^2 +k_3^2}, \\
E_{\sigma} &=& \sqrt{M^2_\sigma+v^2_{\sigma,\perp} \mathbf{k}_\perp^2+k_3^2}.
\end{eqnarray}
The $\pi$-mode is a gapless NG boson, associated with the spontaneous symmetry 
breaking of the chiral symmetry.
The general expressions for the two velocity parameters and the mass of the 
$\sigma$-mode are given by 
\begin{eqnarray}
v_{\pi,\perp}^2 &=&  \frac{F_1^{11}}{ F_1^{00}} , 
\label{v_pi_3+1}
\\
v_{\sigma,\perp} ^2&=&  \frac{F_1^{11}+2F_2^{11}}{ F_1^{00}+2F_2^{00}}, 
\label{v_sigma_3+1}\\
M_\sigma^2 &=&   \frac{V^{\prime\prime}(\bar{\sigma})}{N(F_1^{00}+2F_2^{00})}  .
\label{m_sigma_3+1}
\end{eqnarray}
Note that the longitudinal velocity parameters coincide with the speed of light, i.e.,
$v_{\pi,\parallel}=v_{\sigma,\parallel}=1$. This is consistent with the fact that the 
magnetic field in $3+1$ dimensions does not completely break the $\mathrm{SO}(1,3)$ Lorentz 
group of space-time transformations, but leaves the $\mathrm{SO}(1,1)$ subgroup of the Lorentz 
boosts in the direction of the field unbroken.

Let us start with the discussion of the dispersion relations of the composite $\pi$- and $\sigma$-modes 
in subcritical region, $g<g_c$. As in the case of the $(2+1)$-dimensional NJL model, it is instructive to
analyze the weak coupling and nearcritical regimes separately. 

By making use of the expansion in small $m_{\rm dyn} l$ at weak coupling, $g\ll g_c=1$,  we obtain
the following relations:
\begin{eqnarray}
E_{\pi}&\simeq&\left(\frac{(m_{\rm dyn} l)^2(\Lambda l)^2}{g}
\mathbf{k}^2_{\perp}+k_3^2\right)^{1/2},
\label{30}\\
E_{\sigma} &\simeq& \left(12m_{\rm dyn}^2+\frac{3(m_{\rm dyn} l)^2(\Lambda l)^2}{g}
\mathbf{k}^2_{\perp}+k_3^2\right)^{1/2}.
\label{29}
\end{eqnarray} 
By taking into account the explicit expression for the dynamical mass in Eq.~(\ref{m_dyn_weak_3+1}), we 
find that the transverse velocity $v_{\pi,\perp}$ of the gapless $\pi$-mode is less then 1.

Similarly, in the nearcritical region, we derive 
\begin{eqnarray}
E_{\pi}&\simeq&\left[\left(1-\frac{1}{\ln\pi/
\alpha_Y^{(l)}}\right)\mathbf{k}^2_{\perp}+k_3^2\right]^{1/2},
\label{32}\\
E_{\sigma}&\simeq&\left[4m_{\rm dyn}^2\left(1+\frac{2}{3\ln\pi/
\alpha_Y^{(l)}}\right)+\left(1-\frac{1}{3\ln\pi/\alpha_Y^{(l)}}\right)
\mathbf{k}^2_{\perp}+k_3^2\right]^{1/2},
\label{31}
\end{eqnarray}
where we used the dynamical mass in Eq.~(\ref{24}). We see that, as expected in the case of a 
renormalizable theory, the cutoff $\Lambda$ dropped out from observables $E_{\sigma}$ 
and $E_{\pi}$.

Now, we proceed to the supercritical regime at large coupling, $g>g_c=1$. In this case, the 
analytical expression for the dispersion relations of the composite bosonic fields can be 
easily derived in the weak magnetic field limit ($l\to\infty$). The results read
\begin{eqnarray}
E_{\pi}&\simeq&\left[\left(1-\frac{(eB)^2}{3m_{\rm dyn}^4\ln(\Lambda/
m_{\rm dyn})^2}\right)
\mathbf{k}^2_{\perp}+k_3^2\right]^{1/2},
\label{34} \\
E_{\sigma}&\simeq&\Bigg[4m_{\rm dyn}^2\left(1+\frac{1}{3\ln(\Lambda/
m_{\rm dyn})^2}+\frac{4(eB)^2}{9m_{\rm dyn}^4\ln(\Lambda/m_{\rm dyn})^2}\right)
+\left(1-\frac{11(eB)^2}{45m_{\rm dyn}^4\ln(\Lambda/m_{\rm dyn})^2}\right)
\mathbf{k}^2_{\perp}+k_3^2\Bigg]^{1/2},
\label{33}
\end{eqnarray}
with $m_{\rm dyn}$ is given in Eq.~(\ref{25}). One can see that the magnetic field reduces
the transverse velocity $v_{\pi,\perp}$ of $\pi$ in this case as well. In the nearcritical region 
with $g-g_c\ll 1$, the last expression can be rewritten in terms of the renormalized Yukawa 
coupling $\alpha_Y^{(m_{\rm dyn})}$, related to the scale $\mu=m_{\rm dyn}$. Indeed,
by taking into account that the corresponding renormalized coupling is given by
$\alpha_Y^{(m_{\rm dyn})}\simeq\pi/\ln(\Lambda/m_{\rm dyn})^2$ to leading order
in $1/N$, we find the following dispersion relations in the nearcritical region with $g-g_c\ll 1$:
\begin{eqnarray}
E_{\pi}&\simeq&\left[\left(1-\alpha_Y^{(m_{\rm dyn})}\frac{(eB)^2}{3\pi m_{\rm dyn}^4}\right)
\mathbf{k}^2_{\perp}+k_3^2\right]^{1/2},
\\
E_{\sigma}&\simeq&\Bigg[4m_{\rm dyn}^2\left(1+\alpha_Y^{(m_{\rm dyn})}\left(
\frac{1}{3\pi}+\frac{4(eB)^2}{9\pi m_{\rm dyn}^4}\right)\right)
+\left(1-\alpha_Y^{(m_{\rm dyn})}\frac{11(eB)^2}{45\pi m_{\rm dyn}^4}\right)
\mathbf{k}^2_{\perp}+k_3^2\Bigg]^{1/2}.
\end{eqnarray}

The present model illustrates the general phenomenon in $3+1$ dimensions.
In the infrared region, an external magnetic field reduces the dynamics of fermion
pairing to $(1+1)$-dimensional dynamics (in the lowest Landau level) thus
generating a dynamical mass for fermions even at the weakest attractive
interactions. A concrete realization of dynamical symmetry breaking is of course
different in different models. The detailed analysis of magnetic catalysis in 
$(3 + 1)$-dimensional gauge theories, both quantum electrodynamics and quantum 
chromodynamics, will be considered in Section~\ref{sec:MagCatGauge}. The applications
of the magnetic catalysis for the description of the dynamics in Dirac/Weyl semimetals
will be discussed in Section~\ref{sec:Dirac-Weyl-semimetals}.

\subsubsection{Thermodynamic properties}
\label{sec:NJL3+1Thermo}

Let us now briefly discuss the thermodynamic properties of the $(3+1)$-dimensional 
NJL model in a magnetic field. As expected, just like in $2+1$ dimensions, the 
symmetry will be restored at sufficiently high temperature and/or sufficiently large 
chemical potential.

The leading order approximation for the thermodynamic (effective) potential in the 
$(3+1)$-dimensional NJL model in the $1/N$ expansion can be easily derived in the 
general case of nonzero temperature $T$ and chemical potential $\mu$ \cite{Elmfors:1993wj,
Elmfors:1993bm,Elmfors:1994mq}, see also Appendix~\ref{App:ThermoPot}. The 
corresponding explicit expression reads 
\begin{equation}
V^{(3+1)}_{\beta,\mu}(\rho) = V^{(3+1)}(\rho)
-\frac{N}{2\beta \pi^2l^2}\int_{0}^{\infty}dk_3
\left\{\ln\left[1+e^{-\beta\left(\sqrt{\rho^2+k_3^2}-\mu\right)}\right]
+2\sum_{n=1}^{\infty}
\ln\left[1+e^{-\beta\left(\sqrt{\rho^2+k_3^2+2n/l^2}-\mu\right)}\right]
+\left(\mu\to-\mu\right)
\right\}.
\label{Veff-T-mu_3+1}
\end{equation}
where $\beta=1/T$ and $V^{(3+1)}(\rho)$ is the effective potential at $T=\mu=0$ given in Eq.~(\ref{16}).
The thermodynamic potential in Eq.~(\ref{Veff-T-mu_3+1}) can be studied in the most general 
case by using numerical methods. This is not the goal here. Instead, we would like to show only 
that the main effect of both temperature and chemical potential is to help restoring the chiral
symmetry in the NJL model. 

\begin{figure}[t]
 \includegraphics[width=0.47\textwidth]{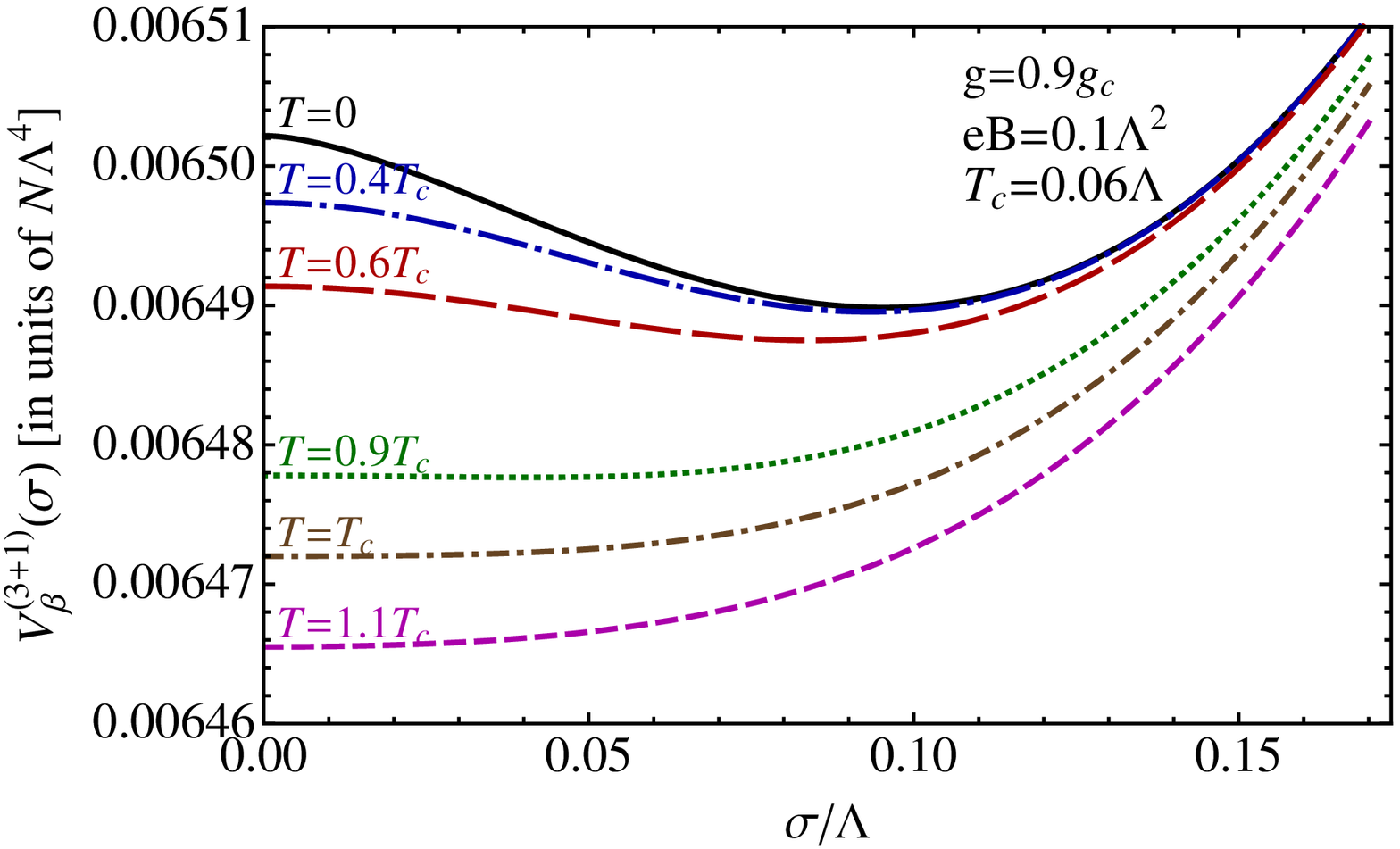}\hspace{0.05\textwidth}
\includegraphics[width=0.47\textwidth]{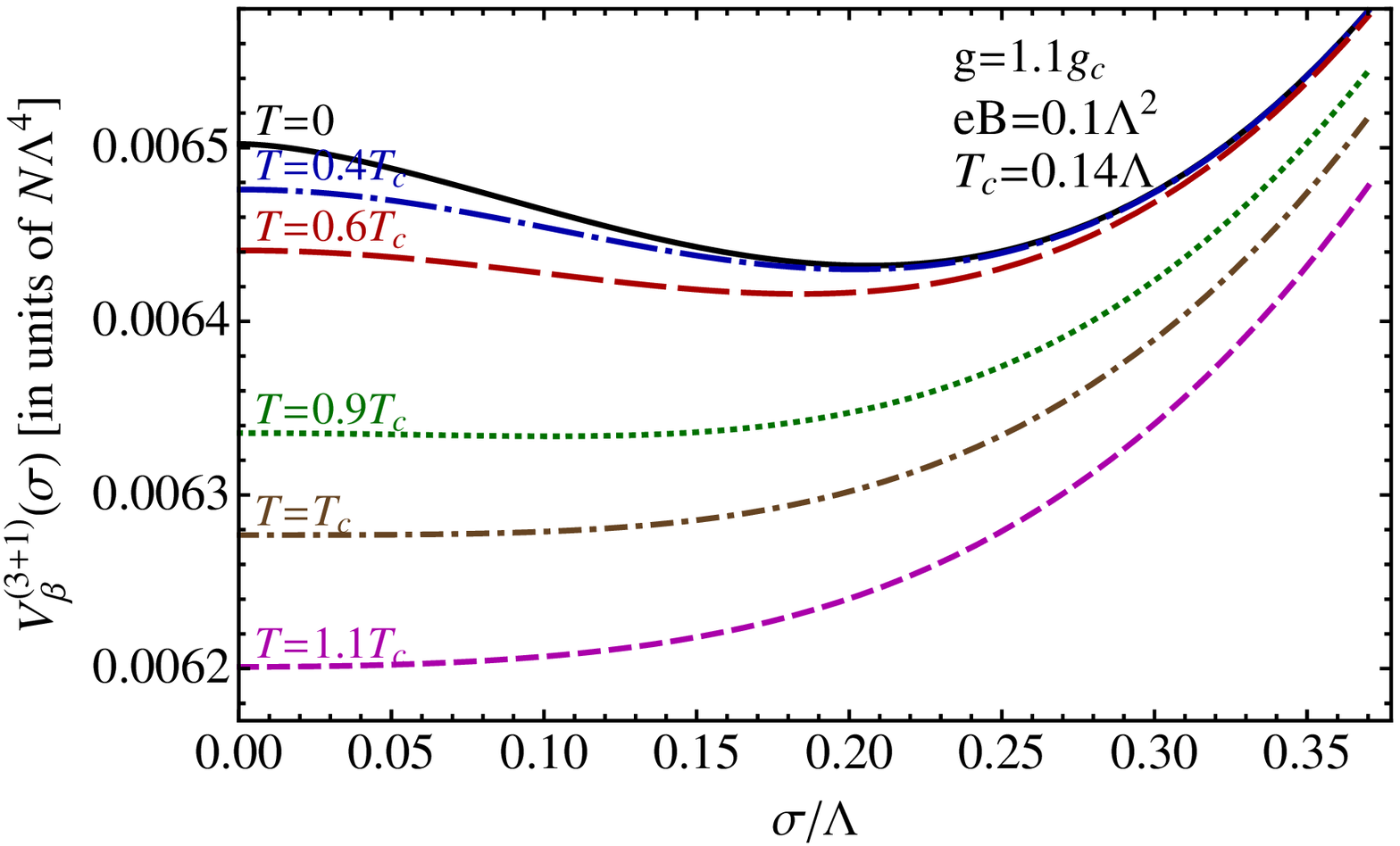}
\caption{(Color online) The thermodynamic potential $V_\beta(\sigma)$ in $3+1$ dimensions 
for several values of temperature and two fixed values of the coupling constant: 
$g=0.9g_c$ (left panel) and 
$g=1.1g_c$ (right panel).}
\label{fig-effVvsSigmaT3+1}
\end{figure}

First let us concentrate on the temperature effects and set $\mu=0$. The representative 
behavior of the thermodynamic potential at several fixed values of temperature are shown in 
Fig.~\ref{fig-effVvsSigmaT3+1}. The corresponding numerical results were calculated for a 
moderately strong magnetic field, $|eB|=0.1\Lambda^2$, and two fixed values of the coupling 
constant: $g=0.9g_c$ in the left panel and $g=1.1g_c$ in the right panel. We see that the 
value of the dynamical mass $m_{\rm dyn}(T) =\bar{\sigma}(T)$, where $\bar{\sigma}(T)$
is the location of the minimum of the potential, decreases with increasing temperature and 
vanishes altogether for $T\geq T_c$. While the chiral symmetry is broken for $T<T_c$, it
is restored for $T\geq T_c$. By noting that $m_{\rm dyn}(T)$ is a continuous function at 
$T=T_c$, the corresponding phase transition is a second order type. The value of the critical temperature $T_c$ 
depends on the coupling constant $g$. For the two cases shown in Fig.~\ref{fig-effVvsSigmaT3+1}, 
we obtain $T_c\approx 0.06 \Lambda$ (at $g=0.9g_c$) and $T_c\approx 0.14 \Lambda$ 
(at $g=1.1g_c$). It is not surprising that the critical temperature $T_c$ grows with increasing
coupling.  

\begin{figure}[t]
 \includegraphics[width=0.47\textwidth]{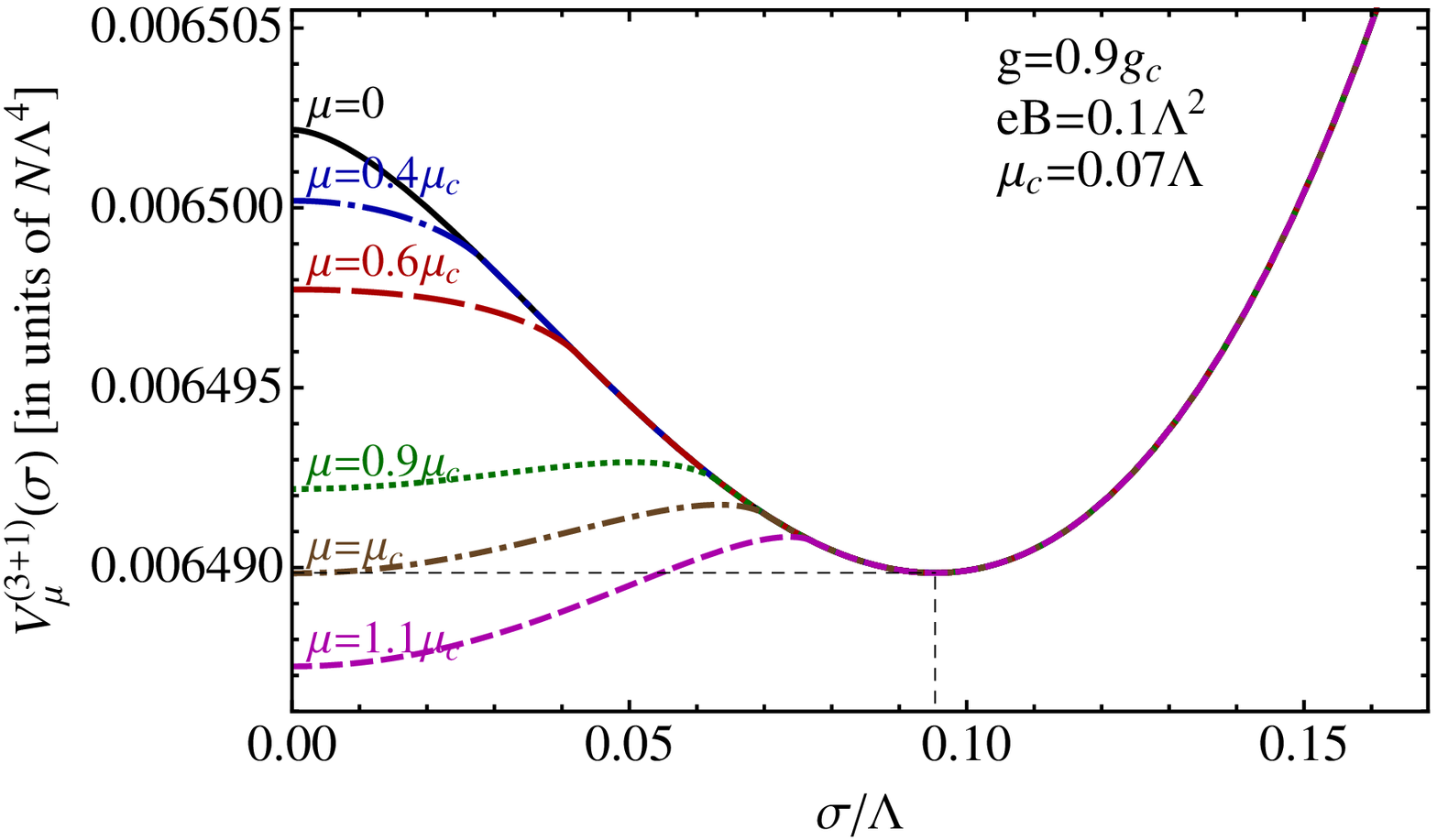}\hspace{0.05\textwidth}
\includegraphics[width=0.47\textwidth]{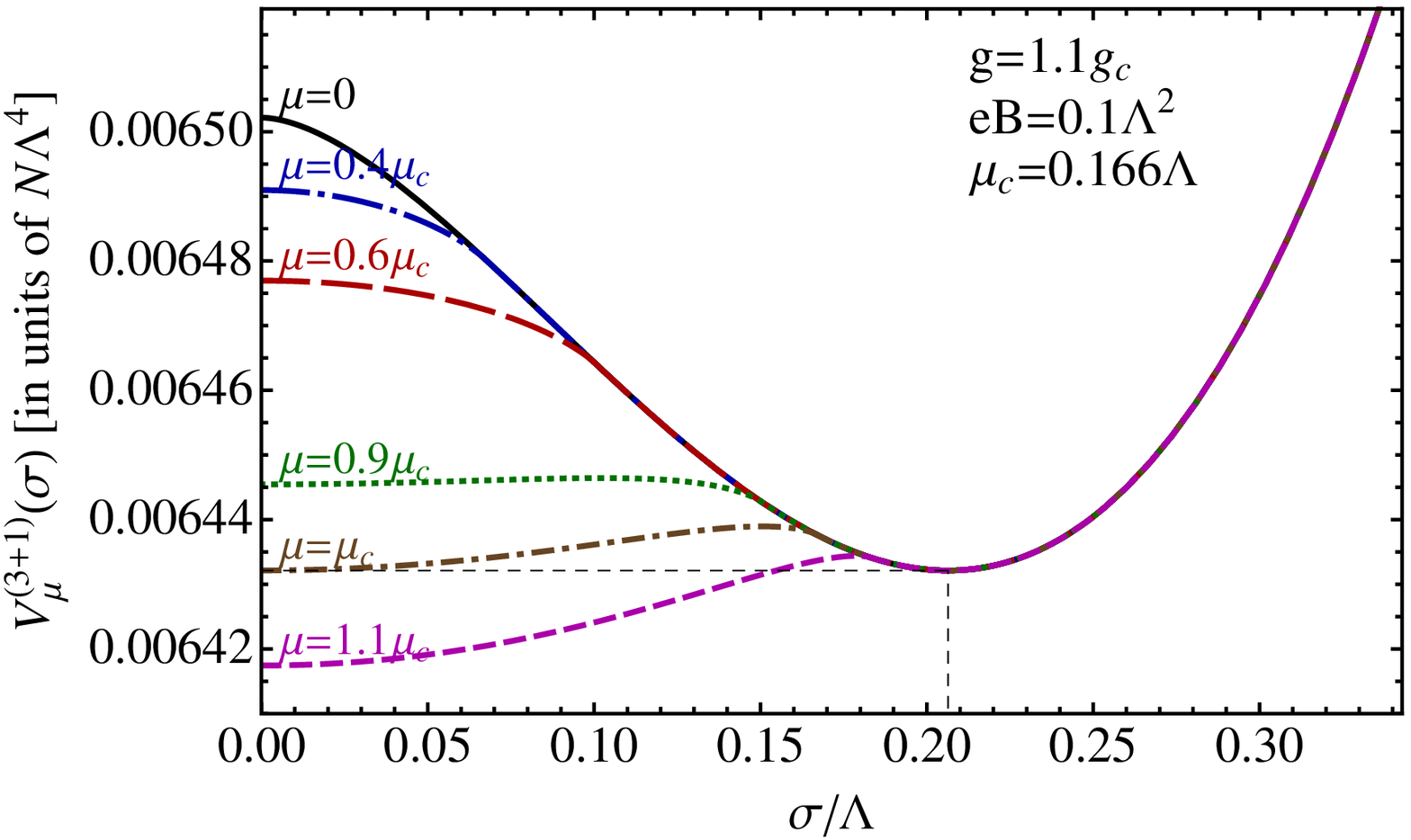}
\caption{(Color online) The thermodynamic potential $V_\mu(\sigma)$ in $3+1$ dimensions 
for several values of the chemical potential and two fixed values of the coupling constant: 
$g=0.9g_c$ (left panel) and 
$g=1.1g_c$ (right panel).}
\label{fig-effVvsSigmaMu3+1}
\end{figure}

At zero temperature (or $\beta\to \infty$), but nonzero chemical potential the effective 
potential takes the following simple form \cite{Persson:1994pz}:
\begin{eqnarray}
V^{(3+1)}_{\mu}(\rho) &=& V^{(3+1)}(\rho)-\frac{N}{4\pi^2l^2}
\left(|\mu|\sqrt{\mu^2-\rho^2}-\rho^2\ln\frac{|\mu|+\sqrt{\mu^2-\rho^2}}{\rho}\right)\theta\left(|\mu|-|\rho|\right)
\nonumber\\
&-&\frac{N}{2\pi^2l^2}\sum^\infty_{n=1}\left(
|\mu|\sqrt{\mu^2-\rho^2-2n/l^2}-\left(\rho^2+2n/l^2\right)
\ln\frac{|\mu|+\sqrt{\mu^2-\rho^2-2n/l^2}}{\sqrt{\rho+2n/l^2}}\right)\theta\left(|\mu|-\sqrt{\rho+2n/l^2}\right).
\nonumber\\
\end{eqnarray}
The latter was studied in detail in Refs.~\cite{Ebert:1999ht,Ebert:2003yk,Inagaki:2003yi,
Boomsma:2009yk,Allen:2013lda}. Here we will discuss only the characteristic 
behavior of the zero temperature effective potential $V^{(3+1)}_{\mu}(\rho)$ in order to support the
claim that the chemical potential tends to restore the chiral symmetry. The potential is shown in 
Fig.~\ref{fig-effVvsSigmaMu3+1} for a moderately strong magnetic field, $|eB|=0.1\Lambda^2$ 
and several values of the chemical potential. In the two plots, we show the numerical results for 
different values of the coupling constant: $g=0.9g_c$ in the left panel and $g=1.1g_c$ in the right 
panel. As expected at $T=0$, the chemical potential $\mu$ affects the behavior of the effective 
potential only in the region with $\sigma<|\mu|$. The chemical potential has no effect on the 
vacuum in the opposite case, $\sigma>|\mu|$, because $\mu$ lies in the energy gap (from 
$E=-\sigma$ to $E=\sigma$) that separates the occupied states in the Dirac see and the 
unoccupied positive energy states.  

Unlike the temperature, the chemical potential generically leads to a first order phase transition
at the critical point. Indeed, as we see from Fig.~\ref{fig-effVvsSigmaMu3+1}, the dynamical 
mass changes discontinuously from $m_{\rm dyn}(\mu_c-0) \neq 0$ to $m_{\rm dyn}(\mu_c+0) = 0$ 
at $\mu=\mu_c$. The value of the critical chemical potential $\mu_c$ is a function of the coupling 
constant $g$. While $\mu_c$ has a general tendency to grow with $g$, the actual dependence 
for small variations of coupling may be more complicated because of an oscillatory behavior 
of the potential, especially at not very large values of the magnetic field. For the two cases 
shown in Fig.~\ref{fig-effVvsSigmaMu3+1}, we find that $\mu_c\approx 0.07 \Lambda$ 
(at $g=0.9g_c$) and $\mu_c\approx 0.166 \Lambda$ (at $g=1.1g_c$). 

\subsection{Bethe-Salpeter equation for Nambu-Goldstone bosons in NJL model}
\label{sec:BS-NJL}

In this subsection, we look at the dynamics of magnetic catalysis in the NJL from 
a slightly different perspective. We study the problem of bound states with the quantum 
numbers of the Nambu--Goldstone (NG) bosons, using the method of a homogeneous 
Bethe-Salpeter equation (for a review, see Ref.~\cite{Miransky:1994vk}). Because of 
the Goldstone theorem \cite{Goldstone:1961eq,Goldstone:1962es,Nambu:1960xd}, 
spontaneous breaking of a continuous global symmetry causes the appearance 
of gapless NG bosons in the low-energy spectrum of the theory. The total number of 
such bosons and their quantum numbers are determined by the broken symmetry 
generators.

Let us start our analysis in $3 + 1$ dimensions. We will assume that the fermions
are weakly interacting and use the ladder approximation, which will be justified in 
this case. The explicit form of the homogeneous Bethe-Salpeter equation for the 
NG bound state $\pi$ is given by \cite{Miransky:1994vk}:
\begin{equation}
\chi_{AB}(u,u^\prime;P) = -i\int d^4u_1d^4u^\prime_1d^4u_2d^4u^\prime_2G_{AA_1}(u,u_1)
K_{A_1B_1;A_2B_2}(u_1u^\prime_1,u_2u^\prime_2) 
\chi_{A_2B_2}(u_2,u^\prime_2;P)G_{B_1B}(u^\prime_1,u^\prime)~, \label{11.6}
\end{equation}
where the Bethe-Salpeter wave function $\chi_{AB}(u,u^\prime;P)=
\langle 0|T\psi_A(u)\bar\psi_B(u^\prime)|P;\pi\rangle$ and the fer\-mi\-on
propagator $G_{AB}(u,u^\prime)=\langle 0|T\psi_A(u)\bar\psi_B(u^\prime)|0\rangle$;
the indices $A=(n\alpha)$ and $B=(m\beta)$ include both Dirac $(n,m)$
and color $(\alpha,\beta)$ indices.  Notice that despite the presence of an
external field $A_\mu$ breaking the conventional translation invariance, the total momentum 
$P$ is a good, conserved, quantum number for neutral bound states such as the gapless 
$\pi$ mode (see Section~\ref{sec:MagneticTranslations}). 
In this problem, it will be convenient to use the symmetric gauge for an external
vector potential, 
\begin{equation}
A^k =  -\frac{B}{2} y \delta^{k}_{1} +\frac{B}{2} x \delta^{k}_{2}
\qquad \mbox{(symmetric gauge)}.
\label{eq:symm_gauge}
\end{equation}
To leading order in $1/N$, the Bethe-Salpeter kernel in the NJL model is given by
\cite{Miransky:1994vk}:
\begin{eqnarray}
K_{A_1B_1;A_2B_2}(u_1u^\prime_1,u_2u^\prime_2) &=& G\Big[
\delta_{A_1B_1}\delta_{B_2A_2}+\delta_{\alpha_1\beta_1}
\delta_{\beta_2\alpha_2}(i\gamma_5)_{n_1m_1}
(i\gamma_5)_{m_2n_2} 
- \delta_{A_1A_2}\delta_{B_2B_1}-\delta_{\alpha_1\alpha_2}
\delta_{\beta_2\beta_1}(i\gamma_5)_{n_1n_2}
(i\gamma_5)_{m_2m_1}\Big] \nonumber\\
&\times& \delta^4(u_1-u^\prime_1)\delta^4(u_1-u_2)
\delta^4(u_1-u^\prime_2) ~. \label{11.8}
\end{eqnarray}
Also, in the ladder approximation,
the full-fermion propagator coincides with the propagator $S(u,u^\prime)$
of a free fermion (with $m=m_{\rm dyn}$) in a magnetic field.

Let us now factorize the Schwinger phase factor in the Bethe-Salpeter wave function:
\begin{equation}
\chi_{AB}(R,r;P) = \delta_{\alpha\beta} \exp
\left[ -  i r^\mu eA_\mu (R)\right] e^{-iPR} \tilde\chi_{nm}(R,r;P)~,
\label{11.8a}
\end{equation}
where we introduced the relative coordinate, $r=u-u^\prime$, and the center
of mass coordinate, $R=(u+u^\prime)/2$. Then, Eq.~(\ref{11.6})
can be rewritten as follows:
\begin{eqnarray}
\tilde\chi_{nm}(R,r;P)&=&-iNG \int d^4R_1
\bar{S}_{nn_1}\left(\frac{r}{2}+R-R_1\right)
\Bigg[\delta_{n_1m_1}~\mathrm{tr}\left[\tilde\chi(R_1,0;P)\right]
-(\gamma_5)_{n_1m_1}\mathrm{tr}\left[\gamma_5\tilde\chi(R_1,0;P)\right]
\nonumber\\
&&
-\frac{1}{N}\tilde\chi_{n_1m_1}(R_1,0;P)  
+\frac{1}{N}
(\gamma_5)_{n_1n_2}\tilde\chi_{n_2m_2}(R_1,0;P)
(\gamma_5)_{m_2m_1}\Bigg]  \nonumber\\
&&\times \bar{S}_{m_1m}\left(\frac{r}{2}-R+R_1\right)
\exp\left[ i r^\mu eA_\mu (R-R_1)\right]
\exp\left[iP(R-R_1)\right]~.   \label{11.8b}
\end{eqnarray}
Let us remind that the effect of translation symmetry breaking by the magnetic 
field in the system of charged fermions is captured by the Schwinger phase 
factor in Eq.~(\ref{11.8a}). The Bethe-Salpeter equation (\ref{11.8b}), however, 
admits a translation invariant solution: $\tilde\chi_{nm}(R,r;P)=\tilde\chi_{nm}(r,P)$. 
This feature is intimately connected with the fact that the corresponding bound 
states are neutral. By rewriting the equation in momentum space, we obtain
\begin{eqnarray}
\tilde\chi_{nm}(p;P) &=&-iNG \int
\frac{d^2q_{\perp}d^2R_{\perp}d^2k_{\perp}d^2k_{\parallel}}{(2\pi)^6}
\exp\left[i(\mathbf{P}_{\perp}-\mathbf{q}_{\perp})
\mathbf{r}_{\perp}\right] \bar{S}_{nn_1}
\left(p_{\parallel}+\frac{P_{\parallel}}{2},\mathbf{p}_{\perp}
- e \mathbf{A} (\mathbf{r}_{\perp}) +
\frac{\mathbf{q}_{\perp}}{2}\right) \nonumber\\
&\times& \left[\delta_{n_1m_1}~\mathrm{tr}\left[\tilde\chi(k;P)\right]-
(\gamma_5)_{n_1m_1}~\mathrm{tr}\left[\gamma_5\tilde\chi(x;P)\right]-
\frac{1}{N} \tilde\chi_{n_1m_1}(k;P) 
+ \frac{1}{N}(\gamma_5)_{n_1n_2}\tilde\chi_{n_2m_2}(k;P)
(\gamma_5)_{m_2m_1}\right] \nonumber\\
&\times &\bar{S}_{m_1m}\left(p_{\parallel}-\frac{P_{\parallel}}{2},
\mathbf{p}_{\perp}
- e\mathbf{A} (\mathbf{r}_{\perp})-\frac{\mathbf{q}_{\perp}}{2}\right)~,
\label{11.9}
\end{eqnarray}
where $p_{\parallel} = (p^0,p^3)$ and $\mathbf{p}_{\perp} = (p^1,p^2)$.  Henceforth, 
we will consider the equation with the total momentum $P_\mu \to 0$, corresponding 
to gapless NG bosons.

We will consider the case of weakly interacting fermions, when the LLL approximation 
for the fermion propagator is justified. Henceforth, for concreteness, we will consider 
the case $eB < 0$ (taking into account that the electron charge is negative,
$e<0$). 
Then,
\begin{equation}
\bar{S}(p) \simeq i \exp\left(- l^2 \mathbf{p}^2_{\perp}\right)
\frac{\hat p_{\parallel}+m_{\rm dyn}}{p_{\parallel}^2-
m^2_{\rm dyn}}\left(1 - i\gamma^1\gamma^2\right) 
\label{tildeSLLL}
\end{equation}
(see Section~\ref{sec:MagCat3+1General}), where $\hat p_{\parallel}=p^0\gamma^0-p^3\gamma^3$,
and Eq.~(\ref{11.9}) transforms into the following one:
\begin{equation}
\rho(p_{\parallel},\mathbf{p}_{\perp}) = \frac{iNG l^2}{(2\pi)^5}
e^{- l^2\mathbf{p}^2_{\perp}} \int d^2A_{\perp}d^2k_{\perp}d^2k_{\parallel}
e^{- l^2\mathbf{A}^2_{\perp}} \left(1 - i\gamma^1\gamma^2\right) 
\hat F\Big[\rho(k_{\parallel},\mathbf{k}_{\perp})\Big]
\left(1 - i\gamma^1\gamma^2\right) ~, 
\label{BSeqLLL}
\end{equation}
where
\begin{equation}
\rho(p_{\parallel},\mathbf{p}_{\perp})=
(\hat p_{\parallel}-m_{\rm dyn})\tilde\chi(p_{\parallel},
\mathbf{p}_{\perp})(\hat p_{\parallel}-m_{\rm dyn}) \label{phoLLL}
\end{equation}
with $\chi(p_{\parallel},\mathbf{p}_{\perp}) \equiv \left.
\chi(p_{\parallel},\mathbf{p}_{\perp};P)\right|_{P=0}$,
and the operator symbol $\hat F[\rho]$ means:
\begin{eqnarray}
\hat F\Big[\rho(k_{\parallel},\mathbf{k}_{\perp})\Big]&=&
\mathrm{tr}\Bigg(\frac{\hat k_{\parallel}+m_{\rm dyn}}{
k^2_{\parallel}-m^2_{\rm dyn}}\rho(k_{\parallel},\mathbf{k}_{\perp})
\frac{\hat k_{\parallel}+m_{\rm dyn}}{k^2_{\parallel}-m^2_{\rm dyn}}\Bigg)
-\gamma_5 \mathrm{tr}\Bigg(\gamma_5\frac{\hat k_{\parallel}+m_{\rm dyn}}{
k^2_{\parallel}-m_{\rm dyn}} \rho(k_{\parallel},\mathbf{k}_{\perp})
\frac{\hat k_{\parallel}+m_{\rm dyn}}{k^2_{\parallel}-m^2_{\rm dyn}}\Bigg)
\nonumber\\
&-&\frac{1}{N} \frac{\hat k_{\parallel}+m_{\rm dyn}}{k^2_{\parallel}-
m^2_{\rm dyn}}
\rho(k_{\parallel},\mathbf{k}_{\perp})
\frac{\hat k_{\parallel}+m_{\rm dyn}}{k^2_{\parallel}-m^2_{\rm dyn}}
+\frac{1}{N} \gamma_5 \frac{\hat k_{\parallel}+m_{\rm dyn}}{k^2_{\parallel}
-m^2_{\rm dyn}}\rho(k_{\parallel},\mathbf{k}_{\perp})
\frac{\hat k_{\parallel}+m_{\rm dyn}}{k^2_{\parallel}-m^2_{\rm dyn}}
\gamma_5~.
\label{hatFopertaor}
\end{eqnarray}
By making use of the representation in Eq.~(\ref{phoLLL}), we find that the 
dependence on the parallel and perpendicular components of the momenta 
is factorized, i.e., 
$\rho(p_{\parallel},\mathbf{p}_{\perp})=\exp(-l^2 \mathbf{p}^2_{\perp})\varphi(p_{\parallel})$, 
where $\varphi(p_{\parallel})$ satisfies the equation:
\begin{equation}
\varphi(p_{\parallel})=\frac{iNG}{32\pi^3l^2} \int
d^2k_{\parallel}\left(1 - i\gamma^1\gamma^2\right) 
\hat F\Big[\varphi(k_{\parallel})\Big]
\left(1 -i\gamma^1\gamma^2\right) ~. \label{phiEqLLL}
\end{equation}
Thus, the Bethe-Salpeter equation has been reduced to a two-dimensional
integral equation.  Of course, this fact reflects the two-dimensional
character of the dynamics of the LLL, that can be explicitly read off
from Eq.~(\ref{tildeSLLL}).

Henceforth, we will use Euclidean space with $k_4=-ik^0$. In order to define
the matrix structure of the wavefunction $\varphi(p_{\parallel})$ of the gapless $\pi$ mode, 
note that, in the magnetic field described by the symmetric gauge (\ref{eq:symm_gauge}),
there remains an unbroken symmetry $\mathrm{SO}(2)\times \mathrm{SO}(2)\times {\cal{P}}$, 
where the $\mathrm{SO}(2)\times \mathrm{SO}(2)$ is connected with rotations in the $xy$ and 
$z x_4$ planes ($x_4=it$) and ${\cal{P}}$ is the inversion transformation $z\to -z$, 
under which a fermion field transforms as $\psi\to i\gamma_5 \gamma_3\psi$.  
This symmetry implies that the function $\varphi(p_{\parallel})$ takes the form:
\begin{equation}
\varphi(p_{\parallel})=\gamma_5(A+i\gamma_1\gamma_2
B+\hat p_{\parallel}C+i\gamma_1\gamma_2 \hat p_{\parallel} D)
\label{phiStructureLLL}
\end{equation}
where $\hat p_{\parallel}=p_3\gamma_3+p_4\gamma_4$ and $A$, $B$, $C$
and $D$ are functions of $p^2_{\parallel}$ (recall that $\gamma_\mu$ are
antihermitian in Euclidean space).  Substituting expansion (\ref{phiStructureLLL})
into Eq.~(\ref{phiEqLLL}), we find that $B=-A$, $C=D=0$,  i.e.,
$\varphi(p_{\parallel})=A\gamma_5\left(1- i\gamma_1\gamma_2\right)$.
The function $A$ satisfies the equation
\begin{equation}
A(p)=\frac{NG}{4\pi^3l^2} \int d^2k
\frac{A(k)}{k^2+m^2_{\rm dyn}}~. 
\label{integralEqA}
\end{equation}
The solution to this equation is $A(p)=\mbox{const}$. Thus, by introducing an 
ultraviolet cutoff $\Lambda$, we obtain the gap equation for the dynamical mass
\begin{equation}
1=\frac{NG}{4\pi^2l^2} \int^{\Lambda^2}_0
\frac{d(k^2)}{k^2+m^2_{\rm dyn}}~. \label{Eq:A=const}
\end{equation}
which leads to the following solution:
\begin{equation}
m^2_{\rm dyn}= \Lambda^2 \exp\Bigg(-\frac{4\pi^2}{NG |eB|}\Bigg) ~ \qquad (D=3+1) .
\label{BSgapMdyn3+1}
\end{equation}
We check that this result agrees with the leading asymptote for the dynamical mass in the weak 
coupling limit derived in Section~\ref{sec:MagCat3+1General}, see Eq.~(\ref{m_dyn_weak_3+1}).
Because of using a less rigorous LLL approximation, however, the subheading pre-exponential 
factor in Eq.~(\ref{BSgapMdyn3+1}) differs from that in Eq.~(\ref{m_dyn_weak_3+1}). 

Let us now show that integral equation (\ref{integralEqA}) can be also rewritten in the 
form of a Schr\"odinger equation. By introducing the wave function
\begin{equation}
\Psi(\mathbf{r})=\int \frac{d^2k}{(2\pi)^2}
\frac{e^{-i\mathbf{k} \cdot\mathbf{r}}}{k^2+m^2_{\rm dyn}} A(k)
\label{Psi-Integral-A}
\end{equation}
and making use of Eq.~(\ref{integralEqA}), we find that $\Psi(\mathbf{r})$ indeed 
satisfies a Schr\"odinger equation in a $2$-dimensional Euclidean space,
\begin{equation}
\left(-\Delta_\mathbf{r}+m^2_{\rm dyn} + V(\mathbf{r})\right) \Psi(\mathbf{r})=0 ,
\label{SchrodingerEq}
\end{equation}
with the potential given by
\begin{equation}
V(\mathbf{r}) = -\frac{NG}{\pi l^2} \delta^2_\Lambda(\mathbf{r})  ,
\label{SchrodingerEqDeltaPot}
\end{equation}
where 
\begin{equation}
\delta^2_\Lambda(\mathbf{r})=\int_\Lambda
\frac{d^2k}{(2\pi)^2} e^{-i\mathbf{k} \cdot\mathbf{r}} 
\label{RegDeltaFunction}
\end{equation}
is a regularized representation of the $\delta$-function. In the Schr\"odinger equation 
(\ref{SchrodingerEq}), $\Delta_\mathbf{r}$ is the Laplace operator in the $2$-dimensional 
Euclidean space, 
\begin{equation}
\Delta_\mathbf{r} = \frac{\partial^2}{\partial x_3^2}+\frac{\partial^2}{\partial x_4^2} , \quad\quad (D=3+1) .
\label{Laplacian}
\end{equation}
where the two coordinates of the Euclidean space are the spatial longitudinal direction 
$x_3=z$ and the imaginary time $x_4=it$. 

By repeating the same derivations in the case of $(2+1)$-dimensional NJL model, 
one finds that the gap equation for the dynamical mass reduces down to 
\begin{equation}
A(p)=\frac{NG}{2\pi^2 l^2} \int^\Lambda_{-\Lambda}dk
\frac{A(k)}{k^2+m^2_{\rm dyn}}~, \label{11.22}
\end{equation}
[compare with Eq.~(\ref{integralEqA})]. In the limit $\Lambda\to\infty$, this yields exactly 
the same expression for $m^2_{\rm dyn}$ as the result at weak coupling in 
Eq.~(\ref{BSgapMdyn2+1}), i.e.,
\begin{equation}
m^2_{\rm dyn}= \frac{N^2G^2|eB|^2}{4\pi^2}~, \qquad (D=2+1) .
\label{BSgapMdyn2+1}
\end{equation}
In $2+1$ dimensions, we can also introduce a wave function analogous to that in Eq.~(\ref{Psi-Integral-A}).
It will satisfy a $1$-dimensional Schr\"odinger equation with the potential
\begin{equation}
V(x_3)= - \frac{NG}{\pi^2} \delta_\Lambda(x_3),
\label{11.21}
\end{equation}
where $\delta_\Lambda(x_3)=\int^\Lambda_{-\Lambda} \frac{dk}{2\pi} e^{-ik x_3}$ is a regularized 
form of the $\delta$-function and $x_3=it$ is the imaginary time coordinate. 

In general, as we see from the NJL model studies, the dimensional reduction $D\to D-2$ 
plays a crucial role in the structure of the homogeneous Bethe-Salpeter equation. The latter  
reduces to a Schr\"odinger equation in a $(D-2)$-dimensional Euclidean space. As will become 
clear in Section~\ref{sec:MagCatGauge}, this is a general feature. Even in gauge theories 
at weak coupling (see Section~\ref{sec:MagCatQED}), when the ladder approximation for 
the Bethe-Salpeter equation is justified, the infrared dynamics of magnetic catalysis reduces 
to a Schr\"odinger problem in $D-2$ dimensions, but with a different attractive potential 
$V(\mathbf{r})$. Unlike in the NJL model, the corresponding potentials in gauge theories 
are generically long-ranged.  

By noting that $-m^2_{\rm dyn}$ plays the role of the energy $E$ in the Schr\"odinger equation 
(\ref{SchrodingerEq}), the problem of dynamical mass generation is reduced to finding the 
spectrum of bound states (with $E=-m^2_{\rm dyn}<0$) in the Schr\"odinger problem with 
an attractive potential $V(\mathbf{r})$. Moreover, by taking into account that only the 
solution with the largest possible value of $m^2_{\rm dyn}$ defines the stable vacuum 
\cite{Miransky:1994vk}, the problem is reduced to finding the lowest energy eigenvalue. 
In this connection, it is useful to mention some known results regarding the properties 
of the Schr\"odinger equation in low-dimensional spaces \cite{Simon:1976un,Simon:1976cx}. 
For our purposes, the most important is the following property: the Schr\"odinger equation 
with an attractive potential in both one-dimensional $(d=1)$ and two-dimensional $(d=2)$ 
spaces always has at least one bound state with a negative energy. The energy of the 
lowest energy bound state $E$ has the form
\begin{equation}
E(G)=-m^2_{\rm dyn}(G)=-|eB|f(G)~, 
\label{Mdyn_vs_weakG}
\end{equation}
where $G$ is a coupling constant. In addition, while the function $f(G)$ is an analytic function 
of the coupling constant at $G\to 0$ for $d=1$, it is non-analytic at $G \to 0$ for $d=2$ 
\cite{Simon:1976un}. 

We see that the general result in Eq.~(\ref{Mdyn_vs_weakG}) agrees with our results 
for the dynamical masses in the $(2+1)$- and $(3+1)$-dimensional NJL models.
More importantly, however, the knowledge of general properties of the Schr\"odinger 
equation provides an additional model-independent insight into the dynamics of chiral 
symmetry breaking in the system of charged fermions in a magnetic field. For example, 
we can now claim that, at $B\neq 0$, the dynamical symmetry breaking will be broken 
in both $D=3+1$ and $D=2+1$ as soon as there is an attractive interaction in the 
particle-antiparticle channel. This statement is a direct consequence of the fact 
that there is always at least one bound state for the one-dimensional ($d=1$) and 
two-dimensional ($d=2$) Schr\"odinger equation with an attractive potential 
\cite{Simon:1976un}. In addition, by taking into account that the energy $E(G)$ 
of the lowest energy bound state is an analytic function of the coupling constant 
at $G\to 0$ for $d=1$ and a non-analytic function for $d=2$ \cite{Simon:1976un},
we can make the same rather general statement about the dependence of the dynamical 
mass squared, $m^2_{\rm dyn}(G)$, on the coupling strength, see Eq.~(\ref{Mdyn_vs_weakG}). 
In the special case of short-range potentials, one can even give the form of the functional
dependence on $G$ as $G\to 0$. In the case of a $(3+1)$-dimensional model (i.e., $d=2$), 
for example, the energy at weak coupling has the form $E(G)\equiv -m_{\rm dyn}^2(G) \propto
-\exp\left(-a/G\right)$, where $a$ is a positive constant \cite{Simon:1976un}.

\subsection{The group of magnetic translations}
\label{sec:MagneticTranslations}

While discussing the magnetic catalysis, we argued that the effective dimensional 
reduction $D\to D-2$ in infrared dynamics plays a profound role in the fermion-antifermion 
pairing responsible for the chiral (flavor) symmetry breaking. In this context, it is 
appropriate to ask whether the dimensional reduction is consistent with spontaneous 
symmetry breaking. According to the Mermin-Wagner-Coleman theorem 
\cite{Mermin:1966fe,Coleman:1973ci}, no spontaneous breakdown of continuous
symmetries are possible in dimensions $d=1+1$ or less. 

In the problem at hand, however, the Mermin-Wagner-Coleman theorem is 
not applicable. The arguments of the theorem are based on the fact that gapless
NG bosons cannot exist in dimensions $d=1+1$ or less. In the problem of 
magnetic catalysis, the reduction $D\to D-2$ takes place only for charged particles,
but not for the NG bosons, which are neutral. In fact, unlike the propagator 
of fermions, the propagator of the NG bosons has a genuine $D$-dimensional 
form, i.e., there is no obstacle for the realization of spontaneous chiral symmetry
breaking in a magnetic field.

The arguments above appear to be intimately connected with the status of the 
spatial translation symmetry in a constant magnetic field $B$. At $B\neq 0$, 
instead of dealing with the usual translations, it is more convenient to introduce 
a generalized group of magnetic translations \cite{Zak:1964zz,1978AnPhy.114..431A}. 
Here we present the key details about such group.

The form of the generators of magnetic translations depends on the gauge choice 
for the background field. The algebra relations are gauge invariant. In order to show
this, we will use interchangeably both the Landau gauge 
(\ref{eq:Landau_gauge}) and the symmetric gauge (\ref{eq:symm_gauge}). 
In the Landau gauge (\ref{eq:Landau_gauge}), the usual translation symmetry 
along the $y$-direction (but not along the $x$-direction) appears to be broken. 
In the symmetric gauge (\ref{eq:symm_gauge}), on the other hand, the translations 
in both $x$ and $y$ directions are broken. 

The generators of the group of magnetic translations are introduced as follows:
\begin{eqnarray}
\hat P_{x}=-i\frac{\partial}{\partial x}~, \quad
\hat P_{y}=-i\frac{\partial}{\partial y}+\hat{Q} Bx~, \quad
\hat P_{z}=-i\frac{\partial}{\partial z} ,
\qquad \mbox{(Landau gauge)}
\label{mtrans1}
\end{eqnarray}
in the Landau gauge, and
\begin{eqnarray}
\hat P_{x}=-i\frac{\partial }{\partial x}-
\frac{\hat{Q}}{2} By~, \quad
\hat P_{y} = -i\frac{\partial}{\partial y}+
\frac{\hat{Q}}{2} Bx~, \quad
\hat P_{z}=-i\frac{\partial}{\partial z}  ,
\qquad \mbox{(symmetric gauge)}
\label{mtrans2}
\end{eqnarray}
in the symmetric gauge. Here $\hat{Q}$ is the charge operator.  One can easily
check that these operators commute with the Hamiltonian of the Dirac
equation in a constant magnetic field.  Also, we get the following commutation 
relations:
\begin{eqnarray}
\left[\hat P_{x},\hat P_{y}\right] =-i \hat{Q}B~, \qquad
\left[\hat P_{x},\hat P_{z}\right] =
\left[\hat P_{y},\hat P_{z}\right] = 0~. \label{algebra}
\end{eqnarray}
For charged states, the perpendicular momenta cannot be defined simultaneously.
If one chooses a basis of quantum states with well defined eigenvalues of the
operator $P_{x}$, they cannot be eigenvalues of $P_{y}$, and vice versa. This 
is exactly what we found by solving the Dirac equation in the Landau gauge 
(\ref{eq:Landau_gauge}).

Now, the situation is qualitatively different for the neutral states. That is because
the charge operator $\hat{Q}$ gives zero when acting on the corresponding states.
Therefore, all commutators of the magnetic translation group vanish, and
the momentum $\mathbf{P}=(P_x,P_y,P_z)$ can be used to describe the
dynamics of the center of mass of neutral states.

Note that, in $2+1$ dimensions, the magnetic translation group consists of only
$\hat P_{x}$ and $\hat P_{y}$. Because, these two generators were the only ones
with a nontrivial commutation relation, i.e., $\left[\hat P_{x},\hat P_{y}\right] =-i \hat{Q}B$, 
essentially the same arguments apply in $(2+1)$-dimensional models.

\subsection{General remarks on magnetic catalysis}
\label{sec:MagCatSummary}

The main conclusion of this section is that a constant magnetic field in $2 + 1$ and $3 + 1$
dimensions is a strong catalyst of dynamical symmetry breaking. It leads to the generation 
of a fermion mass (energy gap) even at the weakest attractive interaction between fermions 
and antifermions. While the ideas were supported by the analysis in the NJL model, it should
be clear that the underlying physics of the magnetic catalysis is universal and should take 
place also in other models with attractive interactions between fermions and antifermions.
(This will be also confirmed by the detailed analysis of magnetic catalysis in QED and QCD 
in Section~\ref{sec:MagCatGauge}.)

As we showed, the underlying reason for the model independent nature of the magnetic 
catalysis phenomenon is directly related to the dynamical reduction $D \to D - 2$ in
infrared fermion-antifermion pairing dynamics in a constant magnetic. Such a pairing 
is dominated by the lowest Landau level. 

Here we consider the dynamics in the presence of a constant magnetic field only. However, 
it may be also interesting to extend the analysis to the case of inhomogeneous electromagnetic 
fields. In connection with that, we note that the present effect in $2 + 1$ dimensions is 
intimately connected with the fact that the massless Dirac equation in a constant magnetic 
field admits an infinite number of normalized solutions with $E=0$ (zero modes). The 
condition for the existence of a nonzero chiral condensate in the case of massless 
fermions in a general external gauge configuration $A_{\mu}(x)$ was established long 
time ago \cite{Banks:1979yr}. It states that the spectral density $\rho(\lambda)$ of an 
Euclidean Dirac massless operator $D[A_{\mu}(x)]$ must be nonzero as $\lambda \to 0$ 
(the Banks-Casher criterion). The physics of this criterion is intimately connected with 
the existence of massless excitations in the spectrum of such an operator. (For a clear 
exposition of this and relating issues, see the review \cite{Smilga:2000ek}.) It would be 
interesting to find examples of inhomogeneous magnetic field configurations that satisfy 
this criterion.
   
The $(3 + 1)$-dimensional case is different. As was shown in Section~\ref{sec:MagCat3+1General}, 
the chiral condensate vanishes for free fermions in a homogeneous magnetic field for $D= 3 + 1$.
The reason is clear: there are no normalized zero modes at the LLL in that case, although at 
$m = 0$ there is a continuous spectrum of eigenvalues with the dispersion relation $E = \pm |k_z|$ 
coinciding with that for massless fermions in $1 + 1$ dimensions. The presence of an 
attractive fermion-antifermion interaction is now necessary for chiral symmetry breaking, 
although it could be even infinitesimally weak.

It may be interesting to ask whether there is magnetic catalysis in higher dimensions, $D = d + 1$ 
with $d > 3$. The NJL model in a magnetic field in higher dimensions was originally studied in 
Ref.~\cite{Gorbar:2000ku}. There it was shown that an analog of the magnetic catalysis indeed 
exists in constant magnetic field configurations with the maximal number $N = [d/2]$ of nonzero 
components of the strength tensor $F^{ab}$, where $a$, $b = 1,2,\ldots,d$ (here $[d/2]$ is the
integer part of $d/2$). We will briefly discuss this problem in Section~\ref{NJL4}.

\section{Magnetic catalysis in gauge theories}
\label{sec:MagCatGauge}

In this section, we will describe the magnetic catalysis effect in QED and QCD. 
These are the gauge theories that play an important role in the Standard Model. 
Because both of them include electromagnetically neutral gauge fields (the photon
and gluon fields, respectively) that mediate long-range interactions, the 
realization of the magnetic catalysis is subtler and more interesting than in 
the NJL model. We will start with the analysis in QED in Section~\ref{sec:MagCatQED}.
In Section~\ref{sec:MagCatQEDreduced} we will consider reduced QED in
a magnetic field \cite{Gorbar:2001qt,Alexandre:2001ps,Gorbar:2002iw}. Finally, we 
will discuss the realization of the magnetic catalysis and some subtleties of the 
low-energy dynamics in QCD in Section~\ref{sec:MagCatQCD}.

\subsection{Magnetic catalysis in massless QED}
\label{sec:MagCatQED}

The magnetic catalysis effect in QED was established in Ref.~\cite{Gusynin:1995gt} by analyzing 
the Bethe-Salpeter equation in a magnetic field in the ladder approximation. This result was 
confirmed in Ref.~\cite{Leung:1996qy}, where the Schwinger-Dyson equation in a magnetic 
field in the rainbow approximation was considered. Because of a smallness of 
the QED coupling in infrared, it looked as if these approximations are justified. However,
as was revealed in Ref.~\cite{Gusynin:1995nb}, this is not the case. Since the magnetic field is the
largest scale in this problem, there are rather large, of order $\alpha|eB|$, contributions in the photon polarization
operator leading to a strong screening of the interactions. In fact, this screening is
a magnetic analog of the Debye screening at nonzero electron density in metals. A consistent
description of the magnetic catalysis effect in QED was done in Refs.~\cite{Gusynin:1998zq,Gusynin:1999pq}.
In this section, following the same strategy as in Section~\ref{sec:MagCatSummary}, 
we will analyze the Bethe-Salpeter equation for the NG modes in QED in a magnetic field 
in the ladder and improved ladder approximations. 

The Lagrangian density of massless QED in a magnetic field is
\begin{equation}
{\cal{L}}=-\frac{1}{4}F^{\mu\nu}F_{\mu\nu}+\frac{1}{2}\left[\bar\psi,
(i\gamma^\mu D_\mu)\psi\right]~, \label{Sec3:Lagrangian_QED}
\end{equation}
where the covariant derivative $D_\mu$ is  
\begin{equation}
D_\mu=\partial_\mu + ie(A_\mu + a_\mu).
\end{equation}
We will use the symmetric gauge (\ref{eq:symm_gauge}) for the external field $A_\mu$. 
The quantum part of the gauge field is denoted by $a_\mu$. 
Besides the Dirac index ($n$), the fermion field carries an additional flavor
index $a=1,2,\ldots,N_f$.  The Lagrangian density (\ref{Sec3:Lagrangian_QED}) is
invariant under the chiral group $\mathrm{SU}(N_f)_{L}\times  \mathrm{SU}(N_f)_{R}\times
\mathrm{U}(1)_{L+R}$. (Here we will not discuss the dynamics related to the anomalous 
singlet axial current  $J_{5\mu}$.)
Since we consider the weak coupling phase of QED, there is no
spontaneous chiral symmetry breaking at $B=0$ \cite{Fomin:1984tv,Miransky:1985aq,Miransky:1994vk}.  We will show
that the magnetic field changes the situation dramatically: at $B\not= 0$
the chiral symmetry $\mathrm{SU}_{L}(N_f)\times \mathrm{SU}_{R}(N_f)$ breaks down to
$\mathrm{SU}_{V}(N_f) \equiv \mathrm{SU}_{L+R}(N_f)$.  As a result, the dynamical
mass $m_{\rm dyn}$ is generated, and $N_f^2-1$ gapless
bosons, composed of fermions and antifermions, appear.

The homogeneous Bethe-Salpeter equation for the $N_f^2-1$ NG bound states takes the
form \cite{Miransky:1994vk}:
\begin{equation}
\chi^\beta_{AB}(u,u^\prime;P) = -i \int d^4u_1 d^4u^\prime_1
d^4u_2d^4u^\prime_2 G_{AA_1}(u,u_1)
K_{A_1B_1;A_2B_2}(u_1u^\prime_1,u_2u^\prime_2)  \chi^\beta_{A_2B_2}(u_2,u^\prime_2;P)
G_{B_1B}(u^\prime_2,u^\prime)~,    \label{31.3}
\end{equation}
where the Bethe-Salpeter wave function $\chi^\beta_{AB}=
\langle0|T\psi_A(u)\bar\psi_B(u^\prime)|P;\beta\rangle$, with $\beta=1,\ldots,
N_f^2-1$, and the fermion propagator $G_{AB}(u,u^\prime)=
\langle0|T\psi_A(u)\bar\psi_B(u^\prime)|0\rangle$. The indices
$A=(na)$ and $B=(mb)$ include both Dirac $(n,m)$ and flavor
$(a,b)$ indices.

As will be shown below, the NG bosons are formed in the infrared region, where 
the QED coupling is weak. This seems to suggest that the Bethe-Salpeter kernel 
in leading order in $\alpha$ should provide a reliable approximation. 
However, because of the $(1+1)$-dimensional form of the fermion propagator 
in the infrared region, there may also be relevant higher order contributions. 
We will return to this problem in Section~\ref{sec:MagCatQCD}, but first, 
we will analyze the Bethe-Salpeter equation with the kernel in leading order 
in $\alpha$.

The Bethe-Salpeter kernel in leading order in $\alpha$ is \cite{Miransky:1994vk}:
\begin{eqnarray}
K_{A_1B_1;A_2B_2}(u_1u^\prime_1,u_2,u^\prime_2)&=&
-4\pi i\alpha \delta_{a_1a_2}\delta_{b_2b_1}\gamma^\mu_{n_1n_2}
\gamma^\nu_{m_2m_1}{\cal{D}}_{\mu\nu}(u^\prime_2-u_2) \delta(u_1-u_2)\delta(u^\prime_1-u^\prime_2)
\nonumber\\
&& +4\pi i\alpha\delta_{a_1b_1}
\delta_{b_2a_2}\gamma^\mu_{n_1m_1}\gamma^\nu_{m_2n_2}
{\cal{D}}_{\mu\nu}(u_1-u_2)   \delta(u_1-u^\prime_1)\delta(u_2-u^\prime_2)~,   
\label{31.4}
\end{eqnarray}
where ${\cal{D}}_{\mu\nu}(u-u^\prime)$ is the photon propagator.

\subsubsection{Magnetic catalysis in QED in the ladder approximation}
\label{QED-ladder}

In the ladder approximation, the photon propagator is given by
\begin{equation}
{\cal{D}}_{\mu\nu}(u-u^\prime)=\frac{-i}{(2\pi)^4} \int d^4k\, e^{ik(u-u^\prime)}
\left(g_{\mu\nu}-\lambda \frac{k_\mu k_\nu }{k^2}\right)
\frac{1}{k^2}                         
\label{photon_D_mu-nu}
\end{equation}
where $\lambda$ is a gauge parameter. 
The first (Fock) term on the
right-hand side of Eq.~(\ref{31.4}) corresponds to the ladder approximation.
The second (Hartree, or annihilation) term does not contribute to the Bethe-Salpeter
equation for NG bosons. This follows from the fact that, due
to the Ward identities for axial currents, the Bethe-Salpeter equation for 
NG bosons can be reduced to the Schwinger-Dyson equation for the fermion
propagator, where there is no contribution of the Hartree term \cite{Miransky:1994vk}.
For this reason, we will omit this term in the following. Then, the Bethe-Salpeter
equation takes the form:
\begin{equation}
\chi^\beta_{AB}(u,u^\prime;P) = -4\pi\alpha \int d^4u_1 d^4u^\prime_1
S_{AA_1}(u,u_1) \delta_{a_1a_2} \gamma^\mu_{n_1n_2}
\chi^\beta_{A_2B_2}(u_1,u^\prime_1;P) 
 \delta_{b_2b_1}\gamma^\nu_{m_2m_1}S_{B_1B}(u^\prime_1,u^\prime)
{\cal{D}}_{\mu\nu}(u^\prime_1-u_1)~, \label{31.6}
\end{equation}
where the full fermion propagator $G_{AB}(u,u^\prime)$ was replaced by the 
free fermion propagator $S(u,u^\prime)$ with the mass $m=m_{\rm dyn}$, see 
Eqs.~(\ref{eq:green}) and (\ref{eq:green-transl-inv}) in Appendix~\ref{App:A-SchwingerProp}.
This is consistent with the leading order in $\alpha$ (ladder) approximation assumed here.

Using the new variables, the center of mass coordinate $R=(u+u^\prime)/2$,
and the relative coordinate $r=u-u^\prime$, Eq.~(\ref{31.6}) can be rewritten as
\begin{eqnarray}
\tilde\chi_{nm}(R,r;P) &=& -4\pi\alpha \int d^4R_1d^4r_1
\bar{S}_{nn_1}\left(R-R_1+\frac{r-r_1}{2}\right)\gamma^\mu_{n_1n_2}\tilde\chi_{n_2m_2}(R_1,r_1;P)
\gamma^\nu_{m_2m_1}
\bar{S}_{m_1m}\left(\frac{r-r_1}{2}-R+R_1\right)\nonumber\\
&\times &
{\cal{D}}_{\mu\nu}(-r_1)\exp\left[ ie(r+r_1)^\mu A_\mu (R-R_1)\right]
\exp\left[iP(R-R_1)\right]~.  \label{31.7}
\end{eqnarray}
Here $\bar{S}(u-u^\prime)$ is the translation invariant part of the propagator and the
function $\tilde\chi_{nm}(R,r;P)$ is defined from the equation
\begin{equation}
\chi^\beta_{AB}(u,u^\prime;P) \equiv
\langle0|T\psi_A(x)\bar\psi_B(y)|P,\beta\rangle
=\lambda^\beta_{ab}e^{-iPR} \exp\left[ - ier^\mu A_\mu^
{\rm ext}(R)\right]\tilde\chi_{nm}(R,r;P) \label{31.8}
\end{equation}
where $\lambda^\beta$ are $N_f^2-1$ flavor matrices
[$\mathrm{tr}(\lambda^\beta \lambda^\gamma)=2\delta_{\beta\gamma}$,
with $\beta,\gamma \equiv 1,\ldots,N_f^2-1$].  The important fact is that,
like in the case of the Bethe-Salpeter equation for the gapless $\pi$ mode 
in the NJL model considered in Section~\ref{sec:BS-NJL}, the effect of translation
symmetry breaking by the magnetic field is factorized in the Schwinger phase 
factor in Eq.~(\ref{31.8}), and Eq.~(\ref{31.7}) admits a translation
invariant solution, $\tilde\chi_{nm}(R,r;P)=\tilde\chi(r;P)$.  Then,
transforming this equation into momentum space, we get
\begin{eqnarray}
\tilde\chi_{nm}(p;P)&=&-4\pi\alpha \int
\frac{d^2q_{\perp} d^2R_{\perp} d^2k_{\perp} d^2k_{\parallel}}{
(2\pi)^6}   \exp\left[i(\mathbf{P}_{\perp}-\mathbf{q}_{\perp})\mathbf{r}_{\perp}\right]
\bar{S}_{nn_1}\left(p_{\parallel}+\frac{P_{\parallel}}{2},\mathbf{p}_{\perp}
- e\mathbf{A} (\mathbf{r}_{\perp})
+\frac{\mathbf{q}_{\perp}}{2}\right) \nonumber\\
&\times& \gamma^\mu_{n_1n_2}\tilde\chi_{n_2m_2}(k,P)
\gamma^\nu_{m_2m_1}\bar{S}_{m_1m}
\left(p_{\parallel}-\frac{P_{\parallel}}{2},
\mathbf{p}_{\perp}
- e\mathbf{A} (\mathbf{r}_{\perp})-
\frac{\mathbf{q}_{\perp}}{2}\right) 
{\cal{D}}_{\mu\nu}\left(k_{\parallel}-p_{\parallel},
\mathbf{k}_{\perp}-\mathbf{p}_{\perp}
+ 2e\mathbf{A} (\mathbf{r}_{\perp})\right)  \nonumber\\
\label{31.9}
\end{eqnarray}
[here $p_{\parallel}\equiv (p^0,p^4)$ and
$\mathbf{p}_{\perp} \equiv (p^1,p^2)$]. Henceforth, we will consider
the equation with the total momentum $P_\mu\to 0$, corresponding to NG bosons.

The crucial point for further analysis will be the assumption that $m_{\rm dyn} \ll  \sqrt{|eB|}$ 
and that the infrared region with $k \lesssim m_{\rm dyn} \ll \sqrt{|eB|}$ is mostly responsible 
for generating the mass. This will allow us to use the LLL approximation (\ref{tildeSLLL}) for 
the propagator $\bar{S}_{nm}(p_{\parallel},p_{\perp})$. As we will see, this assumption is 
self-consistent [see Eq.~(\ref{m_dyn})]. After making use of the LLL approximation
and assuming that $eB<0$, we rewrite Eq.~(\ref{31.9}) in the following form:
\begin{eqnarray}
\rho(p_{\parallel},\mathbf{p}_{\perp}) &=&
\frac{2\alpha l^2}{(2\pi)^4} e^{- l^2\mathbf{p}_{\perp}^2} \int
d^2q_{\perp} d^2k_{\perp} d^2k_{\parallel}
e^{- l^2q_{\perp}^2}\left(1 - i\gamma^1\gamma^2\right) \gamma^\mu \nonumber\\
&& \times \frac{\hat k_{\parallel}+m_{\rm dyn}}{
k^2_{\parallel}-m^2_{\rm dyn}}
\rho(k_{\parallel},\mathbf{k}_{\perp})
\frac{\hat k_{\parallel}+m_{\rm dyn}}{k^2_{\parallel}-m^2_{\rm dyn}}
\gamma^\nu\left(1 - i\gamma^1\gamma^2\right) {\cal{D}}_{\mu\nu}(k_{\parallel}-p_{\parallel},
\mathbf{k}_{\perp}-\mathbf{q}_{\perp})~,     \label{31.10}
\end{eqnarray}
where $\rho(p_{\parallel},\mathbf{p}_{\perp})= 
(\hat p_{\parallel}-m_{\rm dyn})\tilde\chi(p)
(\hat p_{\parallel}-m_{\rm dyn})$. The form of Eq.~(\ref{31.10}) implies that
$\rho(p_{\parallel},\mathbf{p}_{\perp})=\exp(- l^2 \mathbf{p}_{\perp}^2)
\varphi(p_{\parallel})$, where $\varphi(p_{\parallel})$ satisfies
the equation
\begin{equation}
\varphi(p_{\parallel}) =
\frac{\pi\alpha}{(2\pi)^4} \int d^2k_{\parallel}
\left(1 - i\gamma^1\gamma^2\right) \gamma^\mu
\frac{\hat k_{\parallel}+m_{\rm dyn}}{k^2_{\parallel}-m^2_{\rm dyn}}
\varphi(k_{\parallel}) = \frac{\hat k_{\parallel}+m_{\rm dyn}}{k^2_{\parallel}-
m^2_{\rm dyn}} \gamma^\nu\left(1- i\gamma^1\gamma^2\right) D^{\parallel}_{\mu\nu}
(k_{\parallel}-p_{\parallel})~. \label{31.11}
\end{equation}
Here, we introduced the following ``longitudinal" form of the photon propagator  
relevant for the LLL states:
\begin{equation}
D^{\parallel}_{\mu\nu}(k_{\parallel}-p_{\parallel})=
\int d^2k_{\perp} \exp\left(-\frac{ l^2 \mathbf{k}^2_{\perp}}{2}\right)
{\cal{D}}_{\mu\nu}(k_{\parallel}-p_{\parallel},\mathbf{k}_{\perp})~.
\label{31.12}
\end{equation}
Thus, as in the NJL model in Section~\ref{sec:BS-NJL},
the Bethe-Salpeter equation has been reduced to a $2$-dimensional
integral equation.

Henceforth, we will use Euclidean space with $k_4=-ik^0$.
Then, because of the
symmetry $\mathrm{SO}(2)\times \mathrm{SO}(2)\times {\cal{P}}$ in a magnetic field, we
arrive at the matrix structure (\ref{phiStructureLLL}) for $\varphi(p_{\parallel})$:
\begin{equation}
\varphi(p_{\parallel})=\gamma_5(A+i\gamma_1\gamma_2 B+
\hat p_{\parallel}C+i\gamma_1\gamma_2 \hat p_{\parallel} D)
\label{31.13}
\end{equation}
where $A,B,C$ and $D$ are functions of $p^2_{\parallel}$.

We begin the analysis of Eq.~(\ref{31.11}) by choosing the Feynman gauge (the
general covariant gauge will be considered below).  Then,
\begin{equation}
D^{\parallel}_{\mu\nu}(k_{\parallel}-p_{\parallel})=
i\pi \delta_{\mu\nu}  \int^\infty_0
\frac{dx \exp(- l^2 x/2)}{(k_{\parallel}-p_{\parallel})^2+x}~,
\label{31.14}
\end{equation}
and substituting the expression (\ref{31.13}) for $\varphi(p_{\parallel})$
into Eq.~(\ref{31.11}), we find that $B=-A$, $C=D=0$, i.e.,
\begin{equation}
\varphi(p_{\parallel})=A\gamma_5\left(1- i\gamma_1\gamma_2\right)~, \label{31.15}
\end{equation}
and the function $A$ satisfies the following equation:
\begin{equation}
A(p)=\frac{\alpha}{2\pi^2}\int \frac{d^2kA(k)}{k^2+m^2_{\rm dyn}}
\int^\infty_0
\frac{dx \exp(- l^2x/2)}{(\mathbf{k}-\mathbf{p})^2+x} \label{31.16}
\end{equation}
(Henceforth, we will omit the symbol $\parallel$). By making use of the Feynman gauge and 
performing the angular integration, this integral equation can be rewritten in the form of a 
slightly simpler equation for function $A(p^2)$,
\begin{equation}
A(p^2)=\frac{\alpha}{2\pi}\int_0^\infty\frac{dk^2A(k^2)}
{k^2+m^2_{\rm dyn}}K(p^2,k^2)\label{D143}
\end{equation}
where the explicit expression for the kernel is given by
\begin{equation}
K(p^2,k^2)=\int_0^\infty\frac{dz\exp(-z l^2/2)}{\sqrt{
(k^2+p^2+z)^2-4k^2p^2}}.\label{D144}
\end{equation}
The numerical solution of this equation was obtained in Ref.~\cite{Gusynin:1995nb}. 
The corresponding numerical result for the dynamical mass is fitted well by the 
following analytical expression:
\begin{equation}
m_{\rm dyn} \simeq \sqrt{2|eB|} \exp\left[-\frac{\pi}{2}\left(\frac{\pi}{2\alpha}\right)^{1/2}a+b\right],
\label{MdynQED_numerical}
\end{equation}
where $a\approx 1.000059$ and $b\approx -0.283847$ are two fitting parameters. One can also 
find an approximate analytical solution to the integral equation (\ref{D143}) in the Feynman 
gauge. The details of the analysis are presented in Appendix~\ref{GepEqSolutionQED}. 
Here we will quote only the final result for the dynamical mass,
\begin{eqnarray}
m_{\rm dyn}=C\sqrt{|eB|}\exp\left[-\frac{\pi}{2}\left(\frac{\pi}{2\alpha}\right)^{1/2}\right],
\label{m_dyn}
\end{eqnarray}
where the constant $C=O(1)$. This agrees quite well with the numerical solution fitted by 
Eq.~(\ref{MdynQED_numerical}).

A few words are in order about the case of the general covariant gauge in Eq.~(\ref{photon_D_mu-nu}). 
As is known, the ladder approximation is not necessarily gauge invariant. However,
in the case of the magnetic catalysis dynamics, which is due to the weakly coupled 
infrared dynamics in QED, the leading order result for $\ln(m_{\rm dyn}^2 l^2)$ is the 
same in all covariant gauges and is given by $-\pi\sqrt{\pi/2\alpha}$. This is indeed 
supported by the analysis in the general covariant gauge. The corresponding details
of deriving an approximate analytical solution are also given in Appendix~\ref{GepEqSolutionQED}. 

We can see from integral equation (\ref{31.16}) that the dynamical generation of a nonzero
mass is indeed primarily due to the infrared dynamics in QED. Because of the exponent 
$\exp(- l^2x/2)$ on the right hand side of the equation, the natural cutoff in the problem 
is $|eB|$. [The same property is also shared by the integral equations in a general 
covariant gauge, see (\ref{31.24}) and (\ref{31.25}) in Appendix~\ref{GepEqSolutionQED}.] 
The fact that the nonperturbative infrared dynamics decouples from the ultraviolet 
dynamics is reflected also in the asymptotic behavior of function $A(p^2) \sim 1/p^2$ 
[as well as function $C(p^2)$ in the general covariant gauge], which rapidly decreases 
at $p^2\to\infty$. Combining this with the fact that the magnetic field does not affect 
the behavior of the running coupling in QED at $p^2\gg |eB|$ \cite{Dittrich:1985yb}, 
we conclude that the running coupling constant $\alpha$ in the integral equation 
(\ref{31.16}), as well as the generalization in Eqs.~(\ref{31.24}) and (\ref{31.25}), 
has to be interpreted as the value of the running coupling related to the scale 
$\mu^2\sim |eB|$.

Let us now show that integral equation (\ref{31.16}) can be also rewritten in the 
form of a two-dimensional Schr\"odinger equation with a long-range potential. 
By introducing the wave function
\begin{equation}
\Psi(\mathbf{r}) = \int \frac{d^2k}{(2\pi)^2}
\frac{A(k)}{k^2+m^2_{\rm dyn}} e^{i\mathbf{k}\cdot\mathbf{r}}~, \label{31.17}
\end{equation}
and applying the Laplace operator $\Delta_\mathbf{r}$, we find that this function 
indeed satisfies the following Schr\"odinger equation:
\begin{equation}
\left(-\Delta_\mathbf{r}+m^2_{\rm dyn}+V(\mathbf{r})\right) \Psi(\mathbf{r})=0~,
\label{31.18}
\end{equation}
where the potential $V(\mathbf{r})$  is given by
\begin{equation}
V(\mathbf{r})=-\frac{\alpha}{2\pi^2} \int d^2p e^{i\mathbf{p}\cdot\mathbf{r}}
\int^\infty_0 \frac{dx \exp(-x/2)}{ l^2p^2+x}
=-\frac{\alpha}{\pi l^2} \int^\infty_0
dx e^{-x/2} K_0\left(\frac{r}{ l} \sqrt{x}\right) = \frac{\alpha}{\pi l^2} \exp\left(\frac{r^2}{2 l^2}\right)
\mbox{Ei}\left(-\frac{r^2}{2 l^2}\right)   ,      \label{31.19}
\end{equation}
Here $K_0(x)$ is the Bessel function and $\mbox{Ei}(x)$ is the exponential integral function 
\cite{1972hmfw.book.....A}. The essential feature of the potential (\ref{31.19}), which is profoundly 
different from the $\delta$-like potential (\ref{RegDeltaFunction}) in the NJL model, is its long-range
nature. Indeed, by using the asymptotic behavior of $\mbox{Ei}(x)$ \cite{1972hmfw.book.....A}, 
we find
\begin{eqnarray}
V(\mathbf{r}) &\simeq& -\frac{2\alpha}{\pi}\frac{1}{r^2}~, \quad r\to \infty~, \label{31.20a}\\
V(\mathbf{r}) &\simeq& -\frac{\alpha}{\pi l^2} \left(\gamma+\ln \frac{2 l^2}{r^2}\right)~, \quad r\to 0 ~,
\label{31.20b}
\end{eqnarray}
where $\gamma\simeq 0.577$ is the Euler constant. 

Let us recall that, for the Schr\"odinger equation with a short-range potential in two dimensions, 
the ground state energy depends on the coupling constant $\alpha$ at sufficiently weak coupling 
($\alpha\to 0$) as follows: $E(\alpha) \propto -\exp(-a/\alpha)$ where $a>0$ is a model dependent 
constant \cite{Simon:1976un}. This was indeed the case in the NJL model, see Eq.~(\ref{BSgapMdyn3+1}),
but this is not the case in QED, where the asymptote in Eq.~(\ref{31.20a}) clearly demonstrates the 
long-range nature of the potential. 

In fact, the result for the dynamical mass in Eq.~(\ref{m_dyn}) agrees with the formal 
analysis of the Schr\"odinger equation in Ref.~\cite{Perelomov:1970fz}, where the analytic 
properties of the energy eigenvalue $E(\alpha)$ were studied for the long-range 
potentials with the asymptotic behavior $V(\mathbf{r}) \to {\rm const}/r^2$ at $r\to\infty$. 
Such a long-range character of the potential leads to a strong enhancement of the 
dynamical mass. At weak coupling, the result in Eq.~(\ref{m_dyn}) behaves as 
$\propto \exp\left(-a^\prime/\sqrt{\alpha}\right)$ where $a^\prime>0$ is a constant. 
Compared to the behavior $\propto \exp\left(-a /\alpha\right)$, characteristic 
for the short-range potentials \cite{Simon:1976un}, the QED result is much 
larger at weak coupling $\alpha$. (As we will see in the next subsection, 
however, a nonperturbative resummation of all relevant diagrams in QED 
will make the dynamics more similar to the short-range models.)

This concludes the description of spontaneous chiral symmetry breaking by a 
magnetic field in ladder QED. In the next subsection, it will be shown that
the inclusion of screening effects in QED interactions essentially changes
the final result for the dynamical mass $m_{\rm dyn}$.

\subsubsection{Magnetic catalysis in QED beyond ladder approximation}
\label{QED-beyond}

The simplest improvement of the ladder approximation is to replace the free 
photon propagator (\ref{photon_D_mu-nu}) with the one-loop resummed expression. 
The corresponding analysis in the so-called improved rainbow approximation was 
performed in Ref.~\cite{Gusynin:1995nb} and revealed that the result for the 
dynamical mass in all covariant gauges is given by an expression similar to that 
in Eq.~(\ref{m_dyn}), but with $\alpha\to\alpha/2$ replacement. Such a drastic 
change in the result indicates that, despite the smallness of $\alpha$, the 
corresponding diagrammatic expansion of the Bethe-Salpeter equation in powers of 
$\alpha$ is broken in the infrared region. This posed a challenge of defining 
the complete class of diagrams that are relevant for the calculation of the 
dynamical mass in the weakly coupled QED in a magnetic field. The problem
was resolved in Refs.~\cite{Gusynin:1998zq,Gusynin:1999pq} using the framework 
of the Schwinger-Dyson equations. We review the corresponding analysis in this section. 

The Schwinger-Dyson equations in QED in external fields were derived by 
Schwinger and Fradkin (for a review, see Ref.~\cite{Fradkin:1991zq}). The equations 
for the fermion propagator $G(u,u^\prime)$ are
\begin{eqnarray}
&&G^{-1}(u,u^\prime)=S^{-1}(u,u^\prime)+\Sigma(u,u^\prime),\label{G{-1}}\\
&&\Sigma(u,u^\prime)=4\pi\alpha\gamma^\mu\int G(u,v)
\Gamma^\nu (v,u^\prime,v^\prime){\cal D}_{\nu\mu}
(v^\prime,u)d^4vd^4v^\prime.
\label{SD-fer}
\end{eqnarray}
Here $S(u,u^\prime)$ is the free  fermion propagator in the external
field $A_{\mu}$, $\Sigma(u,u^\prime)$ is the fermion mass
operator, and ${\cal D}_{\mu\nu}(u,u^\prime)$, $\Gamma^\nu (u,u^\prime,v)$ are
the full photon propagator and the full amputated vertex.

The full photon propagator satisfies the equations
\begin{eqnarray}
{\cal D}^{-1}_{\mu\nu}(u,u^\prime)&=&D^{-1}_{\mu\nu}(u-u^\prime)
+\Pi_{\mu\nu}(u,u^\prime), \label{SD-pho}\\
\Pi_{\mu\nu}(u,u^\prime)&=&-4\pi\alpha \int d^4 v d^4 v^\prime 
\mathrm{tr} \left[ \gamma_{\mu}
G(u,v)\Gamma_{\nu} (v,v^\prime,u^\prime) G(v^\prime,u) \right],
\label{Pi_munu}
\end{eqnarray}
where $D_{\mu\nu}(u-u^\prime)$ is the free photon propagator and
$\Pi_{\mu\nu}(u,u^\prime)$ is the polarization tensor.

Before proceeding to the main analysis, let us discuss the general structure of the
key Green functions used in the Schwinger-Dyson equation. As we know, the free fermion propagator 
in the presence of a magnetic field is given in the form of a product of the Schwinger 
phase and a translationally invariant part \cite{Schwinger:1951nm}, i.e.,
\begin{equation}
S(u,u^\prime)=\exp\left[i\Phi(\mathbf{r}_\perp,\mathbf{r}_\perp^\prime)\right] \bar{S}(u-u^\prime).
\end{equation}
For the other Green functions, it is not difficult to show directly from the Schwinger-Dyson equations 
that
\begin{eqnarray}
G(u,u^\prime) &=& \exp\left[i\Phi(\mathbf{r}_\perp,\mathbf{r}_\perp^\prime)\right]
\bar{G}(u-u^\prime),
\label{14a}\\
\Gamma(u,u^\prime,v) &=& \exp\left[i\Phi(\mathbf{r}_\perp,\mathbf{r}_\perp^\prime)\right]
\bar{\Gamma}(u-v,u^\prime-v),
\label{14b}\\
{\cal D}_{\mu\nu}(u,u^\prime) &=& \bar{{\cal D}}_{\mu\nu}(u-u^\prime),
\label{14c}\\
\Pi_{\mu\nu}(u,u^\prime) &=& \tilde{\Pi}_{\mu\nu}(u-u^\prime).
\label{14d}
\end{eqnarray}
In other words, in a constant magnetic field, the Schwinger phase
is universal for Green functions containing one fermion field,
one antifermion field, and any number of photon fields, and the
full photon propagator is translation invariant.

Let us show that there exists a gauge in which the bare vertex,
\begin{equation}
\Gamma^{\mu}(u,u^\prime,v)=\gamma^{\mu} \delta(u-u^\prime)\delta(u-v),
\label{bare-ver}
\end{equation}
is reliable for the description of spontaneous chiral symmetry
breaking in a magnetic field in QED.

The following analysis will be reliable in the weak-coupling limit of the gauge theory, when 
the lowest Landau level dominates the dynamics of fermion pairing. In the corresponding 
regime, the higher Landau levels decouple from the infrared dynamics because they are 
separated by the Landau gap of order $\sqrt{|eB|}$, which is larger than the dynamical 
fermion mass $m_{\rm dyn}$. 

The free fermion propagator in the LLL approximation is given by
\begin{equation}
\bar{S}_{\rm LLL}(p)=2i e^{-p_{\perp}^2 l^2}\frac{\hat{p}_{\parallel}+m}
{p^2_{\parallel}-m^2} {\cal P}_{+},
\label{tildeS}
\end{equation}
where $l=|eB|^{-1/2}$ is the magnetic length, $p_{\perp}=(p^1,p^2)$,
$p_{\parallel}=(p^0,p^3)$, and $\hat{p}_{\parallel} =p^0 \gamma^0-p^3 \gamma^3$. The operator 
${\cal P}_{+} \equiv \left[1+ i\gamma^{1}\gamma^{2} \mathrm{sign}(eB)\right]/2$ is 
the projection operator on the fermion states with the spin polarized along the magnetic 
field. The presence of this projection operator implies that the ``effective" bare vertex for 
fermions in the LLL is ${\cal P}_{+} \gamma^\mu {\cal P}_{+} =
{\cal P}_{+} \gamma_{\parallel}^\mu$. As a result, the LLL fermions couple 
only to the longitudinal $0$- and $3$-components of the photon field. In application 
to the polarization tensor, this 
implies that $\Pi^{\mu\nu}(q) \simeq 
\left(q_{\parallel}^{\mu} q_{\parallel}^{\nu}-q_{\parallel}^{2}g_{\parallel}^{\mu\nu}\right)
\Pi (q_{\perp}^2,q_{\parallel}^2)$
in the LLL approximation.

In the one-loop approximation, with fermions from the LLL, the photon propagator takes the 
following form in covariant gauges \cite{Batalin:1971au,Tsai:1974pi,Shabad:1975ik,
1976PhLA...56..151L,Dittrich:1985yb,Calucci:1993fi}:
\begin{eqnarray}
{\cal D}_{\mu\nu}(q)=-i\left[\frac{1}{q^2}g^{\perp}_{\mu\nu}
+\frac{q^{\parallel}_{\mu} 
q^{\parallel}_{\nu}}{q^2 q_{\parallel}^2}
+\frac{1}{q^2+q_{\parallel}^2 \Pi (q_{\perp}^2,q_{\parallel}^2)}
\left(g^{\parallel}_{\mu\nu}-\frac{q^{\parallel}_{\mu}
q^{\parallel}_{\nu}}{q_{\parallel}^2}\right)
-\frac{\lambda}{q^2}\frac{q_{\mu}q_{\nu}}{q^2}\right],
\label{cov-gauge}
\end{eqnarray}
where the symbols $\perp$ and $\parallel$ in $g_{\mu\nu}$ are related to the 
$(1,2)$ and $(0,3)$ components, respectively, and $\lambda$ is a gauge 
parameter. The explicit expression for $\Pi (q_{\perp}^2, q_{\parallel}^2)
=\exp[-(q_{\perp} l)^2/2] \Pi (q_{\parallel}^2)$ is given in Refs.~\cite{Batalin:1971au,
Tsai:1974pi,Shabad:1975ik,1976PhLA...56..151L,Dittrich:1985yb,
Calucci:1993fi}. For our purposes, it is sufficient to know its asymptotes 
in the strong field limit, 
\begin{eqnarray}
\Pi (q_{\parallel}^2)\simeq \frac{\bar{\alpha}}{3\pi}
\frac{|eB|}{m_{\rm dyn}^2}, \quad \mbox{as}
\quad |q_{\parallel}^2| \ll m_{\rm dyn}^2,
\label{Pi-IR}\\
\Pi (q_{\parallel}^2)\simeq -\frac{2\bar{\alpha}}{\pi}
\frac{|eB|}{q_{\parallel}^2} \quad \mbox{as}
\quad |q_{\parallel}^2| \gg m_{\rm dyn}^2,
\label{Pi-UV}
\end{eqnarray}
where $\bar{\alpha}=N_{f}\alpha$.  
(For recent calculations of the polarization tensor beyond the LLL approximation, 
see Refs.~\cite{Hattori:2012je,Hattori:2012ny,Karbstein:2013ufa,Chao:2014wla}.)
Notice that the polarization effects are absent in the transverse components of 
${\cal D}_{\mu\nu}(q)$. This is because the LLL fermions couple only to the 
longitudinal components of the photon field. 

The screening effects in the longitudinal components are rather strong. Indeed, as follows from 
Eq.~(\ref{Pi-UV}), 
\begin{equation}
\frac{1}{q^2+q_{\parallel}^2 \Pi (q_{\perp}^2, q_{\parallel}^2)}
\simeq \frac{1}{q^2-M_{\gamma}^2},
\label{screenedphoton}
\end{equation}
with
\begin{equation}
M_{\gamma}^2= \frac{2\bar{\alpha}}{\pi}|eB|
\label{M_gamma}
\end{equation}
valid for $m^2\ll |q_{\parallel}^2| \ll |eB|$ and $|q_{\perp}^2| \ll |eB|$. This is reminiscent of 
the pseudo-Higgs effect in the $(1+1)$-dimensional {\it massive} QED (massive Schwinger 
model \cite{Schwinger:1962tn,Coleman:1976uz}). It is not the genuine Higgs effect because 
there is no complete screening of the color charge in the infrared region with 
$|k_{\parallel}^2|\ll m_{q}^2$. This can be seen clearly from Eq.~(\ref{Pi-IR}). Nevertheless, 
the pseudo-Higgs effect is manifested in creating a massive resonance and this 
resonance provides the dominant forces leading to chiral symmetry breaking.

We emphasize that the infrared dynamics in this problem is very different from that 
in the Schwinger model. This is because the photon is neutral and, thus, not subject 
to a dimensional reduction in a magnetic field. Accordingly, we see that the complete 
four-dimensional momentum squared, $q^2=q_{\parallel}^2-q_{\perp}^2$, enters the 
photon propagator in Eq.~(\ref{screenedphoton}). However, the tensor and the spinor 
structure of QED in the LLL approximation are exactly the same as in the Schwinger 
model. Indeed, the LLL fermion propagator (\ref{tildeS}) and the vertex ${\cal P}_{+} 
\gamma^\mu {\cal P}_{+} ={\cal P}_{+} \gamma_{\parallel}^\mu$ are 
two-dimensional, and only the longitudinal $(0,3)$ components of a photon field are 
relevant here. This point will be crucial for finding a gauge in which the improved 
rainbow approximation with the bare vertex (\ref{bare-ver}) is reliable. 

We reiterate that despite the smallness of $\alpha$, the expansion in $\alpha$ 
is broken in covariant gauges in massless QED in an external magnetic field 
\cite{Gusynin:1995nb}. The root of the problem is the smallness of $m_{\rm dyn}$ 
in Eq.~(\ref{m_dyn}) as compared to the scale of $\sqrt{|eB|}$ that causes
large logarithmic contributions, or mass singularities of the type 
$\ln(|eB|/m_{\rm dyn}^2) \sim\alpha^{-1/2}\gg 1$ in infrared dynamics. 
The corresponding one-loop corrections to the vertex were discussed in 
Appendix~A of Ref.~\cite{Gusynin:1999pq}. It was found that there are 
contributions of order $\alpha\ln^2(|eB|/m_{\rm dyn}^2)\sim O(1)$ when 
external momenta are of order $m_{\rm dyn}$ or less. They come from the term 
$q_{\mu}^{\parallel} q_{\nu}^{\parallel}/q^2 q_{\parallel}^2$ in ${\cal D}_{\mu\nu}(q)$ 
in Eq.~(\ref{cov-gauge}).

It is natural to ask whether the problem of magnetic catalysis can be solved reliably 
in massless QED if there is no way to avoid large logarithmic corrections in any 
covariant gauge. A resolution of the problem is suggested by the Schwinger model. 
It is known that there is a gauge in which the full vertex is just the bare one \cite{Frishman:1975}. 
It is the gauge with a bare photon propagator
\begin{equation}
D_{\alpha\beta}(k)=-i\frac{1}{k^2}\left(g_{\alpha\beta}-\frac
{k_{\alpha} k_{\beta}}{k^2}\right)
-i d(k^2) \frac{k_{\alpha} k_{\beta}}{(k^2)^2}
\end{equation}
in the (nonlocal) gauge, defined by the specific choice of function $d= 1/(1+\Pi)$. 
In the Schwinger model, the polarization function is $\Pi(k^2)=-e^2/\pi k^2$. Then,
the full propagator is proportional to $g_{\alpha\beta}$,
\begin{eqnarray}
{\cal D}_{\alpha\beta}(k)&=&D_{\alpha\beta}(k)
+ i\left(g_{\alpha\beta}- \frac{k_{\alpha} 
k_{\beta}}{k^2} \right)
\frac{\Pi(k^2)}{k^2\left[1+\Pi(k^2)\right]}
=-i \frac{g_{\alpha\beta}}{k^2\left[1+\Pi(k^2)\right]}.
\label{gauge-S}
\end{eqnarray}
[Here $\alpha,~\beta =0,1$.] The point is that since now ${\cal D}_{\alpha\beta}(k)\sim
g_{\alpha\beta}$ and since the fermion mass $m=0$ in the Schwinger model, all loop 
contributions to the vertex are proportional to 
\begin{equation}
P_{2n+1}\equiv  \gamma_{\alpha} \gamma_{\lambda_{1}} \dots
\gamma_{\lambda_{2n+1}} \gamma^{\alpha}=0
\end{equation}
in the chosen gauge and, therefore, vanish. Note that $P_{2n+1}=0$ results 
from the following identities for the two-dimensional Dirac matrices:
$\gamma_{\alpha} \gamma_{\lambda} \gamma^{\alpha}=0$ and
$\gamma_{\lambda_{i}} \gamma_{\lambda_{i+1}}=g_{\lambda_{i}
\lambda_{i+1}} +\varepsilon_{\lambda_{i} \lambda_{i+1}}
\gamma_{5}$ (here $\gamma_{5}=\gamma_{0} \gamma_{1}$,
$\varepsilon_{\alpha\beta} =-\varepsilon_{\beta\alpha}$, and
$\varepsilon_{01}=1$).

Let us return to the problem of magnetic catalysis in massless QED. 
As emphasized above, the tensor and the spinor structure of the LLL 
dynamics is $(1+1)$-dimensional. Let us take the bare propagator
\begin{equation}
D_{\mu\nu}(q)=-i\frac{1}{q^2}\left(g_{\mu\nu}
-\frac{q_{\mu} q_{\nu}} {q^2}\right)
-i d(q_{\perp}^2, q_{\parallel}^2) \frac{q^{\parallel}_{\mu}
q^{\parallel}_{\nu}}{q^2 q_{\parallel}^2}
\end{equation}
in the (nonlocal) gauge with $d =-q_{\parallel}^2\Pi/[q^2+q_{\parallel}^2\Pi] +
q_{\parallel}^2/q^2$. Then, the full propagator will be given by
\begin{eqnarray}
{\cal D}_{\mu\nu}(q)&=&D_{\mu\nu}(q)
+ i\left(g^{\parallel}_{\mu\nu}-
\frac{q^{\parallel}_{\mu} q^{\parallel}_{\nu}}{q_{\parallel}^2}
\right) \frac{q_{\parallel}^2 \Pi(q_{\perp}^2, q_{\parallel}^2)}
{q^2[q^2+q_{\parallel}^2 \Pi(q_{\perp}^2, q_{\parallel}^2)]}
\nonumber \\
&=&-i\frac{g^{\parallel}_{\mu\nu}}{q^2+q_{\parallel}^2
\Pi(q_{\perp}^2, q_{\parallel}^2)} -i
\frac{g^{\perp}_{\mu\nu}}{q^2}
+i\frac{q^{\perp}_{\mu}q^{\perp}_{\nu} + q^{\perp}_{\mu}
q^{\parallel}_{\nu} + q^{\parallel}_{\mu}q^{\perp}_{\nu}}
{(q^2)^2}.
\label{Dmunu-non-local}
\end{eqnarray}
The crucial point is that, as was pointed out above, the transverse degrees of 
freedom decouple from the LLL dynamics. Therefore, only the first term in 
${\cal D}_{\mu\nu}(q)$, proportional to $g^{\parallel}_{\mu\nu}$, is relevant.

Notice now that dangerous mass singularities in loop corrections to the vertex might 
potentially occur only in the terms containing $\hat{q}^{\parallel}_i 
=q_{i}^{0} \gamma^{0}-q_{i}^{3}\gamma^{3}$ from a numerator 
($\hat{q}^{\parallel}_i+m_{\rm dyn}$) of {\it each} fermion propagator in a diagram 
(all other terms contain positive powers of $m_{\rm dyn}$, coming from at least some 
of the numerators and, therefore, are harmless). However, because of the same 
reasons as in the gauge (\ref{gauge-S}) in the Schwinger model, all those potentially 
dangerous terms disappear in the gauge (\ref{Dmunu-non-local}). Therefore, all the 
loop corrections to the vertex are suppressed by positive powers of $\alpha$ in this 
special gauge. This in turn implies that those loop corrections may result only in a 
change of the overall constant in the expression  for the dynamical mass $C\sim O(1)
\rightarrow \tilde C^\prime\sim O(1)$. In other words, in gauge (\ref{Dmunu-non-local}) 
there exists a {\it consistent} truncation of the Schwinger-Dyson equations and the 
problem is essentially soluble in that gauge. We should also add that, as is shown in 
Ref.~\cite{Gusynin:1999pq}, the gauge (\ref{Dmunu-non-local}) is unique. In other 
gauges, there is an infinite set of diagrams giving relevant contributions to the vertex. 
Therefore, in other gauges, one needs to sum up an infinite set of diagrams to recover 
the same result for the fermion mass. (This implies, in particular, that the approach of 
Refs.~\cite{Leung:2005xz,Leung:2005yq}, where the improved rainbow approximation 
was used in the covariant gauges, is not self consistent.)

It may be appropriate to mention that the dynamical mass is a gauge invariant quantity 
in QED in a magnetic field because there is no confinement of fermions. Therefore, 
{\it any} gauge can be used for the calculations of the mass {\it if} either the calculations 
provide the exact result or a good approximation is used, {\it i.e.}, one can show that 
corrections to the obtained result are small. Below we will find such a gauge in weakly 
coupled massless QED.

Although the above consideration of the mass singularities in the loop corrections 
is general, it is somewhat heuristic. In practice, one has to define more rigorously 
the perturbative expansion for the Schwinger-Dyson equations which is used in this
problem. It is the loop expansion based on the Cornwall-Jackiw-Tomboulis effective 
action $\Gamma(G,{\cal D}_{\mu\nu})$ for composite operators \cite{Cornwall:1974vz} 
(for a review see Ref.~\cite{Miransky:1994vk}). The conditions for extrema of $\Gamma$ 
yield the Schwinger-Dyson equations,
\begin{equation}
\frac{\delta\Gamma}{\delta G(u,u^\prime)}=0, 
\qquad \frac{\delta\Gamma}{\delta
{\cal D}_{\mu\nu}(u,u^\prime)}=0.
\end{equation}
In the loop expansion for $\Gamma$, the {\it full} photon and
fermion propagators are used in two-particle irreducible diagrams
for $\Gamma$. In QED, the problem is essentially reduced to the
loop expansion (with the full photon and fermion propagators) for
the vertex.

The full photon propagator is given by Eq.~(\ref{Dmunu-non-local}), and the
full propagator for fermions from the LLL has the form
\begin{equation}
\bar{G}(p)=2 i e^{-(p_{\perp}l)^2} \frac{ A(p_{\parallel}^2)
\hat{p}_{\parallel} +B(p_{\parallel}^2) }
{A^2(p_{\parallel}^2) p^2_{\parallel} - B^2(p_{\parallel}^2)}
{\cal P}_{+} 
\label{gen-sol}
\end{equation}
[compare with Eq.~(\ref{tildeS}) and see below]. Here $B(p_{\parallel}^2)$ is a 
dynamical mass function of fermions.

The Schwinger-Dyson equations for the fermion propagator in the one-loop and 
two-loop approximations were derived in Ref.~\cite{Gusynin:1999pq}. It was also 
shown that the improved rainbow (one-loop) approximation is reliable in gauge 
(\ref{Dmunu-non-local}) (but not in covariant gauges) and all higher-loop corrections 
are suppressed by positive powers of $\alpha$. Therefore, here we will restrict our 
presentation to the one-loop approximation. 

From Eqs.~(\ref{G{-1}}) and (\ref{SD-fer}) one derives the following equation for the 
fermion propagator:
\begin{equation}
G(u,u^\prime)=S(u,u^\prime)-4\pi \alpha \int d^4u_1 d^4u^\prime_1 
S(u,u_1)\gamma^{\mu}
G(u_1,u^\prime_1)\gamma^{\nu}G(u^\prime_1,u^\prime){\cal D}_{\mu\nu}(u_1-u^\prime_1) .
\label{B1}
\end{equation}
Here $S(u,u^\prime)$ is the free fermion propagator of massless fermions
($m=0$) in  a magnetic field. After extracting the Schwinger phase factors in the 
full and free fermion propagators [see Eq.~(\ref{14a})], the equation for the 
translationally invariant part reads 
\begin{equation}
\bar{G}(u)=\bar{S}(u)
-4\pi \alpha \int d^4 u^{\prime} d^4 u^{\prime\prime} e^{-i e u A(u^{\prime})- ie u^{\prime} A(u^{\prime\prime})}
\bar{S}(u-u^{\prime})\gamma^{\mu}
\bar{G}(u^{\prime}-u^{\prime\prime})\gamma^{\nu}
\bar{G}(u^{\prime\prime}) {\cal D}_{\mu\nu}(u^{\prime}-u^{\prime\prime}),
\label{B3}
\end{equation}
where $A_\mu$ is given in Eq.~(\ref{eq:symm_gauge}) and
the shorthand notation $uA(u^\prime)$ stands for $u^\mu A_\mu(u^\prime)$.

First, let us show that the solution to the above equation, $\bar{G}(u)$, allows 
the factorization of the dependence on the ``parallel" $u_{\parallel}=(t,z)$ and 
perpendicular $\mathbf{r}_{\perp}=(x,y)$ coordinates,
\begin{equation}
\bar{G}(u)=\frac{i}{2\pi l^2}
\exp\left(-\frac{\mathbf{r}_{\perp}^2}{4l^2}\right)
g\left(u_{\parallel}\right){\cal P}_{+} .
\label{B4}
\end{equation}
Notice that this form for $\bar{G}(u)$ is suggested by a
similar expression for the free propagator,
\begin{equation}
\bar{S}(u)=\frac{i}{2\pi l^2}
\exp\left(-\frac{\mathbf{r}_{\perp}^2}{4l^2}\right)
s\left(u_{\parallel}\right){\cal P}_{+} ,
\label{B5}
\end{equation}
with
\begin{equation}
s\left(u_{\parallel}\right)
=\int\frac{d^2 k_{\parallel}}{(2\pi)^2}
e^{-ik_{\parallel}u_{\parallel}}
\frac{\hat{k}_{\parallel}+m}{k_{\parallel}^2-m^2}
\label{B6}
\end{equation}
[see Eq.~(\ref{tildeS}); in the chiral limit, as in the present
problem, the bare mass $m=0$]. In order to perform the
integrations over the perpendicular components of $u_1$ and $u^\prime_1$
in Eq.~(\ref{B3}), it is convenient to make use of the photon
propagator in the momentum representation,
\begin{equation}
{\cal D}_{\mu\nu}(u)=\int\frac{d^2 q_{\parallel} d^2
q_{\perp}}{(2\pi)^4}
e^{-iq_{\parallel}u_{\parallel}+i\mathbf{q}_{\perp}\cdot\mathbf{r}_{\perp}}
{\cal D}_{\mu\nu}\left(q_{\parallel},q_{\perp}\right).
\label{B7}
\end{equation}
After substituting this representation along with those in
Eqs.~(\ref{B4}) and (\ref{B5}) into the Schwinger-Dyson equation (\ref{B3})
and performing the straightforward, though tedious, integrations 
over $\mathbf{r}_{\perp}^{\prime}$ and $\mathbf{r}_{\perp}^{\prime\prime}$, 
we arrive at
\begin{equation}
g\left(u_{\parallel}\right)=s\left(u_{\parallel}\right)
+4\pi \alpha \int \frac{d^4 q}{(2\pi)^4}
d^2 u_{\parallel}^{\prime} d^2 u_{\parallel}^{\prime\prime}
\exp\left(-\frac{(q_{\perp}l)^2}{2}
-iq_{\parallel}(u_{\parallel}^{\prime}-u_{\parallel}^{\prime\prime})\right)
s(u_{\parallel}-u_{\parallel}^{\prime})\gamma^{\mu}_{\parallel}
g(u_{\parallel}^{\prime}-u_{\parallel}^{\prime\prime})\gamma^{\nu}_{\parallel}
g(u_{\parallel}^{\prime\prime})
{\cal D}_{\mu\nu}\left(q_{\parallel},q_{\perp}\right),
\label{B8}
\end{equation}
Since  no dependence on $u_{\perp}$ is left in the equation, 
we conclude that the form of $\bar{G}(u)$ in Eq.~(\ref{B4}) is indeed consistent
with the structure of the Schwinger-Dyson equation.

Regarding this equation, it is necessary to emphasize that the
``perpendicular" components of the $\gamma$-matrices are absent
in it. Indeed, because of the identity  ${\cal P}_{+} 
\gamma_{\perp}^{\mu}{\cal P}_{+} =0$, all those components are killed
by the projection operators coming from the fermion propagators.

Substituting now the photon propagator in the Feynman-like (non-covariant) gauge 
(\ref{Dmunu-non-local}) into the Schwinger-Dyson equation, we see that only the 
first term in Eq.~(\ref{Dmunu-non-local}), proportional to $g^{\parallel}_{\mu\nu}$, 
leads to a nonvanishing contribution. In other words, the photon propagator is 
effectively proportional to $g^{\parallel}_{\mu\nu}$ (justifying the name of the gauge).

\begin{figure}[t]
\begin{center}
\includegraphics[width=0.47\textwidth]{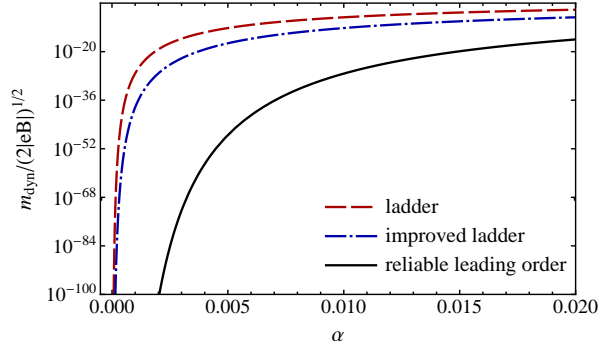}
\caption{(Color online) The dynamical mass as a function of the coupling constant in QED
in various approximations.}
\label{fig:mDynQEDbeyond}
\end{center}
\end{figure}

By switching to the momentum space, we obtain
\begin{eqnarray}
&&g^{-1}\left(p_{\parallel}\right)
=s^{-1}\left(p_{\parallel}\right)
-4\pi \alpha \int \frac{d^4 q}{(2\pi)^4}
\exp\left(-\frac{(q_{\perp}l)^2}{2}\right)
\gamma^{\mu}_{\parallel}
g(p^{\parallel}-q^{\parallel})\gamma^{\nu}_{\parallel}
{\cal D}_{\mu\nu}\left(q_{\parallel},q_{\perp}\right).
\label{B11-text}
\end{eqnarray}
The general solution to this equation is given by the ansatz,
\begin{equation}
g\left(p_{\parallel}\right)=
\frac{A_{p}\hat{p}_{\parallel}+B_{p}}
{A_{p}^2 p_{\parallel}^2-B_{p}^2},
\label{B12-text}
\end{equation}
where $A_{p}=A(p_{\parallel}^2)$ and $B_{p}=B(p_{\parallel}^2)$. By making use of this 
general structure, as well as of the explicit form of the photon propagator in gauge (\ref{Dmunu-non-local}),
we find that $A_p=1$ and function $B_{p}$ satisfies the following equation:
\begin{equation}
B_{p}=-i\frac{\alpha}{2\pi^3} \int
\frac{d^2 q_{\parallel}B_{p-q}}
{\left(p_{\parallel}-q_{\parallel}\right)^2-B_{p-q}^2}
\int \frac{d^2 q_{\perp} \exp\left(-(q_{\perp}l)^2/2\right) }
{q^2+q_{\parallel}^2\Pi\left(q_{\perp}^2,q_{\parallel}^2\right)}.
\label{oneloopB}
\end{equation}
This equation was solved numerically in Refs.~\cite{Gusynin:1998zq,Gusynin:1999pq}. 
For small $\alpha$ ($0.001\leq\alpha\leq 0.1$) and different $N$ ($1\leq N\leq7$) the 
best fit was found to be given by the following expression:
\begin{equation}
m_{\rm dyn} = \sqrt{2|eB|}\, (N\alpha)^{1/3}\exp\left(-\frac{\pi}
{\alpha\ln\left(C_1/N\alpha\right)}\right), 
\label{dynmass}
\end{equation}
where $C_1\simeq 1.82\pm 0.06$. The dynamical mass, described by this 
fit to the numerical solution, is shown in Fig.~\ref{fig:mDynQEDbeyond}. 
For comparison, we also show there the fit to the numerical solution in the 
ladder and improved ladder approximations.

The numerical solution \cite{Gusynin:1998zq,Gusynin:1999pq} reveals that the 
function $B(p_{\parallel}^2)$ is essentially a constant for $p_{\parallel}^2 \ll |eB|$, 
$B(p_{\parallel}^2)=m_{\rm dyn}$, and rapidly decreases for $p_{\parallel}^2\gg |eB|$ 
\cite{Gusynin:1999pq}. Therefore, this approximation is self-consistent: the 
Ward-Takahashi identity for the vertex is satisfied in the relevant kinematic 
region of momenta, and the pole of the fermion propagator appears at 
$p_{\parallel}^2 =m_{\rm dyn}^2$.

It is instructive to review the approximate analytical solution to the gap equation 
(\ref{oneloopB}). In Euclidean space, it takes the form
\begin{equation}
B(p)=\frac{\alpha}{2\pi^2}\int\frac{d^2kB(k)}{k^2+B^2(k)}\int
\limits_0^\infty\frac{dze^{-zl^2/2}}{z+(k-p)^2+\Pi(z)},
\label{masseq}
\end{equation}
where $\Pi(z)=M_\gamma^2e^{-zl^2/2}$ and $M_\gamma^2=2\bar{\alpha} |eB|/\pi$.
Note that we shifted the integration variable and omitted the subscript $\parallel$
in the notation of the longitudinal momentum.

First of all, let us show that the leading singularity,
$1/{\alpha\ln\alpha}$, in $\ln(m_{\rm dyn}^2)$ in
Eq.~(\ref{dynmass}) is induced in the kinematic region with
$m_{\rm dyn}^2\ll|q_{\parallel}^2|\ll|eB|$ and $m_{\rm dyn}^2\ll
M_\gamma^2\lesssim q_\perp^2\ll|eB|$ (in that region, fermions can
be treated as massless).

As follows from the numerical analysis, the approximation with
$B(p_\parallel^2)=m_{\rm dyn}$ for $p_\parallel^2<2|eB|$ and
$B(p_\parallel^2)$ rapidly decreasing for $p_\parallel^2>2|eB|$
is reliable in this problem. Then, taking
$p_{\parallel}^2=0$ in Eq.~(\ref{oneloopB}), we arrive at the equation
\begin{eqnarray}
1&=&\frac{\alpha}{2\pi^2}\int^{2|eB|}\frac{d^2q_\parallel}
{q_\parallel^2+m_{\rm dyn}^2} \int_0^\infty
\frac{dx\exp\left(-xl^2/2\right)}{x+q_\parallel^2+q_\parallel^2
\Pi_E(x,q_\parallel^2)}\nonumber\\
&\simeq&\frac{\alpha}{2\pi^2}\int^{2|eB|}\frac{d^2q_\parallel}
{q_\parallel^2+m_{\rm dyn}^2} \int_0^{2|eB|}
\frac{dx}{x+q_\parallel^2+q_\parallel^2\Pi_E(x, q_\parallel^2)}.
\label{C1}
\end{eqnarray}
Matching now the asymptotes in Eqs.~(\ref{Pi-IR}) and (\ref{Pi-UV}) at
$q_\parallel^2=6m_{\rm dyn}$ in Euclidean space, we get
\begin{eqnarray}
1&\simeq&\frac{\alpha}{2\pi}\int_0^{2|eB|}dx\left(\int_0^{6m_{\rm dyn}^2}\frac{dy}{(y+
m_{\rm dyn}^2)\left(x+y(1+\frac{M_\gamma^2}{6 m_{\rm dyn}^2})\right)}
+\int_{6m_{\rm dyn}^2}^{2|eB|}\frac{dy}{(y+ m_{\rm dyn}^2)\left(x+y+M_\gamma^2
e^{-xl^2/2}\right)}\right).
\label{C2}
\end{eqnarray}
It is clear that, because of $m_{\rm dyn}^2$ in $(y+m_{\rm dyn}^2)$, the
first term in the square bracket on the right hand side of this
equation is of order $O(1)$ and can be neglected: it cannot give
a contribution of order $1/{\alpha\ln\alpha}$ to $m_{\rm dyn}^2$.
Then, we arrive at the estimate,
\begin{equation}
1\simeq\frac{\alpha}{2\pi}\int_{6m_{\rm
dyn}^2}^{2|eB|}\frac{dy}{y+ m_{\rm dyn}^2}
\int_0^{2|eB|}\frac{dx}{x+y+M_\gamma^2e^{-xl^2/2}}.
\label{C3}
\end{equation}
The double logarithmic contribution comes from the region 
$2|eB|\gg y=q_{\parallel}^2\gg m_{\rm dyn}^2$, $2|eB|\gg
x=q_{\perp}^2\gtrsim y+ M_\gamma^2\geq M_\gamma^2$, where $M_\gamma^2
=2\bar\alpha/{\pi l^2}$. Therefore, one can write 
\begin{equation}
1\simeq\frac{\alpha}{2\pi}\int_{6m_{\rm dyn}^2}^{2|eB|}
\frac{dy}{y}
\int_{y+M_\gamma^2}^{2|eB|}\frac{dx}{x}=
\frac{\alpha}{2\pi}\int_{6m_{\rm dyn}^2}^{2|eB|}
\frac{dy}{y}\ln\frac{2|eB|}{y+M_\gamma^2}.
\label{C4}
\end{equation}
To calculate the last integral with  double logarithmic accuracy,
we write
\begin{eqnarray}
1&\simeq&\frac{\alpha}{2\pi}\left(\ln\frac{2eB}{M_\gamma^2}
\int_{6m_{\rm dyn}^2}^{M_\gamma^2}\frac{dy}{y}
+\int_{M_\gamma^2}
^{2|eB|}\frac{dy}{y}\ln\frac{2|eB|}{y}\right)\nonumber\\
&\simeq&\frac{\alpha}{2\pi}
\left(\ln\frac{2eB}{M_\gamma^2}\ln\frac
{M_\gamma^2}{ m_{\rm dyn}^2}+\frac{1}{2}\ln^2\frac{2eB}
{M_\gamma^2}\right)=
\frac{\alpha}{4\pi}\ln\frac{2eB}{M_\gamma^2}
\ln\left(\frac{2|eB| M_\gamma^2}{ m_{\rm dyn}^4}\right).
\label{C5}
\end{eqnarray}
This equation implies that
\begin{equation}
m_{\rm dyn}\sim\sqrt{|eB|}\left(\frac{N\alpha}{\pi}\right)^{1/4}
\exp\left(-\frac{\pi}{\alpha\ln (\pi/N\alpha)}\right).
\label{C6}
\end{equation}
Comparing this expression with Eq.~(\ref{dynmass}), one can see
that this estimate is quite reasonable. The origin of that is a
rather simple form of the fermion mass function
$B(p_\parallel^2)$: $B(p_\parallel^2)\simeq m_{\rm dyn}$ for
$p_\parallel^2\lesssim 2|eB|$ and $B(p_\parallel^2)$ rapidly
decreases for $p_\parallel^2\gtrsim 2|eB|$.

Therefore, the dominant contribution to the Schwinger-Dyson equation
(\ref{oneloopB}) comes from the region with $m_{\rm dyn}^2
\ll|q_{\parallel}^2| \ll|eB|$ and $m_{\rm dyn}^2\ll M_\gamma^2\lesssim
q_\perp^2\ll|eB|$, where fermions can be treated as massless.
This in turn justifies the approximation with the polarization
function $\Pi_E=2\bar\alpha|eB|/{\pi q_\parallel^2}$.

A more rigorous solution to Eq.~(\ref{masseq}) is presented in 
Appendix~\ref{App:BetheSalpeterQEDbeyond}. The corresponding 
final result reads
\begin{equation}
m_{\rm dyn}
\simeq \sqrt{2|eB|} \left(\frac{N\alpha }{\pi}\right)^{1/3}
\exp\left(-\frac{\pi}{\alpha\ln (\pi/N\alpha)}\right).
\end{equation}
This expression for the dynamical mass is close both to the estimate in 
Eq.~(\ref{C6}) and to the numerical solution (\ref{dynmass}). The ratio of 
the values of $C_1$ in the analytical solution and in the numerical one is 
$C_1^{(\rm analyt)}/C_1^{(\rm numer)}\simeq 1.7$. This rather
mild discrepancy reflects the approximations made in the kernel
of the integral equation when it is reduced to the differential equation.

\subsection{Magnetic catalysis in reduced QED}
\label{sec:MagCatQEDreduced}

It had been recognized rather long ago that relativistic field models can
serve as effective theories for the description of long wavelength
excitations in condensed matter systems (for a review, see
Ref.~\cite{Fradkin:1991nr}). In particular, they can be applied to a wide class
of (quasi-) planar systems. In this case, the corresponding relativistic
theories are $(2+1)$-dimensional, i.e., they are formulated in $(2+1)$-dimensional 
Minkowski space with two space like and one time like
coordinates. It is important that amongst these condensed matter systems
are such as high-$T_c$ superconductors and carbon-based materials (for a
list of papers using relativistic field approach to these systems see 
Refs.~\cite{Semenoff:1984dq,Marston:1989zz,Marston:1990bj,Kovner:1990dp,
Dorey:1991kp,Aitchison:1996th,Gonzalez:1999wo,Gonzalez:2001kn,Marinelli:2000xr,
Franz:2000vh,Vafek:2001uk,Ye:2001oq,Vishwanath:2001qp,Liu:1998mg,Farakos:1998pn,
Semenoff:1998bk,Khveshchenko:2001zz,Khveshchenko:2001zza,Ferrer:2001fz,Gorbar:2002iw}.
Here, we review studies of the dynamics in graphite and graphene using 
the framework of the so-called reduced $(3+1)$-dimensional gauge
theories \cite{Gorbar:2001qt,Alexandre:2001ps,Gorbar:2002iw,Teber:2012de,Kotikov:2013eha}. 
These theories share the following common
feature. Their gauge fields (e.g., the electromagnetic field) responsible
for interparticle interaction can propagate in a $3$-dimensional bulk, 
while fermion fields (e.g., describing electron and hole type quasiparticles) 
are localized on a $2$-dimensional plane. Such a model mimics quite 
well the dynamics in graphene and even, to some extend, in high quality
pyrolytic graphite. The underlying reason for this is the well known fact 
that the quasiparticles in such systems are approximately described by 
$2$-dimensional massless Dirac fermions \cite{Wallace:1947wc,McClure:1957hs,Semenoff:1984dq}. 

The Lagrangian density of quasiparticles on a $2$-dimensional plane 
reads 
\begin{equation}
{\cal L}_{0} = 
\bar{\Psi}(t,\mathbf{r})
\left[ i\gamma^{0}(\partial_{t}+i\mu)  
-iv_{F}\left(\gamma^{1} D_{x}
+\gamma^{2} D_{y}\right)\right]\Psi(t,\mathbf{r}),
\label{L-free}
\end{equation}
where $\Psi(t,\mathbf{r})$ is a 4-component spinor, 
$\bar{\Psi}=\Psi^{\dagger}\gamma^{0}$ is the Dirac conjugate spinor, 
$\mathbf{r}=(u,u^\prime)$ is the position vector in the $2$-dimensional plane, 
and the $4\times 4$ Dirac $\gamma$-matrices furnish a reducible 
representation of the Clifford (Dirac) algebra in $2+1$ dimensions 
\cite{Deser:1981wh,Jackiw:1980kv,Pisarski:1984dj,Appelquist:1986fd,Appelquist:1986qw}. 
In order to describe the situation with a finite ``residual"  density of carriers,
here the chemical potential $\mu$, connected with the electric charge,
was introduced. Also, by anticipating the potential application of this 
model to quasirelativistic systems such as graphene, we introduced 
a Fermi velocity $v_F$ instead of the speed of light $c$.

We will consider the case when the fermion fields carry an additional
``flavor" index $i=1,2,\dots,N_f$ (in the example of graphite, $N_{f}=2$,
see Refs.~\cite{Gonzalez:1999wo,Khveshchenko:2001zza}). Following 
the same types of arguments as in Section~\ref{sec:SymBr2+1Free}, we
can show that the symmetry of the Lagrangian (\ref{L-free}) is $\mathrm{U}(2N_{f})$.

The three $4\times4$ $\gamma$-matrices in Eq.~(\ref{L-free}) can be chosen 
in the same form as in Eq.~(\ref{eq:11}). Then, for each four-component spinor, 
there is a global $\mathrm{U}(2)$ symmetry with the generators in Eq.~(\ref{T_i-generators}).
Taking into account that there are $N_f$ fermion flavors, the full symmetry of the action
(\ref{L-free}) is $\mathrm{U}(2N_f)$ with the generators 
\begin{equation}
\frac{\lambda^\alpha}{2},\qquad 
\frac{\lambda^\alpha}{2i}\gamma^3,\qquad 
\frac{\lambda^\alpha}{2} \gamma^5,\qquad\mbox{and}\qquad 
\frac{\lambda^\alpha}{2}\frac{1}{2}[\gamma^3,\gamma^5],
\label{U2Nf-generators}
\end{equation}  
which are obtained by the direct product of the four Dirac generators in Eq.~(\ref{T_i-generators})
and the $N_f^2$ generators of the flavor $\mathrm{U}(N_{f})$ symmetry, i.e., $\lambda^\alpha/2$, with 
$\alpha=0,1,\dots, N_f^2-1$. 
Adding a mass (gap) term $m_{0}\bar\psi\psi$ into the action
(\ref{L-free}) would reduce the $\mathrm{U}(2N_{f})$ symmetry down to 
$\mathrm{U}(N_f)\times \mathrm{U}(N_f)$ with the generators
\begin{equation}
\frac{\lambda^\alpha}{2},\qquad
\frac{\lambda^\alpha}{2}\frac{1}{2}[\gamma^3,\gamma^5],
\end{equation}  
with $\alpha=0,1,\dots, N_f^2-1$. This implies that the dynamical generation of the 
fermion gap leads to the spontaneous breakdown of the $\mathrm{U}(2N_f)$ 
symmetry down to $\mathrm{U}(N_f)\times \mathrm{U}(N_f)$.

The Coulomb interaction between quasiparticles is provided by gauge fields 
which, unlike the quasiparticles themselves, are $3$-dimensional in nature.
Taking into account that, in realistic condensed matter systems, the Fermi 
velocity of gapless fermions $v_F$ is much less than than the speed of light $c$, 
one finds that the static Coulomb forces provide the dominant interactions 
between fermions. The corresponding term in the action takes the following 
form:
\begin{equation}
S_{qp} = 
\int dt \int dt^{\prime} 
\int d^{2} \mathbf{r} \int d^{2} \mathbf{r}^{\prime} 
 \bar{\Psi}(t,\mathbf{r})\gamma^{0}\Psi(t,\mathbf{r}) 
U(t-t^{\prime},|\mathbf{r}-\mathbf{r}^{\prime}|)
\bar{\Psi}(t^{\prime},\mathbf{r}^{\prime})\gamma^{0}
\Psi(t^{\prime},\mathbf{r}^{\prime}), 
\label{Coulomb-action}
\end{equation}  
In general, the potential $U(t,|\mathbf{r}|)$ contains the polarization effects. 
In the simplest case when the polarization effects due to quasiparticles are 
neglected, the Coulomb potential takes the following form:
\begin{equation}
U_{0}(t,|\mathbf{r}|) = \frac{e^2\delta(t)}{\varepsilon_{0}} 
\int \frac{d^{2}\mathbf{k}}{(2\pi)^{2}}  
\exp(i\mathbf{k}\cdot\mathbf{r})\frac{2\pi}{|\mathbf{k}|}
=\frac{e^{2}\delta(t)}{\varepsilon_{0}|\mathbf{r}|},
\end{equation} 
where $\varepsilon_{0}$ is a dielectric constant.
Note, however, that in many cases of interest (e.g., in the case of a
finite temperature and/or a finite density and/or a nonzero magnetic
field), the polarization effects may considerably modify this bare Coulomb
potential. Thus, the interaction should rather be given by
\begin{equation}
U(t,|\mathbf{r}|) = \frac{e^2}{\varepsilon_{0}}
\int\frac{d\omega}{2\pi} \int \frac{d^{2}\mathbf{k}}{2\pi}
\frac{\exp(-i\omega t+i\mathbf{k}\cdot\mathbf{r})}
{|\mathbf{k}|+\Pi(\omega,|\mathbf{k}|)},
\label{retard}
\end{equation}  
where the polarization function $\Pi(\omega,|\mathbf{k}|)$ is proportional (with a factor 
of $2\pi/\varepsilon_{0}$) to the time component of the photon polarization tensor.
In the presence of a strong magnetic field, the polarization effects were calculated 
in Ref.~\cite{Shpagin:1996ch}. After taking them into account, we arrive at the following 
modified interaction:
\begin{equation}
U(t,\mathbf{r}) = \delta(t)
\frac{e^2}{\varepsilon_{0}} \int \frac{d^{2}\mathbf{k}}{2\pi} 
\frac{\exp(i\mathbf{k}\cdot\mathbf{r})}{|\mathbf{k}|(1+a |\mathbf{k}|)} 
= \frac{e^{2}\pi \delta(t)}{2\varepsilon_{0}a}
\left[H_{0}\left(\frac{|\mathbf{r}|}{a}\right)
-N_{0}\left(\frac{|\mathbf{r}|}{a}\right)\right],
\end{equation}
where 
\begin{equation}
a= 2\pi\nu_{0} \frac{e^2 N_{f}}{\varepsilon_{0}v_{F}}\sqrt{\frac{c}{|eB|}},
\end{equation}
and the constant $\nu_{0}$ is given by
\begin{equation}
\nu_{0} \equiv \frac{1}{4\pi\sqrt{\pi}} \int_{0}^{\infty}
\frac{dz}{\sqrt{z}} \left(\frac{\coth(z)}{z}-\frac{1}{\sinh^{2}(z)}\right)
= -\frac{3\zeta(-0.5)}{\sqrt{2}\pi}\approx 0.14. 
\end{equation}
Here we used the following notation: $\zeta(z)$ is the Riemann $\zeta$-function,
$H_{0}(z)$ is the Struve function, and $N_{0}(z)$ is the Bessel function
of the second kind. Let us emphasize that the instantaneous approximation
for the polarization function is justified in this case: the frequency
dependence is suppressed by factors of order $\omega/\sqrt{v_{F}|eB}|$
(which are small in the case of the LLL dominance). This can
be shown directly from the expression for the polarization function in
Ref.~\cite{Shpagin:1996ch}.

Now, the gap equation for the quasiparticle propagator will be the same as in 
the relativistic models of Refs.~\cite{Gusynin:1995gt,Gusynin:1995nb,Gusynin:1998zq,
Gusynin:1999pq}. In the nonrelativistic case at hand, however, it is justified to 
neglect the retardation effects in the interaction potential. 

In the case of a subcritical coupling constant $g \leq g_{c}$, we will distinguish 
two different dynamical regimes. The first regime corresponds to the situation with a
weak coupling $g$, when it is outside the scaling region near the critical
value $g_c$. In this case the LLL dominates and the value of the dynamical
gap $m_{\rm dyn}$ is much less than the gap $\epsilon_{B} \equiv \sqrt{2v_{F}^2|eB|/c}$ 
between the Landau levels. The latter guarantees that the higher Landau levels
decouple from the pairing dynamics and the LLL dominates indeed.

The second, strong coupling, regime is that with a near-critical, although
subcritical, value of $g$. In that case, all Landau levels are relevant
for the pairing dynamics and the value of the dynamical gap $m_{\rm dyn}$ is
of the order of the Landau gap $\epsilon_{B}$.

Let us begin by considering the weak coupled regime. Then, the low-energy
dynamics is dominated by the LLL, and the quasiparticle propagator could
be approximated as follows:
\begin{equation}
\bar{G}(t,\mathbf{r}) = \frac{i|eB|}{4\pi c} 
\exp\left(-\frac{|\mathbf{r}|^{2} |eB|}{4c}\right)
g(t)\left[1+ i\gamma^{1}\gamma^{2}\mbox{sign}(eB)\right],
\label{LLL-app}
\end{equation}
where $g(t)$ is unknown matrix-valued function which should be determined
by solving the Schwinger-Dyson equation. By making use of the LLL ansatz
(\ref{LLL-app}), we derive the following gap equation for the Fourier transform 
of $g(t)$:
\begin{equation}
g^{-1}(\omega) = g_{0}^{-1}(\omega) - i e^{2} \int \frac{d \omega^{\prime}}{2\pi}
\gamma^{0} g(\omega-\omega^{\prime}) \gamma^{0}  \int \frac{d^{2} \mathbf{k}}{(2\pi)^{2}} 
\exp\left(-\frac{c|\mathbf{k}|^{2}}{2|eB|}\right) U(\mathbf{k}).
\label{eq-omega}
\end{equation}
The general structure of the function $g(\omega)$ is suggested by the first (LLL) term 
in the well-known Schwinger propagator, in which the bare gap $m_{0}$ is replaced by 
the dynamical gap function $\Delta_{\omega}$ and additionally the wave function 
renormalization $A_{\omega}$ is introduced. Thus, we have
\begin{equation}
g(\omega) = \frac{A_{\omega}\gamma^{0}\omega +\Delta_{\omega}}
{A_{\omega}^{2}\omega^{2}-\Delta_{\omega}^{2}}.
\end{equation}
The free propagator 
$g_{0}(\omega)$ has a similar structure, but the value of the bare gap is assumed to 
be zero. 

It is easy to check that the integral on the right hand side of
Eq.~(\ref{eq-omega}) is independent of $\omega$. This implies that
$A_{\omega}=1$ and the gap $\Delta_{\omega}$ is independent of $\omega$.
By taking this into account, we straightforwardly derive the solution:
\begin{equation}
m_{\rm dyn} \equiv \Delta_{0} = \frac{g}{\sqrt{2}} \sqrt{\frac{v_{F}^{2}|eB|}{c}}
\int_{0}^{\infty} 
\frac{dk \exp(-k^2)}{1+k \chi_{0}},
\label{gap-T0}
\end{equation}
where $\chi_{0} = 2\sqrt{2}\pi\nu_{0} g N_{f}$. 
In the two limiting cases, $\chi_{0} \ll 1$ and $\chi_{0}\gg 1$, we get the following asymptotes:
\begin{equation}
m_{\rm dyn} \equiv \Delta_{0} \simeq  \frac{g\sqrt{\pi}}{2\sqrt{2}} 
\sqrt{\frac{v_{F}^{2}|eB|}{c}} \left(1-\frac{\chi_{0}}{\sqrt{\pi}} +
\frac{\chi_{0}^{2}}{2}+\cdots\right),
\label{weak}
\end{equation}
(for weak coupling and small $N_{f}$) and
\begin{equation} 
m_{\rm dyn} \equiv\Delta_{0} \simeq \frac{g}{\sqrt{2}} 
\sqrt{\frac{v_{F}^{2}|eB|}{c}} \frac{\ln\chi_{0}}{\chi_{0}}
\equiv \frac{v_{F}}{4\pi\nu_{0}N_{f}}\sqrt{\frac{|eB|}{c}}\ln\chi_{0},
\label{largeN_f}
\end{equation}
(for large $N_{f}$). In accordance with the general conclusion of 
Refs.~\cite{Gusynin:1995gt,Gusynin:1995nb}, in a magnetic field the gap is
generated for any nonzero coupling constant $g=e^2/\varepsilon_{0}v_F$.

One can see that for a sufficiently small $g=e^2/\varepsilon_{0}v_F$ in expression 
(\ref{weak}) and for a sufficiently large $N_f$ in (\ref{largeN_f}), the LLL approximation 
is indeed self-consistent. In both cases, the gap $\Delta_0$ can be made much 
less than the Landau gap (scale) $\epsilon_{B}$. We emphasize that the second 
solution (\ref{largeN_f}), obtained also in Ref.~\cite{Khveshchenko:2001zza},
corresponds to the regime with a large $N_{f}$ and {\it not} to the strong
coupling regime with a large $g$ and $N_f$ of order one. Indeed, taking
$g$ to be large enough in expression (\ref{largeN_f}), one gets the gap
$\Delta_0$ exceeding the Landau scale $\epsilon_{B} $, i.e., for large $g$ the
self-consistency of the LLL dominance approximation is lost. We will
discuss the strong coupling regime below.

What is the energy scale the coupling constant $g$ relates to in this
problem? It is the Landau scale $\epsilon_{B} $. The argument supporting this goes as
follows. There are two, dynamically very different, scale regions in this
problem. One is the region with the energy scale above the Landau scale
$\epsilon_{B}$ and below the ultraviolet cutoff $\Lambda$, defined by the lattice
size. In that region, the dynamics is essentially the same as in the
theory without magnetic field. In particular, the running coupling
decreases logarithmically with the energy scale there \cite{Gonzalez:1999wo}.
Another is the region below the Landau scale $\epsilon_{B}$. In that region, the 
magnetic field dramatically changes the dynamics, in particular, the behavior of
the running coupling constant.  As the analysis of this section shows,
because of the magnetic field, the pairing dynamics (in the particle-hole
channel) is dominated by the infrared region where $\omega \lesssim
m_{\rm dyn} $. Therefore, the scale region above the Landau scale $\epsilon_{B}$
completely decouples from the pairing dynamics in this case. This
manifests itself in expression (\ref{gap-T0}) for the gap: the only
relevant scale is the Landau scale $\epsilon_{B}$ there. Since the effect of the
running of the coupling is taken into account by the polarization function
in the gap equation, we conclude that the coupling $g$ indeed relates to
the Landau scale in this problem. Notice that it can be somewhat smaller
than the bare coupling constant $g(\Lambda)$ related to the scale
$\Lambda$. Taking $\Lambda = 2.4$ eV in graphite (the width of its energy
band) and using the equation for the running coupling from
Ref.~\cite{Gonzalez:1999wo}, one gets that it is smaller by the factors 1.2
and 1.4 than $g(\Lambda)$ for the values of the magnetic field $B = 10$ T
and $B = 0.1$ T, respectively.

Now let us turn to the second dynamical regime at strong coupling. In reduced QED, 
the gap equation in this regime includes the contributions of all Landau levels 
and becomes very formidable. Still, one can estimate the value of the gap in the
strong coupling regime: since there are no small parameters in this regime
for moderate values of $N_f$, the gap should be of the order of the Landau
scale $\epsilon_{B}$. This conclusion is supported by studying the scaling 
regime in a simpler model, a $(2+1)$-dimensional Nambu-Jona-Lasinio model 
\cite{Gusynin:1995gt,Gusynin:1995nb}, considered in Section~\ref{sec:NJL2+1General}. 
In the critical regime, the result for the gap is given by Eq.~(\ref{eq:mdynB}), which can 
be rewritten as $m_{\rm dyn} \simeq 0.315 \epsilon_{B}$, where the Landau scale in the 
relativistic model (with $v_{F}=c$) is $\epsilon_{B} = \sqrt{2c|eB|}$.

\subsection{QCD in a strong magnetic field}
\label{sec:MagCatQCD}

Studies of QCD in external electromagnetic fields had started long time ago 
\cite{Klevansky:1989vi,Suganuma:1990nn,Schramm:1991ex} by
using the Nambu-Jona-Lasinio (NJL) model as a low-energy effective
theory of QCD. Based on these studies, it was concluded that a magnetic
field always enhances the chiral condensate in QCD. Somewhat later, 
based on the property of the asymptotic freedom in QCD, it was suggested 
in Ref.~\cite{Kabat:2002er} that the dynamics in QCD in a magnetic field
is weakly coupled at sufficiently large magnetic fields. This point is
similar to those well known facts that the dynamics in both hot QCD \cite{Gross:1980br} 
and QCD with a large baryon charge \cite{Collins:1974ky,Rajagopal:2000wf,
Rischke:2003mt} are weakly coupled.

In Ref.~\cite{Miransky:2002rp}, the magnetic catalysis in QCD was studied
rigorously, from first principles. In fact, it was shown that, at sufficiently strong magnetic
fields, $|eB| \gg \Lambda_{\rm QCD}^2$, there exists a consistent truncation of the Schwinger-Dyson 
(gap) equation which leads to a reliable asymptotic expression for the quark mass $m_{q}$  
[see Eq.~(\ref{gapQCD})], where $q$ is the electric charge of the $q$-th quark. As we discuss below, because of 
the running of the QCD coupling $\alpha_{s}$, the dynamical mass $m_{q}$ grows very slowly with increasing the value
of the background magnetic field. Moreover, there may exist an intermediate
region of fields where the mass {\it decreases} with increasing the magnetic field.
Another, rather unexpected, consequence is that a strong external magnetic
field can {\it suppress} the chiral vacuum fluctuations leading to the
generation of the usual dynamical mass of quarks $m^{(0)}_{\rm dyn} \simeq 
300~\mbox{MeV}$ in QCD without a magnetic field. In fact, in a wide range of strong magnetic fields 
$\Lambda^{2} \lesssim B \lesssim (10 \mbox{TeV})^{2}$, where $\Lambda$ is the characteristic gap in QCD without the
magnetic field (it can be estimated to be a few times larger than
$\Lambda_{\rm QCD}$), the dynamical mass in a magnetic field remains {\sl smaller}
than $m^{(0)}_{\rm dyn}$. As it will be shown below, this point is
intimately connected with another one: in a strong magnetic field, the
confinement scale, $\lambda_{\rm QCD}$, is much less than the confinement
scale $\Lambda_{\rm QCD}$ in QCD without a magnetic field. 

This picture is very different from that in QED in a magnetic field, which is not surprising:
the QCD and QED dynamics are very different. On the other hand, like in QED in a magnetic field,
one of the central dynamical issues underlying the magnetic catalysis in QCD is the effect of
screening of the gluon interactions in a magnetic field in the region of
momenta relevant for the chiral symmetry breaking dynamics, $m_{q}^2 \ll
|k^2| \ll |eB|$. In this region, gluons acquire a mass $M_{g}$ of order
$\sqrt{N_{f}\alpha_{s}|e_{q}B|}$. This allows us to separate the dynamics
of the magnetic catalysis from that of confinement. More rigorously,
$M_{g}$ is the mass of a quark-antiquark composite state coupled to
the gluon field. As in QED in a magnetic field, the appearance of such 
mass resembles the pseudo-Higgs effect in the $(1+1)$-dimensional {\it massive} 
QED (massive Schwinger model \cite{Schwinger:1962tp,{Coleman:1976uz}}) (see below).

The flavor symmetries in magnetic QCD and magnetic QED are also very different. 
Since the background magnetic field breaks explicitly the global chiral
symmetry that interchanges the up and down quark flavors, the chiral
symmetry in the case of QCD is $\mathrm{SU}(N_{u})_{L}\times \mathrm{SU}(N_{u})_{R} \times
\mathrm{SU}(N_{d})_{L}\times \mathrm{SU}(N_{d})_{R}\times \mathrm{U}^{(-)}(1)_{A}$.
The $\mathrm{U}^{(-)}(1)_{A}$ is connected with the current which is an
anomaly free linear combination of the $\mathrm{U}^{(d)}(1)_{A}$ and $\mathrm{U}^{(u)}(1)_{A}$ currents.
[The $\mathrm{U}^{(-)}(1)_{A}$ symmetry is of course absent if either
$N_d$ or $N_u$ is equal to zero]. The generation of quark masses
breaks this symmetry spontaneously down to $\mathrm{SU}(N_{u})_{V}\times
\mathrm{SU}(N_{d})_{V}$ and, as a result, $N_{u}^{2}+N_{d}^{2}-1$ gapless
Nambu-Goldstone (NG) bosons appear in the spectrum. In Section~\ref{sec:MagCatQCD3}, we derive the
effective action for the NG bosons and calculate their decay constants and
velocities. 

The two main characteristics of the QCD dynamics are of course the properties of asymptotic
freedom and confinement. In connection with that, our second major
result is the derivation of the low-energy effective action for gluons in
QCD in a strong magnetic field [see Eq.~(\ref{gluon-action}) in Section~\ref{sec:MagCatQCDgluo}]. 
The characteristic feature of this action is its anisotropic dynamics. In
particular, the strength of static (Coulomb like) forces along the
direction parallel to the magnetic field is much larger than that in the
transverse directions.  Also, the confinement scale in this theory is much
less than that in QCD without a magnetic field. These features imply a rich
and unusual spectrum of light glueballs in this theory.

A special and interesting case is QCD with a large number of colors, in
particular, with $N_c \to \infty$ (the 't Hooft limit). In this limit, the
mass of gluons goes to zero and the expression for the dynamical quark mass becomes
essentially different [see Eq.~(\ref{m_q}) in Section~\ref{sec:MagCatQCDinfty}]. In fact,
it will be shown that, for any value of an external magnetic field, there
exists a threshold value $N^{\rm thr}_{c}$, rapidly growing with $|eB|$
[e.g., $N^{\rm thr}_c \gtrsim 100$ for $|eB| \gtrsim (1\mbox{ GeV})^2$].
For $N_c$ of the order $N^{\rm thr}_{c}$ or larger, the gluon mass becomes small and irrelevant for
the dynamics of the generation of a quark mass. As a result, expression
(\ref{m_q}) for $m_q$ takes place for such large $N_c$. The confinement
scale in this case is close to $\Lambda_{\rm QCD}$. Still, as is shown in
Section~\ref{sec:MagCatQCDinfty}, the dynamics of chiral symmetry breaking is under
control in this limit if the magnetic field is sufficiently strong.

It is important that, unlike the case of QCD with a nonzero baryon
density, there are no principal obstacles for checking all these results
and predictions in lattice computer simulations of QCD in a magnetic
field. We will discuss the main results of such lattice simulations below 
in Section~\ref{sec:MagCatLattice}.

\subsubsection{Magnetic catalysis}
\label{sec:MagCatQCDgap}
 
We begin the analysis by considering the Schwinger-Dyson (gap) equation for 
the quark propagator. It has the following form:
\begin{equation}
G^{-1}(u,u^\prime) = S^{-1}(u,u^\prime) + 4\pi\alpha_{s} \gamma^{\mu}\int G(u,v)
\Gamma^{\nu}(v,u^\prime,v^{\prime}) {\cal D}_{\nu\mu}(v^{\prime},u) 
d^{4} v d^{4} v^{\prime}, 
\label{SD}     
\end{equation}
where $S(u,u^\prime)$ and $G(u,u^\prime)$ are the bare and full fermion propagators
in an external magnetic field, ${\cal D}_{\nu\mu}(u,u^\prime) $ is the 
full gluon propagator and $\Gamma^{\nu}(u,u^\prime,v)$ is the full amputated
vertex function. Since the coupling $\alpha_s$ related to the scale 
$|eB|$ is small, one might think that the rainbow (ladder) approximation
is reliable in this problem. However, this is not the case.
Because of the $(1+1)$-dimensional form of the fermion propagator
in the LLL approximation, there are relevant higher order 
contributions \cite{Gusynin:1998zq,Gusynin:1999pq}. Fortunately one can solve this problem.
First of all, an important feature of the quark-antiquark pairing dynamics 
in QCD in a strong magnetic field is that this dynamics is essentially 
Abelian. This feature is provided by the form of the polarization 
operator of gluons in this theory. The point is that the dynamics 
of the quark-antiquark pairing is mainly induced 
in the region of momenta $k$ 
much less than $\sqrt{|eB|}$. This implies that the magnetic field 
yields a dynamical ultraviolet cutoff in this problem. On the other
hand, while the contribution of (electrically neutral) gluons
and ghosts in the polarization operator is proportional to
$k^2$, the fermion contribution is proportional to $|e_{q}B|$
\cite{Gusynin:1998zq,Gusynin:1999pq}. As a result, the fermion contribution dominates
in the relevant region with $k^2 \ll |eB|$.

This observation implies that there are three, dynamically very 
different, scale regions in this problem. The first one is the region 
with the energy scale above the magnetic scale $\sqrt{|eB|}$.
In that region, the dynamics is essentially the same as in QCD
without a magnetic field. In particular, the running coupling 
decreases logarithmically with increasing the energy scale there. The
second region is that with the energy scale below the magnetic scale
but much larger than the mass (gap) $m_{q}$. In this region,
the dynamics is Abelian-like and, therefore, the dynamics of the 
magnetic catalysis is similar to that in QED in a magnetic field. 
At last, the third region is the region with the energy scale less 
than the gap. In this region, quarks decouple and an anisotropic confinement 
dynamics for gluons is realized.
 
Let us first consider the intermediate region relevant for the 
magnetic catalysis. As was indicated above, the important ingredient 
of this dynamics is a large contribution of fermions to the 
polarization operator. It is large because of an (essentially)
$(1+1)$-dimensional form of the fermion propagator in a strong magnetic
field. Its explicit form can be obtained by modifying
appropriately the expression for the polarization operator
in QED in a magnetic field \cite{Batalin:1971au,Tsai:1974pi,
Shabad:1975ik,1976PhLA...56..151L,Dittrich:1985yb,Calucci:1993fi}:
\begin{eqnarray}
{\cal P}^{AB,\mu\nu} \simeq \frac{\alpha_{s}}{6\pi}
\delta^{AB} \left(k_{\parallel}^{\mu}
k_{\parallel}^{\nu}-k_{\parallel}^{2}g_{\parallel}^{\mu\nu}\right)
\sum_{q=1}^{N_{f}}\frac{|e_{q}B|}{m^{2}_{q}},
\quad \mbox{for} \quad |k_{\parallel}^2| \ll m_{q}^2,
\label{Pi-IR-QCD}\\
{\cal P}^{AB,\mu\nu} \simeq -\frac{\alpha_{s}}{\pi}
\delta^{AB} \left(k_{\parallel}^{\mu}
k_{\parallel}^{\nu}-k_{\parallel}^{2}g_{\parallel}^{\mu\nu}\right)
\sum_{q=1}^{N_{f}}\frac{|e_{q}B|}{{k_{\parallel}^2}}, 
\quad \mbox{for} \quad m_{q}^2 \ll |k_{\parallel}^2|\ll |eB|  .
\label{Pi-UV-QCD}
\end{eqnarray}
[Compare with Eqs.~(\ref{Pi-IR}) and (\ref{Pi-UV}).]
Here $g_{\parallel}^{\mu\nu}\equiv \mbox{diag}(1,0,0,-1)$ is
the projector onto the longitudinal subspace,
and $k_{\parallel}^{\mu}\equiv g_{\parallel}^{\mu\nu} k_{\nu}$
(the magnetic field is in the $z$ direction).
Similarly, we introduce the orthogonal projector $g_{\perp}^{\mu\nu}\equiv
g^{\mu\nu}-g_{\parallel}^{\mu\nu}=\mbox{diag}(0,-1,-1,0)$ and
$k_{\perp}^{\mu}\equiv g_{\perp}^{\mu\nu} k_{\nu}$ that
we will use below. Notice that quarks in a strong magnetic field
do not couple to the transverse subspace spanned by $g_{\perp}^{\mu\nu}$
and $k_{\perp}^{\mu}$. This is because in a strong magnetic field 
only the quark from the LLL matter and they couple only to the
longitudinal components of the gluon field. The latter property 
follows from the fact that spins of the LLL quarks are polarized 
along the magnetic field \cite{Gusynin:1995gt,Gusynin:1995nb}.  

The expressions (\ref{Pi-IR-QCD}) and (\ref{Pi-UV-QCD}) coincide with 
those for the polarization operator in the massive Schwinger
model if the parameter $\alpha_{s} |e_{q}B|/2$ here is replaced by 
the dimensional coupling $\alpha_{1}$ of QED$_{1+1}$. As in the 
Schwinger model, Eq.~(\ref{Pi-UV}) implies that there is a massive 
resonance in the $k_{\parallel}^{\mu}k_{\parallel}^{\nu}
-k_{\parallel}^{2} g_{\parallel}^{\mu\nu}$ component of the gluon 
propagator. Its mass is
\begin{equation}
M_{g}^2= \sum_{q=1}^{N_{f}}\frac{\alpha_{s}}{\pi}|e_{q}B|=
(2N_{u}+N_{d}) \frac{\alpha_{s}}{3\pi}|eB|.
\label{M_g}
\end{equation}
This is reminiscent of the pseudo-Higgs effect in the 
$(1+1)$-dimensional massive QED. It is not the genuine Higgs effect 
because there is no complete screening of the color charge in the 
infrared region with $|k_{\parallel}^2|\ll m_{q}^2$. This can 
be seen clearly from Eq.~(\ref{Pi-IR-QCD}). Nevertheless, the pseudo-Higgs 
effect is manifested in creating a massive resonance and this 
resonance provides the dominant forces leading to chiral
symmetry breaking.  

Now, after the Abelian-like structure of the dynamics in this 
problem is established, we can use the results of the analysis in 
QED in a magnetic field \cite{Gusynin:1998zq,Gusynin:1999pq} by introducing appropriate 
modifications. The main points of the analysis are: (i) the so
called improved rainbow approximation is reliable in this problem
provided a special nonlocal gauge is used in the analysis, 
and (ii) for a small coupling $\alpha_{s}$ the relevant region of momenta in
this problem is $m_{q}^2 \ll |k^2| \ll |eB|$. We recall 
that in the improved rainbow approximation the vertex 
$\Gamma^{\nu}(u,u^\prime,z)$ is taken to be bare and the gluon propagator 
is taken in the one-loop approximation. Moreover, as we argued 
above, in this intermediate region of momenta, only the contribution
of quarks to the gluon polarization tensor (\ref{Pi-UV}) matters.
It may be appropriate to call this approximation the 
``strong-magnetic-field-loop" improved rainbow approximation.
As to the modifications, they are purely kinematic: the overall 
coupling constant in the gap equation $\alpha$ and the dimensionless
combination $M_{\gamma}^2/|eB|$ in QED have to be replaced by 
$\alpha_s(N_{c}^2-1)/2N_{c}$ and $M_{g}^2/|e_{q} B|$, respectively. 
This leads us to the expression for the dynamical mass (gap),
\begin{equation}
m_{q}^2 \simeq 2 C_{1} |e_{q}B|
\left(c_{q}\alpha_{s}\right)^{2/3}
\exp\left[-\frac{4N_{c}\pi}{\alpha_{s} (N_{c}^{2}-1)
\ln(C_{2}/c_{q}\alpha_{s})}\right],
\label{gapQCD}
\end{equation}
where $e_{q}$ is the electric charge of the $q$-th quark and $N_{c}$
is the number of colors. The numerical factors $C_1$ and $C_2$ equal
$1$ in the leading approximation that we use. Their value, however, can
change beyond this approximation and we can only say that they are 
of order $1$. The constant $c_{q}$ is defined as follows:
\begin{equation}
c_{q} = \frac{1}{6\pi}(2N_{u}+N_{d})\left|\frac{e}{e_{q}}\right|,
\end{equation}
where $N_{u}$ and $N_{d}$ are the numbers of up and down quark 
flavors, respectively. The total number of quark flavors is $N_{f}
=N_{u}+N_{d}$. The strong coupling $\alpha_{s}$ in the last equation is 
related to the scale $\sqrt{|eB|}$, i.e.,
\begin{equation}
\frac{1}{\alpha_{s}} \simeq b\ln\frac{|eB|}{\Lambda_{\rm QCD}^2},
\quad \mbox{ where} \quad
b=\frac{11 N_c -2 N_f}{12\pi}. 
\label{coupling1}
\end{equation}
We should note that in the leading approximation  
the energy scale $\sqrt{|eB|}$ in Eq.~(\ref{coupling1}) 
is fixed only up to a factor of order $1$.

After expressing the magnetic field in terms of the running coupling, 
the result for the dynamical mass takes the following convenient form:
\begin{equation}
m_{q}^2 \simeq 2C_{1}\left|\frac{e_{q}}{e}\right| \Lambda_{\rm QCD}^2 
\left(c_{q}\alpha_{s}\right)^{2/3}
\exp\left[\frac{1}{b\alpha_{s}}-\frac{4N_{c}\pi}{\alpha_{s} 
(N_{c}^{2}-1) \ln(C_{2}/c_{q}\alpha_{s})}\right].
\label{gap-vs-alpha}
\end{equation}
As is easy to check, the dynamical mass of the $u$-quark 
is considerably larger than that of the $d$-quark. It is also
noticeable that the values of the $u$-quark dynamical mass
becomes comparable to the vacuum value 
$m^{(0)}_{\rm dyn}\simeq 300 \mbox{ MeV}$
only when the coupling constant gets as small as $0.05$. 

\begin{figure}[t]
\begin{center}
\includegraphics[width=0.47\textwidth]{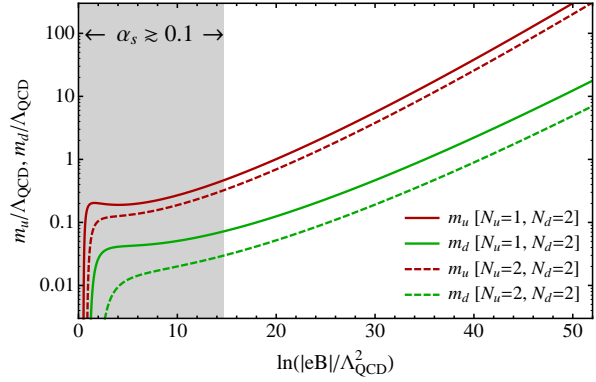}
\caption{(Color online) The dynamical masses of the $u$-quarks (red color) and $d$-quarks (green color) as 
functions of the magnetic field for $N_{c}=3$ and two different choices of the number of flavors: 
(i) $N_{u}=1$ and $N_{d}=2$ (solid lines), and 
(ii) $N_{u}=2$ and $N_{d}=2$ (dashed lines).
The result may not be reliable in the weak magnetic field region (shaded) where 
the running coupling constant becomes strong ($\alpha_s\gtrsim 0.1$). The values of 
masses are given in units of $\Lambda_{\rm QCD} = 250~\mbox{MeV}$.}
\label{fig-DynMassQCD}
\end{center}
\end{figure}

Now, by trading the coupling constant for the magnetic field scale
$|eB|$ using Eq.~(\ref{gap-vs-alpha}), we get the dependence of the dynamical mass 
on the value of the external field. The numerical results 
are presented in  Fig.~\ref{fig-DynMassQCD} [we used 
$C_{1} = C_{2} = 1$ in Eq.~(\ref{gap-vs-alpha})]. 

As one can see in Fig.~\ref{fig-DynMassQCD}, the value of the quark mass
in a wide window of strong magnetic fields, $\Lambda_{\rm QCD}^{2}\ll |eB|
\lesssim (10 \mbox{ TeV})^{2}$, remains smaller than the dynamical mass of
quarks $m^{(0)}_{\rm dyn} \simeq 300 \mbox{ MeV}$ in QCD without a magnetic field.  
In other words, the chiral condensate is partially {\it suppressed} for
those values of a magnetic field.
The explanation of this, rather unexpected, result is
actually simple. The magnetic field leads to the mass $M_g$ (\ref{M_g})  
for gluons. In a strong enough magnetic field,
this mass becomes larger than the characteristic gap
$\Lambda$ in QCD without a magnetic field ($\Lambda$, playing the role of
a gluon mass, can be estimated as a few times larger than
$\Lambda_{\rm QCD}$). This, along with the property of the asymptotic freedom
(i.e., the fact that $\alpha_{s}$ decreases with increasing the magnetic
field), leads to the suppression of the chiral condensate. 

This point also explains why our result for the gap is so different from
that in the NJL model in a magnetic field \cite{Klevansky:1989vi,Suganuma:1990nn}. 
Recall that, in the NJL model, the gap logarithmically (i.e., much faster than in the
present case) grows with a magnetic field. This is the related to the
assumption that both the dimensional coupling constant $G =
g/\Lambda^2$ (with $\Lambda$ playing a role similar to that of the gluon
mass in QCD), as well as the scale $\Lambda$ do not dependent on the value
of the magnetic field. Therefore, in that model, in a strong enough
magnetic field, the value of the chiral condensate is overestimated.

The picture which emerges from this discussion is the following.
For values of a magnetic field $|eB| \lesssim \Lambda^2$ the
dynamics in QCD should be qualitatively similar to that in
the NJL model. For strong values of the field, however, it
is essentially different, as was described above. This in turn
suggests that there should exist an intermediate region of
fields where the dynamical masses of quarks decrease with
increasing the background magnetic field.

\subsubsection{Effective action of NG bosons}
\label{sec:MagCatQCD3}

The presence of the background magnetic field breaks explicitly the global
chiral symmetry that interchanges the up and down quark flavors. This is
related to the fact that the electric charges of the two sets of quarks
are different. However, the magnetic field does not break the global
chiral symmetry of the action completely.  In particular, in the model
with the $N_{u}$ up quark flavors and the $N_{d}$ 
down quark flavors, the action is invariant under the chiral symmetry
$\mathrm{SU}(N_{u})_{L}\times \mathrm{SU}(N_{u})_{R} \times \mathrm{SU}(N_{d})_{L}\times
\mathrm{SU}(N_{d})_{R}\times \mathrm{U}^{(-)}(1)_{A}$. 
The $\mathrm{U}^{(-)}(1)_{A}$ is connected with the current which is an
anomaly free linear combination
of the $\mathrm{U}^{(d)}(1)_{A}$ and $\mathrm{U}^{(u)}(1)_{A}$ currents.
[The $\mathrm{U}^{(-)}(1)_{A}$ symmetry is of course absent if either
$N_d$ or $N_u$ is equal to zero].

The global chiral symmetry of the action is broken spontaneously down to
the diagonal subgroup $\mathrm{SU}(N_{u})_{V}\times \mathrm{SU}(N_{d})_{V}$ when 
dynamical masses of quarks are generated. In agreement with the
Goldstone theorem, this leads to the
appearance of $N_{u}^{2}+N_{d}^{2}-1$ number of the NG gapless excitations
in the low-energy spectrum of QCD in a strong magnetic field. Notice
that there is also a pseudo-NG boson connected with the conventional
(anomalous) $\mathrm{U}(1)_A$ symmetry which can be rather light in a
sufficiently strong magnetic field.
   
Now, in the chiral limit, the general structure of the low-energy action
for the NG bosons could be easily established from the symmetry arguments
alone. First of all, such an action should be invariant with respect to
the space-time symmetry $\mathrm{SO}(1,1)\times \mathrm{SO}(2)$ which is left unbroken by
the background magnetic field [here the $\mathrm{SO}(1,1)$ and the $\mathrm{SO}(2)$ are
connected with Lorentz boosts in the $t$-$z$ hyperplane and
rotations in the $x$-$y$ plane, respectively]. Besides that, the
low-energy
action should respect the original chiral symmetry $\mathrm{SU}(N_{u})_{L}\times
\mathrm{SU}(N_{u})_{R} \times \mathrm{SU}(N_{d})_{L}\times \mathrm{SU}(N_{d})_{R}\times
\mathrm{U}^{(-)}(1)_{A}$. 
These
requirements lead to the following general form of the action:
\begin{eqnarray}
{\cal L}_{NG} &\simeq &\frac{f_{u}^{2}}{4}
\mathrm{tr} \left( g_{\parallel}^{\mu\nu}
\partial_{\mu}\Sigma_{u}\partial_{\nu}\Sigma_{u}^{\dagger}
+v_{u}^{2} g_{\perp}^{\mu\nu}
\partial_{\mu}\Sigma_{u}\partial_{\nu}\Sigma_{u}^{\dagger}\right)
+\frac{f_{d}^{2}}{4}
\mathrm{tr} \left(g_{\parallel}^{\mu\nu}
\partial_{\mu}\Sigma_{d}\partial_{\nu}\Sigma_{d}^{\dagger}
+v_{d}^{2} g_{\perp}^{\mu\nu}
\partial_{\mu}\Sigma_{d}\partial_{\nu}\Sigma_{d}^{\dagger}\right)
\nonumber\\
&+& \frac{\tilde{f}^{2}}{4}
\left(g_{\parallel}^{\mu\nu}
\partial_{\mu}\tilde{\Sigma}\partial_{\nu}
\tilde{\Sigma}^{\dagger}
+\tilde{v}^{2} g_{\perp}^{\mu\nu}
\partial_{\mu}\tilde{\Sigma}\partial_{\nu}\tilde{\Sigma}^{\dagger}\right).
\label{low-e-NG}
\end{eqnarray}
The unitary matrix fields 
$\Sigma_{u}\equiv \exp \left(i\sum_{A=1}^{N_{u}^2-1}\lambda^{A}\pi_{u}^{A}/f_{u} \right)$,
$\Sigma_{d}\equiv \exp \left( i\sum_{A=1}^{N_{d}^2-1}\lambda^{A}\pi_{d}^{A}/f_{d} \right)$, 
and $\tilde{\Sigma} \equiv \exp \left({i\sqrt{2}}\tilde{\pi}/\tilde{f} \right)$ describe the NG 
bosons in the up, down, and $\mathrm{U}^{(-)}(1)_{A}$ sectors of the original theory. 
The decay constants $f_{u}, f_{d}, \tilde{f}$ and transverse velocities $v_{u}, v_{d}, 
\tilde{v}$ can be calculated by using the standard field theory formalism (for a review, 
see for example the book \cite{Miransky:1994vk}). Let us first consider the 
$N_{u}^{2}+N_{d}^{2}-2$ NG bosons in the up and down sectors, assigned 
to the adjoint representation of the 
$\mathrm{SU}(N_{u})_{V}\times \mathrm{SU}(N_{d})_{V}$ symmetry. 
The basic relation is
\begin{equation}
\delta^{AB}P_{q}^{\mu} f_{q} = -i\int 
\frac{d^{4}k}{(2\pi)^{4}}  \mathrm{tr}
\left(\gamma^{\mu}\gamma^{5}\frac{\lambda^A}{2}
\chi^{B}_{q}(k,P)\right),
\label{decay} 
\end{equation}
where $P_{q}^{\mu}= \left(P^{0}, v_{q}^{2}\, \mathbf{P}_{\perp}, P^{3}\right)$
and $\chi^{A}_{q}(k,P)$ is the Bethe-Salpeter wave function of the NG
bosons ($P$ is the momentum of their center of mass). In the weakly
coupled dynamics at hand, one could use an analogue of the Pagels-Stokar 
approximation \cite{Pagels:1979hd,Miransky:1994vk}. 
In this approximation, the Bethe-Salpeter wave function is
determined from the Ward identities for axial currents. In
fact, the calculation of the decay constants and velocities of NG bosons 
resembles closely the calculation in the case of a color superconducting 
dense quark matter \cite{Miransky:2000jt,Miransky:2000bd}. In the LLL approximation, the final result
in Euclidean space is
\begin{eqnarray}
f_{q}^2 &=&  4N_c\int \frac{d^{2}k_{\perp} 
d^{2}k_{\parallel}}{(2\pi)^4}
\exp\left(-\frac{k^{2}_{\perp}}{|e_{q}B|}\right)
\frac{m_{q}^2}{(k^{2}_{\parallel} + m_{q}^2)^2},
\qquad v_{q}=0.
\label{Pagels}
\end{eqnarray}
The evaluation of this integral is straightforward. As a result,
we get 
\begin{eqnarray}
f_{u}^{2} &=& \frac{N_c}{6\pi^{2}}|eB|,\\ 
f_{d}^{2} &=& \frac{N_c}{12\pi^{2}}|eB|.
\end{eqnarray}
The remarkable fact is that the decay constants are nonzero
even in the limit when the dynamical masses of quarks 
approach zero. The reason of that is the $(1+1)$-dimensional
character of this dynamics: as one can see from expression
(\ref{Pagels}), in the limit $m_{q} \to 0$, the infrared
singularity in the integral cancels the mass $m_q$ in the
numerator. A similar situation takes place in color
superconductivity \cite{Beane:2000ms,Son:1999cm,Son:2000tu,
Zarembo:2000pj,Miransky:2000jt,Miransky:2000bd}: in that case the
$(1+1)$-dimensional character of the dynamics is provided by
the Fermi surface. 

Notice that the transverse velocities of the NG bosons are equal to zero.
This is also a consequence of the $(1+1)$-dimensional structure of the quark
propagator in the LLL approximation.  The point is that quarks can move in
the transverse directions only by hopping to higher Landau levels. Taking
into account higher Landau levels would lead to nonzero velocities
suppressed by powers of $|m_{q}|^{2}/|eB|$. In fact, the explicit form of
the velocities was derived in the weakly coupled NJL model in an
external magnetic field [see Eq.~(\ref{30}) in Section~\ref{sec:NJL3+1General}].
It is
\begin{equation}
v_{u,d}^{2} \sim \frac{|m_{u,d}|^{2}}{|eB|}
\ln\frac{|eB|}{|m_{u,d}|^{2}} \ll 1.
\end{equation}
A similar expression should take place also for the transverse 
velocities of the NG bosons in QCD.

Now, let us turn to the NG boson connected with the spontaneous
breakdown of the $\mathrm{U}^{(-)}(1)_{A}$. It is a 
$\mathrm{SU}(N_{u})_{V}\times \mathrm{SU}(N_{d})_{V}$ singlet. Neglecting the
anomaly, we would actually get two NG singlets, connected
with the up and down sectors, respectively. Their decay constants and
velocities would be given by expression (\ref{decay}) in
which $\lambda^A$ has to be replaced by $\lambda^0$.
The latter is proportional to the unit matrix and normalized 
as the $\lambda^A$ matrices:
$\mathrm{tr}[(\lambda^0)^2]=2$. It is clear that their
decay constants and velocities would be the same as for the
NG bosons from the adjoint representation. Now, taking the anomaly
into account, we find that the anomaly free  $\mathrm{U}^{(-)}(1)_{A}$
current is connected with the traceless matrix
$\tilde{\lambda}^{0}/2 \equiv (\sqrt{N_{d}/N_{f}}\lambda^{0}_{u} -
\sqrt{N_{u}/N_{f}}\lambda^{0}_{d})/2$. Therefore, the genuine NG
singlet $|1 \rangle$ is expressed through 
those two singlets, $|1,d \rangle$ and
$|1,u \rangle$, as 
$|1 \rangle = \sqrt{N_{d}/N_{f}}|1,u \rangle - 
\sqrt{N_{u}/N_{f}}|1,d \rangle$. This
implies
that
its decay constant is
\begin{equation}
\tilde{f}^{2} = \frac{(N_{d} f_{u} + 
N_{u} f_{d})^2}{N^{2}_{f}}=
\frac{(\sqrt{2}N_{d} + N_{u})^{2}N_c}{12\pi^{2}N^{2}_{f}}|eB|.
\end{equation}
Its transverse velocity is of course zero in the LLL approximation.

\subsubsection{Low-energy gluodynamics and anisotropic confinement}
\label{sec:MagCatQCDgluo}

Let us now turn to the infrared region with $|k|\lesssim m_d$, where all
quarks decouple (notice that we take here the smaller mass of $d$
quarks). In that region, a pure gluodynamics is realized. However, its
dynamics is quite unusual. The point is that although gluons are
electrically neutral, their dynamics is strongly influenced by an external
magnetic field, as one can see from expression (\ref{Pi-IR-QCD}) for their
polarization operator. In a more formal language, while quarks decouple
and do not contribute into the equations of the renormalization group in
that infrared region, their dynamics strongly influence the boundary
(matching) conditions for those equations at $k \sim m_d$.

A conventional way to describe this dynamics is the method of the low-energy 
effective action. By taking into account the polarization effects
due to the background magnetic field, we arrive at the following quadratic
part of the low-energy effective action for gluons:
\begin{equation}
{\cal L}^{(2)}_{\rm glue, eff} = -\frac{1}{2} \sum_{A=1}^{N_{c}^2-1}
A^{A}_{\mu}(-k) \left[g^{\mu\nu}k^2-k^{\mu}k^{\nu}+\kappa
\left(g_{\parallel}^{\mu\nu}k_{\parallel}^2-k_{\parallel}^{\mu}
k_{\parallel}^{\nu}\right)\right] A^{A}_{\nu}(k),
\label{L-eff-2}
\end{equation}
where
\begin{equation}
\kappa=\frac{\alpha_{s}}{6\pi}
\sum_{q=1}^{N_{f}}\frac{|e_{q}B|}{m^{2}_{q}}=
\frac{1}{12C_{1}\pi} \sum_{q=1}^{N_{f}}
\left(\frac{\alpha_{s}}{c_{q}^{2}}\right)^{1/3}
\exp\left(\frac{4N_{c}\pi}{\alpha_{s} (N_{c}^{2}-1)
\ln(C_{2}/c_{q}\alpha_{s})}\right) \gg 1.
\end{equation}
By making use of the quadratic part of the action, as well
as the requirement of the gauge invariance, we could easily
restore the whole low-energy effective action (including
self-interactions) as follows:
\begin{equation}
{\cal L}_{\rm glue, eff} \simeq  \frac{1}{2} \sum_{A=1}^{N_{c}^2-1}
\left(\mathbf{E}_{\perp}^{A} \cdot \mathbf{E}_{\perp}^{A}
+\epsilon E_{3}^{A} E_{3}^{A}
- \mathbf{B}_{\perp}^{A} \cdot \mathbf{B}_{\perp}^{A}
-B_{3}^{A} B_{3}^{A}\right),
\label{gluon-action}
\end{equation}
where the (chromo-) dielectric constant $\epsilon \equiv 1+\kappa$
was introduced. Also, we introduced the notation for the
chromo-electric and chromo-magnetic fields as follows:
\begin{eqnarray}
E_{i}^{A} &=& \partial_{0} A_{i}^{A} - \partial_{i} A_{0}^{A}
+g f^{ABC} A_{0}^{B} A_{i}^{C}, \\
B_{i}^{A} &=& \frac{1}{2}\varepsilon_{ijk}
\left(\partial_{j} A_{k}^{A} - \partial_{k} A_{j}^{A}
+g f^{ABC} A_{j}^{B} A_{k}^{C}\right).
\end{eqnarray}
This low-energy effective action is relevant for momenta $|k| \lesssim
m_d$. Notice the following important feature of the action: the coupling
$g$, playing here the role of the ``bare" coupling constant related to the
scale $m_d$, coincides with the value of the vacuum QCD coupling related
to the scale $\sqrt{|eB|}$ (and {\it not} to the scale $m_d$). This is
because
$g$ is determined from the matching condition at $|k| \sim m_d$, the lower
border of the intermediate region $m_d \lesssim |k| \lesssim \sqrt{|eB|}$,
where,
because of the pseudo-Higgs effect, the running of the coupling is
essentially frozen. Therefore, the ``bare" coupling $g$ indeed coincides
with the value of the vacuum QCD coupling related to 
the scale $\sqrt{|eB|}$: $g = g_s$. Since this value is much less that
that of the vacuum QCD coupling
related to the scale $m_d$, this implies that the confinement scale
$\lambda_{\rm QCD}$ of the action (\ref{gluon-action}) should be much less
than $\Lambda_{\rm QCD}$ in QCD without a magnetic field.

Actually, this consideration somewhat simplifies the real
situation. Since the
LLL quarks couple to the longitudinal components of the
polarization operator, only the effective coupling connected
with longitudinal gluons is frozen. For transverse gluons, there
should be a logarithmic running
of their effective coupling. It is clear, however, that this
running should be quite different from that in the vacuum QCD.
The point is that the time-like gluons are now massive and 
their contribution in the running in the intermediate region
is severely reduced. On the other hand, 
because of their negative norm, just the time like gluons
are the major players in producing the antiscreening running
in QCD (at least in covariant gauges). 
Since now they effectively decouple,
the running of the effective coupling for the transverse gluons
should slow down. It is even not inconceivable that the
antiscreening running can be transformed into a screening one.  
In any case, one should expect that the value of the transverse
coupling related to the matching scale $m_d$ will be also essentially
reduced in comparison with that in the vacuum QCD. Since the
consideration in this section is rather qualitative, 
we adopt  the simplest scenario with the value of the transverse coupling
at the matching scale $m_d$ also
coinciding with $g_s$.  

In order to determine the new confinement scale $\lambda_{\rm QCD}$, one
should consider the contribution of gluon loops in the perturbative loop expansion 
connected with the {\it anisotropic} action (\ref{gluon-action}), a hard problem that 
could be perhaps solved in the future. [It is interesting to note that the problem of anisotropic 
gluodynamics and the dependence of the confinement scale on the anisotropy could be 
also studied using lattice simulations.] Here we will get an estimate of $\lambda_{\rm QCD}$, 
without studying the loop expansion in detail. Let us start from calculating the interaction 
potential between two static quarks in this theory. It reads
\begin{equation}
V(x,y,z) \simeq \frac{g_{s}^{2}}
{4\pi\sqrt{z^2+\epsilon (x^{2}+y^{2})}}
\label{potent}
\end{equation}
(compare with Problem 2 on p. 56 in Ref.~\cite{Landau-Lifshitz-V8}). 
Because of the dielectric constant, this Coulomb like interaction is
anisotropic in space: it is suppressed by a factor of $\sqrt{\epsilon}$ in
the transverse directions compared to the interaction in the direction of
the magnetic field. The potential (\ref{potent}) corresponds to the
classical (tree) approximation which is good only in the region of
distances much smaller than the confinement radius $r_{\rm QCD} \sim
\lambda^{-1}_{\rm QCD}$. Deviations from this interaction are described by
loop corrections. Let us estimate the value of a fine structure constant
connected with the perturbative loop expansion.

First of all, because of the form of the potential (\ref{potent}), the
effective coupling constants connected with the parallel and transverse
directions are different: while the former is equal to
$g^{\parallel}_{\rm eff} = g_s$, the latter is $g^{\perp}_{\rm eff} =
g_s/\epsilon^{1/4}$. On the other hand, the loop expansion parameter (fine
structure constant) is $g_{\rm eff}^2/4\pi v_g$, where $v_g$ is the velocity
of gluon quanta. Now, as one can notice from Eq.~(\ref{L-eff-2}), while
the velocity of gluons in the parallel direction is equal to the speed
of light $c = 1$, there are gluon quanta with the velocity $v_{g}^{\perp}
= 1/\sqrt{\epsilon}$ in the transverse directions. This seems to suggest
that the fine structure constants may remain the same, or nearly the same,
despite the anisotropy: the factor $\sqrt{\epsilon}$ in
$(g^{\perp}_{\rm eff})^2$ will be canceled by the same factor in
$v_{g}^{\perp}$. Therefore, the fine structure constant can be estimated as
$\alpha_{s} = g^{2}_{s}/4\pi$ (although, as follows from
Eq.~(\ref{L-eff-2}), there are quanta with the velocity $v_{g}^{\perp} =
1$, their contribution in the perturbative expansion is suppressed by the
factor $1/\sqrt{\epsilon}$).

This consideration is of course far from being quantitative. Introducing
the magnetic field breaks the Lorentz group $\mathrm{SO}(3,1)$ down to $\mathrm{SO}(1,1)
\times \mathrm{SO}(2)$, and it should be somehow manifested in the perturbative
expansion. Still, we believe, this consideration suggests that the
structure of the perturbative expansion in this theory can be
qualitatively similar to that in the vacuum QCD, modulo the important
variation: while in the vacuum QCD $\alpha_s$ is related to the scale
$|eB|$, it is now related to much smaller scale $m_d$.

By making use of this observation,
we will approximate the running in   
the low-energy region by a vacuum-like running:
\begin{equation}
\frac{1}{\alpha_{s}^{\prime}(\mu)} = \frac{1}{\alpha_{s}}
+b_{0} \ln\frac{\mu^{2}}{m_{d}^{2}}, \quad \mbox{ where} \quad
b_{0} =\frac{11 N_{c}}{12\pi},
\end{equation}
where the following condition was imposed: 
$\alpha_{s}^{\prime}(m_d)=\alpha_{s}$. From this running law, we estimate
the new confinement scale,
\begin{equation}
\lambda_{\rm QCD} \simeq m_d
\left(\frac{\Lambda_{\rm QCD}}{\sqrt{|eB|}}\right)^{b/b_{0}}.
\label{lambda}
\end{equation}
We emphasize again that expression (\ref{lambda}) is just an
estimate of the new confinement scale. In particular, both
the exponent, taken here to be equal to $b/b_{0}$, and the
overall factor in this expression, taken here to be equal $1$,
should be considered as being fixed only up to a factor of order one.

The hierarchy $\lambda_{\rm QCD} \ll \Lambda_{\rm QCD}$ is intimately 
connected with a somewhat puzzling point that the pairing dynamics decouples 
from the confinement dynamics despite it produces quark masses of order
$\Lambda_{\rm QCD}$ or less [for a magnetic field all the way up to the order
of $(10 \mbox{ TeV})^2$]. The point is that these masses are heavy in units
of the new confinement scale $\lambda_{\rm QCD}$ and the pairing dynamics
is indeed weakly coupled.

Before concluding this section, let us note that recently the authors of 
Ref.~\cite{Bonati:2014ksa} were able to extract results for the 
quark-antiquark potential by making use of lattice QCD simulations
in the background of constant magnetic field. They found that both the 
effective string tension and the Coulomb part of the potential are 
anisotropic. The value of the string tension becomes larger in the 
transverse direction and smaller in the longitudinal direction \cite{Bonati:2014ksa}. 
The absolute value of the Coulomb coupling shows the opposite behavior \cite{Bonati:2014ksa}
and that is consistent with the strong field limit prediction in Eq.~(\ref{potent}).

\subsubsection{Magnetic catalysis in QCD with large number of colors}
\label{sec:MagCatQCDinfty}

In this section, we will discuss the dynamics in QCD in a magnetic field
when the number of colors is large, in particular, we will consider the
('t Hooft) limit $N_c \to \infty$. Just a look at expression (\ref{M_g})
for the gluon mass is enough to recognize that the dynamics in this limit
is very different from that considered in the previous sections. Indeed,
as is well known, the strong coupling constant $\alpha_s$ is proportional
to $1/N_c$ in this limit. More precisely, it rescales as
\begin{equation}
\alpha_s = \frac{\tilde{\alpha}_s}{N_c},
\label{tilde}
\end{equation}
where the new coupling constant $\tilde{\alpha}_s$ remains finite
as $N_c \to \infty$. Then, expression (\ref{M_g}) implies that
the gluon mass goes to zero in this limit. This in turn implies
that the appropriate approximation in this limit is not
the improved rainbow approximation but the rainbow approximation
itself, when {\it both} the vertex and the gluon propagator in
the Schwinger-Dyson equation (\ref{SD}) are taken to be bare.

In order to get the expression for the quark propagator in this case, we can 
use the results of the analysis of the Schwinger-Dyson equation in the rainbow approximation 
in QED in a magnetic field in Section~\ref{QED-beyond}, with the same simple 
modifications as in Section~\ref{sec:MagCatQCDgap}. The result is
\begin{equation}
m_{q}^2 = C |e_{q}B|
\exp\left[-{\pi}
\left(\frac{\pi N_c}{(N_{c}^2-1)\alpha_s}\right)^{1/2}
\right],
\label{m_q}
\end{equation}
where the constant $C$ is of order one. As $N_c \to \infty$, one
gets
\begin{equation}
m_{q,N_c \to \infty}^2 = C |e_{q}B|
\exp\left[-{\pi}
\left(\frac{\pi}{\tilde{\alpha}_s}\right)^{1/2}
\right].
\end{equation}
It is natural to ask how large $N_c$ should be before the expression
(\ref{m_q}) becomes reliable. From
our discussion above, it is clear that the rainbow
approximation may be reliable only when the gluon mass is small, i.e.,
it is of the order of the quark mass $m_q$ or less.
Equating expressions (\ref{M_g}) and (\ref{m_q}), we derive an
estimate for the threshold value of $N_c$:
\begin{equation}
N_{c}^{\rm thr} \sim \frac{2N_u + N_d}{\ln|eB|/\Lambda^{2}_{\rm QCD}}
\exp\left[\frac{\pi}{2\sqrt{3}}
\left(11\ln\frac{|eB|}{\Lambda_{\rm QCD}}\right)^{1/2}
\right].
\label{estimate1}
\end{equation}
Expression (\ref{m_q}) for the quark mass
is reliable for the values of
$N_c$ of the order of $N_{c}^{\rm thr}$ or larger.
Decreasing $N_c$ below $N_{c}^{\rm thr}$, one comes to
expression (\ref{gapQCD}).

It is quite remarkable that one can get a rather
close estimate for $N_{c}^{\rm thr}$ by equating expressions
(\ref{gapQCD}) and (\ref{M_g}):
\begin{equation}
N_{c}^{\rm thr} \sim \frac{2N_u + N_d}{\ln|eB|/\Lambda^{2}_{\rm QCD}}
\exp\left[\left(11\ln\frac{|eB|}{\Lambda_{\rm QCD}}\right)^{1/2}\right]
\label{estimate2}
\end{equation}
[notice that the ratio of the exponents (\ref{estimate1})
and (\ref{estimate2}) is equal to $0.91$].
The similarity of estimates (\ref{estimate1}) and (\ref{estimate2})
implies that, crossing the threshold $N_{c}^{\rm thr}$, expression
(\ref{m_q}) for $m_q$ smoothly transfers into expression (\ref{gapQCD}).

These estimates show that the value of $N_{c}^{\rm thr}$ rapidly grows with
the magnetic field. For example, taking $\Lambda_{\rm QCD} = 250 \mbox{ MeV}$
and $N_{u}=1$, $N_{d}=2$, we find from Eq.~(\ref{estimate1}) that $N^{\rm thr}_c
\sim 10^2$, $10^3$, and $10^4$ for $|eB| \sim (1\mbox{ GeV})^2$, $(10\mbox{
GeV})^2$, and $(100\mbox{ GeV})^2$, respectively.

As was shown in Section~\ref{sec:MagCatQCDgluo}, in the regime with the number
of colors $N_c \ll N_{c}^{\rm thr}$, the confinement scale $\lambda_{\rm QCD}$ in
QCD in a strong magnetic field is essentially smaller than
$\Lambda_{\rm QCD}$. What is the value of $\lambda_{\rm QCD}$ in the regime with
$N_c$ being of the order of $N_{c}^{\rm thr}$ or larger? It is not difficult
to see that $\lambda_{\rm QCD} \simeq \Lambda_{\rm QCD}$ in this case. Indeed, now
the gluon mass and, therefore, the contribution of quarks in the
polarization operator are small (the latter is suppressed by the factor
$1/N_c$ with respect to the contribution of gluons). As a result, the
$\beta$-function in this theory is close to that in QCD without a magnetic
field, i.e., $\lambda_{\rm QCD} \simeq \Lambda_{\rm QCD}$.

Expression (\ref{m_q}) implies that, for a sufficiently strong
magnetic fields, the dynamical mass $m_q$ is much larger than
the confinement scale $\Lambda_{\rm QCD}$. Indeed, expressing the
magnetic field in terms of the running coupling, one gets
\begin{equation}
m_{q}^2 \simeq \left|\frac{e_{q}}{e}\right| \Lambda_{\rm QCD}^2
\exp\left[\frac{1}{b\alpha_{s}} -{\pi}
\left(\frac{\pi N_c}{(N_{c}^2-1)\alpha_s}\right)^{1/2}
\right],
\end{equation}
and for small values of $b\alpha_s \sim N_{c}\alpha_{s} \equiv
\tilde{\alpha}_s$ (i.e., for large values of $|eB|$) the mass $m_q$ is
indeed much larger than $\Lambda_{\rm QCD}$. This point is important for
proving the reliability of the rainbow approximation in this problem.
Indeed, the relevant region of momenta in this problem is $m_{q}^2 \ll
|k^2| \ll |eB|$ \cite{Gusynin:1995gt,Gusynin:1995nb} where,
because of the condition $m_{q}^2 \gg \Lambda_{\rm QCD}^2$ in a strong enough field,
the running coupling is small. Therefore, the rainbow approximation is
indeed reliable for sufficiently strong magnetic fields in
this case.

\subsubsection{Additional remarks about QCD dynamics in a magnetic field}
\label{sec:MagCatQCDdiscuss}

QCD in a strong magnetic field yields an example of a rich, sophisticated
and (that is very important) controllable dynamics. Because of the
property of asymptotic freedom, the pairing dynamics, responsible for
chiral symmetry breaking in a strong magnetic field, is weakly
interacting. The key point why this weakly interacting dynamics manages to
produce spontaneous chiral symmetry breaking is the fact that it is
essentially $(1+1)$-dimensional: in the plane orthogonal to the external field
the motion of charged quarks is restricted to a region of a typical size
of the order of the magnetic length, $ l = 1/\sqrt{|e_{q}B|}$. Moreover,
such a dynamics almost completely decouples from the dynamics of
confinement which develops at very low energy scales in the presence of a
strong magnetic field. 

Here we presented the analysis in the LLL approximation, which 
is expected to be valid in very strong magnetic fields. Beyond such an 
approximation, the role of higher Landau levels in the dynamics of magnetic 
catalysis was recently studied in Ref.~\cite{Mueller:2014tea}. It was found that 
while their contribution lead to quantitative differences in observables, the 
qualitative picture remains similar to that in the LLL approximation.
 
While the pairing dynamics decouples from the dynamics of confinement, the
latter is strongly modified by the polarization effects due to quarks that
have an effective $(1+1)$-dimensional dynamics in the lowest Landau level.
As a result, the confinement scale in QCD in a strong magnetic field
$\lambda_{\rm QCD}$ is much less than the confinement scale $\Lambda_{\rm QCD}$ in
the vacuum QCD. This implies a rich spectrum of light glueballs in this
theory.

This picture changes drastically for QCD with a large number of colors
($N_c \gtrsim 100$ for $|eB| \gtrsim 1 \mbox{ GeV}^2$). In that case a
conventional confinement dynamics, with the confinement scale
$\lambda_{\rm QCD} \sim \Lambda_{\rm QCD}$, is realized.

The dynamics of chiral symmetry breaking in QCD in a magnetic field has both similarities 
and important differences with respect to the dynamics of color superconductivity 
in QCD with a large baryon density \cite{Shovkovy:2004me,Alford:2007xm}.
Both dynamics are essentially $(1+1)$-dimensional. However, while the former
is anisotropic [the rotational $\mathrm{SO}(3)$ symmetry is explicitly broken by 
a magnetic field], the rotational symmetry is preserved in the latter. This fact is connected, 
in particular, with the following fact: while in dense QCD quarks interact via the 
exchange of both chromo-electric and chromo-magnetic gluons 
\cite{Son:1998uk,Schafer:1999jg,Hong:1999fh,Hsu:1999mp,Pisarski:1999tv,Shovkovy:1999mr}, 
in the present theory they interact only via the exchange of the longitudinal components of 
chromo-electric gluons. This in turn leads to very different expressions for the 
dynamical masses of quarks in these two theories.

Another important difference is that while the pseudo-Higgs effect takes place in 
QCD in a magnetic field, the genuine Higgs effect is realized in color superconducting 
dense quark matter. Because of the Higgs effect, the color interactions connected 
with broken generators are completely screened in infrared in the case of color 
superconductivity. In particular, in the color-flavor locked phase of dense QCD 
with three quark flavors, the color symmetry is completely broken and, therefore, 
the infrared dynamics is under control in that case \cite{Alford:1998mk}. As for dense 
QCD with two quark flavors, the color symmetry is only partially broken down to 
$\mathrm{SU}(2)_c$, and there exists an analog of the pseudo-Higgs effect for the 
electric modes of gluons connected with the unbroken $\mathrm{SU}(2)_c$. As a 
result, the confinement scale of the gluodynamics of the remaining $\mathrm{SU}(2)_c$ 
group is much less than $\Lambda_{\rm QCD}$ \cite{Rischke:2000cn}, lresembling the 
present case of QCD in a magnetic field. The essential difference, however, is that, 
unlike QCD in a magnetic field, the infrared dynamics of a color superconductor is isotropic.

Last but not least, unlike the case of QCD with a nonzero baryon density,
there are no principal obstacles for examining all these results and
predictions in lattice computer simulations of QCD in a magnetic field.

\subsection{Lattice studies of QCD in a magnetic field}
\label{sec:MagCatLattice}

In the previous several subsections, we discussed the possibility of magnetic catalysis in gauge theories,
such as QED and QCD. In the case of QED, the corresponding underlying dynamics is weakly 
coupled and, thus, almost completely under theoretical control. As we saw, the only subtle 
complication in the corresponding dynamics originates from the long-range nature of the QED 
interaction. Because of this, in a general gauge, there is no suppression of higher-order diagrams 
and the conventional loop expansion leads to a quantitatively unreliable result for the dynamically 
generated fermion mass. This fact alone could not, however, invalidate the qualitative features and 
consequences of the magnetic catalysis, whose nature is truly universal in the weakly coupled 
regime.

At the same time, it should be emphasized that while the general features of the magnetic catalysis are 
well understood in the weakly interacting regime of quantum field theoretical models, numerous 
questions may arise at strong coupling. This general problem is encountered, in particular, in 
low-energy QCD, in which the strong interaction is responsible for the confinement and other 
nonperturbative phenomena. To circumvent the problem in the analytical studies, in Section~\ref{sec:MagCatQCD}
we had to assume that the background magnetic field is asymptotically strong. In such a 
limit the Landau energy scale $\sqrt{|eB|}$ is much larger than $\Lambda_{\rm QCD}$ and the 
dynamics responsible for the magnetic catalysis becomes effectively weakly interacting. 

The insight into the dynamics at asymptotically strong magnetic fields, however, does not help to fully  
understand the role of moderately strong magnetic fields, $\sqrt{|eB|}\sim\Lambda_{\rm QCD}$
in QCD, when a nontrivial competition between the chiral symmetry breaking and confinement
is expected. In the corresponding strongly-coupled regime of QCD without background fields, 
the use of lattice simulations proved to be invaluable to understand the underlying dynamics at 
zero and finite temperatures. The same techniques have now also been applied to QCD-like gauge 
theories in the presence of a background magnetic field \cite{Buividovich:2008wf,Buividovich:2009bh,
Braguta:2010ej,2010PhRvD..82e1501D,2011PhRvD..83k4028D,Bali:2011qj,Bali:2011uf,Bali:2012zg,
2013LNP...871..181D,Ilgenfritz:2012fw,Ilgenfritz:2013ara,DElia:2013twa,Bruckmann:2013oba,Bruckmann:2013ufa}. 
In this connection, it should be noted that, unlike a baryon chemical potential or electric 
fields, the magnetic field poses no conceptual difficulties in lattice computations.

\subsubsection{Overview of lattice results: catalysis, inverse catalysis and beyond}

At zero temperature, numerous lattice studies in QCD have confirmed that the chiral 
condensate increases with the strength of the magnetic field \cite{Buividovich:2008wf,
Buividovich:2009bh,Braguta:2010ej,2010PhRvD..82e1501D,2011PhRvD..83k4028D,Bali:2011qj,
Bali:2011uf,Bali:2012zg,2013LNP...871..181D,Ilgenfritz:2012fw,Ilgenfritz:2013ara,DElia:2013twa,
Bruckmann:2013oba,Bruckmann:2013ufa}. 
The corresponding computations were performed within several variants of gauge models 
with different values of the pion mass and for magnetic fields up to about $1~\mbox{GeV}^2$. 
The universal conclusion is that the background magnetic field aids chiral symmetry 
breaking in the vacuum. 

\begin{figure}[t]
\begin{center}
\includegraphics[width=0.47\textwidth]{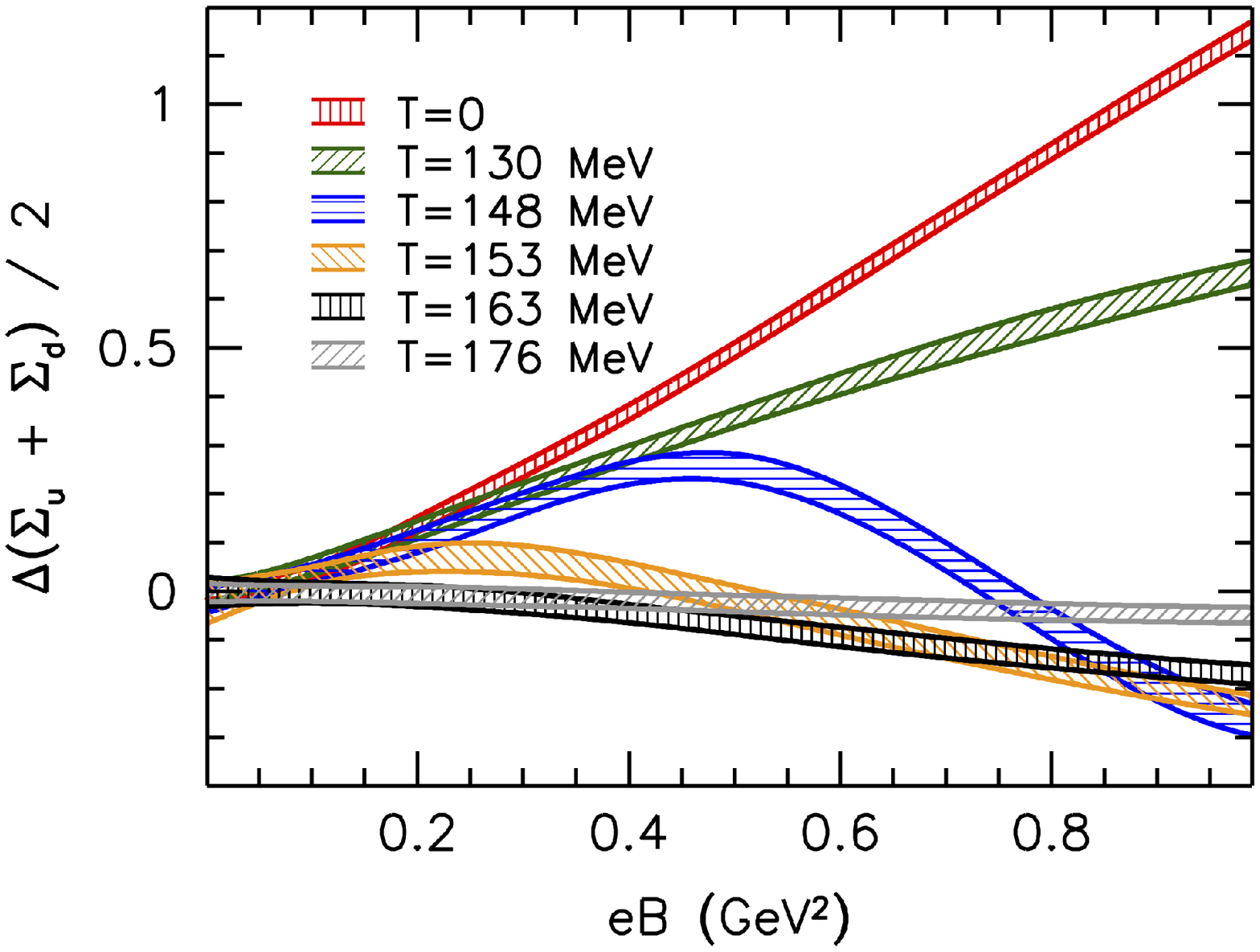}\hspace{0.05\textwidth}
\includegraphics[width=0.45\textwidth]{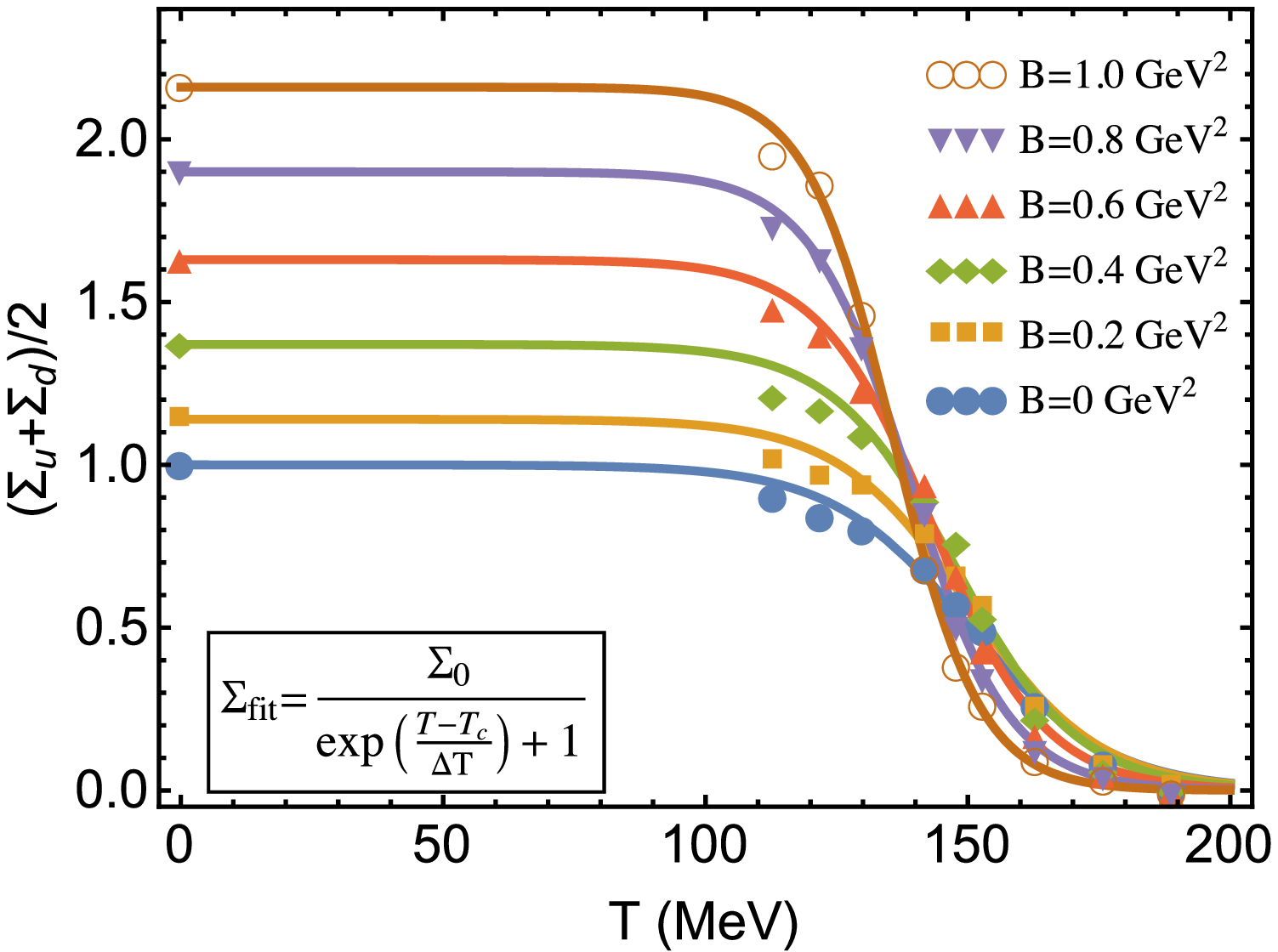}
\caption{(Color online) The change of the average light-quark chiral condensate due to the 
magnetic field (left panel) and the temperature dependence of the condensate (right panel). 
The plot in the left panel is taken from Ref.~\cite{Bali:2012zg}. The temperature dependence 
in the right panel was plotted using the lattice data of Ref.~\cite{Bali:2012zg}.}
\label{fig-cond-vs-T-eB-lattice}
\end{center}
\end{figure}

The representative lattice results for the temperature and magnetic field dependence of the average 
light-quark chiral condensate are shown in Fig.~\ref{fig-cond-vs-T-eB-lattice}. In the left panel,
the change of the average condensate due to the magnetic field is shown at zero, as well as at 
several nonzero values of temperature. This plot is taken from Ref.~\cite{Bali:2012zg}.  
In qualitative agreement with the scenario of the magnetic catalysis, the condensate increases with 
the magnetic field when the temperature is zero, or sufficiently low. Because of the strongly-coupled 
dynamics and confinement, this finding is still very nontrivial.

In contrast and rather surprisingly, the behavior of the condensate drastically changes when 
the temperature is in the near-critical region \cite{Bali:2011qj,Bali:2011uf,Bali:2012zg,
2013LNP...871..181D,Bruckmann:2013oba,Bruckmann:2013ufa}. 
It appears that the magnetic field helps to restore rather than break chiral symmetry. 
In order to understand this better, it is instructive to review the temperature dependence 
of the condensate in some detail. 

The fact that the zero temperature chiral condensate increases with the magnetic field
may suggest that the critical temperature for the chiral symmetry restoration phase 
transition should also increase with the field strength. In the first lattice studies 
\cite{2010PhRvD..82e1501D,Ilgenfritz:2012fw}, it was indeed found that the critical temperature 
rises very slightly as a function of the magnetic field (up to about $|eB|\simeq 0.75~\mbox{GeV}^2$). 
The corresponding exploratory study was not describing the real QCD, however, 
because the pion mass ($200$--$480~\mbox{MeV}$) was considerably larger than 
the physical value. Later, the study of QCD with the physical pion mass 
$m_\pi = 135~\mbox{MeV}$, extrapolated to the continuum limit, has revealed 
that the transition temperature {\it decreases} with the external magnetic field 
\cite{Bali:2011qj,Bali:2011uf,Bali:2012zg,2013LNP...871..181D,Bruckmann:2013oba,
Bruckmann:2013ufa}. This surprising and unexpected property 
was dubbed inverse magnetic catalysis. (Note that the term ``inverse magnetic catalysis" 
is also used by some authors in the context of the chiral symmetry restoration phase 
transition driven by a nonzero chemical potential \cite{Preis:2010cq,Preis:2011sp,Preis:2012fh}.)  

The temperature dependence of the condensate for several fixed values of the magnetic 
field are shown in the right panel of Fig.~\ref{fig-cond-vs-T-eB-lattice}. In additional 
to the lattice data points of Ref.~\cite{Bali:2012zg}, in the same plot we also show simple 
fitting functions (solid lines) for the condensate. The fits are described quite well by the 
following $3$-parameter function:
\begin{equation}
\Sigma_{\rm avr} \equiv \frac{1}{2}\left(\Sigma_u+\Sigma_d\right) 
= \frac{\Sigma_0}{\exp\left[(T-T_c)/\Delta T\right]+1},
\label{fit-condensate}
\end{equation}
where the fit parameters $\Sigma_0$, $T_c$, and $\Delta T$ can be interpreted as the 
approximate values of the zero-temperature condensate, the critical temperature and 
the temperature range of the crossover type transition, respectively. All of them may 
depend on the magnetic field. 

\begin{figure}[t]
\begin{center}
\includegraphics[width=0.47\textwidth]{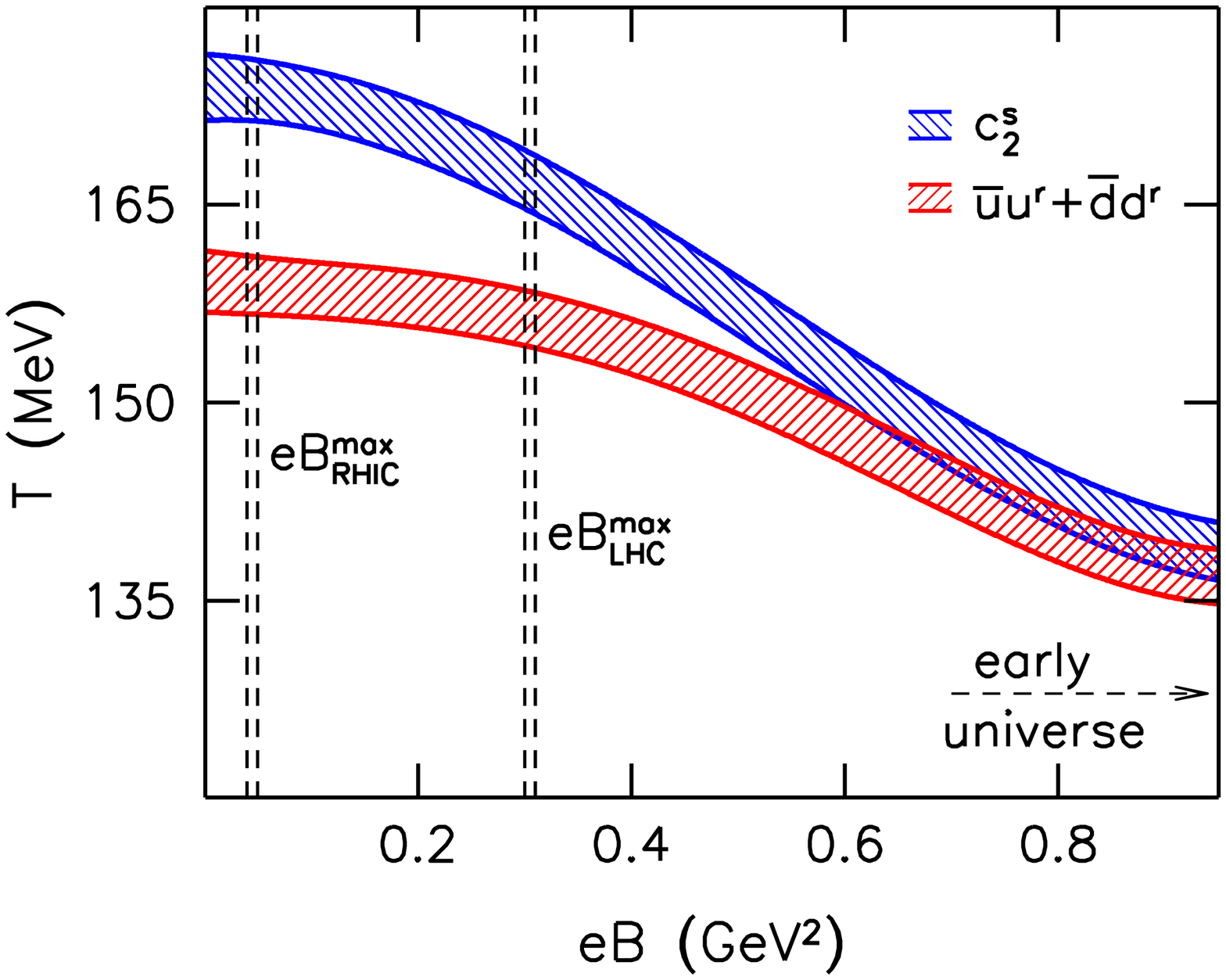}\hspace{0.05\textwidth}
\includegraphics[width=0.45\textwidth]{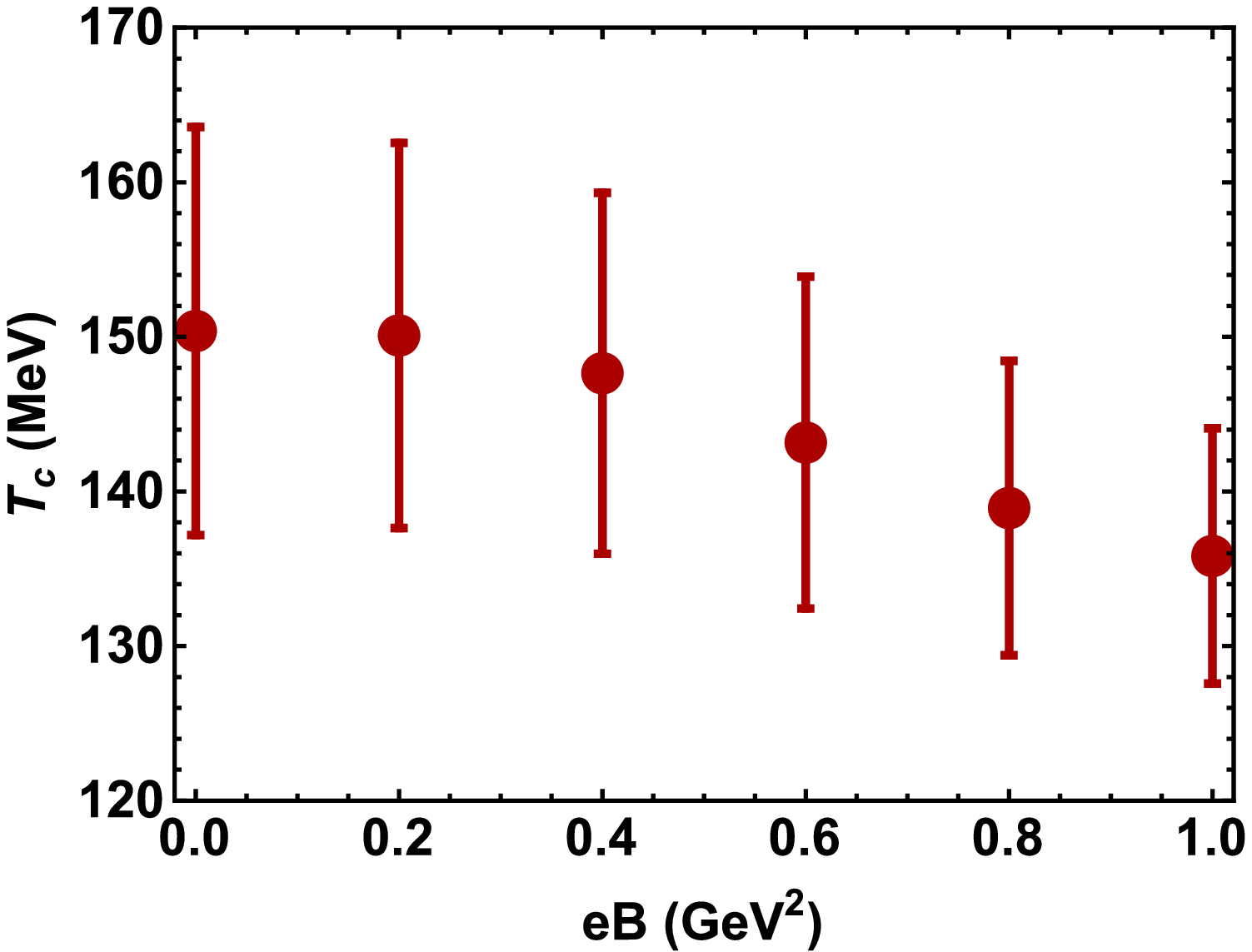}
\caption{(Color online) Left panel: the lattice QCD phase diagram in the magnetic 
field-temperature plane from Ref.~\cite{Bali:2011qj}. Right panel: the QCD phase 
diagram obtained by plotting the model parameters $T_c$ (central value) and $\Delta T$ 
(the vertical range) that fit the data of Ref.~\cite{Bali:2012zg} by the functional dependence 
in Eq.~(\ref{fit-condensate}).}
\label{fig-Tc-lattice}
\end{center}
\end{figure}

As we see from Fig.~\ref{fig-cond-vs-T-eB-lattice}, the normal magnetic catalysis 
hierarchy of the condensates is preserved all the way up to the near-critical temperatures, 
$T\lesssim 140~\mbox{MeV}$. However, a reverse hierarchy sets in above 
$T\gtrsim 140~\mbox{MeV}$. One of the immediate implications of this result 
is the decreasing critical temperature for the chiral symmetry restoration with the 
increasing magnetic field. This is explicitly demonstrated by the lattice phase 
diagram in the left panel of Fig.~\ref{fig-Tc-lattice}, which is taken from Ref.~\cite{Bali:2011qj}. 
As one can see from the right panel in Fig.~\ref{fig-Tc-lattice}, a
qualitatively similar phase diagram is also obtained if one plots the fitting 
parameter $T_c$, used in Eq.~(\ref{fit-condensate}), as a function of the 
magnetic field. Considering that the temperature range of the crossover 
transition is mimicked by the value of $\Delta T$ in the fitting function 
(\ref{fit-condensate}), in the right panel of Fig.~\ref{fig-Tc-lattice} we also 
showed $\Delta T$ as the size of the vertical bar around the central value of $T_c$.

The counterintuitive dependence of the critical temperature on the magnetic field revealed in lattice  
simulations indicates \cite{Bali:2011qj,Bali:2011uf,Bali:2012zg,2013LNP...871..181D,Bruckmann:2013oba,
Bruckmann:2013ufa}, most likely, an important property QCD. This is further supported by the fact 
that no phenomenological models predicted anything like it beforehand. A potentially significant 
piece of information about the observed inverse magnetic catalysis is that it appears to occur only 
in the regime of  rather small current quark masses. Indeed, the lattice studies at sufficiently 
large quark masses \cite{2010PhRvD..82e1501D,Ilgenfritz:2012fw} show the normal behavior 
of the critical temperature consistent with the catalysis. [See however the recent study in 
Ref.~\cite{Bornyakov:2013eya}, which uses chiral lattice fermions and suggests that the
inverse magnetic catalysis may extend also to the models with larger pion masses.] 

An additional insight into the nearcritical  dynamics of chiral symmetry restoration (and 
deconfinement) in QCD in a magnetic field was offered in Refs.~\cite{Bruckmann:2013oba,
Bruckmann:2013ufa}, in which the two competing effects were identified from the 
lattice simulation. The authors separated the effects of ``valence" and ``sea" quark 
contributions in the underlying dynamics. By definition, the valence contribution to the condensate 
is directly connected with the enhancement of the spectral density of the Dirac operator 
in a magnetic field around zero energy. The back-reaction of the quarks on the gauge 
field is not included in the valence part, however. It is accounted in the sea contribution,
which comes from a sampling of the gauge fields modified by the quark determinant.

With such a separation of the two effects, it is not surprising that the valence contribution
always helps to enhance the condensate and drive the magnetic catalysis. The situation 
with the other contribution is more subtle \cite{Bruckmann:2013oba,Bruckmann:2013ufa}:
while the sea also enhances the condensate at small temperatures, it leads to a rather strong 
suppression in the nearcritical region, $T\gtrsim 140~\mbox{MeV}$, and thus drives the 
inverse catalysis. It is natural to ask what is the physical meaning of this lattice result. 
One possible explanation was offered in the same lattice study \cite{Bruckmann:2013oba}, 
where it was found that strong magnetic fields favor large values of the Polyakov loop, 
known to be related to the deconfinement. Large Polyakov loops tend to suppress 
the chiral condensate and the corresponding effect appears to dominate in the transition 
region. 

The temperature dependence of the sea contribution, which appears to be responsible 
for the inverse magnetic catalysis, can be also viewed from a different perspective. It
can be interpreted as the screening effects of the strong interaction. Indeed, as we 
discussed in Section~\ref{sec:MagCatQCDgap}, an external magnetic field has a 
rather strong effect on the gluon interaction in QCD even at zero temperature. For the 
purposes of the pairing dynamics, the gluons behave almost like massive resonances 
with a mass proportional to the field. This has a strong suppressing effect on the chiral 
symmetry breaking and generation of the dynamical mass, but not enough to kill the 
magnetic catalysis at $T=0$. It is quite natural to assume that a nonzero temperature 
will lead to even more screening. Then, the resulting large suppression of the coupling 
constant in the pairing dynamics may become dominant and lead to the inverse catalysis
in the nearcritical region when $T\simeq \Lambda_{\rm QCD}$. 

The corresponding effects can be incorporated very easily in the effective models, by
choosing coupling constant(s) that depend on the magnetic field strength and temperature
\cite{Farias:2014hl,Ferreira:2013tba,Ferreira:2014kpa,Ayala:2014iba,Ayala:2014yla,
Ferrer:2014qka}. The analysis confirms that such models can qualitatively reproduce 
the lattice results for the inverse magnetic catalysis. In all fairness, though, such an
approach has a limited power to provide a deep insight into the subtleties of underlying 
dynamics.

There are also a number of effective model studies that take into account the effects of 
the Polyakov loop on the dynamics of chiral symmetry breaking in a magnetic field background 
\cite{Mizher:2010zb,Gatto:2010qs,Gatto:2010pt,Gatto:2012sp,Ferreira:2013tba,Ferreira:2014kpa,
Andersen:2013swa,Andersen:2014oaa}. There was also an attempt to explain the inverse 
magnetic catalysis using a bag model \cite{Fraga:2012fs}. While
some trends might be captured by such models, the comparison of their predictions with 
the lattice results is still far from perfect. Other model studies try to connect the inverse 
magnetic catalysis to possible condensate suppressing effects due to certain topological 
fluctuations \cite{Chao:2013qpa,Yu:2014sla}. These may indeed suggest some new 
avenues to explore in more details. The existing studies are still heuristic at best and, 
for now, lack a rigorous support from the microscopic studies and/or a more direct confirmation 
from the lattice studies. For a recent review of the model studies of the phase 
diagram of QCD in a magnetic field, see Ref.~\cite{Andersen:2014xxa}.

Additional clues to the puzzle of the inverse magnetic catalysis might be provided by the
observation of the so-called gluonic catalysis and inverse catalysis on the lattice in 
Ref.~\cite{Bali:2013esa}. In essence, instead of the chiral condensate the authors 
consider the gluonic contribution to the interaction measure (trace anomaly) and find
that its dependence on the external magnetic field resembles that of the chiral condensate.
It is enhanced by the magnetic field at small temperatures and is suppressed at high 
temperatures (near and above the transition temperature). This seem to be related 
to another observation in same study \cite{Bali:2013esa}: the chromo-magnetic field 
parallel to the external field is enhanced, while the chromo-electric field in the same direction 
is suppressed, suggesting that the corresponding ``condensates" of color gauge fields 
can be induced spontaneously. Such a possibility is also supported by the analysis of 
a Heisenberg-Euler type effective action that includes not only electromagnetic, but also 
color gauge fields \cite{Bali:2013esa,Ozaki:2013sfa}. In the context of the inverse magnetic 
catalysis, one may speculate that induced chromomagnetic background interferes with 
the dynamics responsible for symmetry breaking. 

Here it might be also appropriate to mention the scenario of the ``magnetic inhibition" 
that was proposed in Ref.~\cite{Fukushima:2012kc} in an attempt to explain the inverse magnetic 
catalysis. The idea is that the symmetry restoration effects are driven by strong fluctuations of the 
composite neutral pions, which have highly nonisotropic dispersion relation in the strong field limit
(or weak coupling regime when the LLL approximation is reliable). Indeed, as we found in 
Section~\ref{sec:NJL3+1KineticTerm}, the transverse velocity of NG bosons can be much smaller 
than the longitudinal one for certain choices of the parameters, e.g., see the weak coupling results 
in Eqs.~(\ref{30}) and (\ref{29}). The problem is that the relevant regime of QCD cannot be well
approximated by the corresponding weak coupling results. At strong coupling, on the other hand, 
the effecting model studies do not show any large anisotropy in the spectrum of composite 
NG bosons. The ``magnetic inhibition" mechanism \cite{Fukushima:2012kc} also appears to be
at odds with the studies based on the functional renormalization group approach 
\cite{Fukushima:2012xw,Kamikado:2013pya}.

The authors of Refs.~\cite{Kojo:2012js,Kojo:2014gha} tried to address the strong coupling 
nonperturbative regime of QCD in a magnetic field in a qualitative manner. They correctly 
observe that the usual weak coupled analysis, which may well be justified in extremely 
strong fields, cannot be directly applied to the regime probed on the lattice. Instead they 
speculate that pairing dynamics responsible for the chiral symmetry breaking is affected 
a lot by the confinement. In the model proposed, the chiral condensate grows linearly 
with the magnetic field, but the dynamical mass stays nearly independent of the magnetic 
field. While the results for the condensate are in qualitative agreement with the lattice 
findings, it remains to be understood whether the critical temperature of a crossover 
type transition can be identified with the zero temperature dynamical mass, which
is not gauge invariant in QCD.  

\subsubsection{Toward the complete phase diagram of QCD in a magnetic field}

In conclusion of this section, let us express our own understanding of the underlying 
dynamics of the magnetic catalysis in QCD. While the case of moderately strong
magnetic field $|eB|\sim \Lambda_{\rm QCD}^2$ is indeed complicated by the 
various nonperturbative effects and confinement, we claim that the regime of 
inverse magnetic catalysis should be replaced by the standard magnetic catalysis
at sufficiently strong magnetic fields, $|eB|\gg \Lambda_{\rm QCD}^2$. In fact, an 
indication of this might have already been seen in the lattice study of 
Ref.~\cite{Ilgenfritz:2013ara}, albeit in a model of chromodynamics with only 
two colors.  

The strong argument in support of the normal magnetic catalysis in QCD at sufficiently 
strong magnetic fields comes from the analytical studied at $B\to \infty$. Even if such a 
regime cannot be realized in nature or happens to be too hard to reproduce in the 
lattice simulations because of a large hierarchy of scales, $|eB|/\Lambda_{\rm QCD}^2\gg 1$, 
this statement is of principle theoretical importance. As we explained in Section~\ref{sec:MagCatQCDgap} 
in detail, in the limit $B\to \infty$, while the pairing dynamics responsible for the chiral symmetry 
breaking is nonperturbative, it is weakly coupled and under theoretical control. 

As follows from the analysis in Section~\ref{sec:MagCatQCDgap}, at $B\to \infty$ and 
zero temperature, the magnetic catalysis should be present and the dynamical quarks 
(or, rather, colorless baryons made out of them) are heavy and decouple from the 
infrared dynamics. The spectrum of the corresponding low-energy theory contains
massless (light) neutral NG bosons, as well as light glueballs, associated with the 
low-energy confined anisotropic gluodynamics. The confinement scale of the latter 
is a function of the magnetic field, $\lambda_{\rm QCD}(B)$, and is suppressed 
compared to usual $\Lambda_{\rm QCD}$ in vacuum at $B=0$, see Section~\ref{sec:MagCatQCDgluo} 
for details.

When the temperature is nonzero, the corresponding theory undergoes a 
deconfinement phase transition at $T^{*}_c(B)\sim \lambda_{\rm QCD}(B)$ 
\cite{Cohen:2013zja}, but remains in the chiral symmetry broken phase in 
the range of temperatures, $T^{*}_c(B) < T < T^{(\chi)}_c(B)$, where 
$T^{(\chi)}_c(B)$ is the critical temperature of the chiral symmetry restoration 
phase transition. It is expected that the deconfinement transition 
at $T^{*}_c(B)$ is first order \cite{Cohen:2013zja}, just like in the conventional 
three-color gluodynamics with the exact $Z_3$ center symmetry \cite{Yaffe:1982qf}.
In their reasoning, the authors of Ref.~\cite{Cohen:2013zja} made a natural 
assumption that the anisotropy should not affect the existence of the phase transition 
itself and should not change its type. It will be interesting to see, however, if the lattice 
simulations of the anisotropic gluodynamics described by the action in Eq.~(\ref{gluon-action}) 
can reconfirm this claim. Also, it will be interesting to know how the anisotropy 
parameter $\epsilon$ (chromo-dielectric constant) affects the critical temperature, 
the interaction potential between static color charges, the spectrum of glueballs, 
the thermodynamic properties, etc. (For a few existing lattice studies of thermodynamic 
properties of QCD in a magnetic field, see Refs.~\cite{Bali:2013txa,Bali:2013esa,Bali:2014kia}.)

\section{Quantum Hall effect in graphene}
\label{sec:QHEgraphene}

\subsection{Generalized magnetic catalysis in graphene}
\label{Sec:MagCatGraphene}

As is well known, the low-energy dynamics of electrons in graphene \cite{2004Sci...306..666N}
is described by the Dirac equation in (2+1) dimensions \cite{Semenoff:1984dq} (for a review,
see Ref.~\cite{CastroNeto:2009zz}). Perhaps the most direct confirmation of the pseudorelativistic 
character of electron motion in graphene is given by the experimental observation 
\cite{2005Natur.438..197N,Zhang:2005zz} of the anomalous quantum Hall (QH) effect 
theoretically predicted in Refs.~\cite{Zheng:2002zz,Gusynin:2005pk,Gusynin:2005iv,Peres:2006zz}. 
The anomalous QH plateaus in the Hall conductivity
are observed at the filling factors $\nu = \pm 4(n + 1/2)$, where $n=0,1,2,\ldots$
is the Landau level (LL) index. The factor 4 in the filling factor is due to a fourfold (spin and
sublattice-valley) degeneracy of each QH state in graphene. The presence of the anomalous
(from the viewpoint of more standard condensed matter systems) term $1/2$ in the filling factor
unmistakably reveals the relativistic-like character of electron motion
in graphene \cite{Semenoff:1984dq,Haldane:1988zza,Khveshchenko:2001zza,
Khveshchenko:2001zz,Gorbar:2002iw,Sharapov:2003te}.

Recall that each plateau in a Hall conductivity corresponds to a gap in the quasiparticle 
energy spectrum. The later experiments in strong magnetic fields ($B \gtrsim 20~\mbox{T}$) 
\cite{2006PhRvL..96m6806Z,2007PhRvL..99j6802J} observed new QH plateaus, with integer 
filling factors  $\nu = 0, \pm 1$, and $\pm 4$. The more recent experiments 
\cite{2009Natur.462..192D,2009Natur.462..196B} discovered additional plateaus, with 
$\nu =\pm 3$ and $\nu = \pm 1/3$. While the latter corresponds to the fractional 
QH effect, the plateaus with $\nu = 0, \pm 1, \pm 4$, and $\nu = \pm 3$ are intimately 
connected with a (quasi-)spontaneous breakdown of the $\mathrm{U}(4)$ symmetry of 
the low-energy effective quasiparticle Hamiltonian in graphene (connected with the spin 
and sublattice-valley degeneracy mentioned above) \cite{Gorbar:2002iw}. It is a 
quasi-spontaneous (and not spontaneous) breakdown because due to the Zeeman 
effect the $\mathrm{U}(4)$ is reduced to a $\mathrm{U}_{\uparrow}(2) \times \mathrm{U}_{\downarrow}(2)$, 
with $\mathrm{U}(2)_s$ being the sublattice-valley symmetry at a
fixed spin ($s=\uparrow$ or $s=\downarrow$). However, taking into account that the
Zeeman interaction is rather weak for realistic magnetic fields, the $\mathrm{U}(4)$ is a good approximate
symmetry guaranteeing that at weak magnetic fields only the QH plateaus with the filling factors
$\nu=\pm 4(n+1/2)$ appear. The observed new QH plateaus $\nu = 0, \pm 1, \pm 4$, and
$\nu = \pm 3$ occur clearly due to the electron-electron interaction leading to (quasi-)spontaneous
$\mathrm{U}(4)$ symmetry breaking that removes the degeneracy of either $n=0$ ($\nu = 0, \pm 1$) 
or $n=1$ ($\nu = \pm 3, \pm 4$) LLs.

For the description of the new QH plateaus, the following two theoretical scenarios were
suggested. One of them is the QH ferromagnetism (QHF)
\cite{2006PhRvL..96y6602N,2006PhRvB..74g5423Y,2006PhRvB..74p1407G,
PhysRevB.74.075422,2007PhRvL..99s6802S}, whose order parameters
are the spin and valley charge densities (the dynamics of a Zeeman spin splitting enhancement
considered in Refs.~\cite{2006PhRvL..96q6803A,2007SSCom.143...77A} is intimately connected 
with the QHF).

The underlying physics of the QHF is connected with the exchange interaction in a many-body system 
with Coulomb repulsion which tends to lift the spin and/or valley degeneracy at half filling
\cite{1995PhRvB..5217366F}. This is similar to the physics behind the Hund's rule 
in atomic physics, in which case the Coulomb energy of the system is lowered by
anti-symmetrizing the coordinate part of the many-body wave function. Because of the 
Fermi statistics of the charge carriers, the corresponding lowest energy state must be 
symmetric in the spin-valley degrees of freedom. In other words, the ground state is
spin and/or valley polarized.

The second scenario is the magnetic catalysis (MC) scenario, whose order parameters 
are excitonic condensates responsible for the generation of the Dirac masses of charge 
carriers \cite{Gusynin:2006gn,Herbut:2006cs,PhysRevB.75.165411,2007PhRvB..76h5432H,
2007PhRvL..98a6803F,2007JPSJ...76i4701E,2007PhyE...40..269E}. The essence of the 
MC effect, discussed in Section~\ref{sec:MagneticCatalysis}, is connected with the 
effective dimensional reduction in the dynamics of charged fermions in an external magnetic field. 
It was first applied for a single layer of graphite in a magnetic field in
Refs.~\cite{Khveshchenko:2001zza,Khveshchenko:2001zz,Gorbar:2002iw}.

One may think that the QH ferromagnetism and magnetic catalysis order parameters 
should compete with each other. However, the analysis in an effective model with the 
local four-fermion interaction performed in Refs.~\cite{Gorbar:2007xh,Gorbar:2008hu}
showed that these two sets of the order parameters necessarily coexist (this feature has been 
recently discussed also in Ref.~\cite{Semenoff:2011ya}). This fact strongly indicates that 
these two sets of the order parameters have a common dynamical origin,  i.e., they are 
two sides of the same coin. For this reason, it is appropriately to call this 
phenomenon a {\it generalized magnetic catalysis}.

\subsection{Quantum Hall effect in model of graphene with short-range interaction}
\label{Sec:MagCatGrapheneLocal}

In this section, we discuss the key features of the dynamics responsible
for lifting the degeneracy of the Landau levels in graphene in the regime 
of the quantum Hall effect in strong magnetic fields. We will use a low-energy
model of graphene similar to that in Section~\ref{sec:MagCatQEDreduced}. 
The quasiparticles are described by Dirac fermions and their interaction is 
point-like. The analysis is done for graphene in the clean limit, neglecting 
all types of possible disorder \cite{2007PhRvB..75c3412K,2007PhRvL..99t6803G,
2007PhRvB..76t5408G,2008PhRvL.100x6806N,2008PhRvL.101c6805J}.
Such a toy model is sufficient to capture the essence of the dynamics 
associated with magnetic catalysis and quantum Hall ferromagnetism. 
Also, by taking into account a considerable improvement in the quality 
of suspended graphene samples \cite{2008SSCom.146..351B,2009PhRvL.102q6804L}, 
it may be expected that even such a toy model in the clean limit provides a reasonable 
qualitative description of some real devices.

\begin{figure}[t]
\begin{center}
\includegraphics[width=0.55\textwidth]{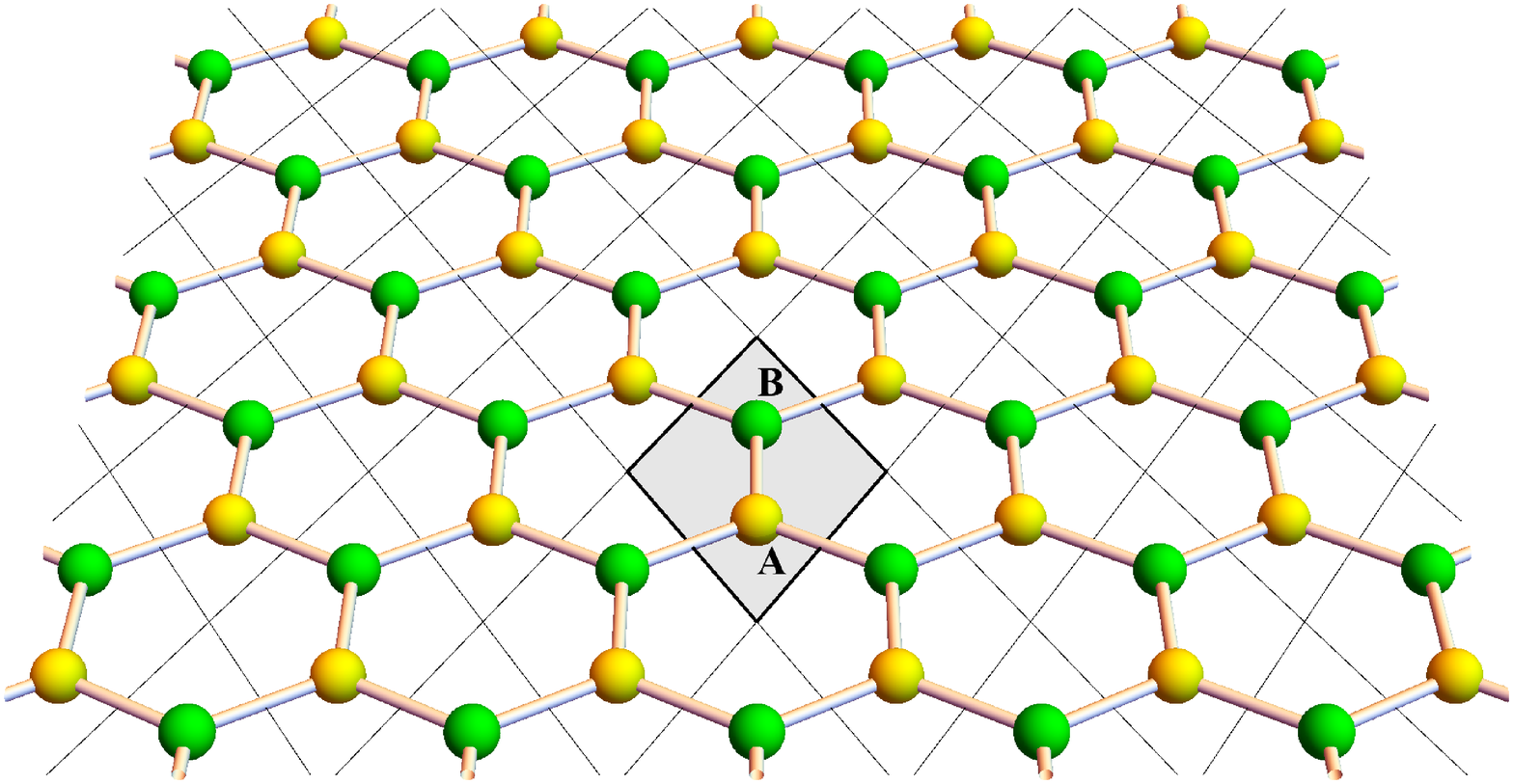}\hspace{0.04\textwidth}
\includegraphics[width=0.4\textwidth]{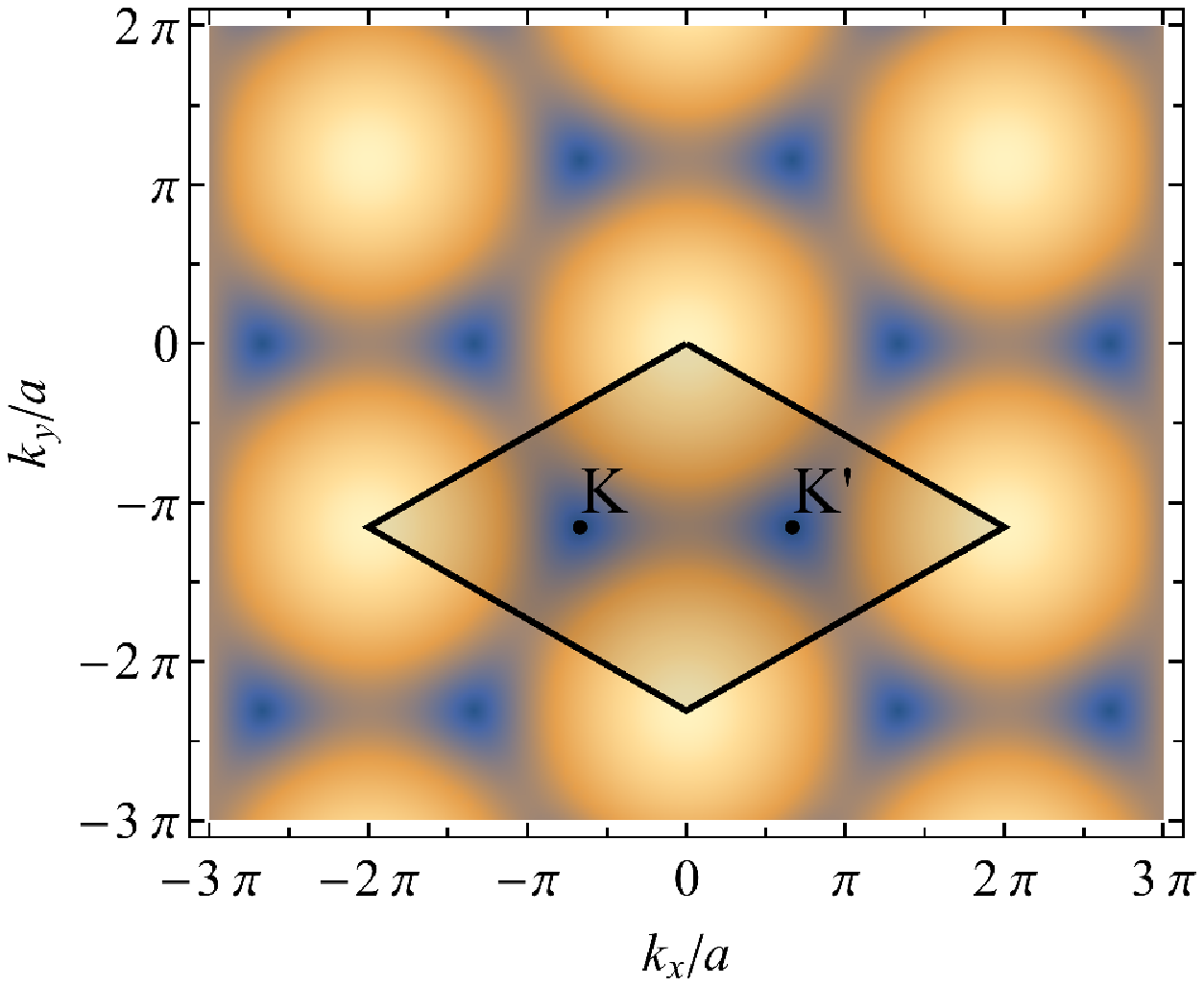}
\caption{(Color online) Left panel: graphene lattice in coordinate space.
Right panel: graphene lattice in reciprocal space.}
\label{graphene_lat}
\end{center}
\end{figure}

The quasiparticles in graphene are described by a four-component Dirac 
spinor field $\Psi_{s}^T = \left( \psi_{KAs},\psi_{KBs}, \psi_{K^\prime Bs},
-\psi_{K^\prime As}\right)$ which combines the Bloch states with spin indices
$s=\pm$ on the two different sublattices ($A$, $B$) of the hexagonal
graphene lattice and with momenta near the two inequivalent valley points
($K$, $K^\prime$) of the two-dimensional Brillouin zone, see Fig.~\ref{graphene_lat}. 
The free quasiparticle Hamiltonian can be recast in a relativistic-like form with the Fermi
velocity $v_F\approx 10^6~\mbox{m/s}$ playing the role of the speed of light:
\begin{equation}
H_0=v_F\int d^2{r}\,\overline{\Psi}\left(\gamma^1\pi_x+\gamma^2\pi_y\right)\Psi,
\label{free-hamiltonian}
\end{equation}
where $\mathbf{r} =(x,y)$ is the position vector in the plane of
graphene and $\overline{\Psi} =\Psi^\dagger\gamma^0$ is the Dirac
conjugated spinor. In Eq.~(\ref{free-hamiltonian}), $\gamma^\nu$
with $\nu=0,1,2$ are $4 \times 4$ gamma matrices belonging to a
reducible representation of the Dirac algebra, namely,
$\gamma^\nu=\tilde{\tau}^3\otimes (\tau^3,i\tau^2,-i\tau^1)$,
where the Pauli matrices $\tilde{\tau}^{i}$ and $\tau^{i}$, with
$i=1,2,3$, act in the subspaces of the valleys ($K$, $K^\prime$)
and sublattices ($A$, $B$), respectively. The matrices
satisfy the usual anticommutation relations
$\left\{\gamma^\mu,\gamma^\nu\right\}=2g^{\mu\nu}$, where
$g^{\mu\nu}=\mbox{diag}\,(1,-1,-1)$ and $\mu,\nu=0,1,2$. The
canonical momentum $\bm{\pi} \equiv (\pi_x, \pi_y)=
-i\hbar\bm{\nabla} - {e\mathbf{A}}/c$ includes the vector potential
$\mathbf{A}$ corresponding to a magnetic field $B_{\perp}$, which
is the component of the external magnetic field $\mathbf{B}$
orthogonal to the $xy$-plane of graphene.

The interaction term has the form
\begin{eqnarray}
H_{C} &=& \frac{1}{2}\int d^2{r} d^2{r}^\prime {\Psi}^{\dagger}(\mathbf{r})
\Psi(\mathbf{r})U_{C}(\mathbf{r}-\mathbf{r}^\prime)
{\Psi}^{\dagger}(\mathbf{r}^\prime) \Psi(\mathbf{r}^\prime),
\label{Coulomb}
\end{eqnarray}
where instead of the actual Coulomb potential we use the contact interaction 
of the following type: $G_{\rm int}\delta^{2}(\mathbf{r})$, 
where $G_{\rm int}$ is a dimensionful coupling constant. The Hamiltonian 
$H = H_0+H_{C}$ possesses a global $\mathrm{U}(4)$ symmetry with the generators
similar to those in Eq.~(\ref{U2Nf-generators}), except that the flavor generators
$\lambda^{\alpha}/2$ are replaced by the spin $\mathrm{SU}(2)$ generators. The
latter are given by the Pauli matrices $\sigma^\alpha$, where $\alpha=0, 1, 2, 3$
and $\sigma^0$ is the $2 \times 2$ unit matrix in the spinor space. In this section, 
we use the representation of Dirac matrices with $\gamma^3$ and $\gamma^5
\equiv i\gamma^0 \gamma^1 \gamma^2 \gamma^3 $ given by 
\begin{equation}
\gamma^3 \equiv i \tilde{\tau}^1\otimes \tau^0= i\left(\begin{array}{cc} 0& I \\ I& 0\end{array}\right),\qquad
\gamma^5 \equiv - \tilde{\tau}^2\otimes \tau^0= i\left(\begin{array}{cc} 0&I\\-I&0\end{array}\right), 
\end{equation}
where $I$ is the $2\times 2$ unit matrix. Note that while the Dirac matrices
$\gamma^0$ and $\pmb \gamma = (\gamma^{1}, \gamma^{2})$ anticommute with
$\gamma^3$ and $\gamma^5$, they commute with the diagonal matrix
$\gamma^3\gamma^5 = - \gamma^5\gamma^3$,
\begin{equation}
\gamma^3\gamma^5 = \left(\begin{array}{cc} I& 0 \\ 0& -I\end{array}\right).
\label{gamma35}
\end{equation}
In the context of the low-energy model at hand, the matrix $\gamma^3\gamma^5$ 
has the meaning of the pseudospin operator.

The electron chemical 
potential $\mu_0$ is introduced by adding the term $-\mu_0 \Psi^{\dagger}\Psi$ 
to the Hamiltonian density (\ref{free-hamiltonian}). This term also preserves the 
$\mathrm{U}(4)$ symmetry. The Zeeman interaction is included by adding the term 
$\mu_{B}B\Psi^{\dagger} \sigma^3 \Psi$, where 
$B \equiv |\mathbf{B}|$ and $\mu_B=|e|\hbar/(2mc)$ is the Bohr
magneton. Here we took into account that the Lande factor for
graphene is $g_L\simeq2$. The spin
matrix $\sigma^3$ has eigenvalue $+1$ ($-1$) for the states with
the spin directed along (against) the magnetic field
$\mathbf{B}$.
Such states will be called spin up (down) states. Because of the Zeeman 
term, the symmetry of the total Hamiltonian
\begin{equation}
H_{\rm tot} \equiv H + \int d^2{r}\,\left(\mu_{B}B\Psi^{\dagger}\sigma^3\Psi -
\mu_{0} \Psi^{\dagger}\Psi\right)
\label{tot}
\end{equation}
is broken down to a symmetry $\mathrm{U}(2)_{+} \times \mathrm{U}(2)_{-}$, where the 
subscript $\pm$ corresponds to spin up and spin down states, respectively. The generators 
of the $\mathrm{U}(2)_{s}$, with $s=\pm$, are $I_4 \otimes P_{s}$, $-i\gamma^3 \otimes P_{s}$,
$\gamma^5 \otimes P_{s}$, and $\gamma^3\gamma^5 \otimes P_{s}$, where 
$P_{\pm}=(1\pm \sigma^3)/2$ are the projectors on spin up and down states. 

It may be appropriate to note that, in addition to the density-density interaction  
in Eq.~(\ref{Coulomb}), preserving the $\mathrm{U}(4)$ flavor symmetry, there are numerous 
short-range interactions that break the symmetry in the complete theory \cite{PhysRevB.76.195415}. 
Despite being rather small, the corresponding interactions could play an important role in 
determining the energetically most stable ground state. The reason for this is the renormalization
of the short-range interactions that greatly enhances the bare values of coupling constants
and makes such interactions even more important than the Zeeman term \cite{KharitonovPRB85.155439}. 
In fact, as was argued in Ref.~\cite{KharitonovPRB85.155439}, the result of such dynamics could 
result in a ground state with almost any symmetry-breaking pattern.

In order to get an insight into the problem of quantum Hall states in graphene, we will 
use the simplest model with a short-range density-density interaction that {\it preserves} the 
$\mathrm{U}(4)$ flavor symmetry. Such an interaction will mimic the long-range Coulomb 
interaction (\ref{Coulomb}). Except for the Zeeman energy, no additional symmetry-breaking 
short-range interactions will be included. The use of such a simplified framework will affect 
our ability to reliably compare the energies of various states, but will be still rather instructive. 
It will allow us to reveal the underlying structure of most (although not all) candidates for the 
ground states of graphene at various filling factors. The power of such an approach lies in the
fact that the dynamics in the real material is largely dominated by the long-range Coulomb 
interaction (\ref{Coulomb}). The small symmetry-breaking terms that help to choose preferred 
directions of the ground state alignment will be considered in Section~\ref{Sec:GrapheneVacuumAlignment}.

The dynamics will be treated in the Hartree-Fock (mean-field) approximation,
which is conventional and appropriate in this case \cite{Khveshchenko:2001zza,Khveshchenko:2001zz,
Gorbar:2002iw,2006PhRvL..96y6602N,2006PhRvB..74g5423Y,2006PhRvB..74p1407G,
Gusynin:2006gn}. The corresponding gap equation takes the following form:
\begin{equation}
G^{-1}(u,u^\prime) = S^{-1}(u,u^\prime)
+ i\hbar G_{\rm int}\gamma^0G(u,u)\gamma^0 \delta^{3}(u - u^\prime) 
- i\hbar G_{\rm int}\gamma^0\, \mathrm{tr}[\gamma^0G(u,u)]\delta^{3}(u- u^\prime),
\label{GapEqLocal}
\end{equation}
where $u \equiv (t,\mathbf{r})$, $t$ is the time coordinate,
$G(u,u^\prime)=\hbar^{-1}\langle 0|T\Psi(u)\bar{\Psi}(u^\prime)|0\rangle$ is
the {\it full} quasiparticle propagator, and
\begin{equation}
iS^{-1}(u,u^\prime)=\left[(i\hbar\partial_t+\mu_0 - \mu_BB\sigma^3)\gamma^0
-v_{F}(\bm{\pi}\cdot\bm{\gamma})\right]\delta^{3}(u- u^\prime)
\label{inversebare}
\end{equation}
is the inverse {\it bare} quasiparticle propagator.
Note that while the second term on the right hand side of Eq.~(\ref{GapEqLocal})
describes the exchange interaction, the third one is the Hartree term describing
the direct interaction. 

It is convenient to introduce the following dimensionless coupling constant 
$\lambda=G_{\rm int} \Lambda/(4\pi^{3/2} \hbar^2 v_{F}^2)$ instead of 
$G_{\rm int}$. Note that $\Lambda$ is the energy cutoff parameter which 
is required when a contact interaction is used. The natural choice for the 
cutoff $\Lambda$ in the problem at hand is the Landau energy scale,
\begin{equation}
\epsilon_{B} \equiv \sqrt{2\hbar|eB_{\perp}|v_{F}^2/c} \simeq
424\sqrt{|B_{\perp}[\mbox{T}]|}~\mbox{K}
\label{Lscale}
\end{equation}
which is in fact the only relevant energy scale in the dynamics with the Coulomb 
interaction. With the choice $\Lambda \sim \epsilon_{B}$, the dimensionful 
coupling constant $G_{\rm int}$ becomes $G_{\rm int} \sim 4\pi^{3/2}\hbar^2 v^{2}_F\lambda/\epsilon_{B}$.

The goal here is to find all solutions of Eq.~(\ref{GapEqLocal}) both with intact and spontaneously
broken $\mathrm{SU}(2)_{s}$ symmetry, where $\mathrm{SU}(2)_{s}$ is the largest non-Abelian 
subgroup of the $\mathrm{U}(2)_{s}$. The Dirac mass term $\tilde{\Delta}_{s}\bar{\Psi}P_{s}\Psi \equiv
\tilde{\Delta}_{s}\Psi^{\dagger}\gamma^{0}P_{s}\Psi$, where $\tilde{\Delta}_{s}$ is a
Dirac gap (mass),\footnote{The energy gap $\tilde{\Delta}_{s}$ is expressed through
the corresponding Dirac mass $\tilde{m}_{s}$ as $\tilde{\Delta}_{s} =
\tilde{m}_{s} v_{F}^2$. In what follows, we will ignore this difference between
them and use the term ``Dirac mass.''} is assigned to the triplet representation of the
$\mathrm{SU}(2)_{s}$, and the generation of such a mass would lead to a spontaneous 
breakdown of the flavor $\mathrm{SU}(2)_{s}$ symmetry down to the $\tilde{\mathrm{U}}(1)_{s}$
with the generator $\gamma^3\gamma^5 \otimes P_{s}$ 
\cite{Khveshchenko:2001zza,Khveshchenko:2001zz,Gorbar:2002iw,Gusynin:2006gn}. 
There is also a Dirac mass term of the form $\Delta_{s}\bar{\Psi}\gamma^3\gamma^{5}P_{s}\Psi$
that is a singlet with respect to $\mathrm{SU}(2)_{s}$, and therefore its generation would
not break this symmetry. On the other hand, while the triplet mass term is even
under time reversal $\cal{T}$, the singlet mass term is $\cal{T}$-odd (for a
review of the transformation properties of different mass terms in
graphene, see Ref.~\cite{Gusynin:2007ix}). Note that the possibility of
a singlet Dirac mass like $\Delta$ in 2D graphite
was first discussed long time ago \cite{Haldane:1988zza}. 

The masses $\Delta_{s}$ and $\tilde{\Delta}_{s}$ are related to the MC order
parameters $\langle {\bar{\Psi}\gamma^3\gamma^{5}P_{s}\Psi} \rangle$ and
$\langle {\bar{\Psi}P_{s}\Psi} \rangle$. In terms of the Bloch components of
the spinors, the corresponding operators take the following forms:
\begin{eqnarray}
\label{singlet_mass}
\Delta_{s}: &\quad&
{\bar{\Psi}\gamma^3 \gamma^5 P_{s} \Psi} =
  \psi_{K  As}^\dagger\psi_{K As}
- \psi_{K^{\prime} As}^\dagger \psi_{K^{\prime}As}
- \psi_{K Bs}^\dagger \psi_{K Bs}
+ \psi_{K^{\prime}Bs}^\dagger \psi_{K^{\prime} Bs},\\
\label{triplet_mass}
\tilde{\Delta}_{s}:&\quad&
{\bar{\Psi} P_{s} \Psi} =
  \psi_{K  As}^\dagger\psi_{K As}
+ \psi_{K^{\prime} As}^\dagger \psi_{K^{\prime}As}
- \psi_{K Bs}^\dagger \psi_{K Bs}
- \psi_{K^{\prime}Bs}^\dagger \psi_{K^{\prime} Bs}.
\end{eqnarray}
The expressions on the right hand side further clarify the
physical meaning of the Dirac mass parameters as the Lagrange
multipliers that control various density imbalances of electrons
at the two valleys and the two sublattices. In particular, the
order parameter (\ref{triplet_mass}), connected with the triplet
Dirac mass, describes a charge density imbalance between
the two sublattices, i.e., a charge density
wave \cite{Khveshchenko:2001zza,Khveshchenko:2001zz,Gusynin:2006gn}.

For quasiparticles of a fixed spin, the full inverse propagator takes the 
following general form [compare with Eq.~(\ref{inversebare})]:
\begin{eqnarray}
iG^{-1}_{s}(u,u^\prime)&=& \left[(i\hbar\partial_t+\mu_{s} +
\tilde{\mu}_{s}\gamma^3\gamma^5)\gamma^0 - v_{F}(\bm{\pi}\cdot\bm{\gamma})
-\tilde{\Delta}_{s} + \Delta_{s}\gamma^3\gamma^5\right]\delta^{3}(u-u^\prime),
\label{full-inverse}
\end{eqnarray}
where the parameters $\mu_{s}$, $\tilde{\mu}_{s}$, $\Delta_{s}$, and $\tilde{\Delta}_{s}$ 
are determined from gap equation (\ref{GapEqLocal}). Note that the {full} electron chemical
potentials $\mu_{\pm}$ include the Zeeman energy $\mp \epsilon_{Z}$, 
which is rather small in magnitude,
\begin{equation}
\epsilon_{Z}=\mu_B B = 0.67B[\mbox{T}]~\mbox{K}.
\label{Zeeman}
\end{equation}
The chemical potential $\tilde{\mu}_{s}$ is related to the density of the
conserved pseudospin charge $\Psi^{\dagger}\gamma^3\gamma^{5}P_{s}\Psi$,
which is assigned to the triplet representation of the $\mathrm{SU}(2)_{s}$. Therefore, unlike
the masses $\Delta_{s}$ and $\tilde{\Delta}_{s}$, the chemical potentials $\mu_3
\equiv (\mu_{+} - \mu_{-})/2$ and $\tilde{\mu}_{s}$ are related to the conventional
QHF order parameters: the spin density $\langle {\Psi^{\dagger}\sigma^3 \Psi} \rangle$ and
the pseudospin density $\langle {\Psi^{\dagger}\gamma^3\gamma^5P_{s}\Psi} \rangle $,
respectively. In terms of the Bloch components, the corresponding operators
take the following forms:
\begin{eqnarray}
\label{singlet_mu}
\mu_3:&\quad&
 {\Psi^{\dagger}\sigma^3 \Psi} =\frac{1}{2}
 \sum_{\kappa=K ,K^{\prime}} \sum_{a =A,B}
\left(  \psi_{\kappa a +}^\dagger\psi_{\kappa a +}
- \psi_{\kappa a-}^\dagger\psi_{\kappa a-}\right),\\
\label{triplet_mu}
\tilde{\mu}_{s}:&\quad&
{\Psi^{\dagger}\gamma^3\gamma^5P_{s}\Psi} =
  \psi_{K  As}^\dagger\psi_{K As}
- \psi_{K^{\prime} As}^\dagger \psi_{K^{\prime}As}
+ \psi_{K Bs}^\dagger \psi_{K Bs}
- \psi_{K^{\prime}Bs}^\dagger \psi_{K^{\prime} Bs}.
\end{eqnarray}
By comparing the last expression with Eq.~(\ref{triplet_mass}), we see that while
the triplet MC order parameter related to $\tilde{\Delta}_{s}$ describes the charge
density imbalance between the two graphene sublattices,
the pseudospin density
(related to $\tilde{\mu}_{s}$)  describes the charge density imbalance between
the two valley points in the Brillouin zone. 
On the other hand, as seen
from Eq.~(\ref{singlet_mu}), $\mu_3$ is related to the conventional ferromagnetic order
parameter $\langle {\Psi^{\dagger}\sigma^3 \Psi} \rangle$ .

The following remark is in order. Because of the relation
$\gamma^5 = i\gamma^0 \gamma^1 \gamma^2 \gamma^3$, the operator in
Eq.~(\ref{triplet_mu}) can be rewritten as
$i{\bar \Psi}\gamma^1 \gamma^2 P_{s}\Psi$. The latter has the same
form as the anomalous magnetic moment operator in Quantum Electrodynamics
(QED). However, unlike QED, in graphene, it describes not the polarization
of the spin degrees of freedom but that of the pseudospin ones,
related to the valleys and sublattices. Because of that, this operator
can be called the anomalous magnetic pseudomoment operator.

Let us describe the breakdown of the $\mathrm{U}(4)$ symmetry down to the $\mathrm{U}(2)_{+}
\times \mathrm{U}(2)_{-}$ flavor symmetry, responsible for a spin gap, in more detail.
Because of the Zeeman term, this breakdown is not spontaneous but explicit.
The point however is that, as was shown in Ref.~\cite{2006PhRvL..96q6803A,2007SSCom.143...77A},
a magnetic field leads to a strong enhancement of the spin gap in graphene.
Such an enhancement is reflected in a large chemical potential
$\mu_3 = (\mu_{+} -\mu_{-})/2\gg \epsilon_{Z}$ and the corresponding QHF order
parameter $\langle {\Psi^{\dagger}\sigma^3 \Psi} \rangle$. But it is not all.
There is also a large contribution to the spin gap connected with
the flavor singlet Dirac mass $\Delta_3 \equiv (\Delta_{+} -\Delta_{-})/2$ and the
corresponding MC order parameter $\langle {\bar{\Psi}\gamma^{3}\gamma^{5}\sigma^3
\Psi} \rangle $. 

The $\mathrm{U}(2)_{+} \times \mathrm{U}(2)_{-}$ is an exact symmetry of the total Hamiltonian
$H_{\rm tot}$ (\ref{tot}) of the continuum effective theory. However, as was pointed out in 
Ref.~\cite{PhysRevB.74.075422} (see also Refs.~\cite{2007arXiv0706.4280L,2007PhRvB..76h5432H,
PhysRevB.75.165411,Herbut:2006cs,PhysRevB.76.195415}), 
it is not exact for the Hamiltonian on
the graphene lattice. In fact there are small on-site repulsion
interaction terms which break the
$\mathrm{U}(2)_{+} \times \mathrm{U}(2)_{-}$ symmetry down to a $\mathrm{U}(1)_{+} \times Z_{2+}
\times \mathrm{U}(1)_{-} \times Z_{2-}$ subgroup, where the elements of the
discrete group $Z_{2s}$ are
$\gamma^5 \otimes P_{s} + I_{4}\otimes P_{-s}$ and the unit matrix.
Unlike a spontaneous breakdown of continuous symmetries, a spontaneous breakdown of
the discrete symmetry $Z_{2\pm}$, with the order parameters
$\langle {\bar{\Psi}P_{\pm}\Psi} \rangle $ and
$\langle\Psi^{\dagger}\gamma^3\gamma^{5}P_{\pm}\Psi\rangle$,
is not forbidden by the Mermin-Wagner theorem at finite
temperatures in a planar system \cite{Mermin:1966fe}. This observation is of relevance
for the description of the ground state responsible for the $\nu = \pm 1$
plateaus.

Thus, there are six order parameters describing the breakdown of
the $\mathrm{U}(4)$ symmetry: the two singlet order parameters connected with $\mu_3$ and
$\Delta_3$ and the four triplet ones connected with $\tilde{\mu}_{\pm}$ and
$\tilde{\Delta}_{\pm}$.

By extracting the location of the poles in the full propagator $G(u,u^\prime)$, 
given in Eq.~(A27) in Appendix A in Ref.~\cite{Gorbar:2008hu},
it is  straightforward to derive the dispersion relations for the quasiparticles 
in graphene. The dispersion relations for LLs with $n \geq 1$ are
\begin{eqnarray}
\hspace{-2mm}\omega^{(\sigma)}_{ns} &=&-\mu_{s} + \sigma\tilde{\mu}_{s}
\pm\sqrt{n \epsilon_{B}^2 + (\tilde{\Delta}_{s}+\sigma\Delta_{s})^2}\,,
\label{higherLLs}
\end{eqnarray}
where $\sigma=\pm 1$ and the two signs in front of the square root
correspond to the energy levels above and below the Dirac point.
In the case of the LLL, which is special, the corresponding
dispersion relations read
\begin{equation}
\omega^{(\sigma)}_{s}= -\mu_{s} + \sigma
\left(\tilde{\mu}_{s}\,\mathrm{sign}(eB_{\perp})+\,\tilde{\Delta}_{s}\right)
+ \Delta_{s}\,\mathrm{sign}(eB_{\perp}).
\label{LLLenergylevels}
\end{equation}
As shown in Appendix A in Ref.~\cite{Gorbar:2008hu}, the parameter
$\sigma$ in Eqs.~(\ref{higherLLs}) and (\ref{LLLenergylevels}) is
connected with the eigenvalues of the diagonal pseudospin matrix
$\gamma_3\gamma_5$ in Eq.~(\ref{gamma35}). For the LLs with $n
\geq 1$, the value $\sigma = \pm 1$ in (\ref{higherLLs})
corresponds to the eigenvalues $\mp 1$ of $\gamma^3\gamma^5$. On
the other hand, for the LLL, the value $\sigma = \pm 1$ in
(\ref{LLLenergylevels}) corresponds to
$\mathrm{sign}(eB_\perp)\times \,(\mp 1)$, with $\mp 1$ being the
eigenvalues of $\gamma^3\gamma^5$.

One can see from Eqs.~(\ref{higherLLs}) and (\ref{LLLenergylevels}) that
at a fixed spin, the terms with $\sigma$ are responsible for splitting of LLs.

\subsubsection{Gap equation: coexistence of QHF and MC parameters}
\label{Sec:Gap_equation}

The equations for the Dirac masses $\Delta_{s}$ and $\tilde{\Delta}_{s}$ and the
chemical potentials $\mu_{s}$ and $\tilde{\mu}_{s}$ follow from the matrix form
of the gap equation in Eq.~(\ref{GapEqLocal}) and expression (\ref{full-inverse}).
Their derivation, while straightforward, is rather tedious. It is considered
in Appendix A in Ref.~\cite{Gorbar:2008hu} in detail. At zero temperature, the equations are
\begin{eqnarray}
\tilde{\Delta}_{s} &=& \frac{A}{2}
\Bigg\{
-\left[\mathrm{sign}(\mu_{s}-\tilde{\mu}_{s})\theta(|\mu_{s}-\tilde{\mu}_{s}|-E_{0s}^{+})
- \mathrm{sign}(\mu_{s}+\tilde{\mu}_{s})\theta(|\mu_{s}+\tilde{\mu}_{s}|-E_{0s}^{-})\right]
\mathrm{sign}(eB_{\perp}) \nonumber \\
&&+\sum_{n=0}^{\infty}\left[\frac{(\tilde{\Delta}_{s}+
\Delta_{s})\theta(E_{ns}^{+}-|\mu_{s}-\tilde{\mu}_{s}|)}{E_{ns}^{+}}
+\frac{(\tilde{\Delta}_{s}-\Delta_{s})\theta(E_{ns}^{-}-|\mu_{s}+\tilde{\mu}_{s}|)}
{E_{ns}^{-}}\right]
[1+\theta(n-1)]\Bigg\},
\label{E1-a}
\end{eqnarray}
\begin{eqnarray}
\Delta_{s} &=& \frac{A}{2}
\Bigg\{
-\left[\mathrm{sign}(\mu_{s}-\tilde{\mu}_{s})\theta(|\mu_{s}-\tilde{\mu}_{s}|-E_{0s}^{+}) +
\mathrm{sign}(\mu_{s}+\tilde{\mu}_{s})\theta(|\mu_{s}+\tilde{\mu}_{s}|-E_{0s}^{-})\right]
\mathrm{sign}(eB_{\perp}) \nonumber \\
&&+ \sum_{n=0}^{\infty}
\left[\frac{(\tilde{\Delta}_{s}+\Delta_{s})\theta(E_{ns}^{+}-|\mu_{s}-\tilde{\mu}_{s}|)}
{E_{ns}^{+}}-\frac{(\tilde{\Delta}_{s}-\Delta_{s})\theta(E_{ns}^{-}
-|\mu_{s}+\tilde{\mu}_{s}|)}{E_{ns}^{-}}\right][1+\theta(n-1)]
\Bigg\},
\label{E2-a}
\end{eqnarray}
\begin{eqnarray}
\tilde{\mu}_{s} &=& \frac{A}{2}
\Bigg\{
\left[\frac{(\tilde{\Delta}_{s}+\Delta_{s})\theta(E_{0s}^{+}
-|\mu_{s}-\tilde{\mu}_{s}|)}{E_{0s}^{+}} +
\frac{(\tilde{\Delta}_{s}-\Delta_{s})\theta(E_{0s}^{-}-
|\mu_{s}+\tilde{\mu}_{s}|)}{E_{0s}^{-}}\right]
\mathrm{sign}(eB_{\perp})\nonumber \\
&&+\sum_{n=0}^{\infty}
\left[-\mathrm{sign}(\mu_{s}-\tilde{\mu}_{s})\theta(|\mu_{s}-\tilde{\mu}_{s}|-E_{ns}^{+}) +
\mathrm{sign}(\mu_{s}+\tilde{\mu}_{s}) \theta(|\mu_{s}+\tilde{\mu}_{s}|-E_{ns}^{-})\right]
 [1+\theta(n-1)]
\Bigg\},
\label{E3-a}
\end{eqnarray}
\begin{eqnarray}
\mu_{s} &=& \bar{\mu}_{s} + X + \frac{A}{2}
\Bigg\{
-\left[\frac{(\tilde{\Delta}_{s}+\Delta_{s})\theta(E_{0s}^{+}-|\mu_{s}-\tilde{\mu}_{s}|)}{E_{0s}^{+}} -
\frac{(\tilde{\Delta}_{s}-\Delta_{s})\theta(E_{0s}^{-}
-|\mu_{s}+\tilde{\mu}_{s}|)}{E_{0s}^{-}}\right]\mathrm{sign}(eB_{\perp})\nonumber \\
&&+\sum_{n=0}^{\infty}
\left[\mathrm{sign}(\mu_{s}-\tilde{\mu}_{s})\theta(|\mu_{s}-\tilde{\mu}_{s}|-E_{ns}^{+}) +
\mathrm{sign}(\mu_{s}+\tilde{\mu}_{s})\theta(|\mu_{s}+\tilde{\mu}_{s}|-E_{ns}^{-})\right]  [1+\theta(n-1)]
\Bigg\}, \label{E4-a}
\end{eqnarray}
where the step function is defined by $\theta(x)= 1$ for $x\geq 0$ and $\theta(x)= 0$
for $x<0$. Regarding the other notation, $\bar{\mu}_{\pm} \equiv \mu_0 \mp \epsilon_{Z}$ is the
bare electron chemical potential which includes the Zeeman energy $\epsilon_{Z} = \mu_{B}B$,
and $E_{ns}^{\pm}=\sqrt{n \epsilon_{B}^2+ (\tilde{\Delta}_{s} \pm \Delta_{s})^2}$
are quasiparticle energies. In {these} equations, we introduced a new
energy scale, $A$, that plays an important role throughout the analysis.
It is determined by the value of the magnetic field and the coupling
constant strength,
\begin{equation}
A\equiv \frac{G_{\rm int}|eB_{\perp}|}{8\pi\hbar c} =
\frac{\sqrt{\pi}\lambda\epsilon_{B}^2}{4\Lambda}.
\label{Apar}
\end{equation}
The second term on the right hand side in Eq.~(\ref{E4-a}) is defined as
follows: 
\begin{equation}
X = \sum_{s=\pm}\,X_{s},
\label{X}
\end{equation}
where
\begin{eqnarray}
X_{s} &=& -2A
\Bigg\{
 -\left[\frac{(\tilde{\Delta}_{s}+\Delta_{s})\theta(E_{0s}^{+}
-|\mu_{s}-\tilde{\mu}_{s}|)}{E_{0s}^{+}} -
\frac{(\tilde{\Delta}_{s}-\Delta_{s})\theta(E_{0s}^{-}
-|\mu_{s}+\tilde{\mu}_{s}|)}{E_{0s}^{-}}\right] \mathrm{sign}(eB_{\perp})
\nonumber \\
&&+\sum_{n=0}^{\infty}
\left[\mathrm{sign}(\mu_{s}-\tilde{\mu}_{s})\theta(|\mu_{s}-\tilde{\mu}_{s}|-E_{ns}^{+}) +
\mathrm{sign}(\mu_{s}+\tilde{\mu}_{s})\theta(|\mu_{s}+\tilde{\mu}_{s}|-E_{ns}^{-})\right]
 [1+\theta(n-1)]
\Bigg\}.
\label{Xs}
\end{eqnarray}
The following comment is in order here. Because of the Hartree term in the gap 
equation (\ref{GapEqLocal}), the equations for the spin up and spin down parameters
do not decouple: they are mixed via the $X$ term in Eq.~(\ref{E4-a}). Fortunately,
it is the only place affected by the Hartree term. This fact strongly simplifies the 
analysis of the system of equations (\ref{E1-a}) -- (\ref{E4-a}).
This point also clearly reflects the essential difference between the roles played
by the exchange and Hartree interactions in the quasiparticle dynamics
of graphene. While the former dominates in producing the QHF and MC order
parameters, the latter participates only in the renormalization of the electron
chemical potential, which is relevant for the filling of LLs.

Since the step functions in the above set of equations depend on $\mu_{s} \pm \tilde{\mu}_{s}$
and $\tilde{\Delta}_{s} \pm \Delta_{s}$, it is more convenient to rewrite the gap
equations for the following set of parameters
\begin{eqnarray}
\Delta_{s}^{(\pm)} = \Delta_{s}\pm\tilde{\Delta}_{s}\,, \qquad
\mu_{s}^{(\pm)} = \mu_{s}\pm\tilde{\mu}_{s}.
\label{newparam}
\end{eqnarray}
In the numerical analysis, we always consider a nonzero temperature. This is
implemented by utilizing the Matsubara formalism. One can check that the 
prescription for modifying Eqs.~(\ref{E1-a}) -- (\ref{E4-a}) at $T \ne 0$ is to
replace
\begin{eqnarray}
\mathrm{sign} (\mu_{s}^{(\pm)}) \theta(|\mu_{s}^{(\pm)}| -
E_{ns}^{\mp}) & \to & \frac{\sinh \frac{\mu_{s}^{(\pm)}}{T}}{\cosh
\frac{E_{ns}^{\mp}}{T} + \cosh \frac{\mu_{s}^{(\pm)}}{T}},
\label{finiteT1}
\\
\theta(E_{ns}^{\pm} - |\mu_{s}^{(\mp)}|) & \to & \frac{\sinh
\frac{E_{ns}^{\pm}}{T}}{\cosh \frac{E_{ns}^{\pm}}{T} + \cosh
\frac{\mu_{s}^{(\mp)}}{T}}.
\label{finiteT2}
\end{eqnarray}
This leads to the following set of equations:
\begin{eqnarray}
\Delta_{s}^{(\pm)} &=& A f_1\left(\Delta_{s}^{(\pm)},\mu_{s}^{(\mp)}\right),
\label{Deltas}
\\
\mu_{s}^{(\pm)} &=& \bar{\mu}_{s} +
A f_2\left(\Delta_{s}^{(\mp)},\mu_{s}^{(\pm)}\right)
+2 A f_2\left(\Delta_{s}^{(\pm)},\mu_{s}^{(\mp)}\right)
+2 A f_2\left(\Delta_{-s}^{(\pm)},\mu_{-s}^{(\mp)}\right)
+2 A f_2\left(\Delta_{-s}^{(\mp)},\mu_{-s}^{(\pm)}\right),
\label{mus}
\end{eqnarray}
where $\Delta^{(\pm)}_{s}$ and $\mu_{s}^{(\pm)}$ are given in
Eq.~(\ref{newparam}), and
\begin{eqnarray} \label{f1}
f_1\left(\Delta^{(\pm)}_{s},\mu^{(\mp)}_{s}\right)&=&
\frac{\sinh\left(\frac{\Delta^{(\pm)}_{s}}{T}\right) -
s_{\perp}\sinh\left(\frac{\mu^{(\mp)}_{s}}{T}\right)}
{\cosh\left(\frac{\Delta^{(\pm)}_{s}}{T}\right)+
\cosh\left(\frac{\mu^{(\mp)}_{s}}{T} \right)}+\sum_{n=1}^{\infty}
\frac{2\Delta^{(\pm)}_{s}\sinh\left(\frac{E_{ns}^{\pm}}{T}\right)}
{E_{ns}^{\pm}\left[\cosh\left(\frac{E_{ns}^{\pm} }{T}\right)+
\cosh\left(\frac{\mu^{(\mp)}_{s}}{T} \right)\right]},\\
\label{f2} f_2\left(\Delta^{(\pm)}_{s},\mu^{(\mp)}_{s}\right)&=&
\frac{s_{\perp}\sinh\left(\frac{\Delta^{(\pm)}_{s}}{T}\right) -
\sinh\left(\frac{\mu^{(\mp)}_{s}}{T}\right)}
{\cosh\left(\frac{\Delta^{(\pm)}_{s}}{T}\right)+
\cosh\left(\frac{\mu^{(\mp)}_{s}}{T} \right)}
-\sum_{n=1}^{\infty}\frac{2\sinh\left(\frac{\mu^{(\mp)}_{s}}{T}\right)}
{\cosh\left(\frac{E_{ns}^{\pm}}{T}\right)+ \cosh\left(\frac{\mu^{(\mp)}_{s}}{T}
\right)},
\end{eqnarray}
with $s_{\perp} \equiv \mathrm{sign}(eB_{\perp})$ and
$E_{ns}^{\pm}= \sqrt{n \epsilon_{B}^2+ \left(\Delta_{s}^{(\pm)}\right)^2}$.

Let us now show that the QHF and MC order parameters should always coexist
in this dynamics. Suppose that Eqs.~(\ref{Deltas}) and (\ref{mus}) have a
solution with some of the chemical potentials $\mu^{\mp}_{s}$ being nonzero
but the Dirac masses being zero, $\Delta^{(\pm)}_{s} = 0$. Then, the
left hand side of Eq.~(\ref{Deltas}) is equal to zero. On the other hand,
taking into account expression (\ref{f1}) for the function $f_1$, we find that for
$\Delta^{(\pm)}_{s} = 0$ the right hand side of this equation takes the form
\begin{equation}
Af_1\left(0,\mu^{(\mp)}_{s}\right) = A\frac{-s_{\perp}
\sinh\left(\frac{\mu^{(\mp)}_{s}}{T}\right)} {1 +
\cosh\left(\frac{\mu^{(\mp)}_{s}}{T} \right)} =-As_{\perp}
\tanh\left(\frac{\mu^{(\mp)}_{s}}{2T}\right),
\end{equation}
and {it} could be zero only if {\it all} chemical potentials 
$\mu^{(\mp)}_{s}$
disappear, in contradiction with our assumption. Therefore, we conclude that the
QHF and MC order parameters in this dynamics necessarily coexist indeed.
This is perhaps one of the central observations in this analysis.

Which factors underlie this feature of the graphene dynamics in a
magnetic field? It is the relativistic nature of the free
Hamiltonian $H_0$ in Eq.~(\ref{free-hamiltonian}) and the special
features of the LLs associated with it. To see this, note that
while the triplet Dirac mass $\tilde{\Delta}_{s}$ multiplies the
unit Dirac matrix $I_4$, the triplet chemical potential
$\tilde{\mu}_{s}$ comes with the matrix $\gamma^3\gamma^5\gamma^0$
in the inverse propagator $G^{-1}_{s}$ in
Eq.~(\ref{full-inverse}). Let us trace how these two structures
are connected with each other. The point is that there are terms
with $i\gamma^1\gamma^2\mathrm{sign}(eB_{\perp})$ matrix in the
expansion of the propagator $G_{s}$ over LLs. Taking into account the
definition $\gamma^5 = i\gamma^0\gamma^1\gamma^2\gamma^3$, we have
$i\gamma^1\gamma^2 = \gamma^3\gamma^5\gamma^0$. Then, through the
exchange term $\sim \gamma^0 G_{s} \gamma^0$ in gap equation
(\ref{GapEqLocal}), the $\tilde{\Delta}_{s}$ term in the inverse
propagator $G^{-1}_{s}$ necessarily induces the term with the
chemical potential $\tilde{\mu}_{s}$. In the same way, the singlet
Dirac mass $\Delta_{s}$ in $G^{-1}_{s}$ is connected with the
singlet chemical potential $\mu_{s}$.

These arguments are based on the kinematic structure of gap equation (\ref{GapEqLocal}),
which is the same as that for equation (\ref{SD}) with the Coulomb interaction.
Taking into account the universality of the MC phenomenon, we conclude that
the coexistence of the QHF and MC order parameters is a robust feature of the
QH dynamics in graphene.

The necessity of the coexistence of the QHF and MC order parameters can be 
clearly seen in the case of the dynamics on the LLL. As one can check from the 
structure of the LLL propagator, it contains only the combinations 
$-\mu_{s} + \Delta_{s}\mathrm{sign}(eB_{\perp})$ and $\tilde{\mu}_{s}
\mathrm{sign}(eB_{\perp}) + \tilde{\Delta}_{s}$. Therefore, in this case, 
the QHF and MC parameters not only coexist but they are not independent,
which in turn reflects the fact that the sublattice and valley degrees of freedom 
are not independent on the LLL. In
particular, by using Eqs.~(\ref{singlet_mass}), (\ref{triplet_mass}), (\ref{singlet_mu}), 
and (\ref{triplet_mu}), one can easily check that, because of the projector 
${\cal{P}_{+}} = [1 + i\gamma^{1}\gamma^{2}\mathrm{sign}(eB_{\perp})]/2$ 
in the LLL propagator, the operators $\Psi^{\dagger}P_{s}\Psi$ and 
$\bar{\Psi}\gamma^3 \gamma^5 P_{s} \Psi$ ($\Psi^{\dagger}\gamma^3\gamma^5 P_{s}\Psi$
and $\bar{\Psi}P_{s} \Psi$), determining the order parameters related to $\mu_{s}$ 
and $\Delta_{s}$ ($\tilde{\mu}_{s}$ and $\tilde{\Delta}_{s}$), coincide up to a sign factor
$\mathrm{sign}(eB_{\perp})$. For example, in the case with $\mathrm{sign}(eB_{\perp}) < 0$, 
one has 
\begin{eqnarray}
\left. \Psi^{\dagger}P_{s}\Psi \right|_{LLL}  &=& \left. \bar{\Psi}\gamma^3 \gamma^5 P_{s} \Psi \right|_{LLL} 
=\psi_{K  As}^\dagger\psi_{K As} +  \psi_{K^{\prime}Bs}^\dagger \psi_{K^{\prime} Bs} ,\\
\left. \Psi^{\dagger}\gamma^3\gamma^5 P_{s}\Psi \right|_{LLL} &=& \left. \bar{\Psi}P_{s} \Psi \right|_{LLL} 
=\psi_{K  As}^\dagger\psi_{K As} - \psi_{K^{\prime}Bs}^\dagger \psi_{K^{\prime} Bs}.
\end{eqnarray}
Similarly, when $\mathrm{sign}(eB_{\perp}) > 0$, the relations are 
\begin{eqnarray}
\left. \Psi^{\dagger}P_{s}\Psi \right|_{LLL} &=& \left. -\bar{\Psi}\gamma^3 \gamma^5 P_{s} \Psi \right|_{LLL}
= \psi_{K^{\prime}  As}^\dagger\psi_{K^{\prime} As} + \psi_{K Bs}^\dagger \psi_{K Bs},\\
\left. \Psi^{\dagger}\gamma^3\gamma^5 P_{s}\Psi \right|_{LLL} &=& \left. -\bar{\Psi}P_{s} \Psi \right|_{LLL}
=-\psi_{K^{\prime}  As}^\dagger\psi_{K^{\prime} As} + \psi_{K Bs}^\dagger \psi_{K Bs}.
\end{eqnarray}
This fact in particular implies that in order to determine all the order parameters, it is 
necessary to analyze the gap equation beyond the LLL approximation.

\subsubsection{Quantum Hall states with LLL filling factors, $|\nu|\leq 2$}
\label{LLLnu} 

As was already discussed in Section~\ref{Sec:MagCatGraphene} , at magnetic fields 
$B \lesssim 10~\mbox{T}$, the plateaus with the filling factors $\nu = \pm 4(n + 1/2)$ 
are observed in the QH effect in graphene \cite{2005Natur.438..197N,Zhang:2005zz}. 
At stronger magnetic fields, new plateaus, with $\nu=0$ and $\nu=\pm 1$ occur: while
the former arises at $B \gtrsim 10~\mbox{T}$, the latter appear at $B \gtrsim 20~\mbox{T}$ 
\cite{2006PhRvL..96m6806Z,2007PhRvL..99j6802J}. In this section, we will describe 
the dynamics underlying these new plateaus, and the plateaus $\nu=\pm 2$ corresponding 
to the gap between the LLL and the $n = 1$ LL, by using the solutions of the gap equation 
presented in the next subsection. We will consider positive $\nu$ and $\mu_0$ (the dynamics 
with negative $\nu$ and $\mu_0$ is related by electron-hole symmetry and will not be discussed
separately). As will be shown below, there is a large number of the solutions corresponding 
to the same $\mu_0$. In order to find the most stable of them, we compare the free energy 
density $\Omega$ for the solutions. The derivation of the expression for $\Omega$ is presented 
in Appendix~\ref{App:FreeEnergyDensity2+1}. (For more details, see also Appendix~C in 
Ref.~\cite{Gorbar:2008hu}.)

The $\nu=0$, $\nu=\pm 1$ and $\nu=\pm 2$ plateaus are connected
with a process of doping of the LLL, which is described by varying
the electron chemical potential $\mu_0$. Therefore, we start our analysis
by reviewing the solutions to the gap equations in the case when $\mu_0$
is much less than the Landau energy scale, i.e.,  $\mu_0\ll \epsilon_{B}$.
At zero temperature the corresponding gap equations are analyzed analytically
in Appendix B in Ref.~\cite{Gorbar:2008hu}. It was concluded there that only the following three
stable solutions are realized:
\vspace{3mm}

(i) The solution with {\it singlet} Dirac masses for both spin up and spin
down quasiparticles,
\begin{equation}
\begin{split}
& \tilde{\Delta}_{+}=\tilde{\mu}_{+}=0,\qquad \mu_{+}=\bar{\mu}_{+}-A,\qquad
\Delta_{+}=s_{\perp}M,
\\
& \tilde{\Delta}_{-}=\tilde{\mu}_{-}=0,\qquad \mu_{-}=\bar{\mu}_{-}+ A,\qquad
\Delta_{-}=-s_{\perp}M.
\end{split}
\label{i}
\end{equation}
[By definition, $M\equiv A/(1-\lambda)$ and $\lambda\equiv 4A\Lambda/(\sqrt{\pi}\epsilon_{B}^2)$].
This solution is energetically most favorable for 
$0 \le \mu_0 < 2A+\epsilon_{Z}$.\footnote{In dynamics in a magnetic field at zero temperature, there is
no one-to-one correspondence between electron density and chemical potential.
As a result, different values of the latter may correspond to the same physics,
as it takes place for this solution.}
It is one of several solutions with nonvanishing singlet Dirac
masses and
we call it the $S1$ solution (here $S$ stands for {\it singlet}). Because
of the opposite signs of both the masses $\Delta_{+}$ and $\Delta_{-}$ and
the chemical
potentials $\mu_{+}$ and $\mu_{-}$, the explicit breakdown of the $\mathrm{U}(4)$
symmetry down to $\mathrm{U}(2)_{+} \times \mathrm{U}(2)_{-}$ by the Zeeman term is
strongly enhanced by the dynamics. Since all triplet order parameters
vanish, the flavor $\mathrm{U}(2)_{+} \times \mathrm{U}(2)_{-}$ symmetry is intact in the
state described by this solution. As will be shown below,
the $S1$ solution corresponds to the $\nu = 0$ plateau.

(ii) The {\it hybrid} solution with a {\it triplet} Dirac mass for spin up and
a {\it singlet} Dirac mass for spin down quasiparticles,
\begin{equation}
\begin{split}
& \tilde{\Delta}_{+} = M,\qquad \tilde{\mu}_{+}=As_{\perp},\qquad \mu_{+} =
\bar{\mu}_{+} - 4A,\qquad \Delta_{+}=0,
\\
& \tilde{\Delta}_{-}=0,\qquad \tilde{\mu}_{-}=0,\qquad \mu_{-}=\bar{\mu}_{-}-3A,\qquad
\Delta_{-}=-s_{\perp}M.
\end{split}
\label{ii}
\end{equation}
It is most favorable for $2A+\epsilon_{Z} \le \mu_0 < 6A+\epsilon_{Z}$. We call it the
$H1$ solution (here $H$ stands for {\it hybrid}, meaning that the
solution is a mixture of the singlet and triplet parameters). In
this case, while the $\mathrm{SU}(2)_{+}\subset \mathrm{U}(2)_{+}$ symmetry
connected with spin up is spontaneously broken down to $\mathrm{U}(1)_{+}$
(whose generator is $\gamma^3\gamma^5 \otimes P_{+}$), the
$\mathrm{SU}(2)_{-}\subset \mathrm{U}(2)_{-}$ symmetry connected with spin down
remains intact. As will be shown below, the $H1$
solution corresponds to the $\nu = 1$ plateau.

(iii) The solution with equal {\it singlet} Dirac masses for both spin up and
spin down quasiparticles
\begin{equation}
\begin{split}
& \tilde{\Delta}_{+}=\tilde{\mu}_{+}=0,\qquad \mu_{+}=\bar{\mu}_{+}-7A,\qquad
\Delta_{+}=-s_{\perp}M,
\\
& \tilde{\Delta}_{-}=\tilde{\mu}_{-}=0,\qquad  \mu_{-}=\bar{\mu}_{-}- 7A,\qquad
\Delta_{-}=-s_{\perp}M.
\end{split}
\label{iii}
\end{equation}
It is most favorable for $\mu_0 > 6A+\epsilon_{Z}$. We call it the $S2$
solution. (Note that the dynamics in the $n=1$ LL will set an
upper limit for the range where the $S2$ solution is the ground
state, see Section~\ref{n=1}.)  In the state given by the $S2$
solution, the $\mathrm{U}(4)$ symmetry is broken down to $\mathrm{U}(2)_{+} \times
\mathrm{U}(2)_{-}$ only by the Zeeman term. Indeed, the singlet masses and
the dynamical contributions to the chemical potentials are of the
same sign for both spin orientations and thus have no effect on
breaking any symmetry. As will be shown below, the
$S2$ solution corresponds to the $\nu = 2$ plateau connected with
the gap between the filled LLL and the empty $n=1$ LL.

\begin{figure}[t]
\begin{center}
\includegraphics[width=.45\textwidth]{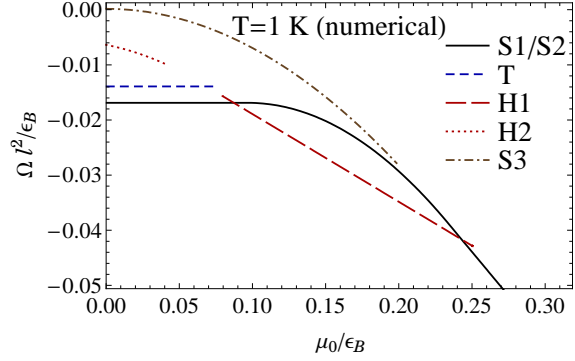}
\caption{(Color online) Free energy density versus the electron chemical potential $\mu_0$
for several different solutions in a range of $\mu_0$ relevant to the dynamics in the lowest
Landau level. The numerical results are shown for a nonzero but small
temperature, $T=1$~K. }
\label{fig.V.eff.T1}
\end{center}
\end{figure}

The free energy densities for the above three solutions are given
by the following expressions (see Appendix~\ref{App:FreeEnergyDensity2+1}
at the end of this review, as well as Appendices~B and C in Ref.~\cite{Gorbar:2008hu}):
\begin{eqnarray}
\label{O1}
\Omega &=& -\frac{|eB_{\perp}|}{2\pi \hbar c} \left( M + A + 2\epsilon_{Z} + h \right),
\quad\mbox{for}\quad 0<\mu_0 < 2A+\epsilon_{Z},\\
\label{O2}
\Omega &=& -\frac{|eB_{\perp}|}{2\pi \hbar c} \left( M - A + \epsilon_{Z} + h + \mu_0 \right),
\quad\mbox{for}\quad 2A+\epsilon_{Z} < \mu_0 < 6A+\epsilon_{Z},\\
\label{O3}
\Omega &=& -\frac{|eB_{\perp}|}{2\pi \hbar c} \left( M - 7A + h + 2\mu_0 \right),
\quad\mbox{for}\quad 6A+\epsilon_{Z} < \mu_0,
\end{eqnarray}
where $h$ is the higher LLs contribution \cite{Gorbar:2008hu}:
\begin{equation}
h \equiv \sum_{n=1}^{\infty}
\frac{2M^4}{\sqrt{n \epsilon_{B}^2+M^2}
\left(\sqrt{n \epsilon_{B}^2+M^2}+
\sqrt{n}\, \epsilon_{B}\right)^2}
\simeq
\frac{M^4}{2\epsilon_{B}^3}
\left[
\zeta\left(\frac32\right)-\zeta\left(\frac52\right) \frac{M^2}{\epsilon_{B}^2}
+O\left(\frac{M^4}{\epsilon_{B}^4}\right)
\right].
\label{h}
\end{equation}
Here $\zeta (x)$ is the Riemann zeta function. We note that
although the parameters of the solutions jump abruptly at the
transition points, $\mu_0=2A+\epsilon_{Z}$ and $\mu_0=6A+\epsilon_{Z}$, their free
energy densities match exactly. We conclude, therefore, that first
order phase transitions take place at these values of the electron
chemical potential $\mu_0$.

The free energy densities in Eqs.~(\ref{O1}) -- (\ref{O3}) are shown as 
functions of the chemical potential $\mu_0$ in Fig.~\ref{fig.V.eff.T1}.
In order to plot the
results, we took $M = 4.84 \times 10^{-2}\epsilon_{B}$ and $A =
3.90 \times 10^{-2}\epsilon_{B}$ which coincide with the values of
the corresponding dynamical parameters in the numerical analysis.
The values of the electron chemical potential are given in units of the 
Landau energy scale $\epsilon_{B}$, and the free energy densities are 
given in units of $\epsilon_{B}/l^2$, where $l=\sqrt{\hbar c/|eB_\perp|}$
is the magnetic length.

As is clear from the numerical results in Fig.~\ref{fig.V.eff.T1},  the singlet-type 
numerical solution, given by the solid line, spans both the $S1$ and $S2$ solutions,
as well as the intermediate (metastable) branch connecting them.
In addition to the $S1$, $H1$, and $S2$ solutions, numerical
results for several other (metastable) solutions are shown. The
metastable solutions will be discussed later.

Let us discuss the key details about the numerical analysis reviewed here.
The default choice of the magnetic field in the numerical
calculations is $B=35~\mbox{T}$. The corresponding Landau energy scale
is $\epsilon_{B}|_{B=35~{\rm T}}\approx 2510~\mbox{K}$. In order to do the
numerical calculations in the model at hand, we use a simple regularization
method that renders the formally defined divergent sum in Eq.~(\ref{f1})
finite. In particular, we redefine the corresponding function as follows:
\begin{equation}
f_1\left(\Delta^{(\pm)}_{s},\mu^{(\mp)}_{s}\right)=
\frac{\sinh\left(\frac{\Delta^{(\pm)}_{s}}{T}\right) -
s_{\perp}\sinh\left(\frac{\mu^{(\mp)}_{s}}{T}\right)}
{\cosh\left(\frac{\Delta^{(\pm)}_{s}}{T}\right)+
\cosh\left(\frac{\mu^{(\mp)}_{s}}{T} \right)}+\sum_{n=1}^{\infty}
\frac{2\Delta^{(\pm)}_{s}\sinh\left(\frac{E_{ns}^{\pm}}{T}\right)
\kappa(\sqrt{n}\,\epsilon_{B},\Lambda)}
{E_{ns}^{\pm}\left[\cosh\left(\frac{E_{ns}^{\pm} }{T}\right)+
\cosh\left(\frac{\mu^{(\mp)}_{s}}{T} \right)\right]},
\label{f1reg}
\end{equation}
where $\kappa(x,\Lambda)$ is a smooth cutoff function defined by
\begin{equation}
\kappa(x,\Lambda)=\frac{\sinh\left({\Lambda}/{\delta\Lambda}\right)}
{\cosh\left({x}/{\delta\Lambda}\right)+\cosh\left({\Lambda}/{\delta\Lambda} \right)}
\label{kappa}
\end{equation}
with $\Lambda=5000$~K and $\delta\Lambda=\Lambda/20=250$~K.
The value of  $\Lambda$ corresponds to an approximate point of the
high-energy cut-off, and the value of $\delta\Lambda$ gives the extent
of the smearing region in either direction from $\Lambda$. (Note
that the energy scale $\Lambda$ is about the same as the energy of the
$n=4$ Landau level at  $B=35~\mbox{T}$.)

{One should emphasize that the specific choice of the cutoff
energy scale $\Lambda$ has little effect on the qualitative as well as
quantitative results of our analysis, provided the dynamical energy scales
$A$ and $M=A/(1-\lambda)$ are kept fixed (see the discussion in the end
of this subsection). Here we assume that the value of the cutoff is
sufficiently large to avoid the reduction of the phase space relevant for
the quasiparticle dynamics at the $n=0$ and $n=1$ LLs.}

Because of the cutoff function $\kappa(x,\Lambda)$ the sum over $n$ on the
right hand side of Eq.~(\ref{f1reg}) is rapidly convergent. In the numerical
calculations, therefore, a sufficiently good accuracy may be achieved by keeping
a finite number of terms in the sum. The optimum choice for the maximum
value of index $n$ is $n_{\rm max}=\left[14\Lambda^2/\epsilon_{B}^2\right]$,
where the square brackets mean the integer number nearest to the result in
the brackets. This choice is large enough to insure a high precision and, at
the same time, it is small enough to make the calculation fast.

While the $f_2$-function in Eq.~(\ref{f2}) is finite, for consistency we redefine
it in the same way as function $f_1$ by smoothly cutting off the contributions
of large-$n$ LLs,
\begin{equation}
f_2\left(\Delta^{(\pm)}_{s},\mu^{(\mp)}_{s}\right)=
\frac{s_{\perp}\sinh\left(\frac{\Delta^{(\pm)}_{s}}{T}\right) -
\sinh\left(\frac{\mu^{(\mp)}_{s}}{T}\right)}
{\cosh\left(\frac{\Delta^{(\pm)}_{s}}{T}\right)+
\cosh\left(\frac{\mu^{(\mp)}_{s}}{T} \right)}
-\sum_{n=1}^{\infty}\frac{2\sinh\left(\frac{\mu^{(\mp)}_{s}}{T}\right)
\kappa(\sqrt{n}\,\epsilon_{B},\Lambda)}
{\cosh\left(\frac{E_{ns}^{\pm}}{T}\right)+ \cosh\left(\frac{\mu^{(\mp)}_{s}}{T}
\right)},
\label{f2reg}
\end{equation}
where $\kappa(x,\Lambda)$ is defined in Eq.~(\ref{kappa}). The numerical
result for the sum in $f_2$ is also approximated by dropping the terms with
$n>n_{\rm max}$ where $n_{\rm max}$ is given above.

By analyzing the solutions to Eqs.~(\ref{Deltas}) and (\ref{mus}) at very
low temperatures, we reproduce all the analytic solutions derived in Appendix~B in 
Ref.~\cite{Gorbar:2008hu}. For the choice of the magnetic field $B=35~\mbox{T}$,
the values of the two dynamical energy parameters $A$ and $M$ are given
by
\begin{equation}
A \approx  98~\mbox{K},\qquad
M \approx  122~\mbox{K}.
\end{equation}
As is easy to check, these correspond to the dimensionless coupling
$\lambda\approx 0.196$. Here one should keep in mind that the
smooth-cutoff regularization used in our numerical calculations is not
the same as in the analytical calculations. Despite this difference,
all analytical results agree very well even quantitatively with the
corresponding numerical ones when expressed in terms of $A$
and $M$ parameters.

The plateau $\nu = 0$ is connected with a range of electron
chemical potentials in the vicinity of the Dirac neutral point
with $\mu_0 = 0$. In this case the $S1$ solution with singlet
Dirac masses of opposite sign for spin up and spin down
quasiparticles, see Eq.~(\ref{i}), is most favorable energetically
and therefore is the ground state solution, provided $\mu_0
<2A+\epsilon_{Z}$.

{From} dispersion relation (\ref{LLLenergylevels}), we find that
while $\omega_{+} =-\mu_0+\epsilon_{Z}+M+A$ is positive for spin up
states, $\omega_{-}= -\mu_0-\epsilon_{Z}-M-A$ is negative for spin down
states, i.e., the LLL is half filled (the energy spectrum in this
solution is $\sigma$ independent). Therefore, there is a nonzero
spin gap $\Delta{E}_{0} = \omega_{+} - \omega_{-}$ associated with
the $\nu = 0$ plateau. The value of this gap is $\Delta{E}_{0} =
2(\epsilon_{Z}+A)+2M$.

\begin{figure}[t]
\begin{center}
\includegraphics[width=.45\textwidth]{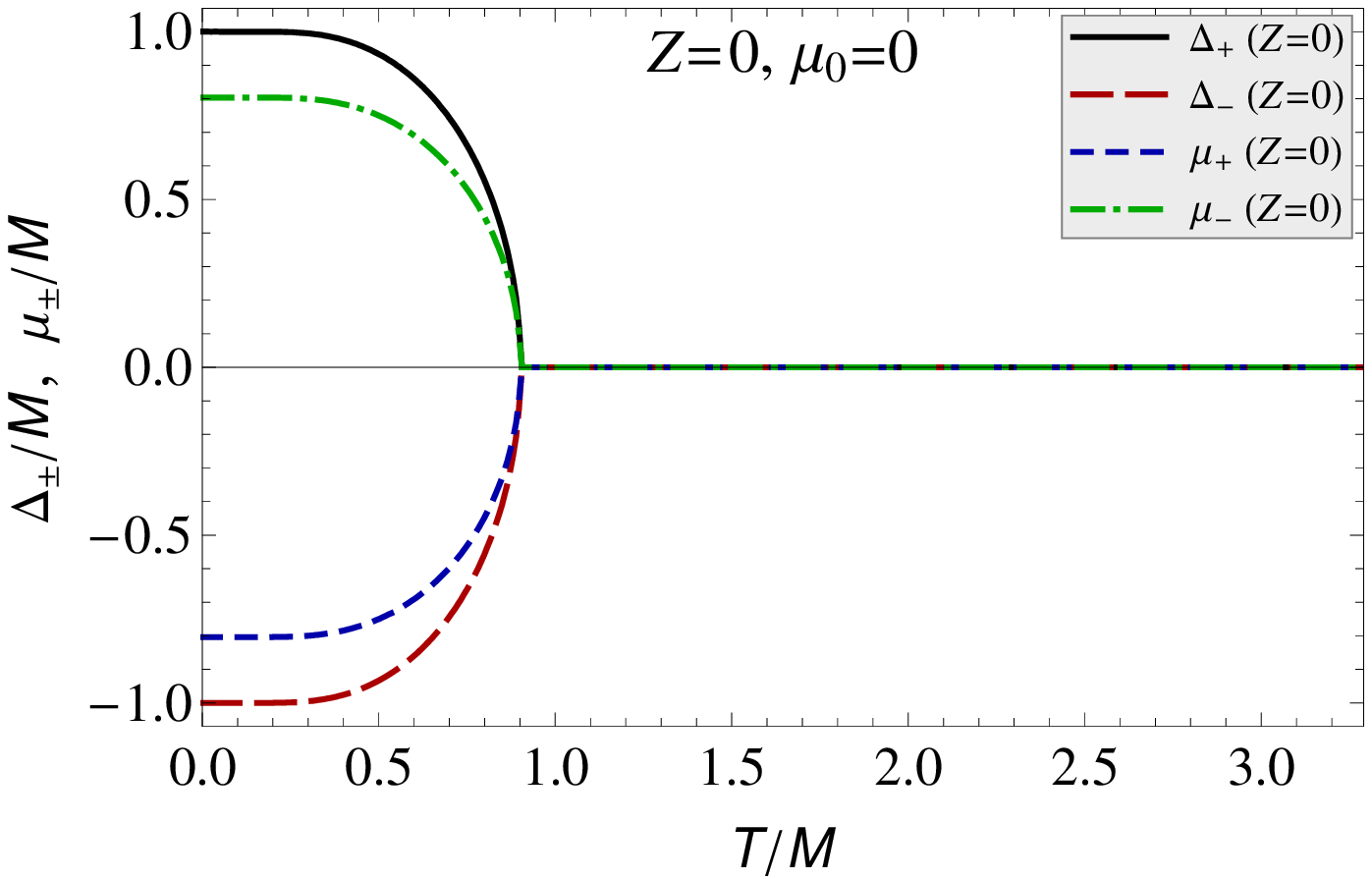}
\hspace{.07\textwidth}
\includegraphics[width=.45\textwidth]{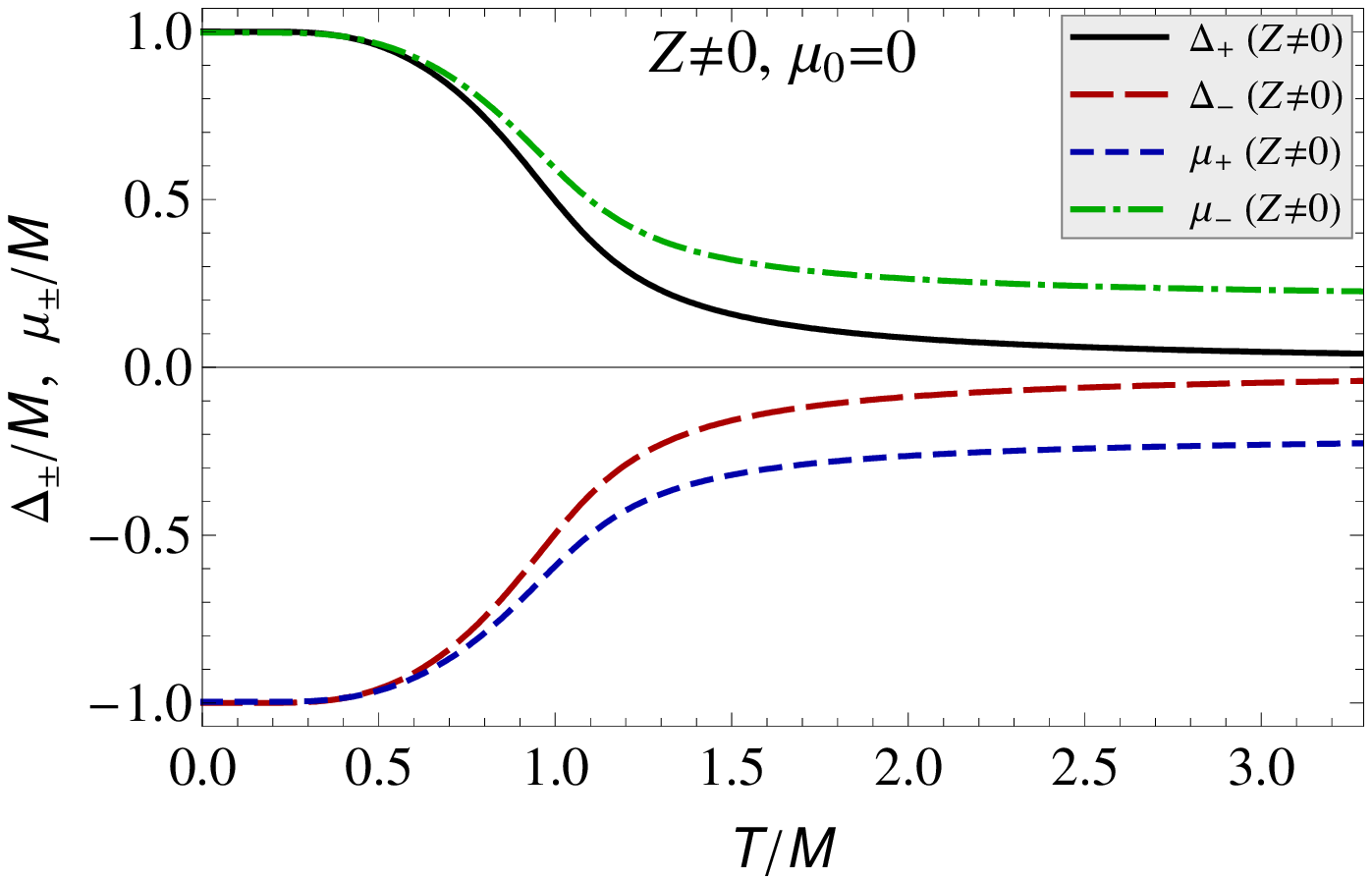}
\caption{(Color online) Temperature dependence of the nontrivial order parameters
in the $\nu=0$ QH state, described by the $S1$ solution. The
results in a model with a vanishing Zeeman energy ($\epsilon_{Z}=0$) are
shown in the left panel, and the results in a realistic model with
a nonzero Zeeman energy ($\epsilon_{Z}\neq 0$) are shown in the right panel.
Note that $\tilde{\mu}_{\pm}=\tilde{\Delta}_{\pm}=0$ in both
cases. The values of the temperature and the order parameters are
given in units of the dynamical scale $M$.}
\label{fig.order_pars_mu0_vs_T}
\end{center}
\end{figure}

While no exact symmetry is broken in the state described by the
$S1$ solution, the explicit spin symmetry breaking by the Zeeman
term $\epsilon_{Z}$ is strongly enhanced by the dynamical contribution $M+A$.
In this case, it is appropriate to talk about the dynamical
symmetry breaking of the approximate spin symmetry. This is also
evident from studying the temperature dependence of the MC and QHF
order parameters in Fig.~\ref{fig.order_pars_mu0_vs_T}. Because of
a nonzero Zeeman energy term ($\epsilon_{Z}\approx 23.51~\mbox{K}$ 
at $B=35~\mbox{T}$), which breaks the spin symmetry explicitly, strictly 
speaking, there is no symmetry restoration phase transition at high 
temperature. Despite that, there is a well pronounced crossover (around 
$T\approx 0.9 M$) between the regimes of low and high temperatures, 
which can be quantified by the relative strength of the bare Zeeman and dynamical
contributions. (It can be also checked that the spontaneous spin-symmetry
breaking occurs in the limit $\epsilon_{Z}=0$ \cite{Gorbar:2008hu}.)

At $T=0$ the solution $S1$ is the ground state for
$\mu_0\lesssim 0.09  \epsilon_{B}$. At sufficiently low temperature, the main
qualitative feature of this solution is that the singlet Dirac masses for spin-up
and spin-down quasiparticles have opposite signs, $\Delta_{+}= -\Delta_{-}$.
This defines the configuration of the MC order parameters that is formally
invariant under the time reversal symmetry. (Of course, the time reversal
symmetry is still explicitly broken by the external magnetic field.) As the
temperature increases, the approximate relation $\Delta_{+}\approx -\Delta_{-}$
may hold at $\mu_0\approx 0$, but deviations from such a relation grow
with increasing $\mu_0$.

It should be emphasized that the solution $S1$ is continuously connected with
the solution $S2$ responsible for the $\nu=2$ QH plateau, see Fig.~\ref{fig.V.eff.T1}.
At low temperatures, the intermediate branch between the $S1$ and $S2$
solutions is metastable. At high temperatures, however, it becomes stable and
the qualitative difference between the two solutions disappears \cite{Gorbar:2008hu}.

In conclusion, the following comment is in order. As one can see from Fig.~\ref{fig.V.eff.T1}, 
besides the $S1$ solution, there is another, triplet ($T$), solution around the Dirac
neutral point. In the $T$ solution, given in Eq.~(B33) in Appendix~B in Ref.~\cite{Gorbar:2008hu},
both spin up and spin down quasiparticle states have a triplet Dirac mass.
Calculating the difference of the free energy densities for these two
solutions, one finds that $\delta\Omega=\Omega_{S1}-\Omega_{T}=- \epsilon_{Z}|eB|/\pi\hbar c$.
Therefore, it is the Zeeman term which makes the $S1$ solution more favorable:
without it, the $S1$ and $T$ solutions would correspond to two degenerate ground states.
The role of the small on-site repulsion interaction terms 
\cite{PhysRevB.74.075422,2007arXiv0706.4280L,2007PhRvB..76h5432H,PhysRevB.75.165411,
Herbut:2006cs,PhysRevB.76.195415} in choosing the genuine ground state in the
present dynamics will be discussed in Section~\ref{Sec:GrapheneRemarks}.

For larger $\mu_0$, the hybrid $H1$ solution (\ref{ii}), with a triplet Dirac
mass for spin up quasiparticles and a singlet Dirac mass for spin
down quasiparticles, is most favorable. It is the ground state for
$2A+\epsilon_{Z} < \mu_0 < 6A+\epsilon_{Z}$. As one can easily check by using
Eq.~(\ref{LLLenergylevels}), while now $\omega_{+}^{(+)} > 0$, the
energies $\omega_{+}^{(-)}$ and $\omega_{-}^{(+)}=
\omega_{-}^{(-)}$ are negative. Consequently, the LLL is now
three-quarter filled and, therefore, the gap $\Delta{E}_{1} =
\omega_{+}^{(+)} - \omega_{+}^{(-)} = 2(M + A)$ corresponds to the
$\nu = 1$ plateau. Notably, the Zeeman term does not enter the
value of the gap. Unlike the $\nu = 0$ state, therefore, the gap
in the $\nu=1$ state is directly related to the spontaneous
breakdown of the flavor symmetry $\mathrm{SU}(2)_{+}$.

The last point regarding the nature of the ground state described
by the $H1$ solution has important consequences for the physical
properties of the $\nu=1$ QH state. {Since the coupling
constant $G_{\rm int}$ in the present model is proportional to
$1/\epsilon_{B}$, Eq.~(\ref{Apar}) implies
that the dynamical parameters $A$ and $M$, and therefore the gap
$\Delta{E}_{1}$, scale with the magnetic field as
$\sqrt{|eB_{\perp}|}$. This fact agrees with the dependence of the
activation energy in the $\nu=1$ state observed in
Ref.~\cite{2007PhRvL..99j6802J}}.

The critical temperature at which the $\mathrm{SU}(2)_{+}$ symmetry is restored,
i.e., when the triplet parameters $\tilde{\mu}_{+}$ and $\tilde{\Delta}_{+}$ vanish,
is $T_c \simeq 0.9 M \simeq 110 K$. The restoration is described by a
conventional second order phase transition.

The temperature dependence of the hybrid $H1$ solution is rather
interesting too: as was shown in Section~\ref{Sec:Gap_equation},
at a $\it fixed$ value of spin $s$ and {\it any} value of
temperature, there are no nontrivial solutions of the gap equation with 
both masses $\Delta_{s}$ and $\tilde{\Delta}_{s}$ being zero. In particular,
at $T > T_c$, when the triplet mass $\tilde{\Delta}_{+}$ vanishes, the mass $\Delta_{+}$ 
is nonzero (in fact, the revival of $\Delta_{+}$ starts already at subcritical $T$).
Note also that in the case of spin down quasiparticles,
the triplet parameters $\tilde{\mu}_{-}$ and $\tilde{\Delta}_{-}$ are identically
zero but the singlet mass $\Delta_{-}$ remains nonzero at all temperatures.

These results are obtained in the mean-field approximation and for the Hamiltonian 
$H_{\rm tot}$ (\ref{tot}), which is symmetric under the 
$\mathrm{U}(2)_{+} \times \mathrm{U}(2)_{-}$. However,
as was already pointed earlier, this symmetry is not
exact for the Hamiltonian on the graphene lattice. In that case,
it is replaced by $\mathrm{U}(1)_{+} \times Z_{2+} \times \mathrm{U}(1)_{-}
\times Z_{2-}$, where the elements of the discrete group $Z_{2\pm}$ are
$\gamma^5 \otimes P_{\pm} + I_{4}\otimes P_{\mp}$ and the unit matrix.
It is important that unlike a spontaneous breakdown of continuous symmetries, 
a spontaneous breakdown of the discrete symmetry $Z_{2\pm}$, with
the order parameters $\langle {\bar{\Psi} P_{\pm}\Psi} \rangle$ and
$\langle\Psi^{\dagger}\gamma^3\gamma^{5}P_{\pm}\Psi\rangle$, is not
forbidden by the Mermin-Wagner theorem at finite temperatures in a
planar system \cite{Mermin:1966fe}. This point strongly suggests that there
exists a genuine phase transition in temperature related to the
$\nu = 1$ state in graphene.

At zero temperature, the $S2$ solution (\ref{iii}) with equal
singlet Dirac masses for spin up and spin down states is most
favorable for $\mu_0 > 6A+\epsilon_{Z}$. It is easy to check from
Eq.~(\ref{LLLenergylevels}) that both $\omega_{+}$ and
$\omega_{-}$ are negative in this case, i.e., the LLL is
completely filled. This solution corresponds to the $\nu = 2$
plateau when the value of the electron chemical potential is in
the gap between the LLL and the $n=1$ LL.

At $T=0$ the solution $S2$ is the ground state for $\mu_0\gtrsim 0.24 \epsilon_{B}$.
As we see, even at high temperatures, the MC order parameters
satisfy the same approximate relation, $\Delta_{+}\approx \Delta_{-}$. Such a
configuration breaks neither spin nor sublattice-valley symmetry of graphene.

In addition to the three {stable} solutions $S1$, $H1$, and $S2$,
describing the $\nu=0$,  $\nu=\pm 1$,  and $\nu=\pm 2$ QH plateaus, the
numerical analysis of the gap equations reveals other, metastable, solutions.

One of such solutions is the $T$ solution with nonzero {\it triplet} Dirac masses
for both spin up and spin down quasiparticles. In the model of graphene used here,
the explicit analytical form of this solution is given in Eq.~(B33) in Appendix~B of 
Ref.~\cite{Gorbar:2008hu}. Note that because there is a contribution of the bare
Zeeman term $\epsilon_{Z}\propto eB$ in the gap $\Delta{E_1}$ for this solution, the 
corresponding activation energy in the $\nu = 1$ state scales with $eB$ differently from
the $\sqrt{|eB|}$ law in the hybrid $H1$ solution.

In addition to the triplet solution, there exist also metastable hybrid ($H2$) and
singlet ($S3$) solutions. Their free energy densities are shown in
Fig.~\ref{fig.V.eff.T1} together with the energy densities of the other solutions.
As seen, neither the $H2$ solution nor the $S3$ one can have sufficiently low free energy
density to become the genuine ground state.

The following remark is in order.
Unlike all the other solutions, the solutions $H2$ and $S3$
cannot be found analytically at $T=0$, see Appendix B in Ref.~\cite{Gorbar:2008hu}. By making use of the
numerical analysis, we find that these two extra solutions are such that $\mu_{s}^{(\pm)}
\approx \pm E_{0s}^{\mp}$. At exactly zero temperature, it is problematic to get such
solutions analytically because Eqs.~(\ref{E1-a}) -- (\ref{E4-a}) contain undetermined
values of the step functions, e.g., $\theta(|\mu_{s}^{(\pm)}|-E_{0s}^{\mp})$. In
contrast, at a nonzero temperature, the step functions are replaced by smooth
expressions, see Eqs.~(\ref{finiteT1}) and (\ref{finiteT2}), and numerical solutions
with $\mu_{s}^{(\pm)}\approx \pm E_{0s}^{\mp}$ are easily found.

\subsubsection{Quantum Hall states with filling factors in the first Landau level, $2\leq |\nu| \leq 6$}
\label{n=1}

In the previous section, we analyzed solutions of the gap equations under the
condition that only states on the LLL can be filled, $|\mu_{s} \pm \tilde{\mu}_{s}|
\ll \epsilon_{B}=\sqrt{2\hbar|eB_{\perp}|v^{2}_F/c}$. Since all the dynamically
generated parameters are much less than $\epsilon_{B}$, this condition implies
that the bare chemical potential $\mu_0$ also has to satisfy a similar inequality,
$\mu_0 \ll \epsilon_{B}$. In this section, we extend the analysis by considering
the dynamics with $\mu_0$ being of the order of the Landau scale $\epsilon_{B}$,
i.e., we study the regime when quasiparticle states on the first Landau level,
$n=1$ LL, can be filled.

We will start from the gap
equations at zero temperature, which are given in Eqs.~(\ref{E1-a}) -- (\ref{E4-a}) in
Section~\ref{Sec:Gap_equation}. In order to get their solutions for $\mu_0 \sim \epsilon_{B}$, we will follow the same steps of the analysis as for the LLL. The corresponding
analysis for the $n=1$ LL, including the calculation of the free energy density for the
solutions, was done in Appendix D in Ref.~\cite{Gorbar:2008hu}. It is shown there
that the following five stable solutions are realized:

\begin{itemize}
\item[(f-i)] The singlet type solution (f-I--f-I)
(here $f$ stands for {\it first}; the nomenclature
used for the $n=1$ LL solutions is defined in Appendix D in Ref.~\cite{Gorbar:2008hu}):

\begin{equation}
\begin{split}
& \tilde{\Delta}_{+}=\tilde{\mu}_{+}=0,
\qquad
\mu_{+} = \bar{\mu}_{+} - 7A,
\qquad
\Delta_{+}=-s_{\perp} M,
\\
& \tilde{\Delta}_{-}=\tilde{\mu}_{-}=0,
\qquad
\mu_{-} = \bar{\mu}_{-} - 7A,
\qquad
\Delta_{-}=-s_{\perp} M
\end{split}
\label{f-i}
\end{equation}
coincides with the solution S2 given in Eq.~(\ref{iii}) 
in the analysis of the LLL. It takes place for
$6A+\epsilon_{Z}<\mu_0 < 7A+\sqrt{\epsilon_{B}^2 +M^2}-\epsilon_{Z}$, and its free energy
density is
\begin{equation}
\Omega=-\frac{|eB_{\perp}|}{2\pi \hbar c}\left(M+2\mu_0-7A+h\right),
\end{equation}
where $h$ is given in Eq.~(\ref{h}). According to SubSection~\ref{LLLnu}, this solution
corresponds to the regime with the filled LLL and the empty $n=1$ LL and is
connected with the $\nu = 2$ plateau.

\item[(f-ii)] The hybrid type solution (f-I--f-II)
\begin{equation}
\begin{split}
& \tilde{\Delta}_{+}=\tilde{\mu}_{+}=0,\qquad  \mu_{+} = \bar{\mu}_{+} -
11A,\qquad  \Delta_{+}=-s_{\perp}\,M,
\\
& \tilde{\Delta}_{-}=\frac{M-M_1}{2},\qquad \tilde{\mu}_{-}=-
As_{\perp},\qquad  \mu_{-} = \bar{\mu}_{-} - 10A,\qquad
\Delta_{-}=-s_{\perp}\, \frac{M+M_1}{2},
\end{split}
\label{f-ii}
\end{equation}
with $M_1$ to be
\begin{equation}
M_1 \simeq \frac{A}{1-\lambda+2\left[1-\zeta(1/2)\right]A/\epsilon_{B}},
\label{m2}
\end{equation}
takes place for $9A+\sqrt{\epsilon_{B}^2+M_1^2}-\epsilon_{Z} < \mu_0 <
11A+\sqrt{\epsilon_{B}^2 +M^2}-\epsilon_{Z}$, and its free energy density is
\begin{equation}
\Omega=-\frac{|eB_{\perp}|}{2\pi \hbar c}\left(\frac{3M+M_1}{4}+
3\mu_0-15A-\epsilon_{B}+\epsilon_{Z}+\frac{3h+h_1}{4}\right),
\end{equation}
where $h_1$ is 
\begin{equation}
h_1 \equiv \sum_{n=2}^{\infty} \frac{2M_1^4}
{\sqrt{n\epsilon_{B}^2 +M_1^2}
\left(\sqrt{n \epsilon_{B}^2+M_1^2} +
\sqrt{n}\epsilon_{B}\right)^2}
\simeq
\frac{M_1^4}{2\epsilon_{B}^3}
\left[ \zeta\left(3/2\right)-1
-\left[\zeta\left(5/2\right)-1\right] \frac{M_1^2 }{\epsilon_{B}^2}
+O\left(\frac{M_1^4}{\epsilon_{B}^4}\right)
\right].
\label{h2}
\end{equation}
As is discussed in Section~\ref{numerical}, this solution corresponds to the $\nu=3$ plateau.
\item[(f-iii)] The singlet type solution (f-I--f-III)
\begin{equation}
\begin{split}
& \tilde{\Delta}_{+}=\tilde{\mu}_{+}=0,\qquad \mu_{+} = \bar{\mu}_{+} -
15A,\qquad \Delta_{+}=-s_{\perp}\,M,
\\
& \tilde{\Delta}_{-}=\tilde{\mu}_{-}=0,\qquad \mu_{-} = \bar{\mu}_{-} -
13A,\qquad  \Delta_{-}=-s_{\perp}\,M_1
\end{split}
\label{f-iii}
\end{equation}
is realized for $13A+\sqrt{\epsilon_{B}^2 +M_1^2}-\epsilon_{Z} < \mu_0 <
15A+\sqrt{\epsilon_{B}^2 +M^2}+\epsilon_{Z}$, and its free energy
density is
\begin{equation}
\Omega=-\frac{|eB_{\perp}|}{2\pi \hbar c}\left(\frac{M+M_1}{2}+4\mu_0-27A-
2\epsilon_{B}+2\epsilon_{Z}+\frac{h+h_1}{2}\right).
\end{equation}
{As is discussed in Section~\ref{numerical}, this solution corresponds to the $\nu=4$ plateau.}
\item[(f-iv)] The hybrid type solution (f-II--f-III)
\begin{equation}
\begin{split}
& \tilde{\Delta}_{+}=\frac{M-M_1}{2},\qquad \tilde{\mu}_{+}=-
As_{\perp},\qquad  \mu_{+} = \bar{\mu}_{+} - 18A,\qquad
\Delta_{+}=-s_{\perp}\,\frac{M+M_1}{2},
\\
& \tilde{\Delta}_{-}=\tilde{\mu}_{-}=0,\qquad  \mu_{-} = \bar{\mu}_{-} -
17A,\qquad \Delta_{-}=-s_{\perp}\,M_1
\end{split}
\label{f-iv}
\end{equation}
takes place for $17A+\sqrt{\epsilon_{B}^2+M_1^2}+\epsilon_{Z} < \mu_0 < 19A+
\sqrt{\epsilon_{B}^2 +M^2}+\epsilon_{Z}$, and its free energy
density is
\begin{equation}
\Omega=-\frac{|eB_{\perp}|}{2\pi \hbar c}\left(\frac{3M_1+M}{4}+
5\mu_0-43A-3\epsilon_{B}+\epsilon_{Z}+\frac{3h_1+h}{4}\right).
\end{equation}
{This solution corresponds to the $\nu=5$ plateau (see 
Section~\ref{numerical}).}
\item[(f-v)] The singlet type solution (f-III--f-III)
\begin{equation}
\begin{split}
& \tilde{\Delta}_{+}=\tilde{\mu}_{+}=0,\qquad  \mu_{+} = \bar{\mu}_{+} -
21A,\qquad \Delta_{+}=-s_{\perp}\,M_1,
\\
& \tilde{\Delta}_{-}=\tilde{\mu}_{-}=0,\qquad  \mu_{-} = \bar{\mu}_{-} -
21A,\qquad  \Delta_{-}=-s_{\perp}\,M_1
\end{split}
\label{f-v}
\end{equation}
is realized for $\mu_0 > 21A+\sqrt{\epsilon_{B}^2 +M_1^2}+\epsilon_{Z}$, and its free energy
density is
\begin{equation}
\Omega=-\frac{|eB_{\perp}|}{2\pi \hbar c}\left(M_1+6\mu_0-63A-
4\epsilon_{B}+h_1\right).
\end{equation}
\end{itemize}
{This solution corresponds to the $\nu=6$ plateau connected with the gap
between the filled $n=1$ LL and the empty $n=2$ LL.}

It should be emphasized that the above analytical solutions do not cover the
whole range of the values of the electron chemical potential around the $n=1$ LL. In
particular, there are no analytical solutions found in the following four intervals:
\begin{eqnarray}
\label{forbidden_region1}
7A+\sqrt{\epsilon_{B}^2+M^2}-\epsilon_{Z}<\mu_0 < 9A+\sqrt{\epsilon_{B}^2 +M_1^2}-\epsilon_{Z},\\
\label{forbidden_region2}
11A+\sqrt{\epsilon_{B}^2 +M^2}-\epsilon_{Z}<\mu_0 < 13A+\sqrt{\epsilon_{B}^2 +M_1^2}-\epsilon_{Z},\\
\label{forbidden_region3}
15A+\sqrt{\epsilon_{B}^2+M^2}+\epsilon_{Z}<\mu_0 < 17A+\sqrt{\epsilon_{B}^2 +M_1^2}+\epsilon_{Z},\\
\label{forbidden_region4}
19A+\sqrt{\epsilon_{B}^2+M^2}+\epsilon_{Z}<\mu_0 < 21A+\sqrt{\epsilon_{B}^2+M_1^2}+\epsilon_{Z}.
\end{eqnarray}
The difficulty in finding analytical solutions at $T=0$ on these
intervals is related to the ambiguities in the definition of some
step functions in gap equations (\ref{E1-a}) -- (\ref{E4-a}). The
same problem, albeit in a weaker form, was also encountered in the
analysis of dynamics at the LLL (see the discussion in the end of
Section~\ref{LLLnu}). As in that case, we remove the ambiguities by
considering a nonzero temperature case. The results at $T=0$ can
then be obtained by taking the limit $T\to 0$. The details of our
numerical analysis are given in the next subsection.

\subsubsection{Numerical analysis, n = 1 LL}
\label{numerical}

\begin{figure}[t]
\begin{center}
\includegraphics[width=.45\textwidth]{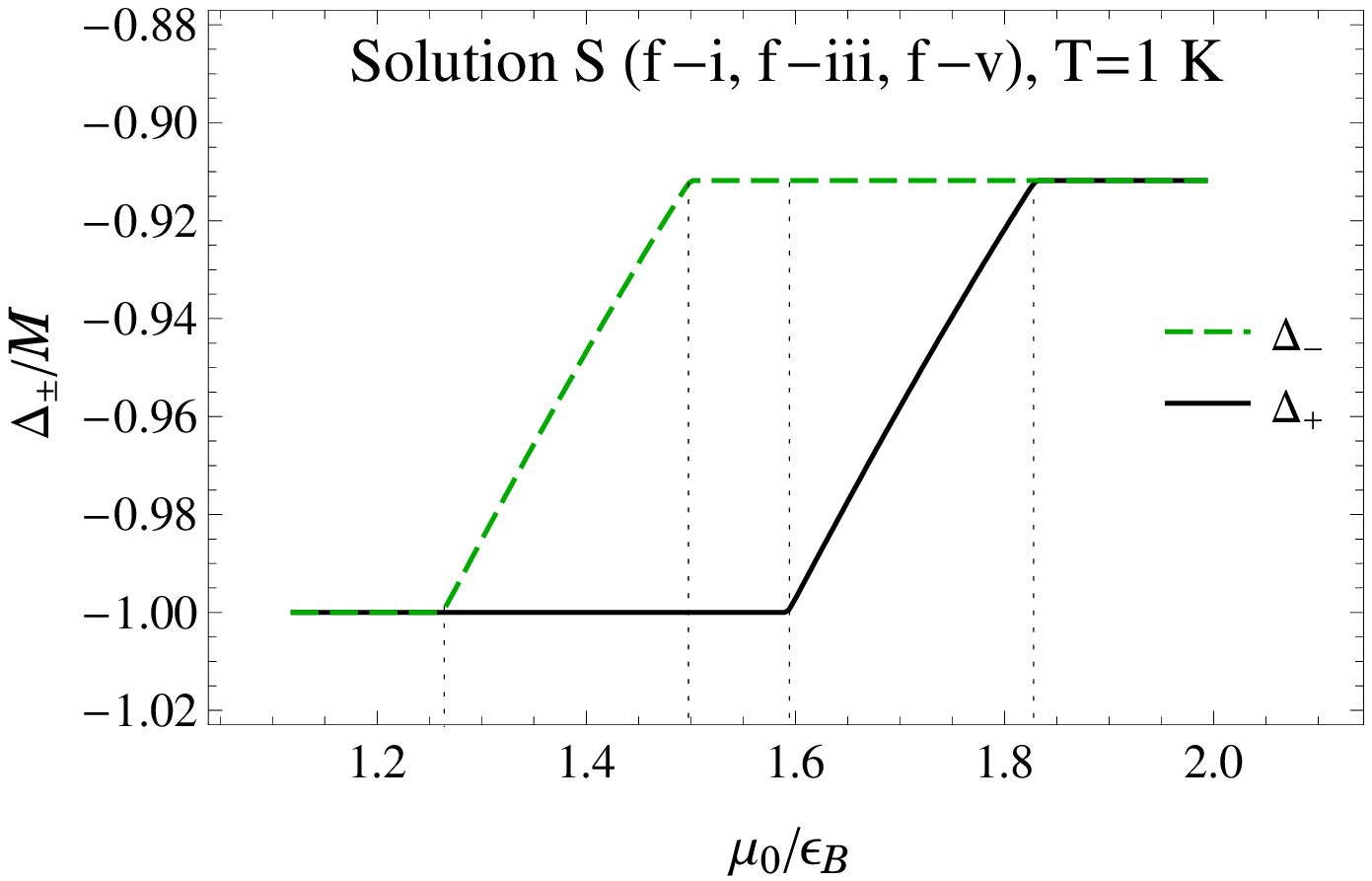}
\hspace{.04\textwidth}
\includegraphics[width=.43\textwidth]{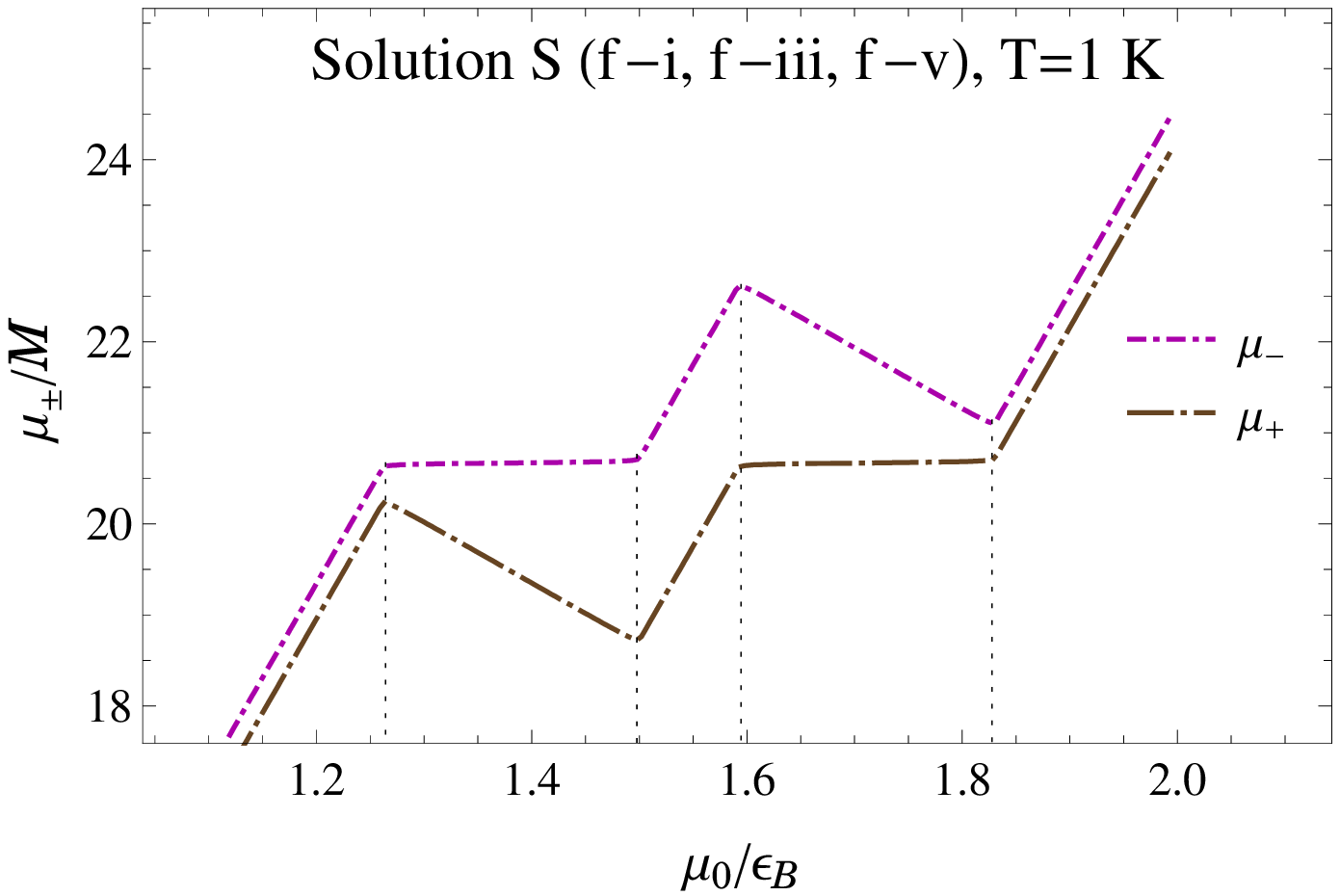}
\caption{Nontrivial order parameters of the $S$-type numerical solution
that contains the analytical solutions (f-i), (f-iii) and (f-v) as parts,
connected by two intermediate solutions.}
\label{fig.sol.S.f-i-iii-v}
\end{center}
\end{figure}

By {performing} a nonzero temperature analysis numerically, we find that the solutions
(f-i), (f-iii), and (f-v), found analytically, are in fact continuously connected. They
are parts of a more general solution $S$ (here $S$ stands for {\it singlet})
that exists at all values of $\mu_0$. At small and intermediate values of $\mu_0$,
this solution includes solutions $S1$ and $S2$.
At larger values of $\mu_0$, relevant for the dynamics of $n=1$ LL, the solution
$S$ is shown in Fig.~\ref{fig.sol.S.f-i-iii-v}.

As seen in Fig.~\ref{fig.sol.S.f-i-iii-v}, the solution $S$
consists of five pieces defined on five adjacent intervals of
$\mu_0$. Three of them are the analytical solutions (f-i),
(f-iii), and (f-v), as defined in the previous subsection. Their
intervals of existence are $\mu_0/\epsilon_{B}\lesssim 1.27$,
$1.5\lesssim\mu_0/\epsilon_{B}\lesssim1.6$ and
$\mu_0/\epsilon_{B}\gtrsim1.83$, respectively. These intervals are
in agreement with the analytical results if one takes $M_1\approx
111~\mbox{K}$, or in terms of the Landau energy scale,
$M_1=4.42\times10^{-2}\epsilon_{B}$. The other two pieces of the
solution $S$ extend the singlet-type analytical solution to the
intermediate intervals.

At $T=0$ the solution $S$ describes the ground state in exactly
the same regions of validity that are found analytically for
solutions (f-i), (f-iii), and (f-v) in the previous subsection.
This can be concluded from the energy consideration: among all
numerical solutions the parts of the solution $S$ have the lowest
free energy density there. {Analyzing the quasiparticle
spectra by using the dispersion relation in Eq.~(\ref{higherLLs}),
we find that the solutions (f-i), (f-iii), and (f-v) describe the
$\nu=2$, $\nu=4$, and $\nu=6$ QH states, respectively.}

{From} the symmetry viewpoint, none of the three parts of the
singlet solution break any exact symmetries in the model. However,
the part (f-iii) of the solution, describing the $\nu=4$ QH state,
corresponds to a quasi-spontaneous breakdown of the $\mathrm{U}(4)$ symmetry
down to the $\mathrm{U}(2)_{+} \times \mathrm{U}(2)_{-}$. {Indeed, by using Eq.
(\ref{higherLLs}), one can check that { the LLL is half filled and
the energy gap} between the pairs of the pseudospin degenerate
spin-up and spin-down states of the $n=1$ LL is given by
$\Delta{E}_{4}\simeq 2(\epsilon_{Z}+A)+(M^{2}-M_{1}^{2})/2\epsilon_{B}$. As we
see, the spin splitting by the Zeeman term $2\epsilon_{Z}$ is strongly enhanced
by the dynamical contribution $2A$.}

This is somewhat similar to the enhancement of the spin splitting
in the $\nu=0$ QH state, discussed in Section~\ref{LLLnu}.
{However, there is an important qualitative difference
between the cases of the LLL and the $n=1$ LL: It is only the
dynamical contribution to the chemical potentials (but not the
Dirac masses) that substantially affects the splitting in the
$\nu=4$ QH state. Indeed, the dynamical contribution due to the
Dirac masses in the gap $\Delta{E}_{4}$, i.e.,
$(M^{2}-M_{1}^{2})/2\epsilon_{B}$, is very small because $M\simeq
M_{1}\ll \epsilon_{B}$.) As a result, the gap $\Delta E_4$ is
substantially smaller than the LLL spin gap $\Delta E_0$ ($\Delta
E_4 \lesssim \Delta E_0/2$).}

Because of having nonvanishing triplet order parameters in the
extended hybrid solutions (f-ii) and (f-iv), the flavor $\mathrm{U}(2)_{+}
\times \mathrm{U}(2)_{-}$ symmetry of graphene is partially broken in the
corresponding ground states. {By using dispersion
relation (\ref{higherLLs}) in the analysis of the quasiparticle
spectra, we find that these solutions describe the $\nu=3$ and
$\nu=5$ plateaus corresponding to the quarter and three-quarter
filled $n=1$ LL, respectively.} In the case of the extended
solution (f-ii), the spin-down flavor subgroup $\mathrm{SU}(2)_{-}\subset
\mathrm{U}(2)_{-}$ is broken down to $\mathrm{U}(1)_{-}$, while the spin-up flavor
subgroup $\mathrm{U}(2)_{+}$ is intact. Similarly, in the case of the
extended solution (f-iv), the spin-up flavor subgroup $\mathrm{SU}(2)_{+}
\subset \mathrm{U}(2)_{+}$ is broken down to $\mathrm{U}(1)_{+}$, while the
spin-down flavor subgroup $\mathrm{U}(2)_{-}$ is intact. {Up to
small corrections due nonzero Dirac masses, the energy gaps
$\Delta E_3$ and $\Delta E_5$ associated with the (f-ii) and
(f-iv) solutions are equal to $2A$. Note that these gaps are
{substantially} smaller than the LLL gap $\Delta E_1$ ($\Delta E_3,
\Delta E_5 \lesssim \Delta E_1/2$).}

\begin{figure}[t]
\begin{center}
\includegraphics[width=.45\textwidth]{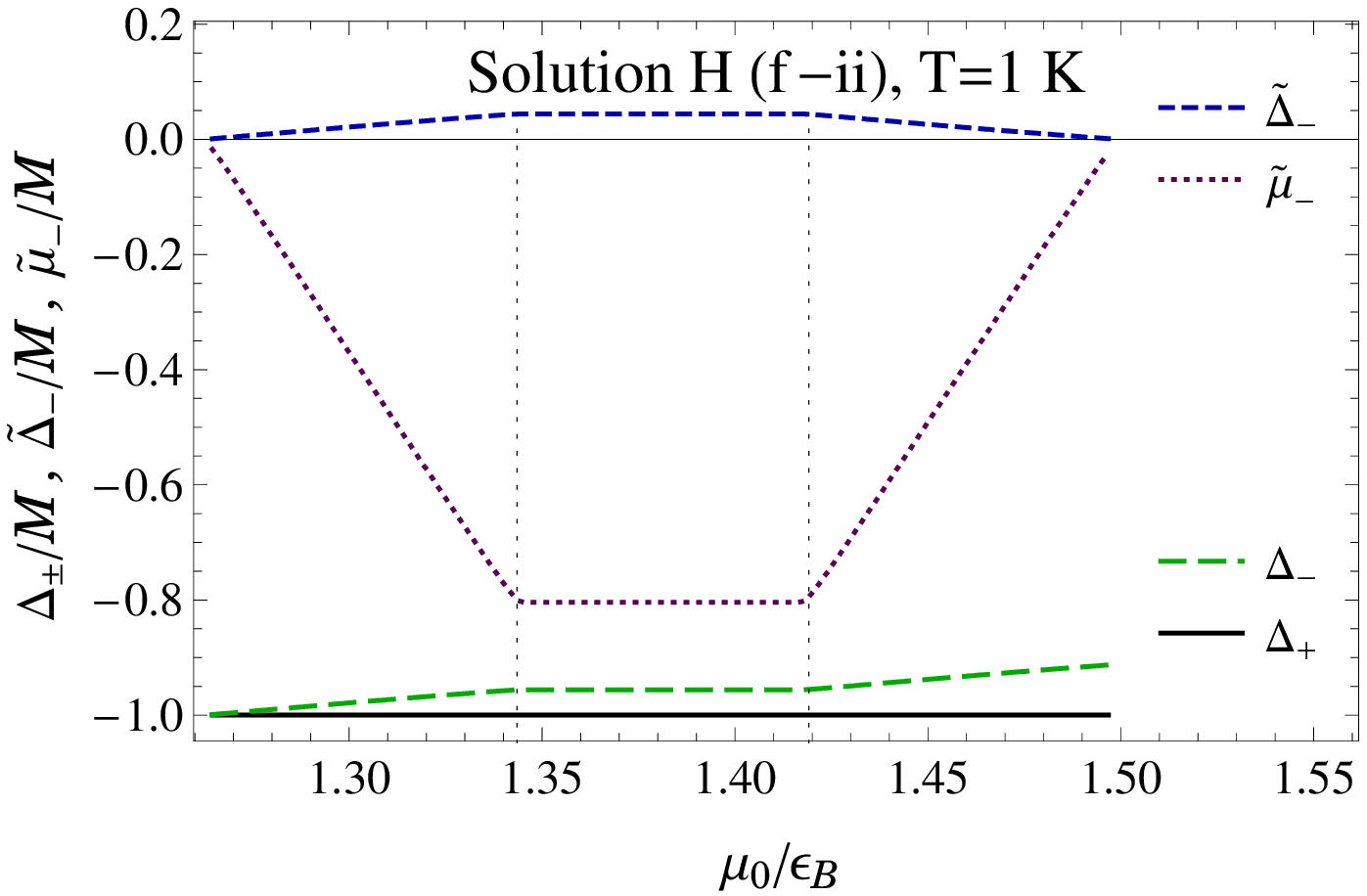}
\hspace{.04\textwidth}
\includegraphics[width=.45\textwidth]{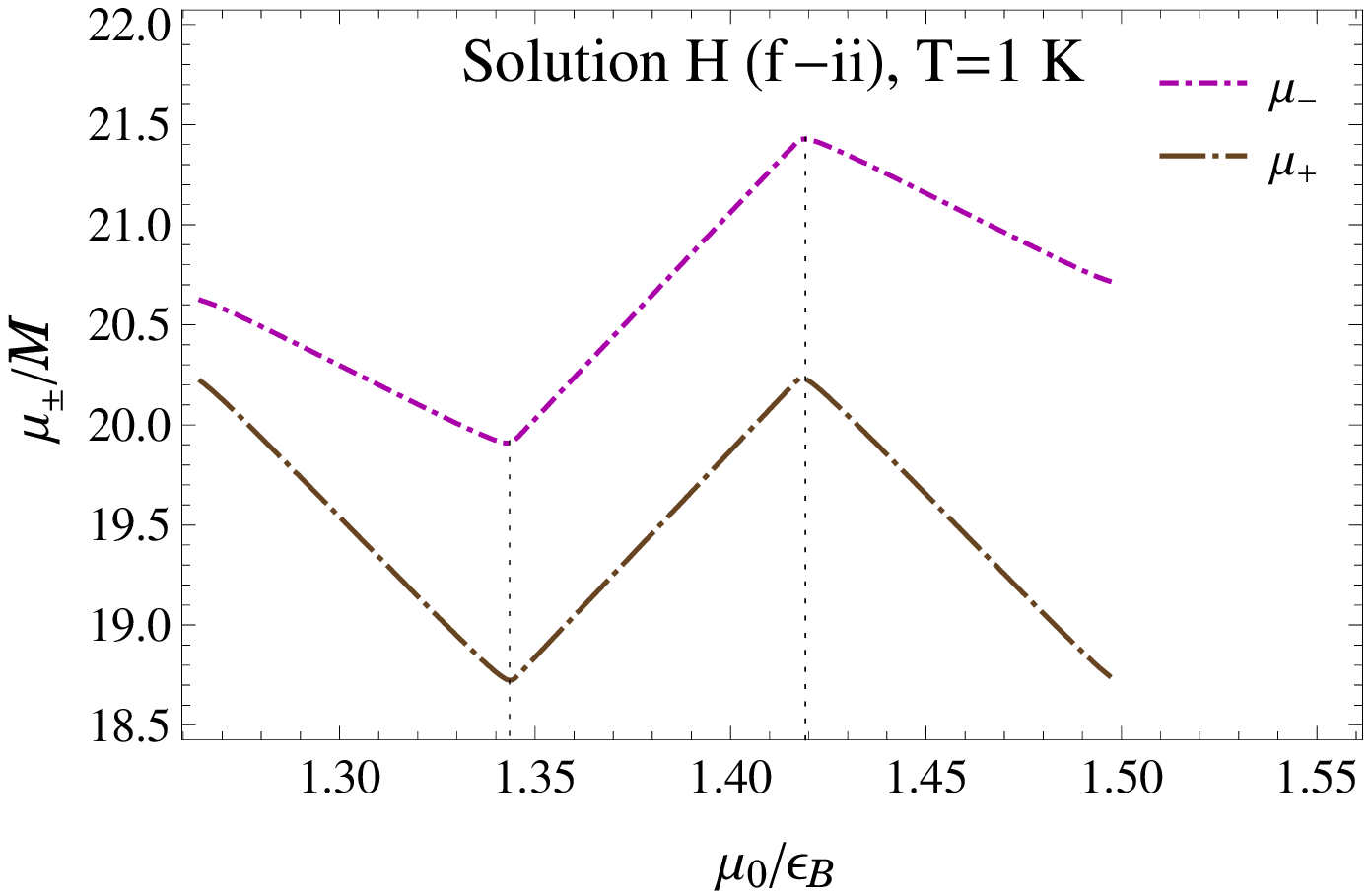}
\caption{Nontrivial order parameters of the extended hybrid solution (f-ii)
which determines the ground state for the $\nu=3$ QH plateau in graphene.}
\label{fig.sol.H.f-ii}
\end{center}
\end{figure}

\begin{figure}[t]
\begin{center}
\includegraphics[width=.45\textwidth]{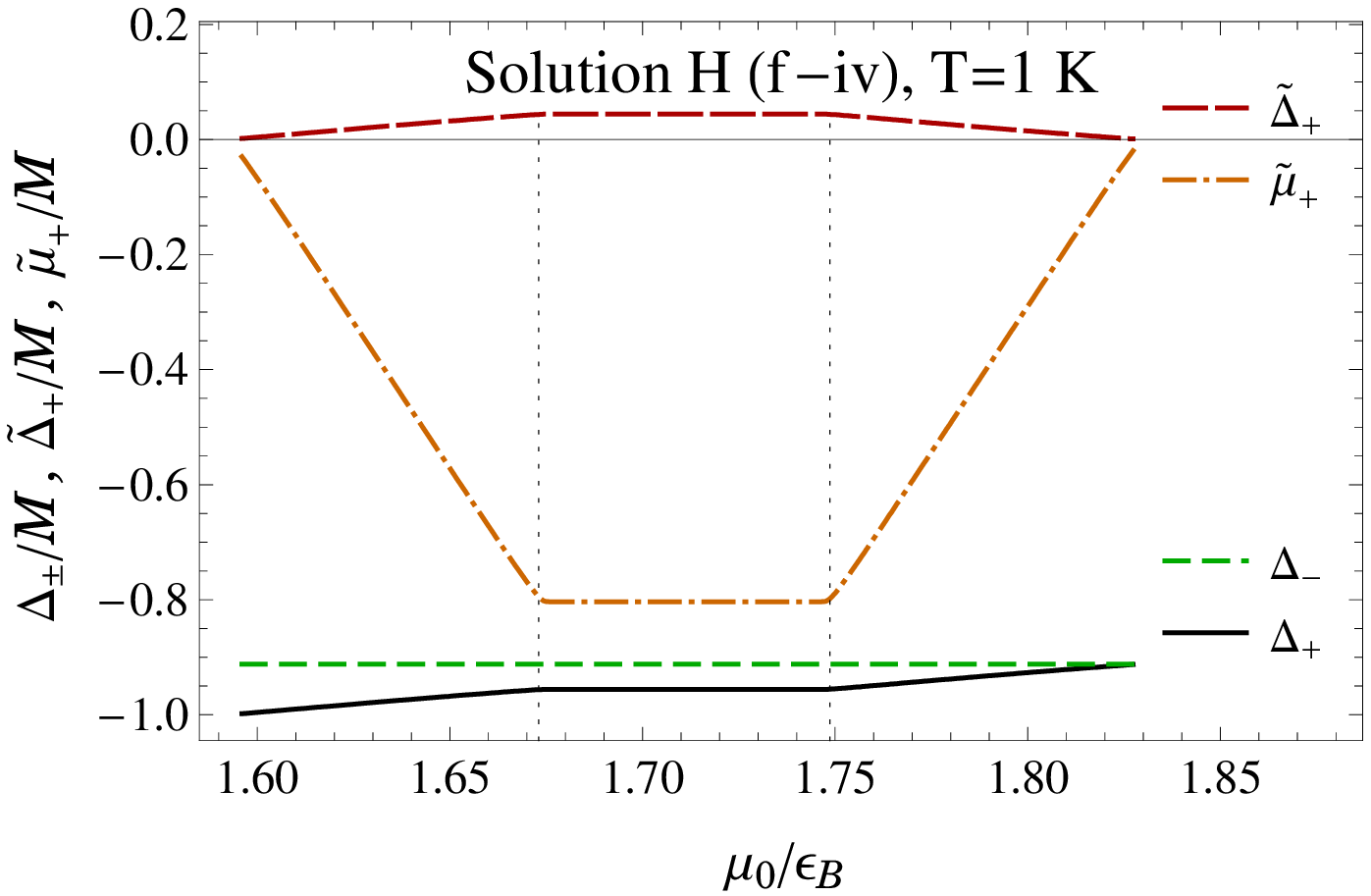}
\hspace{.04\textwidth}
\includegraphics[width=.45\textwidth]{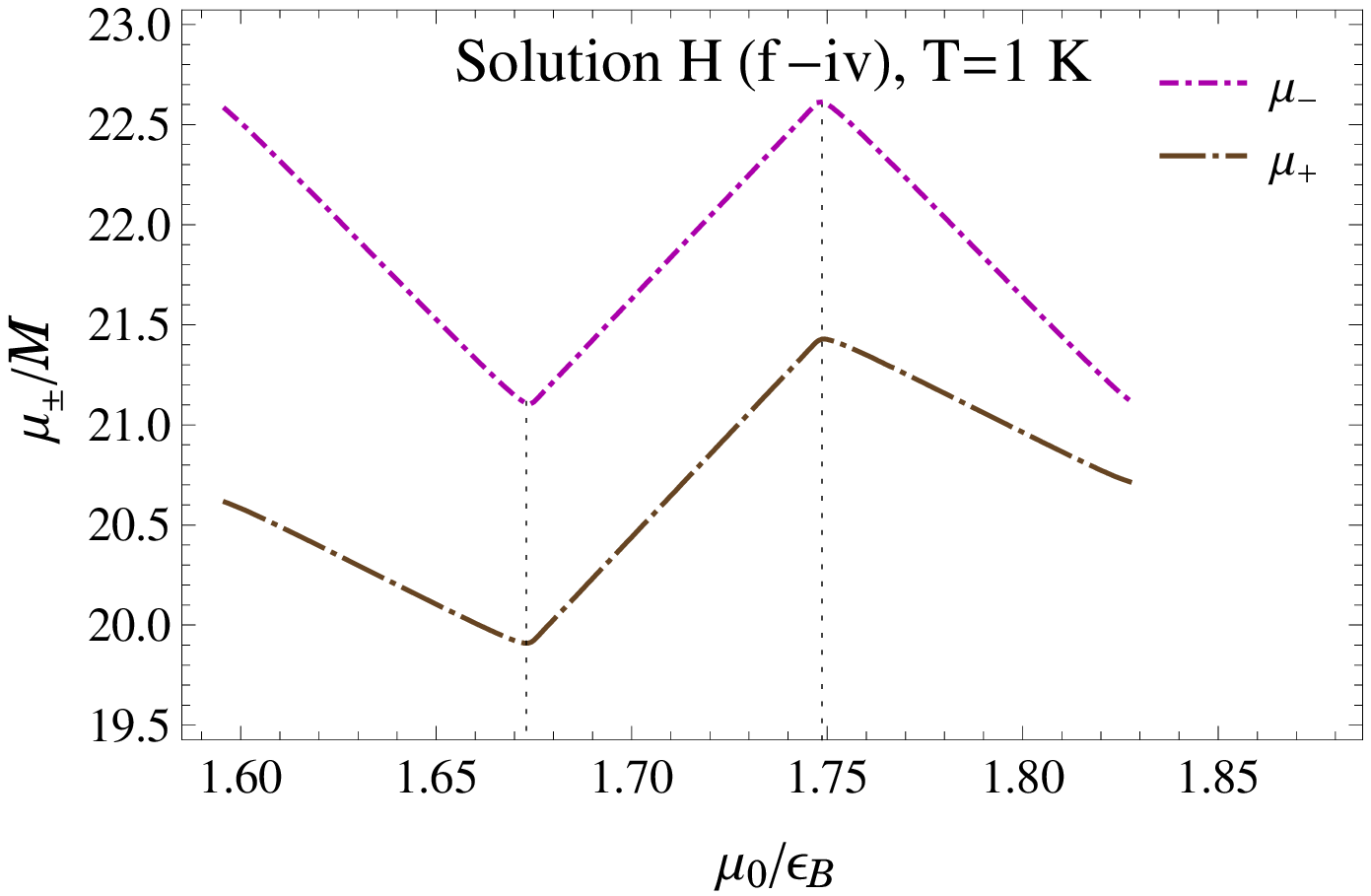}
\caption{Nontrivial order parameters of the extended hybrid solution (f-iv)
which determines the ground state for the $\nu=5$ QH plateau in graphene. }
\label{fig.sol.H.f-iv}
\end{center}
\end{figure}

The analytical hybrid solutions (f-ii) and (f-iv) get continuous extensions to the
left and to the right from their regions of validity found analytically in the previous
subsection. In fact, they extend all the way to cover the neighboring ``forbidden"
regions defined in Eqs.~(\ref{forbidden_region1}) -- (\ref{forbidden_region4}).
The first two ``forbidden" interval are covered by the extension of the solution
(f-ii) to the interval
$7A+\sqrt{\epsilon_{B}^2+M^2}-\epsilon_{Z}<\mu_0 < 13A+\sqrt{\epsilon_{B}^2 +M_1^2}-\epsilon_{Z}$.
The nontrivial Dirac masses and chemical potentials for this numerical solution
are shown in Fig.~\ref{fig.sol.H.f-ii}. The last two ``forbidden" intervals, see
Eqs.~(\ref{forbidden_region3}) and (\ref{forbidden_region4}), are covered by
the extension of the solution (f-iv) to the interval
$15A+\sqrt{\epsilon_{B}^2 +M^2}+\epsilon_{Z}<\mu_0 < 21A+\sqrt{\epsilon_{B}^2 +M_1^2}+\epsilon_{Z}$.
The nontrivial parameters for this solution are shown in Fig.~\ref{fig.sol.H.f-iv}.

\begin{figure}[t]
\begin{center}
\includegraphics[width=.45\textwidth]{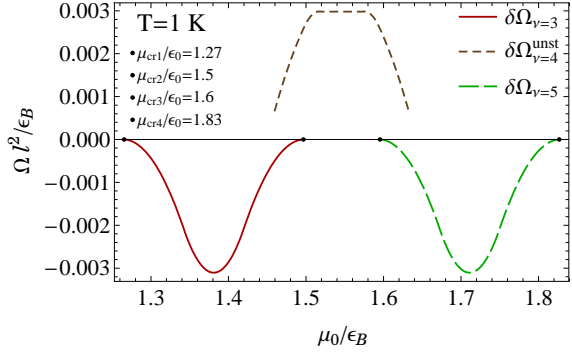}
\caption{(Color online) The difference between the free energy density of three hybrid
type solutions and the free energy density of the $S$-type solution in
the range of $\mu_0$, associated with the dynamics of the $n=1$ LL.}
\label{fig.V.eff.LL1.T1}
\end{center}
\end{figure}

In fact, the extended solutions (f-ii) and (f-iv) are the ground
states in their whole regions of existence. This is seen
in Fig.~\ref{fig.V.eff.LL1.T1}, where we plot the difference
between the free energy density of the hybrid type solutions and
the singlet one. The results for the extended hybrid solutions
(f-ii) and (f-iv) are shown by the solid line and the long-dashed
line, respectively.

In Fig.~\ref{fig.V.eff.LL1.T1} we also show the results for
another hybrid solution that was found numerically. It exists in
the interval of $\mu_0$ that could potentially be relevant for the
$\nu=4$ QH state. However, its free energy density is higher than
that for the solution $S$, and therefore it is unstable.

With increasing the temperature, we find that the extended hybrid
solutions (f-ii) and (f-iv) responsible for the $\nu=3$ and
$\nu=5$ QH states gradually vanish. Their regions of existence
shrink and their free energy densities approach the free energy
density of the singlet solution $S$. At temperatures above
$T^{(\nu=3)}_{\rm cr} \simeq T^{(\nu=5)}_{\rm cr} \simeq 0.4 M
\simeq T^{(\nu=1)}_{\rm cr}/2$, they cease to exist altogether,
and the ground state is described by the singlet solution which
does not break any exact symmetries of the model.

\subsubsection{Discussion of phase diagram in model of graphene with short-range interaction}
\label{Sec:GrapheneToyPhaseDiag}

\begin{figure}[t]
\begin{center}
\includegraphics[width=.45\textwidth]{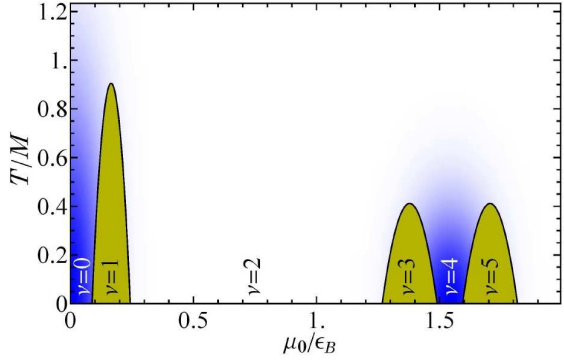}
\caption{(Color online) Schematic phase diagram in the toy model of graphene 
with the point-like density-density interaction. The values of chemical potential are
given in units of the Landau energy scale $\epsilon_{B}$, and the values
of temperature are given in units of the dynamical scale $M$.}
\label{fig-phase-diag}
\end{center}
\end{figure}

By summarizing the numerical results for the ground
states at different temperatures, we obtain the phase diagram of
graphene in the plane of temperature $T$ and electron chemical
potential $\mu_0$ shown in Fig.~\ref{fig-phase-diag}. The areas
highlighted in green correspond to hybrid solutions with a lowered
symmetry in the ground state. These regions are separated from the
rest of the diagram by phase transitions. At the boundary of the
$\nu=1$ region, the transition is of first order at low
temperatures and of second order at higher temperatures. The
transitions to/from the QH $\nu=3$ and $\nu=5$ states are of
second order. It should be kept in mind, however, that here the
analysis is done in the mean-field approximation and in a model
with a simplified contact interaction. Therefore, the predicted
types of the phase transitions may not be reliable. In particular,
the contributions of collective excitations, which are beyond the
mean-field approximation, may change the transitions to first
order type. Also, the types of the transitions may be affected by
the inclusion of disorder and a more realistic long-range Coulomb
interaction. Despite the model limitations, we still expect that
Fig.~\ref{fig-phase-diag} correctly represents the key qualitative
features of the phase diagram of graphene at least in the case of
the highest quality samples. In fact, as will be discussed in
Section~\ref{Sec:GrapheneVacuumAlignment}, the main change in the case of a real 
graphene sample is connected with the $\nu = 0$ state: at small values of a 
longitudinal component of a magnetic field $B_{\parallel}$, it is not in
the S1 (ferromagnetic) phase but in the canted antiferromagnetic (CAF)
one, which is closely connected with the T solution describing a charge density 
wave (CDW) phase. On the other hand, at a sufficiently large $B_{\parallel}$,
the $\nu = 0$ state is ferromagnetic indeed.

In Fig.~\ref{fig-phase-diag} the regions highlighted in blue
correspond to the ground states with a quasi-spontaneous breakdown
of the spin symmetry. In the case of the LLL and the $n=1$ LL,
such are the $\nu=0$ and $\nu=4$ QH states, in which the
quasi-spontaneous breakdown of the approximate $\mathrm{U}(4)$ symmetry
down to $\mathrm{U}(2)_{+}\times \mathrm{U}(2)_{-}$ is enhanced by dynamical
contributions. Because of the explicit breakdown by the Zeeman
term, there is no well-defined order parameter associated with
this symmetry breakdown. Also, there is no well-defined boundary
of the corresponding regions in the diagram. In the plot, this
feature is represented by the fading shades of blue at the
approximate boundaries of the $\nu=0$ and $\nu=4$ regions.

As considered in detail in Section~\ref{LLLnu}, the physical properties of the $\nu=0$ and
$\nu=1$ QH states are determined by the dynamics of the LLL. The corresponding values
of the gaps, $\Delta{E}_{0}=2(\epsilon_{Z}+A+M)$ and $\Delta{E}_{1}= 2(A+M)$, are largely
determined by the dynamical contributions $A$ and $M$ of about equal magnitude.
These two contributions are associated with the QHF and MC order parameters,
respectively.

The results of this study for the LLL qualitatively agree with the experimental data 
\cite{2006PhRvL..96m6806Z,2007PhRvL..99j6802J}. By taking the dimensionless 
coupling $\lambda=4A\Lambda/(\sqrt{\pi} \epsilon_{B}^2)$ to be a free parameter
and utilizing the cutoff $\Lambda$ to be of the order of the Landau scale $\epsilon_{B}$,
we arrive at the following scaling relations: $A  \sim \lambda\sqrt{|eB_{\perp}|}$
and $M \sim \lambda\sqrt{|eB_{\perp}|}$. This implies the same type of scaling
for the gap, $\Delta{E}_1 = 2(A + M) \sim \lambda\sqrt{|eB_{\perp}|}$, associated
with the $\nu =\pm 1$ plateaus. [The square root scaling of the activation
energy in the $\nu=1$ QH state was also obtained in the large-$N$ approximation
in Ref.~\cite{Herbut:2008ui}.] By making use of our results, we find that the
experimental value $\Delta{E}_1 \sim 100~\mbox{K}$ for $B_{\perp}=30~\mbox{T}$
from Ref.~\cite{2007PhRvL..99j6802J} corresponds to $\lambda \sim 0.02$. This estimate,
however, should be taken with great caution: Because interactions with impurities are
ignored and no disorder of any type is accounted for in the present model, it
may not be unreasonable to assume that actual values of $\lambda$ are up to
an order of magnitude larger.

As to the $n=1$ LL, we found that there are the gaps
$\Delta{E}_{3}=\Delta{E}_{5} \simeq 2A$ and $\Delta{E}_{4}\simeq
2(\epsilon_{Z}+A)$ corresponding to the plateaus $\nu=3, 5$ and $\nu=4$,
respectively [the contributions of Dirac masses are suppressed by
a factor of order $(M/\epsilon_{B})^2$ there]. Therefore, the gaps
$\Delta{E}_{3}=\Delta{E}_{5}$ and $\Delta{E}_{4}$ are mostly due
to the QHF type order parameters and are about a factor of two
smaller than the LLL gaps $\Delta{E}_{1}$ and $\Delta{E}_{0}$,
respectively. The first experimental data yield
$\Delta{E}_{4}\simeq 2\epsilon_{Z}$, and no gaps $\Delta{E}_{3}$ and
$\Delta{E}_{5}$ have been observed \cite{2006PhRvL..96m6806Z,2007PhRvL..99j6802J}. The later
experiments \cite{2009Natur.462..192D,2009Natur.462..196B} 
discovered additional plateaus, with $\nu =\pm 3$ and $\nu = \pm 1/3$. 
While the latter corresponds to the fractional QH effect, the plateaus with 
$\nu = 0, \pm 1, \pm 4$, and $\nu = \pm 3$ are intimately connected with a quasi-spontaneous
breakdown of the U(4) symmetry. We think that the discrepancy of the values 
of the theoretical gaps with the experiment could be connected with the fact that unlike $\epsilon_{Z}$, the value of 
the {\it dynamically} generated parameter $A$ corresponding to the $|n| \geq 1$ LLs will be strongly
reduced if a considerable broadening of higher LLs in a magnetic field is taken 
into account \cite{Gusynin:2006gn}. If so, the gap $\Delta{E}_{4}$ will be reduced to $2\epsilon_{Z}$, while the gaps
$\Delta{E}_{3}$ and $\Delta{E}_{5}$ will become much smaller.

In conclusion, we have shown that the QHF and MC order parameters
in graphene are two sides of the same coin and they necessarily
coexist. The present model leads a qualitatively 
reasonable description of the QH plateaus
in graphene in strong magnetic fields. In the rest of this Section,
a more realistic model setup, including the Coulomb interaction between quasiparticles,
will be considered.

\subsection{Quantum Hall effect in model of graphene with Coulomb interaction}
\label{Sec:MagCatGrapheneCoulomb}

In this section, we will extend the analysis of the quantum Hall effect in graphene to the case 
a more realistic long-range Coulomb interaction. While the symmetric structure of the solutions 
describing the quantum Hall states will be similar to those in the model with short-range interaction, 
there will also appear qualitative differences. In particular, as we will see, the long-range 
interaction will be responsible for the ``running'' of all dynamical parameters with the Landau 
level index $n$. Here we will follow the presentation of Ref.~\cite{Gorbar:2011kc}, where 
the corresponding framework of the Landau level representation for the fermion propagator 
in a magnetic field was developed. It should be noted that other approaches of 
studying the effect of the Coulomb interaction on the quantum Hall effect dynamics  in the 
$\nu =0$ state were used in Refs.~\cite{KharitonovPRB85.155439,Sinner:2010ax,Semenoff:2011ya}. 
Here, by making use of the approach of Ref.~\cite{Gorbar:2011kc}, we will be able to easily 
generalize the study to the case of any quantum Hall state with a nonzero filling factor, as 
well as to incorporate the Landau levels mixing effects. As will be clear below, the method 
also allows straightforward generalizations to models with the screening effects and other 
types of interactions (e.g., the symmetry-breaking short-range interactions 
\cite{PhysRevB.76.195415,KharitonovPRB85.155439}). 

The interaction term with the long-range Coulomb interaction is formally given by Eq.~(\ref{Coulomb}),
where $U_{C}(\mathbf{r})$ is the Coulomb potential in a magnetic field. In general, the latter should 
include the polarization effects in the presence of a nonzero density and nonzero magnetic field. 
The corresponding polarization effects were studied in Refs.~\cite{Gorbar:2002iw,Pyatkovskiy:2010xz}.

Before proceeding to the derivation of the gap equation in the problem with the long-range 
interaction, let us discuss the structure of the full fermion propagator that allows its dynamical 
parameters to run with the Landau level index $n$. The corresponding Landau-level representation 
was derived in Ref.~\cite{Gorbar:2011kc} and we will review it here in detail. 

\subsubsection{Fermion propagator with running parameters}
\label{subsec:GrapheneCoulombPropagator}

In order to obtain a convenient form of the gap equation in the model with the long-range 
Coulomb interaction, it is important to have a suitable and, at the same time, rather 
general form of the fermion propagator. Here we present such a propagator in the 
Landau-level representation \cite{Gorbar:2011kc}. 

The general structure of the inverse {\it full} fermion propagator for quasiparticles
of a fixed spin has the following form:
\begin{equation}
i\,G^{-1}(\omega;\mathbf{r},\mathbf{r}^\prime)=\left\{\gamma^0 \omega
+v_F \hat{F}^{+} \, (\bm{\pi}\cdot\bm{\gamma})
+\hat{\Sigma}^{+} \right\}\delta(\mathbf{r}-\mathbf{r}^\prime),
\label{inversefull}
\end{equation}
where $\hat{F}^{+}$ and $\hat{\Sigma}^{+}$ can be viewed as generalized wave-function 
renormalization and self-energy operators, respectively. In the special case of the bare propagator
in Eq.~(\ref{inversebare}), the corresponding functions are $\hat{F}_{\rm bare}^{+}=1$ and
$\hat{\Sigma}_{\rm bare}^{+}=(\mu-\mu_B B\sigma^3)\gamma^0$. 

It should be emphasized that, while the ansatz in Eq.~(\ref{inversefull}) is rather general with 
respect to sublattice and valley degrees of freedom, it does not allow any spin-mixing terms 
in the fermion propagator. For many possible ground states in graphene in a magnetic field, 
this is not a limitation. However, there exist states (e.g., canted ferromagnetism 
\cite{2007PhRvB..76h5432H,KharitonovPRB85.155439}) that cannot be captured by such 
a spin-diagonal ansatz). In principle, the corresponding restriction can be removed by 
replacing functions $\hat{F}^{+}$ and $\hat{\Sigma}^{+}$ with matrices in the spin space. 
We will briefly discuss such a generalization in Section~\ref{Sec:GrapheneVacuumAlignment}. 
In order to avoid numerous technical complications in this section, however, we will assume 
that $\hat{F}^{+}$ and $\hat{\Sigma}^{+}$ are diagonal in spin.

By definition, $\hat{F}^{+}$ and $\hat{\Sigma}^{+}$ are functions of the three mutually commuting
dimensionless operators: $(\bm{\pi}\cdot\bm{\gamma})^{2}\ell^2$, $\gamma^0$ and
$i s_{\perp}\gamma^1\gamma^2$, where $s_{\perp}={\rm sgn}(eB)$ and 
$\ell=\sqrt{\hbar c/|eB|}$ is the magnetic length. Taking into account that
$(\gamma^0)^2=1$ and $(i s_{\perp}\gamma^1\gamma^2)^2=1$, the operators $\hat{F}^{+}$
and $\hat{\Sigma}^{+}$ can be written in the following form:
\begin{eqnarray}
\hat{F}^{+}&=&  f +\gamma^0g
+i s_{\perp}\gamma^1\gamma^2 \tilde{g}+ i s_{\perp}\gamma^0 \gamma^1\gamma^2 \tilde{f},
\label{def-F+TXT}\\
\hat{\Sigma}^{+}&=& \tilde{\Delta} + \gamma^0\mu
+  i s_{\perp}\gamma^1\gamma^2 \tilde\mu+  i s_{\perp}\gamma^0 \gamma^1\gamma^2 \Delta,
\label{def-S+TXT}
\end{eqnarray}
where $f$, $\tilde{f}$, $g$, $\tilde{g}$, $\tilde{\Delta}$, $\Delta$, $\mu$, and
$\tilde\mu$ are functions of only one operator, $(\bm{\pi}\cdot\bm{\gamma})^{2}\ell^2$.
Note that for the functions $\mu$, $\tilde\mu$ and $\tilde{\Delta}$, $\Delta$ we keep the same
notations as for the parameters $\mu$, $\tilde\mu$ and $\tilde{\Delta}$, $\Delta$ in
Eqs.~(\ref{singlet_mu}), (\ref{triplet_mu}) and Eqs.~(\ref{triplet_mass}), (\ref{singlet_mass}),
respectively.

It is obvious from the representations  in Eqs.~(\ref{def-F+TXT}) and (\ref{def-S+TXT})
that $\hat{F}^{+}$ and $\hat{\Sigma}^{+}$
do not necessarily commute with $(\bm{\pi}\cdot\bm{\gamma})$. It is convenient, therefore, to introduce
two other functions $\hat{F}^{-}$ and $\hat{\Sigma}^{-}$, which satisfy the relations:
\begin{eqnarray}
\hat{F}^{+} (\bm{\pi}\cdot\bm{\gamma}) &=& (\bm{\pi}\cdot\bm{\gamma}) \hat{F}^{-},
\label{relation-FplusFminus}\\
\hat{\Sigma}^{+} (\bm{\pi}\cdot\bm{\gamma}) &=& (\bm{\pi}\cdot\bm{\gamma}) \hat{\Sigma}^{-}.
\label{relation-SigmaplusSigmaminus}
\end{eqnarray}
As follows from their definition, the explicit representations of these functions read:
\begin{eqnarray}
\hat{F}^{-}&=&  f -\gamma^0g
-i s_{\perp}\gamma^1\gamma^2 \tilde{g}+ i s_{\perp}\gamma^0 \gamma^1\gamma^2 \tilde{f},
\label{def-F-TXT}\\
\hat{\Sigma}^{-}&=& \tilde{\Delta} - \gamma^0\mu
- i s_{\perp}\gamma^1\gamma^2 \tilde\mu+  i s_{\perp}\gamma^0 \gamma^1\gamma^2 \Delta.
\label{def-S-TXT}
\end{eqnarray}
These are obtained from $\hat{F}^{+}$ and $\hat{\Sigma}^{+}$ by reversing the signs in front 
of the two terms that anticommute with $ (\bm{\pi}\cdot\bm{\gamma}) $.

The physical meaning of the functions $\tilde{\Delta}$, $\Delta$, $\mu$, and $\tilde\mu$ that appear in the
definition of $\hat{\Sigma}^{\pm}$ is straightforward: $\tilde{\Delta}$ is the Dirac mass function, $\Delta$ is
the Haldane (time-reversal odd) mass function, $\mu$ is the charge density chemical potential, and
$\tilde\mu$ is the chemical potential for charge density imbalance between the two valleys in the Brillouin
zone. As for the functions $f$, $\tilde{f}$, $g$, and $\tilde{g}$ that appear in the definition of $\hat{F}^{\pm}$,
they are various structures in the wave function renormalization operator.

It may appear that, in the most general case, the full propagator (\ref{inversefull}) can also include
another wave-function renormalization, multiplying the frequency term $\gamma^0 \omega$. This is
not the case, however, because this Dirac structure is already included in the self-energy
$\hat{\Sigma}^{+}$, which may depend on $\omega$ in general. We note at the same time that the
solution for $\hat{\Sigma}^{+}$ will turn out to be independent of $\omega$ in the instantaneous
approximation utilized in this study. This fact is also one of the reasons that makes it particularly
convenient to separate the term $\gamma^0 \omega$ from the generalized self-energy operator
$\hat{\Sigma}^{+}$ in (\ref{inversefull}).

As mentioned earlier, functions $f$, $\tilde{f}$, $g$, $\tilde{g}$, $\tilde{\Delta}$, $\Delta$, $\mu$, and
$\tilde\mu$ are functions of $(\bm{\pi}\cdot\bm{\gamma})^{2}\ell^2$, whose eigenvalues are nonpositive
even integers: $-2n\equiv-(2N+1 - s_\perp s_{12})$, where $N=0,1,2,\ldots$ is the orbital quantum
number and $s_{12}=\pm 1$ is the sign of the pseudospin projection. [The explicit derivation of this
result can be done by following the same steps as in $(3+1)$-dimensional case in 
Appendix~\ref{CSE:App:Landau-level-rep}.] Therefore, in what follows, it will be convenient to use 
the following eigenvalues of the operators $\hat{F}^{\pm}$ and $\hat{\Sigma}^{\pm}$,
\begin{eqnarray}
F^{s_0,s_{12} }_{n}&\equiv &f_{n} + s_0g_{n} + s_{12} \tilde{g}_{n}+ s_0 s_{12} \tilde{f}_{n}\,,
\label{F-n}\\
 \Sigma^{s_0,s_{12} }_{n}&\equiv &\tilde{\Delta}_{n} + s_0\mu_{n} + s_{12} \tilde{\mu}_{n}+ s_0 s_{12}
 \Delta_{n}\,,
\label{Sigma-n}
\end{eqnarray}
where $f_{n}$, $\tilde{f}_{n}$, $g_{n}$, $\tilde{g}_{n}$, $\tilde{\Delta}_{n}$, $\Delta_{n}$, $\mu_{n}$,
and $\tilde\mu_{n}$ are the eigenvalues of the corresponding coefficient operators in the $n$th
LL state. Further, $s_0=\pm 1$ and $s_{12}=\pm 1$ are the eigenvalues of $\gamma^0$ and
$is_{\perp}\gamma^1\gamma^2$, respectively.

In terms of eigenvalues, the inverse propagator is derived in Ref~\cite{Gorbar:2011kc}.
Its final form reads
\begin{eqnarray}
i\,G^{-1}(\omega;\mathbf{r},\mathbf{r}^\prime) &=& e^{i\Phi(\mathbf{r},\mathbf{r}^{\prime})}
i\,\tilde{G}^{-1}(\omega;\mathbf{r}-\mathbf{r}^\prime)  ,
\label{G-inverse}\\
i\,\tilde{G}^{-1}(\omega;\mathbf{r}) &=&
\frac{e^{-\xi/2}}{2\pi \ell^2}
\sum\limits_{n=0}^{\infty}
\sum_{\sigma=\pm 1}
\sum_{s_0=\pm 1}
\Bigg\{ s_0 \omega L_{n}(\xi)
+\left[s_0 \mu_{n,\sigma}+\tilde{\Delta}_{n,\sigma}  \right]
\left[\delta^{s_{0}}_{+\sigma} L_{n}(\xi)+\delta^{s_{0}}_{-\sigma} L_{n-1}(\xi) \right]
 \nonumber\\
&&
+\frac{i v_F}{\ell^2}(\bm{\gamma}\cdot\mathbf{r})( f_{n,\sigma}-s_0 g_{n,\sigma})L_{n-1}^1(\xi)
\Bigg\}
{\cal P}_{s_0,s_{0}\sigma},
\label{tildeG-inverse}
\end{eqnarray}
where $L^{\alpha}_n$ are Laguerre polynomials ($L^{0}_n \equiv L_n$). 
We also introduced the following short-hand notations:
\begin{eqnarray}
\xi = \frac{(\mathbf{r}-\mathbf{r}^{\prime})^2}{2\ell^2},\qquad
\Phi(\mathbf{r},\mathbf{r}^{\prime})= -s_{\perp} \frac{(x-x^\prime)(y+y^\prime)}{2\ell^2},
\quad \mbox{(Schwinger phase)}
\end{eqnarray}
[here, for the external field the Landau gauge in Eq.~(\ref{eq:Landau_gauge}) is assumed]
and
\begin{eqnarray}
\mu_{n,\sigma} = \mu_{n}+\sigma\tilde\mu_{n},& \quad&
\tilde{\Delta}_{n,\sigma} = \tilde{\Delta}_{n}+\sigma\Delta_{n},
\label{effective-parameters}\\
f_{n,\sigma} = f_{n}+\sigma\tilde{f}_{n},&\quad&
g_{n,\sigma} = g_{n}+\sigma\tilde{g}_{n}.
\end{eqnarray}
Note that, by definition, the Laguerre polynomials $L^{\alpha}_n$ with negative $n$ are identically zero.
Finally, ${\cal P}_{s_0,s_{12}}$ are the projectors in the Dirac space,
\begin{eqnarray}
{\cal P}_{s_0,s_{12}} =\frac{1}{4}(1+s_0\gamma_0)(1+s_{12}i s_{\perp}\gamma^1\gamma^2),
\quad \mbox{with}\quad s_0,s_{12}=\pm 1 .
\label{projDspace}
\end{eqnarray}
Similarly, the expression for the propagator itself reads
\begin{eqnarray}
G(\omega;\mathbf{r},\mathbf{r}^\prime) &=& e^{i\Phi(\mathbf{r},\mathbf{r}^{\prime})}
\tilde{G}(\omega;\mathbf{r}-\mathbf{r}^\prime) ,
\label{G-itself}\\
\tilde{G}(\omega;\mathbf{r}) &=&
i \frac{e^{-\xi/2}}{2\pi \ell^2}
\sum\limits_{n=0}^{\infty}
\sum_{\sigma=\pm 1} \sum_{s_0=\pm 1}
\Big\{
\frac{s_0 (\omega+\mu_{n,\sigma})-\tilde{\Delta}_{n,\sigma}}{(\omega+\mu_{n,\sigma})^2-E_{n,\sigma}^2}
\left[ \delta^{s_{0}}_{+\sigma} L_{n}(\xi)+ \delta^{s_{0}}_{-\sigma} L_{n-1}(\xi)\right]
\nonumber\\
&&+\frac{i v_F}{\ell^2}(\bm{\gamma}\cdot\mathbf{r}) \frac{f_{n,\sigma}-s_0 g_{n,\sigma}}
{(\omega+\mu_{n,\sigma})^2-E_{n,\sigma}^2}L_{n-1}^1(\xi)
\Big\}{\cal P}_{s_0,s_{0}\sigma} ,
\label{propagator-full-expression}
\end{eqnarray}
where the energies in the lowest and higher LLs are
\begin{eqnarray}
E_{0,\sigma} &=& \sigma \, \tilde{\Delta}_{0,\sigma} = \Delta_{0} +\sigma \, \tilde\Delta_{0} ,
\label{eq:E0sigma}
\\
E_{n,\sigma} &=& \sqrt{2n(v_F^2/\ell^2)\left[f_{n,\sigma}^2-g_{n,\sigma}^2\right]+\tilde{\Delta}_{n,\sigma}^2},
\quad\mbox{for}\quad n\geq 1.
\end{eqnarray}
The corresponding energies of quasiparticles are determined by the location of the poles of propagator
(\ref{propagator-full-expression}), i.e.,
\begin{eqnarray}
\omega_{0,\sigma} &=& -\mu_{0,\sigma}+E_{0,\sigma},
\label{LLL-energy}\\
\omega_{n,\sigma}^{\pm} &=& -\mu_{n,\sigma}\pm E_{n,\sigma},\quad\mbox{for}\quad n\geq 1.
\end{eqnarray}
Let us note that $\sigma=\pm 1$ is the eigenvalue of matrix
$i s_{\perp}\gamma^0\gamma^1\gamma^2\equiv  s_{\perp}\gamma_3\gamma_5$,
which up to the overall sign $s_{\perp}$ is the quantum number associated with the valley.
This follows from the explicit representation in Eq.~(\ref{gamma35}) and from our convention
for the four-component Dirac spinor, whose first two components are associated with valley
$K$ and the last two components with valley $K^\prime$.

\subsubsection{Gap equation with long-range interaction}

\begin{figure}[t]
\begin{center}
\includegraphics[width=.75\textwidth]{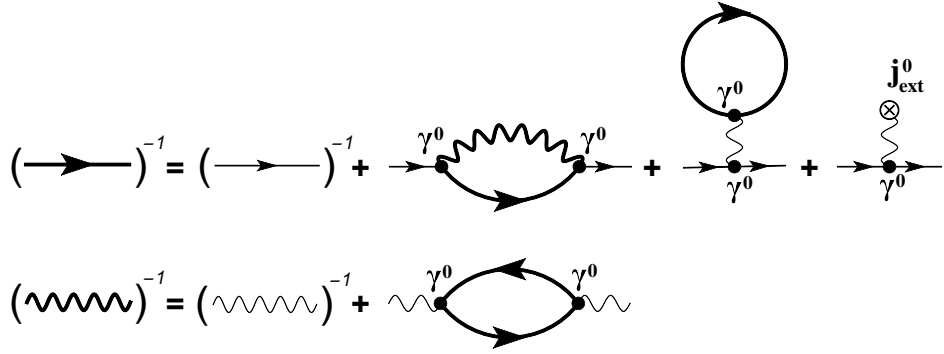}
\caption{The diagrammatic form of the Schwinger--Dyson equations for the electron
and photon propagators in the mean-field approximation.}
\label{fig.SD_eq}
\end{center}
\end{figure}

The Schwinger--Dyson (gap) equation for the fermion propagator in the random-phase
approximation (RPA) is
shown diagrammatically in Fig.~\ref{fig.SD_eq}. Note that in contrast to the naive
mean-field approximation, the RPA Coulomb interaction includes the polarization
(screening) effects, which are not negligible in the dynamics responsible for
symmetry breaking in graphene.

It is important to emphasize that the gap equation for the fermion propagator
in Fig.~\ref{fig.SD_eq}
contains two tadpole diagrams. One of them is connected with the Hartree contribution
due to dynamical
charge carriers, while the other takes into account the background charge from the ions in
graphene and in the substrate. The presence of both tadpoles is essential to insure the
overall neutrality of the sample. Indeed, the equation for the gauge field implies
that the two tadpole contributions should exactly cancel that yields
(the Gauss law):
\begin{equation}
j_{\rm ext}^{0}-e\,\mbox{Tr}\left[\gamma^0 G\right]=0\,.
\end{equation}
As clear from the above arguments, this is directly related to the gauge symmetry in the model. 
(Since the Gauss law does not take place in models with contact interactions \cite{Gorbar:2008hu},
there is an analogue of only one tadpole diagram describing the Hartree interaction, which 
contributes to the gap equation.) Thus, the resulting Schwinger--Dyson
equation for the fermion propagator takes the form
\begin{equation}
G^{-1}(t-t^\prime;\mathbf{r},\mathbf{r}^{\prime}) = S^{-1}(t-t^\prime;\mathbf{r},\mathbf{r}^{\prime})
+ e^2\gamma^0\,G(t-t^\prime;\mathbf{r},\mathbf{r}^{\prime})\gamma^0 D(t^\prime-t;\mathbf{r}^{\prime}-\mathbf{r}),
\label{SD-graphene-Coulomb}
\end{equation}
where $G(t;\mathbf{r},\mathbf{r}^{\prime})$ is the full fermion propagator and
$D(t;\mathbf{r})$ is the  propagator mediating the Coulomb interaction.

At this point it is instructive to compare the cases of a local four-fermion
interaction and a nonlocal Coulomb interaction. In the case of a local four-fermion
interaction, the right-hand side of the Schwinger--Dyson equation contains
$\delta(\mathbf{r}-\mathbf{r}^{\prime})$ and the fermion propagator only at the point
of coincidence $G(0;\mathbf{r},\mathbf{r})$. This means that
the right-hand side of the Schwinger--Dyson equation is a constant in the momentum space and,
hence, it does not renormalize the kinetic $\bm{\pi} \cdot \bm{\gamma}$ part of the fermion
propagator (i.e., $F^{+}=1$). Also, in the case of a local four fermion
interaction, $\Sigma^{+}$ does not depend on the LL index $n$.
Clearly, this simplifies a lot the analysis of the gap equation. The situation changes
in the case of the nonlocal Coulomb interaction.

Following Ref.~\cite{Gorbar:2002iw}, we consider the instantaneous approximation for the
Coulomb interaction by neglecting the dependence of the photon polarization function
$\Pi(\omega,k)$ on $\omega$. Then, in momentum space, the photon propagator takes
the following form:
\begin{equation}
D(\omega,k)\approx D(0,k)=\frac{i}{\epsilon_0 [k+\Pi(0,k)]} ,
\end{equation}
where $\Pi(0,k)$ is the static polarization function and $\epsilon_0$ is a dielectric constant.
In essence, the instantaneous approximation neglects the retardation of the interaction.
This may be a reasonable approximation for graphene, whose charge carriers propagate
much slower than the speed of light. It should be kept in mind, however, that such an
approximation has a tendency to underestimate the strength of the Coulomb
interaction \cite{Gorbar:2002iw,Gamayun:2009em,Gonzalez:2011yb}.

Unfortunately, it is difficult to find exact solutions of Eq.~(\ref{SD-graphene-Coulomb}). Therefore, one has
to use some approximations. Here, we will study the dynamical symmetry breaking in
the model under consideration retaining contributions only of the lowest and several
first LLs. Obviously, this approximation is consistent only if the dynamically
generated gaps are suppressed compared to the Landau scale $\varepsilon_\ell \equiv \sqrt{\hbar v_F^2|eB|/c}$
(it characterizes the energy spectrum of the free theory and is the gap between the lowest and first LLs).

In the instantaneous approximation, the photon propagator reads
\begin{equation}
D(t,\mathbf{r}) = \int\frac{d^2 \mathbf{k}}{(2\pi)^2}  \int\frac{d \omega}{2\pi} D(\omega,k)
e^{-i\omega t+i\mathbf{k}\cdot\mathbf{r}}
= \frac{i}{\epsilon_0} \int_{0}^{\infty} \frac{dk}{2\pi} \frac{k J_{0}(kr)}{k+\Pi(0,k)}
\delta(t)\,.
\label{photon-r}
\end{equation}
By noting that the Schwinger phase on both sides of the gap equation (\ref{SD-graphene-Coulomb}) is
the same, we arrive at the following gap equation for the translation invariant part
of the propagator:
\begin{equation}
i\tilde{G}^{-1}(\omega;\mathbf{r}) = i\tilde{S}^{-1}(\omega;\mathbf{r})
-\frac{e^2}{\epsilon_0} \int_{-\infty}^{\infty} \frac{d\Omega}{2\pi} \int_{0}^{\infty} \frac{dk}{2\pi}
\frac{k J_{0}(kr)}{k+\Pi(0,k)} \gamma^0\,\tilde{G}(\Omega;\mathbf{r})\gamma^0\,.
\label{SD-omega-TXT}
\end{equation}
Multiplying both sides of the gap equation (\ref{SD-omega-TXT}) by either
$e^{-\xi/2}L_{n}(\xi)$ or $e^{-\xi/2}(\bm{\gamma}\cdot\mathbf{r})L_{n}^1(\xi)$, and then integrating
over $\mathbf{r}$, we find that this is equivalent to following set of equations:
\begin{eqnarray}
\left[\mu_{n,\sigma}-\mu-\sigma \tilde{\Delta}_{n,\sigma}\right] \delta^{s_{0}}_{+\sigma}
 &+& \left[\mu_{n+1,\sigma}-\mu+\sigma \tilde{\Delta}_{n+1,\sigma}\right] \delta^{s_{0}}_{-\sigma}
= -\frac{ie^2}{\epsilon_0} \sum_{n^\prime=0}^{\infty}
 \int_{-\infty}^{\infty} \frac{d\Omega}{2\pi} \int_{0}^{\infty} \frac{dk}{2\pi}
\frac{k {\cal L}_{n^\prime,n}^{(0)}(kl)}{k+\Pi(0,k)}
\nonumber \\
&\times&
\left[
\frac{\Omega+\mu_{n^\prime,\sigma}+\sigma \tilde{\Delta}_{n^\prime,\sigma}}{{\cal M}_{n^\prime,
\sigma}} \delta^{s_{0}}_{+\sigma}
+\frac{\Omega+\mu_{n^\prime+1,\sigma}-\sigma \tilde{\Delta}_{n^\prime+1,\sigma}}{{\cal M}_{n^\prime+1,
\sigma}} \delta^{s_{0}}_{-\sigma}
\right],\nonumber \\
\label{SD-algebraic1}
\\
f_{n,\sigma}+s_0 g_{n,\sigma} -1 &=& \frac{ie^2}{n \epsilon_0} \sum_{n^\prime=1}^{\infty}
 \int_{-\infty}^{\infty} \frac{d\Omega}{2\pi} \int_{0}^{\infty} \frac{dk}{2\pi}
\frac{k {\cal L}_{n^\prime-1,n-1}^{(1)}(kl)}{k+\Pi(0,k)} \frac{f_{n^\prime,\sigma}+s_0 g_{n^\prime,\sigma} }{{\cal M}_{n^\prime,\sigma}},
\quad\mbox{for}\quad n\geq1\,,
\label{SD-algebraic2}
\end{eqnarray}
where
\begin{eqnarray}
{\cal L}^{(0)}_{m,n}=\frac{1}{l^{2}}\int_{0}^{\infty}dr\,r\,e^{-\frac{r^{2}}{2\ell^{2}}}
L_{m}\left(\frac{r^{2}}{2\ell^{2}}\right)L_{n}\left(\frac{r^{2}}{2\ell^{2}}\right)J_{0}(k r)
=(-1)^{m+n}e^{-\frac{k^{2}\ell^{2}}{2}}L_{m}^{n-m}\left(\frac{k^{2}\ell^{2}}{2}\right)
L_{n}^{m-n}\left(\frac{k^{2}\ell^{2}}{2}\right),\label{def-L_mn^0}
\end{eqnarray}
\begin{eqnarray}
{\cal L}^{(1)}_{m,n}=\frac{1}{2l^{4}}\int_{0}^{\infty}dr\,r^{3}\,e^{-\frac{r^{2}}{2\ell^{2}}}
L^{1}_{m}\left(\frac{r^{2}}{2\ell^{2}}\right)L^{1}_{n}\left(\frac{r^{2}}{2\ell^{2}}\right)J_{0}(k r)
=(-1)^{m+n} (m+1) \,e^{-\frac{k^{2}\ell^{2}}{2}}L_{m+1}^{n-m}\left(\frac{k^{2}\ell^{2}}{2}\right)
L_{n}^{m-n}\left(\frac{k^{2}\ell^{2}}{2}\right) \nonumber && \\
=(-1)^{m+n} (n+1) \,e^{-\frac{k^{2}\ell^{2}}{2}}L_{m}^{n-m}\left(\frac{k^{2}\ell^{2}}{2}\right)
L_{n+1}^{m-n}\left(\frac{k^{2}\ell^{2}}{2}\right)\,.&&
\label{def-L_mn^1}
\end{eqnarray}
To obtain the results on the right hand sides, we used the  table integral
 7.422 2 in Ref.~\cite{1980tisp.book.....G}.

The gap equations can be equivalently rewritten as follows:
\begin{eqnarray}
\mu_{n,\sigma}-\mu-\sigma \tilde{\Delta}_{n,\sigma} &=& -i \alpha \varepsilon_\ell \sum_{n^\prime=0}^{\infty}
\kappa_{n^\prime,n}^{(0)}   \int_{-\infty}^{\infty}  \frac{d\Omega}{2\pi}
\frac{\Omega+\mu_{n^\prime,\sigma}+\sigma \tilde{\Delta}_{n^\prime,\sigma}}{{\cal M}_{n^\prime,\sigma}} ,
\quad\mbox{for}\quad n\geq0, \label{B11} \\
\mu_{n,\sigma}-\mu+\sigma \tilde{\Delta}_{n,\sigma}&=& -i \alpha \varepsilon_\ell \sum_{n^\prime=1}^{\infty}
\kappa_{n^\prime-1,n-1}^{(0)}  \int_{-\infty}^{\infty}  \frac{d\Omega}{2\pi}
\frac{\Omega+\mu_{n^\prime,\sigma}-\sigma \tilde{\Delta}_{n^\prime,\sigma}}{{\cal M}_{n^\prime,\sigma}} ,
\quad\mbox{for}\quad n\geq1,\label{B12}\\
f_{n,\sigma} &=& 1+ i \alpha \varepsilon_\ell \sum_{n^\prime=1}^{\infty}
\frac{\kappa_{n^\prime-1,n-1}^{(1)} }{n}  \int_{-\infty}^{\infty}  \frac{d\Omega}{2\pi}
\frac{f_{n^\prime,\sigma}  }{{\cal M}_{n^\prime,\sigma}},
\quad\mbox{for}\quad n\geq1, \label{B13}\\
g_{n,\sigma}&=& i \alpha \varepsilon_\ell \sum_{n^\prime=1}^{\infty}
\frac{\kappa_{n^\prime-1,n-1}^{(1)} }{n}   \int_{-\infty}^{\infty}  \frac{d\Omega}{2\pi}
 \frac{ g_{n^\prime,\sigma} }{{\cal M}_{n^\prime,\sigma}},
\quad\mbox{for}\quad n\geq1,\label{B14}
\end{eqnarray}
where $\alpha = e^2/(\epsilon_0 v_F)$, $\varepsilon_\ell = v_F/\ell$, and
\begin{eqnarray}
\kappa_{m,n}^{(\rho)} = \int_{0}^{\infty} \frac{dk}{2\pi}
\frac{k \ell  {\cal L}_{m,n}^{(\rho)}(k\ell)}{k+\Pi(0,k)},\quad \rho=0,1.
\label{kappa_mn^rho}
\end{eqnarray}
When the screening effects are neglected, i.e.,  $\Pi(0,k)=0$, we can use
the explicit form for of ${\cal L}_{m,n}^{(\rho)}$
(with $\rho=0,1$) in Eqs.~(\ref{def-L_mn^0}) and (\ref{def-L_mn^1}) and obtain
the following analytical expressions for $\kappa_{m,n}^{(\rho)}$ \cite{Gorbar:2011kc}:
\begin{eqnarray}
\left. \kappa_{m,n}^{(\rho)}\right|_{\Pi\to0}
&=&\frac{(-1)^{m+n}}{2\sqrt{2}}\sum\limits_{k=0}^{{\rm min}(m,n)}\frac{\Gamma(\rho+1/2+k)}
{(m-k)!(n-k)!\Gamma(1/2-m+k)\Gamma(1/2-n+k)k!} .
\label{kappa0-Pi0} 
\end{eqnarray}
The values of $\kappa_{m,n}^{(\rho)}$ (with $\rho=0,1$) at small values of $m$ and $n$ are given
in Tables~\ref{tab-kappa-0} and \ref{tab-kappa-1}. The leading asymptotes for $n\to \infty$
(at finite $m$) are
\begin{eqnarray}
\left.\kappa_{m,n}^{(0)}\right|_{\Pi\to 0} &\simeq& \frac{1}{2\pi\sqrt{2n}}
+\frac{2m-1}{8\pi(2n)^{3/2}}+O\left(\frac{1}{n^{5/2}}\right)
\quad\mbox{for}\quad n\to\infty ,
\label{kappa-asymptote-0}\\
\left.\kappa_{m,n}^{(1)}\right|_{\Pi\to 0} &\simeq& \frac{(m+1)}{4\pi\sqrt{2n}}
+\frac{(m+1)(3m-1)}{16\pi(2n)^{3/2}}
+O\left(\frac{1}{n^{5/2}}\right)
\quad\mbox{for}\quad n\to\infty .
\label{kappa-asymptote-1}
\end{eqnarray}

\begin{table}[t]
\caption{Values of $\kappa_{m,n}^{(0)}$ when the effects of polarization tensor are neglected}
\begin{center}
\begin{tabular}{l@{\hskip 0.5in}l@{\hskip 0.5in}l@{\hskip 0.5in}l@{\hskip 0.5in}l@{\hskip 0.5in}l@{\hskip 0.5in}l}
\hline
$\kappa_{m,n}^{(0)}$ & $m=0$ & $m=1$ & $m=2$ & $m=3$ & $m=4$ & $m=5$ \\
\hline
$n=0$ & $\frac{1}{2\sqrt{2\pi}}$ & $\frac{1}{4\sqrt{2\pi}}$ & $\frac{3}{16\sqrt{2\pi}}$ & $\frac{5}{32\sqrt{2\pi}}$ & $\frac{35}{256\sqrt{2\pi}}$ & $\frac{63}{512\sqrt{2\pi}}$ \\[2mm]
$n=1$ & $\frac{1}{4\sqrt{2\pi}}$ & $\frac{3}{8\sqrt{2\pi}}$ & $\frac{7}{32\sqrt{2\pi}}$ & $\frac{11}{64\sqrt{2\pi}}$ & $\frac{75}{512\sqrt{2\pi}}$ & $\frac{133}{1024\sqrt{2\pi}}$ \\[2mm]
$n=2$ & $\frac{3}{16\sqrt{2\pi}}$ & $\frac{7}{32\sqrt{2\pi}}$ & $\frac{41}{128\sqrt{2\pi}}$ & $\frac{51}{256\sqrt{2\pi}}$ & $\frac{329}{2048\sqrt{2\pi}}$ & $\frac{569}{4096\sqrt{2\pi}}$ \\[2mm]
$n=3$ & $\frac{5}{32\sqrt{2\pi}}$ & $\frac{11}{64\sqrt{2\pi}}$ & $\frac{51}{256\sqrt{2\pi}}$ & $\frac{147}{512\sqrt{2\pi}}$ & $\frac{759}{4096\sqrt{2\pi}}$ & $\frac{1245}{8192\sqrt{2\pi}}$ \\[2mm]
$n=4$ & $\frac{35}{256\sqrt{2\pi}}$ & $\frac{75}{512\sqrt{2\pi}}$ & $\frac{329}{2048\sqrt{2\pi}}$ & $\frac{759}{4096\sqrt{2\pi}}$ & $\frac{8649}{32768\sqrt{2\pi}}$ & $\frac{11445}{65536\sqrt{2\pi}}$ \\[2mm]
$n=5$ & $\frac{63}{512\sqrt{2\pi}}$ & $\frac{133}{1024\sqrt{2\pi}}$ & $\frac{569}{4096\sqrt{2\pi}}$ & $\frac{1245}{8192\sqrt{2\pi}}$ & $\frac{11445}{65536\sqrt{2\pi}}$ & $\frac{32307}{131072\sqrt{2\pi}}$ \\[2mm]
\hline
\label{tab-kappa-0}
\end{tabular}
\end{center}
\end{table}

\begin{table}[t]
\caption{Values of $\kappa_{m,n}^{(1)}$ when the effects of polarization tensor are neglected}
\begin{center}
\begin{tabular}{l@{\hskip 0.5in}l@{\hskip 0.5in}l@{\hskip 0.5in}l@{\hskip 0.5in}l@{\hskip 0.5in}l@{\hskip 0.5in}l}
\hline
$\kappa_{m,n}^{(1)}$ & $m=0$ & $m=1$ & $m=2$ & $m=3$ & $m=4$ & $m=5$ \\
\hline
$n=0$ & $\frac{1}{4\sqrt{2\pi}}$ & $\frac{1}{8\sqrt{2\pi}}$ & $\frac{3}{32\sqrt{2\pi}}$ & $\frac{5}{64\sqrt{2\pi}}$ & $\frac{35}{512\sqrt{2\pi}}$ & $\frac{63}{1024\sqrt{2\pi}}$ \\[2mm]
$n=1$ & $\frac{1}{8\sqrt{2\pi}}$ & $\frac{7}{16\sqrt{2\pi}}$ & $\frac{15}{64\sqrt{2\pi}}$ & $\frac{23}{128\sqrt{2\pi}}$ & $\frac{155}{1024\sqrt{2\pi}}$ & $\frac{273}{2048\sqrt{2\pi}}$ \\[2mm]
$n=2$ & $\frac{3}{32\sqrt{2\pi}}$ & $\frac{15}{64\sqrt{2\pi}}$ & $\frac{153}{256\sqrt{2\pi}}$ & $\frac{171}{512\sqrt{2\pi}}$ & $\frac{1065}{4096\sqrt{2\pi}}$ & $\frac{1809}{8192\sqrt{2\pi}}$ \\[2mm]
$n=3$ & $\frac{5}{64\sqrt{2\pi}}$ & $\frac{23}{128\sqrt{2\pi}}$ & $\frac{171}{512\sqrt{2\pi}}$ & $\frac{759}{1024\sqrt{2\pi}}$ & $\frac{3495}{8192\sqrt{2\pi}}$ & $\frac{5505}{16384\sqrt{2\pi}}$ \\[2mm]
$n=4$ & $\frac{35}{512\sqrt{2\pi}}$ & $\frac{155}{1024\sqrt{2\pi}}$ & $\frac{1065}{4096\sqrt{2\pi}}$ & $\frac{3495}{8192\sqrt{2\pi}}$ & $\frac{57225}{65536\sqrt{2\pi}}$ & $\frac{67365}{131072\sqrt{2\pi}}$ \\[2mm]
$n=5$ & $\frac{63}{1024\sqrt{2\pi}}$ & $\frac{273}{2048\sqrt{2\pi}}$ & $\frac{1809}{8192\sqrt{2\pi}}$ & $\frac{5505}{16384\sqrt{2\pi}}$ & $\frac{67365}{131072\sqrt{2\pi}}$ & $\frac{261207}{262144\sqrt{2\pi}}$ \\[2mm]
\hline
\label{tab-kappa-1}
\end{tabular}
\end{center}
\end{table}

The zero temperature gap equations (\ref{B11}) through (\ref{B14}) are straightforwardly generalized to the case
of nonzero temperature by making the replacement $\Omega\to i\Omega_{m}\equiv i \pi T (2m+1)$ and using
the Matsubara sums instead of the frequency
integrations,
\begin{eqnarray}
\int\frac{d\Omega}{2\pi}(\ldots) \to i T\sum_{m=-\infty}^{\infty} (\ldots) \, .
\end{eqnarray}
Then, we derive the finite temperature gap equations,
\begin{eqnarray}
\mu_{n,\sigma}-\mu-\sigma \tilde{\Delta}_{n,\sigma} &=&
\frac{\alpha \varepsilon_\ell }{2}\sum_{n^\prime=0}^{\infty}
\kappa_{n^\prime,n}^{(0)}
\Big\{
n_{F}\left(E_{n^\prime,\sigma}-\mu_{n^\prime,\sigma}\right)
-n_{F}\left(E_{n^\prime,\sigma}+\mu_{n^\prime,\sigma}\right)\nonumber\\
&&-\frac{\sigma \tilde{\Delta}_{n^\prime,\sigma}}{E_{n^\prime,\sigma}}
\left[1-n_{F}\left(E_{n^\prime,\sigma}-\mu_{n^\prime,\sigma}\right)
-n_{F}\left(E_{n^\prime,\sigma}+\mu_{n^\prime,\sigma}\right)
\right]
\Big\} ,
\quad\mbox{for}\quad n\geq0,
\label{gap-eq-mu_n-0}\\
\mu_{n,\sigma}-\mu+\sigma \tilde{\Delta}_{n,\sigma}&=&
\frac{\alpha \varepsilon_\ell}{2} \sum_{n^\prime=1}^{\infty}
\kappa_{n^\prime-1,n-1}^{(0)}
\Big\{
n_{F}\left(E_{n^\prime,\sigma}-\mu_{n^\prime,\sigma}\right)
-n_{F}\left(E_{n^\prime,\sigma}+\mu_{n^\prime,\sigma}\right)\nonumber\\
&&+\frac{\sigma \tilde{\Delta}_{n^\prime,\sigma}}{E_{n^\prime,\sigma}}
\left[1-n_{F}\left(E_{n^\prime,\sigma}-\mu_{n^\prime,\sigma}\right)
-n_{F}\left(E_{n^\prime,\sigma}+\mu_{n^\prime,\sigma}\right)
\right]
\Big\} ,
\quad\mbox{for}\quad n\geq1,
\label{gap-eq-mu_n-1}\\
f_{n,\sigma} &=& 1+ \frac{\alpha \varepsilon_\ell}{2} \sum_{n^\prime=1}^{\infty}
\frac{\kappa_{n^\prime-1,n-1}^{(1)} }{n}
\frac{f_{n^\prime,\sigma}  }{E_{n^\prime,\sigma}}
\left[1-n_{F}\left(E_{n^\prime,\sigma}-\mu_{n^\prime,\sigma}\right)
-n_{F}\left(E_{n^\prime,\sigma}+\mu_{n^\prime,\sigma}\right)
\right],
\quad\mbox{for}\quad n\geq1,
\label{gap-eq-f_n}\\
g_{n,\sigma}&=& \frac{\alpha \varepsilon_\ell}{2} \sum_{n^\prime=1}^{\infty}
\frac{\kappa_{n^\prime-1,n-1}^{(1)} }{n}
 \frac{ g_{n^\prime,\sigma} }{E_{n^\prime,\sigma}}
 \left[1-n_{F}\left(E_{n^\prime,\sigma}-\mu_{n^\prime,\sigma}\right)
-n_{F}\left(E_{n^\prime,\sigma}+\mu_{n^\prime,\sigma}\right)
\right],
\quad\mbox{for}\quad n\geq1 ,
\label{gap-eq-g_n}
\end{eqnarray} 
were $n_{F}(x)\equiv 1/(e^x+1)$ is the Fermi-Dirac distribution function.
By taking into account the special status of the LLL states with the energy given by 
Eq.~(\ref{eq:E0sigma}), we find that the LLL parameters enter the gap equations
only in the following two independent combinations:
\begin{eqnarray}
\mu^{\rm eff}_0 =  \mu_0-\Delta_0,\qquad
\tilde{\Delta}^{\rm eff}_{0} =\tilde{\Delta}_{0}-\tilde{\mu}_0 .
\end{eqnarray}
It is the pseudospin polarized nature of the LLL that makes it impossible to determine 
parameters $\mu_0$ and $\Delta_0$ independently. This is also a reflection of the fact
that $\mu^{\rm eff}_0$ is the only combination of $\mu_0$ and $\Delta_0$ with a well 
defined physical meaning.  The same is true for $\tilde{\Delta}_{0}$ and $\tilde{\mu}_0$, 
which give only one observable combination, the effective Dirac mass $\tilde{\Delta}^{\rm eff}_{0}$.

\subsubsection{Solutions to the gap equation}
\label{Sec:GrapheneExamplesSolutions}

In the previous section, we found that the gap equation in the model with a long-range 
interaction is reduced to an infinite set of equations in the Landau-level basis. From physics
viewpoint, however, one may expect that the low-energy dynamics, when expressed in 
terms of renormalized parameters, is not very sensitive to the details at high-energies. 
In the Landau-level basis, then, it may be justified to truncate the corresponding infinite 
set of equations at some large, but finite value of Landau level index $n_{\rm max}$.

In order to concentrate exclusively on the role of the long-range nature of the interaction,
it may be justified here to neglect the screening effects. Thus, we set $\Pi(0,k)=0$
and use the analytical expressions for $\kappa_{m,n}^{(\rho)}$ in Eq.~(\ref{kappa0-Pi0}).
This approximation greatly simplifies the numerical analysis and allows us to get the key 
qualitative features of the dynamics with very few technical complications.

{\it (i) Wave function renormalization.} Let us start by analyzing the simplest case of 
vanishing dynamical mass parameters and no significant splitting of the LLs. This is 
expected to be the case when the magnetic field is not too strong. Even in this case, 
however, there is a nontrivial dynamics responsible for the renormalization of the 
Fermi velocity. This is also interesting from experimental point of view because the 
renormalized value of the Fermi velocity parameter, which is also a function of the 
Landau index $n$, affects the energies of optical transitions \cite{Sadowski:2006ax,
Jiang:2007az,Orlita:2010as}.

When there are no dynamical mass parameters, we are left with a smaller subset of 
the gap equations, involving only the wave-function renormalization $f_n$, see 
Eq.~(\ref{gap-eq-f_n}). However, even this subset contains an infinite number of gap
equations for each choice of spin, i.e.,
\begin{equation}
f_{n} = 1+ \frac{\alpha}{2} \sum_{n^\prime=1}^{\infty}
\frac{\kappa_{n^\prime-1,n-1}^{(1)} }{n \sqrt{2n^\prime}}
\left[1-n_{F}\left(E_{n^\prime}-\mu \right)
-n_{F}\left(E_{n^\prime}+\mu \right)
\right],
\quad\mbox{for}\quad n\geq1,
\label{gap-eq-fn_TXT}
\end{equation}
where we took into account the definition of the coefficients $\kappa_{n^\prime,n}^{(1)}$ in
Eq.~(\ref{kappa0-Pi0}) and used the approximate expression for the LL energies:
$E_{n,\sigma} \approx E_{n} =\sqrt{2n} \varepsilon_\ell f_{n,\sigma}$, which are independent of
the valley quantum number $\sigma\equiv s_0 s_{12}$. Note that the spin index is also omitted, which
is justified especially at weak fields. 

Note that the strength of the Coulomb interaction is characterized by the graphene's ``fine structure 
constant" $\alpha \equiv {e^2}/{\epsilon_0 v_F}$, which is approximately equal to $2.2/\epsilon_0$. 
In the numerical calculations below, we assume that $\epsilon_0=1$, which models the case of 
suspended graphene. Before proceeding to the results of numerical calculations, it is also appropriate 
to clearly determine all relevant energy scales in the problem at hand. By ignoring the large energy 
cutoff due to a finite width of the conductance band, there are essentially only two characteristic 
scales relevant for the low-energy physics: (i) the Landau energy scale $\varepsilon_\ell = 
\sqrt{\hbar v_F^2|eB|/c}$ and (ii) the much smaller Zeeman energy $ \epsilon_{Z} \equiv  \mu_B B $. 
In our numerical calculations, we measure all physical quantities with the units of energy 
in units of $\varepsilon_\ell $. Numerically, these are
\begin{eqnarray}
\varepsilon_\ell = \sqrt{\hbar v_F^2|eB|/c} =26 \sqrt{B[T]}~\mbox{meV}, \qquad
\epsilon_{Z}  = \mu_B B =5.8\times 10^{-2}B[T]~\mbox{meV} ,
\end{eqnarray}
where $B[T]$ is the value of the magnetic field measured in Teslas.
The corresponding temperature scales are $\varepsilon_\ell/k_B=300 \sqrt{B[T]}~\mbox{K}$ and
$ \epsilon_{Z} /k_B=0.67 B[T]~\mbox{K}$. Note that the magnetic length $\ell=\sqrt{ \hbar c/|eB|} $ 
and the Landau energy scale $\varepsilon_\ell $ are related through the Fermi velocity as 
follows: $\ell={\hbar  v_F}/{\varepsilon_\ell } =  {26~\mbox{nm}}/{\sqrt{B[T]}}$ ,
where we used $v_F/c= 1/300$.

It should be noted that the expression on the right hand side of Eq.~(\ref{gap-eq-fn_TXT})
is formally divergent. Indeed, by taking into account the asymptotes of the kernel coefficients
$\kappa_{n^\prime-1,n-1}^{(1)}$ as $n^\prime\to \infty$, see Eq.~(\ref{kappa-asymptote-1}), we
find that the sum over $n^\prime$ on the right hand side of Eq.~(\ref{gap-eq-fn_TXT}) is
logarithmically divergent. From quantum field theoretical point of view, of course, this is just
an indication that the coupling constant $\alpha$ is also subject to a renormalization \cite{Gonzalez:1999wo}.
For our purposes in this study, however, we may simply assume that the sum over the Landau
levels is finite. Indeed, in contrast to actual relativistic models, the effective action for
quasiparticles of graphene is valid only at sufficiently low energies. Moreover, it is also
clear that the energy width of the conducting band of graphene is finite. In fact, it can
be shown that the formal value of the cutoff in the summation over the LL
index $n$ is approximately given by $n_{\rm max}\simeq 10^4/B[T]$ \cite{Roldan:2009az}, where $B[T]$ is the
value of the magnetic field in Teslas. In our numerical calculations, we will use a much
smaller cutoff $n_{\rm max}$. For all practical purposes, when dealing with the observables
in lowest few LLs, such a limitation has little effect on the qualitative and in most cases
even quantitative results. Thus, in the rest of this subsection, we choose the value of the cutoff to
be $n_{\rm max}=100$. (We checked that the numerical results for $f_n$ with the cutoffs 
$n_{\rm max}=50$ and $n_{\rm max}=150$ are qualitatively the same. The values of $f_n$ have
a tendency to grow with increasing $n_{\rm max}$. It is understood, of course, that
such a growth should be compensated by the renormalization of the coupling constant in a
more refined approximation.)

\begin{figure}[t]
\begin{center}
\includegraphics[width=.45\textwidth]{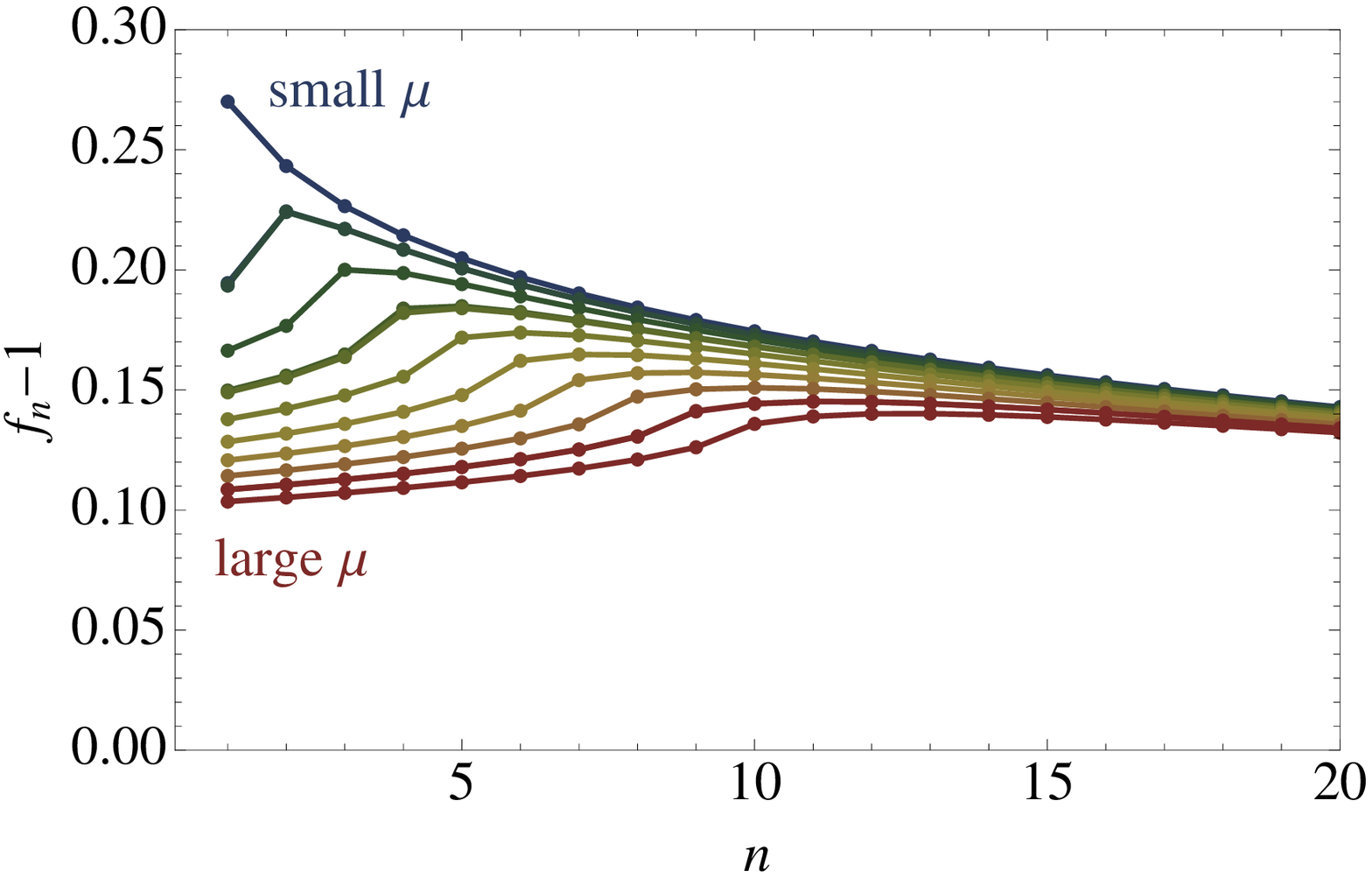}
\hspace{.07\textwidth}
\includegraphics[width=.45\textwidth]{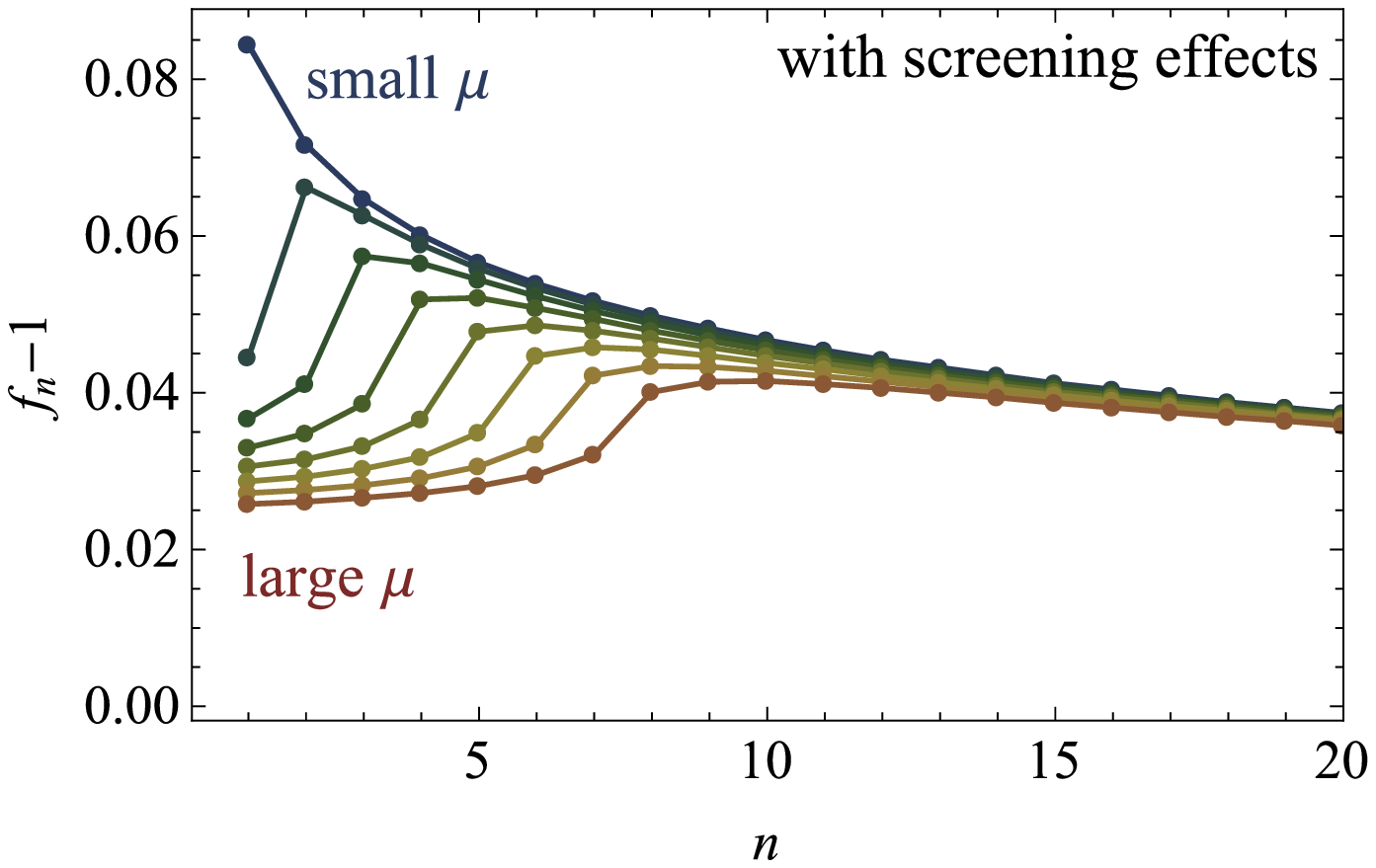}
\caption{(Color online) The numerical values of the wave function renormalization 
coefficients $f_n$ vs. the LL index $n$ for several fixed values of the chemical 
potential. Left panel shows the results for the case of Coulomb interaction without
screening, and the right panel shows the numerical results with the inclusion of
screening effects, as well as the renormalization of parameters $\mu_n$ and 
$\Delta_n$, see Fig.~\ref{fig_Mu_Delta_n_run}.}
\label{fig.fn-vs-n}
\end{center}
\end{figure}

The effective Fermi velocity in the $n$th LL is determined by the following relation:
$\tilde{v}_{F,n} = f_n v_{F}$, where the numerical values of the wave function renormalization
are shown in Fig.~\ref{fig.fn-vs-n}. There we show many sets of the results which correspond to
different values of the chemical potentials. The points are the actual data, while the lines
connecting the points are shown for eye guiding the data for fixed values of the chemical
potentials. The data on the top line corresponds to small values of the chemical potential,
$|\mu|<\sqrt{2}\varepsilon_\ell$. The other lines correspond to the chemical potentials in the
energy gaps between $n$th and $(n+1)$th LL (with $n=0,1,2,\ldots$ from top to bottom).
A part of the same data is also given in Table~\ref{tab-fn}. 

In general, the renormalized Fermi velocity 
$\tilde{v}_{F,n}$ is about $10\%$ to $20\%$ larger than its nonrenormalized value $v_{F}$. These
results seem to be somewhat smaller than the predictions in Ref.~\cite{Iyengar:2007xx}. One should keep 
in mind, however, that there are considerable uncertainties in the theoretical predictions for the
renormalized values of the Fermi velocity. In part, these are associated with a relatively large 
value of the coupling constant in graphene and with the logarithmic running of the wave function 
renormalization itself.

\begin{table}[t]
\caption{Values of the wave function renormalization $f_{n}$ for several values of the chemical potentials.
The results in parenthesis are the results in the case when screening effects, as well as the 
renormalization of parameters $\mu_n$ and $\Delta_n$, are included.}
{\small
\begin{tabular}{c l l l l l l l}
\hline
 & $f_1$ & $f_2$ & $f_3$ & $f_4$ & $f_5$ & $f_6$ \\
\hline
$|\mu|<\sqrt{2}f_{1}\varepsilon_\ell$ 
&  1.270 (1.084) & 1.243 (1.072) & 1.227 (1.065) & 1.214 (1.060) & 1.205 (1.057) & 1.197 (1.054) \\
$\sqrt{2}f_{1}\varepsilon_\ell<\mu<\sqrt{4}f_{2}\varepsilon_\ell$ 
& 1.194 (1.044) & 1.224 (1.066) & 1.217 (1.063) & 1.208 (1.059) & 1.201 (1.056) & 1.194 (1.053)  \\
$\sqrt{4}f_{2}\varepsilon_\ell<\mu<\sqrt{6}f_{3}\varepsilon_\ell$ 
& 1.166 (1.033) & 1.177 (1.035) & 1.200 (1.039) & 1.199 (1.052) & 1.194 (1.052) & 1.189 (1.051)   \\
$\sqrt{6}f_{3}\varepsilon_\ell<\mu<\sqrt{8}f_{4}\varepsilon_\ell$ 
& 1.150 (1.031) & 1.156 (1.032) & 1.165 (1.033) & 1.184 (1.037) & 1.185 (1.048) & 1.182 (1.049)  \\
$\sqrt{8}f_{4}\varepsilon_\ell<\mu<\sqrt{10}f_{5}\varepsilon_\ell$ 
& 1.138 (1.027) & 1.142 (1.028) & 1.148 (1.028) & 1.156 (1.029) & 1.172 (1.031) & 1.174 (1.033)  \\
$\sqrt{10}f_{5}\varepsilon_\ell<\mu<\sqrt{12}f_{6}\varepsilon_\ell$ 
&  1.128 (1.029) & 1.132 (1.029) & 1.136 (1.030) & 1.141 (1.032) & 1.148 (1.035) & 1.162 (1.045) \\
$\sqrt{12}f_{6}\varepsilon_\ell<\mu<\sqrt{14}f_{7}\varepsilon_\ell$ 
&   1.121 (1.027) & 1.123 (1.028) & 1.127 (1.028) & 1.130 (1.029) & 1.135 (1.031) & 1.141 (1.033) \\
\hline
\label{tab-fn}
\end{tabular}}
\end{table}

A convenient quantitative measure of the many-particle effects in the transition energies
is given by the prefactors $C_{n,n^\prime}$, which are introduced as deviations from the
noninteracting carriers in graphene,
\begin{equation}
\Delta E_{n,n^\prime} \equiv E_{n^\prime} \pm E_{n} = \left(\sqrt{2n^\prime}\pm \sqrt{2n}\right)
\varepsilon_\ell +\alpha \varepsilon_\ell C_{n,n^\prime},
\label{DeltaE-scaling}
\end{equation}
where, once again, a small Zeeman splitting of the energy levels is ignored.
Note that both terms in (\ref{DeltaE-scaling}) scale as $\sim\sqrt{B}$ and
experimental data on infrared spectroscopy of LLs of graphene clearly confirm
this behavior \cite{Sadowski:2006ax,Jiang:2007az,Orlita:2010as}. On the other hand, the dependence 
of the coefficients $C_{n,m}$ on the LL pair allows one to get information on many-body effects.
By making use of our notation for the wave-function renormalization, we obtain
\begin{equation}
C_{n,n^\prime} = \frac{\sqrt{2n^\prime}}{\alpha}(f_{n^\prime}-1)\pm \frac{\sqrt{2n}}{\alpha}(f_{n}-1),
\end{equation}
when $n\neq 0$ and $n^\prime\neq 0$. (For transitions from the $n=0$ level and for transitions to the
$n^\prime=0$ level, the LLL never gives any contribution to the corresponding prefactors.)
By making use of our results for $f_{n}$, we obtain the values of prefactors $C_{n,n^\prime}$. For
transitions between several low-lying LLs, the values of some prefactors are 
\begin{equation}
C_{-1,0}= 0.176, \qquad 
C_{-1,1}=0.352, \qquad
C_{-2,0}=0.224, \qquad
C_{-2,1}=0.400, \qquad
C_{-2,2}=0.448 .
\end{equation}
It is clear that the Coulomb interaction contribution to the LL transitions slightly increases the transition 
energies above their noninteracting values in accordance with experimental data (e.g., see Fig.~3a in 
Ref.~\cite{Jiang:2007az}).

{\it (ii) States with LLL filling factors.} 
Let us now perform a simple numerical analysis with dynamical parameters. We will include several 
LLs and account for the spin degree of freedom in the analysis. Because of a large number of the 
dynamical parameters in each Landau level, we will use a rather small number for the cutoff index 
$n_{\rm max}=5$ in the summation over the LLs. 

The gap equations for both spins look the same except for the value of the chemical potential: 
it is $\mu_{\uparrow} \equiv \mu- \epsilon_{Z} $ and $\mu_{\downarrow} \equiv \mu+ \epsilon_{Z} $.
By repeating the same analysis as above, we find that there exist many more solutions around the vanishing
value of $\mu$. Keeping only a subset of several qualitatively different solutions with lowest energies, we
find among them a pure Dirac mass solution, 
four types of the Haldane mass solutions and four types of hybrid solutions (see below).
The pure Dirac mass solution has nonzero Dirac masses for both spins and no Haldane masses [i.e., the order 
parameters are triplets with respect to both $SU_{\uparrow}(2)$ and $SU_{\downarrow}(2)$]. Four Haldane 
mass solutions are determined by four possible sign combinations of the two time-reversal breaking masses:
(i) $\Delta_{0,\uparrow}>0$ and $\Delta_{0,\downarrow}>0$;
(ii) $\Delta_{0,\uparrow}>0$ and $\Delta_{0,\downarrow}<0$;
(iii) $\Delta_{0,\uparrow}<0$ and $\Delta_{0,\downarrow}>0$;
(iv) $\Delta_{0,\uparrow}<0$ and $\Delta_{0,\downarrow}<0$.
All of these are characterized by singlet order parameters with respect to both $SU_{\uparrow}(2)$ 
and $SU_{\downarrow}(2)$ symmetry groups. 
Similarly, four hybrid solutions are determined by the following conditions
(i) $\tilde{\Delta}_{0,\uparrow}\neq 0$ and $\Delta_{0,\downarrow}>0$;
(ii) $\tilde{\Delta}_{0,\uparrow}\neq 0$ and $\Delta_{0,\downarrow}<0$;
(iii) $\Delta_{0,\uparrow}>0$ and $\tilde{\Delta}_{0,\downarrow}\neq 0$;
(iv) $\Delta_{0,\uparrow}<0$ and $\tilde{\Delta}_{0,\downarrow}\neq 0$.
The common feature of the hybrid solutions is that one of their order parameters is a triplet with respect to  
$SU_{\uparrow}(2)$ or $SU_{\downarrow}(2)$ group, while the other order parameter is a singlet with respect 
to the other group.

\begin{figure}[t]
\begin{center}
\includegraphics[width=.48\textwidth]{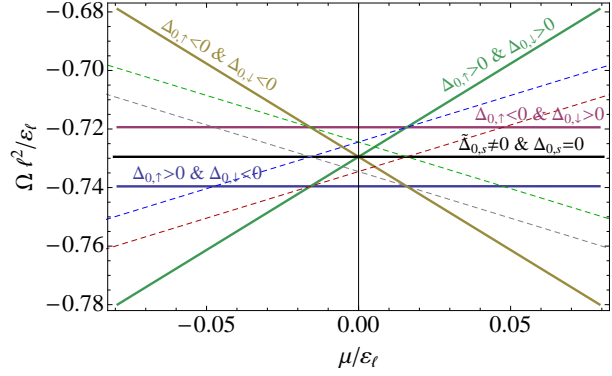}
\caption{Numerical results for free energies of several lowest energy solutions, realized when
the Fermi energy is near $n=0$ LL. The dashed lines correspond to hybrid solutions.}
\label{fig_free-e_LLL_S1S2}
\end{center}
\end{figure}

The free energies of several lowest energy solutions are plotted in Fig.~\ref{fig_free-e_LLL_S1S2}. 
Four singlet type solutions and one triplet solution are shown by solid lines in the figure. Four lowest 
energy hybrid solution are shown by dashed lines.
There are three qualitatively different regions, in which the lowest energy states are different, i.e.,
\begin{eqnarray}
 \mu<- \epsilon_{Z} : && \Delta_{0,\uparrow}>0~\&~ \Delta_{0,\downarrow}>0,\\
- \epsilon_{Z} <\mu< \epsilon_{Z} : &&  \Delta_{0,\uparrow}>0 ~\&~ \Delta_{0,\downarrow}<0,\\
\mu> \epsilon_{Z} : &&  \Delta_{0,\uparrow}<0 ~\&~ \Delta_{0,\downarrow}<0 .
\end{eqnarray}
At the points $\mu=\pm  \epsilon_{Z} $ , there also exist hybrid solutions with the same lowest values of the 
energy as the two Haldane mass solutions.

\begin{itemize}

\item $\bm{\nu=-2}$ (the LLL is empty). This is a {\it singlet} solution, whose free energy as a function of $\mu$ is shown 
by the green solid line in Fig.~\ref{fig_free-e_LLL_S1S2}.  This solution corresponds to the unbroken
$U_{\uparrow}(2) \times U_{\downarrow}(2)$ symmetry and has the lowest energy for $\mu<- \epsilon_{Z} $:
\begin{eqnarray}
\tilde{\Delta}_{0,\uparrow}^{\rm eff}=0, \quad & \mu_{0,\uparrow}^{\rm eff}=\mu- \epsilon_{Z} -\Delta_{0,\uparrow}, &
\quad\Delta_{0,\uparrow}\approx 0.225\varepsilon_{\ell},\\
\tilde{\Delta}_{0,\downarrow}^{\rm eff}= 0, \quad & \mu_{0,\downarrow}^{\rm eff}=\mu+ \epsilon_{Z} -\Delta_{0,\downarrow},&
\quad\Delta_{0, \downarrow}\approx 0.225\varepsilon_{\ell}.
\end {eqnarray}
In this case the LLL quasiparticle energies are
\begin{eqnarray}
\omega_{0,\uparrow} &=& -\mu+ \epsilon_{Z} +|\Delta_{0,\uparrow}| >0 ,\qquad (\times 2),\\
\omega_{0,\downarrow} &=& -\mu- \epsilon_{Z} +|\Delta_{0,\downarrow}|>0,\qquad (\times 2).
\end{eqnarray}
Since none of the LLL sublevels are occupied, this solution corresponds to a $\nu=-2$ state.

\item $\bm{\nu=-1}$ (the LLL is one-quarter filled). This is a {\it hybrid} solution, whose free energy as 
a function of $\mu$ is shown by the red dashed line in Fig.~\ref{fig_free-e_LLL_S1S2}. The symmetry in 
the corresponding state is spontaneously broken down to 
$U_{\uparrow}(2) \times U_{\downarrow}^{(K)}(1) \times U_{\downarrow}^{(K^\prime)}(1)$,
where the two latter factors describe $U(1)$ transformations at a fixed spin and a fixed
valley. At $\mu=- \epsilon_{Z} $, this solution is degenerate in energy with the solutions for the $\nu=-2$ and $\nu=0$ 
cases and is given by
\begin{eqnarray}
\tilde{\Delta}_{0,\uparrow}^{\rm eff}=0, \quad
& \mu_{0,\uparrow}^{\rm eff}=\mu- \epsilon_{Z} -\Delta_{0,\uparrow},
& \quad \Delta_{0,\uparrow}\approx 0.225\varepsilon_{\ell}, \\
\tilde{\Delta}_{0,\downarrow}^{\rm eff}\approx 0.225\varepsilon_{\ell}, \quad
& \mu_{0,\downarrow}^{\rm eff}=\mu+ \epsilon_{Z} ,
& \quad \Delta_{0, \downarrow}=0.
\end{eqnarray}
In this case the LLL quasiparticle energies are
\begin{eqnarray}
\omega_{0,\uparrow} &=& -\mu+ \epsilon_{Z} +|\Delta_{0,\uparrow}| >0 ,\qquad (\times 2),\\
\omega_{0,\downarrow} &=&  -\mu- \epsilon_{Z}   + \tilde{\Delta}_{0,\downarrow}^{\rm eff} >0,\\
\omega_{0,\downarrow} &=&  -\mu- \epsilon_{Z}  - \tilde{\Delta}_{0,\downarrow}^{\rm eff} <0.
\end{eqnarray}
Since only one LLL sublevel is occupied, this solution corresponds to a $\nu=-1$ state.

\item $\bm{\nu=0}$ (the half filled LLL; neutral point). This is a {\it singlet} solution. 
Its free energy as a function of $\mu$ is shown by the blue solid line in Fig.~\ref{fig_free-e_LLL_S1S2}.
The symmetry in the corresponding state is
$U_{\uparrow}(2) \times U_{\downarrow}(2)$, but the Zeeman splitting is dynamically
enhanced. This solution has the lowest energy for $- \epsilon_{Z} <\mu< \epsilon_{Z} $
\begin{eqnarray}
\tilde{\Delta}_{0,\uparrow}^{\rm eff}=0, \quad &
\mu_{0,\uparrow}^{\rm eff}=\mu- \epsilon_{Z} -\Delta_{0,\uparrow},
& \quad \Delta_{0,\uparrow}\approx 0.225\varepsilon_{\ell},\\
\tilde{\Delta}_{0,\downarrow}^{\rm eff}= 0, \quad & \mu_{0,\downarrow}^{\rm eff}=\mu+ \epsilon_{Z} -\Delta_{0,\downarrow},
& \quad \Delta_{0, \downarrow}\approx -0.225\varepsilon_{\ell}.
\end{eqnarray}
In this case the LLL quasiparticle energies are
\begin{eqnarray}
\omega_{0,\uparrow} &=& -\mu+ \epsilon_{Z} +|\Delta_{0,\uparrow}| >0  ,\qquad (\times 2),\\
\omega_{0,\downarrow} &=& -\mu- \epsilon_{Z} -|\Delta_{0,\downarrow}| <0,\qquad (\times 2).
\end{eqnarray}
Since two of the LLL  sublevels are occupied, this solution corresponds to a $\nu=0$ state.

\item $\bm{\nu=1}$ (the LLL is three-quarter filled). Similarly to the solution for the $\nu=-1$ state, 
this is a {\it hybrid} solution. Its free energy as a function of $\mu$ is shown by the gray dashed line 
in Fig.~\ref{fig_free-e_LLL_S1S2}.  The symmetry in the corresponding state is spontaneously broken down to
$U_{\uparrow}^{(K)}(1) \times U_{\uparrow}^{(K^\prime)}(1)\times U_{\downarrow}(2)$.
At $\mu= \epsilon_{Z} $, this solution is degenerate in energy with the solutions for the $\nu=0$ and $\nu=2$ cases
and is given by
\begin{eqnarray}
\tilde{\Delta}_{0,\uparrow}^{\rm eff}\approx0.225\varepsilon_{\ell},\quad
& \mu_{0,\uparrow}^{\rm eff}=\mu- \epsilon_{Z} , & \quad \Delta_{i,\uparrow}=0, \\
\tilde{\Delta}_{0,\downarrow}^{\rm eff}= 0,\quad  & \mu_{0,\downarrow}^{\rm eff}=\mu+ \epsilon_{Z} -\Delta_{0,\downarrow},
& \quad \Delta_{0, \downarrow}\approx-0.225\varepsilon_{\ell}.
\end{eqnarray}
In this case the LLL quasiparticle energies are
\begin{eqnarray}
\omega_{0,\uparrow} &=& -\mu+ \epsilon_{Z}  + \tilde{\Delta}_{0,\uparrow}^{\rm eff}>0,\\
\omega_{0,\uparrow} &=& -\mu+ \epsilon_{Z}  - \tilde{\Delta}_{0,\uparrow}^{\rm eff} <0,\\
\omega_{0,\downarrow} &=& -\mu- \epsilon_{Z} -|\Delta_{0,\downarrow}|<0,\qquad (\times 2).
\end{eqnarray}
Since three of the LLL  sublevels are occupied, this solution corresponds to a $\nu=1$ state.

\item $\bm{\nu=2}$ (the LLL is filled). This is another {\it singlet} solution. 
Its free energy as a function of $\mu$ is shown by the light-brown solid line in Fig.~\ref{fig_free-e_LLL_S1S2}. 
This solution corresponds to the unbroken $U_{\uparrow}(2) \times U_{\downarrow}(2)$ symmetry and has 
the lowest energy for $\mu> \epsilon_{Z} $:
\begin{eqnarray}
\tilde{\Delta}_{0,\uparrow}^{\rm eff}=0, \quad
& \mu_{0,\uparrow}^{\rm eff}=\mu- \epsilon_{Z} -\Delta_{0,\uparrow},
& \quad  \Delta_{0,\uparrow}\approx -0.225\varepsilon_{\ell},\\
\tilde{\Delta}_{0,\downarrow}^{\rm eff}= 0, \quad
& \mu_{0,\downarrow}^{\rm eff}=\mu+ \epsilon_{Z} -\Delta_{0,\downarrow},
& \quad  \Delta_{0, \downarrow}\approx -0.225\varepsilon_{\ell}.
\end{eqnarray}
In this case the LLL quasiparticle energies are
\begin{eqnarray}
\omega_{0,\uparrow} &=& -\mu+ \epsilon_{Z} -|\Delta_{0,\uparrow}| <0 ,\qquad (\times 2),\\
\omega_{0,\downarrow} &=& -\mu- \epsilon_{Z} -|\Delta_{0,\downarrow}| <0,\qquad (\times 2).
\end{eqnarray}
Since all LLL  sublevels are occupied, this solution corresponds to a $\nu=2$ state.

\end{itemize}

\begin{figure}[t]
\begin{center}
\includegraphics[width=.48\textwidth]{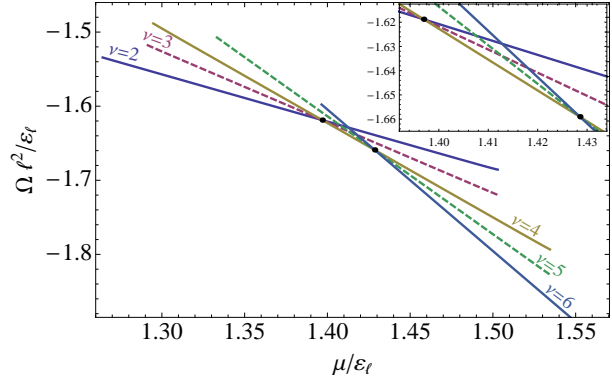}
\caption{(Color online) Numerical results for the free energies of several lowest energy solutions, 
realized when the Fermi energy is near $n=1$ LL.}
\label{fig_free-e_LL1}
\end{center}
\end{figure}

{\it (iii) States with $n=1$ LL filling factors.} Similarly, we can also obtain the lowest 
energy solutions in the $n=1$ LL, when this level is partially or completely filled 
(see Fig.~\ref{fig_free-e_LL1}). In this case, we were also able to identify several 
dozen non-equivalent branches of solutions. The main features of the corresponding 
solutions are described below. Note that filling of the
$n = 1$ LL leads to changing the dispersion relations also in other LLs.
Therefore, although only parameters $\tilde{\Delta}_1$, $\Delta_1$, $\mu_1$,
and $f_1$ determine the QH plateaus at the $n=1$ LL, we write down their values for
two neighbor levels, the LLL and the $n=2$ LL. (Recall that, according to Eqs.~(\ref{F-n})
and (\ref{Sigma-n}), $\tilde{\Delta}_{n}$, $\Delta_{n}$, $\mu_{n}$, $f_{n}$ are functions
of the LL index $n$.)  Physically, the information concerning the gaps in other 
LLs is relevant for experiments connected with transitions between Landau
levels \cite{Jiang:2007az}.

\begin{itemize}

\item $\bm{\nu=3}$ (the $n =1$ LL is one-quarter filled). Hybrid solution, which is valid for
$ \mu \simeq \sqrt{2}\varepsilon_\ell- \epsilon_{Z} $.
The symmetry in this state is spontaneously broken down to 
$U_{\uparrow}(2) \times U_{\downarrow}^{(K)}(1) \times U_{\downarrow}^{(K^\prime)}(1)$:
\begin{eqnarray}
&& \tilde{\Delta}_{0,\uparrow}^{\rm eff}  =0 , \quad
\mu_{0,\uparrow}^{\rm eff} =\mu_{\uparrow}-\Delta_{0,\uparrow} , \quad \Delta_{0,\uparrow} = -0.225\varepsilon_\ell , \\
&& \tilde{\Delta}_{0,\downarrow}^{\rm eff}  =0.052 \varepsilon_\ell , \quad
\mu_{0,\downarrow}^{\rm eff} =\mu_{\downarrow}-\Delta_{0,\downarrow} , \quad \Delta_{0,\downarrow} =-0.277\varepsilon_\ell , \\
&& \tilde{\Delta}_{1,\uparrow}  =0 ,\quad
\mu_{1,\uparrow} =\mu_{\uparrow}+0.053\varepsilon_\ell  ,\quad
\Delta_{1,\uparrow} =-0.067\varepsilon_\ell,\quad
f_{1,\uparrow} = 1.142,\\
&& \tilde{\Delta}_{1,\downarrow}   =-0.018 \varepsilon_\ell , \quad
\mu_{1,\downarrow} =\mu_{\downarrow}+0.148 \varepsilon_\ell,\quad
\Delta_{1,\downarrow} =-0.049\varepsilon_\ell,\quad
f_{1,\downarrow} = 1.103,\\
&& \tilde{\Delta}_{2,\uparrow}  =0 ,\quad
\mu_{2,\uparrow} =\mu_{\uparrow}+0.040\varepsilon_\ell  ,\quad
\Delta_{2,\uparrow} =-0.051\varepsilon_\ell,\quad
f_{2,\uparrow} = 1.111,\\
&& \tilde{\Delta}_{2,\downarrow}   =-0.006 \varepsilon_\ell , \quad
\mu_{2,\downarrow} =\mu_{\downarrow}+0.091 \varepsilon_\ell,\quad
\Delta_{2,\downarrow} =-0.045\varepsilon_\ell,\quad
f_{2,\downarrow} = 1.102.
\end{eqnarray}

\item $\bm{\nu=4}$ (the $n=1$ LL is half filled).
Singlet solution, which is valid for $\sqrt{2}\varepsilon_\ell- \epsilon_{Z} \lesssim \mu\lesssim \sqrt{2}\varepsilon_\ell+ \epsilon_{Z} $.
While formally the symmetry of this state is the same as in the action, $U_{\uparrow}(2) \times U_{\downarrow}(2)$, 
it is characterized by a dynamically enhanced Zeeman splitting:
\begin{eqnarray}
&& \tilde{\Delta}_{0,\uparrow}^{\rm eff}  =0 , \quad
\mu_{0,\uparrow}^{\rm eff} =\mu_{\uparrow}-\Delta_{0,\uparrow} , \quad \Delta_{0,\uparrow} = -0.225\varepsilon_\ell , \\
&& \tilde{\Delta}_{0,\downarrow}^{\rm eff}  =0 , \quad
\mu_{0,\downarrow}^{\rm eff} =\mu_{\downarrow}-\Delta_{0,\downarrow} , \quad \Delta_{0,\downarrow} = -0.328\varepsilon_\ell , \\
&& \tilde{\Delta}_{1,\uparrow}  =0 ,\quad
\mu_{1,\uparrow} =\mu_{\uparrow}+0.053\varepsilon_\ell  ,\quad
\Delta_{1,\uparrow} =-0.067\varepsilon_\ell,\quad
f_{1,\uparrow} = 1.142,\\
&& \tilde{\Delta}_{1,\downarrow}   =0, \quad
\mu_{1,\downarrow} =\mu_{\downarrow}+0.244 \varepsilon_\ell,\quad
\Delta_{1,\downarrow} =-0.031\varepsilon_\ell,\quad
f_{1,\downarrow} = 1.065\\
&& \tilde{\Delta}_{2,\uparrow}  =0 ,\quad
\mu_{2,\uparrow} =\mu_{\uparrow}+0.040\varepsilon_\ell  ,\quad
\Delta_{2,\uparrow} =-0.051\varepsilon_\ell,\quad
f_{2,\uparrow} = 1.111,\\
&& \tilde{\Delta}_{2,\downarrow}   =0, \quad
\mu_{2,\downarrow} =\mu_{\downarrow}+0.142 \varepsilon_\ell,\quad
\Delta_{2,\downarrow} =-0.039\varepsilon_\ell,\quad
f_{2,\downarrow} = 1.092.
\end{eqnarray}

\item $\bm{\nu=5}$ (the $n=1$ LL is three-quarter filled).
Hybrid solution, which is valid for $ \mu \simeq \sqrt{2}\varepsilon_\ell+ \epsilon_{Z} $.
The symmetry in this state is spontaneously broken down to
$U_{\uparrow}^{(K)}(1) \times U_{\uparrow}^{(K^\prime)}(1)\times U_{\downarrow}(2)$:
\begin{eqnarray}
&& \tilde{\Delta}_{0,\uparrow}^{\rm eff}  =0.052 \varepsilon_\ell , \quad
\mu_{0,\uparrow}^{\rm eff} =\mu_{\uparrow}-\Delta_{0,\uparrow} , \quad \Delta_{0,\uparrow} = -0.277\varepsilon_\ell , \\
&& \tilde{\Delta}_{0,\downarrow}^{\rm eff}  =0 , \quad
\mu_{0,\downarrow}^{\rm eff} =\mu_{\downarrow}-\Delta_{0,\downarrow} , \quad \Delta_{0,\downarrow} =-0.328\varepsilon_\ell , \\
&& \tilde{\Delta}_{1,\uparrow}  =-0.018 \varepsilon_\ell  ,\quad
\mu_{1,\uparrow} =\mu_{\uparrow}+0.148\varepsilon_\ell  ,\quad
\Delta_{1,\uparrow} =-0.049\varepsilon_\ell,\quad
f_{1,\uparrow} = 1.103,\\
&& \tilde{\Delta}_{1,\downarrow}   =0, \quad
\mu_{1,\downarrow} =\mu_{\downarrow}+0.244 \varepsilon_\ell,\quad
\Delta_{1,\downarrow} =-0.031\varepsilon_\ell,\quad
f_{1,\downarrow} = 1.065,\\
&& \tilde{\Delta}_{2,\uparrow}  =-0.006 \varepsilon_\ell  ,\quad
\mu_{2,\uparrow} =\mu_{\uparrow}+0.091\varepsilon_\ell  ,\quad
\Delta_{2,\uparrow} =-0.045\varepsilon_\ell,\quad
f_{2,\uparrow} = 1.102,\\
&& \tilde{\Delta}_{2,\downarrow}   =0, \quad
\mu_{2,\downarrow} =\mu_{\downarrow}+0.142 \varepsilon_\ell,\quad
\Delta_{2,\downarrow} =-0.039 \varepsilon_\ell,\quad
f_{2,\downarrow} = 1.092.
\end{eqnarray}

\item $\bm{\nu=6}$ (the $n=1$ LL is filled).
Singlet solution with the unbroken $U_{\uparrow}(2) \times U_{\downarrow}(2)$  symmetry, which is 
valid for $\mu\gtrsim \sqrt{2}\varepsilon_\ell+ \epsilon_{Z}$,
\begin{eqnarray}
&& \tilde{\Delta}_{0,\uparrow}^{\rm eff}  =0 , \quad
\mu_{0,\uparrow}^{\rm eff} =\mu_{\uparrow}-\Delta_{0,\uparrow} , \quad \Delta_{0,\uparrow} = -0.328\varepsilon_\ell , \\
&& \tilde{\Delta}_{0,\downarrow}^{\rm eff}  =0 , \quad
\mu_{0,\downarrow}^{\rm eff} =\mu_{\downarrow}-\Delta_{0,\downarrow} , \quad \Delta_{0,\downarrow} = -0.328\varepsilon_\ell , \\
&& \tilde{\Delta}_{1,\uparrow}  =0 ,\quad
\mu_{1,\uparrow} =\mu_{\uparrow}+0.244\varepsilon_\ell  ,\quad
\Delta_{1,\uparrow} =-0.031\varepsilon_\ell,\quad
f_{1,\uparrow} = 1.065,\\
&& \tilde{\Delta}_{1,\downarrow}   =0, \quad
\mu_{1,\downarrow} =\mu_{\downarrow}+0.244 \varepsilon_\ell,\quad
\Delta_{1,\downarrow} =-0.031\varepsilon_\ell,\quad
f_{1,\downarrow} = 1.065,\\
&& \tilde{\Delta}_{2,\uparrow}  =0 ,\quad
\mu_{2,\uparrow} =\mu_{\uparrow}+0.142\varepsilon_\ell  ,\quad
\Delta_{2,\uparrow} =-0.039\varepsilon_\ell,\quad
f_{2,\uparrow} = 1.092,\\
&& \tilde{\Delta}_{2,\downarrow}   =0, \quad
\mu_{2,\downarrow} =\mu_{\downarrow}+0.142 \varepsilon_\ell,\quad
\Delta_{2,\downarrow} =-0.039\varepsilon_\ell,\quad
f_{2,\downarrow} = 1.092.
\end{eqnarray}

\end{itemize}

It should be emphasized that, while all the results for the QH states, associated with filling the 
$n=0$ and $n=1$ LLs, are qualitatively similar to those obtained in 
Section~\ref{Sec:MagCatGrapheneLocal}. However, the Coulomb long-range interaction 
makes all gaps and other dynamical parameters functions of the LL index $n$.

{\it (iv) States with filling factors $\nu=4(k+1/2)$, where $k$ is an integer.} As the above analysis of the 
special cases of quantum Hall states connected with a partial filling of the lowest and first Landau 
levels suggest, {\it all} integer filling factors $\nu$ are in principle possible. [Note that certain fractional 
filling factors are also possible \cite{2009Natur.462..192D}, but we do not discuss such a possibility in the current framework
of the gap equation.] The simplest quantum Hall states are those with filling factors $\nu=4(k+1/2)$.
Such states do not require any symmetry-breaking order parameters and can be easily observed 
even in weak magnetic fields \cite{2005Natur.438..197N,Zhang:2005zz}. The effect of the Coulomb 
interaction in the corresponding ground states can be seen via the renormalization of the wave-function
as well as the dynamical corrections to the symmetry preserving parameters $\mu_n$ and $\Delta_n$. 

\begin{figure}[t]
\begin{center}
\includegraphics[width=.45\textwidth]{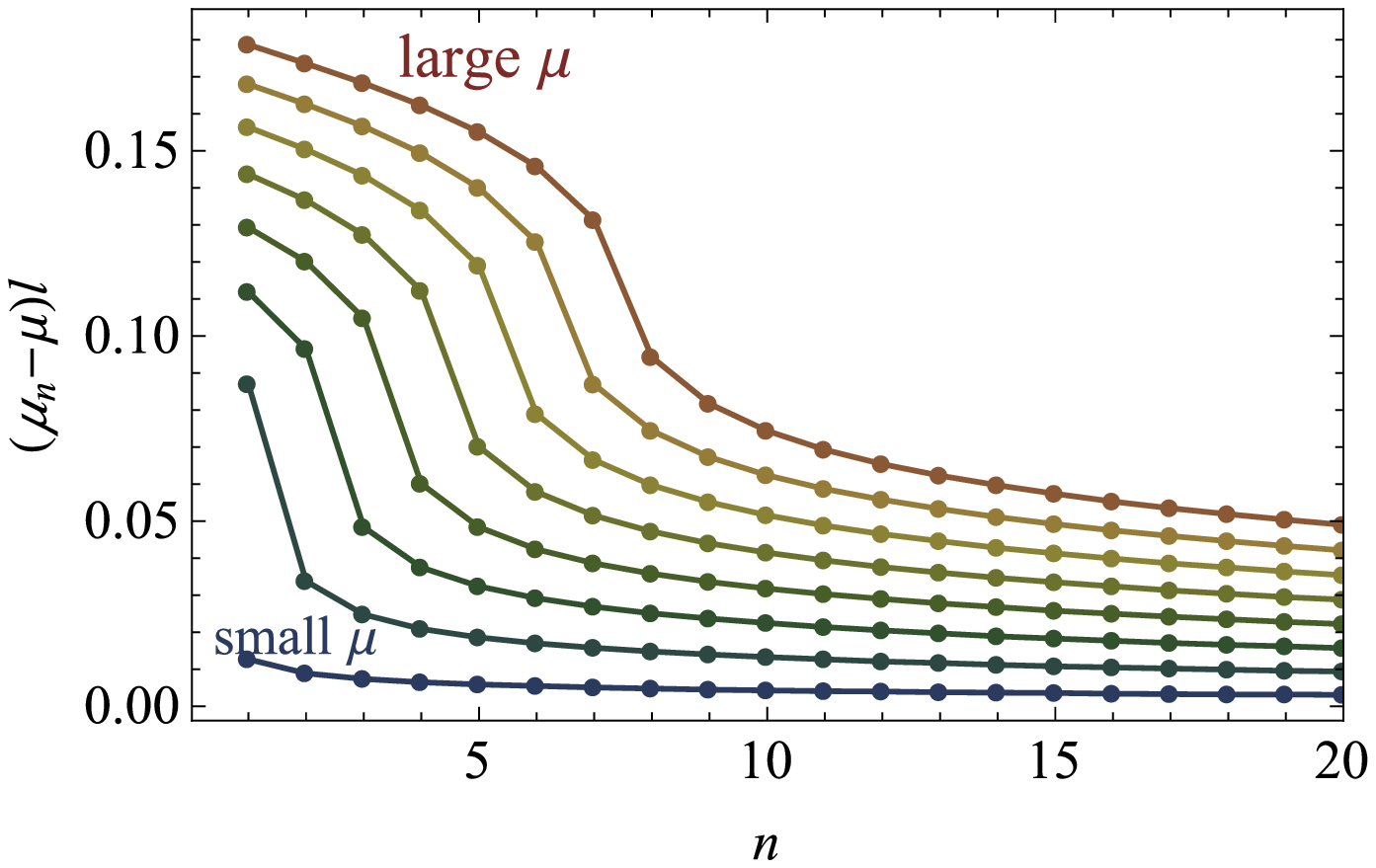}
\hspace{.07\textwidth}
\includegraphics[width=.45\textwidth]{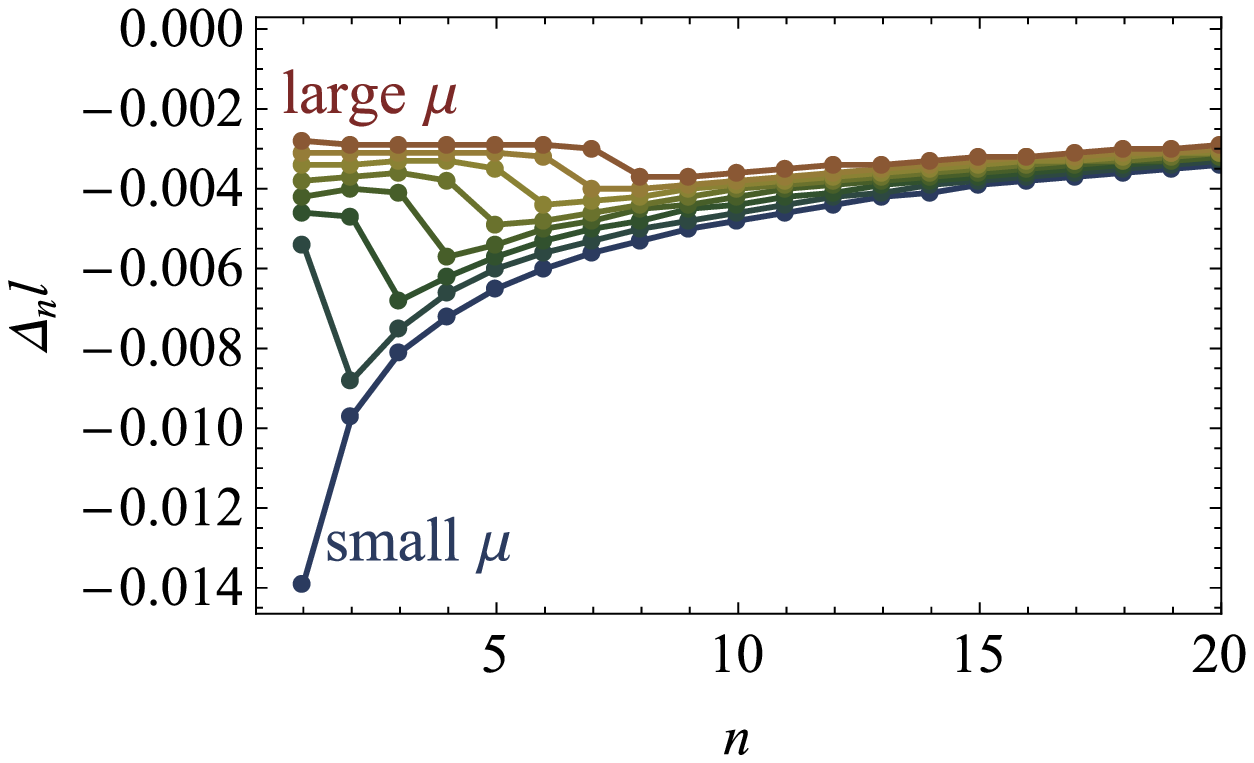}
\caption{(Color online) Numerical results for the parameters $\mu_n$ (left panel) 
and $\Delta_n$ (right panel) as functions of the Landau level index $n$ in the model 
with Coulomb interaction with screening. The corresponding results for the wave-function
renormalization are shown in the right panel of Fig.~\ref{fig.fn-vs-n}.}
\label{fig_Mu_Delta_n_run}
\end{center}
\end{figure}

The coupled set of the corresponding couple set of gap equations follows from the 
Eqs.~(\ref{gap-eq-mu_n-0}) through (\ref{gap-eq-g_n}), in which symmetry-breaking 
parameters are set to be vanishing (i.e., $\tilde{\mu}_n=0$ and $\tilde{\Delta}_n=0$). 
The results of the numerical solutions of the corresponding set of gap equations is
presented in Fig.~\ref{fig_Mu_Delta_n_run}, where we show the dependence of the 
parameters $\mu_n$ (left panel) and $\Delta_n$ (right panel) as functions of the 
Landau level index $n$. The results in Fig.~\ref{fig_Mu_Delta_n_run} correspond 
to a select set of eight different values of the (thermodynamic) chemical potential 
$\mu$ in the gaps between Landau levels, insuring filling of all (nearly degenerate) 
sublevels of the first $k$ Landau level. Lines (points) of different color correspond 
to different $\mu$, describing the first eight states with filling factors $\nu=4(k+1/2)$ 
with integer $k=0,\ldots,7$. [The corresponding results for the wave-function renormalization
were shown in Fig.~\ref{fig.fn-vs-n}.] Note that, in this analysis, we took (static) screening 
effects of the Coulomb interaction into account. The main effect of screening appears 
to be a suppression of the interaction parameters $\kappa_{m,n}^{(\rho)}$, defined in 
Eq.~(\ref{kappa_mn^rho}), by about a factor of 2. 

As we see from all the examples of solutions to the gap equations in the model 
with the long-range Coulomb interaction, the main qualitative feature of the 
corresponding dynamics is a strong dependence of dynamical parameters on the
Landau level index. This is in contrast to the case of the model with a point-like 
interaction discussed in detail in Section~\ref{Sec:MagCatGrapheneLocal}.

\subsubsection{More about quantum Hall effect in graphene}
\label{Sec:GrapheneRemarks}

The model analysis of the dynamics responsible for the quantum Hall effect 
in graphene in a strong magnetic field is quite instructive. From symmetry 
point of view, the transition between different quantum Hall states appear to
be quantum phase transitions, in which the approximate flavor symmetry 
group is broken down to various subgroups. 

The nature of the quantum states with filling factors $\nu=4(k+1/2)$, where $k$ 
is an integer, is simplest because no breaking of the flavor symmetry is needed. 
Of course, this is also the main reason why such states were the first to be discovered 
in experiment \cite{2005Natur.438..197N,Zhang:2005zz}. They are observed 
even at rather weak magnetic fields. The situation with other quantum Hall states
is more complicated. While some small flavor symmetry breaking terms can be 
naturally introduced in the low-energy Hamiltonian, generically they will not be
sufficient to explain the well pronounced properties of states with such filling 
factors as $\nu=0, \pm 1,\pm 3,\pm 4$. They require a substantial enhancement 
of symmetry breaking by a dynamical mechanism. 

In this context, it was rather natural to suggest that the generalized magnetic 
catalysis (i.e., the magnetic catalysis with an admixture of the quantum Hall ferromagnetism) 
could describe the quantum Hall effect in graphene in a strong magnetic field.  
Such a philosophy has a lot of merit even though the underlying dynamics in 
graphene is not exactly the same as the magnetic catalysis in relativistic
quantum field theory, as reviewed in Sections~\ref{sec:MagneticCatalysis} and \ref{sec:MagCatGauge}. 
The analogy is further backed up by the key elements in the dynamics. Just like in the magnetic 
catalysis, the dimensional reduction and finite density of states at the Dirac (neutral) 
point play the crucial role in triggering spontaneous symmetry breaking and 
producing energy gaps in the spectrum in the quantum Hall effect in graphene
even at the weakest repulsive interactions between electrons.

There are, however, also important differences.
Because of numerous complications, that are absent in true relativistic systems, 
the underlying dynamics of graphene appears to be much richer. A few small 
complications were already encountered in the simplified model analysis in 
Sections~\ref{Sec:MagCatGrapheneLocal} and \ref{Sec:MagCatGrapheneCoulomb}. 
Those were connected with the Zeeman term that breaks flavor symmetry as well 
as the fermion interaction that breaks Lorentz symmetry even in the absence of the
external magnetic field. 

One should recall, however, that many important physics effects were neglected 
in the framework of the low-energy Dirac theory used there. One of them is a 
realm of microscopic phenomena (e.g., intrinsic ripples of graphene, phonons, 
impurities due to lattice irregularities and foreign atoms, etc.), responsible for 
the quasiparticle width. Strictly speaking, most of the corresponding phenomena 
are beyond the Dirac theory framework. Taking into account that the language of
Dirac fermions is valid only in the low-energy region, there exist a number of 
higher-order terms that modify the low-energy theory. An important class of 
effects of this type are the generic symmetry breaking short-range interaction
terms, which appear because of the lattice in graphene \cite{PhysRevB.76.195415}. 
We will discuss the effects of short-range interactions in the next subsection.

\subsection{Problem of vacuum alignment and phase diagram of graphene in a magnetic field}
\label{Sec:GrapheneVacuumAlignment}

As mentioned several times before, the Zeeman term is one of the obvious terms 
in the effective Hamiltonian that breaks explicitly the flavor symmetry. It is not 
the only symmetry-breaking term and perhaps even not the most important in
some cases. It appears that there are numerous small, but nonzero four-fermion 
and six-fermion short-range interaction terms \cite{PhysRevB.76.195415} that 
also break flavor symmetry. Some of them are expected to get a rather strong 
renormalization and become sufficiently large in the low-energy region of the
theory to compete with Zeeman term \cite{KharitonovPRB85.155439}. 
Theoretically, however, it is rather hard to predict the outcome of the 
corresponding competitions. Even small uncertainties in the strengths of 
short-range interactions may lead to a rather large variability in the final
predictions for the ground state. This is a typical problem of the vacuum 
alignment when there exists a large approximate symmetry and many 
symmetry breaking terms that prefer different states in a multidimensional 
landscape of nearly degenerate solutions.

In this context, even a rather flexible set of order parameters and a large set 
of ground states introduced in Sections~\ref{Sec:MagCatGrapheneLocal}
and \ref{Sec:MagCatGrapheneCoulomb} is still not able to cover all possible 
solutions with broken symmetry. Therefore, here we will introduce a few 
additional ones, e.g., motivated by some phenomenological models and 
the arguments of renormalization group. We will not make an attempt, however, 
to reanalyze the gap equations with the additional variational parameters. 
To large extend, the solutions and the ground state properties are controlled 
by symmetries and, thus, we will concentrate on the discussion of their symmetry 
properties. 

Also, instead of trying to make definite theoretical predictions about the 
ground states, we will discuss a range of promising possibilities that are 
motivated by general arguments. In the end, the judgment about the true 
ground state should be made by comparing specific predictions for each state 
with experimental observations.

\subsubsection{Antiferromagnetic, Canted Antiferromagnetic and Kekule Distortion phases}
\label{AF-CAF-KD}

As discussed in Sections~\ref{Sec:MagCatGrapheneLocal} and \ref{Sec:MagCatGrapheneCoulomb}, 
there are quite a number of different phases in graphene, see Fig.~\ref{fig.V.eff.T1}. In fact, 
the reality is even more interesting. There are additional, more sophisticated phases, if the 
models analyzed in Sections~\ref{Sec:MagCatGrapheneLocal} and \ref{Sec:MagCatGrapheneCoulomb} 
are properly modified. Let us describe three new phases, which, it seems, can be realized in the 
quantum Hall effect in graphene. For clarity, only the case of the Dirac (neutral) point with 
$\nu = 0$ will be considered.

In the model with a short-range Coulomb interaction in Section~\ref{Sec:MagCatGrapheneLocal}, 
we argued that there are two almost equally good candidates for the description of the $\nu = 0$ 
state: (i) the $S1$ phase with a singlet order parameter and (ii) the $T$ phase with a triplet order 
parameter. [The classifications of the order parameters is given according to their representations 
with respect to the flavor subgroups $\mathrm{SU}(2)_{s}$ with $s=\uparrow,\downarrow$.] 
From the detailed structure of the two phases, it is clear that the $S1$ phase is {\it ferromagnetic} 
(F) and the $T$ phase is a {\it charge-density-wave} (CDW). Because of the Zeeman term, 
the F-phase was more stable than the CDW one. When additional symmetry breaking short-range 
interactions included, it may be expected that these two phases are not the only ones possible. 

In fact, other interesting phases were also suggested in the literature. Among the existing 
ideas, there are proposals for the following three phases:  the {\it antiferromagnetic} (AF) 
phase \cite{PhysRevB.75.165411,Gorbar:2008hu,PhysRevB.80.235417}, the {\it canted antiferromagnetic} (CAF)
\cite{2007PhRvB..76h5432H,KharitonovPRB85.155439,PhysRevB.86.075450,PhysRevB.90.201409}, 
and the {\it Kekule distortion} (KD) one \cite{PhysRevLett.103.216801,PhysRevB.81.075427}.
In principle, each of them can also be a potential candidate for the description of the $\nu = 0$ 
state.

Let us recall that the F-phase is characterized by the following order parameters:
$\langle \bar\Psi\,\sigma_3 \gamma^3\gamma^5 \Psi \rangle$ and 
$\langle\bar\Psi\,\sigma_3 \gamma^0 \Psi \rangle$, where $\sigma_3$ is the Pauli matrix in the 
spin space. Strictly speaking, the actual ferromagnetic order parameter is the magnetization,
given by $\langle\bar\Psi\,\sigma_3 \gamma^0 \Psi \rangle$. As we found in Section~\ref{Sec:MagCatGrapheneLocal}, 
however, the self-consistent solution requires that the other ground state expectation value be
also present. This is understandable because the additional order parameter does not break any
additional symmetries. When the F-phase is realized, the low-energy effective 
Hamiltonian contains the following additional terms: 
\begin{equation}
H^{\rm F} =\int d^2r\,\bar\Psi\,\sigma_3 (\Delta_3\gamma^3\gamma^5-\mu_3\gamma^0)\Psi .
\label{Hgraphene-Fphase}
\end{equation}
In terms of the spin-dependent parameters $\Delta_s$ and $\mu_s$, used in 
Section~\ref{Sec:MagCatGrapheneLocal}, the F-phase is characterized by 
$\Delta_{+}=-\Delta_{-}=\Delta_3$ and $\mu_{+}=-\mu_{-}=\mu_3$. As we 
see from Fig.~\ref{fig.order_pars_mu0_vs_T}, this is exactly the configuration 
obtained for the singlet $S1$ solution. 

The order parameters of the CDW-phase have the following form:
$\langle \bar\Psi \Psi \rangle$ and $\langle\bar\Psi\, i \gamma^1\gamma^2 \Psi \rangle$. 
Among the two expectation values, $\langle \bar\Psi \Psi \rangle$ is directly connected 
with the magnetic catalysis and is the actual order parameter of the charge-density-wave.
It describes a configuration with the imbalance of the charge densities on the two 
sublattices ($A$ and $B$) in the coordinate space. The other expectation value is 
required by the gap equation. As one can check, it describes opposite spin polarizations 
in the valleys $K$ and $K^\prime$. In the CDW-phase, the effective Hamiltonian has the 
following additional contributions \cite{Gorbar:2008hu}: 
\begin{equation}
H^{\rm CDW}=\int d^2r\,\bar\Psi (\tilde{\Delta}_{\rm CDW}- 
\tilde{\mu}_{\rm CDW}\, i \gamma^1\gamma^2)\Psi .
\label{Hgraphene-CDWphase}
\end{equation}
In terms of the spin-dependent parameters $\tilde{\Delta}_s$ and $\tilde{\mu}_s$, used in 
Section~\ref{Sec:MagCatGrapheneLocal}, the CDW-phase is described by a solution with
$\tilde{\Delta}_{+}=\tilde{\Delta}_{-}=\tilde{\Delta}_{\rm CDW}$ and 
$\tilde{\mu}_{+}=\tilde{\mu}_{-}=\tilde{\mu}_{\rm CDW}$. This is indeed the triplet 
solution in the $T$ phase in Section~\ref{Sec:MagCatGrapheneLocal}.

Let us now introduce the three additional phases mentioned above. The order parameters 
of the AF-phase are similar to those in the CDW-phase, but contain an additional spin 
matrix $\sigma^3$: $\langle \bar\Psi \sigma^3 \Psi \rangle$ and 
$\langle\bar\Psi\,\sigma^3  i \gamma^1\gamma^2 \Psi \rangle$. 
The corresponding contributions to the effective Hamiltonian are 
\begin{equation}
H^{\rm AF}=\int d^2r\,\bar\Psi \sigma^3 (\tilde{\Delta}_{\rm AF} -\tilde{\mu}_{\rm AF} \,i  \gamma^1\gamma^2)\Psi .
\label{Hgraphene-AFphase}
\end{equation}
In terms of the spin-dependent parameters $\tilde{\Delta}_s$ and $\tilde{\mu}_s$, used in 
Section~\ref{Sec:MagCatGrapheneLocal}, the AF-phase is described by a solution with
$\tilde{\Delta}_{+}=-\tilde{\Delta}_{-}=\tilde{\Delta}_{\rm AF}$ and 
$\tilde{\mu}_{+}=-\tilde{\mu}_{-}=\tilde{\mu}_{\rm AF}$. It appears that, in the model with 
a short-range Coulomb interaction but without any symmetry breaking short-range terms,
studied in Section~\ref{Sec:MagCatGrapheneLocal}, the free energy density of the AF 
phase is degenerate with that of CDW-phases. Moreover they are not physically distinguishable
in that model (for details, see Appendix~B in Ref.~\cite{Gorbar:2008hu}).

In the CAF-phase, the order parameters are given by a mixture of those in 
the F-phase, see Eq.~(\ref{Hgraphene-Fphase}), and the AF-phase ones, see 
Eq.~(\ref{Hgraphene-AFphase}), but with the spin matrix $\sigma_3$ of the latter
replaced by $\sigma_1$. Then, the effective Hamiltonian will have the following extra terms: 
\begin{equation}
H^{\rm CAF}=H^{\rm F}+\int d^2r\,\bar\Psi\,\sigma_1(\tilde{\Delta}_{\rm CAF}
-\tilde{\mu}_{\rm CAF}\,i \gamma^1\gamma^2)\Psi .
\label{Hgraphene-CAFphase}
\end{equation}
From the physics viewpoint, the CAF-phase is a superposition of the F-order parameter 
along the direction of the magnetic field and the AF-order parameter in the plane perpendicular 
to the magnetic field \cite{2007PhRvB..76h5432H,KharitonovPRB85.155439,PhysRevB.86.075450}.

The KD phase is characterized by a spontaneous formation of a (quasi-long-range)
periodic modulation of the nearest-neighbor hopping amplitude in graphene with the wave 
vector $\mathbf{K}-\mathbf{K}^\prime$ \cite{PhysRevB.81.075427}. Such a dimerization 
pattern of bonds can be also called a ``bond-density-wave" \cite{PhysRevLett.103.216801}.
In the KD phase, the effective Hamiltonian will have the following extra terms: 
\begin{equation}
H^{\rm KD}= \int d^2r\,\bar\Psi \bigl[-(i\gamma^5\cos\theta+\gamma^3\sin\theta)\Delta_{\rm KD}
+\gamma^0(i\gamma^3\cos\theta+\gamma^5\sin\theta)\mu_{\rm KD}\bigr]\Psi .
\label{Hgraphene-KDphase}
\end{equation}
It is the electron-phonon interaction that tends to favor the KD phase. The formation of the 
KD phase is accompanied by an opening of a gap in the spectrum just like in the CDW phase.
It is argued, however, that the low-energy charged excitations in the KD phase are vortices 
and antivortices \cite{PhysRevLett.103.216801,PhysRevB.81.075427}.

\subsubsection{Phase diagram}
\label{Sec:GraphenePhaseDiagram}

As mentioned earlier, there are many nonzero short-range interaction terms that 
break flavor symmetry \cite{PhysRevB.76.195415}. Following the classification of 
Ref.~\cite{PhysRevB.76.195415,KharitonovPRB85.155439}, but reformulating it 
in our relativistic notation, we can write down the most general form of the 
spin-symmetric electron-electron interactions as follows: 
\begin{eqnarray}
H_{\rm e-e}
&=&\frac{g_{\perp\perp}}{2}\int d^2r\sum_{s=\pm} \left(
\bigl[\bar{\Psi}_{s} \gamma^1\gamma^5 \Psi_{s}\bigr]^2
+\bigl[\bar{\Psi}_{s} \gamma^2\gamma^5 \Psi_{s}\bigr]^2
+\bigl[\bar{\Psi}_{s} i\gamma^1\gamma^3\Psi_{s}\bigr]^2
+\bigl[\bar{\Psi}_{s} i\gamma^2\gamma^3\Psi_{s}\bigr]^2
\right)\nonumber\\
&+& \frac{g_{\perp z}}{2}\int d^2r\sum_{s=\pm} \left(
\bigl[\bar{\Psi}_{s} \gamma^3\Psi_{s}\bigr]^2
+\bigl[\bar{\Psi}_{s} i \gamma^5\Psi_{s}\bigr]^2
\right)
\nonumber\\
&+& \frac{g_{0 \perp}}{2}\int d^2r\sum_{s=\pm} \left(
\bigl[\bar{\Psi}_{s} \gamma^1\Psi_{s}\bigr]^2
+\bigl[\bar{\Psi}_{s} \gamma^2\Psi_{s}\bigr]^2
\right)
+\frac{g_{0z}}{2}\int d^2r\sum_{s=\pm} 
\bigl[\bar{\Psi}_{s}\gamma^3\gamma^5 \Psi_{s}\bigr]^2
\nonumber\\
&+&\frac{g_{z \perp}}{2}\int d^2r\sum_{s=\pm} \left(
\bigl[\bar{\Psi}_{s}i\gamma^0\gamma^1\Psi_{s}\bigr]^2
+\bigl[\bar{\Psi}_{s}i\gamma^0\gamma^2\Psi_{s}\bigr]^2
\right)+\frac{g_{zz}}{2}\int d^2r\sum_{s=\pm} 
\bigl[\bar{\Psi}_{s} \Psi_{s}\bigr]^2
\nonumber\\
&+&\frac{g_{\perp 0}}{2}\int d^2r\sum_{s=\pm} \left(
\bigl[\bar{\Psi}_{s} \gamma^0\gamma^5\Psi_{s}\bigr]^2
+\bigl[\bar{\Psi}_{s} i \gamma^0\gamma^3\Psi_{s}\bigr]^2
\right)
+\frac{g_{z0}}{2}\int d^2r\sum_{s=\pm} 
\bigl[\bar{\Psi}_{s} i\gamma^1 \gamma^2 \Psi_{s}\bigr]^2,
\label{H4_X-more}  
\end{eqnarray}
where $g_{\alpha\beta}$ are the eight independent coupling constants identified in
Ref.~\cite{PhysRevB.76.195415,KharitonovPRB85.155439} by using the symmetry 
of the graphene honeycomb lattice. By performing a systematic analysis of the  
energy arising from the short-range interactions, the author of 
Ref.~\cite{KharitonovPRB85.155439} showed that, in the case of the $\nu=0$ 
state, the ground state energy is fully characterized by the Zeeman energy $\epsilon_Z$ 
together with the following two anisotropy energies:
\begin{equation}
u_\perp \equiv \frac{g_{\perp0}+g_{\perp z}}{2\pi l^2}, \qquad
u_z \equiv \frac{g_{z0}+g_{zz}}{2\pi l^2}.
\label{bare_ee_anisotropy_energies}
\end{equation}
Strictly speaking, the energy $u_\perp$ also contains a contribution from the electron-phonon 
interaction, which we omitted here \cite{KharitonovPRB85.155439}.

\begin{figure}[t]
\begin{center}
\includegraphics[width=.49\textwidth]{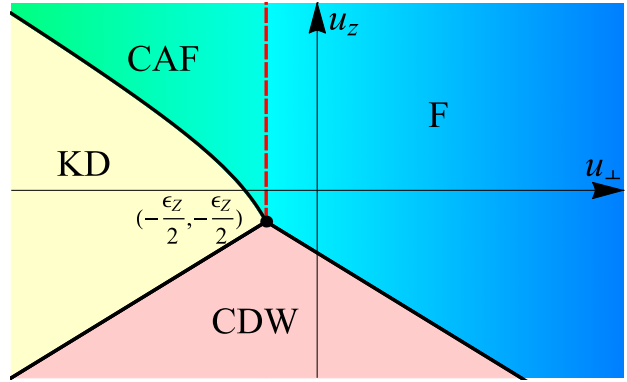}
\caption{(Color online) Phase diagram for the $\nu=0$ ground state of monolayer graphene 
in the plane of two short-range coupling constants that compete with the Zeeman term, based 
on the results from Ref.~\cite{KharitonovPRB85.155439}.}
\label{fig_graphene_PD}
\end{center}
\end{figure}

In the LLL approximation, the energies of the four candidate phases for the $\nu=0$ ground 
state take the following form \cite{KharitonovPRB85.155439}:
\begin{equation}
{\cal E}^{\rm CDW} = u_z , \qquad
{\cal E}^{\rm KD} = u_\perp , \qquad
{\cal E}^{\rm CAF} = -u_z -\frac{\epsilon_Z^2}{2|u_\perp|},  \qquad
{\cal E}^{\rm F} =  -2u_\perp-u_z-2 \epsilon_Z
\end{equation}
[Note that the usual AF-phase with antiparallel spins has the energy ${\cal E}^{\rm AF} = -u_z$ and, 
thus, is always less favored than the CAF-phase.]
Depending on the specific values of the Zeeman energy $\epsilon_Z$ and two anisotropy energies, 
$u_\perp$ and $u_z$, we determine which of the ground states are realized. The result of the
analysis can be summarized as follows:
\begin{itemize}
\item[(i)] F-phase is realized when $u_\perp+u_z+\epsilon_Z>0$ and $u_\perp\geq-\epsilon_Z/2$;
\item[(ii)]  CDW-phase is realized when $u_z<u_\perp<-u_z-\epsilon_Z$;
\item[(iii)]  KD-phase is realized when $u_\perp<u_z<\frac{\epsilon_Z^2}{2u_\perp}-u_\perp$ and $u_\perp<\epsilon_Z/2$;
\item[(iv)]  CAF-phase is realized when $u_z>\frac{\epsilon_Z^2}{2u_\perp}-u_\perp$ and $u_\perp<-\epsilon_Z/2$.
\end{itemize}
The corresponding phase diagram is shown in Fig.~\ref{fig_graphene_PD}. 
The phase transition between the CAF and F phases (which can be induced, for example, 
by changing the in-plane component of the magnetic field) is of the second order, 
while the other phase transitions are of the first order.

There is a large body of experimental studies of monolayer graphene in the 
regime of quantum Hall effect \cite{2006PhRvL..96m6806Z,2007PhRvL..98s6806A,
2007PhRvL..99j6802J,2008PhRvL.100t6801C,CheckelskyPhysRevB.79.115434,
2009Natur.462..192D,2009Natur.462..196B,PhysRevLett.108.106804,
Young:2012aax,2013PhRvB..88k5407A,Young:2014aay}. 
There is a consensus that charge-neutral monolayer graphene is strongly 
insulating at high magnetic fields. While the nature of the insulating state has
been under the debate, its origin is traced to the strong Coulomb interaction. 

The most recent experimental study of the phase diagram in monolayer graphene 
in a tilted magnetic field was done in Ref.~\cite{Young:2014aay} and suggests that 
the F and CAF phases can be realized, depending on the strength and orientation of 
the magnetic field. With increasing the longitudinal component of the field at the fixed 
value of its normal component, a smooth phase transition between an insulator 
phase and a metal phase was found. In both phases, however, the bulk energy gap 
was approximately the same. The measured value $1.8e^2/h$ of the conductance 
in the metal phase is close to $2e^2/h$, which is the predicted contribution due to 
the gapless edge states in the F phase. The fact that the phase transition is continuous 
indicates that the insulator phase is the CAF one, see Fig.~\ref{fig_graphene_PD}.

\subsection{Dynamics and phase diagram of the $\nu = 0$ quantum Hall state in bilayer graphene}
\label{sec:bilayer}

The bilayer graphene \cite{McCann:2006aay,Novoselov:2006aax,Henriksen:2008aa}, 
consisting of two closely connected graphene layers (the distance between the layers is 
$d \simeq 0.3~\mbox{nm}$) is a cousin of the monolayer graphene (for recent 
reviews, see Refs.~\cite{McCann:2012ax,Killi:2012aa}). 

Its properties have attracted great interest. The possibility of inducing and controlling 
the energy gap by gates voltage makes bilayer graphene one of the most active 
research areas with very promising applications in electronic devices. 

The experiments in bilayer graphene \cite{2009NatPh...5..889F,2010PhRvL.104f6801Z} showed 
the generation of gaps in a magnetic field with complete lifting of the eight-fold degeneracy in the
zero energy Landau level, which leads to new quantum Hall states with filling
factors $\nu=0,\pm1,\pm2,\pm3$. Besides that, in suspended bilayer graphene,
Ref.~\cite{2009NatPh...5..889F} reported the observation of an extremely large
magnetoresistance in the $\nu=0$ state due to the energy gap $\Delta E$, which
scales linearly with a magnetic field $B$, $\Delta E \sim 3.5-10.5\,
B[\mbox{T}]~\mbox{K}$, for $B \lesssim 10~\mbox{T}$. This linear scaling is
hard to explain by the standard  mechanisms of the gap generation
used in a monolayer graphene, which lead to large gaps of the order of the
Coulomb energy $e^{2}/l \sim B^{1/2}$, where $l=(\hbar c/eB)^{1/2}$ is the magnetic
length.

In this subsection, we review the dynamics of clean bilayer graphene in a magnetic
field, with the emphasis on the $\nu=0$ state in the quantum Hall effect. It will 
be shown that, as in the case of monolayer graphene (see 
Section~\ref{Sec:MagCatGrapheneLocal} and Ref.~\cite{Gorbar:2008hu}), the dynamics 
in the quantum Hall effect in bilayer graphene is described by the {\it coexisting} the magnetic 
catalysis and quantum Hall ferromagnetism order parameters. The essence of the dynamics
is an effective reduction by two units of the spatial dimension in the
electron-hole pairing in the lowest Landau level with energy $E=0$. 
As we discuss below, there is however an essential difference between the quantum Hall effect 
in these two systems. While the pairing forces in monolayer graphene lead to a
relativistic-like scaling $\Delta E \sim \sqrt{|eB|}$ for the dynamical gap, in
bilayer graphene, such a scaling takes place only for strong magnetic fields,
$B \gtrsim B_{\rm thr}$, $B_{\rm thr} \sim 30-60~\mbox{T}$.
For $B \lesssim B_{\rm thr}$, a nonrelativistic-like scaling $\Delta E\sim |eB|$ is
realized in the bilayer. The origin of this phenomenon is very different forms
of the polarization function in monolayer graphene and bilayer one, which in turn
is determined by the different dispersion relations for quasiparticles in these
two systems. The polarization function is one of the major players in the
quantum Hall effect in bilayer graphene because it leads to a very strong screening of the Coulomb 
interactions there.

The first theoretical studies of the quantum Hall effect in bilayer graphene has been
studied in Refs.~\cite{Barlas:2008xx,Shizuya:2009aa,Nakamura:2009aa,Nandkishore:2009aa,
2010JETPL..91..314G,2010PhRvB..81o5451G}. In particular, the gap equation for the 
quasiparticle propagator including the polarization screening effects has been first studied 
in Refs.~\cite{Nandkishore:2009aa,2010JETPL..91..314G,2010PhRvB..81o5451G}. While
a polarization function with no magnetic field  was used in Ref.~\cite{Nandkishore:2009aa},
the polarization function with a magnetic field was utilized in Refs.~\cite{2010JETPL..91..314G,
2010PhRvB..81o5451G}. The main part of this Section is based on the latter two papers.

\subsubsection{Model Hamiltonian}
\label{bilayer-Hamiltonian}

The free part of the effective low-energy Hamiltonian of bilayer graphene is 
\cite{McCann:2006aay}:
\begin{equation}
H_0 = - \frac{1}{2m}\int d^2\mathbf{r} \, \Psi_{Vs}^{+}(\mathbf{r})
\left( \begin{array}{cc} 0 & (\pi^{\dagger})^2\\ \pi^2 & 0
\end{array} \right)\Psi_{Vs}(\mathbf{r}), 
\label{free-Hamiltonian}
\end{equation}
where $\mathbf{r}=(x,y)$, $\pi=\hat{p}_{x}+i\hat{p}_{y}$ and the canonical 
momentum $\hat{\mathbf{p}} = -i\hbar\bm{\nabla} - {e\mathbf{A}}/c$ includes the
vector potential $\mathbf{A}$ corresponding to the external magnetic field
$\mathbf{B}$. Without magnetic field, this Hamiltonian generates the spectrum
$E=\pm p^2/(2m)$, $m= \gamma_1/2v_{F}^2$, where the Fermi velocity 
$v_F\simeq c/300$ and $\gamma_1 \approx 0.34-0.40~\mbox{eV}$. The two 
component spinor field $\Psi_{Vs}$ carries the valley $(V= K, K^{\prime})$ 
and spin ($s = \pm$) indices. We will use the standard convention:
$\Psi_{Ks}^T=(\psi_A{_1}, \psi_B{_2})_{Ks}$ whereas 
$\Psi_{K^{\prime}s}^T = (\psi_B{_2},-\psi_A{_1})_{K^{\prime}s}$. 
Here $A_1$ and $B_2$ correspond to those sublattices in the layers 1 and 2, 
respectively, which, according to Bernal ($A_2$--$B_1$) stacking, are relevant for 
the low-energy dynamics. [For such Bernal stacking, the degrees of freedom from 
the dimer atoms $A_2$ and $B_1$ lead to a high-energy band and, thus, are irrelevant 
for the low-energy dynamics.] The effective Hamiltonian (\ref{free-Hamiltonian}) is valid 
for magnetic fields $1~\mbox{T} < B < B_{\rm thr}$. For $B < 1~\mbox{T}$, the trigonal 
warping should be taken into account \cite{McCann:2006aay}. For $B> B_{\rm thr}$, 
a monolayer like Hamiltonian with linear dispersion should be used.

The Zeeman and Coulomb interactions plus a top-bottom gates voltage imbalance 
$\tilde{\Delta}_0 = eE_{\perp}d/2$ in bilayer graphene are described by the following 
interaction Hamiltonian (Henceforth, we will omit indices $V$ and $s$ in the field $\Psi_{Vs}$):
\begin{eqnarray}
H_{\rm int} &=& \mu_{B}B \int d^2\mathbf{r} \, \Psi^{+}(\mathbf{r})
\sigma^3\Psi(\mathbf{r})\nonumber +
\frac{1}{2}\int  d^2\mathbf{r}d^2\mathbf{r}^{\prime}\Big[
V(\mathbf{r}-\mathbf{r}^{\prime})\left[\rho_1(\mathbf{r})\rho_1(\mathbf{r}^{\prime})+
\rho_2(\mathbf{r})\rho_2(\mathbf{r}^{\prime})\right]  \\
&+& 2V_{12}(\mathbf{r}-\mathbf{r}^{\prime})\rho_1(\mathbf{r})\rho_2(\mathbf{r}^{\prime})\Big]
+\tilde{\Delta}_0 \int d^2\mathbf{r} \, \Psi^{+}(\mathbf{r})\xi\tau_3\Psi(\mathbf{r})\, .
\label{interaction1}
\end{eqnarray}
(Here $E_{\perp}$ is an electric field orthogonal to the bilayer planes, and 
the Pauli matrices $\sigma^3$ and $\tau^3$ act on the spin and layer indices, respectively.)
The valleys $K$ and $K^{\prime}$ are labeled by $\xi = \pm 1$. 
The Coulomb potential $V(\mathbf{r})$ describes the intralayer interactions and, therefore,
coincides with the bare potential in monolayer graphene whose Fourier transform
is given by $\tilde{V}(k)={2\pi e^2}/{\kappa k}$, where $\kappa$ is the dielectric constant. 
The potential $V_{12}(\mathbf{r})$ describes the interlayer electron interactions. 
Its Fourier transform is $\tilde{V}_{12}(k)=({2\pi e^2}/{\kappa})({e^{-kd}}/{k})$, where 
$d \simeq 0.35~\mbox{nm}$ is the distance between the two layers. The two-dimensional 
charge densities in the two layers, $\rho_1(\mathbf{r})$ and $\rho_2(\mathbf{r})$, are defined
by
\begin{equation}
\rho_1(\mathbf{r})=\Psi^{+}(\mathbf{r})P_1\Psi(\mathbf{r})\,,\qquad 
\rho_2(\mathbf{r})=\Psi^{+}(\mathbf{r})P_2\Psi(\mathbf{r})\,,
\label{density}
\end{equation}
where $P_1=(1+\xi\tau^3)/2$ and $P_2=(1-\xi\tau^3)/2$ are projectors on states in the 
layers 1 and 2, respectively. When the polarization effects are taken into account, the 
potentials $V(\mathbf{r})$ and $V_{12}(\mathbf{r})$ are replaced by effective interactions 
$V_{\rm eff}(\mathbf{r})$ and $V_{12,{\rm eff}}(\mathbf{r})$, respectively, whose Fourier 
transforms are given in Eqs.~(A3) and (A4) in Appendix of Ref.~\cite{2010PhRvB..81o5451G}.

\subsubsection{Symmetries and order parameters}
\label{bilayer-symmetries}

The Hamiltonian $H = H_0 + H_{\rm int}$, with $H_0$ and $H_{\rm int}$ in 
Eqs.~(\ref{free-Hamiltonian}) and (\ref{interaction1}), describes the dynamics
at the neutral point (with no doping). Because of the projectors $P_1$ and
$P_2$ in charge densities (\ref{density}), the symmetry of the Hamiltonian $H$
is essentially lower than the symmetry in monolayer graphene. If both the Zeeman
and $\tilde{\Delta}_0$ terms are ignored, it is 
$\mathrm{U}^{(K)}(2)_S \times \mathrm{U}^{(K^{\prime})}(2)_S\times
Z_{2V}^{(+)}\times Z_{2V}^{(-)}$, where $\mathrm{U}^{(V)}(2)_S$ defines the $\mathrm{U}(2)$ spin
transformations for a fixed valley $V = K, K^{\prime}$, and $Z_{2V}^{(s)}$
describes the valley transformation $\xi \to -\xi$ for a fixed spin $s = \pm$
(recall that in monolayer graphene the symmetry would be $U(4)$).
The Zeeman interaction lowers this symmetry down to $G_2 \equiv \mathrm{U}^{(K)}(1)_{+}
\times \mathrm{U}^{(K)}(1)_{-} \times \mathrm{U}^{(K^{\prime})}(1)_{+} \times \mathrm{U}^{(K^{\prime})}
(1)_{-} \times Z_{2V}^{(+)}\times Z_{2V}^{(-)}$, where $\mathrm{U}^{(V)}(1)_{s}$ is the
$U(1)$ transformation for fixed values of both valley and spin. Recall that the
corresponding symmetry in monolayer graphene is $G_1 \equiv \mathrm{U}^{(+)}(2)_V \times
\mathrm{U}^{(-)}(2)_V$, where $\mathrm{U}^{(s)}(2)_V$ is the $\mathrm{U}(2)$ valley transformations for a
fixed spin. Including the $\tilde{\Delta}_0$ term lowers the $G_2$ symmetry
further down to the $\bar{G}_2 \equiv
\mathrm{U}^{(K)}(1)_{+}\times \mathrm{U}^{(K)}(1)_{-} \times \mathrm{U}^{(K^{\prime}) }(1)_{+} \times
\mathrm{U}^{(K^{\prime})}(1)_{-}$.

Although the symmetries in monolayer and bilayer graphene $G_1$ and $G_2$ are quite 
different, their spontaneous breakdowns are described by essentially the same quantum 
Hall ferromagnetism and magnetic catalysis order parameters. The point is that the symmetry 
groups $G_1$ and $G_2$ define the same four conserved commuting currents whose charge 
densities (and the corresponding four chemical potentials) span the quantum Hall ferromagnetism 
order parameters (we use the notations of Ref.~\cite{Gorbar:2008hu}):
\begin{eqnarray}
\label{mu}
\mu_s:\qquad
\langle{\Psi^\dagger_s\Psi_s}\rangle &=&
\langle{\psi_{K  A_1 s}^\dagger\psi_{K A_1 s}
+\psi_{K^{\prime} A_1 s}^\dagger\psi_{K^{\prime}A_1 s}
 +\psi_{K B_2 s}^\dagger\psi_{K B_2 s}
+ \psi_{K^{\prime}B_2 s}^\dagger\psi_{K^{\prime} B_2 s}}\rangle\,,\\
\label{mu-tilde}
\tilde{\mu}_{s}:\qquad
\langle{\Psi^\dagger_s\xi\Psi_s}\rangle &=&
\langle{\psi_{K  A_1 s}^\dagger\psi_{K A_1 s}
- \psi_{K^{\prime} A_1 s}^\dagger\psi_{K^{\prime}A_1 s}
 + \psi_{K B_2 s}^\dagger\psi_{K
B_2 s} - \psi_{K^{\prime}B_2 s}^\dagger\psi_{K^{\prime} B_2 s}}\rangle\,.
\end{eqnarray}
The order parameter (\ref{mu}) is the charge density for a fixed spin
whereas the order parameter (\ref{mu-tilde}) determines the charge-density
imbalance between the two valleys. The corresponding chemical potentials
are $\mu_s$ and $\tilde{\mu}_s$, respectively. While the former order parameter
preserves the $G_2$ symmetry, the latter completely breaks its discrete subgroup 
$Z_{2V}^{(s)}$.

The magnetic catalysis order parameters read 
\begin{eqnarray}
\label{delta}
\Delta_s: \qquad
\langle{\Psi^\dagger_s\tau_3\Psi_s}\rangle &=&
\langle{\psi_{K  A_1 s}^\dagger\psi_{K A_1 s}
-\psi_{K^{\prime} A_1 s}^\dagger\psi_{K^{\prime}A_1 s}
 -\psi_{K B_2 s}^\dagger\psi_{K
B_2 s}+ \psi_{K^{\prime}B_2 s}^\dagger\psi_{K^{\prime} B_2 s}}\rangle\,,\\
\label{delta-tilde}
\tilde{\Delta}_{s}:\qquad
\langle{\Psi^\dagger_s\xi\tau_3\Psi_s}\rangle &=&
\langle{\psi_{K  A_1 s}^\dagger\psi_{K A_1 s}
+\psi_{K^{\prime} A_1 s}^\dagger\psi_{K^{\prime}A_1 s}
 -\psi_{K B_2 s}^\dagger\psi_{K
B_2 s} - \psi_{K^{\prime}B_2 s}^\dagger\psi_{K^{\prime} B_2 s}}\rangle\,.
\end{eqnarray}
These order parameters can be rewritten in the form of Dirac mass
terms \cite{Gorbar:2008hu}. The corresponding masses are $\Delta_s$ and
$\tilde{\Delta}_{s}$, respectively. While the order parameter (\ref{delta})
preserves the $G_2$, it is odd under time reversal \cite{Haldane:1988zza}. 
On the other hand, the order parameter (\ref{delta-tilde}), connected with the 
conventional Dirac mass $\tilde{\Delta}$, determines the charge-density 
imbalance between the two layers. Like the quantum Hall ferromagnetism 
order parameter (\ref{mu-tilde}), this mass term completely breaks the $Z_{2V}^{(s)}$
symmetry and is even under $\cal{T}$. Let us emphasize that unlike a
spontaneous breakdown of continuous symmetries, a spontaneous breakdown 
of the discrete valley symmetry $Z_{2V}^{(s)}$, with the order parameters
$\langle{\Psi^\dagger_s\xi\Psi_s}\rangle$ and 
$\langle{\Psi^\dagger_s\xi\tau_3\Psi_s}\rangle$,
is not forbidden by the Mermin-Wagner theorem at finite
temperatures in a planar system \cite{Mermin:1966fe}.

Note that because of the Zeeman interaction, the $\mathrm{SU}^{(V)}(2)_S$ is explicitly 
broken, leading to a spin gap. This gap could be dynamically strongly enhanced 
\cite{2006PhRvL..96q6803A}. In that case, a quasispontaneous breakdown of the 
$\mathrm{SU}^{(V)}(2)_S$ takes place. The corresponding ferromagnetic (Fr) phase is described 
by the chemical potential $\mu_3 = (\mu_{+} - \mu_{-})/2$, corresponding to the
quantum Hall ferromagnetism  order parameter $\langle\Psi^\dagger\sigma_3\Psi\rangle$, 
and by the mass $\Delta_3 = (\Delta_{+} - \Delta_{-})/2$ corresponding to the magnetic catalysis 
order parameter $\langle\Psi^\dagger\tau_3\sigma_3\Psi\rangle$ \cite{Gorbar:2008hu}.
Thus, the physical picture behind the symmetry breaking in the LLL in bilayer
graphene is quite similar to that in the LLL in monolayer one.

\subsubsection{Solutions of the gap equations}
\label{bilayer-solutions}

In Section~\ref{bilayer-symmetries}, we found the order parameters connected with two
phases. One of them is the ferromagnetic (F) phase that is described by the chemical 
potential $\mu_3 = (\mu_{+} - \mu_{-})/2$ and the mass $\Delta_3 = (\Delta_{+} - \Delta_{-})/2$, 
connected with the order parameters $\langle\Psi^\dagger\tau_3\sigma_3\Psi\rangle$ and 
$\langle\Psi^\dagger\sigma_3\Psi\rangle$, respectively. The second phase is the layer 
polarized (LP) phase with the charge-density imbalance between the two layers. It is 
described by the chemical potential $\tilde{\mu}_s$ and the Dirac mass $\tilde{\Delta}_s$, 
relating to the order parameters $\langle{\Psi^\dagger_s\xi\Psi_s}\rangle$ and 
$\langle{\Psi^\dagger_s\xi\tau_3\Psi_s}\rangle$, respectively. 

It is clear that these two phases play in bilayer graphene the role similar to that 
of the ferromagnetic (F) phase and the charge density wave (CDW) in monolayer 
graphene, discussed in Section~\ref{Sec:GrapheneVacuumAlignment}. The latter are 
basic in monolayer graphene: the other three phases are given by simple deformations 
of the CDW phase. Because of that, one can expect that the F and LP phases are 
basic in bilayer graphene. 

The solutions of the gap equation describing these two phases in the LLL were 
obtained in the framework of the Baym-Kadanoff formalism \cite{Baym:1961zz} 
(known also as the Cornwall-Jackiw-Tomboulis formalism \cite{Cornwall:1974vz} 
in relativistic field theory) in Refs.~\cite{2010JETPL..91..314G,2010PhRvB..81o5451G}.
The crucial point was including the polarization function in the gap equation that was calculated 
in the random phase approximation (RPA). Here we will describe the main results of that analysis. 

 Recall that in bilayer graphene, the LLL includes both the $n=0$ and $n=1$ Landau levels (LLs), 
if the Coulomb interaction is ignored \cite{McCann:2006aay}. Therefore there are sixteen parameters
$\mu_{s}(n)$, $\Delta_{s}(n)$, $\tilde{\mu}_{s}(n)$, and $\tilde{\Delta}_{s}(n)$ with $n=0, 1$, 
which describe the solutions.

Let us start from the description of the polarization function. As was already mentioned in the end
of Section~\ref{bilayer-Hamiltonian}, the polarization effects change the potentials: 
$V(\mathbf{r}) \to V_{\rm eff}(\mathbf{r})$ and $V_{12}(\mathbf{r}) \to V_{12,{\rm eff}}(\mathbf{r})$. 
As is shown in Ref.~\cite{2010PhRvB..81o5451G}, the Fourier transform $\tilde{V}_{\rm eff}(k)$ of 
$V_{\rm eff}(\mathbf{r})$, describing the exchange interactions, is
\begin{equation}
\tilde{V}_{\rm eff}(k)=\frac{2\pi e^2}{\kappa}\,\frac{1}{ k +\frac{4\pi e^2}{\kappa}\Pi (k^2)}
\label{V-D}
\end{equation}
with $\Pi(k^{2})\equiv \Pi_{11}(k)+\Pi_{12}(k)$, where the polarization function $\Pi_{ij}$ describes 
electron densities correlations on the layers $i$ and $j$ in a magnetic field. Below, following 
Ref.~\cite{2010PhRvB..81o5451G}, the static approximation with no dependence on frequency 
$\omega$ will be used. (For the analysis of the gap equations with the dynamically screened 
Coulomb interaction and Landau level mixing, see Ref.~\cite{Gorbar:2012jc}.)

As to the Hartree interactions (interlayer) interaction 
$V_{\rm IL}(\mathbf{r}) \equiv V_{12,{\rm eff}}(\mathbf{r}) - V_{\rm eff}(\mathbf{r})$,
its Fourier transform is
\begin{equation}
\tilde{V}_{\rm IL}(\omega=0,{k}=0)=-\frac{2\pi e^{2}d}{\kappa}.
\label{interlayer-interaction1}
\end{equation}
Let us emphasize that this Hartree contribution is given in terms of the bare interlayer 
potential \cite{Gorbar:2012jc}.

\begin{figure}[t]
\begin{center}
\includegraphics[width=6.0cm]{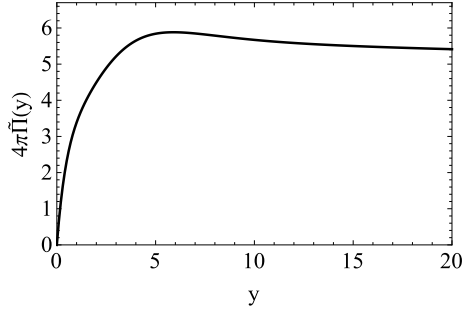}
\caption{The static polarization function $4\pi\tilde{\Pi}(y)$. }
\label{bilayer-polarization}
\end{center}
\end{figure}

It is convenient to rewrite the static polarization $\Pi(k^2)$ in the form $\Pi = (m/{\hbar}^2)\tilde{\Pi}(y)$, 
where both $\tilde{\Pi}(y)$ and $y \equiv (kl)^2/2$ are dimensionless. The function $\tilde{\Pi}(y)$ 
can be expressed in terms of the sum over all the Landau levels and can be analyzed both analytically 
and numerically, for details see Appendix~of Ref.~\cite{2010PhRvB..81o5451G}. At $y\ll 1$,
$\tilde{\Pi}(y) \simeq 0.55y$ and its derivative $\tilde{\Pi}'(y)$ changes from $0.55$ at $y=0$ to $0.12$ 
at $y=1$. At large $y$ it approaches a zero magnetic field value, $\tilde{\Pi}(y)\simeq \ln4/\pi$,
see Fig.~\ref{bilayer-polarization}.

At the neutrality point, $\mu_0 = 0$, there are two competing solutions of the gap equation: 
(i) a ferromagnetic (F) solution, and (ii) a layer polarized (LP) solution. The energy of the LLL states 
of the F solution equals:
\begin{equation}
E^{\rm (F)}_{\xi ns}=
s\left(\epsilon_{Z}+\frac{\hbar^{2}}{2ml^2}F_{n}(x)\right) - \xi\tilde{\Delta}_0 \,, 
\label{solution-I}
\end{equation}
where $\epsilon_{Z} = \mu_{B}B$, $F_{0}(x)=I_{1}(x)+I_{2}(x)$, $F_{1}(x)=I_{2}(x)+I_{3}(x)$, 
and the functions $I_i(x)$ are defined by the following integral expression:
\begin{equation}
I_i(x)=\int_0^{\infty} \frac{dy\,f_i(y)\,e^{-y}}{\kappa \sqrt{xy} + 4\pi
\tilde{\Pi}(y)}
\label{integralsI}
\end{equation}
with $f_i(y)=1,\,y,\,(1-y)^2$ for $i=1,2,3$, respectively. Here, by definition, 
$x=2\hbar^{4}/e^{4}m^{2}l^{2}=(4\hbar\omega_{c}/\alpha^{2}\gamma_{1})
(v_{F}/c)^{2}\simeq 0.003 B[\mbox{T}]$, where $\alpha=1/137$ is the
fine-structure constant and the values of the parameters $\gamma_{1}=0.39~\mbox{eV}$, 
$\hbar\omega_{c}=\hbar^{2}/ml^{2}=2.19 B[\mbox{T}]~\mbox{meV}$, and
$v_{F}=8.0\times10^{5}~\mbox{m/s}$ are the same as in Ref.~\cite{McCann:2006aay}.

The solution exists for $\tilde{\Delta}_0<\epsilon_{Z}+\frac{\hbar^{2}}{2m l^{2}}F_1(x)$.
The Hartree interaction does not contribute in $E^{\rm (F)}_{\xi ns}$. Note that the
dynamical term $(\hbar^{2}/2ml^2)F_{n}(x)$ in Eq.~(\ref{solution-I}) can be rewritten as
$(\hbar|eB|/2mc) F_{n}(x)$, where $F_{n}(x)$ depends on $B$ logarithmically for $x \ll 1$.

The energies of the LLL states for the LP solution are given by the following expression:
\begin{equation}
 E^{\rm (LP)}_{\xi ns} = s \epsilon_{Z} - \xi\left(\tilde{\Delta}_0 -
\frac{\hbar^{2}}{2ml^2}F_{n}(x) - \frac{2e^2 d}{\kappa l^2}\right)\,. \label{solution-II}
\end{equation}
The last term in the parenthesis is the Hartree one, and the solution exists for
$\tilde{\Delta}_0>\frac{2e^{2}d}{\kappa l^{2}}+\epsilon_{Z}-\frac{\hbar^{2}}{2m l^{2}}F_1(x)$. 
For illustration, here we assume the case of a suspended bilayer graphene with $\kappa\simeq 1$.

\begin{figure}[t]
\begin{center}
\includegraphics[width=7.0cm]{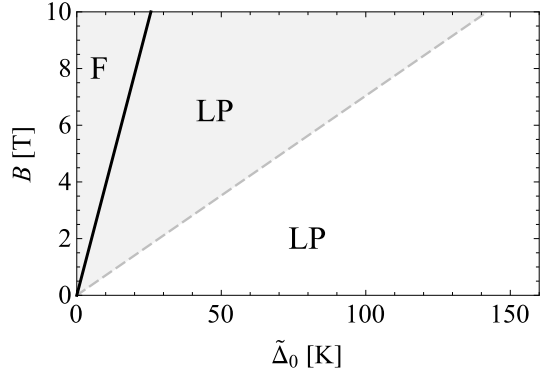}
\caption{The phase diagram for the $\nu=0$ ground state of bilayer graphene 
 in the plane of $\tilde{\Delta}_0$ and $B$ (at $B_\parallel =0$). The two solutions 
coexist in the shaded region.}
\label{bilayer-diagram}
\end{center}
\end{figure}

In Fig.~\ref{bilayer-diagram}, we show the phase diagram in the plane of $\tilde{\Delta}_0$ 
and $B$, taking $B_\parallel =0$ (recall that, by definition, $\tilde{\Delta}_0 = eE_{\perp}/2$). 
The areas labeled F and LP correspond to the regions of the parameter space where the F and LP 
phases are the ground states, respectively. The shaded area shows the region where both types of 
solutions exist. The black line is the phase transition line between the two phases. 
Since these two phases coexist around the critical line, it was suggested in 
Refs.~\cite{2010JETPL..91..314G,2010PhRvB..81o5451G} that the phase transition 
between the F and LP phases is of the first order.

\begin{figure}[t]
\begin{center}
\includegraphics[width=.47\textwidth]{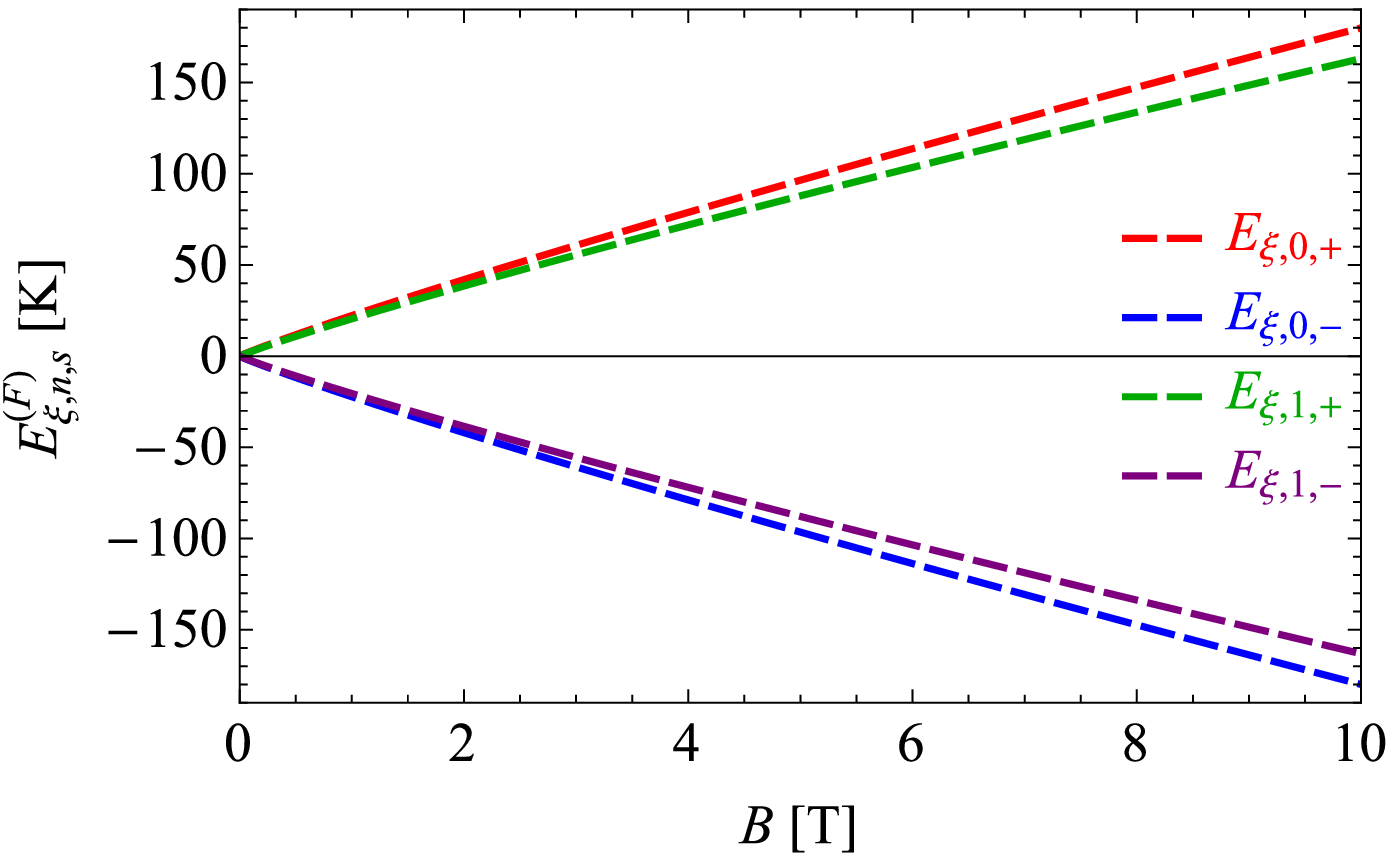}\hspace{.03\textwidth}
\includegraphics[width=.47\textwidth]{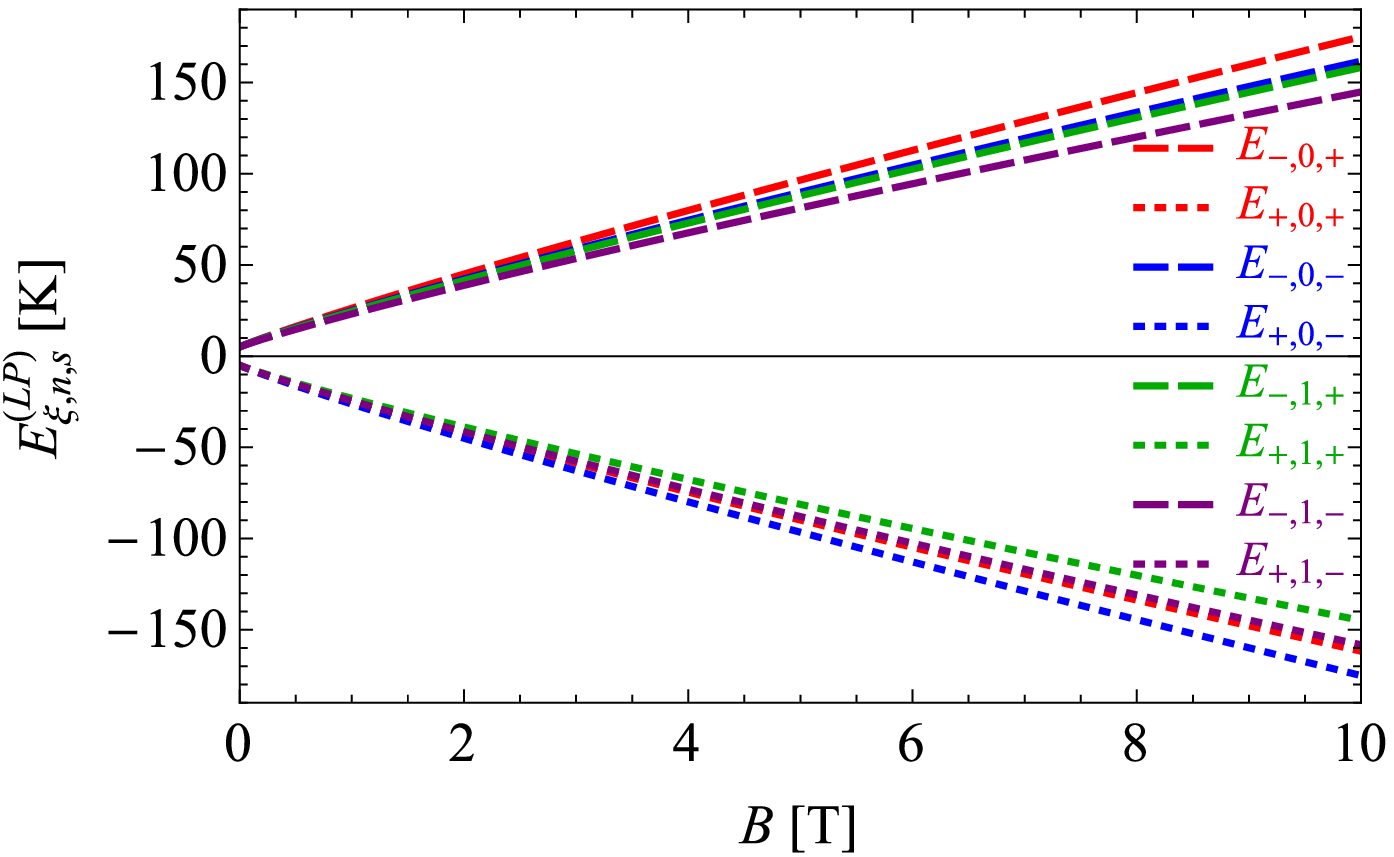}
\caption{The LLL energies of the F (left panel) and LP (right panel) solutions
as functions of $B$ at $B_{\parallel}=0$. The results for the F and LP solutions
are plotted for $\tilde{\Delta}_{0}=0$ and $\tilde{\Delta}_{0}=5~\mbox{K}$, 
respectively.}
\label{bilayer-spectrum}
\end{center}
\end{figure}

For the F solution at $\tilde{\Delta}_0 = 0$, the dependence of the LLL energies
$E^{\rm (F)}_{\xi ns}$ on $B$ (at $B_{\parallel}=0$) is shown on the left panel in 
Fig.~\ref{bilayer-spectrum} (the LLL states with opposite $\xi$ remain degenerate 
in this solution). Nearly perfectly linear forms of the energy dependences
on the field are evident. Also, the degeneracy between the states of the $n=0$ LL 
and those of the $n =1$ LL is removed. The energy gap corresponding to the $\nu = 0$ 
plateau is $\Delta E^{\rm (F)} = (E_{\xi1 + }^{\rm (F)} - E_{\xi1 - }^{\rm (F)})/2 
\simeq 16.3 B [\mbox{T}]~\mbox{K}$.

In the right panel of Fig.~\ref{bilayer-spectrum}, the dependence of the LLL energies on 
$B$ (at $B_{\parallel}=0$) is shown for the LP solution at $\tilde{\Delta}_0= 5~\mbox{K}$.
The energy dependences on the field are nearly perfectly linear again. Unlike the 
F solution, the LLL degeneracy is now completely removed. As to the energy gap corresponding 
to the $\nu = 0$ plateau, it is $\Delta E^{\rm (LP)} = (E_{- 1 - }^{\rm (LP)} - E_{+ 1 + }^{\rm (LP)})/2  
\simeq 5~\mbox{K} + 14.0 B [\mbox{T}]~\mbox{K}$.

In conclusion we would like to note that the theoretical description of the broken 
symmetry states at higher plateaus with $\nu = \pm1, \pm2, \pm3$ was given in 
Ref.~\cite{Gorbar:2011ce}. The agreement of the theory and experiment 
\cite{2010Sci...330..812W} is satisfactory.

\subsubsection{Comparison with experiment}
\label{Sec:Bilayerexperiment}

The first experiments in bilayer graphene in a magnetic field \cite{Novoselov:2006aax,Henriksen:2008aa} 
revealed quantum Hall states with the filling factor $\nu= \pm 4n$, $n= 1, 2...$ predicted 
in the framework of the one electron problem in Ref.~\cite{McCann:2006aay}. No traces 
of lifting the eightfold degeneracy of the LLL and the fourfold degeneracy of higher LLs 
were observed.

Later experiments in bilayer graphene \cite{2009NatPh...5..889F,2010PhRvL.104f6801Z} 
showed the generation of energy gaps in a magnetic field resulting in complete lifting the 
eightfold degeneracy in the LLL, which leads to new quantum Hall states with filling factors 
$\nu=0,\pm1,\pm2,\pm3$. While in Ref.~\cite{2009NatPh...5..889F} suspended bilayer 
graphene was used, bilayer graphene samples deposited on Si${\mbox O_2}$/Si substrates 
were used in Ref.~\cite{2010PhRvL.104f6801Z}. Because suspended bilayer graphene is 
much cleaner than that on a substrate, the new quantum Hall states in the former start to 
develop at considerably smaller magnetic fields than in the latter. Also, the energy gaps 
corresponding to these states are considerably larger in suspended samples than in those 
on substrates. Both these experiments clearly showed that the $\nu = 0$ state is an 
insulating one.

Here the dynamics of the $\nu = 0$ state is analyzed in the case of clean bilayer graphene.
Therefore, it is appropriate to compare the model results with the experimental data for 
suspended graphene. The central result concerning the $\nu = 0$ state in 
Ref.~\cite{2009NatPh...5..889F} is the observation of an extremely large magnetoresistance 
in the $\nu=0$ state due to the energy gap $\Delta E$, which scales linearly with a magnetic
field $B$. For $B_{\perp} \lesssim 10~\mbox{T}$, the value of the gap is approximately given by 
$\Delta E \sim 3.5-10.5B_{\perp}[{\mbox T}]~{\mbox K}$. This experimental value of the gap 
appears to be in a satisfactory agreement with the expressions for both gaps $\Delta E^{\rm (F)}$ and 
$\Delta E^{\rm (LP)}$ obtained in Section~\ref{bilayer-solutions}.

The observed large magnetoresistance implies that the ground state cannot be ferromagnetic. 
The reason is that such a state has gapless edge states and, thus, is not an insulator.
One might think, then, that the $\nu = 0$ state is described by the LP phase. In fact, this guess 
is not very far from the correct one: the $\nu = 0$ state is the canted antiferromagnet (CAF) state 
\cite{Kharitonov:2012PRL,Kharitonov:2012PRB}, which is closely connected with the LP one 
(similarly to the close connection between the CDW phase and CAF one in monolayer graphene, 
see Section~\ref{AF-CAF-KD}). We will return to this issue in Section~\ref{Sec:BilayerPhaseDiagram}.

Let us now turn to the phase diagram in Fig.~\ref{bilayer-diagram}. The phase transition 
between two phases in bilayer graphene was reported in several experimental studies
\cite{2010Sci...330..812W,PhysRevLett.107.016803,2012NatNa...7..156V,Maher:2013aa}. 
The theoretical prediction for the form of the critical line in the present model is a straight 
line with the maximum value of its slope about $4.7~\mbox{mV}\, \mbox{nm}^{-1}\mbox{T}^{-1}$, 
corresponding to the smallest permissible value $\kappa = 1$. On the other hand, the value 
of the slope in the experiments of Refs.~\cite{2010Sci...330..812W,PhysRevLett.107.016803,
2012NatNa...7..156V,Maher:2013aa} is about $10~\mbox{mV}\, \mbox{nm}^{-1}\mbox{T}^{-1}$. 
The discrepancy may have its roots in disorder. Indeed, an external electric field in a clean 
sample (considered in the present model) is more effective and, therefore, its critical value 
should be smaller than in a real sample with charged impurities.

In conclusion we would like to note that unlike monolayer graphene, there is a 
nonzero energy gap $\Delta_0 \sim 1~\mbox{meV}$ at the charge neutral point in bilayer 
graphene even at zero value of a magnetic field \cite{2012PhRvL.108g6602F,2012NatNa...7..156V}. 
This fact reflects the nonrelativistic dispersion relation $E \sim p^2$ for quasiparticles in 
bilayer graphene that leads to a nonzero quasiparticle density at the charge neutral point.

\subsubsection{Phase diagram}
\label{Sec:BilayerPhaseDiagram}

Here we will describe the phase diagram of bilayer graphene derived in 
Refs.~\cite{Kharitonov:2012PRL,Kharitonov:2012PRB}. It is quite remarkable that 
this phase diagram and its analysis are very similar to those in monolayer graphene
considered in Section~\ref{Sec:GrapheneVacuumAlignment}. Because of that, here we 
will briefly describe only the final result.

As in the case of monolayer graphene, in addition to the F and LP phases, there are additional 
candidates for the $\nu=0$ ground state: the counted antiferromagnetic (CAF) phase and the 
partially layer polarized (PLP) one, which is an analog of the Kekule distortion (KD) phase in 
monolayer graphene. (Like in monolayer, the antiferromagnetic phase is always less favorable 
than the CAF phase in bilayer graphene, and, therefore, it is not considered below.) The new 
feature in bilayer is that beside an external magnetic field, one can also use an electric field 
$E_{\perp}$.

In the LLL approximation, the energies of the four candidate phases for the $\nu=0$ ground 
state take the following form \cite{Kharitonov:2012PRL}:
\begin{equation}
{\cal E}^{\rm LP} = u_z  - 2\epsilon_V, \qquad
{\cal E}^{\rm PLP} = u_\perp -\frac{\epsilon_V^2}{u_z+|u_\perp|}, \qquad
{\cal E}^{\rm CAF} = -u_z -\frac{\epsilon_Z^2}{2|u_\perp|},  \qquad
{\cal E}^{\rm F} =  -2u_\perp-u_z-2\epsilon_Z ,
\end{equation}
where $\epsilon_V \approx E_{\perp} d/2$. Note that the perpendicular electric field affects only the
layer-polarized phases. 

\begin{figure}[t]
\begin{center}
\includegraphics[width=.49\textwidth]{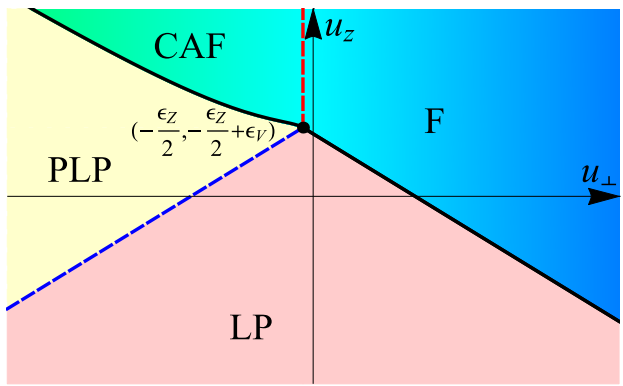}
\caption{(Color online) Phase diagram for the $\nu=0$ ground state of bilayer graphene in 
the plane of two short-range coupling constants, based on the results from 
Ref.~\cite{Kharitonov:2012PRL}.}
\label{fig_bilayer_PD}
\end{center}
\end{figure}

The phase diagram is shown in Fig.~\ref{fig_bilayer_PD}. As one can see, in agreement with 
the conclusion of Refs.~\cite{2010JETPL..91..314G,2010PhRvB..81o5451G} (see also 
Section~\ref{bilayer-solutions}), there is a first order phase transition between the LP and F phases. 
On the other hand, there is a smooth phase transition between the CAF and F phases. This 
yields a potential signature for the CAF phase: increasing a longitudinal component of a 
tilted magnetic field at its fixed normal component, the CAF phase smoothly transfers into the F one.
Another useful signature is connected with electric field $E_{\perp}$. When $E_{\perp}$ increases,
the CAF phase transfers (through a first order phase transition) into the PLP phase and then
the latter smoothly transfers into the LP one (note that like the KD phase in monolayer graphene,
the PLP one cannot directly transfer into the F phase). 

The experiment \cite{Maher:2013aa} confirmed the two signatures described above. Moreover, 
in agreement with the presence of four gapless edge states in the F phase, it was found that the 
conductance of the metallic state, to which the insulator state transfers with increasing $B_{\parallel}$, 
is close to $4e^2/h$, thus yielding an additional support of the phase diagram in Fig.~\ref{fig_bilayer_PD}.

\section{Chiral asymmetry in magnetic field}
\label{sec:CSEandCME}

\subsection{Introduction}

As should be clear by now, one of the main recurrent themes throughout this review is 
symmetry. Symmetries play an important role in quantum field theory. They provide general 
tools to classify fundamental as well as emergent (composite) particles, allow unambiguous 
classification of different vacua (ground states), impose constraints on the equations 
of motion and explain various conservation laws. Symmetries and their breaking patterns 
help to explain the observed degeneracies of particle states and mass hierarchies, 
provide deeper insights into the elementary processes observed in nature, and allow a 
systematic approach in model building of the ultimate theory of elementary particles.

\subsubsection{Anomalous symmetries}
\label{subsec:AnomalousSym}

It appears, however, that sometimes the absence of certain symmetries may play an 
equally profound role too. In particular, this is the case when the symmetry is present  
in the classical Lagrangian, but is absent in the full quantum field theory. The 
corresponding types of ``missing" symmetries are called anomalous. It should be
emphasized that there is a profound difference between such anomalous symmetries 
and the spontaneously broken ones. The latter are broken in the ground state of the 
system, but remain exact symmetries of the quantum mechanical action. The former 
are not even the symmetries of the action. Also, the consequences of anomalous 
symmetries are very different. 

The history of anomalous symmetries started with the problem of understanding the 
decay rate of the neutral pion into a pair of gammas, $\pi^0\to2\gamma$, which is the
primary decay mode. The measured value of the corresponding decay rate was in 
apparent contradiction with the approximate axial-vector current conservation. 
Quantitatively, the measured value $\Gamma(\pi^0\to \gamma\gamma)\approx 7.76~\mathrm{eV}$ 
($\tau\approx 8.38 \times 10^{-17}~\mathrm{s}$) for the decay width (mean lifetime) is about 
three orders of magnitude larger (smaller) than the prediction naively following 
from the assumed presence of the axial $\mathrm{U}(1)_{A}$ symmetry in the field 
equations of motion, for a review see \cite{Bernstein:2011bx}. The discovery of the 
quantum anomaly for this symmetry offered an elegant resolution of the puzzle \cite{Bell:1969ts,Adler:1969gk}.  

Another use of the $\mathrm{U}(1)_{A}$ anomalous symmetry is the underlying nature 
of the $\eta^\prime$
meson and the explanation of its rather large mass. Compared to the eight pseudoscalar 
mesons, associated with breaking of the (approximate) $\mathrm{SU}(3)_{A}$ chiral symmetry, 
$\eta^\prime$ is considerably more massive. The root of this can be traced again to the 
fact that the axial $\mathrm{U}(1)_A$ transformations are not part of the (approximate) 
chiral symmetry in the hadronic physics \cite{Veneziano:1979ec,Witten:1979vv}. 

The absence of the chiral $\mathrm{U}(1)_A$ symmetry in a quantum system can be 
formally stated as the nonvanishing divergence of the axial current. In particular, in the
massless QED, the electromagnetic anomaly of the axial current takes the form of the 
following operator relation \cite{Bell:1969ts,Adler:1969gk}:
\begin{equation} 
\partial_{\mu} j_{5}^{\mu} = -\frac{e^2}{16 \pi^2}
\epsilon^{\kappa\lambda\mu\nu}F_{\kappa\lambda}F_{\mu\nu}.
\label{intro-anomaly}
\end{equation} 
where $j_{5}^{\mu} = \bar\psi \gamma_5 \gamma^{\mu} \psi $ is the axial current density 
and $F_{\mu\nu}$ is the field strength tensor in QED. Interestingly, this operator relation 
receives no radiative corrections \cite{Adler:1969er} and is, therefore, correct to all orders in 
perturbation theory. The simplest (semi-classical) interpretation of the anomaly relation is that 
the numbers of left-handed and right-handed fermions are not separately conserved when background 
electromagnetic fields are such that $\epsilon^{\kappa\lambda\mu\nu}F_{\kappa\lambda}F_{\mu\nu}\neq 0$.
A similar relation can be also obtained in the non-Abelian $\mathrm{SU}(3)_{c}$ gauge 
theory of strong interactions,
\begin{equation} 
\partial_{\mu} j_{5}^{\mu} = -\frac{g^2 N_f}{32 \pi^2}
\epsilon^{\kappa\lambda\mu\nu}  F^{a}_{\kappa\lambda}F^{a}_{\mu\nu}.
\label{intro-anomaly-non-Abelian}
\end{equation} 
where $j_{5}^{\mu} = \bar{q}_{i} \gamma_5 \gamma^{\mu} q_{i}$ is the axial current density 
constructed from quark fields with the color index $i=1,2,3$, and $F^{a}_{\mu\nu}$ is the gluon 
field strength in QCD.

In general, there exist many well-known examples of anomalous symmetries that play an
important role in particle physics and beyond. They are covered extensively in textbooks, 
see for example Refs.~\cite{Peskin:1995ev,Weinberg:1996kr}, and monographs, see for 
example Refs.~\cite{Bertlmann:book,Fujikawa:book}. However, it is not our goal to review 
them here. Instead, we will discuss several relatively recent ideas regarding the chiral 
properties in magnetized relativistic matter at nonzero density that were inspired by anomalies.
An interesting feature of the corresponding types of matter is a range of possible anomalous 
transport phenomena that they can display. 

It should be mentioned that, strictly speaking, certain elements of chiral anomalous transport 
have been known from 1980s \cite{Vilenkin:1980fu}. However, it was only within the last 
five years or so that numerous applications of the corresponding phenomena started to 
attract a widespread attention. In part, this was driven by the real possibility of testing
them in experiments and observational data. The chiral effects of this type may affect the 
physics of heavy-ion collisions \cite{2013LNP...871..181D,Kharzeev:2013ffa,Liao:2014ava,
Abelev:2009ac,Abelev:2012pa,Adamczyk:2014mzf}, compact stars \cite{Charbonneau:2009hq,Charbonneau:2009ax,
Ohnishi:2014uea}, the early Universe \cite{Giovannini:1997eg,Boyarsky:2011uy,Tashiro:2012mf}, 
and perhaps even Dirac/Weyl semimetals \cite{2013arXiv1301.0330T,Vafek:2013vn,Li:2014aax}.

The current growing list of related anomalous phenomena includes a number of chiral effects: 
chiral separation \cite{Vilenkin:1980fu,Son:2004tq,Metlitski:2005pr,Newman:2005as,
Gorbar:2009bm,Gorbar:2011ya,Gorbar:2013upa}, 
chiral magnetic  \cite{Kharzeev:2004ey,Kharzeev:2007tn,Kharzeev:2007jp,Fukushima:2008xe},
chiral vortical \cite{Vilenkin:1979ui,Kharzeev:2007tn,Erdmenger:2008rm,Banerjee:2008th,Son:2009tf}, 
chiral electric \cite{Neiman:2011mj}, 
chiral electric separation \cite{Huang:2013iia,Pu:2014cwa}, 
chiral charge generation  \cite{Bhattacharyya:2012xi,Jimenez-Alba:2014pea} effects.
Also the existence of several new types of anomalous collective excitations, e.g., the 
chiral magnetic \cite{Kharzeev:2010gd,Burnier:2011bf} 
and chiral electric \cite{Huang:2013iia,Pu:2014fva} waves, were predicted 
for relativistic plasmas in external fields. 
For the recent reviews, see Refs.~\cite{Kharzeev:2013ffa,Liao:2014ava}

We will primarily concentrate on the chiral separation effect and the chiral magnetic
effect, which are of particular interest when there is a background magnetic field. 
These effects are connected, although in a nontrivial way, to the chiral anomaly.

\subsubsection{Chiral separation and chiral magnetic effects}
\label{subsec:CSE-CME-general}

The analysis of the spectrum of $(3+1)$-dimensional free fermions in a constant 
magnetic field in Section~\ref{sec:MagCat3+1General} demonstrates that the lowest
Landau level is completely spin-polarized. In the case of massless fermions, each 
state should also have a well defined chirality. Let us now introduce a nonzero chemical 
potential $\mu$ and discuss the chirality of occupied LLL states with $|k_z|\leq \mu$. 
Because of the special nature of the LLL, the spin direction of all occupied states 
is the same as (opposite to) the magnetic field when the fermions are 
positively (negatively) charged. This also means that there is a one-to-one 
correspondence between the sign of the longitudinal momentum $k_z$ and the 
chirality. All states with negative longitudinal momenta, $-\mu < k_z<0$, are left (right) 
handed, while all states with positive longitudinal momenta, $0 < k_z<\mu$, are 
right (left) handed. In other words, there is a nonzero axial current density, which
is determined by the total number density of the occupied LLL states  (recall that 
the speed of light is $c=1$). The formal expression for the axial current density 
reads \cite{Vilenkin:1980fu,Metlitski:2005pr}
\begin{equation}
\mathbf{j}_5=\frac{e\mathbf{B}}{2\pi^2} \mu .
\label{CSEdefinition}
\end{equation}
This result is known in the literature as the chiral separation effect (CSE). 
It is curious to note that the CSE appears already in the free theory, in 
which only the interaction with the background field is accounted for, but 
otherwise the fermions remain noninteracting. Also, it appears that only the lowest 
Landau level contributes to the axial current in such a free theory \cite{Metlitski:2005pr}.
This is significant because in the presence of an external magnetic field the chiral anomaly 
is known to be generated entirely on the lowest Landau level \cite{Ambjorn:1983hp}. 

The above mentioned connection of the axial current density (\ref{CSEdefinition}) to 
the anomaly relation can be made more explicit. Indeed, let us assume that a 
constant magnetic field points in the $z$ direction and consider the corresponding 
relation in the case of a spatially varying chemical potential $\mu(z)$. This can be 
modeled, for example, by a classical electric potential $\Phi(z) = \mu(z)/e$. In response 
to the electric potential $\Phi(z)$, the system will induce a spatial variation of the axial 
current with the following nonzero divergence: 
\begin{equation}
\partial_z j^{5}_{z}=\frac{e B_z }{2\pi^2} \partial_z \left(e \Phi \right) 
= - \frac{e^2}{2\pi^2} B_z E_z,
\label{CSEdefinitionA}
\end{equation}
where we used the definition of the electric field, $E_z=-\partial_z \Phi$. 
As is easy to check, this is a special case of the chiral anomaly relation in 
Eq.~(\ref{intro-anomaly}). (Note that $E_z=F_{03}$ and $B_z=-F_{12}$.)

A closely related phenomenon can be realized also when a nonzero density is replaced by the 
chiral charge density. It is called the chiral magnetic effect (CME) \cite{Kharzeev:2007jp,
Fukushima:2008xe}. The needed chiral asymmetry can be semi-rigorously described by a 
nonvanishing chiral chemical potential $\mu_5 \ne 0$. The essence of CME is that $\mu_5$ 
causes a nondissipative electric current 
\begin{equation}
\mathbf{j}=\frac{e\mathbf{B}}{2\pi^2}  \mu_5 . 
\label{CMEdefinition}
\end{equation}
Both the CSE and CME have been studied analytically in various quantum field 
theoretical models \cite{Vilenkin:1980fu,Son:2004tq,Metlitski:2005pr,Newman:2005as,
Kharzeev:2007jp,Fukushima:2008xe,Gorbar:2009bm,Fukushima:2009ft,Fukushima:2010fe,
Fukushima:2010zza,Rubakov:2010qi,Gorbar:2011ya,Hou:2011ze,Fukushima:2012vr,Basar:2012gm,Gorbar:2013upa} 
as well as in holographic models \cite{Yee:2009vw,Rebhan:2009vc,Gorsky:2010xu,Gynther:2010ed,Hoyos:2011us,
Kalaydzhyan:2011vx,Gahramanov:2012wz,Jimenez-Alba:2014iia}. They were reproduced in kinetic theory 
\cite{Son:2012wh,Son:2012zy,Stephanov:2012ki,Gao:2012ix,Chen:2012ca,Satow:2014lva} 
and, with a varying degree of success, they were also tested in lattice simulations 
\cite{Buividovich:2009wi,Buividovich:2009zj,Buividovich:2009zzb,Abramczyk:2009gb,Braguta:2010ej,
Yamamoto:2011gk,Yamamoto:2011ks,Yamamoto:2012bi,Bali:2014vja,Chang:2014dga}. 
(See, however, the recent holographic study in Ref.~\cite{Jimenez-Alba:2014iia}, 
which points some fundamental differences between the realization of the CME and CSE.)

The combination of the chiral separation and chiral magnetic effects can in turn give 
rise to a collective gapless excitation, known as the chiral magnetic wave (CMW) 
\cite{Kharzeev:2010gd}. The corresponding mode can be thought of as a propagating
perturbation of alternating electric and chiral charge density fluctuations. A local 
fluctuation of the electric charge density induces via CSE a local fluctuation of the 
axial current. The resulting fluctuation of the chiral chemical potential produces via the CME
a local fluctuation of the electric current. The latter in turn leads again to a local fluctuation 
of electric charge density and thus provides a self-sustaining mechanism for the propagation
of a chiral magnetic wave. In heavy-ion collisions, such a wave may lead to an observational 
feature such as the quadrupole correlations of charged particles \cite{Gorbar:2011ya,Burnier:2011bf}
(for more details, see the discussion in Section~\ref{ChiralShiftPhys}).

Let us recall that the anomalous operator relations like the one in Eq.~(\ref{intro-anomaly}) 
are exact \cite{Adler:1969er}. If such relations cannot get any higher-order radiative 
corrections, one may suggest that the earlier mentioned one-loop relations for the axial 
and electromagnetic current densities are exact as well \cite{Metlitski:2005pr,Newman:2005as}. 
As we discuss in detail below, this is not exactly true. The following simplified argument may explain why
this is not the case. The point is that these anomalous transport relations for the currents are meaningful when given in terms of the corresponding ground state expectation values, rather than the formal 
operators. Therefore, there is no guarantee that interaction corrections will be absent, 
especially when the ground state expectation values are calculated at nonzero temperature 
or density, which break the Lorentz symmetry.

Below, we will address the issue of the interaction effects within the framework of a
Nambu-Jona-Lasinio model as well as quantum electrodynamics. We will mostly 
concentrate on the CSE effects, but make numerous comments about CME too. 
The benefit of the NJL model is the simplicity of the analysis that allows us to capture 
almost all qualitative features of the CSE effect without complications present in 
the gauge theory.  However, we will make the case that many details of the 
underlying physics will remain essentially the same in QED.

\subsubsection{The essence of chiral asymmetry}
\label{subsec:intro-essence}

As we argued in the previous subsection, the lowest Landau level has a built-in 
chiral asymmetry which is the consequence of the spin-polarized nature of the LLL 
states. In the case of massless or ultrarelativistic particles, this translates into 
a current of chirality (or helicity), which is formally captured by the relation in 
Eq.~(\ref{CSEdefinition}). 

Notably, higher Landau levels do not contribute to the axial current density 
(\ref{CSEdefinition}) because they are not spin-polarized in the free theory.
It is logical to ask if the same will remain true in interacting theory. As we will 
argue below, the situation changes: the chiral asymmetry of LLL gets promoted 
to all Landau levels. This was first found in the framework of the NJL model 
at nonzero density \cite{Gorbar:2009bm,Fukushima:2010zza,Gorbar:2010kc},
and later generalized to gauge theories as well \cite{Gorbar:2013uga,Gorbar:2013upa}.

The effect of the chiral asymmetry in higher Landau levels is quantified by a dynamically 
generated chiral shift parameter $\Delta$. For a magnetic field directed along the 
positive $z$-axis, the corresponding low-energy effective action receives a new term, 
$\Delta\bar{\psi}\gamma^3\gamma^5\psi$. The physical meaning of the chiral shift 
$\Delta$ is most transparent in the chiral limit. It determines the relative shift of the 
longitudinal momenta in the dispersion relations of opposite chirality fermions 
$k_z \to k_z \pm \Delta$. Just like the external magnetic field, the chiral shift 
is symmetric under parity transformations ${\cal P}$, breaks time reversal 
${\cal T}$ and the rotational symmetry $\mathrm{SO}(3)$ [down to $\mathrm{SO}(2)$ 
subgroup of rotations about the direction of the magnetic field]. Also, the $\Delta$ term 
is even under charge conjugation ${\cal C}$. It breaks ${\cal CPT}$, but this discrete 
symmetry is already broken by a nonzero chemical potential. We conclude, therefore, 
that the absence of the chiral shift parameter is not protected by any symmetry. This, 
in turn, suggests that a nonzero chiral shift could be dynamically generated even in 
perturbation theory \cite{Gorbar:2009bm,Gorbar:2011ya}. 

Interaction between fermions not only induces a chiral shift, but also leads to a 
dynamical correction to the axial current in Eq.~(\ref{CSEdefinition}). Unlike the 
anomalous contribution, which is exclusively due to LLL, the dynamical one will 
come from all Landau levels. This observation brings up an important question 
about the connection between the structure of the induced axial current and the 
chiral anomaly. As is well known, the operator relation for the divergence of the 
axial current (\ref{intro-anomaly}) receives no radiative corrections \cite{Adler:1969er}. 
Should then the axial current also remain the same as in the free theory in a 
magnetic field \cite{Metlitski:2005pr}? This was indeed a predominant belief in 
earlier studies, see Refs.~\cite{Metlitski:2005pr,Gorbar:2009bm,Rebhan:2009vc,
Fukushima:2010zza,Rubakov:2010qi}. As we argue below and then support the 
arguments with explicit calculations in QED, the ground state expectation value 
of the axial current receives nonzero radiative corrections. At the same time, 
the chiral shift does not affect the exact operator form of the chiral anomaly 
(\ref{intro-anomaly}).

\subsection{Chiral asymmetry in NJL model in a magnetic field}
\label{ChiralAsymNJL}

In this section, we study chiral asymmetry within the framework the NJL model 
with a single fermion flavor. The simplicity of such a model allows us to illustrate the 
dynamics underlying the dynamical generation of the chiral shift $\Delta$ in the 
simplest form. The only serious drawback of this study is the nonrenormalizable  
nature of the model. As will become clear later, this is not a critical limitation for qualitative 
findings regarding the (perturbative) generation of the chiral shift. However, it will 
naturally raise questions regarding the validity of the NJL results for the axial current, 
which is thought to be intimately connected with the exact chiral anomaly relation
that should not receive any radiative corrections. We will discuss the chiral anomaly 
relation and its implications on the validity of the key results in 
Section~\ref{CSE:ChiralAnomaly}. As we will show, the chiral anomaly relation for 
the divergence of the axial vector current (\ref{intro-anomaly}) remains exact even 
when the chiral shift is generated dynamically. The same will not be true for the 
defining relation of the chiral separation effect in Eq.~(\ref{CSEdefinition}). The corresponding 
expression for the axial current will in general receive radiative corrections. This will 
be also demonstrated by explicit two-loop calculations in QED in Section~\ref{ChiralAsymQED}.

\subsubsection{Gap equations}
\label{subsec:ChiralShift}

In order to illustrate the dynamics underlying the dynamical generation of the chiral 
shift in the simplest form, we will start here from the analysis of the NJL model with 
a single fermion flavor. The Lagrangian density of the corresponding model is given
by
\begin{eqnarray}
{\cal L} &=& \bar\psi \left(iD_\nu+\mu_0\delta_{\nu}^{0}\right)\gamma^\nu \psi
-m_{0}\bar\psi \psi  +\frac{G_{\rm int}}{2}\left[\left(\bar\psi \psi\right)^2
+\left(\bar\psi i\gamma^5\psi\right)^2\right],
\label{CSE-CME:NJLmodel}
\end{eqnarray}
where $m_{0}$ is the bare fermion mass and $\mu_0$ is the chemical potential. 
By definition, $\gamma^5\equiv i\gamma^0\gamma^1\gamma^2\gamma^3$. The 
covariant derivative is defined as before, $D_{\nu}=\partial_\nu + i e A_{\nu}$, and 
includes the external gauge field $A_{\nu}$, which is assumed to be in the Landau 
gauge, $A^{k}= - B y \delta_{1}^{k}$. This corresponds to the external magnetic field 
$\mathbf{B}$ pointing in the $z$-direction. In addition to the $(3+1)$-dimensional 
Lorentz symmetry, which is broken down to the $\mathrm{SO}(2)$ symmetry of 
rotations, the discrete symmetries ${\cal C}$,  ${\cal T}$, ${\cal CP}$, ${\cal CT}$, 
${\cal PT}$, and ${\cal CPT}$ are broken as well. The parity ${\cal P}$ is the only 
discrete symmetry that is left unbroken. 

In the chiral limit, $m_{0}=0$, this model possesses the chiral $\mathrm{U}(1)_L\times 
\mathrm{U}(1)_R$ symmetry. In the vacuum state ($\mu_0=0$), as discussed in 
Section~\ref{sec:NJL3+1General}, this chiral symmetry is spontaneously broken 
because of the magnetic catalysis. At a sufficiently large chemical potential, however,
the chiral symmetry is expected to be restored. As we will see below, this is indeed 
the case, but the corresponding normal ground state (with restored chiral symmetry) 
is characterized by a nonzero chiral shift parameter $\Delta$. 

As follows from the structure of the Lagrangian density in Eq.~(\ref{CSE-CME:NJLmodel}), the
tree level fermion propagator in coordinate space is determined by
\begin{equation}
iS^{-1}(u,u^\prime) =\left[(i\partial_t+\mu_0)\gamma^0 -(\bm{\pi}\cdot\bm{\gamma}) -m_{0}\right]
\delta^{4}\left(u- u^\prime\right),
\label{CSE-CME:sinverse}
\end{equation}
where $u=(t,\mathbf{r})$ and $\pi^{k} \equiv i \partial^k - e A^k$ are the spatial component 
of the canonical momenta.

We will study the properties of model (\ref{CSE-CME:NJLmodel}) in the mean-field approximation,
which is reliable at weak coupling, $G_{\rm int}\to 0$. In this case, the effect of the wave function 
renormalization is negligible and the (inverse) full propagator takes the form 
\cite{Gorbar:2009bm}:
\begin{equation}
iG^{-1}(u,u^\prime) = \Big[(i\partial_t+\mu)\gamma^0 -
(\bm{\pi}\cdot\bm{\gamma})
+ i\tilde{\mu}\gamma^1\gamma^2
+\Delta\gamma^3\gamma^5
-m\Big]\delta^{4}(u- u^\prime),
\label{ginverse}
\end{equation}
This propagator contains two new types of dynamical parameters that are absent
at tree level in Eq.~(\ref{CSE-CME:sinverse}): $\tilde{\mu}$ and $\Delta$. From its
Dirac structure, it should be clear that $\tilde{\mu}$ plays the role of an anomalous
magnetic moment. As for $\Delta$, it is the chiral shift parameter already mentioned
in Section~\ref{subsec:intro-essence}. Note that in $2 + 1$ dimensions (without $z$ coordinate),
$\Delta\gamma^3\gamma^5$ plays the role of a time-reversal breaking mass term.
This mass is responsible for inducing the Chern-Simons term in the effective action
for gauge fields \cite{Niemi:1983rq,Redlich:1983kn,Redlich:1983dv}, and it plays an 
important role in the quantum Hall effect in graphene \cite{Gorbar:2007xh,Gorbar:2008hu},
see Section~\ref{sec:QHEgraphene}. (In condensed matter literature, the corresponding 
time-reversal breaking mass is called the Haldane mass \cite{Haldane:1988zza}.)

It should be emphasized that the Dirac mass and the chemical potential terms in the 
(inverse) full propagator (\ref{ginverse}) are determined by $m$ and $\mu$ that may differ
from their tree level counterparts, $m_0$ and $\mu_0$. While $m_0$ is the bare fermion
mass, $m$ has the physical meaning of a dynamical mass that, in general, depends on
the density and temperature of the matter, as well as on the strength of interaction.
Concerning the chemical potentials, it is $\mu_0$ that is the chemical potential in the
thermodynamic sense. The value of $\mu$, on the other hand, is an ``effective" chemical
potential that determines the quasiparticle dispersion relations in interacting theory.

In order to determine the values of the parameters $m$, $\mu$, $\Delta$ and $\tilde{\mu}$
in the model at hand, we will use the Schwinger-Dyson (gap) equation for the full
fermion propagator. By making use of the effective action for composite operators 
\cite{Luttinger:1960ua,Baym:1961zz,Cornwall:1974vz}, see Appendix~\ref{App:EffActionThermoEnergy}, 
we obtain the following equation in the mean-field approximation:
\begin{eqnarray}
G^{-1}(u,u^\prime) = S^{-1}(u,u^\prime)
- i G_{\rm int} \left\{ G(u,u) -  \gamma^5 G(u,u) \gamma^5 
- \mathrm{tr}[G(u,u)] +  \gamma^5\, \mathrm{tr}[\gamma^5G(u,u)]\right\}
\delta^{4}(u- u^\prime). 
\label{CSE-CME:gap}
\end{eqnarray}
While the first two terms in the curly brackets describe the exchange (Fock) interaction,
the last two terms describe the direct (Hartree) interaction. By making use of the 
ansatz (\ref{ginverse}) for the full fermion propagator, we rewrite this gap equation 
in the following equivalent form:
\begin{equation}
(\mu-\mu_0)\gamma^0 +i\gamma^1\gamma^2 \tilde{\mu}
+i\Delta\gamma^0\gamma^1\gamma^2 -m+m_0 = -\frac{1}{2}G_{\rm int}
\left(\gamma^0  \langle j^0\rangle +\gamma^3\gamma^5 \langle j_5^3\rangle\right) 
+G_{\rm int}  \langle\bar\psi \psi\rangle ,
\label{CSE-CME:gap-more}
\end{equation}
where $\langle \bar\psi \psi\rangle$ is the chiral condensate, $\langle j^0\rangle$ is the 
fermion density, and $\langle j_5^3\rangle$ is the axial current density. The corresponding 
ground state expectation values are defined in terms of the full propagator, 
\begin{eqnarray}
\langle \bar\psi \psi\rangle &=& -\mathrm{tr}\left[G(u,u)\right],
\label{condensate} \\
\langle j^0\rangle &=& -\mathrm{tr}\left[\gamma^0G(u,u)\right],
\label{charge}  \\
\langle j_5^3\rangle &=& -\mathrm{tr}\left[\gamma^3\gamma^5G(u,u)\right].
\label{current} 
\end{eqnarray}
We note that there is no Dirac structure of the anomalous magnetic moment type 
on the right hand side of the matrix gap equation (\ref{CSE-CME:gap-more}). This 
implies that only the zero value of parameter $\tilde{\mu}$ is consistent with the
equation. This is an artifact of the mean-field approximation in the model at hand 
\cite{Gorbar:2011ya}. In the remainder of this subsection, therefore, we will set 
$\tilde{\mu} = 0$. We should mention, however, that $\tilde{\mu}$ may be 
nonzero in models with other types of interactions \cite{Gorbar:2007xh,
Gorbar:2008hu,Ferrer:2008dy} and probably even in the NJL model used, 
but beyond the leading order mean-field approximation.

The matrix form of the gap equation in Eq.~(\ref{CSE-CME:gap-more}) can be 
equivalently rewritten as a coupled set of three equations for the dynamical 
parameters $m$, $\mu$, and $\Delta$, i.e.,
\begin{eqnarray}
m &=& m_0 - G_{\rm int}   \langle \bar\psi \psi\rangle  ,
\label{CSE-CME:gap-m-text}\\
\mu &=&\mu_0 -\frac{1}{2}G_{\rm int} \langle j^0\rangle ,
\label{CSE-CME:gap-mu-text}  \\
\Delta &=& -\frac{1}{2}G_{\rm int} \langle j_5^3\rangle  .
\label{CSE-CME:gap-Delta-text} 
\end{eqnarray}
As we see, the gap equations depend only on the full fermion propagator $G(u,u^\prime)$ 
at $u^\prime=u$. This fact greatly simplifies the analysis. Of course, it is related to the fact that 
we use the model with point-like interaction. This feature will be lost in more realistic models with 
long-range interactions.

The explicit expression for the fermion propagator in the Landau-level representation is calculated 
in Appendix~\ref{CSE:App:Landau-level-rep}. In the case when $m$, $\mu$ and $\Delta$ are 
the only nonzero parameters in the propagator, the result takes the form:
\begin{equation} 
G(u,u) =
\frac{i}{2\pi l^2}\sum_{n=0}^{\infty} \int\frac{d\omega d k^{3}}{(2\pi)^2}
\frac{{\cal K}_{n}^{+}{\cal P}_{+}+{\cal K}_{n}^{-}{\cal P}_{-}\theta(n-1)}{U_n},
\label{CSE-CME:full-propagator}
\end{equation}
where $l=1/\sqrt{|eB|}$ is the magnetic length, $\theta(n-1)\equiv 1$ for $n\geq 1$, 
and $\theta(n-1)\equiv0$ for $n\leq 0$. We also use the standard spin projectors:
\begin{equation}
{\cal P}_{\pm} =\frac12\left(1\pm is_{\perp}\gamma^1\gamma^2\right),
\label{CSE-CME:projectorP}
\end{equation}
with the shorthand notation $s_{\perp}\equiv \mathrm{sign}(e B )$. 
The explicit form of functions ${\cal K}_{n}^{\pm}$ can be extracted from the more
general expression in Eq.~(\ref{inverse-M-2neB}), and function $U_n$ is given in 
Eq.~(\ref{U_inv-simple}). For the reader's convenience, let us quote them in the special 
case, which is of the main interest for us here, i.e., when $m$, $\mu$ and $\Delta$ are 
the only nonzero parameters,
\begin{eqnarray}
{\cal K}_{n}^{\pm}&=& \left[(\omega+\mu \mp s_{\perp}\Delta)\gamma^0
+ m  -k^{3}\gamma^3\right]\left[
(\omega+\mu)^2  - m^2 - \Delta^2 - (k^{3})^2 - 2n|eB|
\mp 2 s_{\perp}\Delta \left(m + k^{3} \gamma^3 \right)\gamma^0
\right],
\label{CSE-CME:K_n^pm_text}
\\
U_n &=&
 \left[(\omega+\mu)^2 - 2n|eB| -\left(s_\perp \Delta-\sqrt{m^2+(k^{3})^2}\right)^2\right]
 \left[(\omega+\mu)^2 - 2n|eB| -\left(s_\perp \Delta+\sqrt{m^2+(k^{3})^2}\right)^2\right].
\label{CSE-CME:U_n_text}
\end{eqnarray}
By making use of these expression, the $n$th Landau level contribution to the
fermion propagator can be written in the following form:
\begin{equation}
\frac{{\cal K}_{n}^{\pm} {\cal P}_{\pm}}{U_n} =  \gamma^0 \left[
\frac{\omega+\mu -i s_{\perp} \gamma^1\gamma^2 \left(s_{\perp} \Delta - \sqrt{m^2+(k^{3})^2} \right)}
{(\omega+\mu)^2 - \left(s_{\perp}\Delta-\sqrt{m^2+(k^{3})^2} \right)^2 - 2n|eB|}{\cal H}^{-}
+
\frac{\omega+\mu -i s_{\perp} \gamma^1\gamma^2 \left(s_{\perp} \Delta + \sqrt{m^2+(k^{3})^2} \right)}
{(\omega+\mu)^2 - \left(s_{\perp}\Delta+\sqrt{m^2+(k^{3})^2} \right)^2 - 2n|eB|}{\cal H}^{+}
\right]{\cal P}_{\pm},
\label{CSE-CME:K_P_U}
\end{equation}
where
\begin{equation}
{\cal H}^{\pm}=\frac{1}{2}\left(1 \pm s_{\perp}\frac{ m  +k^{3}\gamma^3}{\sqrt{m^2+(k^{3})^2} }\gamma^5\gamma^3\right)
\label{CSE-CME:projectorH}
\end{equation}
are the projectors on the quasiparticle states, whose energies are given in terms of either the sum 
or the difference of $s_{\perp}\Delta$ and $\sqrt{m^2+(k^{3})^2}$. The projectors take a particularly simple form 
in the massless limit: ${\cal H}_{m=0}^{\pm}=\frac{1}{2}\left[1 \pm s_{\perp}\mathrm{sign}(k^3)\gamma^5 \right]$. 
In this case, ${\cal H}_{m=0}^{\pm}$ almost coincide (up to the sign of the longitudinal momentum) 
with the chirality projectors ${\cal P}_{5}^{\pm}=\frac{1}{2}\left(1 \pm s_{\perp} \gamma^5 \right)$. 
For each choice of the signs of $eB$ and $k^3$, the chirality of the states 
can be easily determined from the following relation between the two sets of projectors:
\begin{equation}
{\cal H}_{m=0}^{\pm}=\frac{1 \mp \mathrm{sign}(k^3)}{2} {\cal P}_{5}^{-}+\frac{1 \pm \mathrm{sign}(k^3)}{2} {\cal P}_{5}^{+}.
\label{CSE-CME:Hpm-P5pm}
\end{equation}
By making use of this relation and taking into account that $ \sqrt{m^2+(k^{3})^2}\to |k^{3}|$ when $m\to0$, 
it is straightforward to check that the propagator in Eq.~(\ref{CSE-CME:full-propagator}) in the massless limit takes 
the form \cite{Gorbar:2009bm}: 
\begin{equation}
G(u,u)=G_0^{+}{\cal P}_{+}
+\sum_{n=1}^{\infty}\left(G_{n}^{+}{\cal P}_{+}+G_{n}^{-}{\cal P}_{-}\right),
\end{equation}
where 
\begin{equation}
G_{n}^{\pm} =
\frac{i|eB|\gamma^0}{2\pi}\int\frac{d\omega d k^3}{(2\pi)^2}
 \Big[
 \frac{\omega+\mu \pm[k^3 - s_{\perp} \Delta ]}
      {(\omega+\mu)^2 - 2n|eB| - [k^3- s_{\perp} \Delta]^2 }{\cal P}_{5}^{-}
+\frac{\omega+\mu \mp[k^3+  s_{\perp} \Delta]}
      {(\omega+\mu)^2 - 2n|eB| - [k^3+s_{\perp} \Delta]^2 }{\cal P}_{5}^{+}
\Big].
\label{prop:G_n_pm}
\end{equation}   
The opposite chirality fermions, described by such a propagator, are characterized by a relative 
shift of the longitudinal momenta, $k^{3}\to k^{3}\pm s_{\perp}\Delta$, in their dispersion relations.

Unlike the higher Landau level terms in the propagator, the LLL contribution is rather simple,
\begin{eqnarray}
\frac{{\cal K}_{0}^{+} {\cal P}_{+}}{U_0} 
&=&\gamma^0  \left[\frac{{\cal H}^{-}}{\omega+\mu + s_{\perp} \Delta -  \sqrt{m^2+(k^{3})^2} } 
+ \frac{{\cal H}^{+}}{\omega+\mu + s_{\perp} \Delta +  \sqrt{m^2+(k^{3})^2} }\right]{\cal P}_{+}.
\label{CSE-CME:K_P_U-LLL}
\end{eqnarray}
In the LLL, $s_{\perp} \Delta$ is a part of the effective chemical potential $\mu - s_{\perp} \Delta$,
and the two terms in Eq.~(\ref{CSE-CME:K_P_U-LLL}) can be associated with the antiparticle (negative energy) and 
particle (positive energy) contributions, respectively. In order to avoid a potential confusion, let us also mention 
that, as seen from Eq.~(\ref{CSE-CME:K_P_U}), the connection between ${\cal H}^{\pm}$ and the particle/antiparticle 
states is not preserved in the higher Landau levels. 

As follows from Eq.~(\ref{CSE-CME:K_P_U}) and Eq.~(\ref{CSE-CME:K_P_U-LLL}), the poles of the full fermion 
propagator are at
\begin{equation}
\omega_0 = -\mu - s_{\perp} \Delta\pm\sqrt{m^2+(k^{3})^2},
\label{CSE-CME:omega0}
\end{equation}
for the lowest Landau level, and at
\begin{equation}
\omega_n = -\mu\pm \sqrt{\left(s_{\perp}\Delta\pm\sqrt{m^2+(k^{3})^2}\right)^2 + 2n|eB| },
\label{CSE-CME:omegan}
\end{equation}
for higher Landau levels, $n\geq 1$. Note that all four combinations of signs are possible for
the latter.

Let us first discuss the normal phase at zero temperature and in the chiral limit, when 
$m = m_0 = 0$ and $\langle \bar\psi \psi\rangle = 0$. This will be realized when the 
chemical potential $\mu_0 > m_{\rm dyn}/\sqrt{2}$ \cite{Gorbar:2009bm}, where $m_{\rm dyn}$ 
is a dynamical fermion mass in a magnetic field at zero chemical potential and zero 
temperature. We can solve Eqs.~(\ref{CSE-CME:gap-mu-text}) and (\ref{CSE-CME:gap-Delta-text}) 
in weak coupling perturbation theory. To the zeroth order in coupling, the fermions are noninteracting 
and we have $\mu=\mu_0$ and $\Delta=0$. The fermion density $\langle j^0\rangle$ 
and the axial current density $\langle j_5^3\rangle$ are nonzero, however. They are 
\begin{eqnarray}
\langle j^0\rangle_0 &=& \frac{\mu_0|eB|}{2\pi^2} +\frac{\mathrm{sign}(\mu_0)|eB|}{\pi^2}
\sum_{n=1}^{\infty}
\sqrt{\mu_{0}^2-2n|eB|}\,
 \theta\left(|\mu_0|-\sqrt{2n|eB|}\right),
\label{j}
\\
\langle j^3_5\rangle_0 &=& \frac{eB}{2\pi^2}\mu_0\, ,
\label{MZ}
\end{eqnarray}
respectively. By making use of Eq.~(\ref{CSE-CME:gap-Delta-text}), to the first order in coupling, 
we then obtain 
\begin{equation}
\Delta \simeq -\frac{1}{2}G_{\rm int} \langle j_5^3\rangle_0 
= -G_{\rm int} \frac{eB}{4\pi^2}\mu_0 \neq 0,
\end{equation}
which shows that a nonzero shift parameter $\Delta$ is {\it necessarily} present in the normal phase of
the theory. This was one of the main results of Ref.~\cite{Gorbar:2009bm}. We emphasize that 
$\Delta$ is generated by perturbative dynamics, which should not be surprising because the 
vanishing $\Delta$ is not protected by any symmetry of the theory. [Recall that ${\cal C} =+1$, 
${\cal P} =+1$, and ${\cal T} = -1$ for the axial current density $j_5^3$, and beside parity 
${\cal P}$, all the discrete symmetries are broken in model (\ref{CSE-CME:NJLmodel}).] Similarly, 
from Eq.~(\ref{CSE-CME:gap-mu-text}) we find that $\mu - \mu_0 \propto G_{\rm int}\langle j^0\rangle_0 \neq 0$, 
i.e., $\mu$ and $\mu_0$ are different. The origin of this difference can be traced to the Fock 
terms in the gap equation, see Eqs.~(\ref{CSE-CME:gap}) and (\ref{CSE-CME:gap-more}).

This result was obtained for the case of zero temperature. As will be shown below in 
Section~\ref{ChiralAsymNJLMCshift}, the chiral shift parameter is rather insensitive to the value 
of the temperature in the regime of cold dense matter appropriate for potential applications in 
stars. In the case of hot matter, which is relevant for heavy ion collisions, a temperature larger 
than the chemical potential may even play an important role in enhancing the chiral shift 
parameter.

By noting from Eq.~(\ref{CSE-CME:gap-Delta-text}) that the chiral shift $\Delta$ is induced 
by the axial current, it is naturally to ask whether $\Delta$ itself affects this current. The 
answer to this question is affirmative \cite{Gorbar:2009bm}. Another natural question is 
whether the divergence of this modified current satisfies the conventional anomaly equation 
\cite{Adler:1969gk,Bell:1969ts}. As will be shown in the next subsection, the answer to this 
question is also affirmative.

In conclusion, let us briefly discuss the following issue. One may think that when
the fermion mass $m$ is zero, the term with the chiral shift $\Delta$ is unphysical
because it could be formally removed by the gauge transformation 
$\psi\to e^{iz\gamma_5 \Delta}\psi$, $\bar{\psi} \to \bar{\psi}e^{iz\gamma_5\Delta}$.  
However, the point is that this transformation is singular (anomalous). It follows 
from the following two facts: (i) as was already pointed out above, in the LLL, $s_{\perp}\Delta$ 
is a part of the chemical potential, see Eq.~(\ref{CSE-CME:K_P_U-LLL}),
and (ii) this happens because the LLL dynamics is $1 + 1$-dimensional \cite{Gusynin:1994xp,Gusynin:1995nb}.
It is well known that in $1 + 1$ dimensions this transformation, which formally varies the value of the 
chemical potential, is anomalous (for a recent thorough discussion of this transformation, 
see Ref.~\cite{Kojo:2009ha}).

\subsubsection{Chiral shift versus magnetic catalysis}
\label{ChiralAsymNJLMCshift}

To set up a benchmark for the numerical results, it is instructive to start the analytical analysis 
of gap equations (\ref{CSE-CME:gap-mu-text}), (\ref{CSE-CME:gap-m-text}), and (\ref{CSE-CME:gap-Delta-text}) at zero 
temperature. We will use two regularization schemes: (i) the gauge noninvariant one, with 
a sharp momentum cutoff, $|k^{3}|\leq \Lambda$, in the integrals over $k^{3}$ (which are 
always performed first) and a smooth cutoff in the sums over the Landau levels (which are 
performed last), and (ii) the gauge invariant proper-time regularization. It will be shown that 
the results in these two regularizations are qualitatively the same.

Let us first analyze the equations using the momentum cutoff regularization. 
The smoothing function in the sums over the Landau levels is taken in the following form:
\begin{equation}
\kappa(n)= \frac{\sinh(\Lambda/\delta\Lambda)}
{\cosh\left((\Lambda/\delta\Lambda)\sqrt{n/n_{\rm cut}}\right)+\cosh(\Lambda/\delta\Lambda)},
\label{CSE-CME:kappa}
\end{equation}
where the cutoff value $n_{\rm cut}$ is determined by the number of the Landau levels below 
the energy scale set by $\Lambda$, i.e., $n_{\rm cut} \equiv \left[\Lambda^2/2|eB|\right]$ with 
the square brackets denoting the integer part. The width of the energy window in which the 
cutoff is smoothed is determined by the ratio $\delta\Lambda/\Lambda$. When the value of 
the latter goes to zero, the function $\kappa(n)$ approaches the step function. (In the numerical 
calculations below we use $\delta\Lambda/\Lambda=1/20$.) 

Also, instead of using a dimensionful coupling constant $G_{\rm int}$, let us introduce the 
following dimensionless coupling constant
\begin{equation}
g \equiv \frac{G_{\rm int}\Lambda^2}{4\pi^2},
\label{CSE-CME:g}
\end{equation}
The weakly coupled limit corresponds to $g \ll 1$. Note that the coupling constant $g$ is 
defined in such a way that the critical value for generating a fermion dynamical mass in 
the NJL model without magnetic field is $g_{\rm cr} =1$.

There exist two qualitatively different types of solutions to the gap equation. 
The first of them describes a zero density symmetry broken state where the dynamics 
is dominated by the magnetic catalysis. The other is a normal state with restored 
symmetry at finite density. Let us describe both of these solutions in detail, following 
the original studies in Refs.~\cite{Gorbar:2009bm,Gorbar:2011ya}.

{\it Solution of type I}.
The first solution type corresponds to $m\neq 0$ and $\Delta=0$ in accordance 
with the magnetic catalysis scenario in the vacuum discussed in Section~\ref{sec:MagneticCatalysis}.
For $m\neq 0$ and $\Delta=0$, the zero temperature expressions for the chiral condensate 
$\langle \bar\psi \psi\rangle$, and the vacuum expectation values of the fermion density 
$j^0$ and the axial current density $j_5^3$, take the following form \cite{Gorbar:2011ya}:
\begin{eqnarray}
\langle \bar\psi \psi\rangle&\simeq & -\frac{m}{2(\pi l)^2}\left[\ln\frac{2\Lambda}{|m|}
-\ln\frac{|\mu|+\sqrt{\mu^2-m^2}}{|m|}\theta\left(|\mu|-|m|\right) \right]
-\frac{m}{(\pi l)^2}\sum_{n=1}^{\infty} \left[\ln\frac{\Lambda+\sqrt{\Lambda^2+m^2+2n|eB|}}{\sqrt{m^2+2n|eB|}}\right.
\nonumber \\
&&\left.
-\ln\frac{|\mu|+\sqrt{\mu^2-m^2-2n|eB|}}{\sqrt{m^2+2n|eB|}}
\theta\left(|\mu|-\sqrt{m^2+2n|eB|}\right)\right],
\label{CSE-CME:BT0}
\\
\langle j^0\rangle&=& \frac{\mathrm{sign}(\mu)}{2(\pi l)^2}\sqrt{\mu^2-m^2} \theta\left(|\mu|-|m|\right)
+\frac{\mathrm{sign}(\mu)}{(\pi l)^2}\sum_{n=1}^{\infty} \sqrt{\mu^2-2n|eB|-m^2}\,
\theta\left(|\mu|-\sqrt{m^2+2n|eB|}\right),
\label{CSE-CME:AT0}
\\
\langle j_5^3\rangle&\simeq & s_\perp\frac{\mathrm{sign}(\mu)}{2(\pi l)^2}\sqrt{\mu^2-m^2} \theta\left(|\mu|-|m|\right).
\label{CSE-CME:DT0}
\end{eqnarray}
Note that the expression for $\langle j_5^3\rangle$ is proportional to the LLL contribution to the fermion density
and, as a result, vanishes when $|\mu|<|m|$. In this case, a solution with $\Delta=0$ is consistent with
gap equation (\ref{CSE-CME:gap-Delta-text}). Then, the other two gap equations reduce down to $\mu =\mu_0$
and
\begin{equation}
m = m_0 + \frac{2 g m}{(\Lambda l)^2}
\left(\ln\frac{2\Lambda}{|m|} 
+2\sum_{n=1}^{\infty}\kappa(n)\ln\frac{\Lambda+\sqrt{\Lambda^2+m^2+2n|eB|}}{\sqrt{m^2+2n|eB|}}
\right),
\label{CSE-CME:gap-m-Delta0}
\end{equation}
where we utilized the smooth cutoff function (\ref{CSE-CME:kappa}) in the sum over the Landau levels. 

{\it Solution of type II}.
In the chiral limit, in addition to the solution with a nonzero Dirac mass $m$, the gap equation
also allows a solution with $m=0$ and a nonzero chiral shift parameter $\Delta$. To see this,
we derive the functions that appear on the right hand sides of the gap equations for this
special case:
\begin{eqnarray}
\langle \bar\psi \psi\rangle&=& 0,
\label{CSE-CME:BT0m0}
\\
\langle j^0\rangle&=& \frac{ \mu-s_{\perp}\Delta }{2(\pi l)^2} +\frac{\mathrm{sign}(\mu)}{(\pi l)^2} \sum_{n=1}^{N_B}
\sqrt{\mu^2-2n|eB|}  ,
\label{CSE-CME:AT0m0}
\\
\langle j_5^3\rangle&=& \frac{1}{2(\pi l)^2}\left(s_{\perp} \mu - \Delta
-2 \Delta \sum_{n=1}^{\infty}\kappa(n)\right),
\label{CSE-CME:DT0m0}
\end{eqnarray}
where $N_B$ is the integer part of $\mu^2/(2|eB|)$.
(Here it might be appropriate to note that the above result for $\langle j_5^3\rangle$ remains
unchanged also at nonzero temperatures!) The fact that
now $\langle \bar\psi \psi\rangle= 0$ is in agreement with Eq.~(\ref{CSE-CME:gap-m-text})
and the assumption $m=m_0=0$. The remaining two equations, (\ref{CSE-CME:gap-mu-text}) and (\ref{CSE-CME:gap-Delta-text}),
reduce down to
\begin{equation}
\mu =\mu_0+  \frac{g}{(\Lambda l)^2}\left( s_{\perp}\Delta -\mu -2 \mathrm{sign}(\mu)\sum_{n=1}^{N_B}\sqrt{\mu^2-2n|eB|}\right)
\label{CSE-CME:TypeII-mu-equation}
\end{equation}
and
\begin{equation}
\Delta = \frac{g }{(\Lambda l)^2}\left(s_{\perp} \mu - \Delta
-2 \Delta \sum_{n=1}^{\infty}\kappa(n)\right),
\label{CSE-CME:TypeII-Delta-equation}
\end{equation}
respectively. To leading order in the coupling constant, the solutions for $\mu$ and $\Delta$
are straightforward,
\begin{eqnarray}
\mu & \simeq & \frac{\mu_0}{1+g/(\Lambda l)^2},   
\label{CSE-CME:mu-cutoff}\\
\Delta &=& -\frac{g s_{\perp} \mu}{(\Lambda l)^2 +g \left[1 +2\sum_{n=1}^{\infty}\kappa(n)\right]},
\label{CSE-CME:Delta-cutoff}
\end{eqnarray}
which are derived under the assumption that the chemical potential $\mu_0$ is not large enough for the
first Landau level to start filling up, i.e., $|\mu|\lesssim \sqrt{2 |eB|}$. When the chemical potential
becomes larger, the result for $\mu$ will get corrections, but the expression for $\Delta$ in terms of
$\mu$ will keep the same form. 
Note that with the function $\kappa(n)$ given in Eq.~(\ref{CSE-CME:kappa}), one 
finds that 
\begin{equation}
\sum_{n=1}^{\infty} \kappa\left(\sqrt{2n|eB|},\Lambda\right) = a \frac{\Lambda^2}{|eB|} ,
\label{CSE-CME:sumkappa}
\end{equation}
where $a=O(1)$. [For the chosen value of the regulating parameter 
$\delta\Lambda/\Lambda=1/20$, the numerical results for $a$ are fitted well by 
the following expression: $a\approx 0.504+0.492|eB|/\Lambda^2$, where the 
subleading term in the inverse powers of the cutoff parameter is also included.]

Let us now redo the analysis using the gauge invariant proper-time 
regularization. As expected, we find the same two types of the solutions.

{\it Solution of type I}. 
In the regime of magnetic catalysis, we have shown above that the dynamical mass parameter 
satisfies Eq.~(\ref{CSE-CME:gap-m-Delta0}) in the momentum cutoff regularization scheme. 
Now, let us show that this is consistent with the result obtained in the proper-time regularization. 
The expression for the vacuum part of function $\langle \bar\psi \psi\rangle$ in this regularization
is given in Eq.~(\ref{Condensate3+1Free}). The corresponding gap equation in the chiral limit, 
$m_0=0$, was analyzed in Section~\ref{NJL3+1:EffPot}. To leading order in coupling, it is equivalent 
to Eq.~(\ref{CSE-CME:gap-m-Delta0}). The approximate analytical solution for $m_{\rm dyn}$ was 
given in Eq.~(\ref{m_dyn_weak_3+1}). The corresponding solution exists for $|\mu_0| < m_{\rm dyn}$. 
As we will discuss below, however, it corresponds to the ground state only in a part of this parameter
range, i.e., $|\mu_0| \lesssim m_{\rm dyn}/\sqrt{2}$.

{\it Solution of type II}. 
Now let us consider the chiral limit and search for a solution with $m=0$ and a nonzero chiral shift 
parameter $\Delta$ using the proper-time representation. In this case, $\langle \bar\psi \psi\rangle=0$, 
and the expressions for $\langle j^0\rangle$ and $\langle j_5^3\rangle$ are given by the same 
formal expressions as in Eqs.~(\ref{CSE-CME:AT0m0}) and (\ref{CSE-CME:DT0m0}). Function 
$\langle j_5^3\rangle$ contains ultraviolet divergences. The corresponding regularized expression 
is derived in Ref.~\cite{Gorbar:2011ya},  
\begin{equation}
\langle j_5^3\rangle= \frac{\sqrt{\pi}\Lambda}{2(2\pi l)^2} e^{-(\Delta/\Lambda)^2 }
\mbox{erfi}\left(\frac{\Delta}{\Lambda} \right)\coth\left(\frac{eB}{\Lambda^2} \right)
-\frac{2 s_{\perp}\mu}{(2\pi l)^2},
\end{equation}
where $\mbox{erfi}(x) \equiv -i\,\mbox{erf}(ix)$ 
is the imaginary error function. By expanding the expression for 
$\langle j_5^3\rangle$ in inverse powers of $\Lambda$, we arrive at the following approximate result:
\begin{equation}
\langle j_5^3\rangle\simeq -\frac{1}{2(\pi l)^2}\left( s_{\perp}\mu-\frac{(\Lambda l)^2}{2}\Delta\right),
\label{CSE-CME:DT0m0-proper-time}
\end{equation}
which is in agreement with the result in Eq.~(\ref{CSE-CME:DT0m0}) after making the identification
$\frac{1}{2}(\Lambda l)^2 \equiv 1 +2\sum_{n=1}^{\infty}\kappa(n) \simeq 2a(\Lambda l)^2$, where
the parameter $a$ is defined in Eq.~(\ref{CSE-CME:sumkappa}). As we see, $a= 1/4$ in the proper-time regularization.

The gap equation for $\mu$ is insensitive to the ultraviolet dynamics and coincided with Eq.~(\ref{CSE-CME:TypeII-mu-equation}). 
By making use of the approximation in Eq.~(\ref{CSE-CME:DT0m0-proper-time}), we arrive at the following 
equation for $\Delta$:
\begin{equation}
\Delta  = \frac{g }{(\Lambda l)^2}\left(s_{\perp} \mu - \frac{(\Lambda l)^2}{2}\Delta \right),
\end{equation}
which is equivalent to Eq.~(\ref{CSE-CME:TypeII-Delta-equation}) after the same identification of the regularization schemes 
is made. Also, the proper-time solution,
\begin{eqnarray}
\mu & \simeq & \frac{\mu_0}{1+g/(\Lambda l)^2},\\
\Delta &=& \frac{g s_{\perp} \mu}{(\Lambda l)^2 +\frac{1}{2}g (\Lambda l)^2},
\label{CSE-CME:Delta-proper-time}
\end{eqnarray}
is equivalent to the solution in Eqs.~(\ref{CSE-CME:mu-cutoff}) and (\ref{CSE-CME:Delta-cutoff}).

As should be clear from the above discussion, in the region $|\mu_0|< m_{\rm dyn}$, the two
inequivalent solutions coexist. In order to decide which of them describes the ground state, one
has to compare the corresponding free energies. The general expression for the free energy density
is derived in Appendix~\ref{App:FreeEnergyDensity3+1}. [For more details, see Ref.~\cite{Gorbar:2011ya}.]
For the two cases of interest here, the corresponding results read 
\begin{equation}
\Omega_{m} \simeq  -\frac{m_{\rm dyn}^2 |eB|}{2(2\pi)^2}
\left(1+\frac{m_{\rm dyn}^2}{2|eB|} \ln\frac{\Lambda^2}{|eB|}\right)
\label{CSE-CME:Omega0}
\end{equation}
and
\begin{equation}
\Omega_{\Delta} \simeq -\frac{\mu_0^2 |eB|}{(2\pi)^2}\left(1-g\frac{|eB|}{\Lambda^2}\right),
\label{CSE-CME:OmegaDelta0}
\end{equation}
respectively. In deriving the last expression, we used the approximate relations
$\mu\simeq \mu_0$ and $\Delta \simeq g\mu_0 eB/\Lambda^2$. By comparing the free
energies in Eqs.~(\ref{CSE-CME:Omega0}) and (\ref{CSE-CME:OmegaDelta0}), we see that the ground
state with a nonzero $\Delta$ becomes favorable when $\mu_0\gtrsim m_{\rm dyn}/\sqrt{2}$.
This is analogous to the Clogston relation in superconductivity \cite{Clogston:1962zz}.

\subsubsection{Numerical solution of gap equations}
\label{CSE-CME:Numerical}

In order to solve numerically the set of gap equations (\ref{CSE-CME:gap-mu-text}), 
(\ref{CSE-CME:gap-m-text}), and (\ref{CSE-CME:gap-Delta-text}), we have to regulate 
the divergences that appear in the integrals over the longitudinal momentum $k^{3}$ 
and the sums over the Landau levels in the expressions for the chiral condensate 
$\langle \bar\psi \psi\rangle$ and the axial current density $\langle j_5^3\rangle$. 
In Section~\ref{ChiralAsymNJLMCshift} we used two regularizations: (i) with a sharp 
momentum cutoff, $|k^{3}|\leq \Lambda$, in the integrals over $k^{3}$ (which are 
always performed first) and a smooth cutoff in the sums over the Landau levels, and 
(ii) the proper-time regularizations. By taking into account that the zero-temperature 
results in these two regularizations are qualitatively similar, we perform a detailed 
numerical analysis of the gap equations at nonzero temperature by using the first 
regularization only, which is technically much simpler to implement.

The form of the smoothing function $\kappa(n)$ in this regularization is given in Eq.~(\ref{CSE-CME:kappa}).
The width of the energy window in which the cutoff is smoothed is determined
by the ratio $\delta\Lambda/\Lambda$. When the value of this ratio goes to zero, $\kappa(n)$
approaches a step function, $\theta(n_{\rm cut}-n)$, corresponding to the case of a sharp cutoff at
$n_{\rm cut} = \left[\Lambda^2/2|eB|\right]$. We note, however, that taking a very sharp cutoff in the sums 
over the Landau levels may
result in some unphysical discontinuities in the physical properties of the model as a function of the
magnetic field. This is because of the discontinuities in the dependence of the function $n_{\rm cut}(|eB|)$,
which defines the number of the dynamically accessible Landau levels. In our numerical calculations,
we choose a reasonably large value $\delta\Lambda/\Lambda =1/20$.

In order to keep our model study as general as possible, we specify all energy/mass parameters in units
of the cutoff parameter $\Lambda$. In the numerical calculations below, we use the following values of
the coupling constant and the magnetic field:
\begin{eqnarray}
g &=& \frac{G_{\rm int}\Lambda^2}{4\pi^2} =0.25,
\label{CSE-CME:g-value}\\
|eB| &=& 0.125\Lambda^2 .
\label{CSE-CME:eB-value}
\end{eqnarray}
The above choice of coupling is sufficiently weak to justify the approximations used in the analysis. 
In real dense or hot quark matter, however, the actual value of dimensionless coupling may be stronger. 
In the degenerate electron gas in the interior of compact stars, on the other hand, it is still much
weaker. Our purpose here, however, is to perform a qualitative analysis of the model and
reveal the general features of the dynamics relevant for the generation of the chiral shift
parameter. Therefore, our ``optimal" choice of $g$ is sufficiently weak to make the analysis
reliable, while not too weak to avoid a very large hierarchy of the energy scales which would
make the numerical analysis too difficult. Similar reasoning applies to the choice of the
magnetic field in Eq.~(\ref{CSE-CME:eB-value}). This is a sufficiently strong field that makes it
easier to explore and understand the qualitative features of the dynamics behind both the
magnetic catalysis and the generation of the chiral shift parameter. In applications related
to compact stars, the actual fields might be considerably weaker. However, this value may in
fact be reasonable for applications in heavy ion collisions \cite{Kharzeev:2004ey,Kharzeev:2007tn,Kharzeev:2007jp,Fukushima:2008xe,Nam:2010nk,Skokov:2009qp,Bzdak:2011yy,Voronyuk:2011jd,Deng:2012pc}.

The gap equation is solved by multiple iterations of the gap equations. The
convergence is checked by measuring the following error function:
\begin{equation}
\epsilon_n =\sqrt{\sum_{i=1}^{3}\frac{(x_{i,n}-x_{i,n-1})^2}{\mbox{max}(x_{i,n}^2,x_{i,n-1}^2)}},
\end{equation}
where ${x}_i=\mu,\Delta,m$ for $i=1,2,3$, respectively. (In the case when both $x_{i,n}$
and $x_{i,n-1}$ vanish, the corresponding $i$th contribution to $\epsilon_n$ is left out.)
When the value of $\epsilon_n$ becomes less than $10^{-4}$ (at $T\neq 0$) or $10^{-5}$
(at $T=0$), the current set of $\mu_{n}$, $\Delta_{n}$ and $m_{n}$ is accepted as an
approximate solution to the gap equation. Usually, the convergence is achieved after
several dozens of iterations. In some cases, even as few as five iterations suffices
to reach the solution with the needed accuracy. This is often the case when we
automatically sweep over a range of values of some parameter (e.g., the temperature
or the chemical potential) and use the solution obtained at the previous value of the
parameter as the starting guess to solve the equation for a new nearby value of the same
parameter. However, even in this approach, the required number of iterations may
sometimes be in the range of hundreds. This is usually the case when the dynamically
generated $\Delta$ and $m$ have a steep dependence on the model parameters,
which is common, e.g., in the vicinity of phase transitions.

\begin{figure}[t]
\begin{center}
\includegraphics[width=.45\textwidth]{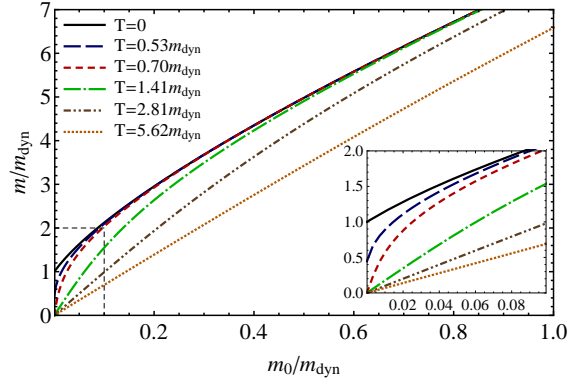}
\caption{(Color online) The dependence of the constituent mass $m$ on the bare
mass $m_0$  at $\mu_0=0$ for several fixed values of temperature:
$T=0$ (black), 
$T=3.75\times 10^{-4}\Lambda=0.53m_{\rm dyn}$ (blue), 
$T=5\times 10^{-4}\Lambda=0.70m_{\rm dyn}$ (red), 
$T=10^{-3}\Lambda=1.41m_{\rm dyn}$ (green),
$T=2\times 10^{-3}\Lambda=2.81m_{\rm dyn}$ (dark brown), and 
$T=4\times 10^{-3}\Lambda=5.62m_{\rm dyn}$ (light brown). 
The inset shows the details in the rectangular area around the origin.} 
\label{CSE-CME:figMass-vs-m0}
\end{center}
\end{figure}

In order to set up the reference point for the nonzero chemical potential calculations,
let us start by presenting the results for the constituent fermion mass as a function
of the bare mass $m_0$ at $\mu_0=0$. The corresponding zero temperature dependence is
shown by the black line in Fig.~\ref{CSE-CME:figMass-vs-m0}. As expected, the mass 
approaches the value of $m_{\rm dyn}$ in the chiral limit ($m_0 \to 0$). 
For the model parameters used here, it reads:
\begin{equation}
m_{\rm dyn} \approx 7.1 \times 10^{-4} \, \Lambda.
\end{equation}
In the same figure, we also
plotted the results for several nonzero values of temperature. These results show that
the value of the dynamical mass in the chiral limit gradually vanishes with increasing
the temperature. Within our numerical accuracy, the corresponding value of the critical
temperature is consistent with the Bardeen-Cooper-Schrieffer theory relation, $T_c \approx 0.57 m_{\rm dyn}$.
We also note that the results for $\mu$ and $\Delta$ are trivial at all temperatures
when $\mu_0=0$.

\begin{figure}[t]
\includegraphics[width=.45\textwidth]{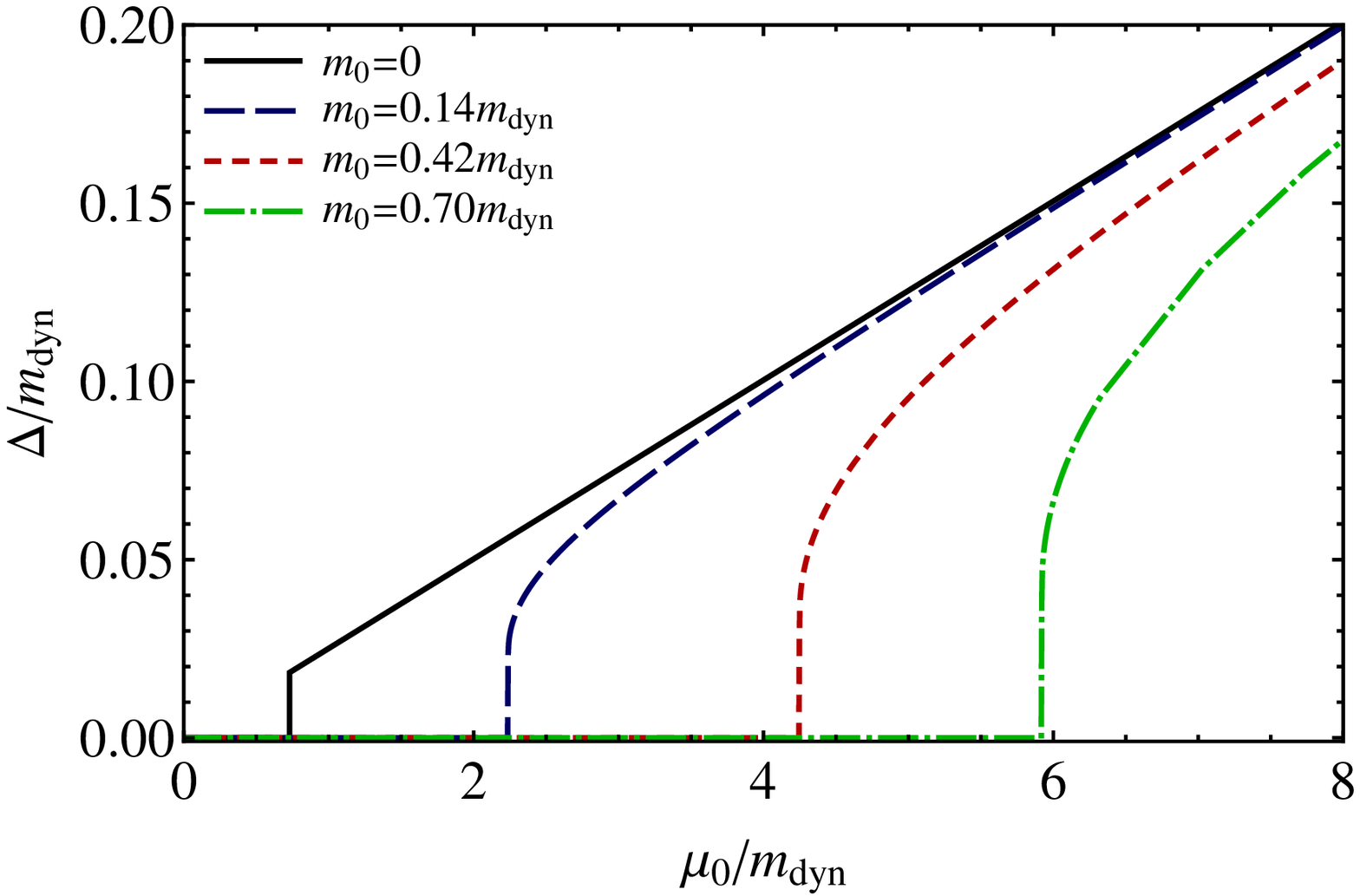}
\hspace{10pt}
\includegraphics[width=.45\textwidth]{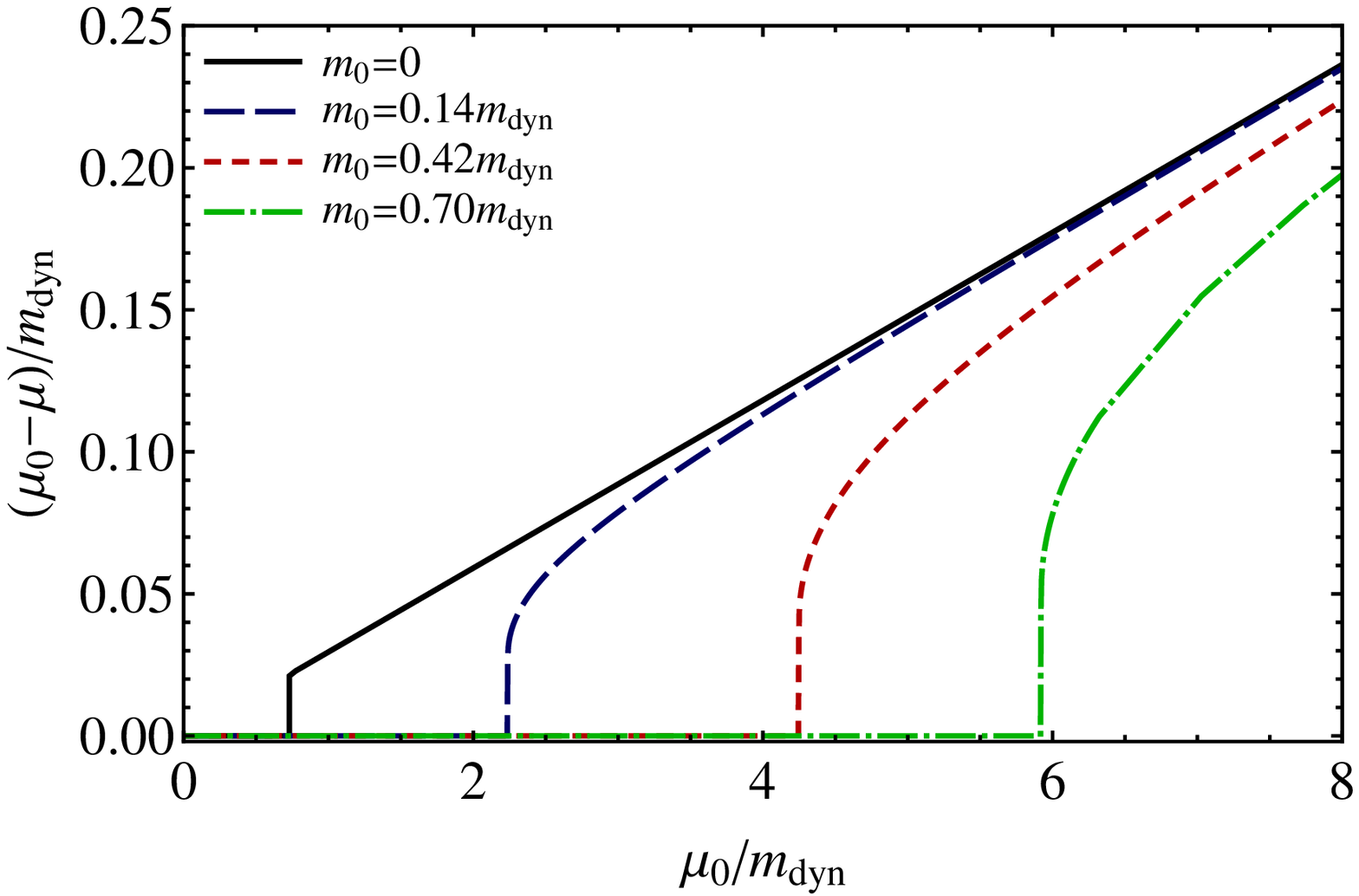}\\
\hspace*{10pt}
\includegraphics[width=.425\textwidth]{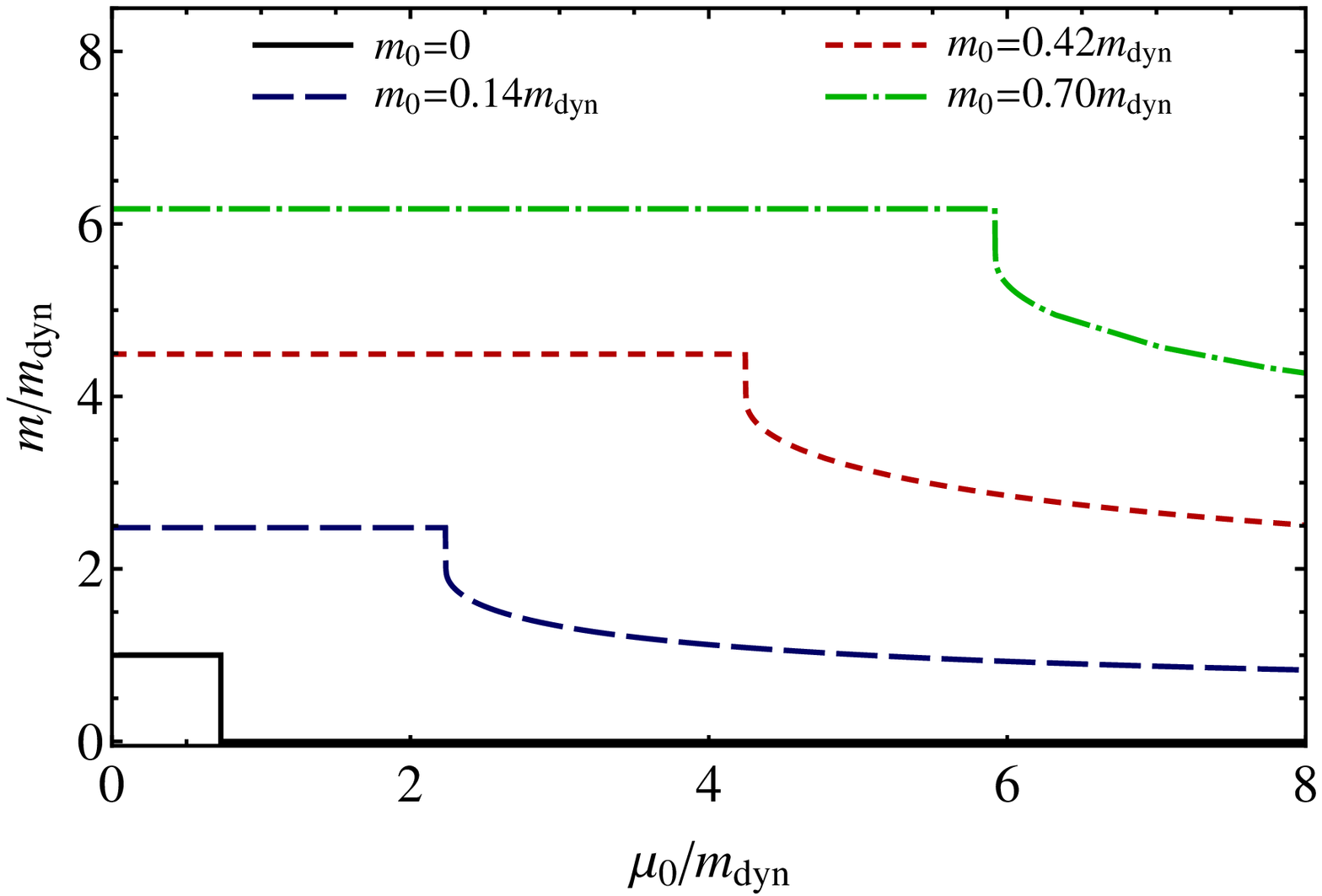}
\hspace{10pt}
\includegraphics[width=.45\textwidth]{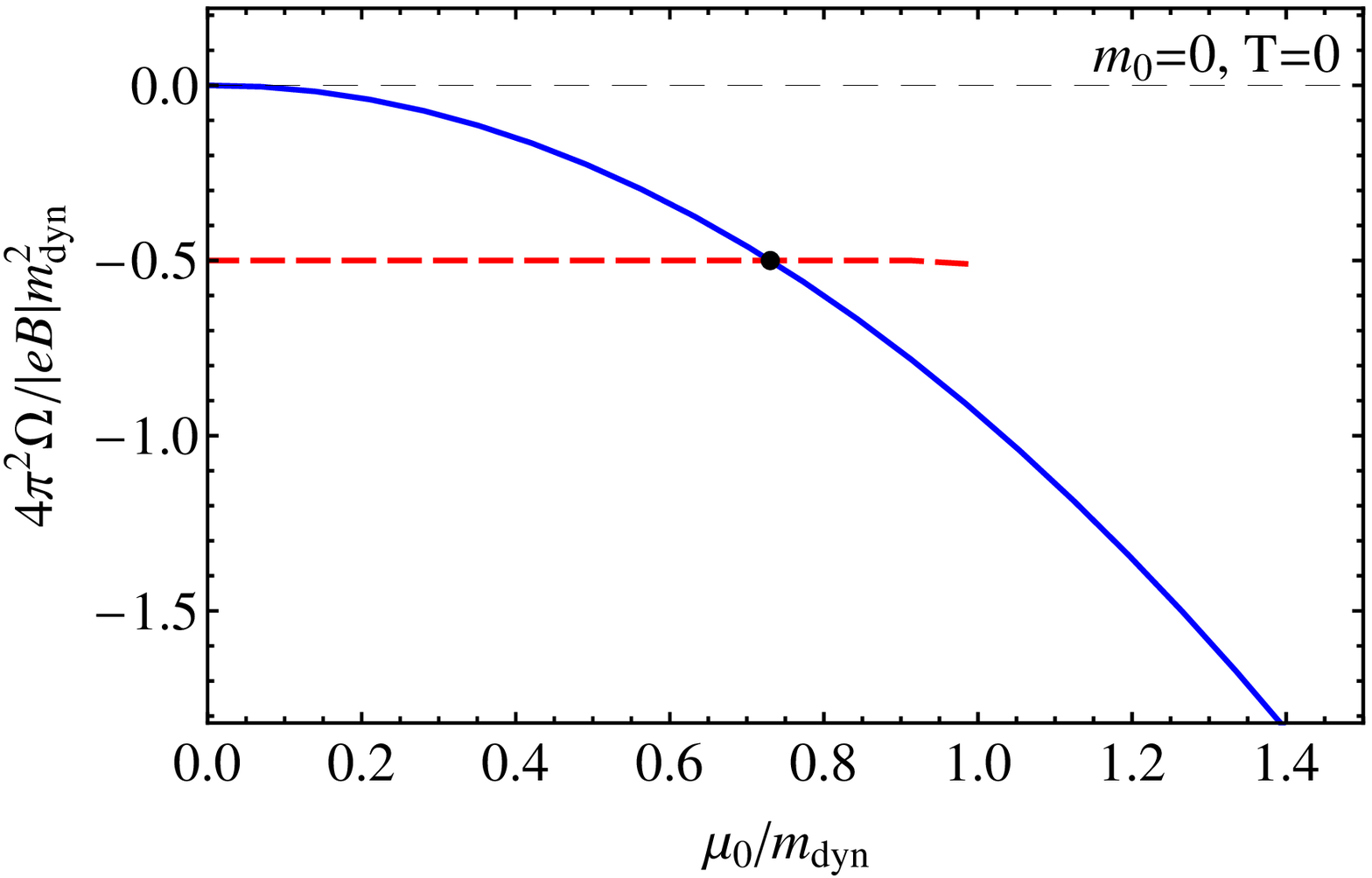}
\caption{(Color online) The zero temperature results for the chiral shift
parameter $\Delta$ (upper left panel), the mass $m$ (lower left panel), and the
difference $\mu_0-\mu$ (upper right panel) on the chemical potential $\mu_0$
for several fixed values of the bare mass: $m_0=0$ (black solid line), 
$m_0=10^{-4}\Lambda=0.14m_{\rm dyn}$ (blue long-dashed line), 
$m_0=3\times 10^{-4}\Lambda=0.42m_{\rm dyn}$ (red short-dashed line), 
$m_0=5\times 10^{-4}\Lambda=0.70m_{\rm dyn}$ (green dash-dotted line). 
The lower right panel shows the free energies for the solution with a nonzero 
dynamical mass (red dashed line) and the solution with a chiral shift (blue solid 
line) in the chiral limit, $m_0=0$, at zero temperature.} 
\label{CSE-CME:fig-T0multi-m0}
\end{figure}

The solutions of $\Delta$ vs $\mu_0$ and $m$ vs $\mu_0$, obtained by the iterating the set
of gap equations, are shown in upper and lower left panels of Fig.~\ref{CSE-CME:fig-T0multi-m0},
respectively. There the solutions for several different values of bare masses $m_0$ are shown.
It might be appropriate to note here that, in the vicinity of the phase transitions, we
had obtained a pair of solutions for each fixed value of $m_0$. The two different
solutions are obtained by sweeping over the same range of the chemical potentials $\mu_0$
in two different directions: (i) from left to right and (ii) from right to left. When such
a pair of solution, forming a small hysteresis loop, is observed, a first order phase
transition is expected somewhere within the loop. To determine the location of such a phase
transition, the comparison of free energy densities for the corresponding pairs of solutions
is required. The general expression for the free energy is derived in Appendix~\ref{App:FreeEnergyDensity3+1}.
By calculating the corresponding expression for each of the solutions around the hysteresis loop,
we could point the position of the actual phase transition. An example of such a calculation
is presented in the lower right panel of Fig.~\ref{CSE-CME:fig-T0multi-m0}. There the free energies
of the pair of solutions in the case $m_0=0$ are shown. The blue solid line correspond to the
solution with $m=0$ and
$\Delta\neq 0$. The free energy shown by the red dashed line represents the other solution, in
which the Dirac mass $m$ is nonzero and $\Delta$ is zero at small $\mu_0< m_{\rm dyn}$. The free
energies of the two solutions become equal at $\mu_{0,{\rm cr}}\approx 0.73 m_{\rm dyn}$. This
is where the first order phase transition occurs. Note that the numerical value of
$\mu_{0,{\rm cr}}$ is within several percent of the analytical estimate $m_{\rm dyn}/\sqrt{2}$.

Concerning the solution for $\mu$ vs $\mu_0$, the results are always such that 
$\mu\approx \mu_0$ to within a few percent. Therefore, the corresponding plot 
would give little information. In order to get a deeper insight into the deviation of 
$\mu$ from $\mu_0$, we find it instructive to plot the result for the difference 
$\mu_0-\mu$ instead. Note that as follows from Eq.~(\ref{CSE-CME:gap-mu-text}), 
the latter is proportional to the fermion charge density. The result is presented 
in the upper right panel of Fig.~\ref{CSE-CME:fig-T0multi-m0}. We see that 
$\mu_0-\mu$ is always positive, meaning that the value of $\mu$ is slightly 
smaller than $\mu_0$. 

By comparing that graph for $\mu_0-\mu$ with the dependence of the chiral shift
parameter $\Delta$ on $\mu_0$ in the upper left panel of the same figure, we 
observe that they have the same qualitative behaviors. In particular, $\Delta$ 
is nonzero in the ground state only if the fermion charge density is also nonzero 
there. In other words, the shift parameter is a manifestation of dynamics in a 
system with matter. Note that when the numerical values of the model parameters 
are used, the results for $\mu$ and $\Delta$ in the normal phase are approximately 
given by $\mu\simeq \mu_0-0.0295 \mu_0$ and $\Delta \simeq 0.0252\mu_0$ in 
the chiral limit ($m_0 = 0$).

\begin{figure}[t]
\includegraphics[width=.45\textwidth]{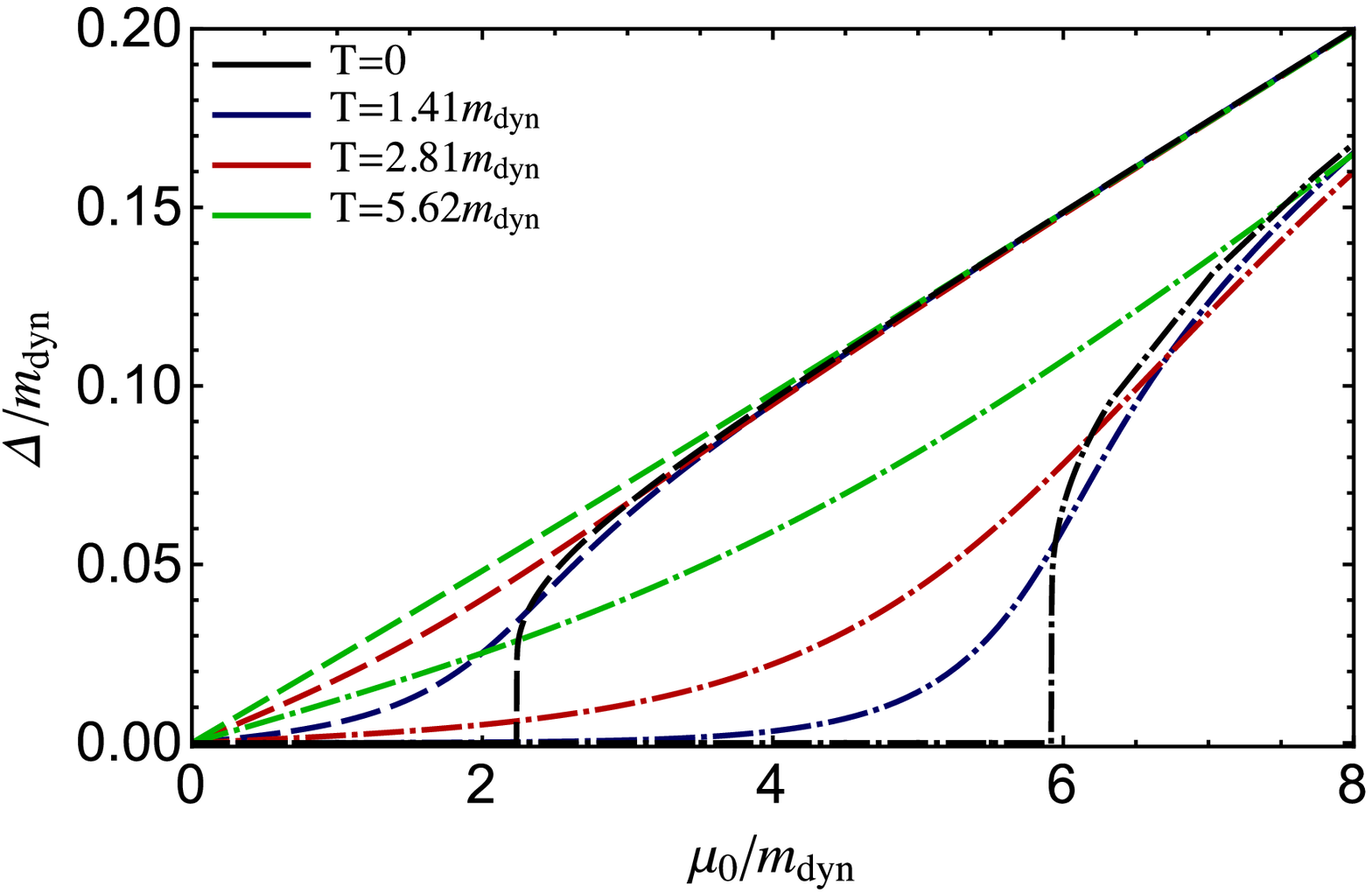}
\hspace*{10pt}
\includegraphics[width=.425\textwidth]{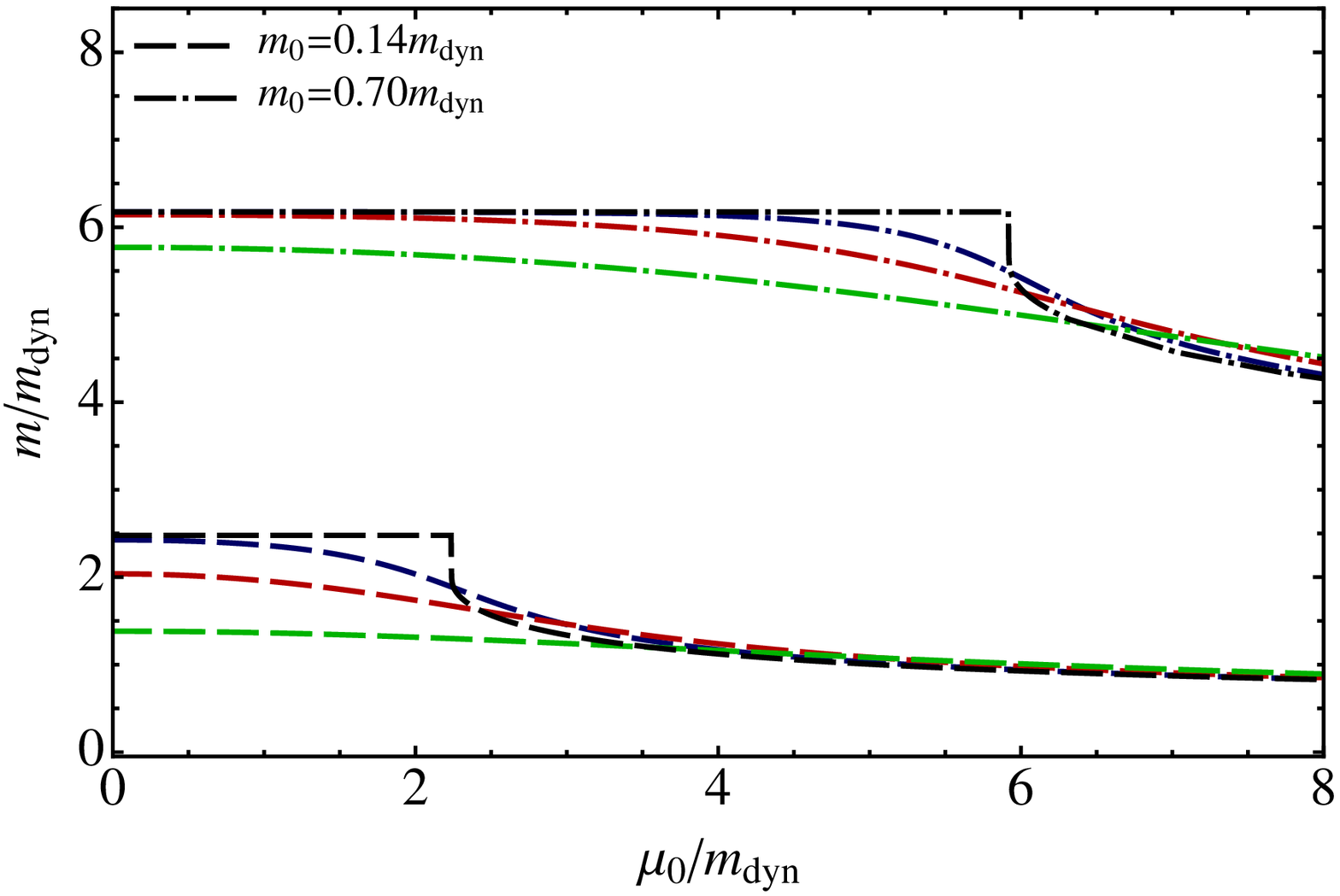}
\caption{(Color online) The nonzero temperature results for the chiral shift parameter
$\Delta$ (left panel) and the mass $m$ (right panel) as function of the chemical potential
$\mu_0$ for several fixed values of the temperature,
$T=0$ (black),
$T=1.41m_{\rm dyn}$ (blue lines),
$T=2.81m_{\rm dyn}$ (red lines), and
$T=5.62m_{\rm dyn}$ (green lines), and two values of the bare mass,
$m_0=0.14m_{\rm dyn}$ (solid lines) and
$m_0=0.70m_{\rm dyn}$ (short-dashed lines).}
\label{CSE-CME:fig-Tmulti-mo}
\end{figure}

Let us now proceed with the numerical solution of the gap equation at nonzero temperature.
At vanishing value of $\mu_0$, several results for the constituent mass have already been
presented in Fig.~\ref{CSE-CME:figMass-vs-m0}. The other two parameters, $\mu$ and $\Delta$, 
were identically zero in that special case. Here we extend the solutions to nonzero values of
$\mu_0$. The numerical results for $\Delta$ and $m$ as functions of $\mu_0$ are presented
in Fig.~\ref{CSE-CME:fig-Tmulti-mo}. Note that the dependence of $\mu - \mu_0$ on $\mu_0$ 
(not shown in Fig.~\ref{CSE-CME:fig-Tmulti-mo}) is similar to the dependence of $\Delta$ 
on $\mu_0$ at all temperatures. As should be expected, temperature suppresses the 
dynamical fermion mass (see the right panel of this figure). However,
the situation is quite different for the chiral shift parameter. As one can see in the left
panel of the figure, $\Delta$ is rather insensitive to temperature when $T \ll \mu_0$, and 
{\it increases} with $T$ when $T > \mu_0$. This property reflects the fact that higher 
temperature leads to higher matter density, which is apparently a more favorable environment 
for generating the chiral shift $\Delta$. While the first regime with $T \ll \mu_0$ is appropriate 
for stellar matter, the second one with $T > \mu_0$ (actually, $T \gg \mu_0$) is realized in 
heavy ion collisions. As we discuss in Section~\ref{ChiralShiftPhys}, the generation 
of $\Delta$ may have important implications for both stellar matter and (less likely)
heavy ion collisions.

\subsubsection{Axial current and charge density}
\label{ChiralAsymNJLaxial}

It is instructive to calculate the ground state expectation value of the axial current density,
\begin{eqnarray}
\langle j_5^3\rangle &=& -\mathrm{tr}\left[\gamma^3\gamma^5 G(u,u)\right].
\label{CSE-CME:axial-general}
\end{eqnarray}
In the case of the vanishing Dirac mass, $m=0$, an explicit expression for $\langle j_5^3\rangle$ within 
the momentum cutoff and the proper-time regularization schemes were presented in Eqs.~(\ref{CSE-CME:DT0m0}) 
and (\ref{CSE-CME:DT0m0-proper-time}), respectively. Both expressions can be written in the same form:
\begin{equation}
\langle j_5^3\rangle \simeq \frac{eB}{2\pi^2}\left[\mu -2 a s_{\perp}\Delta (\Lambda l)^2 \right],
\label{CSE-CME:axial}
\end{equation}
where $a$ is a dimensionless constant of order $1$, determined by the specific regularization scheme. 
When the proper time is used, we find from Eq.~(\ref{CSE-CME:DT0m0-proper-time}) that $a=1/4$. In the case 
of the cutoff regularization, it is defined by Eq.~(\ref{CSE-CME:sumkappa}). Note that qualitatively the same 
result is also obtained in the point-splitting regularization \cite{Gorbar:2010kc}.

The first term in the parenthesis in Eq.~(\ref{CSE-CME:axial}) is the same topological term that was derived in the free 
theory in Ref.~\cite{Metlitski:2005pr}, while the second term is an outcome of interactions \cite{Gorbar:2009bm}.
It is interesting to note that, by making use of the gap equation (\ref{CSE-CME:gap-Delta-text}) for $\Delta$,
the result for the axial current can be also rewritten in an alternative form:
\begin{equation}
\langle j_5^3\rangle =-\frac{2\Delta}{G_{\rm int} } =- \frac{\Delta}{2\pi^2}\frac{\Lambda^2}{g}.
\label{CSE-CME:axial-Delta}
\end{equation}
While this may not be very convenient in the free theory, in which 
both the coupling constant $g$ and the chiral shift $\Delta$ vanish,
and the cutoff is formally infinite, it is helpful to get a deeper insight in interacting theory.

Formally, the results for the axial current either in Eq.~(\ref{CSE-CME:axial}) or in Eq.~(\ref{CSE-CME:axial-Delta}) appear
to be quadratically divergent when $\Lambda\to \infty$. It should be noticed, however, that the solution to 
the gap equation, see Eq.~(\ref{CSE-CME:Delta-cutoff}) in the case of cutoff regularization and  Eq.~(\ref{CSE-CME:Delta-proper-time}) 
in the case of proper-time regularization, is inversely proportional to $\Lambda^2$, i.e., 
$\Delta\sim g \mu eB/\Lambda^2$. Taking this into account, we see that the axial current density is 
actually finite in the continuum limit $\Lambda\to \infty$,
\begin{equation}
\langle j^{3}_5 \rangle \simeq \frac{eB}{2\pi^{2}} \mu - a \frac{\Lambda^2}{\pi^{2}}  \Delta
\simeq \frac{eB}{2\pi^{2}}\frac{\mu}{\left(1+ 2 a g\right)} .
\label{CSE-CME:a}
\end{equation}

Before concluding this section, let us also note the following expression for fermion number density:
\begin{eqnarray}
\langle j^0\rangle &=& -\mathrm{tr}\left[\gamma^0 G(u,u)\right]  .
\end{eqnarray}
The explicit form of the function $\langle j^0\rangle$ can be derived from the general form 
of the propagator in Appendix~\ref{CSE:App:Landau-level-rep}. The corresponding
result is complicated and adds no new information when the solution to the gap equation is available.
Indeed, the fermion number density can be conveniently rewritten in a simpler form by making use of
the gap equation (\ref{CSE-CME:gap-mu-text}) for $\mu$,
\begin{equation}
\langle j^0\rangle = 2\frac{\mu_0-\mu}{G_{\rm int} } = \frac{\mu_0-\mu}{2\pi^2}\frac{\Lambda^2}{g} .
\end{equation}
This shows that the result for this density is proportional to $\mu_0-\mu$ presented earlier.

\subsubsection{Chirality dependent asymmetry of the Fermi surface}
\label{CSE:ChiralFermiSurface}

In the case of a strong magnetic field, when the LLL approximation is appropriate, we find that the 
chiral shift parameter (and, thus, the axial current density) and the fermion number density are 
proportional to each other. This is apparent in Fig.~\ref{CSE-CME:fig-T0multi-m0}. The underlying 
reason for this proportionality is the same (up to a sign)  LLL contribution to both functions 
$\langle j^0\rangle$ and $\langle j_5^3\rangle$. Moreover, this property seems to be at least 
approximately valid in a general case. In turn, this suggests that a fermion number density 
and the chiral shift parameter are two closely connected characteristics of the normal phase 
of magnetized relativistic matter.

\begin{figure}[t]
\begin{center}
\includegraphics[width=.3\textwidth]{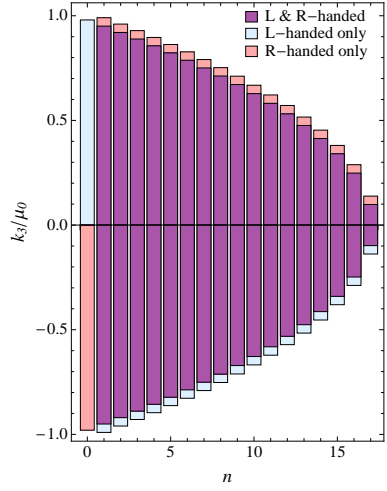}
\caption{(Color online) A schematic distribution of (negatively charged) 
particles in the ground state of dense relativistic matter in a magnetic field (pointing 
in the positive $z$ direction). The values of the quantum number $n$ (Landau levels) 
are shown along the horizontal axis, while the longitudinal momenta are shown along 
the vertical axis. The colored bars indicate the filled states of given chirality.}
\label{CSE-CME:figFermiSurface}
\end{center}
\end{figure}

The immediate implication of a nonzero chiral shift parameter in dense magnetized matter is the 
modification of the quasiparticle dispersion relations, see Eqs.~(\ref{CSE-CME:omega0}) and (\ref{CSE-CME:omegan}).  
These relations can be used to define the ``Fermi surface" in the space of the longitudinal momentum 
$k^3$ and the Landau index $n$. Note that the quantity $2n |eB|$ plays the role analogous 
to the square of the transverse momentum $k_\perp^2\equiv (k^1)^2+(k^2)^2$ in the absence of 
the magnetic field. Following the standard philosophy, we define the Fermi surface as the hypersurface 
in the space of quantum numbers $n$ and $k^3$, which correspond to quasiparticles with zero energy, i.e.,
\begin{eqnarray}
n=0: &\quad & k^{3}=\pm\sqrt{(\mu + s_\perp \Delta)^2-m^2},   \label{CSE-CME:Fermi-k3-1}\\
n>0: &\quad & k^{3}=\pm\sqrt{\left(\sqrt{\mu^2-2n|eB|}\pm s_\perp \Delta\right)^2-m^2}. \label{CSE-CME:Fermi-k3-2}
\end{eqnarray}
(In the last equation, all four combinations of signs are possible.) In order to better 
understand the nature of quasiparticles at the Fermi surface, described by Eqs.~(\ref{CSE-CME:Fermi-k3-1}) 
and (\ref{CSE-CME:Fermi-k3-2}), we recall that there are two types of quasiparticles. The Dirac structures of 
their wave functions are obtained by applying the projection operators in Eq.~(\ref{CSE-CME:projectorH}). 
In relativistic dense matter ($\mu\gg m$), the corresponding states at the Fermi surface can be 
approximately characterized by their chiralities. This follows from the fact that $|k^3|\gg m$ for 
a large fraction of the Fermi surface in Eq.~(\ref{CSE-CME:Fermi-k3-2}), except for the limiting values of $n$ 
around $n_{\rm max} \equiv \left[\mu^2/(2|eB|)\right]$. At such large values of the relative momentum, 
the projection operators in Eq.~(\ref{CSE-CME:projectorH}) are very closely related to the chiral projectors. 
Indeed, for $|k^3|\gg m$, the relation between the two sets of projectors is approximately the same 
as in the {\it massless} case in Eq.~(\ref{CSE-CME:Hpm-P5pm}). Taking this into account, it is possible to 
define quasiparticles at the Fermi surface, which are predominantly left-handed or right-handed.
Without loss of generality, let us assume that $\mathrm{sign}(eB)<0$. Then, the Fermi surface for the 
{\it predominantly left-handed} particles is determined by 
\begin{eqnarray}
n=0: &\quad& k^{3}=+\sqrt{(\mu+s_\perp \Delta)^2-m^2},   \label{CSE-CME:Fermi-k3-1L}\\
n>0: &\quad& k^{3}=+\sqrt{\left(\sqrt{\mu^2-2n|eB|}+ s_\perp \Delta\right)^2-m^2}, \label{CSE-CME:Fermi-k3-2L} \\
        &\quad& k^{3}=-\sqrt{\left(\sqrt{\mu^2-2n|eB|}- s_\perp \Delta\right)^2-m^2}, \label{CSE-CME:Fermi-k3-3L}
\end{eqnarray}
and the Fermi surface for the {\it predominantly right-handed} particles is determined by 
\begin{eqnarray}
n=0: &\quad& k^{3}=-\sqrt{(\mu+s_\perp \Delta)^2-m^2},   \label{CSE-CME:Fermi-k3-1R}\\
n>0: &\quad& k^{3}=-\sqrt{\left(\sqrt{\mu^2-2n|eB|}+ s_\perp \Delta\right)^2-m^2}, \label{CSE-CME:Fermi-k3-2R}\\
        &\quad & k^{3}=+\sqrt{\left(\sqrt{\mu^2-2n|eB|}- s_\perp \Delta\right)^2-m^2}. \label{CSE-CME:Fermi-k3-3R}
\end{eqnarray}
In the massless case, of course, this correspondence becomes exact. Then, we find that the Fermi surface 
for fermions of a given chirality is asymmetric in the direction of the magnetic field. In Fig.~\ref{CSE-CME:figFermiSurface}, 
we show a schematic distribution of negatively charged fermions and take into account that the parameter 
$s_\perp \Delta$ has the same sign as the chemical potential, see Eqs.~(\ref{CSE-CME:Delta-cutoff}) or (\ref{CSE-CME:Delta-proper-time}). 
(A similar distribution is also valid for positively charged fermions, but the left-handed and right-handed fermions 
will interchange their roles.) For the fermions of a given chirality, the LLL and the higher Landau levels give 
opposite contributions to the overall asymmetry of the Fermi surface. For example, the left-handed electrons in 
the LLL occupy only the states with {\it positive} longitudinal momenta (pointing in the magnetic field direction). The 
spins of the corresponding LLL electrons point against the magnetic field direction. In the higher Landau levels, 
while the left-handed electrons can have both positive and negative longitudinal momenta (as well as both spin 
projections), there are more states with {\it negative} momenta occupied, see Fig.~\ref{CSE-CME:figFermiSurface}. 
If there are many Landau levels occupied, which is the case when $\mu\gg \sqrt{|eB|}$, the relative contribution 
of the LLL to the whole Fermi surface is small, and the overall asymmetry is dominated by higher Landau levels.
In the opposite regime of superstrong magnetic field (if it can be realized in compact stars at all), only the LLL is 
occupied and, therefore, the overall asymmetry of the Fermi surface will be reversed. In the intermediate regime 
of a few Landau levels occupied, one should expect a crossover from one regime to the other, where the 
asymmetry goes through zero.

\subsection{Chiral anomaly relation in theory with chiral asymmetry}
\label{CSE:ChiralAnomaly}

By making use of the gauge invariant point-splitting regularization, let us study the influence
of the shift parameter $\Delta$ on the form of the axial current and the axial anomaly (for
reviews of this regularization, see Refs.~\cite{Peskin:1995ev,Ioffe:2006ww}). The analysis is 
model independent and is based only on the form of the fermion propagator with $\Delta$ in 
an external electromagnetic field. Our main conclusion is that while including $\Delta$ essentially
changes the form of the axial current, it does not modify the axial anomaly. Moreover, while
the contribution of the chemical potential in the axial current is generated in the infrared kinematic
region (at the LLL) \cite{Metlitski:2005pr}, the contribution of $\Delta$ in the current is
mostly generated in ultraviolet (at all Landau levels).

We consider the case of a constant electromagnetic field. Since it is known that the axial anomaly
is insensitive to chemical potential \cite{Son:2004tq,Gavai:2009vb}, the latter will be omitted. We 
will use the Schwinger representation \cite{Schwinger:1951nm} for the fermion propagator. In the 
case of a constant magnetic field, it is given in Appendix~\ref{App:A-SchwingerProp}. A similar 
representation will be also valid in the general case of a constant electromagnetic field. In fact, 
the only property of the propagator that we need to know is that it is given in the form of a 
product of the Schwinger phase,
\begin{equation}
\Phi(u,u^{\prime})=- e\int_{u^{\prime}}^u dv^{\nu}A_{\nu}(v),
\label{phase}
\end{equation}
and a translation invariant part $\bar{G}(u - u^{\prime})$ that depends only on the field strength 
$F_{\mu\nu}$. In the normal phase with $m=0$, the inverse propagator (\ref{ginverse})
(with $\tilde{\mu}=0$) can be rewritten as
\begin{equation}
iG^{-1} = iD_{\nu}\gamma^{\nu}+\Delta\gamma^3\gamma^5
=\left(iD_{\nu}\gamma^{\nu}-\Delta s_\perp\gamma^3\right) {\cal P}^{-}_5+
\left(iD_{\nu}\gamma^{\nu}+\Delta s_\perp\gamma^3\right) {\cal P}^{+}_5\,,
\end{equation}
where $s_\perp \equiv \mathrm{sign}(eB)$, $D_{\nu}=\partial_{\nu}+ ieA_{\nu}$, and
${\cal P}^{\pm}_5=(1 \pm  s_\perp \gamma^5)/2$. This equation implies that the
effective electromagnetic vector potential equals $\tilde{A}_{\nu}^{-}=A_{\nu}
+ (s_\perp \Delta /e)\delta_{\nu}^{3} $ and $\tilde{A}_{\nu}^{+}=A_{\nu} -
(s_\perp \Delta /e)\delta_{\nu}^{3}$ for the $-$ and $+$ chiral fermions,
respectively. Since the field strength $F_{\mu\nu}$ for $\tilde{A}_{\nu}^{\mp}$
is the same as for $A_{\nu}$, $\Delta$ affects only the Schwinger phase
(\ref{phase}):
\begin{eqnarray}
\Phi^{-}_{\Delta}(u,u^{\prime}) = \Phi(u,u^{\prime}) - s_{\perp}\Delta (u^3-u^{\prime\,3}),\\
\Phi^{+}_{\Delta}(u,u^{\prime}) = \Phi(u,u^{\prime}) + s_{\perp}\Delta (u^3-u^{\prime\,3}).
\end{eqnarray}
Thus, we find
\begin{equation}
G(u,u^{\prime}) = 
  \exp[- is_{\perp}\Delta (u^3-u^{\prime\,3})]\,{\cal P}^{-}_5\,G_0(u,u^{\prime})
+\exp[ i s_{\perp}\Delta (u^3-u^{\prime\,3})]\,{\cal
P}^{+}_5\,G_0(u,u^{\prime})\,, \label{propagator-transformed}
\end{equation}
where $G_0$ is the propagator with $\Delta=0$. Note that $\Delta$ appears now only in the phase factors.

According to Eq.~(\ref{current}), the axial current density is equal to
\begin{equation}
\langle j^{\mu}_5(u)\rangle=-\mathrm{tr}
\left[\gamma^{\mu}\gamma^5\,G(u,u+\epsilon)\right]_{\epsilon \to 0}. 
\label{density1}
\end{equation}
On the other hand, the fermion propagator in an electromagnetic field has the
following singular behavior for $u^{\prime}-u=\epsilon \to 0$
\cite{Peskin:1995ev,Ioffe:2006ww}:
\begin{equation}
G_0(u,u+\epsilon)\simeq
\frac{i}{2\pi^2}\left[\frac{\hat{\epsilon}}{\epsilon ^4} 
-\frac{1}{16\epsilon ^2}eF_{\mu\nu} \left(\hat{\epsilon}\sigma^{\mu\nu}
+\sigma^{\mu\nu}\hat{\epsilon } \right)\right], \label{asymptotics}
\end{equation}
where $\hat{\epsilon }=\gamma_{\mu}\epsilon^{\mu}$. Then, by using
Eqs.~(\ref{propagator-transformed}) -- (\ref{asymptotics}), we find
\begin{equation}
\langle j^{\mu}_5 \rangle_{\rm singular} = \left.
\frac{i\epsilon^{\mu}s_\perp}{\pi^2\epsilon^4} \left(e^{  is_\perp
\Delta \epsilon^3}- e^{- is_\perp  \Delta \epsilon^3}\right) +
\frac{ieF_{\lambda\sigma}\epsilon_{\beta}\epsilon^{\beta\mu\lambda\sigma}}
{8\pi^2\epsilon^2} \left(e^{ + i s_\perp \Delta \epsilon^3}
+e^{- i s_\perp \Delta \epsilon^3}\right)\right|_{\epsilon \to 0}\,.
\label{axialcurrent}
\end{equation}
Taking into account that the limit $\epsilon \to 0$ should be taken in this
equation symmetrically \cite{Peskin:1995ev,Ioffe:2006ww},  i.e.,
$\epsilon^{\mu}\epsilon^{\nu}/\epsilon^2 \to \frac{1}{4}g^{\mu\nu}$, and the
fact that its second term contains only odd powers of $\epsilon$, we arrive at
\begin{equation}
\langle j^{\mu}_5 \rangle_{\rm singular} =
 \frac{\Delta}{2\pi^2\epsilon^2} \delta^{\mu}_{3} \sim
\frac{\Lambda^2 \Delta }{2\pi^2}\delta^{\mu}_{3} \,. \label{chiral-current}
\end{equation}
It is clear that $- 1/\epsilon^2$ plays the role of a Euclidean ultraviolet cutoff
$\Lambda^2$. This feature of expression (\ref{chiral-current}) agrees with the
results obtained in Sections~\ref{ChiralAsymNJLMCshift} and \ref{CSE-CME:Numerical}.
As was shown there, while the contribution of each Landau level into the axial current 
density $\langle j^{\mu}_5 \rangle$ is finite at a fixed $\Delta$, their total contribution 
is quadratically divergent. However, the important point is that since the
solution of gap equation (\ref{CSE-CME:gap-Delta-text}) for the dynamical shift $\Delta$
yields $\Delta \sim g\mu\, eB/\Lambda^2 $, the axial current density is
actually finite.

We conclude that interactions leading to the shift parameter $\Delta$ essentially change
the form of the induced axial current in a magnetic field. It is important to mention that
unlike the topological contribution in $\langle j^{\mu}_5 \rangle$ \cite{Metlitski:2005pr},
the dynamical one is generated by all Landau levels.

It is straightforward to show that the shift parameter $\Delta$ does not affect
the axial anomaly. By using again the gauge invariant point-splitting regularization 
scheme, the divergence of the axial current in the massless theory is given by 
\cite{Peskin:1995ev,Ioffe:2006ww}
\begin{equation}
\partial_{\mu}j^{\mu}_5(u)=ie\epsilon^{\alpha}\bar{\psi}(u +\epsilon)
\gamma^{\mu}\gamma^5\psi(u)\left. F_{\alpha\mu}\right|_{ \epsilon \to 0}\,. 
\label{divergence}
\end{equation}
Then, calculating the vacuum expectation value of the divergence of the axial current, we find
\begin{equation}
\langle \partial_{\mu}j^{\mu}_5(u) \rangle
=-ie\epsilon^{\alpha}F_{\alpha\mu}\mathrm{tr}\left[\gamma^{\mu}\gamma^5
G(u,u+\epsilon) \right]_{\epsilon \to 0} =
ie\epsilon^{\alpha}F_{\alpha\mu} \langle j^{\mu}_5(u) \rangle\,,
\label{divergence1}
\end{equation}
where $G(u,u^{\prime})$ is the fermion propagator in Eq.~(\ref{propagator-transformed}).
Let us check that the presence of $\Delta$ in $G(u,u^{\prime})$, which modifies the axial
current, does not affect the standard expression for the axial anomaly.

We start by considering the first term in the axial current density in Eq.~(\ref{axialcurrent}):
\begin{equation}
\frac{i\epsilon^{\mu}s_{\perp}}{\pi^2\epsilon^4} \left(
e^{ i s_{\perp} \Delta \epsilon^3}
-e^{-i s_{\perp} \Delta \epsilon^3}\right) 
\simeq -\frac{2\Delta\epsilon^{\mu}\epsilon^3}{\pi^2\epsilon^4}
\left(1-\frac{\Delta^2\epsilon^2_3}{6}+...\right)\,.
\label{chiral-current-series}
\end{equation}
Its contribution to the right-hand side of Eq.~(\ref{divergence1}) is
\begin{equation}
-\frac{2i\Delta\,\epsilon^{\alpha}\epsilon^{\mu}\epsilon^3}{\pi^2\epsilon^4}
\left(1-\frac{\Delta^2\epsilon^2_3}{6}+...\right)eF_{\alpha \mu} \,.
\label{new-term}
\end{equation}
Since this expression contains only odd powers of $\epsilon$, it gives zero contribution after
symmetric averaging over space-time directions of $\epsilon$.

Thus, only the second term in  Eq.~(\ref{axialcurrent}) is relevant for the divergence of
axial current in Eq.~(\ref{divergence1}):
\begin{equation}
\langle \partial_{\mu}j^{\mu}_5(u) \rangle =
-\frac{e^2\epsilon^{\beta\mu\lambda\sigma}F_{\alpha\mu}F_{\lambda\sigma}
\epsilon^{\alpha}\epsilon_{\beta}}{8\pi^2\epsilon^2} \left(
e^{+ i s_{\perp} \Delta \epsilon^3}
+e^{-i s_{\perp} \Delta \epsilon^3}\right) \to
-\frac{e^2}{16 \pi^2}
\epsilon^{\beta\mu\lambda\sigma}F_{\beta\mu}F_{\lambda\sigma}
\label{divergence-final}
\end{equation}
for $\epsilon\to 0$ and symmetric averaging over space-time directions of $\epsilon$. Therefore,
the presence of the shift parameter $\Delta$ does not affect the axial anomaly indeed.

\subsection{Chiral asymmetry in QED in a magnetic field}
\label{ChiralAsymQED}

The NJL model studies of the chiral asymmetry in Section~\ref{ChiralAsymNJL} revealed that the interaction
effects induce a nonzero chiral shift  $\Delta$ in the magnetized relativistic matter. In turn, the
chiral shift leads to an additional dynamical contribution into the axial current. Unlike the topological
contribution due to the LLL, the dynamical one affects the fermions in all Landau levels, including those
around the whole Fermi surface. 

Taking into account that the NJL model is nonrenormalizable and recalling the fact 
the chiral anomaly is intimately connected with ultraviolet divergencies, one may question 
whether the corresponding toy model analysis is reliable in application to more realistic gauge 
theories. In order to address the problem rigorously, below we consider the problem of radiative 
corrections within the framework of a renormalizable model \cite{Gorbar:2013upa}. In order to 
make the problem manageable, we will assume that the magnetic field $\mathbf{B}$ is weak. 
This will allow us to calculate leading order radiative corrections to the axial current in QED,
using the expansion in powers of $\mathbf{B}$ up to linear order.

The Lagrangian density of QED in a magnetic field is given by
\begin{equation}
{\cal L} = -\frac{1}{4}F^{\mu\nu}F_{\mu\nu}+\bar{\psi}\left( i\gamma^{\nu}{\cal D}_{\nu}+\mu
\gamma^0-m\right)\psi +\delta_2\bar{\psi}( i \gamma^{\nu} \partial_{\nu}+\mu\gamma^{0}+e A _{\nu} \gamma^{\nu}
)\psi-\delta_m\bar{\psi}\psi,
\label{CSE:Lagrangian}
\end{equation}
where $\mu$ is the fermion chemical potential, the last two terms are counterterms (we use the 
notation of Ref.~\cite{Peskin:1995ev}), and the covariant derivative is ${\cal D}_{\mu} 
=\partial_{\mu}+i e A_{\mu}+i e a _{\mu}$. Without the loss of generality, we assume that the
external magnetic field $\mathbf{B}$ points in the $+z$ direction and is described by the 
vector potential in the Landau gauge, $\mathbf{A} =\left(-yB,0,0\right)$. Note that the 
counterterms include the chemical potential $\mu$ and the external field $A _{\mu}$.

\subsubsection{Fermion self-energy in QED in a magnetic field}
\label{ChiralAsymQEDstrongB}

In order to resolve the limitations of the weak-field expansion, let us now investigate  
the fermion self-energy and chiral asymmetry in cold magnetized QED plasma in a
strong magnetic field. To this end, we will use a generalized form of the Landau level 
representation.

Before presenting the details of the formalism, let us recall that the Ritus method is  
one of the standard approaches to treat quantum field theory of charged particles in 
an external magnetic field \cite{1972AnPhy..69..555R}. Here, however, we advocate 
a different approach that was developed in Ref.~\cite{Gorbar:2011kc} in a 
study of graphene in a magnetic field. From the technical viewpoint, the key 
difference between the two methods lies in the use of the complete sets of eigenfunctions 
of different operators. In the Ritus method, one uses the eigenfunctions of the operator 
$(\bm{\pi}\cdot\bm{\gamma})$ with a nontrivial Dirac structure and, thus, treats both the 
orbital and spinor parts of the fermion kinematics in a uniform fashion. In the method of 
Ref.~\cite{Gorbar:2011kc}, on the other hand, the eigenfunctions of the scalar operator 
$\bm{\pi}^2$ are used. This operator includes only the orbital part of the kinematics and, 
thus, requires one to treat the spin part separately. The seeming inconvenience of dealing with 
the spinor part separately in the second method, in fact, appears to offer many advantages, 
ranging from a much more transparent interpretation of various Dirac structures 
in the propagator and self-energy to significant technical simplifications in calculations.
  
To leading order in the coupling constant $\alpha=e^2/(4\pi)$, the fermion self-energy in QED is 
given by 
\begin{equation}
\Sigma(u,u^\prime)=-4i\pi\alpha\gamma^\mu S(u,u^\prime) \gamma^\nu D_{\mu\nu}(u-u^\prime),
\label{CSE:self-energy}
\end{equation}
where $S(u,u^\prime)$ is the free fermion propagator in magnetic field and $D_{\mu\nu}(u-u^\prime)$ is the free 
photon propagator. 

As is well known, the fermion propagator $S(u,u^\prime)$ in the presence of an external magnetic field is
not translation invariant. It can be written, however, in a form of an overall Schwinger phase 
(breaking the translation invariance) and a translation invariant function \cite{Schwinger:1951nm}, 
i.e.,
\begin{equation}
S(u,u^\prime) = \exp\left[i\Phi(\mathbf{r}_\perp,\mathbf{r}_\perp^\prime)\right]\bar{S}(u-u^\prime),
\label{CSE:S-Schwinger-phase}
\end{equation}
where the Schwinger phase equals 
$\Phi(\mathbf{r}_\perp,\mathbf{r}_\perp^\prime)=-eB(x-x^\prime)(y+y^\prime)/2$ in the Landau gauge (\ref{eq:Landau_gauge}). 
The Fourier transform of $\bar{S}(u-u^\prime)$ is presented in Eq.~(\ref{CSE:Fourier-tranlation-inv-S}) in 
Appendix~\ref{CSE:App:proper-time-rep-mu}. The expression in Eq.~(\ref{CSE:self-energy}) implies that 
the self-energy $\Sigma(u,u^\prime)$ has an analogous representation
\begin{equation}
\Sigma(u,u^\prime) = \exp\left[i\Phi(\mathbf{r}_\perp,\mathbf{r}_\perp^\prime)\right]\bar{\Sigma}(u-u^\prime),
\label{CSE:Sigma-Schwinger-phase}
\end{equation}
with the same Schwinger phase as in the propagator. 
  
In the rest of this section, we give a detailed derivation of the expansion of
the fermion self-energy over Landau levels. We will start by writing down the self-energy
with all possible Dirac structures in Eq.~(\ref{CSE:self-energy}) in coordinate 
space. Thus, we have
\begin{equation}
\Sigma(u,u^\prime) =
\left[-\gamma^0 {\delta\mu} 
+\pi^3\gamma^3  \delta v_{3}
+(\bm{\pi}_{\perp}\cdot\bm{\gamma}_{\perp}){\delta v_{\perp}} 
+{\cal M} 
- i\gamma^1\gamma^2\tilde{\mu}
-\gamma^3\gamma^5\Delta
-\gamma^0\gamma^5\mu_{5} \right] \delta^{4}(u-u^\prime),
\label{weakQED:self-energy-coordinate-space}
\end{equation}
where $\mathbf{\pi}_{\perp}$ is the canonical transverse momentum operator, which 
includes the vector potential. Here, $\delta\mu$, $\delta v_{3}$, $\delta v_{\perp}$, ${\cal M}$, 
$\tilde{\mu}$, $\Delta$, and $\mu_5$, are functions of the operators $-i\partial_0$ and $\pi^3$, 
as well as the operator $(\bm{\pi}_{\perp}\cdot\bm{\gamma}_{\perp})^2 l^2$.

The eigenvalues of the operator $\bm{\pi}_{\perp}^2 l^2$ are positive odd integers,
$2N+1$, where $N=0,1,2,\ldots$ is the orbital quantum number \cite{1930ZPhy...64..629L}. The corresponding 
eigenfunctions $\psi_{Np}(\mathbf{r}_{\perp})$ are well known and have the following explicit form: 
\begin{equation}
\psi_{Np}(\mathbf{r}_{\perp}) 
=\frac{1}{\sqrt{2\pi  l}}\frac{1}{\sqrt{2^NN!\sqrt{\pi}}}
H_N\left(\frac{y}{l}+pl\,\mathrm{sign}(eB)\right) e^{-\frac{1}{2l^2}\left[y+pl^2\,\mathrm{sign}(eB)\right]^2} e^{i p x},
\end{equation}
where $H_N(x)$ are Hermite polynomials \cite{1980tisp.book.....G}.
By making use of the completeness of these eigenfunctions 
\begin{equation}
\delta^2(\mathbf{r}_{\perp}-\mathbf{r}^{\prime}_{\perp})=
\sum_{N=0}^{\infty}\int^{+\infty}_{-\infty} dp\,\,\psi_{Np}(\mathbf{r}_{\perp})\psi^{*}_{Np}(\mathbf{r}^{\prime}_{\perp}),
\label{weakQED:completeness}
\end{equation}
one can rewrite the self-energy (\ref{weakQED:self-energy-coordinate-space}) in the form
\begin{eqnarray}
\Sigma(u,u^\prime)&=&\sum_{N=0}^{\infty}\sum_{s=\pm}\int\frac{dp_0 dp^3 dp}{(2\pi)^2}
e^{-ip_0(x_0-y_0)+ip^3(x^3-y^3)}
\nonumber\\
&&\times\left[-\gamma^0\delta\mu+p^3\gamma^3\delta v_3
+(\bm{\pi}_{\perp}\cdot\bm{\gamma}_{\perp}){\delta v_{\perp}} 
+{\cal M}
- i\gamma^1\gamma^2\tilde{\mu}-\gamma^3\gamma^5\Delta-\gamma^0\gamma^5\mu_{5}
\right] \,{\cal P}_s\,\psi_{Np}(\mathbf{r}_{\perp}) \psi_{Np}^{*}(\mathbf{r}_{\perp}^\prime),
\label{weakQED:self-energy-coordinate-levels}
\end{eqnarray}
where $\delta\mu$, $\delta v_3$, $\ldots$, $\mu_5$ are now functions of $p_0$, $p^3$, and the operator 
$(\bm{\pi}_{\perp}\cdot\bm{\gamma}_{\perp})^2 l^2$.  In Eq.~(\ref{weakQED:self-energy-coordinate-levels}), we 
also inserted the unit matrix in the form of the sum of spin projectors, i.e., $1=\sum_{s=\pm}{\cal P}_s$. 
It is easy to see that any function of $(\bm{\pi}_{\perp}\cdot\bm{\gamma}_{\perp})^2 l^2$ acting on 
${\cal P}_s\,\psi_{Np}$ reduces to a constant in the $n$th Landau level. This is a consequence of 
the following identity:
\begin{equation}
(\bm{\pi}_{\perp}\cdot\bm{\gamma}_{\perp})^2 l^2\,{\cal P}_s\,\psi_{Np}
=-(\bm{\pi}^2_{\perp}- i\gamma^1\gamma^2 eB ) l^2\,{\cal P}_s\,\psi_{Np}
=-2 n\,{\cal P}_s\,\psi_{Np},
\end{equation}
where $n\equiv N+(1+s)/2$ is the standard Landau level quantum number and $s=\pm 1$. 
This allows us to rewrite Eq.~(\ref{weakQED:self-energy-coordinate-levels}) as follows:
\begin{eqnarray}
\Sigma(u,u^\prime)&=&\sum_{N=0}^{\infty}\sum_{s=\pm}\int\frac{dp_0 dp^3 dp}{(2\pi)^2}
e^{-ip_0(x_0-y_0)+ip^3(x^3-y^3)} \Big[-\gamma^0\delta\mu_n+p^3\gamma^3\delta v_{3,n}
+(\bm{\pi}_{\perp}\cdot\bm{\gamma}_{\perp}){\delta v_{\perp,n}} 
+{\cal M}_n\nonumber\\
&&
- i\gamma^1\gamma^2\tilde{\mu}_n-\gamma^3\gamma^5\Delta_n-\gamma^0\gamma^5\mu_{5,n}
\Big] \,{\cal P}_s\,\psi_{Np}(\mathbf{r}_{\perp}) \psi_{Np}^{*}(\mathbf{r}_{\perp}^\prime),
\label{weakQED:SigmaXY_Eigenstates}
\end{eqnarray}
where the coefficient functions $\delta\mu_n$, $\delta v_{3,n}$, etc., depend on energy $p_0$ 
and longitudinal momentum $p^3$. Now, by taking into account the relation
\begin{equation}
(\bm{\pi}_{\perp}\cdot\bm{\gamma}_{\perp})  l \, \psi_{Np}(\mathbf{r}_{\perp}) = 
i \gamma^2  \left[
\sqrt{2(N+1)}{\cal P}_{-} \psi_{N+1,p}(\mathbf{r}_{\perp}) 
-\sqrt{2N}{\cal P}_{+} \psi_{N-1,p}(\mathbf{r}_{\perp}) 
\right]
\label{weakQED:pi-gamma-relation}
\end{equation}
and using formula 7.377 from Ref.~\cite{1980tisp.book.....G},
\begin{equation}
\int\limits_{-\infty}^\infty\,e^{-x^2}H_m(x+y)H_n(x+z)dx
=2^n\pi^{1/2}m!z^{n-m}L_m^{n-m}(-2yz),
\label{weakQED:HnHm-integral}
\end{equation}
we can perform the integration over $p$ in Eq.~(\ref{weakQED:SigmaXY_Eigenstates}). As expected, the result takes 
the form of a product of the Schwinger phase and a translationally invariant function, i.e.,
\begin{equation}
\Sigma(u,u^\prime) = e^{i\Phi(\mathbf{r}_\perp,\mathbf{r}_\perp^\prime)} \bar{\Sigma}(u-u^\prime),
\end{equation}
where the latter is given by 
\begin{eqnarray}
\bar{\Sigma}(x)&=&\frac{e^{-\xi/2}}{2\pi l^2}\sum_{n=0}^{\infty}
\int \frac{dp_0 dp^3 }{(2\pi)^2}  e^{-ip_0x_0+ip^3x^3} 
\Big\{
\left(-\gamma^0 \delta\mu_{n} 
+p^3\gamma^3  \delta v_{3,n}
- i\gamma^1\gamma^2\tilde{\mu}_{n}
-\gamma^3\gamma^5\Delta_{n}
-\gamma^0\gamma^5\mu_{5,n}
+{\cal M}_{n} \right)\nonumber\\
&&\times \left[ L_{n}(\xi){\cal P}_{+}+L_{n-1}(\xi){\cal P}_{-}\right]
-\frac{i}{ l^2}(\mathbf{r}_{\perp}\cdot\bm{\gamma}_{\perp}){\delta v_{\perp,n}} L^{1}_{n-1}(\xi)
\Big\},
\end{eqnarray}
and $\xi=\mathbf{r}^2_{\perp}/(2 l^2)$. Here $L_{-1}(\xi)=0$ by definition. Performing the 
Fourier transform, we finally find the sought expansion of the fermion self-energy over the 
Landau levels:
\begin{eqnarray}
\bar{\Sigma}(p)&=& 2  e^{-p_{\perp}^2  l^2} 
\sum_{n=0}^{\infty}(-1)^n\Big\{
\left(-\gamma^0 \delta\mu_{n} 
+p^3\gamma^3  \delta v_{3,n}
- i\gamma^1\gamma^2\tilde{\mu}_{n}
-\gamma^3\gamma^5\Delta_{n}
-\gamma^0\gamma^5\mu_{5,n}
+{\cal M}_{n} \right) \nonumber\\
&&\times \left[ L_{n}(2p_{\perp}^2  l^2){\cal P}_{+} - L_{n-1}(2p_{\perp}^2  l^2){\cal P}_{-}\right]
-2 (\bm{\gamma}_\perp\cdot \bm{p}_{\perp}){\delta v_{\perp,n}} L^{1}_{n-1}(2p_{\perp}^2  l^2)
\Big\}.
\label{weakQED:self-energy-LL}
\end{eqnarray} 

In what follows, we will drop the $\delta v_{\perp}$-type corrections to the self-energy. For the 
purposes of this study, this is justified, because such terms neither break the chiral symmetry
nor modify the chiral asymmetry of the ground state. On the other hand, it is necessary to 
keep the terms with ${\delta\mu(p)}$ and ${\delta v_3(p)}$. The reason for this becomes 
obvious after noticing that, when restricted to the subspaces of fixed spin projections, 
${\delta\mu(p)}$ and ${\delta v_3(p)}$ mix up with $\Delta(p)$ and $\mu_5(p)$, respectively. 
The argument can be made explicit by making use of the following identities: 
$\gamma^0 {\cal P}_{\pm} =\pm \mbox{sign}(eB)\gamma^3\gamma^5 {\cal P}_{\pm}$ and 
$\gamma^3 {\cal P}_{\pm} =\pm \mbox{sign}(eB)\gamma^0\gamma^5 {\cal P}_{\pm}$. Note 
that a similar argument also necessitates the inclusion of the anomalous magnetic 
moment function $\tilde{\mu}(p)$ whenever the mass function ${\cal M}(p)$ is present.

Equation (\ref{weakQED:self-energy-LL}) defines the expansion of the fermion self-energy over Landau 
levels. On the other hand, in the leading order of perturbation theory, the self-energy is given 
by Eq.~(\ref{weakQED:self-energy-momentum}). By combining these equations, it is not difficult to express 
the dynamical functions $\delta\mu_n$, $\delta v_{3,n}$, etc., projected onto specific Landau 
levels, through the self-energy (\ref{weakQED:self-energy-momentum}). Multiplying these two equivalent 
expressions for the fermion self-energy by 
$ l^2\pi^{-1}(-1)^{n^\prime}e^{-p^2_{\perp} l^2}L_{n^\prime}(2p^2_{\perp} l^2){\cal P}_{\pm}$ 
and integrating over the perpendicular momentum $\bm{p}_{\perp}$, we arrive at the following set 
of equations:
\begin{eqnarray}
\left[-\gamma^0 \delta\mu_{n} 
+p^3\gamma^3  \delta v_{3,n}
+{\cal M}_{n}  
-\mbox{sign}(eB)\left(\tilde{\mu}_{n}
+ \gamma^0\Delta_{n}
+ \gamma^3\mu_{5,n}\right)
\right] {\cal P}_{+} &=& I_n{\cal P}_{+}\,,
\label{weakQED:equation-spin-1}
\\
\left[ -\gamma^0 \delta\mu_{n} 
+p^3\gamma^3  \delta v_{3,n}
+{\cal M}_{n}  
+\mbox{sign}(eB)\left(\tilde{\mu}_{n}
+ \gamma^0\Delta_{n}
+ \gamma^3\mu_{5,n}\right)
\right]{\cal P}_{-} &=& I_{n-1}{\cal P}_{-}\,,
\label{weakQED:equation-spin-2}
\end{eqnarray}
where
\begin{equation}
I_n = -4i (-1)^n \alpha l^2 \int\frac{d^2k_{\parallel}d^2k_{\perp}d^2p_{\perp}}{(2\pi)^4}
e^{-p_{\perp}^2 l^2}\,L_n(2p^2_{\perp} l^2)\,
\gamma^{\mu}\,\bar{S}(k)\gamma^{\nu}
D_{\mu\nu}(p-k).
\label{weakQED:I_n_definition_1}
\end{equation}
When the free fermion propagator in Eq.~(\ref{weakQED:I_n_definition_1}) is replaced by the full propagator, which itself 
is a function of $\delta\mu_{n}$, $\delta v_{3,n}$, etc., the above set of equations will become an infinite set of 
the Schwinger-Dyson equations for the dynamical functions.

From Eqs.~(\ref{weakQED:equation-spin-1}) and (\ref{weakQED:equation-spin-2}), we obtain the following relations which express 
the dynamical functions projected onto specific Landau levels through the self-energy (\ref{weakQED:self-energy-momentum}):
\begin{eqnarray}
\delta\mu_{n}  &=&  -\frac{1}{4}\mathrm{Tr}\left[\gamma^0\left( I_{n}{\cal P}_{+}+ I_{n-1}{\cal P}_{-}\right)\right],
\label{weakQED:eq-n-1} \\
\Delta_{n}    &=& -\frac{\mbox{sign}(eB)}{4}\mathrm{Tr}\left[\gamma^0\left( I_{n}{\cal P}_{+}- I_{n-1}{\cal P}_{-}\right)\right] ,
\label{weakQED:eq-n-2}\\
{\cal M}_{n}  &=& \frac{1}{4}\mathrm{Tr} \left( I_{n}{\cal P}_{+}+ I_{n-1}{\cal P}_{-}\right) ,
\label{weakQED:eq-n-3}\\
\tilde{\mu}_{n}   &=&  -\frac{\mbox{sign}(eB)}{4}\mathrm{Tr} \left( I_{n}{\cal P}_{+}- I_{n-1}{\cal P}_{-}\right)  ,
\label{weakQED:eq-n-4}\\
p^3 \delta v_{3,n} &=& -\frac{1}{4}\mathrm{Tr}\left[\gamma^3\left( I_{n}{\cal P}_{+}+ I_{n-1}{\cal P}_{-}\right)\right]   ,
\label{weakQED:eq-n-5}\\
\mu_{5,n} &=& +\frac{\mbox{sign}(eB)}{4}\mathrm{Tr}\left[\gamma^3\left( I_{n}{\cal P}_{+}- I_{n-1}{\cal P}_{-}\right)\right]    .
\label{weakQED:eq-n-6}
\end{eqnarray}
The special role of LLL ($n=0$) should be noted here. By taking into account that $I_{-1}=0$, 
we find the following relations between the pairs of parameters: 
$\Delta_{0} =+ \mbox{sign}(eB)\delta\mu_{0} $, 
$\tilde{\mu}_{0} =-\mbox{sign}(eB){\cal M}_{0} $, and 
$\mu_{5,0} =-\mbox{sign}(eB)p^3 \delta v_{3,0} $; i.e.,  only half of them remain independent in LLL. 
From the physics viewpoint, this reflects the spin-polarized nature of the lowest energy level. 

The dynamical functions $\Delta_n$ and $\mu_{5,n}$ for $n \ge 1$ define chiral asymmetry 
in higher Landau levels. Therefore, these functions are of the prime interest for us here.
In terms of the self-energy (\ref{weakQED:self-energy-momentum}), we can represent 
$I_n$ in Eq.~(\ref{weakQED:I_n_definition_1}) as follows:
\begin{equation}
I_n =  (-1)^n\frac{ l^2}{\pi}\int d^2p_{\perp}
e^{-p_{\perp}^2 l^2}\,L_n(2p^2_{\perp} l^2)\,
\bar{\Sigma}(p).
\label{weakQED:I_n_definition_2}
\end{equation}
Using it, we may rewrite Eqs.~(\ref{weakQED:eq-n-2}) and (\ref{weakQED:eq-n-6}) in the following perhaps more transparent form:
\begin{eqnarray}
\Delta_n 
&=&  -\frac{(-1)^{n}}{8} \frac{ l^2}{\pi} \mbox{sign}(eB) \int d^2p_{\perp}
e^{-p_{\perp}^2 l^2}\left[L_n(2p^2_{\perp} l^2)+L_{n-1}(2p^2_{\perp} l^2)\right]
\mathrm{Tr}\left[\gamma^0 \bar{\Sigma}(p)\right]\nonumber\\
&& -  \frac{(-1)^{n}}{8} \frac{ l^2}{\pi}  \int d^2p_{\perp}
e^{-p_{\perp}^2 l^2}\left[L_n(2p^2_{\perp} l^2)-L_{n-1}(2p^2_{\perp} l^2)\right]
\mathrm{Tr}\left[\gamma^3\gamma^5 \bar{\Sigma}(p)\right],
\label{strongB:Delta-n}
\end{eqnarray}
\begin{eqnarray}
\mu_{5,n} 
&=&+\frac{(-1)^{n}}{8} \frac{ l^2}{\pi} \mbox{sign}(eB) \int d^2p_{\perp}
e^{-p_{\perp}^2 l^2}\left[L_n(2p^2_{\perp} l^2)+L_{n-1}(2p^2_{\perp} l^2)\right]
\mathrm{Tr}\left[\gamma^3 \bar{\Sigma}(p)\right]\nonumber\\
&& + \frac{(-1)^{n}}{8} \frac{ l^2}{\pi}  \int d^2p_{\perp}
e^{-p_{\perp}^2 l^2}\left[L_n(2p^2_{\perp} l^2)-L_{n-1}(2p^2_{\perp} l^2)\right]
\mathrm{Tr}\left[\gamma^0\gamma^5 \bar{\Sigma}(p)\right],
\label{strongB:mu-5-n}
\end{eqnarray}
where $\bar{\Sigma}(p)$ is given by Eq.~(\ref{weakQED:self-energy-momentum}). These expressions will 
in principle allow us to determine the chiral asymmetry for fermions in higher Landau levels. 
In the general case, however, the calculation of these parameters can be done only with the 
help of numerical methods.

Before proceeding to the numerical analysis of the chiral asymmetry functions $\Delta_{n}(p_3)$ 
and $\mu_{5,n}(p_3)$, it is instructive to discuss the implications of the well known ultraviolet (UV)
divergency in the fermion self-energy function in QED \cite{Peskin:1995ev}. In the Pauli-Villars 
regularization scheme, the self-energy contains the following logarithmically divergent 
contribution \cite{Peskin:1995ev}:
\begin{equation}
\Sigma^{\rm (div)} (p) = \frac{\alpha}{4\pi} \left[-\gamma^\nu (p_\nu+\mu\delta_{\nu}^{0}) +4m \right]
\ln\frac{\Lambda^2}{m^2} .
\end{equation}
Note that the only effect of a nonzero chemical potential here is to shift $p_0\to p_0+\mu$ 
in the vacuum expression \cite{Gorbar:2013upa,Freedman:1976xs}. Of course, the 
above divergency cannot be affected by the magnetic field. When projected onto 
Landau levels as prescribed by Eqs.~(\ref{strongB:Delta-n}) and (\ref{strongB:mu-5-n}), this result 
leads to the following contributions to the chiral asymmetry functions:
\begin{eqnarray}
\Delta^{\rm (div)}_n(p_3) & = & - \delta_{n}^{0}\frac{\alpha (p_0+\mu)  }{8\pi} \mathrm{sign}(eB) \ln\frac{\Lambda^2}{m^2},\\
\mu^{\rm (div)}_{5,n}(p_3) & = &  \delta_{n}^{0} \frac{\alpha p_3 }{8\pi} \mathrm{sign}(eB)  \ln\frac{\Lambda^2}{m^2} .
\end{eqnarray}
As we see, the corresponding functions are free of the UV divergences in all, but the lowest Landau level 
($n=0$). As explained in detail in Ref.~\cite{Gorbar:2013uga}, the LLL ($n=0$) is very special also because of 
its spin-polarized nature. As a consequence, the LLL chiral shift is indistinguishable from the correction 
to the chemical potential, and the LLL axial chemical potential is indistinguishable from the correction 
to the longitudinal velocity. It was concluded, therefore, that the novel type of the chiral asymmetry is determined 
exclusively by the dynamical functions $\Delta_n$ and $\mu_{5,n}$ with $n \ge 1$. These functions 
are of the prime interest for us here. In the next subsection, we take into account that all 
these functions are free from the UV divergences and study them using numerical methods.

\subsubsection{Fermion self-energy and chiral asymmetry in QED in a weak field}
\label{ChiralAsymQEDweakB}

After dropping the common Schwinger phase factor on both sides of Eq.~(\ref{CSE:self-energy}), 
we will arrive at a translation invariant form of the equation for the self-energy. The Fourier 
transform of the corresponding relation reads
\begin{equation}
\bar{\Sigma} (p) = -4i\pi\alpha \int\frac{ d^4 k}{(2\pi)^4}
\gamma^{\mu}\,\bar{S} (k)\gamma^{\nu} D_{\mu\nu}(k-p), 
\label{CSE:self-energy-momentum}
\end{equation}
where $\bar{S}(k)$ is the Fourier transform of the translation invariant part of the 
fermion propagator and $D_{\mu\nu}(q)$ is the photon propagator. 

Note that, despite the appearance, $\bar{S} (k)$ is not a conventional momentum-space 
representation of the fermion propagator. Strictly speaking, with the Schwinger phase missing,
$\bar{S}(u-u^\prime)$, as well as its Fourier transform, cannot even be interpreted as an 
actual propagator. This should be also clear from the physics viewpoint because the transverse 
components $\mathbf{k}_\perp$ of four-vector $k^\mu$ are not good quantum numbers for 
classifying fermionic states in a magnetic field. To avoid a potential confusion, in what follows, we 
call $\mathbf{k}_\perp$ a pseudomomentum. It is instructive to mention though that, in the limit of 
large pseudomomentum or weak magnetic field (i.e., $\mathbf{k}_\perp^2\gg |eB|$), the 
effect of the Schwinger phase is negligible and the pseudomomentum can be interpreted as 
an approximate (or ``quasiclassical") fermion's momentum. 

In the weak magnetic field limit we use the photon propagator in the Feynman gauge. 
In momentum space, it reads
\begin{equation}
D_{\mu\nu}(q)=-i\frac{g_{\mu\nu}}{q^2_{\Lambda}} \equiv  
-i\left(\frac{g_{\mu\nu}}{q_0^2-\mathbf{q}^2-m_\gamma^2+i\epsilon}
-\frac{g_{\mu\nu}}{q_0^2-\mathbf{q}^2-\Lambda^2+i\epsilon} \right).
\label{CSE:photon-propagator}
\end{equation}
Here we introduced a nonzero photon mass $m_{\gamma}$ which serves as an infrared regulator 
at the intermediate stages of calculations. Of course, none of the physical observables should 
depend on this parameter (see, however, Section~\ref{ChiralAsymQEDaxial}). [As is 
well know since the classical paper of Stueckelberg \cite{Stueckelberg:1957zz}, introducing a 
photon mass causes no problems in an Abelian gauge theory, such as QED. The same is not 
true in non-Abelian theories.] In general, radiative corrections to the self-energy and axial current 
density are expected to have logarithmic divergencies in the ultraviolet region. The corresponding 
divergencies will be compensated by the vacuum counterterms. As in Ref.~\cite{Adler:1969er}, 
we find that the Feynman regularization of the photon propagator (\ref{CSE:photon-propagator})
with ultraviolet regularization parameter $\Lambda$ presents the most convenient way of
regularizing the theory.

By taking into account the Dirac structure of the fermion propagator, it is straightforward 
to show that the resulting representation of the self-energy (\ref{CSE:self-energy-momentum}) 
has the following Dirac structures:
\begin{equation}
\bar{\Sigma} (p) =
- \gamma^0{\delta\mu(p)} 
+ p^3\gamma^3\,{\delta v_{3}(p)} 
+ (\bm{\gamma}_{\perp}\cdot\mathbf{p}_{\perp}){\delta v_{\perp}(p)} 
+ {\cal M}(p) 
- i \gamma^1 \gamma^2 \tilde{\mu}(p)
- \gamma^3\gamma^5 \Delta(p)
- \gamma^0\gamma^5 \mu_5(p).
\label{CSE:self-energy-general}
\end{equation}
[Compare with Eq.~(\ref{weakQED:self-energy-LL}).]
The first four Dirac structures in Eq.~(\ref{CSE:self-energy-general}) are standard and are present 
in the fermion self-energy also when the magnetic field is absent. The functions ${\delta\mu(p)}$, 
${\delta v_{3}(p)}$, and $\delta v_{\perp}(p)$ define the wave-function renormalization and the 
modification of the (longitudinal and transverse) fermion velocities. Note that in the absence of 
a magnetic field $\delta v_3=\delta v_{\perp}$. The contribution with the unit matrix ${\cal M} (p)$ 
gives a correction to the fermion mass function. As for the last three terms in the self-energy 
(\ref{CSE:self-energy-general}), they are present only if there is a magnetic field. The terms 
with $\tilde{\mu} (p)$ and $\Delta(p)$ are the anomalous magnetic moment function and chiral 
shift, respectively. They were already generated in the study of the NJL model in 
Section~\ref{ChiralAsymNJLMCshift}. Here, due to the long-range character of the QED 
interaction, however, these functions generally depend on the momentum.

The last term in the self-energy (\ref{CSE:self-energy-general}) presents a qualitatively new 
type of contribution in QED. As we show below, it has the form $\mu_5(p)\equiv p_3 f(p)$, 
where $f(p)$ is a dimensionless function. This new contribution comes as a result of the long-range QED interaction 
and, thus, has no analog in the NJL model. If $\mu_5(p)$ were a constant and did not depend on pseudomomentum, 
it would be identical with the chiral chemical potential $\mu_5$ \cite{Fukushima:2008xe,Kharzeev:2009fn}
and would, therefore, break parity. Considering that neither the external magnetic field nor the electromagnetic 
interaction breaks parity, the genuine chiral chemical potential cannot 
be generated in perturbation theory. Instead, we find that the function $\mu_5(p)$ in the self-energy 
(\ref{CSE:self-energy-general}) is an odd function of the $p^3$-component of momentum and, 
therefore, is even under parity.

To linear order in $B$, the translation invariant part of the free fermion propagator in the 
pseudomomentum representation has the following structure:
\begin{equation}
\bar{S}(k) = \bar{S}^{(0)}(k) +\bar{S}^{(1)}(k) +\cdots,
\end{equation}
where $\bar{S}^{(0)}$ is the free fermion propagator in the absence of magnetic field
and $\bar{S}^{(1)}$ is the linear in the magnetic field part. Both of them are derived 
in Appendix~\ref{CSE:App:proper-time-rep-mu} by making use of a generalized Schwinger 
parametrization when the chemical potential is nonzero. The final expressions for
$\bar{S}^{(0)}$  and $\bar{S}^{(1)}$ can be also rendered in the following equivalent form:
\begin{equation}
\bar{S}^{(0)}(k) = i \frac{(k_0+\mu)\gamma^0- \mathbf{k}\cdot\bm{\gamma}+m}
{(k_0+\mu+i\epsilon\, \mathrm{sign}(k_0))^2-\mathbf{k}^2-m^2}
\label{CSE:free-term}
\end{equation}
and 
\begin{equation}
\bar{S}^{(1)}(k) = \gamma^1\gamma^2 eB \frac{(k_0+\mu)\gamma^{0}-k_3 \gamma^3+m}
{\left[ (k_0+\mu+i\epsilon\, \mathrm{sign}(k_0))^2-\mathbf{k}^2 -m^2\right]^2 }.
\label{CSE:linear-term}
\end{equation}
The self-energy at zero magnetic field
\begin{equation}
\bar{\Sigma}^{(0)}(p)=-4i\pi \alpha \int\frac{d^4k}{(2\pi)^4}\gamma^\mu \bar{S}^{(0)}(k) \gamma^\nu D_{\mu\nu}(p-k)
\label{CSE:self-energy-momentum-space-B0}
\end{equation}
determines the counterterms $\delta_2$ and $\delta_m$ in Eq.~(\ref{CSE:Lagrangian}). 
To calculate the self-energy (\ref{CSE:self-energy-momentum-space-B0}),
we will use the generalized Schwinger parametrization of the fermion propagator $\bar{S}^{(0)}(k)$,
see Eq.~(\ref{CSE:S0-vac-plus-matter}) in Appendix~\ref{CSE:App:proper-time-rep-mu}. Such a representation allows a natural  
separation of the propagator (as well as the resulting self-energy) into the ``vacuum" and ``matter" parts. The 
former is very similar to the usual vacuum self-energy in QED in the one-loop approximation. The only difference 
will be the appearance of $p_0+\mu$ instead of $p_0$. The matter part is an additional contribution that comes
from the $\delta$-function contribution in Eq.~(\ref{CSE:S-prop-time-eB0}). Unlike the vacuum part, the matter one has no 
ultraviolet divergences and vanishes when $|\mu|<m$. 

The explicit expression for the vacuum part reads
\begin{equation}
\bar{\Sigma}^{(0)}_{\rm vac} (p) =\frac{\alpha}{2\pi}
\int_0^1 dx \left\{2m-x\left[(p_0+\mu)\gamma^{0}-\mathbf{p}\cdot\bm{\gamma}\right]\right\}
\ln\frac{x \Lambda^2}{(1-x) m^2 + x m_\gamma^2 - x(1-x)\left[(p_0+\mu)^2-\mathbf{p}^2\right]}.
\label{CSE:self-energy-vacuum}
\end{equation}
Note that, while the integral over $x$ can be easily calculated, we keep the result in this more 
compact form. We see that the self-energy (\ref{CSE:self-energy-vacuum}) becomes identical with 
the well-known vacuum self-energy in QED in the Feynman gauge after performing the 
substitution $p_0+\mu\to p_0$ \cite{Peskin:1995ev}. Further, using Eq.~(\ref{CSE:self-energy-vacuum}), 
we find that the counterterms in (\ref{CSE:Lagrangian}) are defined as follows \cite{Peskin:1995ev}:
\begin{eqnarray}
\delta_2 &=& \frac{d\bar{\Sigma}^{(0)}_{\rm vac}(p)}{d\!\! \not{\!P}}\Big|_{\not{P}=m}= -\frac{\alpha}{2\pi}\left(\frac{1}{2}\ln\frac{\Lambda^2}{m^2}+\ln\frac{m_\gamma^2}{m^2}+\frac{9}{4}\right),
\label{CSE:wave-function-renormalization-constant} \\
\delta m &=& m-m_0 = \bar{\Sigma}^{(0)}_{\rm vac}(p)\Big|_{\not{P}=m} =\frac{3\alpha}{4\pi} m \left(\ln\frac{\Lambda^2}{m^2}+\frac{1}{2}\right),
\label{CSE:mass-renormalization-constant}
\end{eqnarray}
where $P=(p_0+\mu,\mathbf{p})$. Note that the fermion wave function renormalization constant is defined 
as follows: $Z_2=1+\delta_2$.

For completeness, let us calculate the additional matter part of the self-energy due to the filled fermion states given by
\begin{equation}
\bar{\Sigma}^{(0)}_{\rm mat} (p) = -\frac{i\alpha}{\pi^2} \int_{-\mu}^{0} d k_0 \int  d^3\mathbf{k}\, 
\frac{(k_0+\mu)\gamma^{0}-\mathbf{k}\cdot\bm{\gamma}-2m}{(k_0-p_0)^2-(\mathbf{k}-\mathbf{p})^2}\,\,
\delta \left[(k_0+\mu)^2-\mathbf{k}^2 -m^2 \right].
\end{equation}
After performing the integration over the energy and spatial angular coordinates, we find
\begin{eqnarray}
\bar{\Sigma}^{(0)}_{\rm mat} (p) &=&
-\frac{\alpha}{\pi}\int_{0}^{\sqrt{\mu^2-m^2}}\frac{k dk}{|\mathbf{p}|}\Bigg\{\frac{1}{2}\left(\gamma^0-\frac{2m}{\sqrt{k^2+m^2}}\right) 
\ln\frac{(p_0+\mu-\sqrt{m^2+k^2})^2-(k-|\mathbf{p}|)^2}{(p_0+\mu-\sqrt{m^2+k^2})^2-(k+|\mathbf{p}|)^2}
\nonumber\\
&-&\frac{k (\mathbf{p}\cdot\bm{\gamma})}{|\mathbf{p}|\sqrt{m^2+k^2}} 
\left(1+\frac{k^2+\mathbf{p}^2-(p_0+\mu-\sqrt{m^2+k^2})^2}{4k|\mathbf{p}|} 
\ln\frac{(p_0+\mu-\sqrt{m^2+k^2})^2-(k-|\mathbf{p}|)^2}{(p_0+\mu-\sqrt{m^2+k^2})^2-(k+|\mathbf{p}|)^2}\right)
\Bigg\}.
\end{eqnarray}
While the remaining integral over the absolute value of the momentum $k$ can be also performed, 
the result will take a rather complicated form that will not add any clarity.

The linear in the magnetic field correction to the translation invariant part of the fermion self-energy in a magnetic field reads
\begin{equation}
\bar{\Sigma}^{(1)}(p)=-4i\pi \alpha \int\frac{d^4k}{(2\pi)^4}\gamma^\mu \bar{S}^{(1)}(k) \gamma^\nu D_{\mu\nu}(p-k).
\label{weakQED:self-energy-momentum}
\end{equation}
This correction, which in particular contains a chiral shift parameter term, is analyzed 
below. We will also use this expression in the derivation of leading radiative corrections  
to the axial current density in Section~\ref{ChiralAsymQEDaxial}.

To linear order in the magnetic field, the vacuum part of the self-energy is given by
Eq.~(\ref{weakQED:self-energy-momentum}). As shown in Appendix~\ref{CSE:App:proper-time-rep-mu}, 
the explicit form of the linear order contribution to the fermion propagator reads \cite{Gorbar:2013upa}
\begin{equation}
\bar{S}^{(1)}(k) = - e B \Bigg\{
\int_{0}^{\infty} s ds\, e^{i s [(k_0+\mu)^2-m^2-\mathbf{k}^2+i\epsilon]} 
+2i\pi  \theta(|\mu|-|k_0|) \theta(-k_0\mu) 
\delta^{\prime}\left[(k_0+\mu)^2-m^2-\mathbf{k}^2 \right]
\Bigg\}  \left[(k_0+\mu)\gamma^{0}-k^3\gamma^3+m\right]\gamma^1\gamma^2.
\label{weakQED:S-prop-time-mu3}
\end{equation}

Note that $\bar{S}^{(1)}(k)$ splits naturally into the ``vacuum" and ``matter" parts, with the latter 
containing the $\delta$ function. [Note that the vacuum part is not precisely reflecting the 
nature of the first contribution, because it depends on the chemical potential.] It is 
convenient to treat the two pieces separately in the calculation of the self-energy.

To linear order in the magnetic field, the vacuum part of the self-energy is given by
\begin{eqnarray}
\bar{\Sigma}_{\rm vac}^{(1)} (p) &=& - 8 i\pi\alpha eB  \int_{0}^{\infty} d\tau \int_{0}^{\infty} s ds\, 
\int\frac{ d^4 k}{(2\pi)^4} e^{i s \left[(k_0+\mu)^2-m^2-\mathbf{k}^2\right]+i\tau \left[(p_0-k_0)^2-
(\mathbf{p}-\mathbf{k})^2\right]}
\left[(k_0+\mu)\gamma^{0}-k^3\gamma^3\right]\gamma^1\gamma^2\nonumber\\
&=& \frac{\alpha eB}{2\pi} \left[(p_0+\mu)\gamma^{0}-p^3\gamma^3\right]\gamma^1\gamma^2 
\int_{0}^{\infty} \int_{0}^{\infty} \frac{s \tau ds d\tau}{(s+\tau)^3}
e^{-ism^2+i\frac{s\tau}{s+\tau}\left[(p_0+\mu)^2-\mathbf{p}^2 \right]} 
\nonumber\\
&=& \frac{\alpha eB}{2\pi}\,i\gamma^1\gamma^2 \frac{(p_0+\mu)\gamma^{0}-p^3\gamma^3}
{(p_0+\mu)^2-\mathbf{p}^2}
\left[1+\frac{m^2}{(p_0+\mu)^2-\mathbf{p}^2}\left(\ln\frac{|m^2+\mathbf{p}^2-(p_0+\mu)^2|}{m^2}-i
\pi \theta[...] \right) \right],
\label{weakQED:self-energy-vacuum}
\end{eqnarray}
where the imaginary part is nonzero when $(p_0+\mu)^2-\mathbf{p}^2>m^2$. Note that this 
expression simplifies a lot in the chiral limit.

To the same linear order in the magnetic field, the matter part of the self-energy is given by 
\begin{equation}
\bar{\Sigma}_{\rm mat}^{(1)} (p) = \frac{\alpha eB}{\pi^2}\,i\gamma^1\gamma^2
\int d^4 k \frac{(k_0+\mu)\gamma^{0}-k^3\gamma^3}{(p_0-k_0)^2-(\mathbf{p}-\mathbf{k})^2}\, 
\theta(|\mu|-|k_0|) \theta(-k_0\mu) 
\delta^{\prime}\left[(k_0+\mu)^2-m^2-\mathbf{k}^2 \right].
\label{weakQED:self-energy-matter}
\end{equation}
We would like to emphasize that Eqs.~(\ref{weakQED:self-energy-vacuum}) and 
(\ref{weakQED:self-energy-matter}) imply that only the chiral asymmetric structures 
$\Delta$ and $\mu_5$ are generated in the fermion self-energy (\ref{CSE:self-energy-general})
in the linear in $B$ approximation. The vacuum contributions to the chiral shift and 
the chiral chemical potential terms are
\begin{equation}
\Delta_{\rm vac}(p)=\frac{i \alpha (p_0+\mu) eB }{2\pi \left[(p_0+\mu)^2 -\mathbf{p}^2\right]
^2}\left(
(p_0+\mu)^2 -\mathbf{p}^2
+m^2\ln\frac{|m^2+\mathbf{p}^2-(p_0+\mu)^2 |}{m^2}-i\pi \, m^2 \, \theta \left[(p_0+\mu)^2 -
\mathbf{p}^2-m^2\right]
\right),
\label{weakQED:Delta-vac}
\end{equation}
and 
\begin{equation}
\mu^{\rm vac}_5(p) = -\frac{p^3}{p_0+\mu}\Delta_{\rm vac},
\label{weakQED:parameters-vacuum}
\end{equation}
respectively. The corresponding matter contributions are given by
\begin{eqnarray}
\Delta_{\rm mat}(p)&=&
\frac{\alpha e B }{4\pi|\mathbf{p}|}\ln\frac{p_0^2-(|\mathbf{p}|-\sqrt{\mu^2-m^2})^2}{p_0^2-(|\mathbf{p}|+\sqrt{\mu^2-m^2})^2}
-  \frac{\alpha e B \sqrt{\mu^2-m^2}}{2\pi [\mathbf{p}^2-(p_0+\mu)^2]}
- \frac{\alpha e B m^2 (p_0+\mu)}{2\pi \left[(p_0+\mu)^2-\mathbf{p}^2\right]^2}
\ln\frac{\mu-\sqrt{\mu^2-m^2}}{\mu+\sqrt{\mu^2-m^2}}\nonumber\\
&- &\frac{\alpha e B }{8\pi|\mathbf{p}|}\left(\frac{(p_0+\mu+|\mathbf{p}|)^2-m^2}{(p_0+\mu+|\mathbf{p}|)^2}
\ln\frac{p_0+|\mathbf{p}|-\sqrt{\mu^2-m^2}}{p_0+|\mathbf{p}|+\sqrt{\mu^2-m^2}}
-\frac{(p_0+\mu-|\mathbf{p}|)^2-m^2}{(p_0+\mu-|\mathbf{p}|)^2}
\ln\frac{p_0-|\mathbf{p}|-\sqrt{\mu^2-m^2}}{p_0-|\mathbf{p}|+\sqrt{\mu^2-m^2}}\right), \nonumber \\
\label{weakQED:Delta-mat}
\end{eqnarray}
and 
\begin{eqnarray}
\mu^{\rm mat}_5(p)&=&\frac{\alpha eBp^3 }{2\pi\mathbf{p}^2}\Bigg(
\frac{(p_0+\mu)\sqrt{\mu^2-m^2}}{\mathbf{p}^2-(p_0+\mu)^2} 
+\frac{\mu^2+p_0(p_0+\mu)-\mathbf{p}^2+m^2}{2p\mu } 
\ln\frac{p_0^2-(|\mathbf{p}|-\sqrt{\mu^2-m^2})^2}{p_0^2-(|\mathbf{p}|+\sqrt{\mu^2-m^2})^2}
\nonumber\\
&+& 
\frac{3(p_0+\mu)(p_0+\mu+|\mathbf{p}|)^2+m^2(p_0+\mu+2|\mathbf{p}|)}{4|\mathbf{p}|(p_0+\mu+|\mathbf{p}|)^2}
\ln\frac{p_0+|\mathbf{p}|-\sqrt{\mu^2-m^2}}{p_0+|\mathbf{p}|+\sqrt{\mu^2-m^2}}\nonumber\\
&-&\frac{3(p_0+\mu)(p_0+\mu-|\mathbf{p}|)^2+m^2(p_0+\mu-2|\mathbf{p}|)}{4|\mathbf{p}|(p_0+\mu-|\mathbf{p}|)^2}
\ln\frac{p_0-|\mathbf{p}|-\sqrt{\mu^2-m^2}}{p_0-|\mathbf{p}|+\sqrt{\mu^2-m^2}}
\Bigg).
\label{weakQED:parameters-matter}
\end{eqnarray}
Eqs.~(\ref{weakQED:parameters-vacuum}) and (\ref{weakQED:parameters-matter}) show that
$\mu_5(p)$ is indeed an odd function of $p^3$, and, therefore, it does not break parity.

It is useful to consider some particular limits of the obtained expressions. The first interesting 
case corresponds to the behavior of the chiral shift and chiral chemical potential on the Fermi 
surface, i.e., for $p_0\to 0$ and $|\mathbf{p}|\to p_F\equiv \sqrt{\mu^2-m^2}$. We have
\begin{eqnarray}
\Delta=\Delta_{\rm mat}+\Delta_{\rm vac} &\simeq & - \frac{\alpha e B \mu }{\pi m^2}
\left(\ln\frac{m^2}{2\mu\left(|\mathbf{p}|-p_F\right)}-1\right),\\
\mu_5=\mu^{\rm mat}_5+\mu^{\rm vac}_5 &\simeq & \frac{\alpha e B \mu \cos\theta}{\pi m^2}
\left(\ln\frac{m^2}{2\mu\left(|\mathbf{p}|-p_F\right)}-1\right),
\end{eqnarray}
where $\cos\theta=p^3/p$; i.e., $\theta$ is the angle between the magnetic field and momentum. 
Furthermore, Eqs.~(\ref{weakQED:Delta-vac}) through
(\ref{weakQED:parameters-matter}) simplify strongly in the chiral limit, where, for $\mu>0$,
\begin{eqnarray}
\bar{\Sigma}^{(1)} (p) &\simeq & 
\frac{\alpha e B }{2\pi}\gamma^3\gamma^5\left[\frac{p_0+2\mu}{(p_0+\mu)^2-\mathbf{p}^2}
-\frac{1}{4|\mathbf{p}|}\ln\frac{p_0^2-(|\mathbf{p}|+\mu)^2}{p_0^2-(|\mathbf{p}|-\mu)^2}\right]
\nonumber\\
&- &\frac{\alpha e B }{2\pi}\gamma^0\gamma^5\frac{p^3}{\mathbf{p}^2}\left[
\frac{\mu(p_0+\mu)+\mathbf{p}^2}{(p_0+\mu)^2-\mathbf{p}^2}
+\frac{\mu-p_0}{4 |\mathbf{p}|}\ln\frac{p_0^2-(|\mathbf{p}|-\mu)^2}{p_0^2-(|\mathbf{p}|+\mu)^2}
\right].
\end{eqnarray}

The dispersion relations for fermion quasiparticles in a weak magnetic field can be formally 
obtained by considering the location of the poles of the fermion propagator. In the limit 
of large pseudomomentum or weak magnetic field (i.e., $\mathbf{k}^2_{\perp} \gg |eB|$), the effects 
of the Schwinger phase can be neglected and pseudomomentum can be interpreted as 
an approximate (or ``quasiclassical") fermion's momentum. Then, the poles of the fermion propagator
are defined by the following equation:
\begin{equation}
\mbox{det}\left[i\bar{S}^{-1}(p)-\Sigma(p)\right] =0.
\label{weakQED:det=0}
\end{equation}

To determine the dispersion relations from Eq.~(\ref{weakQED:det=0}), we should define the inverse free propagator 
in the pseudomomentum representation. This is not difficult to do by following the procedure given in 
Section~\ref{ChiralAsymQEDstrongB}. The inverse free propagator in the coordinate space is defined as follows:
\begin{eqnarray}
iS^{-1}(u,u^\prime) &=& \left(i\gamma^{\nu}{\cal D}_{\nu}+\mu\gamma^0-m\right)\delta^4(u-u^\prime).
\label{weakQED:inverse-S-coordinate-space}
\end{eqnarray}
By making use of Eq.~(\ref{weakQED:completeness}), one can rewrite the inverse free propagator (\ref{weakQED:inverse-S-coordinate-space}) 
in the form
\begin{equation}
iS^{-1}(u,u^\prime) =\sum_{N=0}^{\infty} \int\frac{dp_0 dp^3 dp\,\,e^{-ip_0(x_0-y_0)+ip^3(x^3-y^3)}}{(2\pi)^2}
\left[(p_0+\mu)\gamma^0-p^3\gamma^3
-(\bm{\pi}_{\perp}\cdot\bm{\gamma}_{\perp})-m
\right] \psi_{Np}(\mathbf{r}_{\perp}) \psi_{Np}^{*}(\mathbf{r}_{\perp}^\prime).
\end{equation}
After taking into account the identity in Eq.~(\ref{weakQED:pi-gamma-relation}) and the table 
integral in Eq.~(\ref{weakQED:HnHm-integral}), we can easily perform the integration over the quantum 
number $p$. Just like in the case of the self-energy, the result takes the form of a 
product of the standard Schwinger phase and a translationally invariant function, i.e.,
\begin{equation}
iS^{-1}(u,u^\prime) = e^{i\Phi(\mathbf{r}_\perp,\mathbf{r}_\perp^\prime)} i\bar{S}^{-1}(u-u^\prime) .
\end{equation}
The translationally invariant function is given by
\begin{equation}
i\bar{S}^{-1}(x) =\frac{e^{-\xi/2}}{2\pi l^2}
\sum_{n=0}^{\infty}\int \frac{dp_0 dp^3 }{(2\pi)^2}  e^{-ip_0x_0+ip^3x^3} \Big\{
\left[(p_0+\mu)\gamma^0 -p^3\gamma^3-m\right] \left[ L_{n}(\xi){\cal P}_{+}+L_{n-1}(\xi){\cal P}_{-}\right]
+\frac{i}{ l^2}(\mathbf{r}_{\perp}\cdot\bm{\gamma}_{\perp}) L^{1}_{n-1}(\xi)
\Big\},
\end{equation}
where $\xi = \mathbf{r}_{\perp}^2/(2 l^2)$. By performing the Fourier transform, we finally arrive at 
the following expansion of the translation invariant part of the inverse free propagator over Landau levels 
[compare with the corresponding expansion of the self-energy in Eq.~(\ref{weakQED:self-energy-LL})]:
\begin{equation}
i \bar{S}^{-1}(p) = 2  e^{-p_{\perp}^2  l^2}  \sum\limits_{n=0}^{\infty} (-1)^n\Big\{
\left[(p_0+\mu)\gamma^0 - p^3\gamma^3 -m\right] 
\left[{\cal P}_{+} L_n(2p_{\perp}^2  l^2)-{\cal P}_{-} L_{n-1}(2p_{\perp}^2  l^2) \right] 
+2 (\bm{\gamma}_\perp\cdot \bm{p}_{\perp})
L^1_{n-1}(2p_{\perp}^2  l^2)\Big\} .
\label{weakQED:inverse-bare-propagator}
\end{equation}
Interestingly, by performing the summation over Landau levels in this expression using formula
(\ref{sum-Laguerre}), we obtain
\begin{equation}
i\bar{S}^{-1}(p) =(p_0 +\mu)\gamma^0 -(\bm{\gamma}_\perp\cdot \mathbf{p}_{\perp})- p^3\gamma^3 -m .
\label{weakQED:inverse-free-momentum}
\end{equation}
This is a remarkable result, because it means that the translation invariant part of the inverse free propagator in 
a magnetic field is identical to the inverse free propagator in the absence of a magnetic field. Therefore, for the 
inverse free propagator, only the Schwinger phase contains information about the presence of a magnetic field.
 
For the free propagator in the weak field limit, the dependence on the Landau level index [which is 
the eigenvalue of the operator $-\frac{1}{2}(\bm{\pi}_\perp\cdot\bm{\gamma}_\perp)^2 l^2$] can be 
unambiguously replaced by the square of the transverse momentum, i.e., $2n|eB| \to \mathbf{p}_\perp^2$. 
Therefore, when using the pseudomomentum representation in Eq.~(\ref{weakQED:det=0}), we can interpret 
$\mathbf{p}_\perp^2$ as a convenient shorthand substitution for $2n|eB|$. Indeed, this is natural in 
the weak field limit, when the quantization of Landau levels is largely irrelevant. This implies the 
standard dispersion relation $p_0=-\mu \pm \sqrt{\mathbf{p}_{\perp}^2+p_3^2+m^2}$, or equivalently 
$p_0=-\mu \pm \sqrt{2n|eB|+p_3^2+m^2}$ after the substitution $\mathbf{p}_\perp^2\to 2n|eB|$.

\begin{figure}[t]
 \includegraphics[width=0.45\textwidth]{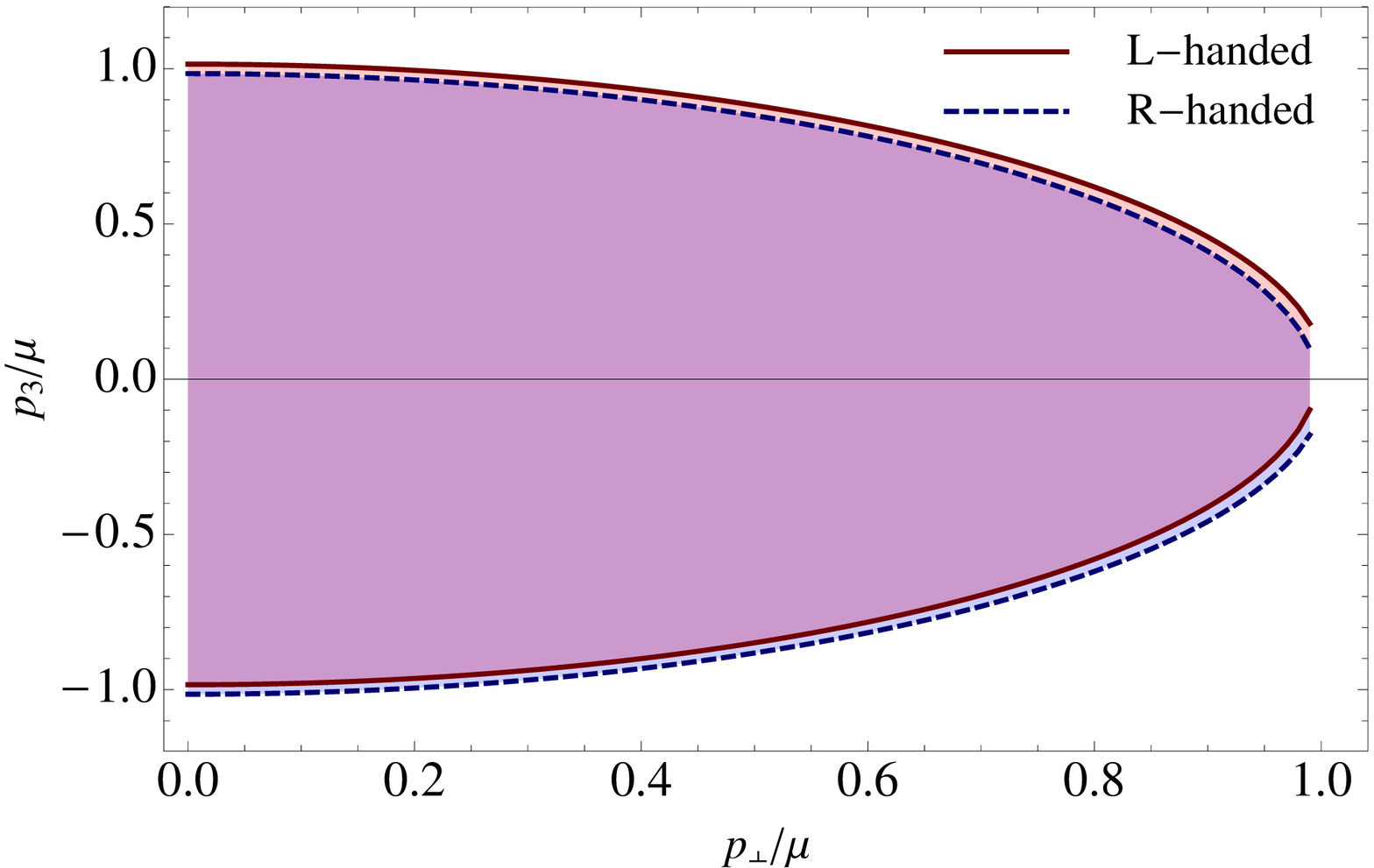}\hspace{0.05\textwidth}
\includegraphics[width=0.45\textwidth]{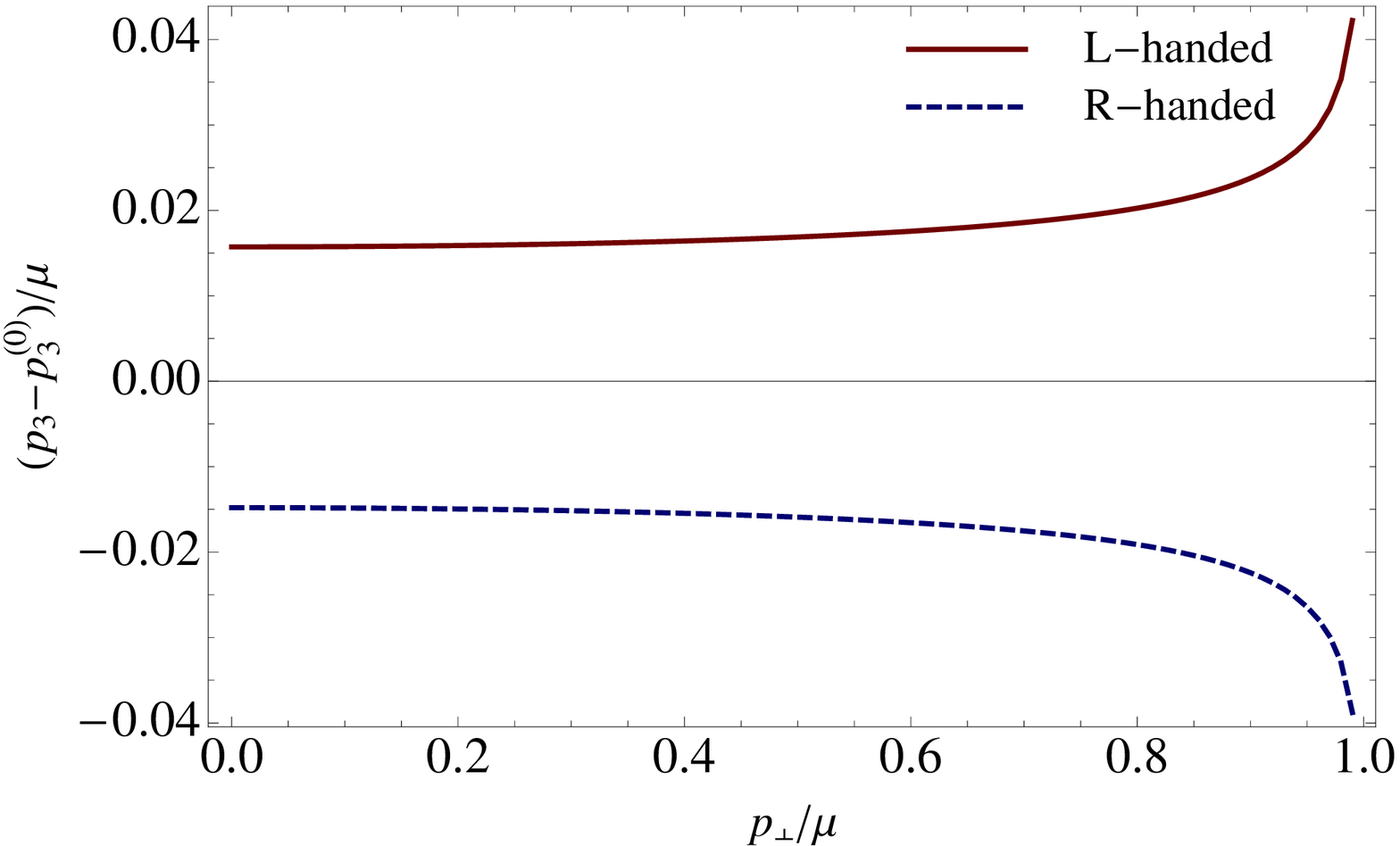}
\caption{Asymmetry of the Fermi surface for predominantly left-handed and right-handed particles 
for $|eB|=0.1\mu^2$ and $\alpha=1/137$.}
\label{weakQED:fig-fermi-surf}
\end{figure}

By making use of the chiral representation of the Dirac $\gamma$ matrices, the inverse free propagator 
(\ref{weakQED:inverse-free-momentum}), and the self-energy in the weak magnetic field limit, Eq.~(\ref{weakQED:det=0}) 
can be rewritten in the following equivalent form:
\begin{equation}
\mbox{det}\left(
\begin{array}{cc}
p_0 +\mu -(\bm{\sigma}_\perp\cdot \mathbf{p}_{\perp})+(\Delta- p^3)\sigma^3 +\mu_5& m \\
m & p_0 +\mu +(\bm{\sigma}_\perp\cdot \mathbf{p}_{\perp})+ (\Delta+ p^3)\sigma^3-\mu_5
\end{array}
\right)=0,
\label{weakQED:det=0-more1}
\end{equation}
where $\bm{\sigma}$ are Pauli matrices. Calculating the determinant, we obtain
\begin{equation}
\left[(p_0 +\mu-\mu_5)^2-\mathbf{p}_{\perp}^2-(p^3+\Delta)^2\right]
\left[(p_0 +\mu+\mu_5)^2-\mathbf{p}_{\perp}^2-(p^3-\Delta)^2\right]
-2m^2\left[(p_0 +\mu)^2+\Delta^2-\mathbf{p}_{\perp}^2-p_3^2-\mu_5^2\right]+m^4=0.
\label{weakQED:det=0-more2}
\end{equation}
This expression can be factorized to produce two equations for predominantly left-handed 
and predominantly right-handed particles:
\begin{eqnarray}
(p_0 +\mu)^2-\mathbf{p}_{\perp}^2-p_3^2-m^2-\Delta^2+\mu_5^2
-2\sqrt{(p^3\Delta +\mu_5(p_0 +\mu))^2+m^2(\Delta^2-\mu_5^2)}
=0,
\label{weakQED:det=0-more3}\\
(p_0 +\mu)^2-\mathbf{p}_{\perp}^2-p_3^2-m^2-\Delta^2+\mu_5^2
+2\sqrt{(p^3\Delta +\mu_5(p_0 +\mu))^2+m^2(\Delta^2-\mu_5^2)}
=0.
\label{weakQED:det=0-more4}
\end{eqnarray}
By making use of the analytical results for the self-energy obtained in Eqs.~(\ref{weakQED:Delta-vac}) --
(\ref{weakQED:parameters-matter}) and the dispersion relations that follow from Eqs.~(\ref{weakQED:det=0-more3}) and (\ref{weakQED:det=0-more4}),
we can easily write down the equations for the Fermi surfaces of both types of particles. Namely, we take $p^0=0$ and 
solve for $p^3$ as a function of $p_\perp$. The results are shown in the left panel of Fig.~\ref{weakQED:fig-fermi-surf} in the
case of the physical value of the fine structure constant ($\alpha=1/137$) and the magnetic field $|eB|=0.1\mu^2$. 
In order to clearly demonstrate the magnitude of the effect, in the right panel of Fig.~\ref{weakQED:fig-fermi-surf} we 
also plot the difference between the longitudinal momenta with and without the inclusion of the interaction 
induced chiral asymmetry.

As the results in Fig.~\ref{weakQED:fig-fermi-surf} demonstrate, the Fermi surface of the predominantly 
left-handed particles is slightly shifted in the direction of the magnetic field, while the Fermi surface of 
the predominantly right-handed particles is slightly shifted in the direction opposite of the magnetic field. 
This is in qualitative agreement with the finding in the NJL model \cite{Gorbar:2011ya}. In the case of QED
with its long-range interaction, however, the chiral asymmetry of the Fermi surfaces comes not only from 
the $\Delta$ function, but also from the new function $\mu_5(p)\equiv p_3 f(p)$. Also, unlike in the NJL 
model, both of these functions have a nontrivial dependence on the particles' momenta. In particular, they
reveal a logarithmic enhancement of the asymmetry near the Fermi surface.

As we see, the chiral asymmetry in QED is in qualitative agreement with the earlier results in the 
NJL model. In both models, the Fermi surfaces of the left- and right-handed fermions are shifted 
relative to each other in momentum space in the direction of the magnetic field. Also, in both cases,
the chiral asymmetry in the ground state of magnetized matter is the source of the radiative corrections 
to the axial current density.

\subsubsection{Radiative corrections to axial current density}
\label{ChiralAsymQEDaxial}

The renormalization group invariant axial current density, which is a quantity of 
the principal interest here, is given by
\begin{equation}
\langle j_{5}^{3}\rangle = -Z_2\mathrm{tr}\left[\gamma^3\gamma^5 G(x,x)\right],
\label{CSE:j53-general}
\end{equation}
where $G(u,u^\prime)$ is the full fermion propagator and $Z_2 = 1 + \delta_2$ is the wave function 
renormalization constant of the fermion propagator, cf. Eq.~(\ref{CSE:Lagrangian}).

To the first order in the coupling constant $\alpha=e^2/(4\pi)$, the propagator reads
\begin{equation}
G(u,u^\prime) = S(u,u^\prime)+i\int d^4 u d^4 v S(x,u) \Sigma(u,v) S(v,y)  + i\int d^4 u d^4 v S(x,u) \Sigma_{\rm ct}(u,v) S(v,y),
\label{CSE:gxy-expansion}
\end{equation}
where $S(u,u^\prime)$ is the free fermion propagator in the magnetic field, $\Sigma(u,v)$ is 
the one-loop fermion self-energy, and $\Sigma_{\rm ct}(u,v)$ is the counterterm
contribution to the self-energy. The structure of the counterterm contribution is 
determined by the last two terms in the Lagrangian density (\ref{CSE:Lagrangian}). 

\begin{figure}[t]
\begin{center}
\includegraphics[width=0.45\textwidth]{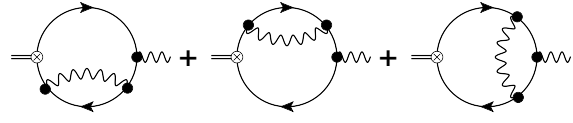}
\caption{The leading radiative corrections to the axial current in the linear 
in magnetic field approximation. Solid and wavy lines correspond to the fermion and photon 
propagators, respectively. Double solid lines describe the axial current insertions and the external 
wavy lines attached to the fermion loops indicate the insertions of the external gauge field.}
\label{CSE:fig-correlator}
\end{center}
\end{figure}

Below, we make use of the weak magnetic field expansion in the calculation 
of the axial current density. Such an expansion is straightforward to obtain from the 
general expression in Eq.~(\ref{CSE:j53-general}) and the representation (\ref{CSE:gxy-expansion}) 
for the fermion propagator. For the fermion propagator to linear in $B$ order, we have
\begin{equation}
S(u,u^\prime) = \bar{S}^{(0)}(u-u^\prime)-  i e\int d^4z\,\bar{S}^{(0)}(x-z)\gamma^{\nu}\bar{S}^{(0)}(z-y)A _{\nu}(z).
\label{CSE:free-propagator-magnetic-1}
\end{equation}
Further, by making use of Eq.~(\ref{CSE:free-propagator-magnetic-1}), the weak field expansion of the 
self-energy follows from the definition in Eq.~(\ref{CSE:self-energy}). (Note that the photon 
propagator is independent of the magnetic field to this order.) Combining all pieces 
together, we can find the complete expression for the leading radiative corrections 
to the axial current (\ref{CSE:j53-general}) in the approximation linear in the magnetic field. 
In this framework, the diagrammatical representation for the leading radiative corrections 
to the axial current is shown in Fig.~\ref{CSE:fig-correlator} (for simplicity, we do not display 
the contributions due to counterterms).

As follows from Gauss's law, the consistent description of the 
dynamics in QED at finite density requires the overall neutrality of the system. The neutralizing 
background can be provided, for example, by protons (in neutron stars) or nuclei (in white dwarfs).
This implies that all tadpole diagrams should cancel. In the diagrammatic form of the expression 
for the axial current in Fig.~\ref{CSE:fig-correlator}, this corresponds to removing the one-particle 
reducible diagram (not shown in Fig.~\ref{CSE:fig-correlator}) that is made of two fermion loops 
connected by a photon line.
   
Instead of using the expansion for the free propagator in Eq.~(\ref{CSE:free-propagator-magnetic-1}), we find it 
much more convenient to utilize the Schwinger form of the fermion propagator (\ref{CSE:S-Schwinger-phase}), 
which consists of a simple phase, that breaks the translation invariance, and a translation invariant function. 
Taking into account that the Schwinger phase $\Phi(\mathbf{r}_\perp,\mathbf{r}_\perp^\prime)$ is linear in magnetic field, we arrive at the following 
alternative form of the weak field expansion of the fermion propagator in the linear in $B$ approximation: 
\begin{equation}
S(u,u^\prime) = \bar{S}^{(0)}(u-u^\prime)+i\Phi(\mathbf{r}_\perp,\mathbf{r}_\perp^\prime)\bar{S}^{(0)}(u-u^\prime)+\bar{S}^{(1)}(u-u^\prime),
\label{CSE:free-propagator-magnetic-2}
\end{equation}
where $\bar{S}^{(0)}(u-u^\prime)$ and $\bar{S}^{(1)}(u-u^\prime)$ are the zeroth- and first-order terms in powers of $B$
in the translation invariant part of the propagator. [For the explicit forms of their Fourier transforms see 
Eqs.~(\ref{CSE:free-term}) and (\ref{CSE:linear-term}) above.] Of course, the representations in 
Eqs.~(\ref{CSE:free-propagator-magnetic-1}) and (\ref{CSE:free-propagator-magnetic-2}) are equivalent. One can 
check this explicitly, for example, by making use of the Landau gauge for the external field 
$A _{\nu}$.

Furthermore, Eq.~({\ref{CSE:self-energy}) implies that a similar expansion takes place also for the fermion self-energy
\begin{equation}
\Sigma(u,v)= \bar{\Sigma}^{(0)}(u-v)+i\Phi(u,v)\bar{\Sigma}^{(0)}(u-v)+ \bar\Sigma^{(1)}(u-v).
\end{equation}
The Fourier transforms of the self-energies $\bar{\Sigma}^{(0)}(u-u^\prime)$ and $\bar{\Sigma}^{(1)}(u-u^\prime)$ are given by
Eqs.~(\ref{CSE:self-energy-momentum-space-B0}) and (\ref{weakQED:self-energy-momentum}), respectively.

Omitting the noninteresting zeroth order in $B$ contribution in Eq.~(\ref{CSE:gxy-expansion}), we arrive at the following 
linear in $B$ contribution to the propagator: 
\begin{eqnarray}
G^{(1)}(x,x)&=&\bar{S}^{(1)}(x,x) 
+i\int  d^4 u d^4 v\left[\bar{S}^{(1)}(x-u)\bar{\Sigma}^{(0)}(u-v)\bar{S}^{(0)}(v-x)+\bar{S}^{(0)}(x-u)\bar{\Sigma}^{(0)}(u-v) 
\bar{S}^{(1)}(v-x)\right]\nonumber\\
&+&i\int  d^4 u d^4 v \left[\bar{S}^{(0)}(x-u)\bar\Sigma^{(1)}(u-v)\bar{S}^{(0)}(v-x) \right]\nonumber\\
&-&\int  d^4 u d^4 v \left[\Phi(x,u)+\Phi(u,v)+\Phi(v,x)\right]\bar{S}^{(0)}(x-u)\bar{\Sigma}^{(0)}(u-v)\bar{S}^{(0)}(v-x).
\label{CSE:Mat-propagator-linear-magnetic}
\end{eqnarray}
Noting that $\Phi(x,u)+\Phi(u,v)+\Phi(v,x) = \frac{eB}{2}\left[(x_1-u_1)(v_2-x_2)-(v_1-x_1) (x_2-u_2)\right]$ 
is a translation invariant function, it is convenient to switch to the momentum space on the right-hand 
side of Eq.~(\ref{CSE:Mat-propagator-linear-magnetic}). The result reads 
\begin{eqnarray}
G^{(1)}(x,x)&=& \int \frac{d^4 p}{(2\pi)^4}\bar{S}^{(1)}(p)
+i\int \frac{d^4 p}{(2\pi)^4}\left[ 
\bar{S}^{(1)}(p)\bar{\Sigma}^{(0)}(p)\bar{S}^{(0)}(p) 
+\bar{S}^{(0)}(p)\bar{\Sigma}^{(0)}(p)\bar{S}^{(1)}(p) 
+\bar{S}^{(0)}(p)\bar\Sigma^{(1)}(p)\bar{S}^{(0)}(p) \right]\nonumber\\
&+ &\frac{eB}{2}\int\frac{d^4 p}{(2\pi)^4}\left[\frac{\partial\bar{S}^{(0)}(p) }{\partial p_1}
\bar{\Sigma}^{(0)}(p)\frac{\partial\bar{S}^{(0)}(p) }{\partial p_2}
-\frac{\partial\bar{S}^{(0)}(p)}{\partial p_2}\bar{\Sigma}^{(0)}(p)\frac{\partial\bar{S}^{(0)}(p) }{\partial p_1} \right].
\label{CSE:linear-propagator}
\end{eqnarray}
By substituting this into the definition in Eq.~(\ref{CSE:j53-general}), we obtain the following expression 
for the axial current density:
\begin{eqnarray}
\langle j_{5}^{3}\rangle=\langle j_{5}^{3}\rangle_0+\langle j_{5}^{3}\rangle_{\alpha},
\label{CSE:axial-current-complete}
\end{eqnarray}
where
\begin{eqnarray}
\langle j_{5}^{3}\rangle_0= - \int \frac{d^4 p}{(2\pi)^4}\mathrm{tr} \left[ \gamma^3\gamma^5 \bar{S}^{(1)}(p) \right]
\end{eqnarray}
is the contribution to the axial current in the free theory and
\begin{eqnarray}
\langle j_{5}^{3}\rangle_{\alpha} &=&
- \frac{eB}{2} \int \frac{d^4 p}{(2\pi)^4}\mathrm{tr} \Bigg[ \gamma^3\gamma^5 
\frac{\partial \bar{S}^{(0)}(p) }{\partial p_1}\bar{\Sigma}^{(0)}(p)  \frac{\partial \bar{S}^{(0)}(p) }{\partial p_2}
- \gamma^3\gamma^5  \frac{\partial \bar{S}^{(0)}(p) }{\partial p_2}\bar{\Sigma}^{(0)}(p)\frac{\partial \bar{S}^{(0)}
(p)}{\partial p_1} \Bigg]
\nonumber\\
&-&i\int \frac{d^4 p}{(2\pi)^4} \mathrm{tr} \Bigg[ 
\gamma^3\gamma^5 \bar{S}^{(1)}(p)\bar{\Sigma}^{(0)}(p)\bar{S}^{(0)}(p) 
+\gamma^3\gamma^5 \bar{S}^{(0)}(p)\bar{\Sigma}^{(0)}(p) \bar{S}^{(1)}(p) 
+ \gamma^3\gamma^5 \bar{S}^{(0)}(p) \bar{\Sigma}^{(1)}(p)\bar{S}^{(0)}(p) \Bigg]
+\langle j_{5}^{3}\rangle_{\rm ct}\nonumber\\
\label{CSE:axial-current-initial}
\end{eqnarray}
defines the leading radiative corrections to the axial current. The counterterm contribution $\langle j_{5}^{3}\rangle_{\rm ct}$ in
Eq.~(\ref{CSE:axial-current-initial}) contains all the contributions with $\delta_2$ and $\delta_m$. 
Its explicit form will be given later, see Eq.~(\ref{CSE:counterterm-contribution-general}).

It is instructive to start from investigating the structure of Eq.~(\ref{CSE:axial-current-complete}) in the 
free theory (i.e., to the zeroth order in $\alpha$). By making use of the explicit form of 
$\bar{S}^{(1)}(k) $ in Eq.~(\ref{CSE:linear-term}), we straightforwardly derive the following 
contribution to the axial current density:
\begin{equation}
\langle j_{5}^{3}\rangle_0
=\frac{eB\,\mathrm{sign}(\mu)}{4\pi^3}\int d^3\mathbf{k}\,\delta(\mu^2-\mathbf{k}^2-m^2)
=\frac{eB\,\mathrm{sign}(\mu)}{2\pi^2}\sqrt{\mu^2-m^2},\,
\label{CSE:axial-current-topological}
\end{equation}
which coincides, of course, with the very well-known topological contribution \cite{Metlitski:2005pr}. 
Note that in contrast to the approach using the expansion over the Landau levels, where the 
contribution to $\langle j_{5}^{3}\rangle_0$ comes only from the filled LLL states, the origin 
of the same topological contribution in the formalism of weak magnetic fields is quite different.
As Eq.~(\ref{CSE:axial-current-topological}) implies, it comes from the Fermi surface and, therefore,
provides a dual description of the topological contribution in this formalism. (Interestingly, 
the origin of the topological contribution in the weak field analysis above may have some similarities with 
the Wigner function formalism \cite{Gao:2012ix,Chen:2012ca}.)

By substituting the propagators (\ref{CSE:free-term}) and (\ref{CSE:linear-term}) into Eq.~(\ref{CSE:axial-current-initial}), 
we find the following leading radiative corrections to the axial current:
\begin{eqnarray}
\langle j_{5}^{3}\rangle_{\alpha} &=& - 32 \pi \alpha eB \int\frac{d^4 p\,d^4 k}{(2\pi)^8}  \frac{1}{(P-K)^2_{\Lambda}}
\Bigg[\frac{(k_0+\mu)  [(p_0+\mu)^2+p_\perp^2- p_3^2-m^2 ] -2(p_0+\mu)(p_1k_1+p_2k_2) }
{(P^2-m^2)^3 (K^2-m^2)}
\nonumber\\
&&-2\frac{(p_0+\mu)  (p_1k_1+p_2k_2+2k_3p_3+4m^2)-(k_0+\mu)[(p_0+\mu)^2+p_3^2+m^2]}
{(P^2-m^2)^3 (K^2-m^2)}
\nonumber\\
&&-\frac{(k_0+\mu) [(p_0+\mu)^2-p_\perp^2+ p_3^2+m^2] - 2 (p_0+\mu) p_3k_3}{(P^2-m^2)^2 
(K^2-m^2)^2}
\Bigg]+\langle j_{5}^{3}\rangle_{\rm ct}\nonumber\\
&=&- 32 \pi \alpha eB \int\frac{d^4 p\,d^4 k}{(2\pi)^8}  \frac{1}{(P-K)^2_{\Lambda}}
\Bigg[\frac{(k_0+\mu)  [3(p_0+\mu)^2+\mathbf{p}^2+m^2 ] -4(p_0+\mu)(\mathbf{p}\cdot\mathbf{k}
+2m^2) }{(P^2-m^2)^3 (K^2-m^2)}
\nonumber\\
&& - \frac{(k_0+\mu) [3(p_0+\mu)^2-\mathbf{p}^2+3m^2] - 2 (p_0+\mu) (\mathbf{p}\cdot \mathbf
{k})}
{3(P^2-m^2)^2 (K^2-m^2)^2}
\Bigg]+\langle j_{5}^{3}\rangle_{\rm ct}.
\label{CSE:j53-1}
\end{eqnarray}
Here we use the shorthand notation $K^2=[k_0+\mu+i\epsilon\, \mathrm{sign}(k_0)]^2-\mathbf{k}^2$
and $P^2=[p_0+\mu+i\epsilon\, \mathrm{sign}(p_0)]^2-\mathbf{p}^2$. As for the definition 
of $(P-K)^2_{\Lambda}$, it follows Eq.~(\ref{CSE:photon-propagator}). Furthermore, the following replacements 
have been made in the integrand: $p_\perp^2\to \frac{2}{3}\mathbf{p}^2$, $p_3^2\to \frac{1}{3}\mathbf{p}^2$, 
and $p_3k_3 \to \frac{1}{3}(\mathbf{p}\cdot \mathbf{k})$. These replacements are allowed by the 
rotational symmetry of the other parts of the integrand.

It is convenient to represent Eq.~(\ref{CSE:j53-1}) as follows:
\begin{eqnarray}
\langle j_{5}^{3}\rangle_{\alpha}
&=&- 32 \pi \alpha eB \int\frac{d^4 p d^4 k}{(2\pi)^8}  \frac{1}{(P-K)^2_{\Lambda}}
\Bigg[\frac{4(p_0+\mu)[(k_0+\mu) (p_0+\mu)  -\mathbf{p}\cdot\mathbf{k}-2m^2] }{(P^2-m^2)^3 (K^2-m^2)}
-\frac{(k_0+\mu)}{(P^2-m^2)^2 (K^2-m^2)}
\nonumber\\
&& - \frac{(k_0+\mu) [3(p_0+\mu)^2-\mathbf{p}^2+3m^2- 2 (\mathbf{p}\cdot \mathbf{k})]}
{3(P^2-m^2)^2 (K^2-m^2)^2}\Bigg]+\langle j_{5}^{3}\rangle_{\rm ct}.
\label{CSE:j53-2}
\end{eqnarray}
Since the denominators of the integrand in this expression contain the factors $(P^2-m^2)^n$ and $(K^2-m^2)^n$, 
with $n = 2, 3$, which vanish on the Fermi surface, the integrand in (\ref{CSE:j53-2}) is singular there. Therefore, 
one should carefully treat the singularities in the calculation of the axial current. For this, we find it very 
convenient to use the following identity valid for all integers $n\geq 1$:
\begin{eqnarray}
\frac{1}{ \left[ [k_0+\mu +i\epsilon\, \mathrm{sign}(k_0)]^2-\mathbf{k}^2-m^2 \right]^n}&=&
 \frac{1}{\left[(k_0+\mu )^2-\mathbf{k}^2-m^2+i\epsilon \right]^n}\nonumber\\
&+&\frac{2\pi i(-1)^{n-1}}{(n-1)!}\theta(|\mu |-|k_0|) \theta(-k_0 \mu )\delta^{(n-1)}\left[(k_0+\mu )^2-\mathbf{k}^2-m^2\right],
\label{CSE:1overK^2-n}
\end{eqnarray}
which can be obtained from Eq.~(\ref{CSE:pole-identity}) in Appendix~\ref{CSE:App:proper-time-rep-mu}  
by differentiating it $n-1$ times with respect to $m^2$. Since the first term on the right-hand side has 
the pole prescription as in the theory without the filled fermion states, we call it the ``vacuum" part. 
The second term in this expression takes care of the filled fermion states, and we call it the ``matter"
part.

One can also obtain another useful relation by differentiating Eq.~(\ref{CSE:1overK^2-n}) 
with respect to energy $k_0$,
\begin{eqnarray}
&& \frac{\partial}{\partial k_0}\left(
\frac{1}{ \left[ [k_0+\mu+i\epsilon\, \mathrm{sign}(k_0)]^2-m^2-\mathbf{k}^2 \right]^{n}} \right) 
= -\frac{2n(k_0+\mu)}{ \left[ [k_0+\mu+i\epsilon\, \mathrm{sign}(k_0)]^2-m^2-\mathbf{k}^2 \right]^{n+1}} \nonumber \\
&&\hspace{2.5in}+ \frac{2\pi i (-1)^n \mathrm{sign}(\mu)}{(n-1)!} \delta^{(n-1)} \left[(k_0+\mu)^2-\mathbf{k}^2-m^2\right]
\left[\delta(k_0) - \delta(k_0+\mu) \right], \nonumber\\
\label{CSE:1overK^2-n-der}
\end{eqnarray}
where we made use of Eq.~(\ref{CSE:1overK^2-n}) the second time, albeit with $n\to n+1$, in order to render 
the result on the right-hand side in the form of the $(n+1)$th order pole with the conventional 
$i\epsilon$ prescription at nonzero $\mu$. In addition, we used the following easy to derive result:
\begin{equation}
\frac{\partial}{\partial k_0}\left[\theta(|\mu |-|k_0|) \theta(-k_0 \mu )\right] = \mathrm{sign}(\mu)\left[\delta(k_0+\mu) - \delta(k_0) \right].
\end{equation}
We note that $\delta(k_0+\mu)$ in the last term on the right-hand side of Eq.~(\ref{CSE:1overK^2-n-der}) 
never contributes. Indeed, this $\delta$ function is nonvanishing only when $k_0+\mu=0$. It multiplies, however, 
another $\delta$ function, which is nonvanishing only when $(k_0+\mu)^2-\mathbf{k}^2-m^2=0$. Since the two 
conditions cannot be simultaneously satisfied, the corresponding contribution is trivial. After taking this into account, 
we finally obtain
\begin{eqnarray}
\frac{\partial}{\partial k_0}\left(
\frac{1}{ \left[ [k_0+\mu+i\epsilon\, \mathrm{sign}(k_0)]^2-m^2-\mathbf{k}^2 \right]^{n}} 
\right)&=&-\frac{2n(k_0+\mu)}{ \left[ [k_0+\mu+i\epsilon\, \mathrm{sign}(k_0)]^2-m^2-\mathbf{k}^2 \right]^{n+1}} \nonumber \\
&+& \frac{2\pi i (-1)^n \mathrm{sign}(\mu)}{(n-1)!} \delta^{(n-1)} \left(\mu^2-\mathbf{k}^2-m^2\right) \delta(k_0).
\label{CSE:1overK^2-n-derivative}
\end{eqnarray}
Now, by making use of the above identities, we can proceed to the calculation of $\langle j_{5}^{3}\rangle_{\alpha}$ 
in Eq.~(\ref{CSE:j53-2}). We start by simplifying the corresponding expression using integrations by parts. 
Note that the $\Lambda$-regulated representation has nice convergence properties in the ultraviolet and, 
therefore, all integrations by parts in the analysis that follows will be perfectly justified.

The first term in $\langle j_{5}^{3}\rangle_{\alpha}$ in Eq.~(\ref{CSE:j53-2}) is proportional to $p_0+\mu$ 
and contains $(P^2-m^2)^3$ in the denominator. Therefore, we use identity (\ref{CSE:1overK^2-n-derivative}) 
with $n=2$ and $k\to p$, i.e.,
\begin{equation}
\frac{4(p_0+\mu)}{\left(P^2-m^2\right)^3}
=-\frac{\partial}{\partial p_0}\left(\frac{1}{(P^2-m^2)^2}\right)
+2i\pi \delta^{\prime}\left[\mu^2-m^2-\mathbf{p}^2\right]\delta(p_0).
\label{CSE:Mat-identity}
\end{equation}
Using it, we rewrite the first term in the integrand of Eq.~(\ref{CSE:j53-2}) as follows:
\begin{eqnarray}
\mbox{1st}&=& f_1 -  32 \alpha eB\int\frac{d^4 p d^4 k}{(2\pi)^8}  
\frac{ (k_0+\mu) (p_0+\mu)  -\mathbf{p}\cdot\mathbf{k}-2m^2 }{(P-K)^2_{\Lambda}(K^2-m^2)}
\frac{\partial }{\partial p_0}\left(\frac{-1}{(P^2-m^2)^2}\right) \nonumber\\
&=&f_1 -  32 \alpha eB\int\frac{d^4 p d^4 k}{(2\pi)^8}  \frac{1}{(P^2-m^2)^2}\frac{\partial }{\partial p_0}
\left(\frac{ (k_0+\mu) (p_0+\mu)  -\mathbf{p}\cdot\mathbf{k}-2m^2 }{(P-K)^2_{\Lambda}(K^2-m^2)}\right) 
\nonumber\\
&=&f_1 -  32 \alpha eB\int\frac{d^4 p d^4 k}{(2\pi)^8}  \frac{1}{(P^2-m^2)^2}
\left(\frac{ (k_0+\mu)}{(P-K)^2_{\Lambda}(K^2-m^2)}
+\frac{ (k_0+\mu)(p_0+\mu)  -\mathbf{p}\cdot\mathbf{k}-2m^2 }{(K^2-m^2)}\frac{\partial }{\partial p_0}\frac{ 1}{(P-K)^2_{\Lambda}}
\right),\nonumber\\
\label{CSE:Mat-first-term}
\end{eqnarray}
where the singular ``matter" term, containing the derivative of a $\delta$ function at the Fermi surface, 
was separated into a new function, 
\begin{equation}
f_1= - 64 i \pi^2 \alpha eB  \int\frac{d^4 p d^4 k}{(2\pi)^8}  
\frac{ (k_0+\mu) (p_0+\mu) -\mathbf{p}\cdot\mathbf{k}-2m^2 }{(P-K)^2_{\Lambda}(K^2-m^2)}
\delta^{\prime}\left[\mu^2-m^2-\mathbf{p}^2\right]\delta(p_0).
\label{CSE:Mat-f-1}
\end{equation}
We note that the first term in the parentheses in Eq.~(\ref{CSE:Mat-first-term}) cancels with the second 
term in the integrand of Eq.~(\ref{CSE:j53-2}). Then, by using 
\begin{equation}
\frac{\partial}{\partial p_0}\frac{ 1}{(P-K)^2_{\Lambda}}=-\frac{\partial }{\partial k_0}\frac{ 1}{(P-K)^2_{\Lambda}}
\end{equation}
and integrating by parts, we find that the sum of the first and second terms in the integrand of Eq.~(\ref{CSE:j53-2}) is equal to
\begin{eqnarray}
I_{1,2}&=& f_1 -  32 \alpha eB \int\frac{d^4 p d^4 k}{(2\pi)^8}  \frac{ 1}{(P-K)^2_{\Lambda}(P^2-m^2)^2}
\left(
\frac{p_0+\mu}{(K^2-m^2)}
+\left[(k_0+\mu)(p_0+\mu)  -\mathbf{p}\cdot\mathbf{k}-2m^2\right]\frac{\partial }{\partial k_0}
\frac{ 1 }{(K^2-m^2)}
\right)
\nonumber\\
& =& f_1+f_2- 32 \alpha eB \int\frac{d^4 p d^4 k}{(2\pi)^8}  \frac{ 1}{(P-K)^2_{\Lambda}}
\left( \frac{(p_0+\mu) }{(P^2-m^2)^2(K^2-m^2)}
-2(k_0+\mu)\frac{(k_0+\mu)(p_0+\mu)  -\mathbf{p}\cdot\mathbf{k}-2m^2}{(P^2-m^2)^2(K^2-m^2)^2} \right).\nonumber\\
\label{CSE:Mat-two-terms}
\end{eqnarray}
Note that here we used the identity
\begin{equation}
\frac{\partial}{\partial k_0}\left(\frac{1}{K^2-m^2}\right)
=\frac{-2(k_0+\mu)}{\left(K^2-m^2\right)^2}-2i\pi \delta \left(\mu^2-m^2-\mathbf{k}^2\right)\delta(k_0),
\label{CSE:Mat-derivative}
\end{equation}
which follows from Eq.~(\ref{CSE:1overK^2-n-derivative}) with $n=1$, and introduced another 
function, which contains the leftover contribution with the $\delta$ function, 
\begin{equation}
f_2= 64 i\pi^2 \alpha eB \int\frac{d^4 p d^4 k}{(2\pi)^8} \frac{(k_0+\mu)(p_0+\mu)  
-\mathbf{p}\cdot\mathbf{k}-2m^2}{(P-K)^2_{\Lambda}(P^2-m^2)^2}
\delta \left(\mu^2-m^2-\mathbf{k}^2\right)\delta(k_0).
\label{CSE:Mat-f-2}
\end{equation}
It is convenient to make the change of variables $p \to k$ and $k \to p$ in the first term in Eq.~(\ref{CSE:Mat-two-terms}). 
Then, the two terms in the integrand can be combined, resulting in
\begin{equation}
I_{1,2} = f_1 +f_2- 32 \alpha eB\int\frac{d^4 p d^4 k}{(2\pi)^8}  \frac{(k_0+\mu)
\left[-(p_0+\mu)^2  -\mathbf{p}^2+2\mathbf{p}\cdot\mathbf{k}+3m^2\right]}{(P-K)^2_{\Lambda}(P^2-m^2)^2(K^2-m^2)^2}.
\label{CSE:two-terms-1}
\end{equation}
Finally, by combining the result in Eq.~(\ref{CSE:two-terms-1}) with the last term in the integrand of Eq.~(\ref{CSE:j53-2}), we obtain
\begin{equation}
\langle j_{5}^{3}\rangle_{\alpha} =  f_1 +f_2 + \frac{64}{3} \pi \alpha eB \int\frac{d^4 p d^4 k}{(2\pi)^8}
\frac{(k_0+\mu) }{(P-K)_{\Lambda}^2}
\frac{3(P^2-m^2)+ 4\mathbf{p}\cdot(\mathbf{p}-\mathbf{k})}{(P^2-m^2)^2(K^2-m^2)^2}+\langle j_{5}^{3}\rangle_{\rm ct}.
\end{equation}
Using the identity in Eq.~(\ref{CSE:Mat-derivative}) once again, we rewrite the last expression as follows:
\begin{equation}
\langle j_{5}^{3}\rangle_{\alpha} =  f_1 +f_2+f_3  +\langle j_{5}^{3}\rangle_{\rm ct}
- \frac{64}{3} \pi \alpha eB \int\frac{d^4 p d^4 k}{(2\pi)^8} 
\frac{(k_0-p_0)}{(P-K)_{\Lambda}^4}\left(\frac{3}{(P^2-m^2)(K^2-m^2)}
+\frac{4\mathbf{p}\cdot(\mathbf{p}-\mathbf{k})}{(P^2-m^2)^2(K^2-m^2)}  
\right),
\label{CSE:j53-3}
\end{equation}
where
\begin{equation}
f_3 = - \frac{64i \pi^2 \alpha eB }{3}\int\frac{d^4 p d^4 k}{(2\pi)^8}  
\frac{3(P^2-m^2)+ 4\mathbf{p}\cdot(\mathbf{p}-\mathbf{k})}{(P-K)_{\Lambda}^2(P^2-m^2)^2}
\delta \left(\mu^2-m^2-\mathbf{k}^2\right)\delta(k_0) .
\label{CSE:Mat-f-3}
\end{equation}
Since the first term of the integrand in Eq.~(\ref{CSE:j53-3}) is odd under the exchange 
$p \leftrightarrow k$, its contribution vanishes, and we obtain
\begin{equation}
\langle j_{5}^{3}\rangle_{\alpha} = f_1 +f_2+f_3+\langle j_{5}^{3}\rangle_{\rm ct}
- \frac{64}{3} \pi \alpha eB \int\frac{d^4 p d^4 k}{(2\pi)^8} \frac{(k_0-p_0)}{(P-K)_{\Lambda}^4} 
\frac{4\mathbf{p}\cdot(\mathbf{p}-\mathbf{k})}{(P^2-m^2)^2(K^2-m^2)}.
\end{equation}
Finally, by making use of the identity 
\begin{equation}
\frac{\mathbf{p}\cdot(\mathbf{p}-\mathbf{k})}{(P-K)_{\Lambda}^4} 
=\frac{1}{2}\mathbf{p}\cdot\bm{\nabla}_\mathbf{k}\frac{-1}{(P-K)_{\Lambda}^2}
\end{equation}
and integrating by parts, we derive
\begin{eqnarray}
\langle j_{5}^{3}\rangle_{\alpha}
&=& f_1 +f_2+f_3+\langle j_{5}^{3}\rangle_{\rm ct}- \frac{64}{3} \pi \alpha eB \int\frac{d^4 p d^4 k}{(2\pi)^8} 
\frac{2(k_0-p_0)}{(P^2-m^2)^2(K^2-m^2)}  
\mathbf{p}\cdot\bm{\nabla}_\mathbf{k}\frac{-1}{(P-K)_{\Lambda}^2}
\nonumber\\
&=& f_1 +f_2+f_3+\langle j_{5}^{3}\rangle_{\rm ct}- \frac{64}{3} \pi \alpha eB \int\frac{d^4 p d^4 k}{(2\pi)^8} 
\frac{2(k_0-p_0)}{(P-K)_{\Lambda}^2(P^2-m^2)^2}  
\mathbf{p}\cdot\bm{\nabla}_\mathbf{k}\frac{1}{(K^2-m^2)}
\nonumber\\
&=& f_1 +f_2+f_3 +\langle j_{5}^{3}\rangle_{\rm ct}- \frac{64}{3} \pi \alpha eB \int\frac{d^4 p d^4 k}{(2\pi)^8} 
\frac{4(k_0-p_0)\mathbf{p}\cdot\mathbf{k}}{(P-K)_{\Lambda}^2(P^2-m^2)^2(K^2-m^2)^2}  \nonumber\\
&=& f_1 +f_2+f_3 +\langle j_{5}^{3}\rangle_{\rm ct},
\label{CSE:chiral-current-new} 
\end{eqnarray}
where the last integral term in the second to the last line of Eq.~(\ref{CSE:chiral-current-new}) vanishes 
because it is odd under the exchange $p \leftrightarrow k$. Collecting together all contributions, 
i.e., $f_1$ in Eq.~(\ref{CSE:Mat-f-1}), $f_2$ in Eq.~(\ref{CSE:Mat-f-2}) and $f_3$ in Eq.~(\ref{CSE:Mat-f-3}),
we have the following leading radiative corrections to the axial current:
\begin{eqnarray}
\langle j_{5}^{3}\rangle_{\alpha}&=&- 64i\pi^2\alpha eB \int\frac{d^4 p d^4 k}{(2\pi)^8}\Bigg[ 
\frac{ (k_0+\mu) (p_0+\mu) -\mathbf{p}\cdot\mathbf{k}-2m^2 }{(P-K)^2_{\Lambda}(K^2-m^2)}
\delta^{\prime}\left[\mu^2-m^2-\mathbf{p}^2\right]\delta(p_0) \nonumber\\
&+& \frac{3(p_0+\mu)^2-3(k_0+\mu)(p_0+\mu)+\mathbf{p}^2-\mathbf{p}\cdot\mathbf{k}+3m^2}{3(P-K)^2_{\Lambda}(P^2-m^2)^2}
\delta \left(\mu^2-m^2-\mathbf{k}^2\right)\delta(k_0)\Bigg]+\langle j_{5}^{3}\rangle_{\rm ct},
\label{CSE:j53-by-parts}
\end{eqnarray}
where the first term in the integrand comes from $f_1$, while the second term comes from 
the sum $f_2+f_3$. The result in Eq.~(\ref{CSE:j53-by-parts}) is quite remarkable for several reasons. 
From a technical viewpoint, it reveals that the integration by parts allowed us to reduce the original 
two-loop expression in Eq.~(\ref{CSE:j53-2}) down to a much simpler one-loop form. Indeed, after the 
integration over one of the momenta in Eq.~(\ref{CSE:j53-by-parts}) is performed using the $\delta$ functions 
in the integrand, the expression will have an explicit one-loop form. Such a simplification will turn out to 
be extremely valuable, allowing us to obtain an analytic result for the leading radiative corrections to 
the axial current.

In addition, the result in Eq.~(\ref{CSE:j53-by-parts}) reveals important physics details
about the origin of the radiative corrections to the axial current. It shows that all nonzero corrections
come from the regions of the phase space, where either $p$ or $k$ momentum is restricted 
to the Fermi surface. This resembles the origin of the topological contribution in 
Eq.~(\ref{CSE:axial-current-topological}). In both cases, the presence of the singular ``matter" 
terms in identities like (\ref{CSE:Mat-identity}) and (\ref{CSE:Mat-derivative}) was crucial for 
obtaining a nonzero result. Moreover, by tracing back the derivation of the result in 
Eq.~(\ref{CSE:j53-by-parts}), we see that all nonsingular terms are gone after the 
integration by parts. This makes us conclude that the nonzero radiative corrections 
to the axial current are intimately connected with the precise form of the singularities in the fermion 
propagator at the Fermi surface, that separates the filled fermion states with energies 
less than $\mu$ and empty states with larger energies.

The calculation of the axial current in Eq.~(\ref{CSE:j53-by-parts}) is still technically quite involved. 
However, it is relatively straightforward to show (see Appendix~B in Ref.~\cite{Gorbar:2013upa}) that 
the right-hand side in (\ref{CSE:j53-by-parts}) without the counterterm has a logarithmically 
divergent contribution when $\Lambda\to \infty$, i.e., 
\begin{equation}
- \frac{\alpha eB(2\mu^2+m^2)}
{4 \pi^3 \sqrt{\mu^2-m^2}}\ln\frac{\Lambda}{m}.
\label{CSE:j53-loop-UV}
\end{equation}
To cancel this divergence, we should add the contribution due the
counterterms in Lagrangian (\ref{CSE:Lagrangian}). The Fourier transform
of the translational invariant part of the counterterm contribution to the self-energy reads
\begin{equation}
\bar{\Sigma}^{(0)}_{\rm ct}(p)= \delta_2[(p_0+\mu)\gamma^0-\mathbf{p}\cdot\bm{\gamma}] -  \delta_m,
\end{equation}
where $\delta_2$ was defined in Eq.~(\ref{CSE:wave-function-renormalization-constant}), while 
$\delta_m = Z_2 m_0-m \simeq m\delta_2-\delta m$ and $\delta m$ was defined in 
Eq.~(\ref{CSE:mass-renormalization-constant}).

We find the following leading order contributions to the axial current density due to counterterms:
\begin{eqnarray}
\langle j_{5}^{3}\rangle_{\rm ct} &=&
-\delta_2\langle j_{5}^{3}\rangle_0
+ 4ieB\int\frac{d^4 p}{(2\pi)^4}\frac{\delta_2(p_0+\mu)}{(P^2-m^2)^2}
+ 8ieB\int\frac{d^4 p}{(2\pi)^4}\frac{(p_0+\mu) 
\left[ \delta_2 ( (p_0+\mu)^2-\mathbf{p}^2+m^2 ) -2 m\delta_m \right]}{(P^2-m^2)^3}
\nonumber\\
&=&
8ieB\int\frac{d^4 p}{(2\pi)^4}\frac{(p_0+\mu) \left[ \delta_2 ( P^2-m^2) +2 m(m\delta_2-\delta_m) \right]}{(P^2-m^2)^3}
\nonumber\\
&=&
8ieB\delta_2 \int\frac{d^4 p}{(2\pi)^4}\frac{p_0+\mu}{(P^2-m^2)^2}
+ 8 i m \left(m\delta_2-\delta_m\right)eB\frac{\partial }{\partial (m^2)}\int\frac{d^4 p}{(2\pi)^4}\frac{p_0+\mu}{(P^2-m^2)^2}
\nonumber\\
&=&\frac{eB}{\pi^2}\sqrt{\mu^2-m^2}\delta_2
- \frac{eB m \left(m\delta_2-\delta_m\right) }{2\pi^2\sqrt{\mu^2-m^2}}.
\label{CSE:j35-ct}
\end{eqnarray}
Here we used the same result of integration as in the topological term, see Eq.~(\ref{CSE:axial-current-topological}).

By making use of the explicit form of the counterterms (\ref{CSE:wave-function-renormalization-constant}) and 
(\ref{CSE:mass-renormalization-constant}), we obtain
\begin{equation}
\langle j_{5}^{3}\rangle_{\rm ct} =
\frac{\alpha eB}{2\pi^3}\sqrt{\mu^2-m^2}\left(\frac{1}{2}\ln\frac{\Lambda^2}{m^2}+\ln\frac{m_\gamma^2}{m^2}+\frac{9}{4}\right)  
+ \frac{3\alpha eB m^2}{4\pi^3\sqrt{\mu^2-m^2}}\left(\frac{1}{2}\ln\frac{\Lambda^2}{m^2} +\frac{1}{4}\right).
\label{CSE:counterterm-contribution-general}
\end{equation} 
For $m\ll|\mu|$, it reduces to 
\begin{equation}
\langle j_{5}^{3}\rangle_{\rm ct} \simeq
\frac{\alpha eB \mu}{2\pi^3} \left(\frac{1}{2}\ln\frac{\Lambda^2}{m^2}+\ln\frac{m_\gamma^2}{m^2}+\frac{9}{4}\right)  
+ \frac{\alpha eB m^2}{2\pi^3\mu}\left(\frac{1}{2}\ln\frac{\Lambda^2}{m_\gamma^2} -\frac{3}{4}\right).
\label{CSE:counterterm-contribution-general-small-m}
\end{equation} 
By combining this counterterm contribution with the matter contribution $\langle j_{5}^{3}\rangle_{\rm mat} 
\equiv f_1 +f_2+f_3$, we will obtain the complete expression for the leading radiative corrections to the axial 
current, see Eq.~(\ref{CSE:chiral-current-new}). The matter part is calculated in Appendix~B of 
Ref.~\cite{Gorbar:2013upa}. In the limit $m \ll |\mu|$, which is of main interest here, the corresponding 
result reads
\begin{equation}
\langle j_{5}^{3}\rangle_{\rm mat} \equiv  f_1+f_2+f_3 
=  - \frac{\alpha eB\mu}{2\pi^3} \left(\ln\frac{\Lambda}{2\mu}+\frac{11}{12}\right)
- \frac{\alpha eBm^2}{2 \pi^3 \mu} \left(\ln\frac{\Lambda}{2^{3/2}\mu}+\frac{1}{6}\right).
\end{equation}
Note that this expression has the correct ultraviolet logarithmic divergencies (when $\Lambda\to\infty$)
that will cancel exactly with those in the counterterm (\ref{CSE:counterterm-contribution-general-small-m}). 
Combining the two results, we finally obtain the following leading 
radiative corrections to the axial current in the case $m \ll |\mu|$:
\begin{equation}
\langle j_{5}^{3}\rangle_{\alpha}=\frac{\alpha eB \mu}{2\pi^3}\left(\ln\frac{2\mu}{m}+\ln\frac{m_\gamma^2}{m^2}+\frac{4}{3}\right)  
+ \frac{\alpha eB m^2}{2\pi^3\mu} \left(\ln\frac{2^{3/2}\mu}{m_\gamma} -\frac{11}{12}\right).
\label{CSE:j53-small-m-final}
\end{equation}
As expected, this result is independent of the ultraviolet regulator $\Lambda$. It does contain,
however, the dependence on the fictitious photon mass $m_\gamma$. This is the only infrared 
regulator left in our result. Its origin can be easily traced back to the infrared singularity
of the wave function renormalization $Z_2$ in the Feynman gauge used. As we discuss 
in the next subsection, this singularity is typical for a class of QED observables, obtained by 
perturbative methods. As we will explain below, in the complete physical expression for 
the axial current, obtained by going beyond the simplest double expansion in the coupling 
constant and magnetic field, the regulator $m_\gamma^2$ will likely be replaced by a physical
scale, e.g., such as $|eB|$ or $\alpha\mu^2$.

In summary, the main results of this subsection is that the chiral separation effect in QED 
receives nonvanishing radiative corrections. To leading order, these corrections are shown 
to be directly connected with the Fermi surface singularities in the fermion propagator at 
nonzero density. This interpretation is strongly supported by another observation: had we 
ignored the corresponding singular terms in the fermion propagator, the calculation of the 
two-loop radiative corrections would give a vanishing result.

The final result for the leading radiative corrections to the axial current density is presented 
in Eq.~(\ref{CSE:j53-small-m-final}). This is obtained by a direct calculation of all relevant
contributions to linear order in $\alpha$ and to linear order in the external magnetic field 
(strictly speaking, linear in $eB$ because the field always couples with the charge). 
The result in Eq.~(\ref{CSE:j53-small-m-final}) is presented in terms of renormalized (physical) 
parameters. As expected, it is independent of the ultraviolet regulator $\Lambda$, used 
at intermediate stages of calculations. This is a nontrivial statement since the original 
two-loop expression for the leading radiative corrections contains ultraviolet divergencies. 
In fact, the divergencies are unavoidable because the corresponding diagrams contain the 
insertions of the one-loop self-energy and vertex diagrams, which are known to have 
logarithmic divergencies. However, at the end of the day, all such divergencies 
are canceled exactly with the contributions due to the counterterms.

Our analysis shows that the matter contribution, $f_1+f_2+f_3$, to the axial current density 
(calculated in the Feynman gauge) has no additional singularities. While functions $f_1$ 
and $f_2+f_3$ separately do have additional infrared singularities, the physically relevant 
result for the sum $f_1+f_2+f_3$ is finite, see Appendix~B in Ref.~\cite{Gorbar:2013upa} for details. 
As we see from Eq.~(\ref{CSE:j53-small-m-final}), however, the final result depends on the photon 
mass $m_\gamma$, which was introduced as the conventional infrared regulator. This 
feature deserves some additional discussion. 

It is straightforward to trace the origin of the $m_\gamma$ dependence in Eq.~(\ref{CSE:j53-small-m-final}) 
to the calculation of the well-known result for the wave function renormalization constant 
$\delta_2$, presented in Eq.~(\ref{CSE:wave-function-renormalization-constant}). In fact, 
this infrared problem is common for dynamics in external fields in QED (for a thorough 
discussion, see Sec.~14 in Ref.~\cite{Weinberg:1995mt}). The most famous example is 
provided by the calculation of the Lamb shift, when an electron is in a Coulomb field. 
The point is that even for a light nucleus with $Z\alpha \ll 1$, one cannot consider the 
Coulomb field as a weak perturbation in deep infrared. The reason is that this field 
essentially changes the dispersion relation for the electron at low energies and momenta. 
As a result, its four-momenta are not on the electron mass shell, where the infrared divergence
is generated in the renormalization constant $Z_2$. Because of that, this infrared divergence 
is fictitious. The correct approach is to consider the Coulomb interaction perturbatively only 
at high energies, while to treat it nonperturbatively at low energies. The crucial point is 
matching those two regions that leads to replacing the fictitious parameter $m_{\gamma}$ 
by a physical infrared scale. This is the main subtlety that makes the calculation of the 
Lamb shift quite involved \cite{Weinberg:1995mt}.

In the case of the Lamb shift, the infrared scale is related to the atomic binding energy, or equivalently 
the inverse Bohr radius. For smaller energies and momenta, the electron wave functions cannot possibly 
be approximated with plane waves, which is the tacit assumption of the weak field approximation. 
Almost exactly the same line of arguments applies in the present problem of  QED in an external 
magnetic field. In particular, the fermion momenta perpendicular to the magnetic field cannot be 
defined with a precision better than $\sqrt{|eB|}$, or equivalently the inverse magnetic length. This
implies that the contribution to the axial current, which comes from the low-energy photon exchange 
between the fermion states near the Fermi surface, should be treated nonperturbatively. Just like in the 
Lamb shift problem \cite{Weinberg:1995mt}, we can anticipate that a proper nonperturbative treatment will 
result in a term proportional to $\ln(|eB|/m_\gamma^2)$, with a coefficient such as to cancel the 
$m_\gamma$ dependence in Eq.~(\ref{CSE:j53-small-m-final}). 

The additional complication in the problem at hand, which is absent in the study of the Lamb shift, 
is a nonzero density of matter. While doing the expansion in $\alpha$ and keeping only the leading 
order corrections, we ignored all screening effects, which formally appear to be of higher order. 
It is understood, however, that such effects can be very important at nonzero density. In particular, 
they could replace the unphysical infrared regulator $m_\gamma^2$ with a physical screening 
mass, i.e., the Debye mass $\sqrt{\alpha}\mu$.

In contrast to the physics underlying the Lamb shift, where the nonperturbative result 
can be obtained with the logarithmic accuracy by simply replacing $m_\gamma$ with the only 
physically relevant infrared scale in the problem, the same is not possible in the problem of the 
axial current at hand. The major complication here comes from the existence of two different 
physical regulators that can replace the unphysical infrared scale $m_\gamma$. One of 
them is $\sqrt{|eB|}$ and the other is $\sqrt{\alpha}\mu$. Because of the use of a double 
perturbative expansion in the analysis controlled by the small parameters $|eB|/\mu^2$ 
and $\alpha$, it is not possible to unambiguously resolve (without performing a direct 
nonperturbative calculation) which one of the two scales (or their combination) will 
cure the singularity in Eq.~(\ref{CSE:j53-small-m-final}).

Another natural question to address is the chiral limit, $m \to 0$. As one can see from
Eq.~(\ref{CSE:j53-small-m-final}),  the current $\langle j_{5}^{3}\rangle_{\alpha}$ is singular in this limit.
This point reflects the well-known fact that massless QED possesses new types of infrared singularities: 
beside the well-known divergences connected with soft photons, there are also divergences connected 
with the emission and absorption of collinear fermion-antifermion pairs \cite{Kinoshita:1962ur,Lee:1964is}. 
In addition, because of a Gaussian infrared fixed point in massless QED, the renormalized electric charge 
of massless fermions is completely shielded. One can show that this property is also intimately related to 
the collinear infrared divergences \cite{Fomin:1976am,Miransky:1977jt}. The complete screening of the renormalized 
electric charge makes this theory very different from massive QED. It remains to be examined whether 
there is a sensible way to describe the interactions with external electromagnetic fields in massless 
QED.

In massless QED without external electromagnetic fields, fermions interact through neutral 
vector currents \cite{Fomin:1976am,Miransky:1977jt}, despite the vanishing 
renormalized electric charge. In fact, massless QED yields the simplest example of an 
unparticle gauge field theory \cite{Georgi:2007ek}, in which the infrared fixed point
is Gaussian. There are no one-particle asymptotic states in unparticle theories. 
Instead, the asymptotic states are described by fermionic and bosonic jets 
\cite{Lee:1964is,Fomin:1976am,Miransky:1977jt,Georgi:2007ek}.

In addition to the quantitative study of the nonperturbative low-energy contributions and the 
effect of screening, there remain several other interesting problems to investigate in the future. 
Here we will mention only the following three. (i) It is of special interest to clarify the exact
connection of the nontrivial radiative corrections to the axial current density (\ref{CSE:j53-small-m-final}) 
with the generation of the chiral shift parameter in dense QED. The analysis of the chiral 
shift in the weak field limit in Section~\ref{ChiralAsymQEDweakB} suggests that a connection may exist, 
but it is probably much more complicated than in the NJL model \cite{Gorbar:2011ya,Gorbar:2009bm,Gorbar:2010kc}.
(ii) In order to make a contact with the physics of heavy-ion collisions, it would be interesting
to generalize the study of radiative corrections to the case of a nonzero temperature. The 
corresponding study in the NJL model \cite{Gorbar:2011ya} suggests that the temperature 
dependence of the axial current density should be weak.  
(iii) The analysis made in the NJL model shows a lot of similarities between the structure of the 
axial current in the CSE effect \cite{Gorbar:2011ya,Gorbar:2009bm,Gorbar:2010kc} with that 
of the electromagnetic current in the CME one \cite{Fukushima:2010zza}. On the other hand, 
the arguments of Ref.~\cite{Rubakov:2010qi} may suggest that the dynamical 
part of the result for the electromagnetic current should vanish, while the topological contribution 
(which needs to be added as part of the modified conserved axial current) will have no radiative 
corrections. It remains to be seen if these expectations will be supported by direct calculations of 
the induced electromagnetic current in the CME effect in QED with a chiral chemical 
potential $\mu_5$.

We also note that the radiative corrections to the chiral vorticity conductivity connected 
with the chiral vortical effect were calculated in Refs.~\cite{Golkar:2012kb,Hou:2012xg}. 
The role of the interaction effects and radiative corrections in various chiral anomalous 
effects in magnetized relativistic matter were recently discussed in 
Refs.~\cite{Jensen:2013vta,Kirilin:2013fqa}.

\subsubsection{Chiral asymmetry in a strong field}
\label{strongB:Sec:NumericalResults}

Since we will be using numerical methods in this section, there is no real advantage 
in utilizing the free photon propagator in the analysis. Moreover, the corresponding 
approximation is not reliable. It was problematic even in the weak magnetic field limit. 

Since we are dealing with the properties of cold QED matter at nonzero density, which 
is a great conductor, the screening  effects are extremely important. To have all the 
approximations under control, we will assume that the fermion number density is large, 
i.e., the corresponding value of the chemical $\mu$ is much larger than other energy scales in the problem.
In particular, we assume that $\mu\gg \sqrt{|eB|}\gg m$, which is a reasonable hierarchy, for
example, in the case of electron plasma in magnetars. One of the most important effects associated 
with the nonzero density is the screening of the one-photon exchange interaction. Even at 
weak coupling, such screening is strong and plays an important role in the dynamics. The 
well known scheme that captures the corresponding effects is called the hard-dense-loop (HDL) 
approximation \cite{Vija:1994is,Manuel:1995td}. The explicit form of the HDL photon propagator is 
given by 
\begin{equation} 
 D_{\mu\nu}(q) \simeq i\left(\frac{|\bm{q}| }{|\bm{q}|^3+\frac{\pi}{4}m_D^2|q_{4}|}O^{\rm (mag)}_{\mu\nu}
+ \frac{O^{\rm (el)}_{\mu\nu}}{q_{4}^2+|\bm{q}|^2+m_D^2}\right),
\label{strongB:D-HDL}
\end{equation}
where $q_4\equiv iq_0$ and $m_D^2=2\alpha \mu^2/\pi$ is the Debye screening mass.
In the Coulomb gauge assumed here, the Lorentz space projectors onto the electric and 
magnetic modes are defined as follows:
\begin{eqnarray} 
O^{\rm (mag)}_{\mu\nu} &=&  g_{\mu\nu}-u_{\mu} u_{\nu}
+\frac{\bm{q}_{\mu}\bm{q}_{\nu}}{|\bm{q}|^{2}},\\
O^{\rm (el)}_{\mu\nu} &=&  u_{\mu} u_{\nu},
\end{eqnarray}
where $u_{\mu}=(1,0,0,0)$. 

By making use of the definition in Eqs.~(\ref{strongB:Delta-n}) and (\ref{strongB:mu-5-n}), 
as well as the explicit form of the HDL photon propagator in Eq.~(\ref{strongB:D-HDL}), 
we derive the following results for the two coefficient functions of interest:
\begin{eqnarray}
\Delta_n(p_3)  & =  & (-1)^n e^2 l^2 \mathrm{sign}(eB) 
                        \int \frac{ dk_3 d^2 \mathbf{k}_{\perp}d^2\mathbf{p}_{\perp}}{(2\pi)^4} e^{-k_{\perp}^2l^2 - p^2_{\perp}l^2}  \displaystyle\sum_{N=0}^{\infty}  (-1)^N  \nonumber \\
&  &                       \Bigg\{ \left[  L_n(2p_{\perp}^2l^2) + L_{n-1}(2p_{\perp}^2l^2) \right] 
                        \left[ L_N(2k_{\perp}^2l^2) - L_{N-1}(2k_{\perp}^2l^2) \right] \left( - \mathcal{D}^{\rm (mag)} + \frac{1}{2} \mathcal{D}^{\rm (el)} \right) \nonumber \\ 
          &    &  +   \left[  L_n(2p_{\perp}^2l^2) - L_{n-1}(2p_{\perp}^2l^2) \right] 
                        \left[ L_N(2k_{\perp}^2l^2) + L_{N-1}(2k_{\perp}^2l^2) \right]  \left( \frac{ q_3^2 }{|\bm{q}|^{2}} \mathcal{D}^{\rm (mag)} + \frac{1}{2} \mathcal{D}^{\rm (el)} \right) \Bigg\}, 
\end{eqnarray}
\begin{eqnarray}
\mu_{5,n}(p_3)  & =  &  (-1)^n e^2 l^2 \mathrm{sign}(eB) 
                        \int \frac{ dk_3 d^2 \mathbf{k}_{\perp}d^2\mathbf{p}_{\perp}}{(2\pi)^4} e^{-k_{\perp}^2l^2 - p^2_{\perp}l^2}  \displaystyle\sum_{N=0}^{\infty} 
                         (-1)^N    \nonumber \\
&  &                       \Bigg\{ - \left[  L_n(2p_{\perp}^2l^2) + L_{n-1}(2p_{\perp}^2l^2) \right] \left[ L_N(2k_{\perp}^2l^2) - L_{N-1}(2k_{\perp}^2l^2) \right] k_3 \left( \frac{ q_3^2 }{|\bm{q}|^{2}} \mathcal{F}^{\rm (mag)} 
			+ \frac{1}{2} \mathcal{F}^{\rm (el)}  \right) \nonumber \\
         &    &    +   \left[  L_n(2p_{\perp}^2l^2) - L_{n-1}(2p_{\perp}^2l^2) \right] 
                        \left[ L_N(2k_{\perp}^2l^2) + L_{N-1}(2k_{\perp}^2l^2) \right] k_3 \left( \mathcal{F}^{\rm (mag)} - \frac{1}{2} \mathcal{F}^{\rm (el)} \right)  \nonumber \\
&  &               	+ 4  L^1_{N-1}(2k_{\perp}^2l^2) \left[  L_n(2p_{\perp}^2l^2) + L_{n-1}(2p_{\perp}^2l^2) \right] \frac{ q_3(k_1 q_1 + k_2 q_2 ) }{|\bm{q}|^{2}} \mathcal{F}^{\rm (mag)}  \Bigg\},
\end{eqnarray}
where the explicit expressions for the functions $\mathcal{D}^{\rm (mag)}$, $\mathcal{D}^{\rm (el)}$, $\mathcal{F}^{\rm (mag)}$, and 
$\mathcal{F}^{\rm (el)}$ are obtained after the integrations over $k_0$ performed, i.e.,  
\begin{eqnarray}
\mathcal{D}^{\rm (mag)}
& = & \frac{i}{\pi} |\bm{q}| \int_{-\infty} ^\infty \frac{ ( \omega_E - i \mu ) d\omega_E  }{ \left[ (\omega_E - i\mu )^2 + \mathcal{M}^2_N \right] \left( |\bm{q}|^3 + \frac{\pi}{4} m_D^2 |\omega_E + i p_0 | \right)  }\nonumber \\
& = &   \frac{  |\bm{q}|^4   \mathrm{sign}(\mu) \mathrm{sign}(\mathcal{M}_N^2 - \mu^2) }   { 2 \left[ |\bm{q}|^6 + (\frac{\pi}{4}m_D^2)^2 \left( \mathcal{M}_N-|\mu|  \right)^2 \right]} 
     - |\bm{q}| \mathrm{sign}(\mu)  \frac{ \frac{1}{4}m_D^2  \left(  \mathcal{M}_N- |\mu| \right)  \ln {\frac{|\bm{q}|^3}{ \frac{\pi}{4}m_D^2 | \mathcal{M}_N-|\mu|  | }   } } 
      {|\bm{q}|^6 + (\frac{\pi}{4}m_D^2)^2 \left( \mathcal{M}_N-|\mu|  \right)^2 }  \nonumber \\
& -&  \frac{  |\bm{q}|^4  \mathrm{sign}(\mu) }  { 2 \left[ |\bm{q}|^6 + (\frac{\pi}{4}m_D^2)^2 \left( \mathcal{M}_N+|\mu|  \right)^2 \right] }
            +  |\bm{q}|  \mathrm{sign}(\mu) \frac{ \frac{1}{4}m_D^2  \left(  \mathcal{M}_N+ |\mu| \right)  \ln {\frac{|\bm{q}|^3}{ \frac{\pi}{4}m_D^2 \left( \mathcal{M}_N+|\mu|  \right) }   } } 
      {|\bm{q}|^6 + (\frac{\pi}{4}m_D^2)^2 \left( \mathcal{M}_N+|\mu|  \right)^2 } ,
\end{eqnarray}
\begin{eqnarray}
\mathcal{D}^{\rm (el)}
& = & \frac{1}{\pi} \int_{-\infty} ^\infty \frac{ ( i \omega_E + \mu ) d\omega_E  }{ \left[ (\omega_E - i\mu )^2 + \mathcal{M}^2_N \right] \left[ (\omega_E-ip_0)^2 + |\bm{q}|^2+m_D^2  \right]  } \nonumber \\
& = & \frac{\mu \Theta[\mathcal{M}_N^2 - \mu^2 ] } {\sqrt{|\bm{q}|^2+m_D^2} \left[ \left( \sqrt{|\bm{q}|^2+m_D^2} + \mathcal{M}_N \right)^2 - \mu^2 \right] } 
       - \frac{ \mathrm{sign}(\mu) \Theta \left[ \mu^2 - \mathcal{M}_N^2 \right] \left( \sqrt{|\bm{q}|^2+m_D^2} + |\mu| \right)  } {\sqrt{|\bm{q}|^2+m_D^2 }
           \left[ \left( \sqrt{|\bm{q}|^2+m_D^2} + |\mu| \right)^2 - \mathcal{M}_N^2  \right]},
\end{eqnarray}
and
\begin{eqnarray}
\mathcal{F}^{\rm (mag)}
& = & \frac{1}{\pi} |\bm{q}| \int_{-\infty} ^\infty \frac{ d\omega_E  }{ \left[ (\omega_E - i\mu )^2 + \mathcal{M}^2_N \right] \left( |\bm{q}|^3 + \frac{\pi}{4} m_D^2 |\omega_E + i p_0 | \right)  }\nonumber \\
& = &  \frac{1}{\mathcal{M}_N} \Bigg( \frac{ |\bm{q}|^4   \mathrm{sign}(\mathcal{M}_N^2 - \mu^2) }   {2 \left[ |\bm{q}|^6 + (\frac{\pi}{4}m_D^2)^2 \left( \mathcal{M}_N-|\mu|  \right)^2 \right] } 
     - |\bm{q}|  \frac{ \frac{1}{4}m_D^2  \left(  \mathcal{M}_N- |\mu| \right)  \ln {\frac{|\bm{q}|^3}{ \frac{\pi}{4}m_D^2 | \mathcal{M}_N-|\mu|  | }   } } 
      {|\bm{q}|^6 + (\frac{\pi}{4}m_D^2)^2 \left( \mathcal{M}_N-|\mu|  \right)^2 }  \nonumber \\
&    &   + \frac{ |\bm{q}|^4  } {2 \left[ |\bm{q}|^6 + (\frac{\pi}{4}m_D^2)^2 \left( \mathcal{M}_N+|\mu|  \right)^2 \right] }
            -  |\bm{q}|   \frac{ \frac{1}{4}m_D^2  \left(  \mathcal{M}_N+ |\mu| \right)  \ln {\frac{|\bm{q}|^3}{ \frac{\pi}{4}m_D^2 \left( \mathcal{M}_N+|\mu|  \right) }   } } 
      {|\bm{q}|^6 + (\frac{\pi}{4}m_D^2)^2 \left( \mathcal{M}_N+|\mu|  \right)^2 }  \Bigg),    
\end{eqnarray}
\begin{eqnarray}
\mathcal{F}^{\rm (el)}
& = & \frac{1}{\pi} \int_{-\infty} ^\infty \frac{  d\omega_E  }{ \left[ (\omega_E - i\mu )^2 + \mathcal{M}^2_N \right] \left[ (\omega_E-ip_0)^2 + |\bm{q}|^2+m_D^2  \right]  } \nonumber \\
& = & - \frac{ \Theta[ \mu^2 - \mathcal{M}_N^2 ] } {\sqrt{|\bm{q}|^2+m_D^2} \left[ \left( \sqrt{|\bm{q}|^2+m_D^2} + |\mu| \right)^2 - \mathcal{M}_N^2 \right] } 
       + \frac{ \Theta \left[ \mathcal{M}_N^2 -\mu^2 \right] \left( \sqrt{|\bm{q}|^2+m_D^2} + \mathcal{M}_N \right)  } { \sqrt{|\bm{q}|^2+m_D^2 } \mathcal{M}_N
           \left[ \left( \sqrt{|\bm{q}|^2+m_D^2} + \mathcal{M}_N \right)^2 - \mu^2  \right]} ,
\end{eqnarray}
where 
$\mathcal{M}_N^2 = k_3^2 + 2N|eB| + m^2$ and 
$|\bm{q}|^2 = | k_3 - p_3 |^2 + k_\perp^2 + p_\perp^2 - 2 k_\perp p_\perp \cos \phi$.

While performing the numerical analysis, it is convenient to render the above expressions in a dimensionless
form. Therefore, we introduce the following dimensionless functions: $\bar{\Delta}_n \equiv  \Delta_n/\mu$ and 
$\bar{\mu}_{5,n} \equiv  \mu_{5,n}/\mu$, as well as the following dimensionless variables: 
$x = p_{\perp}/\mu$, $y \equiv  k_{\perp}/\mu$,  $ x_{3} \equiv p_{3}/\mu$, and $y_{3} \equiv k_{3}/\mu$.
By using this new notation, we have 
\begin{eqnarray}
\bar{\Delta}_n  & =  & (-1)^n \frac {e^2}{b} \mathrm{sign}(eB) 
                        \int \frac{ dy_3 dy dx d\phi}{(2\pi)^3} e^{-(x^2+y^2)/b }  \displaystyle\sum_{N=0}^{\infty}  (-1)^N xy  \nonumber \\
& \times &                       \Bigg\{ \left[  L_n(2x^2/b) + L_{n-1}(2x^2/b) \right] 
                        \left[ L_N(2y^2/b) - L_{N-1}(2y^2/b) \right] \left( \mathcal{ - \bar{D}}^{\rm (mag)} + \frac{1}{2} \mathcal{\bar{D}}^{\rm (el)} \right) \nonumber \\ 
          &  -  &    \left[  L_n(2x^2/b) - L_{n-1}(2x^2/b) \right] 
                        \left[ L_N(2y^2/b) + L_{N-1}(2y^2/b) \right]  \left( \frac{ (x_3-y_3)^2 \mathcal{\bar{D}}^{\rm (mag)} }{(x_3-y_3)^2+x^2+y^2-2xy\cos\phi}  + \frac{1}{2} \mathcal{\bar{D}}^{\rm (el)} \right) \Bigg\},
\label{strongB:Delta-n-dimless}
\end{eqnarray}
and 
\begin{eqnarray}
\bar{\mu}_{5,n}  & =  & (-1)^n \frac {e^2}{b} \mathrm{sign}(eB) 
                        \int \frac{ dy_3 dy dx d\phi}{(2\pi)^3} e^{-(x^2+y^2)/b }  \displaystyle\sum_{N=0}^{\infty}  (-1)^N xy y_3 \nonumber \\
& \times &                       \Bigg\{ - \left[  L_n(2x^2/b) + L_{n-1}(2x^2/b) \right] 
                        \left[ L_N(2y^2/b) - L_{N-1}(2y^2/b) \right]  \left( \frac{ (x_3-y_3)^2 \mathcal{ \bar{F}}^{\rm (mag)}  }{(x_3-y_3)^2+x^2+y^2-2xy\cos\phi} + \frac{1}{2} \mathcal{\bar{F}}^{\rm (el)} \right) \nonumber \\ 
          &  +  &  \left[  L_n(2x^2/b) - L_{n-1}(2x^2/b) \right] 
                        \left[ L_N(2y^2/b) + L_{N-1}(2y^2/b) \right]  \left( \mathcal{\bar{F}}^{\rm (mag)} - \frac{1}{2} \mathcal{\bar{F}}^{\rm (el)} \right) \Bigg\},
\label{strongB:mu-5-n-dimless}
\end{eqnarray}
where 
\begin{eqnarray}
\bar{\mathcal{D}}^{\rm (mag)}
& = &   \frac{   |\bm{\bar{q}}|^4   \mathrm{sign}(\mu) \mathrm{sign}(\mathcal{\bar{M}}_N^2 - 1) }   {2 \left[ |\bar{\bm{q}}|^6 + (\frac{\pi d^2}{4})^2 \left( \mathcal{\bar{M}}_N- \mathrm{sign}(\mu)  \right)^2 \right] } 
     - |\bm{\bar{q}}| \mathrm{sign}(\mu)  \frac{ \frac{d^2}{4}  \left(  \mathcal{\bar{M}}_N- \mathrm{sign}(\mu) \right)  \ln {\frac{|\bm{\bar{q}}|^3}{ \frac{\pi d^2}{4} | \mathcal{\bar{M}}_N- \mathrm{sign}(\mu) | }   } } 
      {|\bm{\bar{q}}|^6 + (\frac{\pi d^2}{4})^2 \left( \mathcal{M}_N-\mathrm{sign}(\mu)  \right)^2 }  \nonumber \\
&    &   - \frac{   |\bm{\bar{q}}|^4  \mathrm{sign}(\mu) }  { 2 \left[ |\bar{\bm{q}}|^6 + (\frac{\pi d^2}{4})^2 \left( \mathcal{\bar{M}}_N+\mathrm{sign}(\mu)  \right)^2 \right] }
            +  |\bm{\bar{q}}|  \mathrm{sign}(\mu) \frac{ \frac{d^2}{4}  \left(  \mathcal{\bar{M}}_N+ \mathrm{sign}(\mu) \right)  \ln {\frac{|\bm{\bar{q}}|^3}{ \frac{\pi d^2}{4} \left( \mathcal{\bar{M}}_N+\mathrm{sign}(\mu) \right) }   } } 
      {|\bm{\bar{q}}|^6 + (\frac{\pi d^2}{4})^2 \left( \mathcal{\bar{M}}_N+\mathrm{sign}(\mu)  \right)^2 }    ,
\end{eqnarray}
\begin{eqnarray}
\bar{\mathcal{D}}^{\rm (el)}
& = & \frac{ \Theta[\mathcal{\bar{M}}_N^2 - 1 ] } {\sqrt{|\bar{\bm{q}}|^2+d^2} \left[ \left( \sqrt{|\bar{\bm{q}}|^2+d^2} + \mathcal{\bar{M}}_N \right)^2 - 1 \right] } 
       - \frac{ \mathrm{sign}(\mu) \Theta \left[ 1 - \mathcal{\bar{M}}_N^2 \right] \left( \sqrt{|\bar{\bm{q}}|^2+d^2} + \mathrm{sign}(\mu) \right)  } {\sqrt{|\bar{\bm{q}}|^2+d^2 }
           \left[ \left( \sqrt{|\bar{\bm{q}}|^2+d^2} + \mathrm{sign}(\mu) \right)^2 - \mathcal{\bar{M}}_N^2  \right]},
\end{eqnarray}
\begin{eqnarray}
\bar{\mathcal{F}}^{\rm (mag)}
& = &   \frac{1}{\mathcal{\bar{M}}} \Bigg\{ \frac{   |\bm{\bar{q}}|^4   \mathrm{sign}(\mathcal{\bar{M}}_N^2 - 1) }  
        {2 \left[ |\bar{\bm{q}}|^6 + (\frac{\pi d^2}{4})^2 \left( \mathcal{\bar{M}}_N- \mathrm{sign}(\mu)  \right)^2 \right] } 
     - |\bm{\bar{q}}|  \frac{ \frac{d^2}{4}  \left(  \mathcal{\bar{M}}_N- \mathrm{sign}(\mu) \right)  \ln {\frac{|\bm{\bar{q}}|^3}{ \frac{\pi d^2}{4} | \mathcal{\bar{M}}_N- \mathrm{sign}(\mu) | }   } } 
      {|\bm{\bar{q}}|^6 + (\frac{\pi d^2}{4})^2 \left( \mathcal{M}_N-\mathrm{sign}(\mu)  \right)^2 }  \nonumber \\
&    &   + \frac{   |\bm{\bar{q}}|^4  }  { 2 \left[ |\bar{\bm{q}}|^6 + (\frac{\pi d^2}{4})^2 \left( \mathcal{\bar{M}}_N+\mathrm{sign}(\mu)  \right)^2 \right] }
            -  |\bm{\bar{q}}|  \frac{ \frac{d^2}{4}  \left(  \mathcal{\bar{M}}_N+ \mathrm{sign}(\mu) \right)  \ln {\frac{|\bm{\bar{q}}|^3}{ \frac{\pi d^2}{4} \left( \mathcal{\bar{M}}_N+\mathrm{sign}(\mu) \right) }   } } 
      {|\bm{\bar{q}}|^6 + (\frac{\pi d^2}{4})^2 \left( \mathcal{\bar{M}}_N+\mathrm{sign}(\mu)  \right)^2 } \Bigg\},       
\end{eqnarray}
\begin{eqnarray}
\bar{\mathcal{F}}^{\rm (el)}
& = & - \frac{ \Theta[ 1 - \mathcal{ \bar{M}}_N^2 ]} {\sqrt{|\bar{\bm{q}}|^2+d^2} \left[ \left( \sqrt{|\bar{\bm{q}}|^2+d^2} + 1 \right)^2 - \mathcal{\bar{M}}_N^2 \right] } 
    + \frac{ \Theta \left[ \mathcal{\bar{M}}_N^2 - 1 \right] \left( \sqrt{|\bar{\bm{q}}|^2+d^2} + \mathcal{\bar{M}}_N \right)  } {\mathcal{\bar{M}}_N \sqrt{|\bar{\bm{q}}|^2+d^2 }
           \left[ \left( \sqrt{|\bar{\bm{q}}|^2+d^2} +\mathcal{\bar{M}}_N \right)^2 - 1 \right]},
\end{eqnarray}
with $\mathcal{\bar{M}}_N^2 = y_3^2 + 2Nb + a^2$ and $|\bar{\bm{q}}|^2 = ( y_3 - x_3 )^2 + y^2 + x^2 - 2 xy \cos \phi$.

It is instructive to note that the function under the integral in the expression for $\bar{\mu}_{5,n}$ contains 
an overall factor of $y_3$ in the numerator. Clearly, such a dependence on $y_3$ is not very helpful for 
the numerical convergence of the integral. By taking into account, however, that the rest of the integrand 
depends on $y_3$ only via $( y_3 - x_3 )^2$ and $y_3^2$ combinations, the convergence can be 
substantially improved by using the following identity:
\begin{equation}
\int_{-\infty}^{\infty} dy_3 \, y_3 \, F\left((y_3-x_3)^2, y_3^2 \right) 
= \int_{-\infty}^{\infty}  dy_3  \, \frac{y_3}{2}  \left[ F\left((y_3-x_3)^2, y_3^2 \right) - F\left((y_3+x_3)^2, y_3^2 \right) \right].
\end{equation}
In order to calculate such an integral, it is convenient to use the importance sampling Monte 
Carlo method \cite{Weinzierl:2000wd}. The details of the method are presented in Ref.~\cite{Xia:2014wla}.

In order to analyze numerically the two chiral asymmetry functions in dense QED, we need 
to fix several model parameters (i.e., the strength of magnetic field, the value of the chemical potential,
and the fermion mass). In principle, when using the dimensionless description, the value of the 
chemical potential $\mu$ may be left unspecified. Keeping in mind, however, that the value of 
the fermion (electron) mass has to be measured in units of $\mu$, we will assume that the 
default choices of the chemical potential and the magnetic field are $\mu=420~\mbox{MeV}$ 
and $B=10^{18}~\mbox{G}$. Then, the two dimensionless model parameters used in the 
calculations will be
\begin{eqnarray}
a & = & \frac{m}{\mu} \approx 1.22\times10^{-3} \frac{m}{m_e} \frac{420~\mbox{MeV}}{\mu}, \\
b & = & \frac{ |eB| }{\mu^2} \approx \frac{1}{30} \left( \frac{B}{10^{18}~\mbox{G}} \right) \left( \frac{420~\mbox{MeV}}{\mu} \right)^2.
\end{eqnarray}
Note that, in agreement with the assumption made earlier, the chosen value of the magnetic field 
strength is rather small compared to the chemical potential scale $\mu^2$. By taking into account 
the definition of the Debye mass and the QED fine structure constant, it is also convenient to  
introduce the following short-hand notation for the dimensionless Debye mass:
\begin{equation}
d = \frac{m_D}{\mu} \equiv \sqrt{ \frac{2\alpha}{\pi} } \approx 6.816\times10^{-2}.
\end{equation}
In the final expressions for the chiral asymmetry functions, there will be a need to sum over  
an infinite number of Landau levels. In the numerical calculations, however, the sums will 
be truncated at $n_{\rm max}=200$. 

The numerical results for the chiral shift are summarized in Fig.~\ref{strongB:fig_Delta-n}.
In the left panel, we show the dependence of the chiral shift $\Delta_{n}$ on the 
longitudinal momentum $y_3=p_3/\mu$ for several low-lying Landau levels. 
Since obtaining the complete functional dependence on $p_3$ is rather 
expensive numerically, we used only a moderately large number of sampling
points, $N=2\times 10^8$ and calculated the results only for the first four lowest 
lying Landau levels. The common feature of the corresponding functions 
is the appearance of a maximum at an approximate location of the Fermi surface. 
In the free (weakly interacting) theory, this is determined by the following value of
the longitudinal momentum: $p_3/\mu = \sqrt {1 - 2 nb - a^2}$. In agreement with 
this expression, the location of the maximum of the chiral shift function in the $n$th 
Landau level $\Delta_{n}(p_3)$ decreases with increasing $n$. At large values of 
the momentum $p_3$, the chiral shift function decreases and gradually approaches 
zero as expected. 

\begin{figure}[t]
\begin{center}
\includegraphics[width=0.47\textwidth]{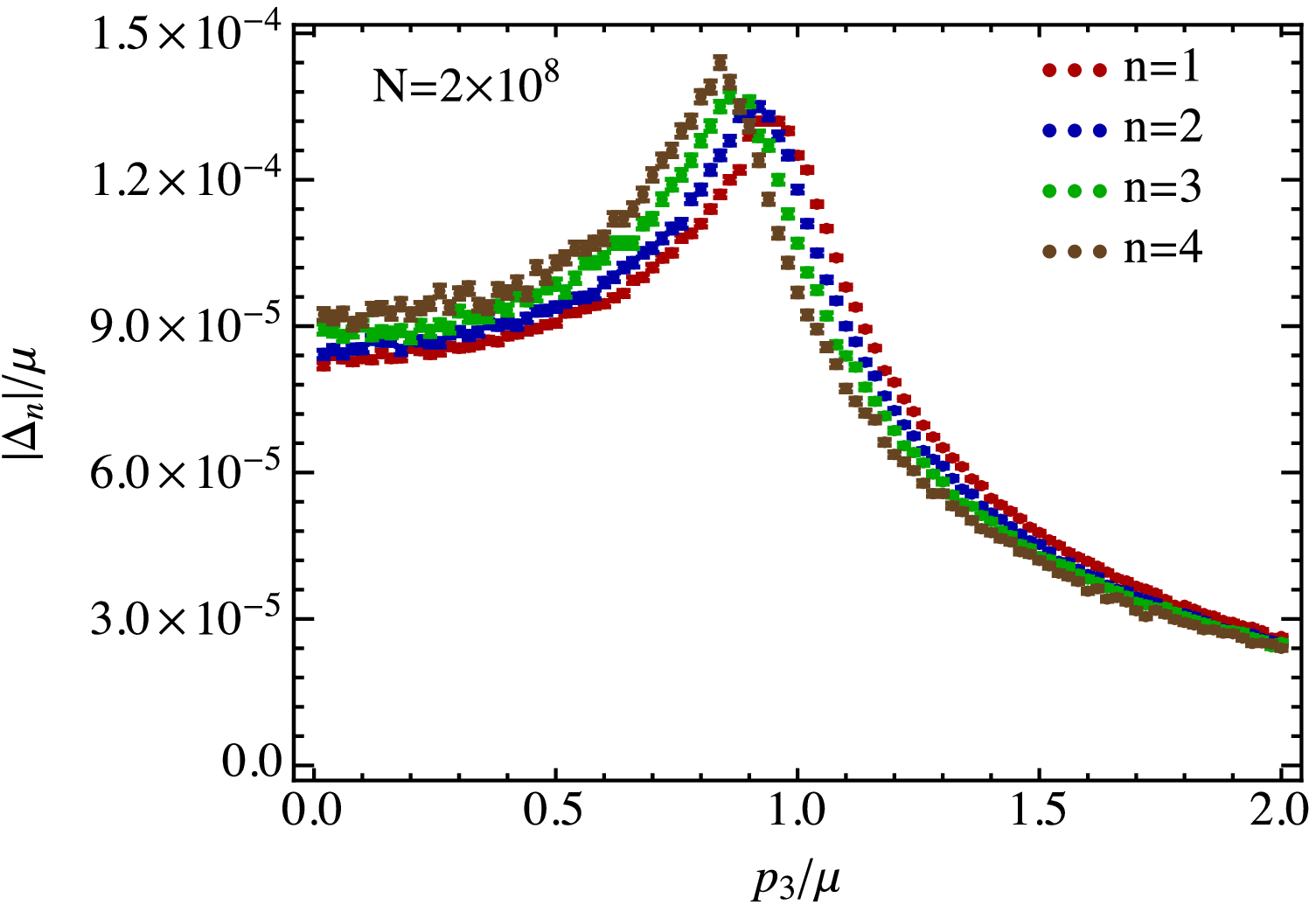}\hspace{0.03\textwidth}
\includegraphics[width=0.47\textwidth]{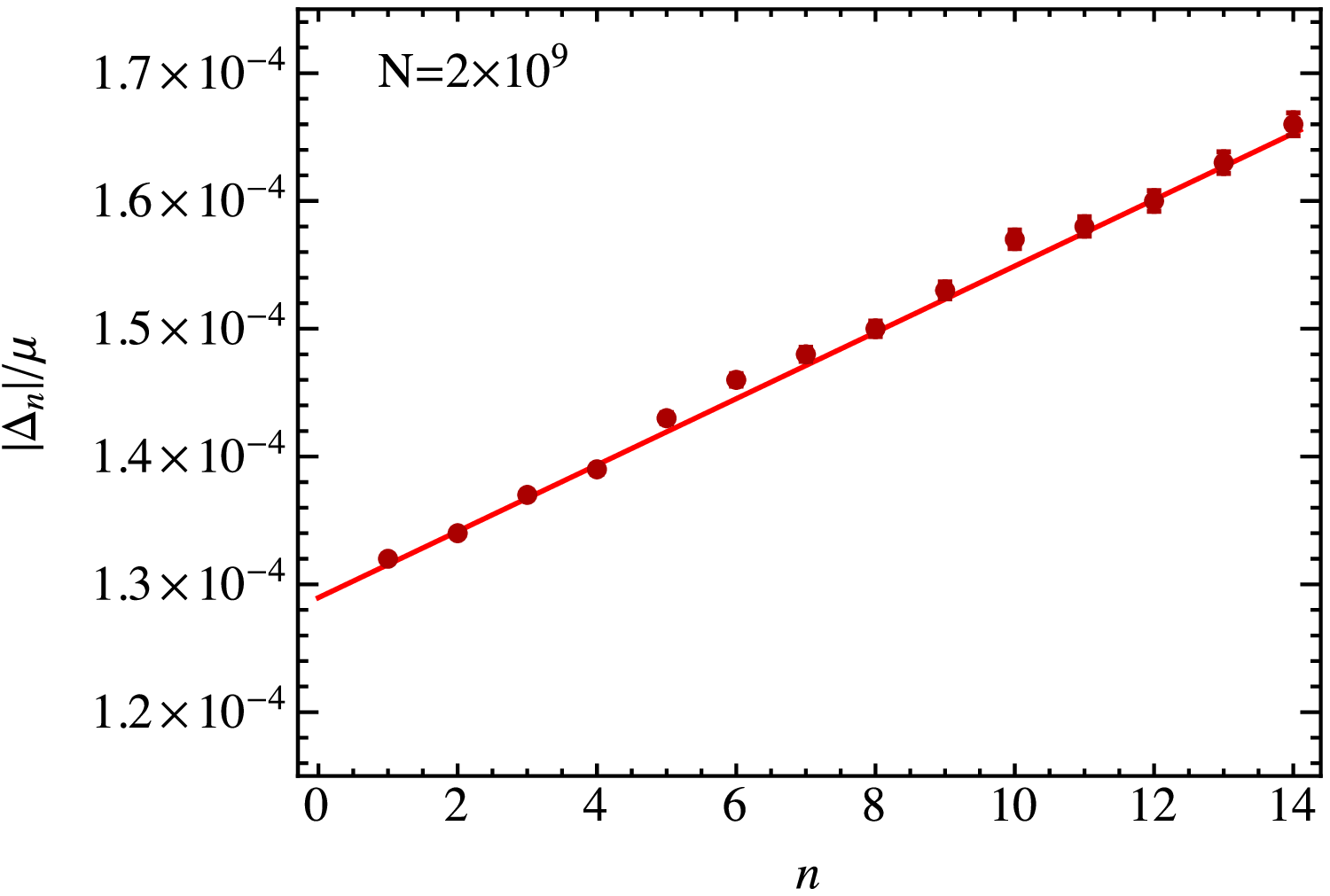}
\caption{(Color online) 
Left panel: the chiral shift $\Delta_n$ as a function of the longitudinal 
momentum $p_3$ for $n=1$ (red), $n=2$ (blue), $n=3$ (green), and $n=4$ (brown) Landau 
levels. Right panel: the values of the chiral shift $\Delta_n$ at the Fermi surface.}
\label{strongB:fig_Delta-n}
\end{center}
\end{figure}

From the viewpoint of the low-energy physics, it is most important to know the chiral 
shift at the Fermi surface. The corresponding results are presented in the right 
panel of Fig.~\ref{strongB:fig_Delta-n}. By assumption, the location of the Fermi surface
is determined by the perturbative expression, $p_3/\mu = \sqrt {1 - 2 nb - a^2}$. 
In this calculation, we used a larger number of sampling points, $N=2\times 10^9$. 
As we see, the Fermi surface values of $\Delta_{n}$ grow with the Landau-level 
index $n$. (The corresponding numerical values are also given in the first 
column of Table~\ref{strongB:Table-Fermi-surface}.) This growth is somewhat surprising, 
but is in agreement with the general behavior of functions $\Delta_{n}(p_3)$ 
shown in the left panel of Fig.~\ref{strongB:fig_Delta-n}. The corresponding dependence 
on the Landau-level index can be fitted quite well by a linear function. 

It is easy to check that the numerical results for the chiral shift in Fig.~\ref{strongB:fig_Delta-n} 
are of the same order of magnitude as $\alpha |eB|/\mu$. Taking into account that 
$\Delta_{n}$ is one of the structures in the fermion self-energy, induced by a nonzero 
magnetic field, it is indeed quite natural that the corresponding function is proportional 
to the coupling constant and the magnetic field strength. As for the chemical potential 
in the denominator, it is the only other relevant energy scale in the problem that can 
be used to render the result for $\Delta_{n}$ with the correct energy units. 
(Formally, the fermion mass is yet another energy scale, but it is unlikely 
to play a prominent role at the Fermi surface in the high density and strong-field 
limit.) The linear fit for the chiral shift at the Fermi surface is shown by the solid 
line in the right panel of Fig.~\ref{strongB:fig_Delta-n}. The corresponding function can 
be presented in the following form:
\begin{equation}
\Delta_{n} \simeq -\frac{\alpha |eB|}{\mu}\left(0.53+0.32 \frac{|eB|n}{\mu^2}\right),
\label{strongB:Delta-n-fit}
\end{equation}
where we took into account that the numerical results in Fig.~\ref{strongB:fig_Delta-n} were 
obtained for the magnetic field $|eB|=\mu^2/30$ and $\alpha =1/137$. The result in 
Eq.~(\ref{strongB:Delta-n-fit}) should be contrasted with a very different parametric dependence 
obtained in the weak-field limit in Ref.~\cite{Gorbar:2013uga}, i.e., $\Delta_{n} \propto 
\alpha |eB|\mu/m^2$, which is a factor of $(\mu/m)^2$ larger. Such a large factor is 
quite natural in the weak-field limit, where it is an artifact of the expansion in powers of 
$|eB|/m^2$. In contrast, one does not expect anything like that in the case of a strong 
magnetic field.

\begin{figure}[t]
\centering
\includegraphics[angle=0,width=0.47\textwidth]{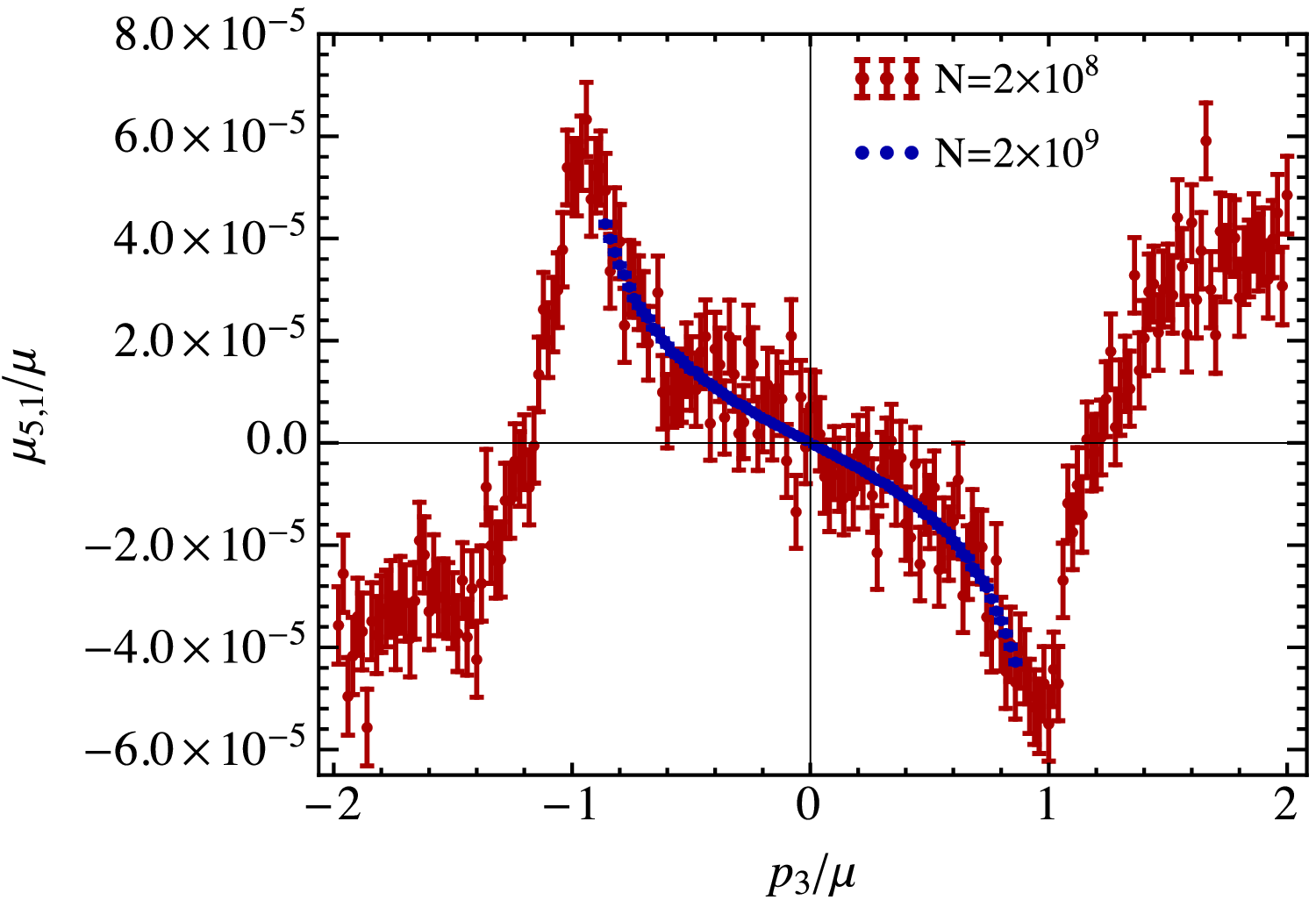}\hspace{0.03\textwidth}
\includegraphics[width=0.47\textwidth]{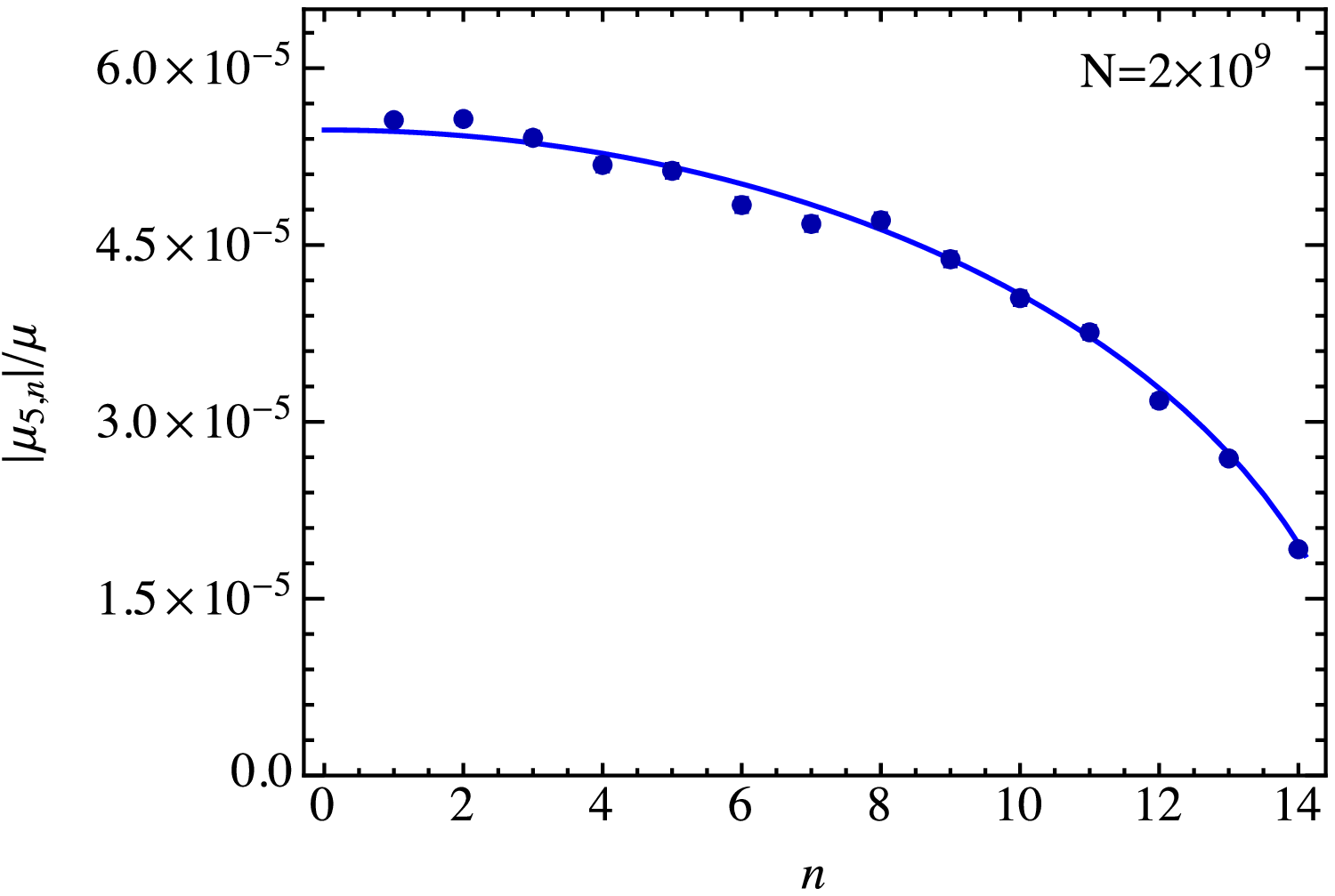}
\caption{(Color online) 
Left panel: the chiral chemical potential $\mu_{5,n}$ as a function of the longitudinal 
momentum $p_3$ for $n=1$ Landau level. Right panel: the values of the chiral 
chemical potential $\mu_{5,n}$ at the Fermi surface.}
\label{strongB:fig_mu-5-n}
\end{figure}

The numerical results for the chiral chemical potential $\mu_{5,n}$ are summarized in 
Fig.~\ref{strongB:fig_mu-5-n}. In the left panel, we present the chiral chemical potential in the 
$n=1$ Landau level as a function of the longitudinal momentum $p_3$. (The results 
for larger $n$ are expected to have the same qualitative dependence on $p_3$.) 
The red and blue points represent the results for two different numbers of sampling 
points, $N=2\times 10^8$ and $N=2\times 10^{9}$, respectively. The numerical 
results confirm that $\mu_{5,n}$ is an odd function of $p_3$ and, as such, it does 
not break parity. The dependence of $\mu_{5,n}$ on $p_3$ also reveals a pair of 
sharp peaks on the Fermi surface at $p_3/\mu \simeq \pm\sqrt {1 - 2 nb - a^2}$. 
In the context of the low-energy physics, it is these values of $\mu_{5,n}$ on the 
Fermi surface that are of main importance.

The numerical results for the chiral chemical potential at the Fermi surface are shown 
in the right panel of Fig.~\ref{strongB:fig_mu-5-n}. In the corresponding calculation, we again 
assumed that the location of the Fermi surface is determined by the perturbative 
expression, $p_3/\mu = \pm \sqrt {1 - 2 nb - a^2}$, and used the Monte Carlo integration
algorithm with $N=2\times 10^9$ sampling points. We find that the values of $\mu_{5,n}$ 
decrease with the Landau-level index $n$. (The corresponding numerical values are 
given in the second column of Table~\ref{strongB:Table-Fermi-surface}.) The order of magnitude 
of the obtained results is similar to those for the chiral shift function. Following the same
arguments, therefore, we can assume that $\mu_{5,n}$ is also proportional to the 
coupling constant and the magnetic field strengths, i.e., $\mu_{5,n} \propto \alpha |eB|/\mu$.
(Let us emphasize again that this dependence is quite different from the weak-field 
limit in Ref.~\cite{Gorbar:2013uga}.) In order to fit the numerical results, we could 
try to use a polynomial function of $n$. However, by following a trial and error approach 
instead, we found that the following simple function approximates our numerical results 
really well: 
\begin{equation}
\mu_{5,n} \simeq - 0.225 \frac{\alpha |eB|}{\mu}\sqrt{1-\left(\frac{2n|eB|}{\mu^2}\right)^2},
\label{strongB:mu5-n-fit}
\end{equation}
where we took into account that $|eB|=\mu^2/30$ and $\alpha =1/137$. The corresponding 
function is shown by the solid line in the right panel of Fig.~\ref{strongB:fig_mu-5-n}.

By making use of the numerical results and the analytical expression for the fermion 
propagator with the chiral asymmetry parameters included, we can now straightforwardly 
determine the interaction-induced deviations of the Fermi momenta $(p_3-p_3^{(0)})/\mu$ for 
the predominantly left-handed and right-handed fermions in the considered 
ultrarelativistic limit $\mu\gg m$. Here $p_3^{(0)}$ is the value of the Fermi 
momentum in the absence of the chiral asymmetry (i.e., $\Delta_{n}=0$ and 
$\mu_{5,n}=0$). Such 
deviations can be viewed as the actual measure of the chiral asymmetry at the 
Fermi surface. The numerical results for $(p_3-p_3^{(0)})/\mu$ in each occupied 
Landau level are shown in Fig.~\ref{strongB:fig_pFermi}. (The corresponding numerical 
values are also presented in the last two columns of Table~\ref{strongB:Table-Fermi-surface}.) 
This is a generalization of the analogous results in the weak-field limit, obtained in 
Ref.~\cite{Gorbar:2013uga}. 

\begin{figure}[t]
\centering
\includegraphics[width=0.47\textwidth]{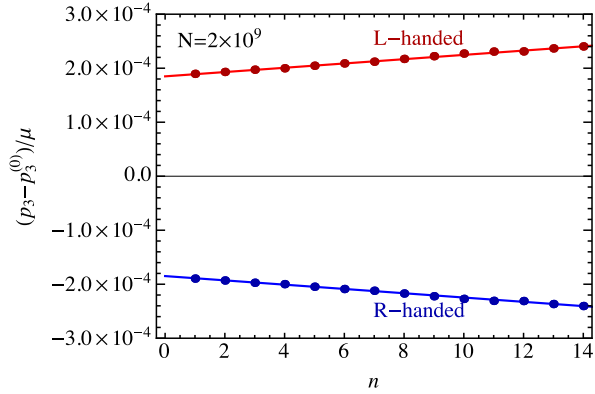}
\caption{Asymmetry of the Fermi surface for predominantly left-handed (red)
and right-handed (blue) particles as a function of the Landau-level index $n$.}
\label{strongB:fig_pFermi}
\end{figure}

We find that the results for $(p_3-p_3^{(0)})/\mu$ in Fig.~\ref{strongB:fig_pFermi} are well approximated by 
linear functions of $n$. When written in the same form as the chiral shift and the chiral 
chemical potential functions, the corresponding linear fits take the following form:
\begin{equation}
p_3-p_3^{(0)} \simeq \pm \frac{\alpha |eB|}{\mu}\left(0.76+0.49 \frac{|eB|n}{\mu^2}\right).
\label{strongB:p3-p30}
\end{equation}
As is easy to check, the values of these Fermi momenta shifts are of the order of
$10$--$100~\mbox{keV}$ and, thus, are not very large in the context of compact stars, 
even though we already assumed an extremely strong value of the magnetic field, 
$B=10^{18}~\mbox{G}$. One should keep in mind, however, that here we used the 
model of a dense QED plasma, whose coupling constant $\alpha$ is extremely small. 
This conclusion could change drastically in the case of dense quark matter, where the 
relevant coupling constant $\alpha_{s}$ is about two orders of magnitude stronger. Indeed, 
by taking into account that the estimate for the Fermi momenta shift in Eq.~(\ref{strongB:p3-p30}) 
is proportional to the coupling, we conclude that the chiral asymmetry should be of the 
order of $1$--$10~\mbox{MeV}$ in dense quark matter. Such a large asymmetry may 
in turn produce observable consequences for protoneutron stars \cite{Gorbar:2011ya}.

\begin{table}
\begin{center}
\caption{Data for the chiral asymmetry functions $\Delta_n$, $\mu_{5,n}$, and 
$(p_3-p_3^{(0)})/\mu$ at the Fermi surface.} 
\label{strongB:Table-Fermi-surface}
\begin{tabular}{c@{\hskip 0.5in}c@{\hskip 0.5in}c@{\hskip 0.5in}c}
\hline 
$n$ & $\Delta_n/\mu$ & $\mu_{5,n}/\mu$ & $(p_3-p_3^{(0)})/\mu$  \\
\hline 
  1  &  $-1.32\times 10^{-4} \pm 1.84\times 10^{-7} $  &  $-5.56\times 10^{-5} \pm 3.30\times 10^{-7}$  &  $\pm 1.90\times 10^{-4}$ \\
  2  &  $-1.34\times 10^{-4} \pm 2.56\times 10^{-7} $  &  $-5.57\times 10^{-5} \pm 4.12\times 10^{-7}$  &  $\pm 1.93\times 10^{-4}$    \\
  3  &  $-1.37\times 10^{-4} \pm 3.21\times 10^{-7}$  &  $-5.41\times 10^{-5} \pm 4.64\times 10^{-7} $  &  $\pm 1.97\times 10^{-4}$    \\
  4  &  $-1.39\times 10^{-4} \pm 3.76\times 10^{-7} $  &  $-5.18\times 10^{-5} \pm 5.01\times 10^{-7}$  &  $\pm 2.00\times 10^{-4}$    \\
  5  &  $-1.43\times 10^{-4} \pm 4.34\times 10^{-7}$  &  $-5.13\times 10^{-5} \pm 5.27\times 10^{-7}$  &  $\pm 2.05\times 10^{-4}$    \\
  6  &  $-1.46\times 10^{-4} \pm 4.90\times 10^{-7}$  &  $-4.84\times 10^{-5} \pm 5.43\times 10^{-7}$  &  $\pm 2.09\times 10^{-4}$    \\
  7  &  $-1.48\times 10^{-4} \pm 5.46\times 10^{-7}$  &  $-4.68\times 10^{-5} \pm 5.52\times 10^{-7}$  &  $\pm 2.12\times 10^{-4}$    \\
  8  &  $-1.50\times 10^{-4} \pm 5.97\times 10^{-7}$  &  $-4.71\times 10^{-5} \pm 5.52\times 10^{-7}$  &  $\pm 2.17\times 10^{-4}$    \\
  9  &  $-1.53\times 10^{-4} \pm 6.54\times 10^{-7}$  &  $-4.38\times 10^{-5} \pm 5.42\times 10^{-7}$  &  $\pm 2.22\times 10^{-4}$    \\
 10  &  $-1.57\times 10^{-4} \pm 7.12\times 10^{-7} $  &  $-4.05\times 10^{-5} \pm 5.23\times 10^{-7}$  &  $\pm 2.27\times 10^{-4}$    \\
 11  &  $-1.58\times 10^{-4} \pm 7.53\times 10^{-7}$  &  $-3.76\times 10^{-5} \pm 4.94\times 10^{-7}$  &  $\pm 2.31\times 10^{-4}$    \\
 12  &  $-1.60\times 10^{-4} \pm 8.01\times 10^{-7}$  &  $-3.18\times 10^{-5} \pm 4.48\times 10^{-7}$  &   $\pm 2.31\times 10^{-4}$    \\
 13  &  $-1.63\times 10^{-4} \pm 8.47\times 10^{-7}$  &  $-2.69\times 10^{-5} \pm 3.83\times 10^{-7}$  &  $\pm 2.37\times 10^{-4}$    \\
 14  &  $-1.66\times 10^{-4} \pm 8.96\times 10^{-7}$  &  $ -1.92\times 10^{-5} \pm 2.84\times 10^{-7}$  &   $\pm 2.40\times 10^{-4}$    \\
\hline 
\end{tabular}
\end{center}
\end{table}

As we see from the above results, 
the values of the chiral shift $\Delta_{n}$ and the chiral chemical potential $\mu_{5,n}$ 
at the Fermi surface appear to be of order $\alpha |eB|/\mu$. This differs from the
corresponding weak-field prediction $\alpha |eB|\mu/m^2$ by a rather large factor 
$(\mu/m)^2$. Such a difference is not surprising and, in fact, should have been 
expected in the ultrarelativistic limit when $|eB|/m^2$ is not a good expansion 
parameter. While the dependence of $\Delta_{n}$ on the Landau-level index $n$ 
shows a weak growth, $\mu_{5,n}$ decreases with $n$. 

The interaction induced deviations of the Fermi momenta $(p_3-p_3^{(0)})/\mu$ 
for the predominantly left-handed and right-handed fermions provide a formal 
measure of the chiral asymmetry at the Fermi surface. Because of the smallness 
of the electromagnetic coupling constant, the corresponding values 
appear to be rather small in the case of dense QED matter even at extremely 
large densities and extremely strong magnetic fields. It is suggested, however, 
that the results can be substantial in the case of quark matter.

\subsection{Physics phenomena due to chiral effects in magnetic fields}
\label{PhysPhenChiralMag}

\subsubsection{Pulsar kicks due to chiral shift}
\label{ChiralShiftPhys}

The asymmetry with respect to longitudinal momentum $k^{3}$ of the opposite chirality fermions 
in the ground state of dense magnetized matter may have important physical consequences 
\cite{Gorbar:2009bm,Gorbar:2011ya}. In particular, the manifestation of the asymmetry could 
be rather dramatic in the processes that rely exclusively on the weak interaction. This is due to 
the fact that only left-handed fermions participate in weak processes. For example, the Fermi 
surface asymmetry of the left-handed charged particles (electrons or quarks) could dramatically 
affect the distribution of neutrinos inside dense magnetized matter. 

Dense relativistic matter (i.e., electron or quark plasma) in rather strong magnetic fields is 
naturally present in neutron stars. Also, in the context of neutron star, neutrinos may play 
an important role in determining some of the observed stellar properties. One of the interesting 
phenomena that we would like to mention here is the pulsar kicks \cite{Lyne:1994az,
Cordes:1997mm,Hansen:1997zw,Fryer:1997nu,Arzoumanian:2001dv,Chatterjee:2005mj}.

It has been known for a long time that typical spatial velocities of pulsars are an order of 
magnitude larger than those of their progenitors \cite{1968AJ.....73..535T,Lyne:1994az}. 
Taking into account the violent conditions at the pulsar birth, this may not be so surprising. 
Even a small asymmetry in the supernova explosion may result in a kick velocity of order  
$100~\mbox{km/s}$. However, there are some pulsars whose velocities are considerably 
higher, e.g., of the order of $1000~\mbox{km/s}$ \cite{Wang:2005jg}. While many mechanisms
were previously proposed, we suggest that a qualitatively new mechanism for the pulsar 
kicks could be provided by the chiral asymmetry associated with the induced chiral shift $\Delta$  
in the stellar plasma of charged particles (electrons or quarks) \cite{Gorbar:2009bm,Gorbar:2011ya}. 

The kick could develop during the early stages of the (proto-)neutron star evolution when a
large number of neutrinos are trapped in dense and hot stellar matter. The corresponding state
lasts about a minute. During this time the neutrinos experience multiple scatterings and 
come in near perfect thermal equilibrium with dense magnetized plasma. More precisely,
they interact and equilibrate with the {\it left-handed} sector of the plasma. As we argued earlier, 
however, the latter is characterized by an asymmetric distribution with nonequal absolute 
values of the longitudinal momenta on the opposite sides of the Fermi surface. Therefore, 
this asymmetry should result in an asymmetric momentum distribution of the trapped
neutrinos \cite{Gorbar:2009bm,Gorbar:2011ya}. After gradually diffusing through the bulk 
of dense matter, these neutrinos will carry a nonzero momentum away from the protoneutron 
star. In view of the conservation of the total momentum, the star will receive a kick in the 
opposite direction.  

The crucial detail of this mechanism is the asymmetric Fermi surface of the left-handed charged 
fermions in the magnetized plasma. Such a plasma is a thermally equilibrated, but nonisotropic 
ground state. Because of this and in sharp contrast to the commonly assumed diffusion through 
an {\it isotropic} hot matter \cite{Kusenko:1998yy,Sagert:2006yn,Sagert:2007as,Sagert:2007ug}, 
the asymmetry of the neutrino distribution is build up rather than wash out by interaction with 
the stellar plasma. 

It appears also very helpful for the proposed pulsar kick mechanism that the chiral shift parameter 
is not affected a lot even by moderately high temperatures, $5~\mbox{MeV} \lesssim T\lesssim 
50~\mbox{MeV}$, present during the earliest stages of protoneutron stars \cite{Page:2006ud}. 
Indeed, as our findings show, the value of $\Delta$ is primarily determined by the chemical 
potential and has a weak/nonessential temperature dependence when $\mu\gg T$. In the 
stellar context, this ensures the feasibility of the proposed mechanism even at the earliest 
stages of the protoneutron stars, when there is a sufficient amount of thermal energy 
to power the strongest (with $v\gtrsim 1000~\mbox{km/s}$) pulsar kicks observed \cite{Lyne:1994az,Cordes:1997mm,Hansen:1997zw,Fryer:1997nu,Arzoumanian:2001dv,Chatterjee:2005mj}.
Alternatively, the constraints of the energy conservation would make it hard, if not impossible, 
to explain any sizable pulsar kicks if the interior matter is cold ($T\lesssim 1~\mbox{MeV}$) 
\cite{Charbonneau:2009hq,Charbonneau:2009ax}.

Let us also mention that the robustness of the chiral shift in hot magnetized matter may be useful to 
provide an additional neutrino push to facilitate successful supernova explosions as suggested in 
Ref.~\cite{Fryer:2005sz}. The specific details of such a scenario are yet to be worked out.

\subsubsection{Magnetic instability in neutron stars}
\label{MagInstabilityStars}

A rather unusual application of the chiral magnetic and chiral separation effects in the compact 
stars was proposed recently in Ref.~\cite{Ohnishi:2014uea}. It was suggested that a strong and 
stable magnetic field could be produced inside the neutron stars spontaneously as the result of 
the chiral plasma instability \cite{Akamatsu:2013pjd}. [Similar instabilities were also proposed 
earlier in different contexts \cite{Niemi:1985ir,Redlich:1984md,Rubakov:1986am,Tsokos:1985ny,
Rubakov:1985nk,Joyce:1997uy,Boyarsky:2011uy}.] According to Ref.~\cite{Ohnishi:2014uea}, 
the seed of the corresponding instability is provided by a chirality imbalance of electrons. Such 
an imbalance is argued to be produced during the core collapse of supernova via parity-violating 
weak processes. (The same origin of the chirality imbalance was also discussed in 
Refs.~\cite{Charbonneau:2009hq,Charbonneau:2009ax}.) The maximal magnetic field due 
to this instability was estimated to be of order $10^{18}~\mbox{G}$ at the stellar core.
The beneficial byproduct of the proposed mechanism \cite{Ohnishi:2014uea} is the production of
a stable field configuration with a large magnetic helicity. If valid, this could be a candidate 
mechanism for explaining the origin of extremely large magnetic field fields in magnetars 
\cite{Kouveliotou:1998ze,Rea:2011fa,Olausen:2013bpa}.

Clearly, one of the key ingredients for the development of the proposed instability is the chirality 
imbalance of electrons. Presumably, a sufficiently large imbalance could be produced via electron 
capture inside a collapsing star, i.e., $p+e^{-}_{L}\to n+\nu_{e,L}$. The argument utilizes the fact that 
only left-handed electrons are captured in such a process, leading to an excess of the right-handed   
chirality in the electron plasma. The authors of Ref.~\cite{Ohnishi:2014uea} estimated that the 
difference of the chemical potentials of the right- and left-handed electrons could be as large as 
$\mu_5 =(\mu_R-\mu_L)/2 \sim 200~\mbox{MeV}$. (As we point below, this estimate is  
questionable when the role of a nonzero electron mass is taken into account 
\cite{Grabowska:2014efa}.)

If a sufficiently large chiral imbalance $\mu_5$ is indeed created in the electron plasma, it will 
induce an electric current along the direction of the magnetic field via the chiral magnetic effect, 
see Eq.~(\ref{CMEdefinition}). Such a current itself will induce a toroidal magnetic field that, in
turn, also produces a toroidal electric current again via the chiral magnetic effect. Finally, the 
toroidal current will induce a magnetic field in the same direction as the original magnetic field
and, thus, provide a positive feedback to the exponential growth of the magnetic field and 
currents. In the process of the magnetic field generation, the chiral imbalance will gradually 
decrease and the process will turn off eventually \cite{Akamatsu:2014yza}. 

The validity of the spontaneous magnetic field generation scenario of Ref.~\cite{Ohnishi:2014uea} 
was criticized, however, in Ref.~\cite{Grabowska:2014efa}. In particular, the authors of 
Ref.~\cite{Grabowska:2014efa} calculated the rate of helicity changing scattering of electrons 
off the ambient protons and showed that it is many orders of magnitude larger than the rate of 
the electron capture weak process responsible for producing the chirality imbalance. It was 
concluded, therefore, that it is impossible to generate the seed chiral imbalance needed to 
drive the instability during the core collapse of a supernova. 

A slightly different mechanism to explain the generation of strong magnetic fields in magnetars 
based on the chiral magnetic instability was proposed in Ref.~\cite{Dvornikov:2014uza}. However, 
its theoretical validity is yet to be tested.

\subsubsection{Chiral magnetic effect in heavy ion collisions}
\label{ChiralMagEffHIC}

It is natural to ask whether the chiral shift parameter can have any interesting implications in the regime 
of relativistic heavy ion collisions. As was recently discussed in the literature, hot relativistic matter 
in a magnetic field may have interesting properties even in the absence of the chiral shift parameter. 
The examples of the recently suggested phenomena, that appear to be closely related to the generation 
of the chiral shift, are the chiral magnetic effect \cite{Kharzeev:2004ey,Kharzeev:2007tn,
Kharzeev:2007jp,Fukushima:2008xe,Nam:2010nk,Rebhan:2009vc,Fukushima:2010zza,Bzdak:2012ia}, 
the chiral magnetic spiral \cite{Basar:2010zd,Kim:2010pu,Frolov:2010wn}, and the chiral magnetic wave 
\cite{Kharzeev:2010gd,Burnier:2012ae,Taghavi:2013ena}.

As we discussed in Section~\ref{CSE-CME:Numerical}, at high temperatures, i.e., in the regime relevant 
for relativistic heavy ion collisions, the chiral shift parameter is generated for any nonzero chemical 
potential. This is seen from the results presented in Fig.~\ref{CSE-CME:fig-Tmulti-mo}. However, its 
role is not as obvious as in the case of stellar matter. At high temperatures, the Fermi surface and 
the low-energy excitations in its vicinity are not very useful concepts any more. Instead, it is the 
axial current itself that is of interest. The chiral shift should induce a correction to the topological 
axial current (\ref{MZ}). As seen from Eq.~(\ref{CSE-CME:axial}), the corresponding correction in 
the NJL model is proportional to the chiral shift parameter $\Delta$, multiplied by a 
factor $(\Lambda l)^2$, where $\Lambda$ is the ultraviolet cutoff. Formally, the product of $\Delta$ 
and $(\Lambda l)^2$ is finite and is proportional to the chemical potential. However, unlike the 
topological term, which is also proportional to the chemical potential, the dynamical one contains 
an extra factor of the coupling constant. Therefore, only at relatively strong coupling, which can be 
provided by QCD interactions, the effect of the chiral shift parameter on the axial current can be 
substantial. The corresponding effect is still likely to be just a quantitative change of 
the overall constant in relation (\ref{CSEdefinition}) and, thus, is not easy to measure.

On the other hand, as suggested in Refs.~\cite{Gorbar:2011ya,Burnier:2011bf}, 
the interplay of the chiral separation and chiral magnetic effects can lead to a modified 
version of the latter, the quadrupole CME, which does not rely on the initial topological 
charge fluctuations. This can presumably be realized as follows \cite{Gorbar:2011ya},
see Fig.~\ref{CSE-CME:figQuadCME}. An initial axial 
current generates an excess of opposite chiral charges around the polar regions of the 
fireball. Then, these chiral charges trigger two ``usual" chiral magnetic effects with 
opposite directions of the vector currents at the opposite poles. The inward flows of 
these electric currents will diffuse inside the fireball, while the outward flows will lead 
to a distinct observational signal: an excess of same sign charges going back-to-back. 
This picture can also be interpreted as the generation of the chiral magnetic wave 
\cite{Kharzeev:2010gd}, a collective gapless excitation already mentioned in 
Section~\ref{subsec:CSE-CME-general}. It should be also emphasized that the 
quadrupole CME does not rely on the existence of topological charge fluctuations in 
the hot quark-gluon plasma. It is triggered by the conventional chemical potential 
$\mu$, rather than by the chiral chemical potential $\mu_5$, which is not as well defined 
as $\mu$ \cite{Jimenez-Alba:2014iia}.

\begin{figure}[t]
\begin{center}
\includegraphics[width=.6\textwidth]{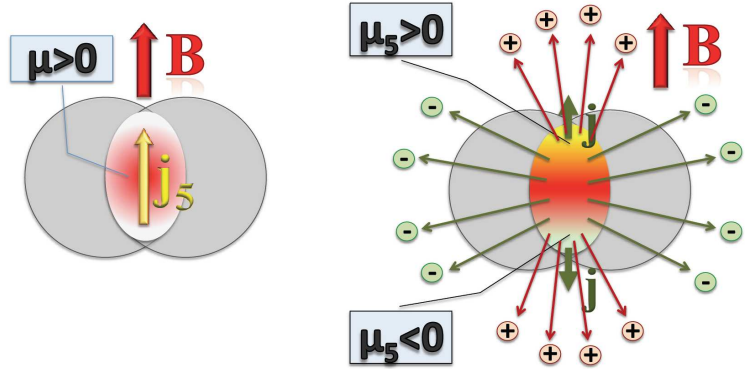}
\caption{(Color online) The schematic mechanism behind the quadrupole chiral magnetic 
effect in heavy ion collisions, when an axial current is initially driven by a nonzero baryon 
chemical potential that results in a pair of back-to-back electric currents produced by the 
axial charges in the polar regions.}
\label{CSE-CME:figQuadCME}
\end{center}
\end{figure}

One of the observable implications of the quadrupole CME is the splitting 
of the elliptic flows of positive and negative pions, i.e., $v^{\pi^{-}}_2-v^{\pi^{+}}_2=r_eA$, 
where $A$ is the net charge asymmetry $A=(\bar{N}^{+}-\bar{N}^{-})/(\bar{N}^{+}+\bar{N}^{-})$ 
and $r_e > 0$ is the slope parameter \cite{Burnier:2011bf}. Such a splitting was observed 
by the STAR collaboration \cite{Adamczyk:2013kcb,Wang:2012qs,Ke:2012qb} 
and appears to be in qualitative agreement with the theoretical predictions.

Concerning the regime of hot relativistic matter, let us also mention that it will be of 
interest to extend the present analysis of magnetized relativistic matter to address 
the properties of collective modes, similar to those presented in Ref.~\cite{Kharzeev:2010gd}, 
by studying various current-current correlators.

\subsubsection{Magnetic instability in the early Universe}
\label{MagInstabilityUniverse}

Let us recall that, historically, the idea of the chiral separation/magnetic effect came from 
cosmological studies \cite{Vilenkin:1980fu}. Early on, it was also proposed that the parity 
violation in weak interactions could potentially explain the origin of cosmic magnetic fields
\cite{Vilenkin:1982pn}. Recently similar ideas and their developments became very popular
again \cite{Semikoz:2007ti,Semikoz:2009ye,Semikoz:2012ka,Boyarsky:2011uy,Boyarsky:2012ex,
Tashiro:2012mf,Long:2013tha,Dvornikov:2013bca}. This renewed interest is driven by the 
substantial progress in the observational cosmology in the recent decade. Among other 
characteristics, intergalactic magnetic fields may provide useful probes of the early 
Universe \cite{Durrer:2013pga}. 

By making use of the generalized set of Maxwell equations that take into account 
the chiral chemical potential in relativistic plasma with $T\gg m$, the authors of 
Ref.~\cite{Boyarsky:2011uy} analyzed a self-consistent time evolution of magnetic 
fields. They found that the magnetic fields survive much longer than the time defined 
by magnetic diffusion and that the magnetic helicity is transferred from short to long 
scales. Using ideas closely related to the chiral magnetic effect, it was also suggested
that the parity-violating weak interactions in a plasma with nonzero density of lepton or 
baryon numbers may drive a generation a horizon-scale helical magnetic fields 
\cite{Boyarsky:2012ex}. 

By supplementing the leptogenesis scenario with the analysis of magnetic field evolution 
in a turbulent medium, in Ref.~\cite{Long:2013tha} it was proposed that the combination
of the chiral magnetic and chiral vortical effects can give rise to a right-handed helical 
magnetic field that is coherent on astrophysical length scales. A range of other related 
ideas that lead to the generation of the primordial magnetic fields were given in 
Refs.~\cite{Tashiro:2012mf,Semikoz:2007ti,Semikoz:2009ye,Semikoz:2012ka,
Kahniashvili:2012uj,Dvornikov:2013bca}. One common theme in all of them is the
key role played by the chiral anomaly, e.g., via the chiral separation and chiral
magnetic effects, see Eqs.~(\ref{CSEdefinition}) and (\ref{CMEdefinition}), or
their generalizations.

\section{Dirac semimetals in magnetic fields}
\label{sec:Dirac-Weyl-semimetals}

After the discovery of graphene and the appreciation of its unique electronic and transport 
properties, a lot of effort was invested in the search of condensed matter materials with 
three-dimensional Dirac-like spectra of low-energy quasiparticles. Historically, bismuth and 
some of its alloys were the first materials in which the electron states in a part of the Brillouin 
zone were described by 3D Dirac fermions \cite{1960PMag....5..115C,1964JPCS...25.1057W,
Falkovsky:1968,1976AdPhy..25..555E,1977SvPhU..20..819E}. Moreover, in an alloy of bismuth 
with a small admixture of antimony, Bi$_{1-x}$Sb$_x$, it was found that the Dirac 
mass could be controlled by changing the concentration of antimony. At the concentration of about 
$x \approx 0.03-0.04$, the mass vanishes and the alloy turns into a massless Dirac semimetal, 
albeit a highly anisotropic one \cite{Lenoir199689,PhysRevB.78.045426}. 

Although other materials with 3D Dirac fermions can be obtained by fine tuning the strength of 
the spin-orbital coupling or chemical composition \cite{2007NJPh....9..356M,2010NJPh...12f5013Z,
2011Sci...332..560X,2011NatPh...7..840S,2013PhRvB..88c5444D}, it is difficult to 
control such realizations. An interesting idea was expressed in Ref.~\cite{Young:2012kx}, 
where it was shown that the formation of Dirac points can be protected by a crystal symmetry, and
metastable $\beta$-cristobalite BiO$_2$ was suggested as a specific example of a massless 
Dirac material. Later, by using first-principles calculations and effective model analysis, the authors 
of Refs.~\cite{Wang:2012uq,Wang:2013fk} predicted that $A_3$Bi ($A$ = Na, K, Rb) and Cd$_3$As$_2$ should
be Dirac semimetals with bulk 3D Dirac points protected by crystal symmetry. The experimental
discovery of the 3D Dirac fermions in Na$_3$Bi and Cd$_3$As$_2$ was reported in 
Refs.~\cite{Liu:2013uq} and \cite{Neupane:2013vn,Borisenko:2013kx}, respectively. The Dirac nature 
of the quasiparticles was confirmed by investigating the electronic structure of these materials 
with angle-resolved photoemission spectroscopy. For a recent review of 3D Dirac semimetals, see
Ref.~\cite{Gibson:2014aa}.
 
The Dirac four-component spinors are composed of a pair of (left-handed and 
right-handed) two-component spinors. The latter satisfy the Weyl equation and describe 
fermionic quasiparticles of a fixed (left-handed or right-handed) chirality. If the 
requirement of inversion or time-reversal symmetry is relaxed, the degeneracy 
of the dispersion relations of the left- and right-handed Weyl modes can be lifted, 
transforming the Dirac semimetal into a Weyl one. While pyrochlore iridates 
\cite{2011PhRvB..83t5101W}, as well as some heterostructures of topological and 
normal insulators \cite{PhysRevLett.107.127205}, are conjectured to be Weyl semimetals 
(for a review, see Refs.~\cite{PhysRevB.84.235126,2013arXiv1301.0330T,Vafek:2013vn}), 
no material at present is experimentally confirmed to be a Weyl semimetal. Since 
magnetic field breaks time reversal symmetry, one may engineer a Weyl semimetal from a 
Dirac one by applying the external magnetic field. One such mechanism was originally 
proposed in the context of high-energy physics in Ref.~\cite{Gorbar:2009bm} and then extended 
to Dirac semimetals in Ref.~\cite{Gorbar:2013qsa}. The essence of the corresponding idea
relies on the dynamical generation of the chiral shift that was reviewed in Section~\ref{sec:CSEandCME}. 
The extension of the same mechanism to semimetals is reviewed in the next subsection.
It is expected that such a mechanism can be naturally realized in the recently discovered 3D Dirac 
semimetals Na$_3$Bi and Cd$_3$As$_2$ \cite{Liu:2013uq,Neupane:2013vn,Borisenko:2013kx} 
(in addition to the magnetic field, a necessary condition for this mechanism to operate is 
a nonzero density of charge carriers).

\subsection{Chiral shift in Dirac semimetals}
\label{sec:ChiralShiftDiracSemimetals}

Following the ideas similar to those in Section~\ref{sec:CSEandCME}, here we will show that 
an external magnetic field turns a Dirac semimetal with a nonzero density of charge carriers into 
a Weyl semimetal with a pair of Weyl nodes for each of the original Dirac nodes \cite{Gorbar:2013qsa}. 
The corresponding transformation is induced by the electron-electron interaction and is 
intimately connected with the topological nature of the spin-polarized lowest Landau level 
in the Dirac semimetal. As expected, the induced momentum separation between the Weyl 
nodes will be determined by the chiral shift parameter $\mathbf{b}$. (This is the same as the 
chiral shift $\Delta$ in Section~\ref{sec:CSEandCME}, but here we use the notation conventional
in solid state physics literature.) The latter is a pseudovector with the direction along the magnetic 
field and the magnitude proportional to the quasiparticle charge density, the strength of the magnetic 
field, and the strength of the interaction.

\subsubsection{Model}
\label{secDW:model}

Let us start by writing down the general form of the low-energy Hamiltonian for a Weyl semimetal, 
\begin{equation}
H^{\rm (W)}=H^{\rm (W)}_0+H_{\rm int},
\label{secDW:Hamiltonian-model-Weyl}
\end{equation}
where
\begin{equation}
H^{\rm (W)}_0=\int d^3r \left[\,v_F \psi^{\dagger} (\mathbf{r})\left( 
\begin{array}{cc} \bm{\sigma}\cdot(-i\bm{\nabla}-\mathbf{b})  & 0\\ 0 & 
-\bm{\sigma}\cdot(-i\bm{\nabla}+\mathbf{b}) \end{array} 
\right)\psi(\mathbf{r})-\mu_{0}\, \psi^{\dagger} (\mathbf{r})\psi(\mathbf{r})
\right]
\label{secDW:free-Hamiltonian}
\end{equation}
is the Hamiltonian of the free theory, which describes two Weyl nodes of opposite (as required by 
the Nielsen--Ninomiya theorem \cite{Nielsen:1983rb}) chirality separated by vector $2\mathbf{b}$ 
in momentum space. It should be clear that $\mathbf{b}$ is the same as the chiral shift parameter
in Section~\ref{sec:CSEandCME}. The rest of the notations used in the definition of the low-energy 
Hamiltonian are: $v_F$ is the Fermi velocity, $\mu_{0}$ is the chemical potential, and 
$\bm{\sigma}=(\sigma_x,\sigma_y,\sigma_z)$ are Pauli matrices associated with the 
conduction-valence band degrees of freedom in a generic low-energy model \cite{Zyuzin:2012ab}. 
Because of the similarity of the latter to the spin matrices in the relativistic Dirac equation, 
we will call them pseudospin matrices in this section.

The $H_{\rm int}$ part of the Hamiltonian describes the electron-electron Coulomb interaction 
$U(\mathbf{r}-\mathbf{r}^{\prime})$. In general, it has the following nonlocal form:
\begin{equation}
H_{\rm int} = \frac{1}{2}\int d^3rd^3r^{\prime}\,\psi^{\dagger}(\mathbf{r})\psi(\mathbf{r})U(\mathbf{r}-\mathbf{r}^{\prime})
\psi^{\dagger}(\mathbf{r}^{\prime})\psi(\mathbf{r}^{\prime}). 
\label{secDW:int-Hamiltonian}
\end{equation} 
In order to describe the underlying physics in the simplest possible form, below we will utilize a simple
model with a contact four-fermion interaction,
\begin{equation}
U(\mathbf{r}) = \frac{e^2}{\kappa |\mathbf{r}|} \rightarrow g\, \delta^3(\mathbf{r}),
\label{secDW:model-interaction}
\end{equation}
where $g$ is a dimensionful coupling constant. Such a model interaction should at least be 
sufficient for a qualitative description of the effect of the dynamical generation of the chiral 
shift parameter by a magnetic field in Dirac semimetals.

Before proceeding further with the analysis of the underlying dynamics, it is convenient to 
rewrite the model in terms of the four-component Dirac spinors. To achieve this, let us 
introduce the following Dirac matrices in the chiral representation:
\begin{equation}
\gamma^0 = \left( \begin{array}{cc} 0 & -I\\ -I & 0 \end{array} \right),\qquad
\bm{\gamma} = \left( \begin{array}{cc} 0& \bm{\sigma} \\  - \bm{\sigma} & 0 \end{array} \right),
\label{secDW:Dirac-matrices}
\end{equation}
where $I$ is the two-dimensional unit matrix. Then, the original Hamiltonian takes the following 
relativistic form:
\begin{equation}
H^{\rm (W)} = \int d^3 r\, \bar{\psi} (\mathbf{r})\left[
-i v_F (\bm{\gamma}\cdot \bm{\nabla})-(\mathbf{b}\cdot \bm{\gamma})\gamma^5-\mu_{0}\gamma^0 
\right]\psi(\mathbf{r})
+\frac{g}{2}\int d^3r\,\rho(\mathbf{r})\rho(\mathbf{r}).
\label{secDW:free-Hamiltonian-Weyl-rel}
\end{equation}
Here, by definition, $\bar{\psi} \equiv \psi^{\dagger}\gamma^0$ is the Dirac conjugate spinor field, 
$\rho(\mathbf{r})\equiv \bar{\psi}(\mathbf{r}) \gamma^0 \psi (\mathbf{r})$ is the charge density 
operator, and matrix $\gamma^5$ is defined as usual, i.e., 
\begin{equation}
\gamma^5 \equiv i\gamma^0\gamma^1\gamma^2\gamma^3 
= \left( \begin{array}{cc} I & 0\\ 0 & -I \end{array} \right).
\end{equation}
By comparing the structure of this matrix with that of the chiral shift term in the free Hamiltonian 
(\ref{secDW:free-Hamiltonian}), we see that the eigenvalues of $\gamma^5$ correspond to the 
node degrees of freedom.

As should be clear from the definition, a Dirac semimetal is a special case of the Weyl one, 
and its low-energy Hamiltonian is the same as in Eq.~(\ref{secDW:free-Hamiltonian-Weyl-rel}), 
but with $\mathbf{b}=0$, i.e.,
\begin{equation}
H^{\rm (D)} = H^{\rm (W)}\Big|_{\mathbf{b} = 0}.
\label{secDW:free-Hamiltonian-Dirac}
\end{equation}
Unlike Weyl semimetals, Dirac semimetals are invariant under the time reversal symmetry. 
While the most general Dirac model also allows a nonzero mass term $m_0\bar{\psi} 
(\mathbf{r})\psi(\mathbf{r})$, we do not include it in Eq.~(\ref{secDW:free-Hamiltonian-Weyl-rel}).
This is justified by the fact that the recently discovered Dirac semimetals \cite{Liu:2013uq,
Neupane:2013vn,Borisenko:2013kx} appear to have no mass gaps at the Dirac points.

In the presence of an external magnetic field, the $\bm{\nabla}$ operator in the Hamiltonian
should be replaced by the covariant derivative $\bm{\nabla}-ie\mathbf{A}/c$, where 
$\mathbf{A}$ is the vector potential and $c$ is the speed of light. Therefore, the model 
Hamiltonian for the Dirac semimetal in an external magnetic field has the following form:
\begin{equation}
H^{\rm (D)}_{\rm mag} = \int d^3 r\, \bar{\psi} (\mathbf{r})\left[
-i v_F \left( \bm{\gamma}\cdot(\bm{\nabla}-ie\mathbf{A}/c) \right)-\mu_{0}\gamma^0 
\right]\psi(\mathbf{r})
+\frac{g}{2}\int d^3r\,\rho(\mathbf{r})\rho(\mathbf{r}).
\label{secDW:Hamiltonian-Dirac-magnetic}
\end{equation}
In the absence of the Dirac gap, both this Hamiltonian and the Weyl semimetal Hamiltonian in 
Eq.~(\ref{secDW:free-Hamiltonian-Weyl-rel}) are invariant under the chiral $\mathrm{U}(1)_{+}\times \mathrm{U}(1)_{-}$ 
symmetry, where $+$ and $-$ correspond to the node states with $+1$ and $-1$ eigenvalues 
of the $\gamma_5$ matrix, respectively. Strictly speaking, the currents connected with the 
$\mathrm{U}(1)_{+}$ and $\mathrm{U}(1)_{-}$ symmetries are anomalous. However, because these symmetries 
are Abelian, one can still introduce conserved charges for them \cite{Jackiw:1986dr}.

\subsubsection{Gap equation}
\label{secDW:gapEq}

In this section, we will derive the gap equation for the fermion propagator in the 
Dirac semimetal model (\ref{secDW:Hamiltonian-Dirac-magnetic}) and show that at a nonzero 
charge density, a nonzero $\mathbf{b}$ {\it necessarily} arises in the normal phase as 
soon as a magnetic field is turned on. 

In model (\ref{secDW:Hamiltonian-Dirac-magnetic}), we easily find the following free fermion 
propagator:
\begin{equation}
iS^{-1}(u,u^\prime) = \left[(i\partial_t+\mu_0)\gamma^0
-v_F(\bm{\pi}\cdot\bm{\gamma})\right]\delta^{4}(u- u^\prime),
\label{secDW:sinverse}
\end{equation}
where $u=(t,\mathbf{r})$ and $\bm{\pi} \equiv -i \bm{\nabla} - e\mathbf{A}/c$ 
is the canonical momentum.The vector potential is chosen in
the Landau gauge, $\mathbf{A}= (0, x B,0)$, where $B$ is a strength of the external 
magnetic field pointing in the $z$ direction.

An ansatz for the full fermion propagator can be written in the following form (we will see 
that this ansatz is consistent with the Schwinger--Dyson equation for the fermion propagator 
in the mean-field approximation):
\begin{equation}
iG^{-1}(u,u^\prime)= \Big[(i\partial_t+\mu )\gamma^0 - v_F(\bm{\pi}\cdot\bm{\gamma}) 
+\gamma^0(\bm{\tilde{\mu}}\cdot\bm{\gamma})\gamma^5
+v_F (\mathbf{b} \cdot \bm{\gamma})\gamma^5
-m\Big]\delta^{4}(u- u^\prime).
\label{secDW:ginverse}
\end{equation}
This propagator contains dynamical parameters $\tilde{\bm{\mu}}$, $\mathbf{b}$, and $m$ 
that are absent at tree level in Eq.~(\ref{secDW:sinverse}). Here $m$ plays the role of the dynamical 
Dirac mass and $\mathbf{b}$ is the chiral shift \cite{Gorbar:2009bm,Gorbar:2011ya}.
By taking into account the Dirac structure of the $\tilde{\bm{\mu}}$ term, we see that it is 
related to the anomalous magnetic moment $\mu_{\rm an}$ (associated with the pseudospin) 
as follows: $\tilde{\bm{\mu}}\equiv \mu_{\rm an}\mathbf{B}$.
It should be also emphasized that the dynamical parameter $\mu$ in the full propagator may 
differ from its tree-level counterpart $\mu_0$ [see Eq.~(\ref{secDW:gap-mu-text})].

In order to determine the values of these dynamical parameters, we will use the 
Schwinger--Dyson (gap) equation for the fermion propagator in the mean-field 
approximation, i.e.,
\begin{equation}
iG^{-1}(u,u^\prime) = iS^{-1}(u,u^\prime) - g \left\{ 
\gamma^0 G(u,u) \gamma^0 - \gamma^0\, \mathrm{tr}[\gamma^0G(u,u)]\right\}
\delta^{4}(u- u^\prime). 
\label{secDW:gap}
\end{equation}
The first term in the curly brackets describes the exchange (Fock) interaction and
the last term presents the direct (Hartree) interaction. 

Separating different Dirac structures in the gap equation, we arrive at the following set
of equations:
\begin{eqnarray}
\mu - \mu_0  &=& -\frac{3}{4}\, g \, \langle j^{0}\rangle ,
\label{secDW:gap-mu-text}  \\
\mathbf{b} &=& \frac{g}{4v_F } \, \langle\mathbf{j}_5\rangle   ,
\label{secDW:gap-Delta-text} \\
m &=& - \frac{g}{4} \,  \langle \bar{\psi}\psi\rangle  ,
\label{secDW:gap-m-text}\\
\bm{\tilde{\mu}} &=& \frac{g}{4} \, \langle \bm{\Sigma} \rangle  .
\label{secDW:gap-tilde-mu-text}
\end{eqnarray}
The fermion charge density, the axial current density, the chiral condensate, and the anomalous magnetic moment 
condensate on the right hand side of the above equations are determined through the full fermion propagator as follows:
\begin{eqnarray}
\langle j^{0}\rangle &\equiv  &-\mathrm{tr}\left[\gamma^0G(u,u)\right], 
 \label{secDW:density-text}\\
\langle \mathbf{j}_5\rangle &\equiv  &-\mathrm{tr}\left[\bm{\gamma}\,\gamma^5G(u,u)\right],
\label{secDW:axial-current-text}\\
\langle \bar{\psi}\psi\rangle&\equiv  & -\mathrm{tr}\left[G(u,u)\right] ,
\label{secDW:chi-condensate-text}\\
\langle \bm{\Sigma} \rangle &\equiv  &-\mathrm{tr}\left[ \gamma^0\bm{\gamma}\,\gamma^5G(u,u)\right] . 
 \label{secDW:pseudospin-condensate-text}
\end{eqnarray}
Note that the right-hand sides in Eqs.~(\ref{secDW:chi-condensate-text}) and (\ref{secDW:pseudospin-condensate-text})
differ from those in Eqs.~(\ref{secDW:density-text}) and  (\ref{secDW:axial-current-text}) by the inclusion of an 
additional $\gamma^0$ matrix inside the trace. Since, according to Eq.~(\ref{secDW:Dirac-matrices}), 
the $\gamma^0$ matrix mixes quasiparticle states {from different Weyl nodes, we conclude 
that while $\langle j^{0}\rangle$ and $\langle \mathbf{j}_5\rangle$ describe the charge density
and the axial current density, the chiral condensate $\langle \bar{\psi}\psi\rangle$ and the 
anomalous magnetic moment condensate $\langle \bm{\Sigma} \rangle$ describe internode 
coherent effects.

As is known \cite{Vilenkin:1980fu,Metlitski:2005pr,Newman:2005as}, in the presence of a fermion charge density and 
a magnetic field, the axial current $\langle \mathbf{j}_5\rangle$ is generated even in the free theory. 
Therefore, according to Eq.~(\ref{secDW:gap-Delta-text}), the chiral shift $\mathbf{b}$ is induced already in the lowest 
order of the perturbation theory. As a result, a Dirac semimetal is {\it necessarily} transformed into a Weyl one, 
as soon as an external magnetic field is applied to the system (see also a discussion in Section~\ref{secDW:perturb}).

In order to derive the propagator $G(u,u^\prime)$ in the Landau-level representation, we invert 
$G^{-1}(u,u^\prime)$ in Eq.~(\ref{secDW:ginverse}) by using the approach described in 
Appendix~\ref{CSE:App:Landau-level-rep}. For our purposes here, the expression 
for the propagator in the coincidence limit $u^\prime\to u$ is sufficient [cf. Eq.~(A26) in 
Ref.~\cite{Gorbar:2011ya}]:
\begin{equation}
G(u,u)= \frac{i}{2\pi l^2}\sum_{n=0}^{\infty} \int\frac{d\omega d k^{3}}{(2\pi)^2}
\frac{{\cal K}_{n}^{+ }{\cal P}_{+ }+{\cal K}_{n}^{- }{\cal P}_{- }\theta(n-1)}{U_n},
\label{secDW:full-propagator}
\end{equation}
where ${\cal P}_{\pm}\equiv \frac12 \left(1\pm i s_\perp \gamma^1\gamma^2\right)$ are the pseudospin 
projectors, $l=\sqrt{c/|eB|}$ is the magnetic length, and  $s_\perp=\mathrm{sign} (eB)$. Also, by definition, $\theta(n-1)=1$ for $n\geq 1$ 
and $\theta(n-1)=0$ for $n < 1$. The functions ${\cal K}_{n}^{\pm}$ and $U_n$ with $n\geq 0$ are given by
\begin{eqnarray}
{\cal K}_{n}^{\pm}&=& \left[\left(\omega+\mu \mp s_{\perp}v_F b\right)\gamma^0
\pm s_{\perp}\tilde\mu + m -v_Fk^{3}\gamma^3\right]\Big\{
(\omega+\mu)^2 + \tilde{\mu}^2 -  m^2 - (v_F b)^2 - (v_Fk^{3})^2 - 2nv^2_F|eB|/c \nonumber \\
&& \mp 2s_{\perp}\left[\tilde\mu (\omega+\mu)+v_F b m \right]\gamma^0
\pm 2s_{\perp} (\tilde\mu + v_F b\gamma^0)v_Fk^{3} \gamma^3\Big\}
\label{secDW:K_n^pm}
\end{eqnarray}
and
\begin{equation}
U_n =\left[(\omega+\mu)^2 + \tilde{\mu}^2 - m^2 - (v_F b)^2 - (v_Fk^{3})^2 -2nv^2_F|eB|/c\right]^2
- 4\left[\left(\tilde{\mu}\,(\omega+\mu) + v_F b m\right)^2
        +(v_Fk^{3})^2\left((v_F b)^2-\tilde{\mu}^2\right)\right],
\label{secDW:U_n}
\end{equation}
where we took into account that the only nonvanishing components of the axial vectors 
$\mathbf{b}$ and $\bm{\tilde{\mu}}$ are the longitudinal projections $b$ and $\tilde{\mu}$ on the
direction of the magnetic field. Note that the zeros of the function $U_n$ determine the dispersion 
relations of quasiparticles.

\subsubsection{Perturbative solution}
\label{secDW:perturb}

In order to obtain the leading order perturbative solution to the gap equations, we can use the free propagator 
on the right-hand side of Eqs.~(\ref{secDW:density-text}) through (\ref{secDW:pseudospin-condensate-text}), i.e.,
\begin{equation}
S(u,u)= \frac{i}{2\pi l^2}\sum_{n=0}^{\infty} \int\frac{d\omega d k^{3}}{(2\pi)^2}
\frac{ \left[\left(\omega+\mu_{0}\right)\gamma^0 -v_Fk^{3}\gamma^3\right]
\left[{\cal P}_{+}+{\cal P}_{-}\theta(n-1)\right]}{(\omega+\mu_{0})^2 - (v_F k^{3})^2 -2nv^2_F|eB|/c}.
\label{secDW:full-free-propagator}
\end{equation}
Note that unlike the high Landau levels with $n\geq 1$, where both spin projectors ${\cal P}_{+}$ 
and ${\cal P}_{-}$ contribute, the lowest Landau level (LLL) with $n=0$ contains only one projector ${\cal P}_{+}$.
The reason of this is well known. The Atiyah-Singer theorem connects the number of the zero energy 
modes (which are completely pseudospin polarized) of the two-dimensional part of the Dirac operator 
to the total flux of the magnetic field through the corresponding plane. This theorem states that  
LLL is topologically protected (for a discussion of the Atiyah-Singer theorem in the 
context of condensed matter physics, see Ref.~\cite{Katsnelson:2012GrapheneBook}).

By making use of Eq.~(\ref{secDW:full-free-propagator}), we straightforwardly calculate the zeroth order result for the 
charge density,
\begin{equation}
\langle j^{0}\rangle_0=\frac{\mu_0}{2v_F(\pi l)^2}
+\frac{\mathrm{sign}(\mu_0)}{v_F(\pi l)^2}\sum_{n=1}^{\infty} \sqrt{\mu_0^2-2nv^2_F|eB|/c}\,\,
\theta\left(\mu_0^2-2nv^2_F|eB|/c\right),
\label{secDW:perturbative-charge}
\end{equation}
and the axial current density
\begin{equation}
\langle \mathbf{j}_5\rangle_0 = \frac{e\mathbf{B}\mu_0}{2\pi^2 v_Fc}.
\label{secDW:perturbative-axial-current}
\end{equation}
As to the chiral condensate and the anomalous magnetic moment condensate, they vanish, 
i.e., $\langle \bar{\psi}\psi\rangle_0 = 0$ and $\langle\bm{\Sigma} \rangle_0 = 0$.
This is not surprising because both of them break the chiral $\mathrm{U}(1)_{+} \times \mathrm{U}(1)_{-}$ 
symmetry of the Dirac semimetal Hamiltonian (\ref{secDW:Hamiltonian-Dirac-magnetic}).
Then, taking into account Eqs.~(\ref{secDW:gap-m-text})} and (\ref{secDW:gap-tilde-mu-text}), 
we conclude that both the Dirac mass $m$ and the parameter $\tilde{\bm{\mu}}$ 
are zero in the perturbation theory.

After taking into account the gap equations (\ref{secDW:gap-mu-text}) and (\ref{secDW:gap-Delta-text}),
the results in Eqs.~(\ref{secDW:perturbative-charge}) and (\ref{secDW:perturbative-axial-current}) imply that     
there is a perturbative renormalization of the chemical potential and a dynamical generation 
of the chiral shift. Of special interest is the result for the axial current density given by 
Eq.~(\ref{secDW:perturbative-axial-current}). This is generated already in the free theory and 
known in the literature as the topological contribution \cite{Son:2004tq,Metlitski:2005pr}.
Its topological origin is related to the following fact: in the free theory, only the LLL contributes 
to both the axial current and the axial anomaly. By combining Eqs.~(\ref{secDW:gap-Delta-text}) and 
(\ref{secDW:perturbative-axial-current}), we find
\begin{equation}
\mathbf{b} = \frac{ge\mathbf{B}\mu_0}{8\pi^2 v_F^2  c}.
\label{secDW:bvalue}
\end{equation}
This is our principal result, which reflects the simple fact that at $\mu_0 \neq 0$ (i.e., nonzero
charge density), the absence of the chiral shift is not protected by any additional symmetries 
in the normal phase of a Dirac metal in a magnetic field. Indeed, in the presence of a homogeneous 
magnetic field pointing in the $z$ direction, the rotational $\mathrm{SO}(3)$ symmetry in the model is 
explicitly broken down to the $\mathrm{SO}(2)$ symmetry of rotations around the $z$ axis. The 
dynamically generated chiral shift parameter $\mathbf{b}$ also points in the same direction 
and does  not break the leftover $\mathrm{SO}(2)$ symmetry. The same is true for the discrete symmetries: 
while the parity ${\cal P}$ is preserved, all other discrete symmetries, charge conjugation 
${\cal C}$, time reversal ${\cal T}$, ${\cal CP}$, ${\cal CT}$, $PT$, and ${\cal CPT}$, are broken.
Last but not least, the chiral shift does not break the chiral $\mathrm{U}(1)_{+}\times \mathrm{U}(1)_{-}$ symmetry 
considered in Section~\ref{secDW:model}. This implies that the dynamical chiral shift is necessarily generated 
in the normal phase of a Dirac semimetal in a magnetic field, and the latter is transformed into 
a Weyl semimetal.

In the phase with a dynamically generated chiral shift $\mathbf{b}$, the quasiparticle dispersion 
relations, i.e., $\omega_{n,\sigma} = -\mu + E_{n,\sigma}$, are determined by the following Landau-level 
energies (recall that we assume that the magnetic field points in the $z$ direction):
\begin{eqnarray}
E_{0, \sigma} &=& v_F \left( s_\perp b +\sigma k^3\right),\qquad n=0, \\
E_{n,\sigma} &=& \pm v_F \sqrt{\left(s_{\perp}b+\sigma k^3\right)^2+2n|eB|/c},\qquad n\geq 1,
\end{eqnarray}
where $\sigma=\pm$ corresponds to different Weyl nodes. The corresponding dispersion relations
are shown graphically in Fig.~\ref{secDW:fig:Fermi}. Note a qualitatively different character (compared to higher 
Landau levels) of the dispersion relations in the LLL given by the straight lines whose signs of slope 
correlate with the Weyl nodes. Of course, this correlation is due to the complete polarization of 
quasiparticle pseudospins in the LLL. As seen from Fig.~\ref{secDW:fig:Fermi}, the effect of the chiral shift
is not only to shift the relative position of the Weyl nodes in momentum space, but also to induce 
a chiral asymmetry of the Fermi surface \cite{Gorbar:2009bm,Gorbar:2011ya}.

\begin{figure}[t]
\begin{center}
\includegraphics[width=.45\textwidth]{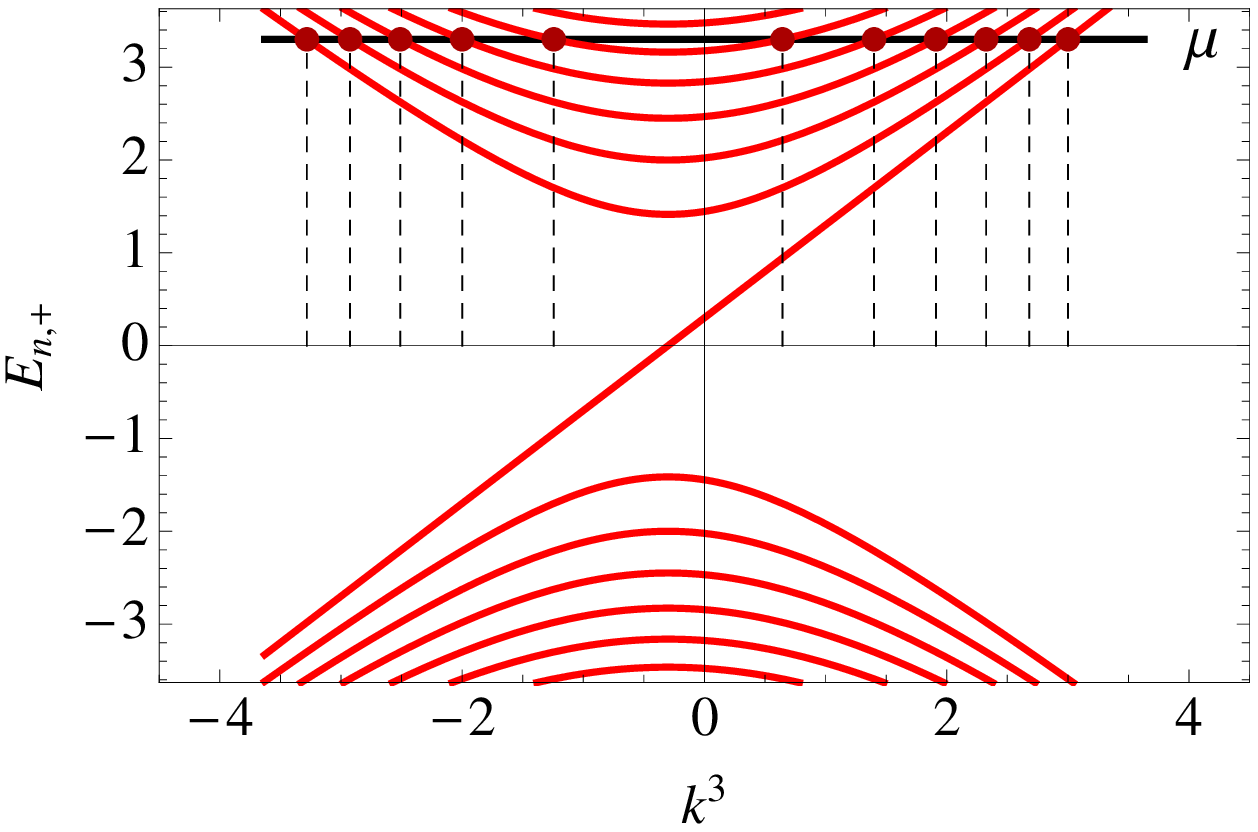}
\hspace{.04\textwidth}
\includegraphics[width=.45\textwidth]{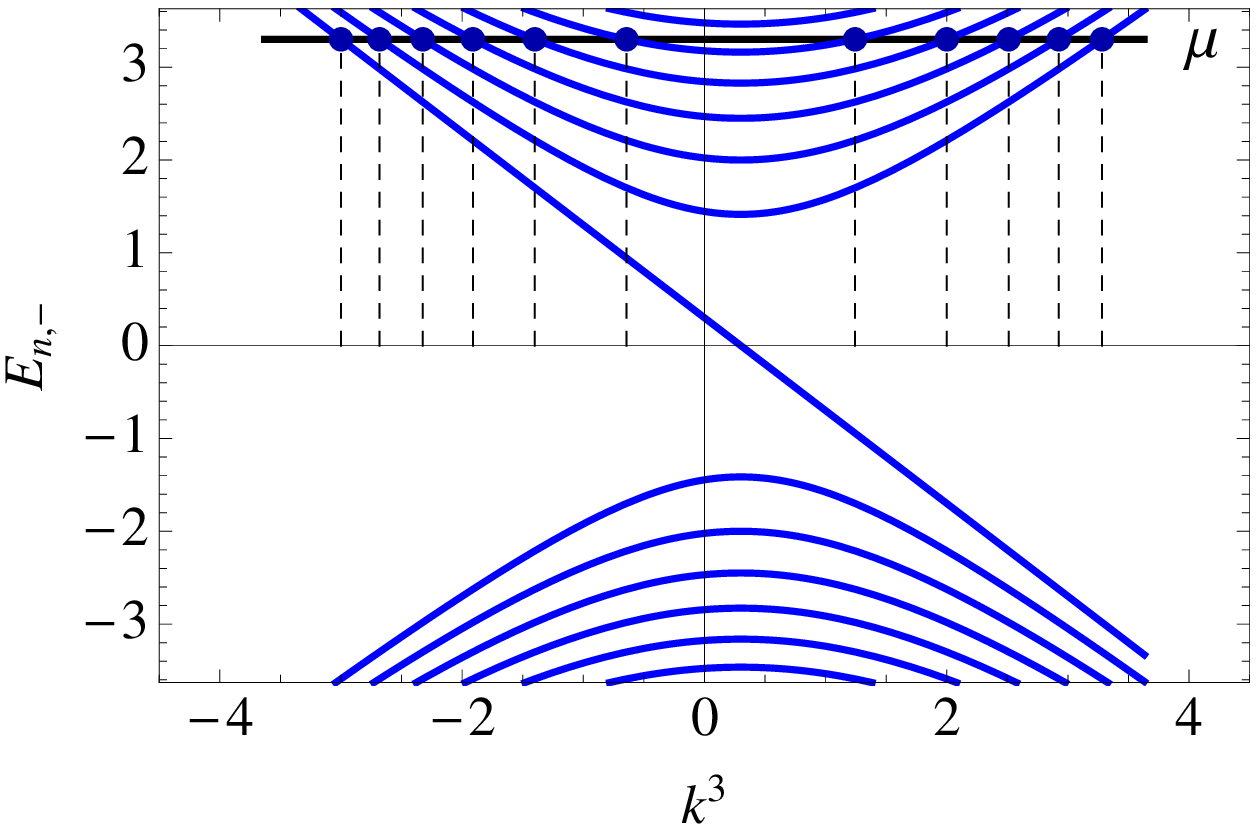}
\caption{Dispersion relations of the quasiparticles from different Weyl nodes and 
their Fermi surfaces. The two dispersion relations are the mirror images of 
each other.}
\label{secDW:fig:Fermi}
\end{center}
\end{figure}

\subsubsection{Nonperturbative solution: Phase transition}
\label{secDW:nonperturb}

The magnetic catalysis phenomenon, which was reviewed in detail in 
Sections~\ref{sec:MagneticCatalysis} and \ref{sec:MagCatGauge}, implies that at vanishing 
$\mu_0$, the ground state in the model at hand is characterized by a nonzero Dirac mass 
$m$ that spontaneously breaks chiral symmetry.

Such a vacuum state can withstand a finite stress due to a nonzero chemical potential. However, as 
we discuss below, when $\mu_0$ exceeds a certain critical value $\mu_{\rm cr}$, the chiral symmetry 
restoration and a new ground state are expected. The new state is characterized by a nonvanishing 
chiral shift parameter $\mathbf{b}$ and a nonzero axial current in the direction of the magnetic field. 
Since no symmetry of the theory is broken, this state is the {\it normal} phase of the magnetized 
matter that happens to have a rather rich chiral structure. 

Let us describe this transition in more detail. The value of the dynamical Dirac mass $m$ in the vacuum 
state can be easily calculated following the same approach as in Ref.~\cite{Gusynin:1994xp,Gusynin:1995nb}. At weak 
coupling, in particular, we can use the following expression for the chiral condensate: 
\begin{equation}
\langle \bar{\psi}\psi\rangle \simeq -\frac{m}{4\pi^2 v_F} 
\left(\Lambda^2 +\frac{1}{l^2}\ln\frac{v_F^2}{\pi m^2 l^2}  \right),
\label{secDW:condensate3+1}
\end{equation}
obtained in the limit of a small mass (which is consistent with the weak coupling approximation), using 
the gauge invariant proper-time regularization. Here the ultraviolet momentum cutoff $\Lambda$ can be related, for 
example, to the value of lattice spacing $a$ as follows: $\Lambda \simeq \pi/a$. Finally, by taking into 
account gap equation (\ref{secDW:gap-m-text}), we arrive at the solution for the dynamical mass,
\begin{equation}
m \simeq \frac{v_F}{\sqrt{\pi}l}\exp\left(-\frac{8 \pi^2 v_F l^2}{g}+\frac{(\Lambda l)^2}{2}\right).
\label{secDW:DiracMass}
\end{equation}
This zero-temperature, nonperturbative solution exists for $\mu_0< m$.

The free energies of the two types of states, i.e., the nonperturbative state with a dynamically 
generated Dirac mass (and no chiral shift) and the perturbative state with a nonzero chiral 
shift (and no Dirac mass) become equal at about $\mu_0\simeq m/\sqrt{2}$. This is analogous 
to the Clogston relation in superconductivity \cite{Clogston:1962zz}.

At the critical value $\mu_{\rm cr} \simeq m/\sqrt{2}$, a first order phase transition takes place.
Indeed, both these solutions coexist at $\mu_0< m$, and while for $\mu_0 < \mu_{\rm cr}$
the nonperturbative (gapped) phase with a chiral condensate is more stable, the normal 
(gapless) phase becomes more stable at  $\mu_0 > \mu_{\rm cr}$. Note that at 
$\mu_0 < m$ the chemical potential is irrelevant in the gapped phase: the charge 
density is absent there. On the other hand, at any nonzero chemical potential, there is a 
nonzero charge density in the normal (gapless) phase. Therefore, at 
$\mu_{\rm cr} \simeq m/\sqrt{2}$, there is a phase transition with a jump in the charge 
density, which is a clear manifestation of a first order phase transition.

\subsubsection{Dirac semimetals vs. graphene in a magnetic field}
\label{secDW:graph}

It is instructive to compare the states in the magnetized Dirac semimetals and graphene. 
First of all, we would like to point out that the chiral shift is a three dimensional
analog of the Haldane mass \cite{Haldane:1988zza,Niemi:1983rq,Redlich:1983kn}, which plays an important role in the 
dynamics of the quantum Hall effect in graphene. Indeed, in the formalism of the 
four-component Dirac fields in graphene, the Haldane mass condensate is described 
by the same vacuum expectation value as that of the axial current in three 
dimensions \cite{Gorbar:2008hu}:
\begin{equation}
\langle \bar{\psi}\gamma^{3}\gamma^{5}\psi\rangle =  -\mathrm{tr}\left[\gamma^3\,\gamma^5G(u,u)\right]
\end{equation}
for a magnetic field pointing in the $z$ direction, which is orthogonal to the graphene plane,
cf. Eq.~(\ref{secDW:axial-current-text}). Moreover, similarly to the solution with the chiral shift,
the solution with the Haldane mass (with the same sign for both spin-up and 
spin-down quasiparticles) describes the normal phase: it is a singlet with respect to the 
$\mathrm{SU}(4)$ symmetry, which is a graphene analog of the chiral group in Dirac and Weyl 
semimetals. 

Also, in graphene, there is a phase transition similar to that described in Section~\ref{secDW:nonperturb}. 
It happens when the LLL is completely filled \cite{Gorbar:2008hu}. In other words, the quantum 
Hall state with the filling factor $\nu = 2$ in graphene is associated with the normal phase 
containing the Haldane mass. 

As is well known, the Haldane mass leads to the Chern-Simons term in an external electromagnetic 
field \cite{Haldane:1988zza,Niemi:1983rq,Redlich:1983kn}. This feature reflects a topological nature of the state with the filling factor 
$\nu = 2$ in a graphene. As was recently shown in Ref.~\cite{Grushin:2012mt}, the chiral shift term 
\begin{equation}
\bar{\psi}(\mathbf{b} \cdot \bm{\gamma})\gamma^5\psi
\end{equation}
leads to an induced Chern-Simons term of the form 
$\frac{1}{2}b_\mu\epsilon^{\mu\nu\rho\sigma}F_{\rho\sigma}A_\nu$ in Weyl semimetals 
[here $b_\mu$ is a four-dimensional vector $(0,\mathbf{b})$]. Therefore, it should also be 
generated in Dirac semimetals in a magnetic field. Note however the following principle 
difference between them: while in Weyl semimetals the chiral shift $\mathbf{b}$ is present  
in the free Hamiltonian, it is dynamically generated in the normal phase of Dirac semimetals 
in a magnetic field [see Eq.~(\ref{secDW:bvalue})]. Like in graphene, the generation of the 
Chern-Simons term implies a topological nature of the normal state in this material.
Note that the generation of the Chern-Simons term in these materials is intimately 
connected with their anisotropy. A possibility of the generation of an induced Chern-Simons 
term in a relativistic isotropic hot matter was considered in Ref.~\cite{Redlich:1984md}.

In summary, the normal phase of a Dirac semimetal at a nonzero charge density and in a 
magnetic field has a nontrivial chiral structure: it is transformed into a Weyl semimetal with 
a pair of Weyl nodes for each of the original Dirac points. The nodes are separated by a 
dynamically induced chiral shift that is directed along the magnetic field. The phase 
transition between the normal phase and the phase with chiral symmetry breaking is revealed, 
and the rearrangement of the Fermi surface accompanying this phase transition is 
described.

Although we used a simple model with a contact four-fermion interaction, the same 
qualitative results are expected also in more realistic models. The studies in the 
relativistic Nambu-Jona-Lasinio (NJL) model (with a contact interaction) on the one hand 
\cite{Gorbar:2009bm,Gorbar:2011ya} and in QED on the other \cite{Gorbar:2013upa} 
strongly support the validity of this statement. Namely, the dynamical generation of the 
chiral shift in a magnetic field and at a nonzero fermion density is a universal phenomenon. 

\subsection{Observational implications}
\label{sec:ObservationDiracSemimetals}

It was suggested in the literature that negative longitudinal magnetoresistivity in Weyl 
semimetals \cite{Nielsen:1983rb,Aji:2011aa,Son:2012bg} is a consequence of the 
chiral anomaly \cite{Adler:1969gk,Bell:1969ts}
and is a fingerprint of a Weyl semimetal phase. In particular, the anomaly is responsible 
for pumping the electrons between the nodes of opposite chirality 
at a rate proportional to the scalar product of the applied electric and magnetic fields 
$\mathbf{E}\cdot\mathbf{B}$. Recently, a negative longitudinal magnetoresistivity
was observed in Bi$_{1-x}$Sb$_x$ alloy with $x \approx 0.03$ in moderately strong magnetic 
fields \cite{Kim:2013ys} and could be interpreted as an experimental signature 
of the presence of a Weyl semimetal phase, where a single Dirac point splits into two Weyl 
nodes with opposite chiralities and the separation between the nodes in momentum space is 
proportional to the applied field. (Similar measurement were also reported in other 
materials \cite{Liang:2014aa,Li:2014aax}, but they were interpreted differently.)
As we will show below, however, the observation of the negative longitudinal 
magnetoresistivity is also expected in Dirac semimetals. 
Therefore, negative magnetoresistivity alone may not be sufficient to unambiguously 
distinguish between the Dirac and Weyl semimetals. Note that a nonlocal transport 
can be another way of probing the chiral anomaly in Weyl semimetals \cite{Parameswaran:2013aa}.

In Refs.~\cite{Nielsen:1983rb,Aji:2011aa,Son:2012bg}, the magnetoresistivity in Weyl semimetals 
was studied by using the semiclassical Boltzmann kinetic equation. Since negative longitudinal 
magnetoresistivity is one of the key characteristics of Weyl semimetals intimately connected 
with the chiral anomaly, below we derive magnetoresistivity using the microscopic 
Kubo formalism, which takes into account quantum effects. (In a special class of gapless 
Dirac semiconductors with a small carrier concentration, transverse magnetoresistivity 
was previously studied in Ref.~\cite{1998PhRvB..58.2788A}.) We found that the negative 
longitudinal magnetoresistivity takes place not only in Weyl semimetals, but also in 
Dirac ones. 

As we argue in Section~\ref{sec:MagnetoresistanceDiracSemimetals}, the origin of the negative magnetoresistivity 
is intimately connected with the spatial dimensional reduction $3 \to 1$ in the low-energy 
dynamics dominated by the LLL. Such a dimensional reduction is a universal phenomenon, 
taking place in the dynamics of charged fermions in a magnetic field \cite{Gusynin:1994xp,Gusynin:1995nb}. The 
low-energy quasiparticles are described by the spin-polarized LLL states and effectively 
have one-dimensional dispersion relations, which depend only on the longitudinal 
momentum $k_z$ and do not contain the magnetic field at all.
The physics behind this phenomenon is the following. As is well known, the 
transverse momenta $\mathbf{k}_\perp$ are not good quantum numbers for quasiparticles 
in a magnetic field. In the dispersion relations, such momenta are replaced by a single 
discrete quantum number $n$, labeling the Landau levels (which have a degeneracy 
proportional to the value of the magnetic field).

The consequences of the dimensional reduction are rather dramatic in the case 
of relativistic-like massless fermions because of their chiral nature: such fermions disperse 
only one way in the longitudinal direction for each chirality \cite{Hosur:2013aa}. The existence of massless 
chiral fermions and their high degeneracy in the presence of a magnetic field are topologically 
protected by the index theorem \cite{Aharonov:1978gb}. We find that it is this unique nature 
of the low-energy states that is responsible for the main contribution (growing linearly with 
the field) to the longitudinal conductivity in Dirac/Weyl semimetals. In fact, as we will see 
in the following, the special nature of the LLL plays a profound role also in the 
anomalous Hall contribution to the transverse conductivity.

Finally, we would like to mention that electric transport in Weyl semimetals in the absence 
of magnetic field was studied in Refs.~\cite{Hosur:2011aa,Rosenstein:2013aa}. The magneto-optical 
conductivity of Weyl semimetals was investigated in Ref.~\cite{Ashby:2013aa}. Recent 
developments in transport phenomena in Weyl semimetals are reviewed in 
Ref.~\cite{Hosur:2013aa} focusing on signatures connected with the chiral anomaly.

\subsubsection{Magnetoresistance}
\label{sec:MagnetoresistanceDiracSemimetals}

According to the Kubo linear response theory, the direct current conductivity tensor
\begin{equation}
\sigma_{ij}  = \lim_{\Omega\to 0}\frac{\mbox{Im}\,\Pi_{ij}(\Omega+i0 ;\mathbf{0})}{\Omega}
\label{sigma-ij}
\end{equation}
is expressed through the Fourier transform of the current-current correlation function
\begin{equation}
\Pi_{ij}(\Omega;\mathbf{0}) =  e^2 v_F^2 T \sum_{k=-\infty}^{\infty} 
\int \frac{d^3 \mathbf{p}}{(2\pi)^3} \mbox{tr} \Big[ \gamma^i \bar{G}(i\omega_k ;\mathbf{p})  
\gamma^j \bar{G}(i\omega_k-\Omega; \mathbf{p})\Big].
\label{Pi_Omega_k}
\end{equation}
Note that this function is given in terms of the translation invariant part of the quasiparticle Green's function.
By making use of the spectral representation for the Green's function
\begin{equation}
\bar{G}(i\omega_k ;\mathbf{p})  = \int_{-\infty}^{\infty} \frac{d\omega A (\omega ;\mathbf{p}) }{i\omega_k+\mu-\omega},
\label{spectral-fun}
\end{equation}
we obtain the following standard representation for the current-current correlation function:
\begin{equation}
\Pi_{ij}(\Omega+i 0;\mathbf{0}) =
e^2 v_F^2  \int  d\omega \int d \omega^\prime 
\frac{n_F(\omega)-n_F(\omega^\prime)}{\omega -\omega^\prime-\Omega-i 0}
\int \frac{d^3 \mathbf{k}}{(2\pi)^3} 
\mbox{tr} \left[ \gamma^i A(\omega ;\mathbf{k}) \gamma^j A(\omega^\prime; \mathbf{k})\right], 
\label{Pi_ij}
\end{equation}
where $n_F(\omega)= 1/\left[e^{(\omega-\mu)/T}+1\right]$ is the Fermi distribution function. 

In the normal phase (with $m=0$), the spectral function can be decomposed into the sum 
of two separate chiral contributions \cite{Gorbar:2013dha}, i.e.,
\begin{equation}
A(\omega;\mathbf{k}) =  \sum_{\chi=\pm} A^{(\chi)}(\omega;\mathbf{k}) {\cal P}_{5}^{(\chi)} ,
\label{spectral-A}
\end{equation}
where  
\begin{eqnarray}
A^{(\chi)}(\omega;\mathbf{k}) &=& \frac{ie^{-k_\perp^2 l^{2}}}{\pi}\sum_{\lambda=\pm}\sum_{n=0}^{\infty} 
 \frac{(-1)^n}{E_n^{(\chi)}}\Bigg\{
 \left[E_n^{(\chi)} \gamma^{0} 
 -\lambda v_F (k_3-\chi b)\gamma^3\right]\left[{\cal P}_{+}L_n\left(2 k_\perp^2 l^{2}\right)
-{\cal P}_{-}L_{n-1}\left(2 k_\perp^2 l^{2}\right)\right] \nonumber\\
&& + 2\lambda v_F(\mathbf{k}_\perp\cdot\bm{\gamma}_\perp) L_{n-1}^1\left(2 k_\perp^2 l^{2}\right)
 \Bigg\}\frac{\Gamma_n}{\left(\omega -\lambda E_n^{(\chi)} \right)^2+\Gamma_n^2}.
 \label{sp-funct-Gamma}
\end{eqnarray}
This model spectral function assumes that quasiparticles in the $n$th Landau level have a nonzero 
decay width $\Gamma_n$, which is determined by the disorder and/or interactions. The functional 
form of $\Gamma_n$ is needed for quantitative studies of the transport properties, but will not be 
addressed here. By aiming only at the key features of the magneto-transport characteristics in 
Weyl and Dirac semimetals, we will assume that $\Gamma_n$ take constant values. Because of the 
special nature of LLL, we will allow that the decay width in the LLL can be smaller than the decay 
widths in higher Landau levels. 

In the expression for the diagonal components of the current-current correlation function (\ref{Pi_ij}), 
the traces in the integrand are real. 
Therefore, in order to extract the imaginary part of $\Pi_{ii}(\Omega+i 0;\mathbf{0})$, we can use 
the identity
\begin{equation}
\frac{1}{\omega -\omega^\prime-\Omega-i 0} = {\cal P}\frac{1}{\omega -\omega^\prime-\Omega}
+i \pi \delta\left(\omega -\omega^\prime-\Omega\right) .
\label{prin-value}
\end{equation}
Taking this into account in Eq.~(\ref{Pi_ij}) and using the definition in Eq.~(\ref{sigma-ij}),
we derive a much simpler and more convenient expression for the diagonal components of the conductivity 
tensor:
\begin{eqnarray}
\sigma_{ii} &=& -\pi e^2 v_F^2  \sum_{\chi=\pm} \int \frac{d \omega }{4T\cosh^2\frac{\omega-\mu}{2T}}
\int \frac{d^3 \mathbf{k}}{(2\pi)^3} 
\mbox{tr} \left[ \gamma^i A^{(\chi)}(\omega ;\mathbf{k}) \gamma^i A^{(\chi)}(\omega; \mathbf{k})
 {\cal P}_{5}^{(\chi)}\right].
\label{sigma-ii-AA-chi}
\end{eqnarray}
(Here there is no sum over index $i$.)

The calculation of the off-diagonal components of the transverse conductivity $\sigma_{12}=-\sigma_{21}$ 
is complicated by the fact that the corresponding traces in Eq.~(\ref{Pi_ij}) are imaginary. In this case, 
it is convenient to rewrite the expression for the current-current correlation 
function as follows:
\begin{eqnarray}
\Pi_{ij}(\Omega+i 0;\mathbf{0}) 
&=& e^2 v_F^2    \sum_{\chi=\pm} \int \frac{d^3 \mathbf{k}}{(2\pi)^3}  \int  d\omega n_F(\omega)
\mbox{tr} \Big[ \gamma^i A^{(\chi)}(\omega ;\mathbf{k}) \gamma^j \bar{G}^{(\chi)}_{\mu=0}(\omega-\Omega-i0; \mathbf{k})
 {\cal P}_{5}^{(\chi)}\nonumber\\
&&
 +\gamma^i \bar{G}^{(\chi)}_{\mu=0}(\omega +\Omega+i0 ;\mathbf{k}) \gamma^j A^{(\chi)}(\omega; \mathbf{k}) {\cal P}_{5}^{(\chi)}\Big],
\label{correlation-function}
\end{eqnarray}
where we used Eq.~(\ref{spectral-fun}) at $\mu=0$ in order to eliminate one of the energy integrations.
By substituting this result into 
Eq.~(\ref{sigma-ij}) and taking the limit $\Omega\to0$, we obtain
\begin{eqnarray}
\sigma_{ij} &=& e^2 v_F^2   \sum_{\chi=\pm} \mbox{Im} 
\int \frac{d^3 \mathbf{k}}{(2\pi)^3} \int  d\omega  n_F(\omega)
\mbox{tr} \Bigg[ 
\gamma^i  \frac{d\bar{G}^{(\chi)}_{\mu=0}(\omega+i0 ;\mathbf{k})}{d\omega} \gamma^j A^{(\chi)}(\omega;\mathbf{k}) 
 {\cal P}_{5}^{(\chi)}\nonumber\\
&&
 -\gamma^i A^{(\chi)}(\omega;\mathbf{k}) \gamma^j \frac{d\bar{G}^{(\chi)}_{\mu=0}(\omega-i0 ;\mathbf{k})}{d\omega} 
 {\cal P}_{5}^{(\chi)} \Bigg].
 \label{sigma-ij-AA-chi}
\end{eqnarray}
In principle this is valid for both the diagonal and off-diagonal components. In the case of the diagonal components, 
however, this is equivalent to the much simpler expression in Eq.~(\ref{sigma-ii-AA-chi}). In order to show their 
equivalency explicitly, one needs to integrate the expression in Eq.~(\ref{sigma-ij-AA-chi}) by parts and use the 
definition for the spectral function in Eq.~(\ref{spectral-A}). In the calculation of the off-diagonal components 
$\sigma_{ij}$, only the representation in Eq.~(\ref{sigma-ij-AA-chi}) is valid. 

It may be appropriate to mention that the analysis of the conductivity here does not take into 
account the effect of weak localization/antilocalization \cite{1980PThPh..63..707H,1980PhRvB..22.5142A}. (For a recent 
study of weak localization and antilocalization in 3D Dirac semimetals, see Ref.~\cite{2012PhRvB..86c5422G,Lu:2014bb}.) 
The corresponding quantum interference effects play an important role in weak magnetic fields and can 
even change the qualitative dependence of the conductivity/resistivity on the magnetic field. This expectation 
is also supported by the analysis of the experimental results \cite{Kim:2013ys}, where the 
signs of weak antilocalization are observed in weak magnetic fields. While the physics behind 
this effect is very interesting, it is not of prime interest for the purposes of our study here. 
Indeed, in the case of moderately strong magnetic fields considered, the effect of the weak
antilocalization is not expected to modify the qualitative behavior of the magnetoresistance. 
(The localization effects in interacting Weyl semimetals in a magnetic field was recently studied in 
Ref.~\cite{Lu:2014bb}.)

The longitudinal conductivity is of special 
interest in Weyl semimetals because, as first suggested in Ref.~\cite{Nielsen:1983rb}, 
it may reveal a unique behavior characteristic for these materials. Using Eq.~(\ref{sigma-ii-AA-chi}), we
find that the longitudinal conductivity is given by 
\begin{eqnarray}
\sigma_{33} &=&
\frac{e^2 v_F^2 }{2^4\pi^3 l^2 T}
\sum_{\chi} \sum_{n=0}^{\infty}
\int \frac{d \omega d k_3 }{\cosh^2\frac{\omega-\mu}{2T}}
\frac{\Gamma_n^2\left[\left(\omega-s_\perp\chi v_F(k_3-\chi b)\right)^2+2n\epsilon_{L}^2+\Gamma_n^2\right]^2}
{\left[\left(\omega -E_n^{(\chi)} \right)^2+\Gamma_n^2\right]^2
\left[ \left(\omega +E_{n}^{(\chi)} \right)^2+\Gamma_{n}^2\right]^2} \nonumber \\
&+&  \frac{e^2 v_F^2 }{2^4\pi^3 l^2 T}
\sum_{\chi} \sum_{n=1}^{\infty}
\int \frac{d \omega d k_3 }{\cosh^2\frac{\omega-\mu}{2T}}
\frac{\Gamma_n^2\left[\left(\omega+s_\perp\chi v_F(k_3-\chi b)\right)^2+2n\epsilon_{L}^2+\Gamma_n^2\right]^2}
{\left[\left(\omega -E_n^{(\chi)} \right)^2+\Gamma_n^2\right]^2
\left[ \left(\omega +E_{n}^{(\chi)} \right)^2+\Gamma_{n}^2\right]^2}\nonumber \\
&-&  \frac{e^2 v_F^2 }{\pi^3 l^2 T}
\sum_{\chi}\sum_{n=1}^{\infty}
\int \frac{d \omega d k_3 }{\cosh^2\frac{\omega-\mu}{2T}}
\frac{\Gamma_n^2 \omega^2  n\epsilon_{L}^2}
{\left[\left(\omega -E_n^{(\chi)} \right)^2+\Gamma_n^2\right]^2
\left[ \left(\omega +E_{n}^{(\chi)} \right)^2+\Gamma_{n}^2\right]^2},
\label{conductivity-longitudinal}
\end{eqnarray}
where $\epsilon_{L} \equiv v_F/l \equiv v_F\sqrt{|eB|/c}$ is the Landau energy scale. 

Before analyzing the complete expression, it is instructive to extract the LLL contribution
 $\sigma_{33}^{\rm (LLL)}$ to the  longitudinal conductivity. It is given by the following exact 
result:
\begin{eqnarray}
\sigma_{33}^{\rm (LLL)} &=&\frac{e^2 v_F^2 }{2^4\pi^3 l^2 T}
\sum_{\chi} 
\int \frac{d \omega d k_3 }{\cosh^2\frac{\omega-\mu}{2T}}
\frac{\Gamma_0^2}{\left[\left(\omega+s_\perp\chi v_F(k_3-\chi b)\right)^2+\Gamma_0^2\right]^2} 
=\frac{e^2 v_F}{4 \pi^2 l^2 \Gamma_0}
=\frac{e^2 v_F |eB|}{4 \pi^2 c \Gamma_0}.
\label{sigma33LLL}
\end{eqnarray}
This is a {\it topological contribution} associated with the chiral anomaly, which is generated entirely 
on the LLL in the presence of a magnetic field \cite{Ambjorn:1983hp}. It is completely independent of the 
temperature and the chemical potential. This result agrees also with the corresponding result obtained by 
using the semiclassical Boltzmann kinetic equation in Refs.~\cite{Nielsen:1983rb,Aji:2011aa,Son:2012bg}. By comparing the
expression in Eq.~(\ref{sigma33LLL}) with those in Refs.~\cite{Nielsen:1983rb,Aji:2011aa,Son:2012bg}, we see that the 
quasiparticle width $\Gamma_0$ is related to the collision time as follows: $\Gamma_0=\hbar/\tau$.

It is interesting that the origin of the topological contribution in Eq.~(\ref{sigma33LLL}) is 
intimately connected with the spatial dimensional reduction $3 \to 1$ in the LLL dynamics \cite{Gusynin:1994xp,Gusynin:1995nb}. 
The dimensional reduction of the LLL states can be made explicit by noting that the 
propagator of the corresponding quasiparticles of given chirality $\chi$ (Weyl node) is given by
\begin{equation}
\bar{G}^{(\chi)}_{LLL}(\omega,\mathbf{k})=ie^{-k^2_{\perp}l^2}\frac{(\omega+\mu)\gamma^0-v_F(k_3-\chi b)\gamma^3}
{(\omega+\mu)^2-v^2_F(k_3-\chi b)^2}\,(1+is_{\perp}\gamma^1\gamma^2).
\label{LLLprop}
\end{equation}
This propagator implies that the LLL modes are characterized by a one-dimensional form of the 
relativistic-like dispersion relation $\omega^{(\chi)}=-\mu \pm v_F(k_3-\chi b)$, which is independent 
of the magnetic field. The final expression for the topological contribution is proportional to the magnetic 
field only because the LLL density of states is determined by the strength of the field.

The remaining higher Landau level (HLL) contribution to the  longitudinal conductivity is given 
by the following expression:
\begin{eqnarray}
\sigma_{33}^{\rm (HLL)} &=&\frac{e^2 v_F^2 }{4\pi^3 l^2 T}
\sum_{n=1}^{\infty} 
\int \frac{d \omega d k_3 }{\cosh^2\frac{\omega-\mu}{2T}}
\frac{\Gamma_n^2\left[
\left(\omega^2+E_n^2+\Gamma_n^2\right)^2- 4n \epsilon_{L}^2\omega^2 \right]}
{\left[\left(\omega -E_n  \right)^2+\Gamma_n^2\right]^2
\left[ \left(\omega +E_{n}  \right)^2+\Gamma_{n}^2\right]^2},
\label{sigma33HLL}
\end{eqnarray}
where $E_n=v_F \sqrt{k_3^2 + 2 n |eB|/c}$. Note that the integration over $k_3$ in the last 
expression can be performed analytically. Moreover, in the limit of zero temperature, the remaining 
integration over $\omega$ can be performed as well. The corresponding explicit results are presented 
in Appendix~B of Ref.~\cite{Gorbar:2013dha}.

\begin{figure}[t]
\begin{center}
\includegraphics[width=.45\textwidth]{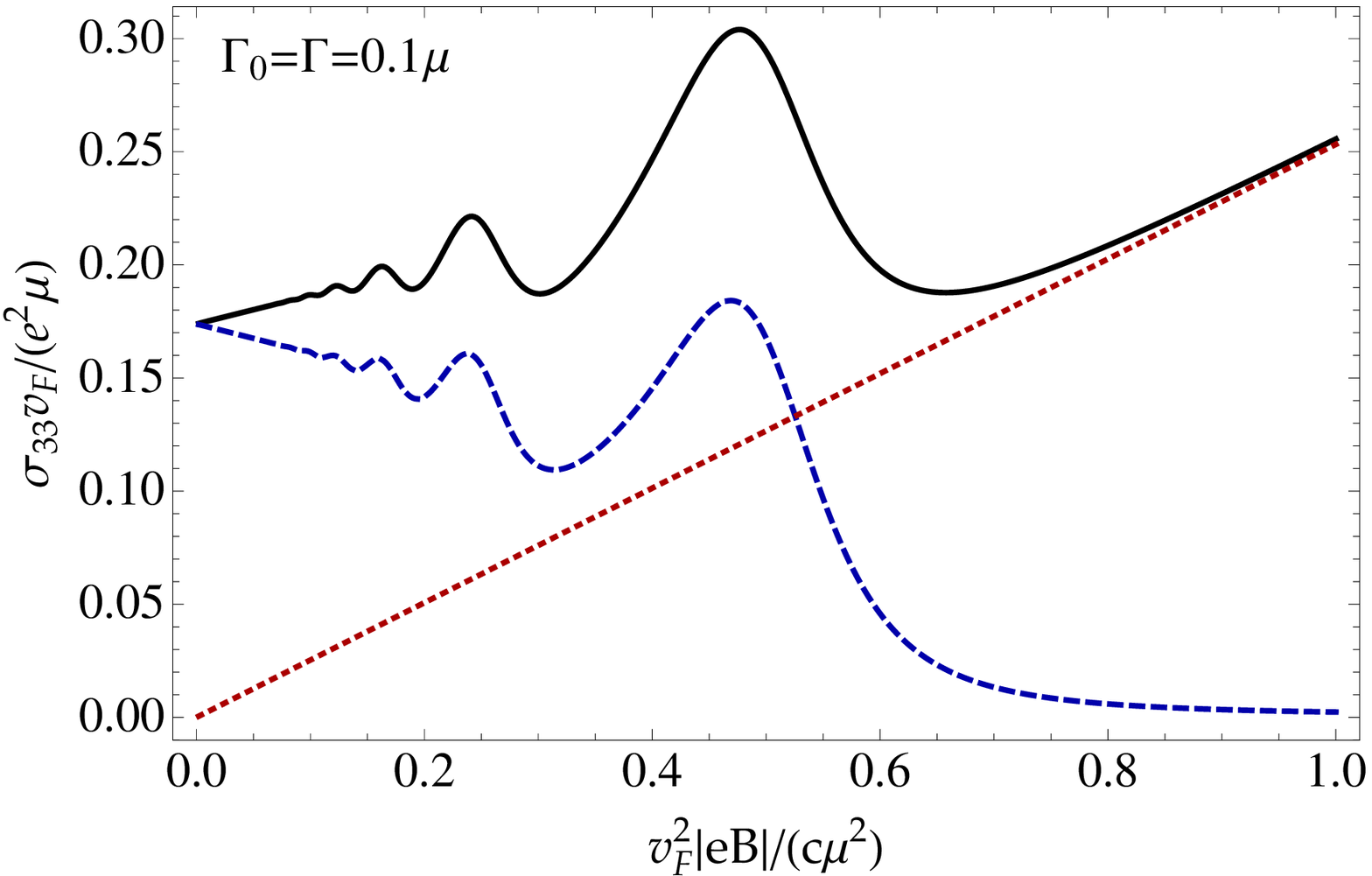}
\hspace{5mm}
\includegraphics[width=.45\textwidth]{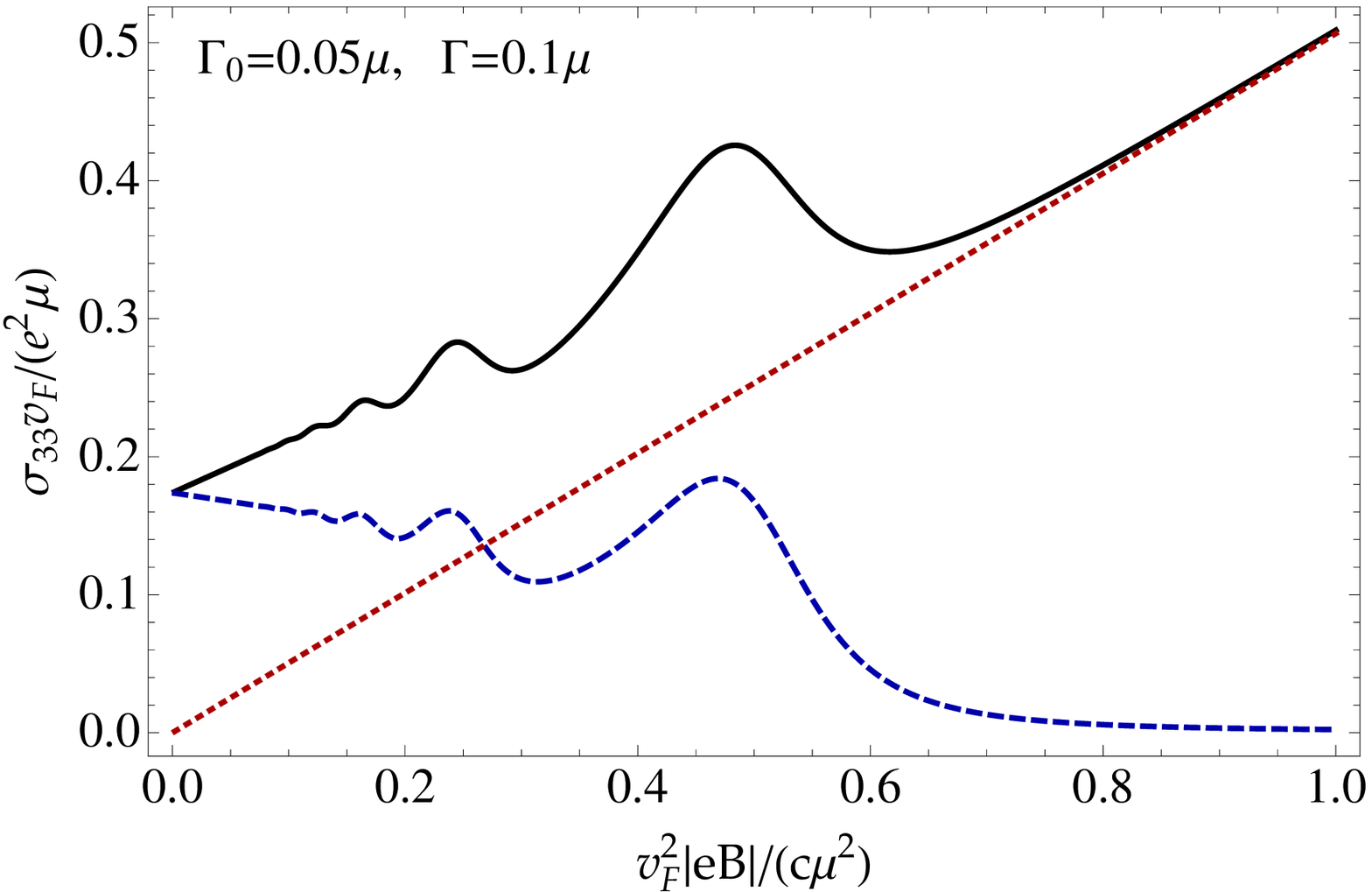}
\caption{(Color online) Longitudinal conductivity $\sigma_{33}$ at zero temperature as a function of the magnetic field. 
The solid line shows the complete result, the dashed line shows the contribution without the lowest 
Landau level, and the dotted line shows the topological contribution of the lowest Landau level alone.
The quasiparticle width in higher Landau levels is $\Gamma=0.1\mu$.
The LLL quasiparticle width is the same or half the width in higher Landau levels.}
\label{fig:Fermi}
\end{center}
\end{figure}

The numerical results for the longitudinal magnetoconductivity as functions of $v^2_F|eB|/\mu^2c$ are 
plotted in Fig.~\ref{fig:Fermi} for two fixed values of the quasiparticle widths in the higher Landau 
levels, i.e., $\Gamma=0.1\mu$ (left panels) and $\Gamma=0.2\mu$ (right panels), and with the two 
possible choices of the LLL quasiparticle width $\Gamma_0$, i.e., the same (upper panels) or two 
times smaller (lower panels) than the width in the higher Landau levels. The LLL contribution is 
shown by the red dotted line, the HLL contribution is shown by the blue dashed line, and the 
complete expression for the longitudinal magnetoconductivity, 
$\sigma_{33}=\sigma_{33}^{\rm (LLL)}+\sigma_{33}^{\rm (HLL)}$, is shown by the black solid line.
Leaving aside the characteristic Shubnikov-de Haas oscillations, we see that the HLL contribution has an 
overall tendency to decrease with increasing the field. In spite of that, the total longitudinal magnetoconductivity, 
which also includes the linearly increasing topological LLL contribution, has the opposite tendency.

\begin{figure}[t]
\begin{center}
\includegraphics[width=.45\textwidth]{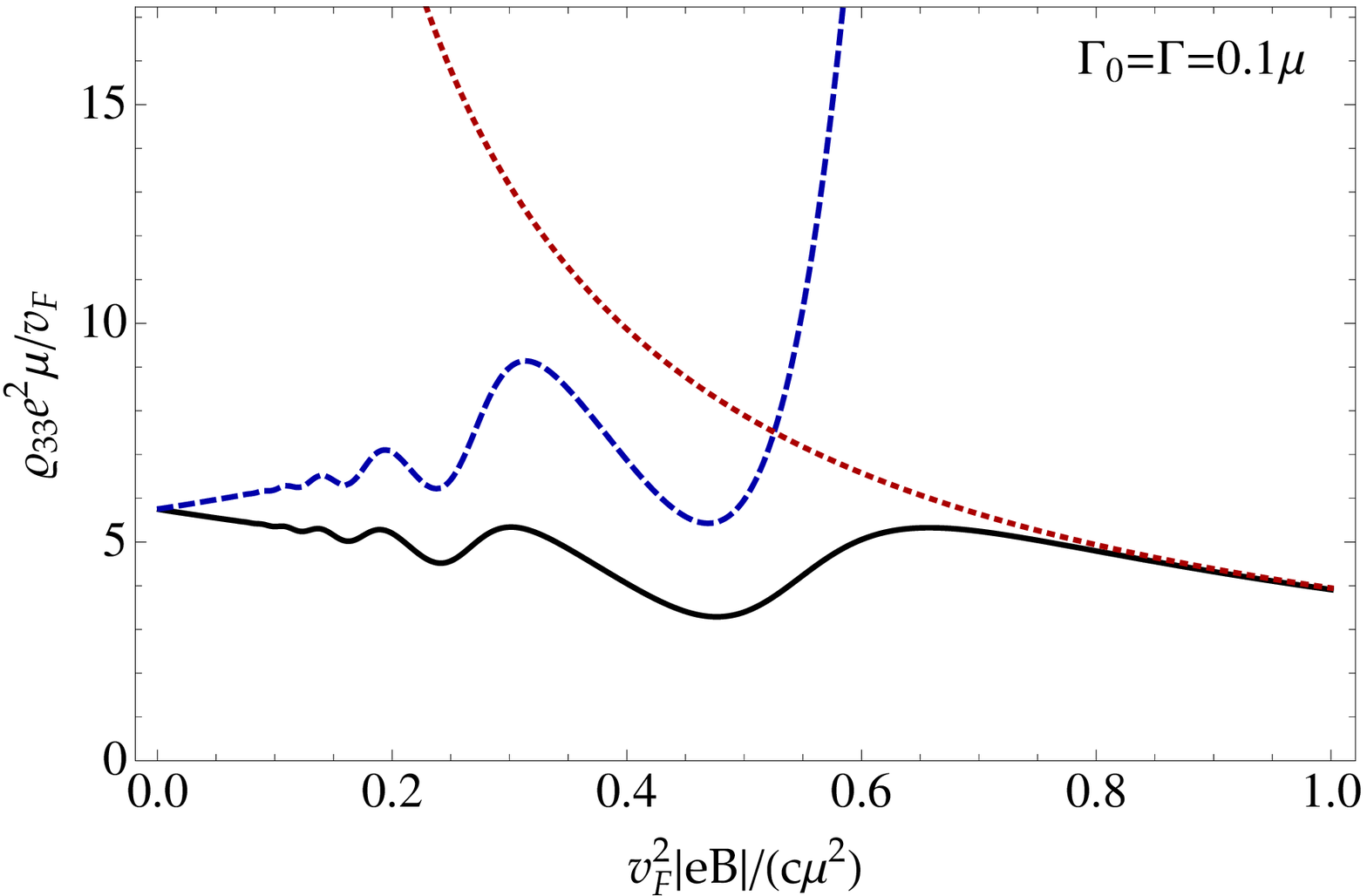}
\hspace{5mm}
\includegraphics[width=.45\textwidth]{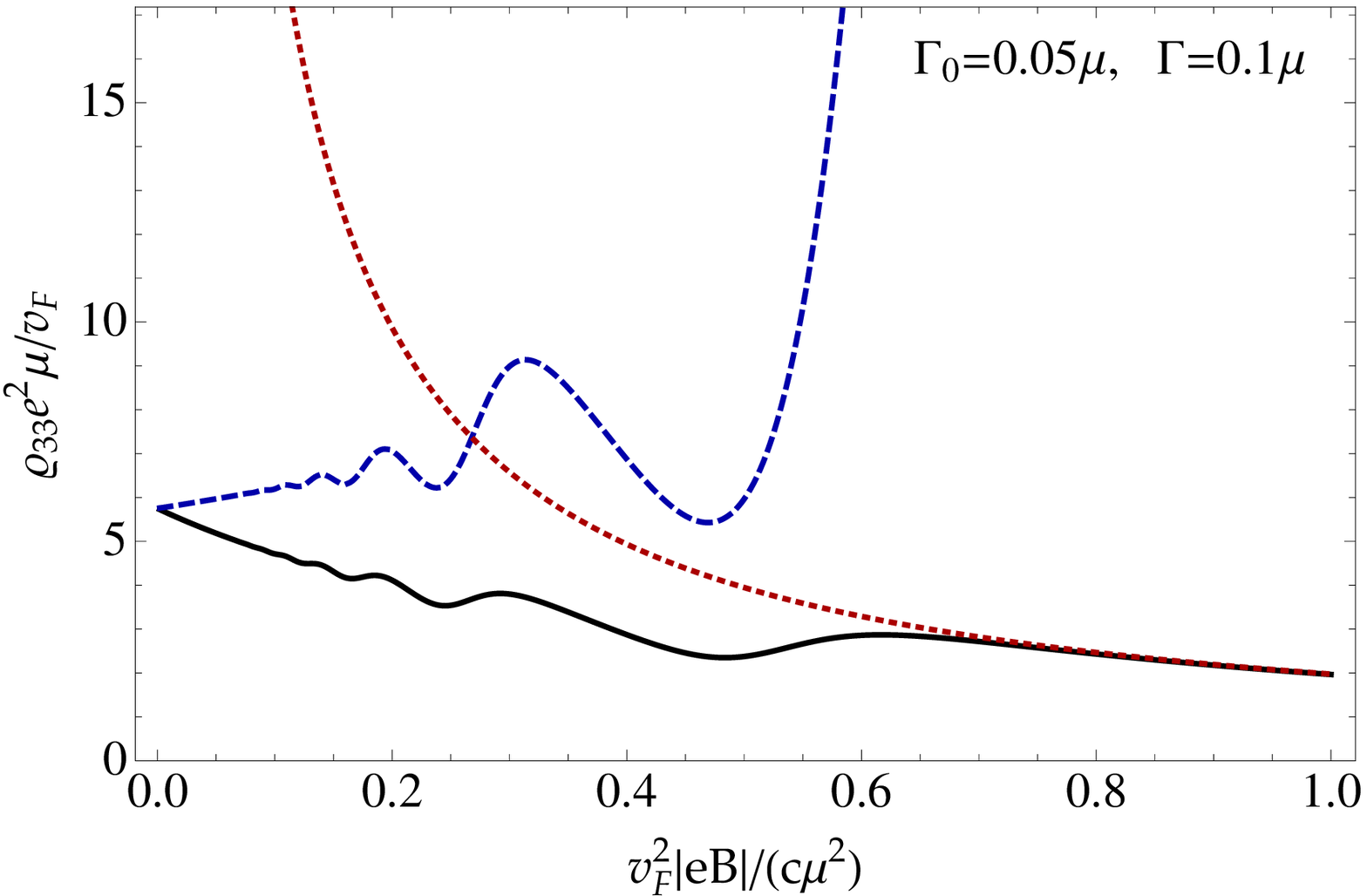}
\caption{(Color online) Longitudinal resistivity $\rho_{33}$ at zero temperature as a function of the magnetic field. 
The solid line shows the complete result, the dashed line shows the contribution without the lowest 
Landau level, and the dotted line shows the topological contribution of the lowest Landau level alone.
The quasiparticle width is $\Gamma=0.1\mu$.
The LLL quasiparticle width is the same or half the width in higher Landau levels.}
\label{fig:longitudinal}
\end{center}
\end{figure}

Taking into account that $\sigma_{13}=\sigma_{31}=\sigma_{23}=\sigma_{32}=0$ and 
using $\sigma_{33}$ calculated above, we also find the longitudinal magnetoresistivity. 
It is given by $\rho_{33}=1/\sigma_{33}$. The corresponding numerical results are plotted in 
Fig.~\ref{fig:longitudinal} as functions of $v^2_F|eB|/\mu^2c$. Oscillations of 
magnetoresistivity connected with the Shubnikov-de Haas effect are clearly seen 
in the left panels in Fig.~\ref{fig:longitudinal}, which show the results for a smaller value 
of the quasiparticle width $\Gamma=0.1 \mu$ in higher Landau levels. The 
oscillations in the case of twice as large width, $\Gamma=0.2 \mu$, are not as well 
pronounced. The longitudinal magnetoresistivity in the case with the LLL
quasiparticle width two times smaller than the width of higher Landau levels is plotted in 
the two lower panels. Overall, the longitudinal 
magnetoresistivity decreases as the magnetic field grows. As we mentioned in the Introduction, 
this phenomenon is known in the literature as negative magnetoresistivity. As is clear from our results 
in Fig.~\ref{fig:longitudinal}, the negative longitudinal magnetoresistivity is exclusively due to 
the LLL contribution\cite{Nielsen:1983rb} which in turn is connected with the chiral anomaly \cite{Adler:1969gk,Bell:1969ts}.

We would like to emphasize that we did not assume in our calculations that $\Gamma_0$ 
is much less than the quasiparticle width in higher Landau levels. This assumption was made in 
semiclassical calculations in Refs.~\cite{Nielsen:1983rb,Aji:2011aa,Son:2012bg} due to the fact that the 
quasiparticle width $\Gamma_0$ in the LLL is not equal to zero only because of the internode 
scatterings. This is unlike the quasiparticle width in higher Landau levels where intranode scatterings 
contribute too. Since Weyl nodes are separated by the distance $2b$ in momentum 
space in Weyl semimetals, internode scattering processes are less efficient compared to intranode 
ones. Therefore, it is usually assumed that $\Gamma_0$ is much less than $\Gamma_n$ in higher 
Landau levels $n \ge 1$. Although we did not make this assumption, we still observe the negative 
longitudinal magnetoresistivity. It is also important to emphasize another point. After the change 
of the integration variable $k_3  \to  k_{\rm new}^3\equiv k_3-\chi b$, the chiral shift $b$ does not 
enter in the longitudinal magnetoconductivity (\ref{conductivity-longitudinal}) and affects the result 
only indirectly through the quasiparticle width \cite{Nielsen:1983rb}. Since our results show that the 
negative longitudinal magnetoresistivity takes place even when the LLL quasiparticle width 
$\Gamma_0$ is comparable to the width $\Gamma_n$ in the higher Landau levels, we conclude that 
this phenomenon is quite robust and will also take place in Dirac semimetals as well.

Now, let us proceed to the calculation of the diagonal $\sigma_{11}=\sigma_{22}$ components 
of the transverse conductivity. From Eq.~(\ref{sigma-ii-AA-chi}), one obtains the following
result:
\begin{eqnarray}
\sigma_{11} &=&   \frac{e^2 v_F^2 }{4\pi^3 l^2 T}
\sum_{n=0}^{\infty}
\int \frac{d \omega d k_3 }{\cosh^2\frac{\omega-\mu}{2T}}
 \frac{\Gamma_{n+1} \Gamma_{n}
 \left[ \left(\omega^2+E_{n}^2+\Gamma_{n}^2 \right) \left(\omega^2+E_{n+1}^2+\Gamma_{n+1}^2 \right)-4(v_F k_3)^2\omega^2\right]}
{\left[\left(E_{n}^2+\Gamma_{n}^2 -\omega^2\right)^2+4\omega^2 \Gamma_{n}^2 \right]
\left[\left(E_{n+1}^2+\Gamma_{n+1}^2 -\omega^2\right)^2+4\omega^2 \Gamma_{n+1}^2 \right]}.  
\label{sigma11_text}
\end{eqnarray}
In the limit of zero temperature, we can easily integrate over $\omega$ and $k_3$. The corresponding 
analytical result is presented in Appendix~B of Ref.~\cite{Gorbar:2013dha}.

The numerical results for the transverse diagonal conductivity $\sigma_{11}$ 
as a function of $v^2_F|eB|/(\mu^2c)$ are shown in the left panel of Fig.~\ref{fig:sigma11-12} 
for three different values of the quasiparticle 
width. Just as in the case of longitudinal conductivity, the Shubnikov-de Haas oscillations are clearly seen for smaller 
values of the width, but gradually disappear when the width becomes larger. In all cases, however, the transverse 
diagonal conductivity has an overall tendency to decrease with increasing the field. 

\begin{figure}[t]
\begin{center}
\includegraphics[width=.45\textwidth]{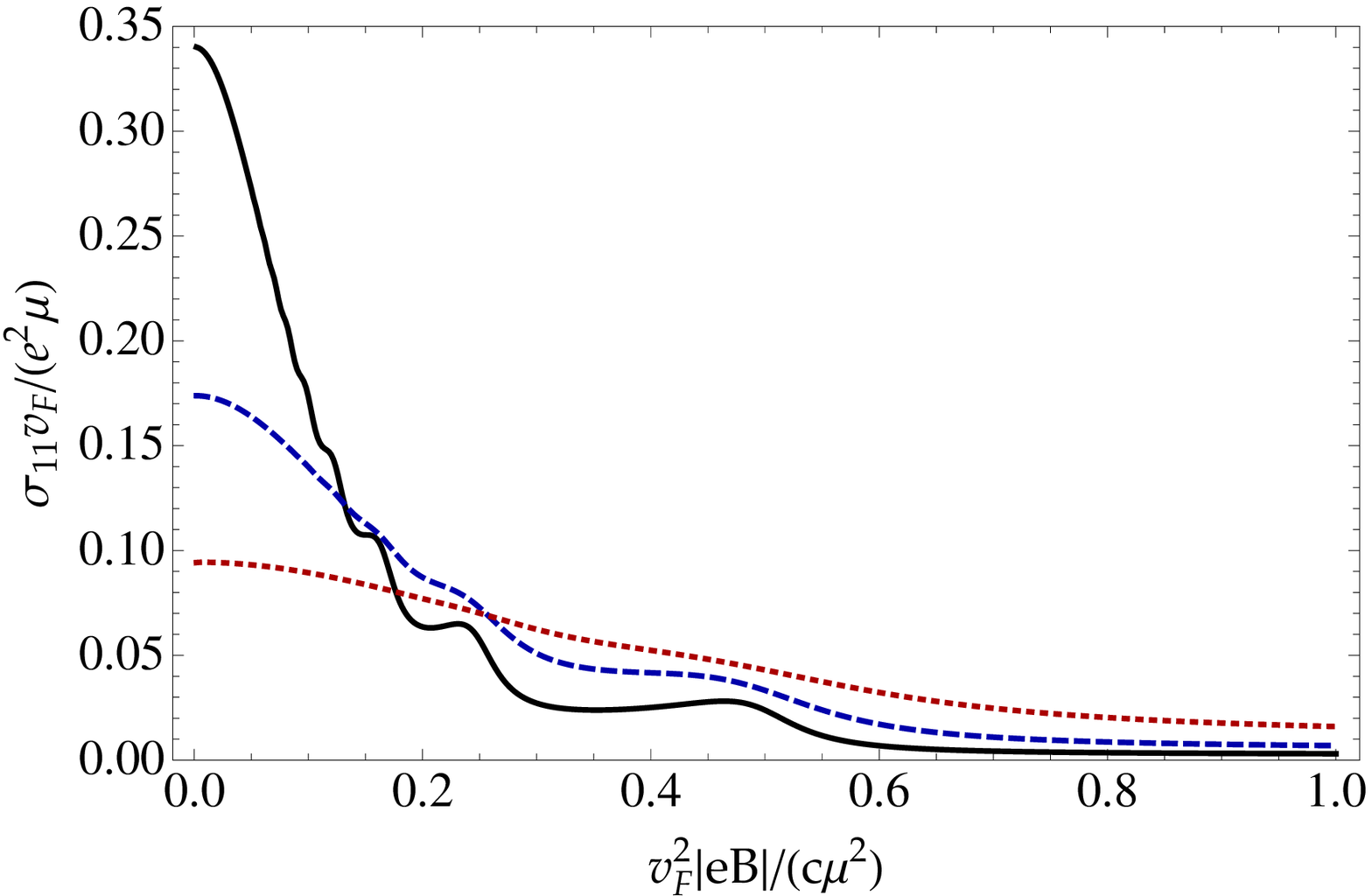}
\hspace{5mm}
\includegraphics[width=.45\textwidth]{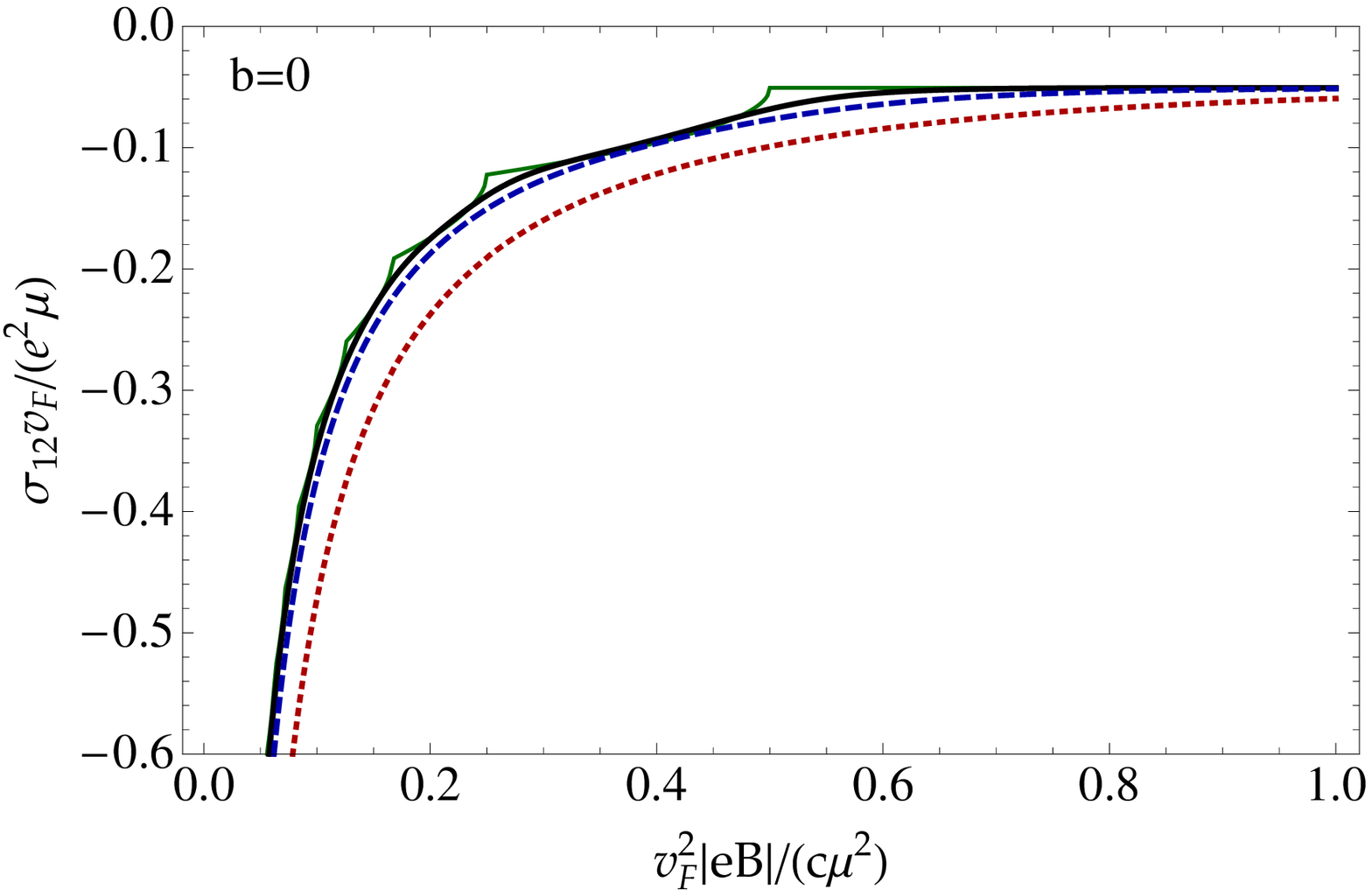}
\caption{(Color online) 
Left panel: Diagonal components of the transverse conductivity $\sigma_{11}=\sigma_{22}$ 
at zero temperature as a function of the magnetic field. The quasiparticle width is 
$\Gamma=0.05\mu$ (black solid line), $\Gamma=0.1\mu$ (blue dashed line), and 
$\Gamma=0.2\mu$ (red dotted line). The sum over Landau levels includes 
$n_{\rm max} =10^{4}$ levels.
Right panel: Off-diagonal components of the transverse conductivity $\sigma_{12}=-\sigma_{21}$ 
as a function of the magnetic field for the vanishing chiral shift, $\mathbf{b}=0$. The results are shown 
for $\Gamma = T=0$ (green thin solid line), $\Gamma \to T=0.05 \mu$ (black solid line), 
$\Gamma \to T=0.1\mu$ (blue dashed line), and $\Gamma \to T=0.2\mu$ (red dotted line). 
If $b \ne 0$, the conductivity will simply shift by $e^2 b/(2\pi^2)$.}
\label{fig:sigma11-12}
\end{center}
\end{figure}

In order to calculate the off-diagonal components of the transverse conductivity, we use Eq.~(\ref{sigma-ij-AA-chi}).
Let us start from the simplest case when $\Gamma_n\to 0$. In this limit, the Lorentzian in the spectral function (\ref{sp-funct-Gamma}) is replaced by a $\delta$ function and the analysis greatly simplifies. The corresponding result reads as
\begin{eqnarray}
\sigma_{12} &=& - \frac{e^2 v_F^2 s_\perp}{4\pi^2 l^2} 
\sum_{\lambda,\lambda^\prime=\pm}
\sum_{n}
\int d k_3   
\frac{n_F\left(\lambda^\prime E_{n}\right)-n_F\left(\lambda^\prime E_{n+1}\right)}{\left(E_{n} -\lambda E_{n+1}\right)^2}
\left(1- \frac{\lambda \left(v_F k_{3}\right)^2 }{ E_{n+1} E_{n}} \right)\nonumber\\
&&+\frac{e^2 v_F^2}{8\pi^2 l^2} \sum_{\chi=\pm} \sum_{\lambda,\lambda^\prime=\pm}
\sum_{n,n^\prime}
\int d k_3 \frac{n_F\left(\lambda E_{n}^{(\chi)} \right)}{ E_{n}^{(\chi)} E_{n^\prime}^{(\chi)}}\frac{\chi v_F(k_3-\chi b)}
{ \lambda^\prime E_{n}^{(\chi)} - \lambda E_{n^\prime}^{(\chi)} }\left(\delta_{n-1,n^\prime}+\delta_{n,n^\prime-1}\right)
\nonumber\\
&= &- \frac{e^2 s_\perp}{4\pi^2} 
\sum_{n}\alpha_{n}\int d k_3 \frac{ \sinh\frac{\mu}{T}}{\cosh\frac{E_n}{T}+\cosh\frac{\mu}{T}}
-\frac{e^2}{8\pi^2} \sum_{\chi=\pm}\chi\int dk_3\frac{\sinh\frac{v_F(k_3-\chi b)}{T}}{\cosh\frac{v_F(k_3-\chi b)}{T}+\cosh\frac{\mu}{T}},
\label{sigma12-T}
\end{eqnarray}
where $\alpha_{n}=2-\delta_{n,0}$ is the spin degeneracy of the Landau levels. The first 
term in the last line is associated with a nonzero density of charge carriers. It comes from the occupied 
Landau levels and, as expected, depends on the temperature, chemical potential, and magnetic 
field. In contrast, the last term in Eq.~(\ref{sigma12-T}) is a topological vacuum contribution (which 
is present even at $\mu=0$) and comes exclusively from the lowest Landau level. Such a contribution
is a specific feature of Weyl semimetals and is directly related to the 
anomalous Hall effect \cite{Haldane:1988zza}, which is produced by the dynamical Chern-Simons term in 
Weyl semimetals \cite{PhysRevLett.107.127205,Grushin:2012ac,Zyuzin:2012ab,Vazifeh:2013fk,Goswami:2012aa}. This topological (anomalous) 
contribution is independent of the temperature, chemical potential, and magnetic field and equals
\begin{eqnarray}
\sigma_{12,{\rm anom}} &=& 
-\frac{e^2}{8\pi^2 v_F} T \left.
\ln\frac{\cosh\frac{v_F(k_3-b)}{T}+\cosh\frac{\mu}{T} }{\cosh\frac{v_F(k_3+ b)}{T}+\cosh\frac{\mu}{T} }
\right|_{k_3=-\infty}^{k_3=\infty}
 =\frac{e^2 b}{2\pi^2}.
\label{anomaly-contribution}
\end{eqnarray}
As usual in calculations of anomalous quantities, the integral form of the topological contribution 
in the last term in Eq.~(\ref{sigma12-T}) should be treated with care. Indeed, while 
separate left- and right-handed contributions appear to be poorly defined because of a 
linear divergency, the sum of both chiralities results in a convergent integral. 

It should be noted that there is no interference between the topological contribution and the 
remaining contribution due to the finite density of charge carriers. We should also emphasize that the anomalous
contribution (\ref{anomaly-contribution}) will be present even in Dirac semimetals in a magnetic 
field because, as we discussed in the Introduction, $\mathbf{b} \ne 0$ is generated in Dirac 
semimetals by the Zeeman interaction or dynamically \cite{Gorbar:2013qsa}. The anomalous contribution 
(\ref{anomaly-contribution}) unambiguously distinguishes a Weyl semimetal from a Dirac one only in the absence 
of a magnetic field. In such a case, nonzero $\mathbf{b}$ breaks time reversal symmetry in Weyl 
semimetals and provides finite $\sigma_{12}$ unlike the case of Dirac semimetals where $\mathbf{b}$ 
is absent and, therefore, time reversal symmetry is preserved and $\sigma_{12}$ vanishes. 

In the limit of zero temperature, the complete expression for the off-diagonal conductivity is given by the 
following analytical expression:
\begin{eqnarray}
\sigma_{12} &=&\frac{e^2 b}{2\pi^2}
- \frac{e^2 s_\perp \mbox{sgn}(\mu)}{4\pi^2}  \sum_{n} \alpha_{n} \int dk_3  
 \theta\left(|\mu|-|E_{n}|\right)
 =\frac{e^2 b}{2\pi^2}
 - \frac{e^2 s_\perp \mbox{sgn}(\mu)}{2\pi^2 v_F}  \sum_{n=0}^{n_{\rm max}} \alpha_{n}\sqrt{\mu^2-2nv_F^2|eB|/c},
\label{sigma12}
\end{eqnarray}
where $n_{\rm max}$ is given by the integer part of $\mu^2/(2\epsilon_L^2)$ and has the meaning of
the Landau level index in the highest occupied Landau level. The off-diagonal component of the 
conductivity is plotted in the right panel of Fig.~\ref{fig:sigma11-12} (green thin solid line).

It may be appropriate to note here that the expression for the off-diagonal component of the
conductivity in the case of quasiparticles with nonzero widths, modeled by the Lorentzian 
distribution in Eq.~(\ref{sp-funct-Gamma}), is not as convenient or even useful as the above expression. 
In fact, unlike the similar expressions for the diagonal components of the conductivity, 
off-diagonal component $\sigma_{12}$ contains a formally divergent sum over the Landau levels
when $\Gamma_n\neq 0$. This can be checked by first explicitly calculating the integrals 
over the energy and the longitudinal momentum, and then examining the contributions of 
the Landau levels with large values of Landau index $n$. The corresponding contributions 
are suppressed only as $1/\sqrt{n}$ when $n\to\infty$ and, therefore, cause a divergence in the 
sum. From the physics viewpoint, the origin of the problem is rooted in the use of the simplest Lorentzian 
model (\ref{sp-funct-Gamma}) for the quasiparticle spectral function with nonzero quasiparticle widths. 
The corresponding distribution falls off too slowly as a function of the energy. As a result, the 
Landau levels with very large $n$, which are completely empty and should not have much 
of an effect on the conductivity, appear to give small individual contributions (suppressed 
only as $1/\sqrt{n}$) that collectively cause a divergence.  

In order to illustrate the problem in the simplest possible mathematical form, we can mimic 
the result of the integration by the following approximate form:
\begin{eqnarray}
\sigma_{12} &\simeq & - \frac{e^2 s_\perp}{4\pi^3} \sum_{n}\alpha_n \int d k_3
\left[\arctan\frac{E_n+\mu}{\Gamma}-\arctan\frac{E_n-\mu}{\Gamma}\right]\nonumber\\
&= &- \frac{e^2 s_\perp}{\sqrt{2}  \pi^2 v_F}  \sum_{n}\alpha_n
\frac{\Gamma\mu}{\sqrt{2n\epsilon_{L}^2+\Gamma^2-\mu^2+\sqrt{(2n\epsilon_{L}^2+\Gamma^2-\mu^2)^2+4\Gamma^2\mu^2}}},
\end{eqnarray}
which correctly captures the zero quasiparticle width approximation on the one hand and 
shares the same problems as the exact result obtained from the expression in the model with
the Lorentzian quasiparticle widths. 

Ideally, in order to better incorporate the effects of finite widths of quasiparticles in the 
calculation of the off-diagonal component of the conductivity, one has to use a more 
realistic model for the spectral function. A simple model approach to incorporate 
the effects of finite widths of quasiparticles that we will follow here is suggested by the 
finite-temperature expression in Eq.~(\ref{sigma12-T}). We will assume that a nonzero 
but small width $\Gamma$ could be mimicked by the effects of a small temperature 
$T\simeq \Gamma$. Then, by making use of the expression in Eq.~(\ref{sigma12-T}) with the
corresponding replacement, we can roughly estimate the effect of a small nonzero width. The 
corresponding numerical results for $\Gamma\to T=0.05\mu$, $\Gamma\to T=0.1\mu$, and 
$\Gamma\to T=0.2\mu$ are shown in the right panel of Fig.~\ref{fig:sigma11-12} as the solid black line, the blue 
dashed line, and the red dotted line, respectively.

By making use of the transverse conductivity, we calculate all remaining nonzero 
components of the resistivity tensor, i.e.,
\begin{eqnarray}
\rho_{11} &=& \rho_{22} =\frac{\sigma_{11}}{\sigma_{11}^2+\sigma_{12}^2},\\
\rho_{12} &=&-\rho_{21}=-\frac{\sigma_{12}}{\sigma_{11}^2+\sigma_{12}^2}.
\end{eqnarray}
Using the conductivity results at zero temperature, we calculate $\rho_{11}$ and 
$\rho_{12}$ numerically. The corresponding diagonal and off-diagonal components of 
resistivity are shown as functions of $v^2_F|eB|/(\mu^2c)$ in Fig.~\ref{fig:resistivity} 
for $b=0$ (upper panels) and $b=0.3 \mu$ (lower panels).

\begin{figure}[t]
\begin{center}
\includegraphics[width=.45\textwidth]{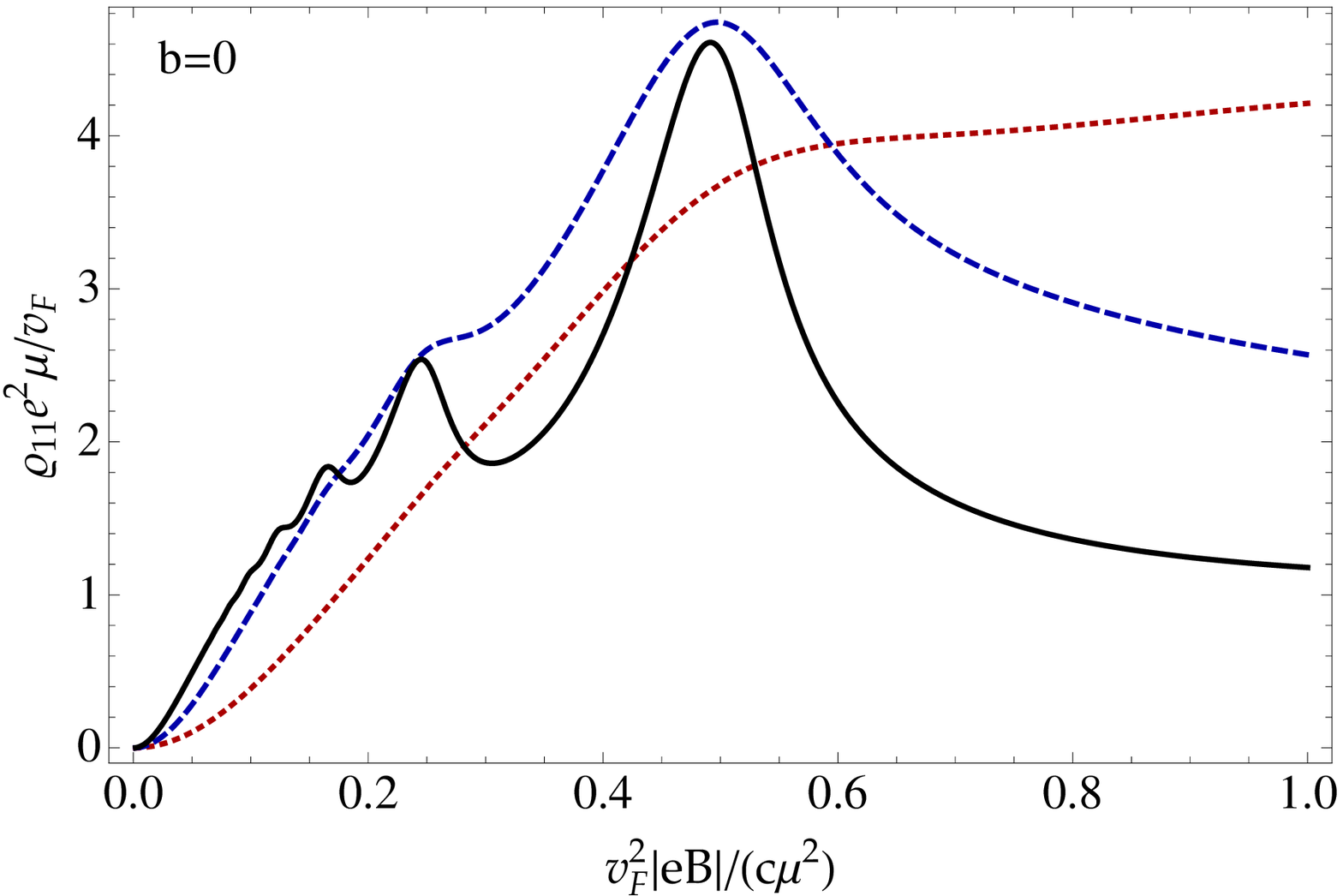}
\hspace{.05\textwidth}
\includegraphics[width=.45\textwidth]{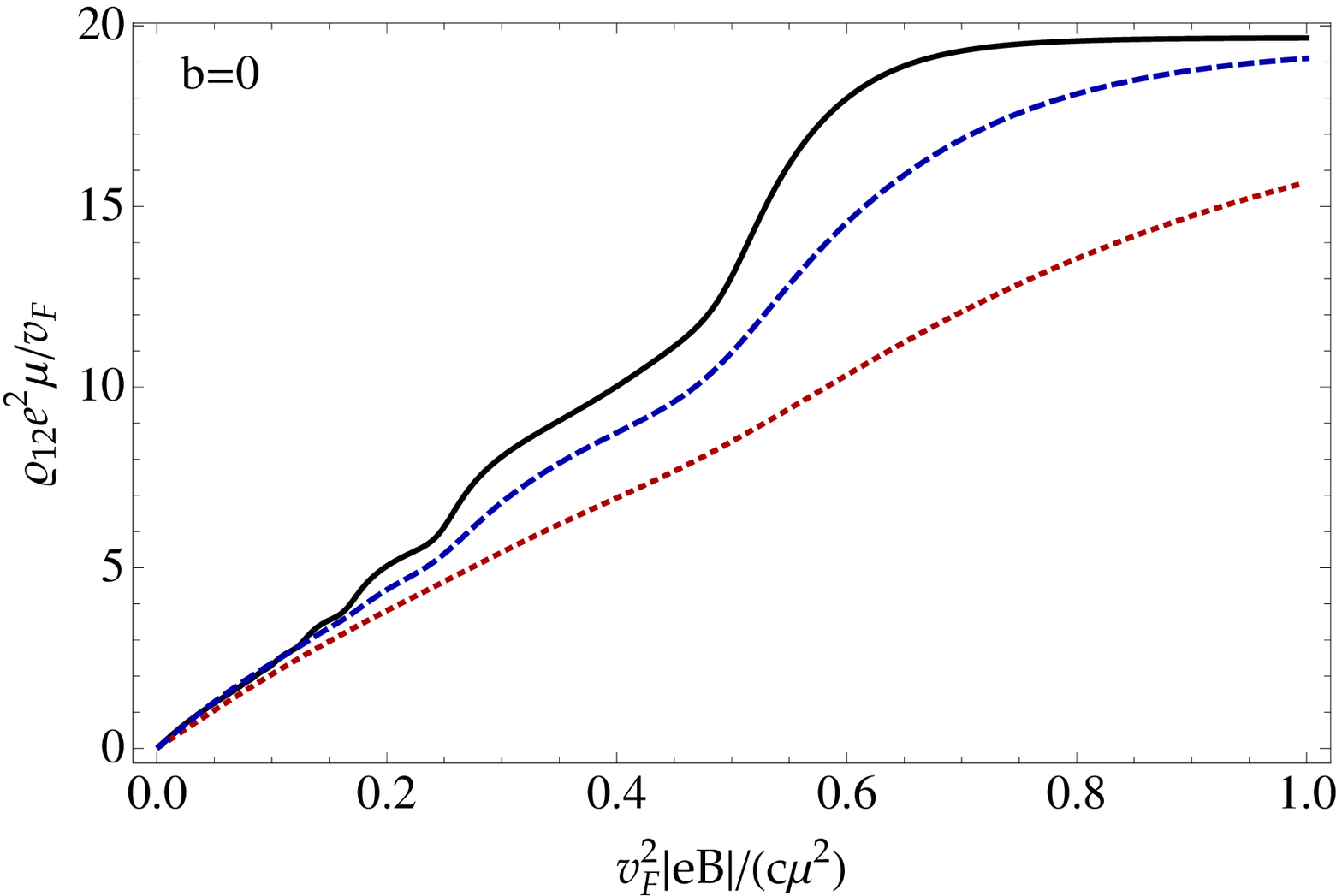}\\
\includegraphics[width=.45\textwidth]{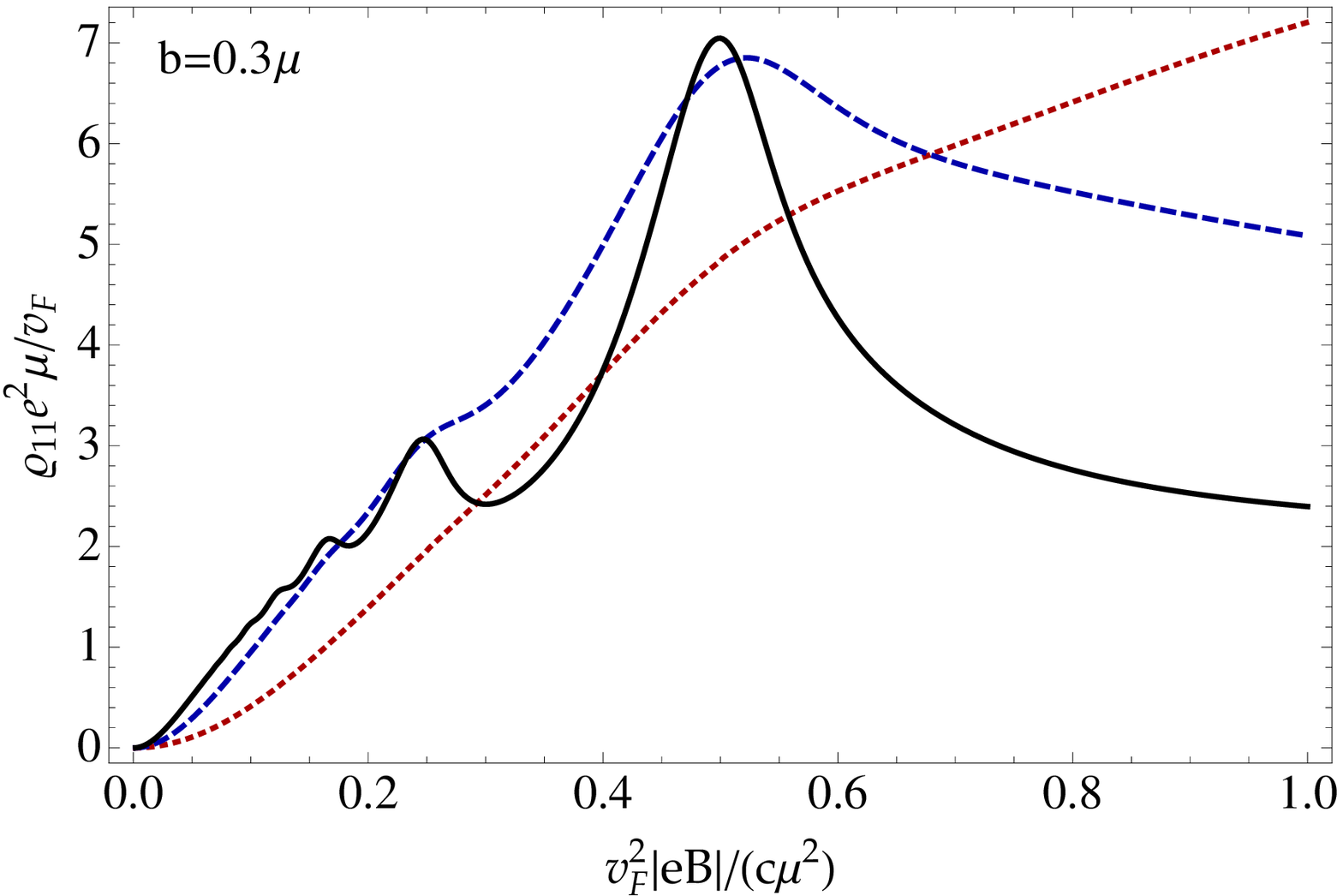}
\hspace{.05\textwidth}
\includegraphics[width=.45\textwidth]{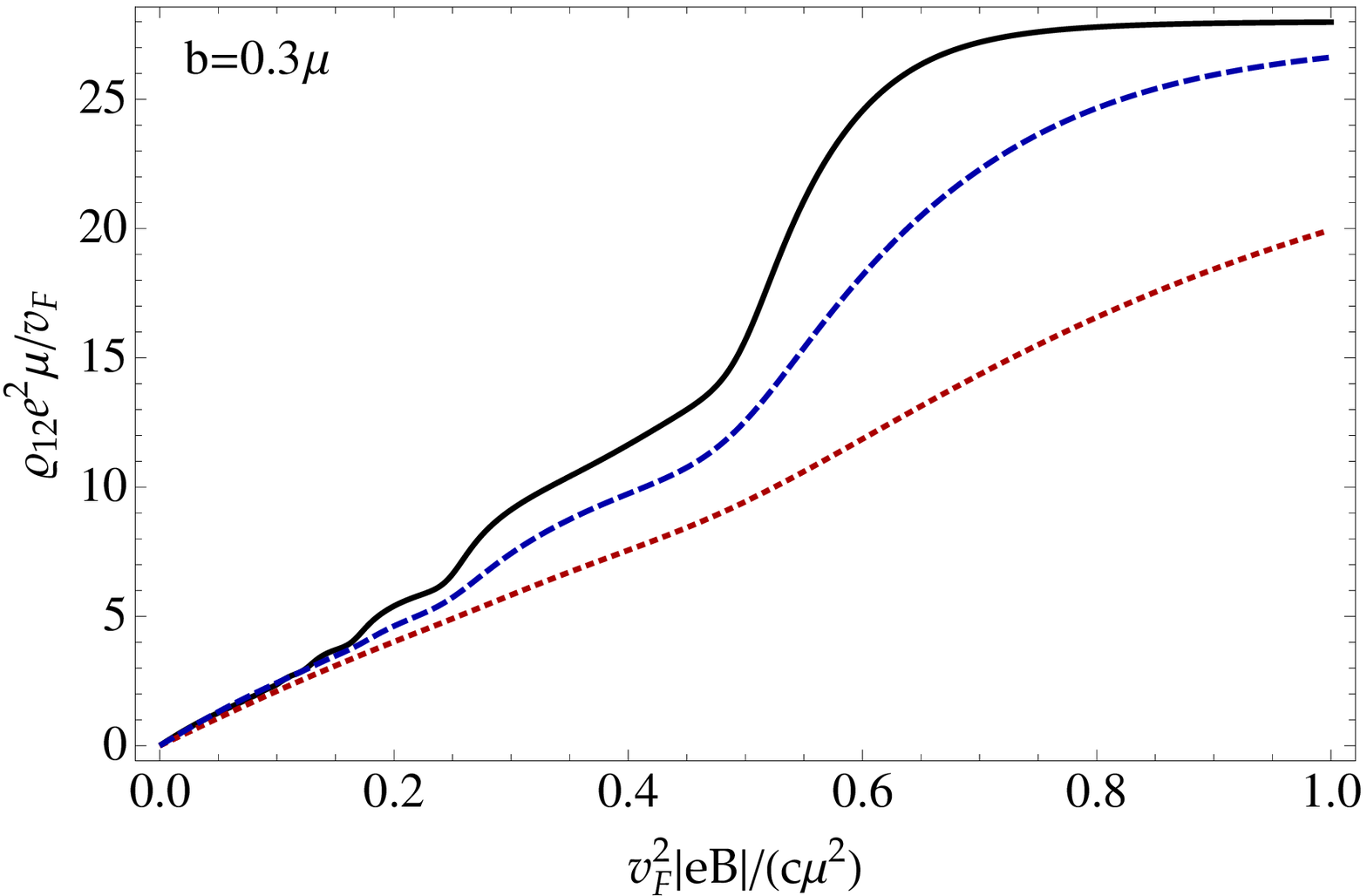}\\
\caption{(Color online) Transverse components of resistivity $\rho_{11}$ and $\rho_{12}$ at zero temperature 
as functions of the magnetic field for $b=0$ (upper panels) and $b=0.3 \mu$ (lower panels).
The quasiparticle width is $\Gamma=0.05\mu$ (black solid line), 
$\Gamma=0.1\mu$ (blue dashed line), and $\Gamma=0.2\mu$ (red dotted line). The sum over 
Landau levels includes $n_{\rm max} =10^{4}$ levels.}
\label{fig:resistivity}
\end{center}
\end{figure}

As our calculations show, the negative magnetoresistivity in the longitudinal 
conductivity occurs solely due to the lowest Landau level.  This contribution has a 
topological origin and is associated with the chiral anomaly. It is also intimately
connected with the dimensional reduction $3 \to 1$ in the dynamics of the LLL
in three-dimensional relativistic-like systems. While the dispersion relation of the 
LLL quasiparticles is independent of the magnetic field, the longitudinal conductivity 
$\sigma_{33}$ grows linearly with the magnetic field because it is proportional to the 
LLL density of states, i.e., $\propto |eB|$. In essence, this growth is the main mechanism 
behind the negative longitudinal magnetoresistivity.

The present results qualitatively agree with the quasiclassical results obtained in 
Refs.~\cite{Nielsen:1983rb,Aji:2011aa,Son:2012bg} using the Boltzmann equation. In general, however,
the quasiclassical results are not sufficient because the quantum corrections due to higher 
Landau levels are quantitatively important in the complete result, especially in the regime 
of moderately strong magnetic fields when a few Landau levels are occupied.

We found that the longitudinal conductivity does not explicitly depend on the value of the shift 
$\mathbf{b}$ between the Weyl nodes. A potential indirect dependence may enter, however,
through the corresponding dependence of the widths of quasiparticles \cite{Nielsen:1983rb,Aji:2011aa,Son:2012bg}. 
This is in contrast to the transverse transport which does reveal an explicit dependence on the 
chiral shift $\mathbf{b}$. Specifically, the off-diagonal transverse component of conductivity 
$\sigma_{12}$ has an anomalous contribution directly proportional to the chiral shift, but 
independent of the temperature, chemical potential and magnetic field. From our analysis,
we see that this anomalous part of conductivity is determined exclusively by the LLL 
quasiparticles. It is also interesting to point out that this contribution has exactly the 
same form as in a Weyl semimetal (with an intrinsic $\mathbf{b}\neq0$) without an external 
magnetic field. It is manifested via the anomalous part of the electric current $\mathbf{j}_{\rm anom}
=e^2/(2\pi^2)\mathbf{b}\times\mathbf{E}$ which is perpendicular to the applied 
electric field \cite{PhysRevLett.107.127205,Grushin:2012ac,Zyuzin:2012ab,Vazifeh:2013fk,Goswami:2012aa}.

In both Dirac and Weyl semimetals, the chiral shift $\mathbf{b}$ should receive
dynamical corrections proportional to the magnetic field. It would be very interesting to observe 
such corrections experimentally. This is not easy when Landau levels are partially filled and 
an ordinary Hall effect, associated with a nonzero density of charge carriers, is superimposed
over the anomalous Hall conductivity. However, as was demonstrated in Ref.~\cite{Liu:2013uq}
in the case of Na$_{3}$Bi, such a problem can be circumvented by tuning the chemical 
potential to the Dirac or Weyl points and, thus, eliminating the contribution due to the ordinary Hall 
effect. This can be done by using surface $K$-doping \cite{Liu:2013uq}. If this works, it may also allow us to 
observe the dependence of the chiral shift $\mathbf{b}$ on the magnetic field through 
the measurements of the off-diagonal transverse conductivity.

It is also interesting to mention that an experimental observation of a transition 
from a Dirac to Weyl semimetal driven by a magnetic field has been recently reported in 
Ref.~\cite{Kim:2013ys} (see also Refs.~\cite{Liang:2014aa,Li:2014aax} where
similar measurements were done for other materials, the interpretation of the results 
differs). By applying moderately strong magnetic fields to the Bi$_{1-x}$Sb$_x$ 
alloy with the antimony concentration of about $x \approx 0.03$ 
(i.e., the regime of a massless Dirac semimetal), the authors observed negative
longitudinal magnetoresistivity and interpreted it as an unambiguous signature of 
the anomaly contribution [see Eq.~(\ref{sigma33LLL})]. As our current study indicates, 
such an observation is indeed the consequence of the anomaly, but not necessarily 
of a Weyl semimetal. In fact, the only direct indication of the Weyl nature of a semimetal
is present in the off-diagonal component of the transverse conductivity $\sigma_{12}$ 
[see Eq.~(\ref{anomaly-contribution})]. Extracting such a contribution from the experimental 
data may be quite challenging, however, because the value of the chiral shift $b$ itself 
is expected to depend on the magnetic field and the density of charge carriers \cite{Gorbar:2013qsa}.

\subsubsection{Fermi Arcs}
\label{sec:FermiArcsDiracSemimetals}

As argued in Refs.~\cite{PhysRevB.83.205101,Haldane:2014aa,Okugawa:2014aa}, the topological nature of Weyl nodes 
should lead to the existence of the surface Fermi arc states that connect Weyl nodes of 
opposite chirality. These surface states are topologically protected and are well defined 
at momenta away from the Weyl nodes because there are no bulk states with the same 
energy and momenta. Taken together the two Fermi arc states on opposite surfaces form 
a closed Fermi surface.

As we discussed in Section~\ref{sec:ChiralShiftDiracSemimetals}, interaction effects lead to dramatic consequences in a 
Dirac semimetal at nonzero charge density in a magnetic field, making it a Weyl semimetal 
with a pair of Weyl nodes for each of the original Dirac points. The Weyl nodes are separated 
in momentum space by a dynamically generated chiral shift $\mathbf{b}$ directed along the 
magnetic field. The magnitude of the momentum space separation between the Weyl nodes 
is determined by the quasiparticle charge density, the strength of the magnetic field,
and the strength of the interaction. 

A remarkable new way to determine experimentally the momentum space separation of 
Weyl nodes was suggested in Ref.~\cite{Potter:2014aa}. Although the surface states of Weyl 
semimetals consist of disjoint Fermi arcs, it was shown that there exist closed magnetic orbits 
involving the surface Fermi arcs. These orbits produce periodic quantum oscillations of the 
density of states in a magnetic field. If observed, this unconventional Fermiology of surface 
states would provide a clear fingerprint of the Weyl semimetal phase. Since, according to
Ref.~\cite{Gorbar:2013qsa}, the interaction effects change the separation of Weyl nodes
in momentum space, they also affect the quantum oscillations in Weyl semimetals \cite{Gorbar:2014qta}.

Let us here briefly review the influence of interactions on the oscillation of the density of states associated
with the Fermi arc modes in the Weyl semimetals. We will start by first reproducing the results obtained
in Ref.~\cite{Potter:2014aa} and then study the influence of interactions on the period of oscillations. In the
{\it laboratory frame} of reference, see Fig.~\ref{fig:ref-frames}, the Weyl semimetal slab is characterized by the
Weyl nodes located at $\pm\mathbf{b}_{0}=\pm(0, b_{0,y}, b_{0,z})$, where $b_{0,y}$ ($b_{0,z}$) is the component of the
chiral shift perpendicular (parallel) to the slab. As the explicit analysis in Ref.~\cite{Gorbar:2014qta} reveals,
the quasiparticle velocity of the Fermi arc modes is $\mathbf{v}_{\rm b}=(v_F, 0, 0)$
on the bottom surface ($y=0$) and is $\mathbf{v}_{\rm t}=(-v_F, 0, 0)$ on the top surface ($y=L$).
The direction of the magnetic field $\mathbf{B}=B\, \hat{\mathbf{n}}$ is specified by the unit
vector $\hat{\mathbf{n}} = (\sin\theta\sin\varphi,\cos\theta,\sin\theta\cos\varphi)$, where $\theta$ is
the angle between the $y_{\rm lab}$ axis and the direction of the magnetic field, and $\varphi$ is the angle between
the $z_{\rm lab}$ axis and the projection of $\mathbf{B}$ onto the $z_{\rm lab}x_{\rm lab}$ plane,
see the left panel in Fig.~\ref{fig:ref-frames}.

\begin{figure}[t]
\begin{center}
\includegraphics[width=0.45\textwidth]{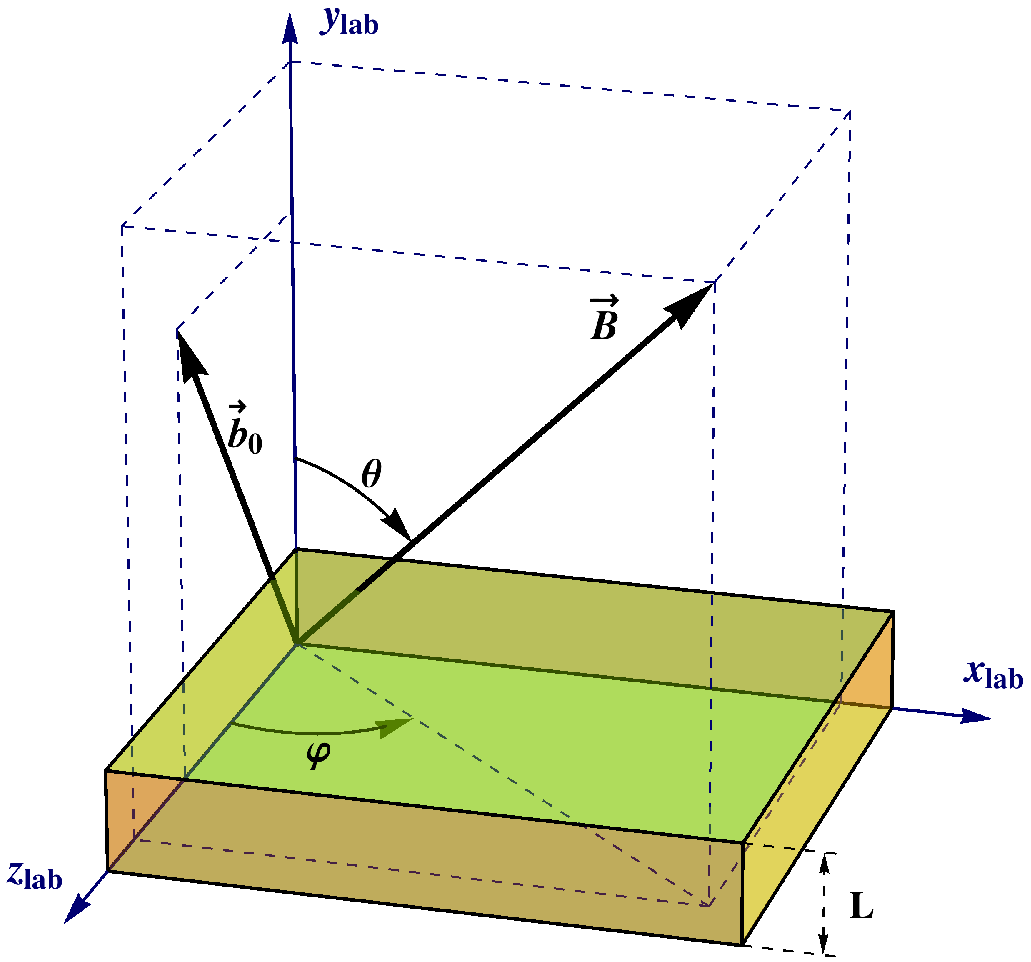}\hspace{0.075\textwidth}
\includegraphics[width=0.45\textwidth]{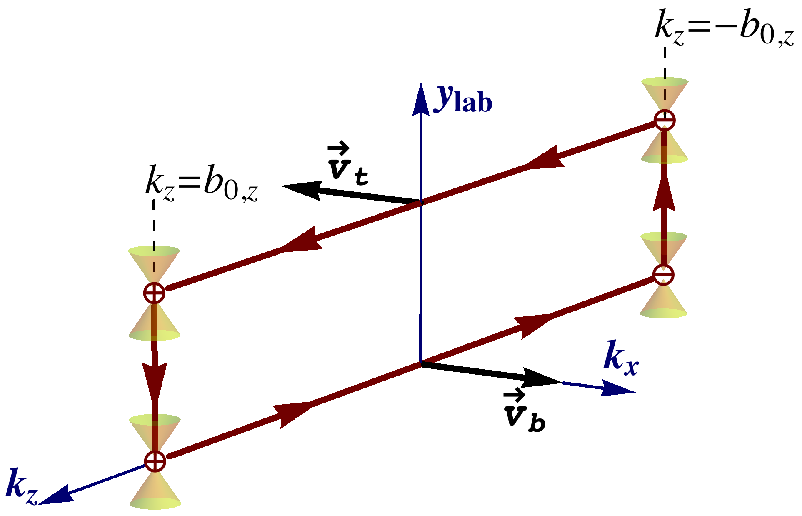}
\caption{(Color online) The model setup and the choice of the laboratory reference frame
used in the current study (left panel) and a schematic representation of the closed
quasiparticle orbits in a magnetic field involving the Fermi arcs (right panel).}
\label{fig:ref-frames}
\end{center}
\end{figure}

The Fermi arc states are localized on the surfaces of the semimetal and are characterized by
wave vectors $k_x$ and $k_z$. Since the velocities of these modes are parallel to the $x$ axis,
only the $z$ component of the quasiclassical equation of motion in the magnetic field is
nontrivial; i.e.,
\begin{eqnarray}
&\partial_t k_{z} = -e\left[\mathbf{v}_{\rm b}\times\mathbf{B}\right]_{z}=-e v_F B_y , &\qquad \mbox{(bottom surface)},
\label{k_z-bottom}\\
&\partial_t k_{z} = -e\left[\mathbf{v}_{\rm t}\times\mathbf{B}\right]_{z}=e v_F B_y , &\qquad \mbox{(top surface)}.
\label{k_z-top}
\label{arc-momentum}
\end{eqnarray}
Therefore, the corresponding quasiparticles slide along the bottom (top) Fermi arc from the right-handed
(left-handed) node at $k_z=b_{0,z}$ ($k_z=-b_{0,z}$) to the left-handed (right-handed) node at $k_z=-b_{0,z}$
($k_z=b_{0,z}$), where $b_{0,z}$ is the component of the chiral shift parallel to the surface of the
semimetal. The surface parts of the quasiparticle orbits are connected with each other via the gapless
bulk modes of fixed chirality. The corresponding closed orbits are schematically shown in Fig.~\ref{fig:ref-frames},
where we use a mixed coordinate--wave vector representation combining the out-of-plane $y$ axis with the in-plane
$k_x$  and $k_z$ axes.

The semiclassical quantization condition for the closed orbits involving the Fermi arc modes reads
\cite{Potter:2014aa}
\begin{equation}
E_n t = 2\pi(n+\gamma),
\label{semiclassical-condition}
\end{equation}
where $n$ is an integer and $\gamma$ is an unknown phase shift that can be determined only via the
rigorous quantum analysis. The total time $t$ includes the time of quasiparticle propagation through the bulk 
$t_{\rm bulk}=2 L/(v_F \cos{\theta})$, where the presence of $\cos{\theta}$ is related to the fact that the 
movement through the bulk occurs only along the magnetic field, and the time it takes for the surface 
Fermi arcs to evolve from one chirality node to the other. The latter can be estimated from 
Eqs.~(\ref{k_z-bottom}) and (\ref{k_z-top}), i.e., $t_{\rm arcs}=2k_0/(v_F e B_y)$, where $k_0$ 
is the arc length in the reciprocal space. The latter is determined by the component of the (full) chiral shift
parallel to the surface of the semimetal \cite{Potter:2014aa,Gorbar:2014qta}. Therefore, 
Eq.~(\ref{semiclassical-condition}) implies that
\begin{equation}
E_n = \frac{\pi v_F(n+\gamma)}{L/\cos{\theta}+k_0/(e B_y)}.
\label{semiclassical-condition-01}
\end{equation}
If the Fermi energy $\mu$ is fixed and the magnetic field is varied, the density of states, associated
with the corresponding close quasiclassical orbits, will be oscillating. The peaks of such oscillations
occur when the Fermi energy crosses the energy levels in Eq.~(\ref{semiclassical-condition-01}).
Taking into account that $B_y = B\cos\theta$, we derive the following discrete values of the magnetic
field that correspond to the maxima of the density of state oscillations:
\begin{equation}
\frac{1}{B_n} = \frac{e}{k_0} \left(\frac{\pi v_F \cos{\theta}}{\mu}(n+\gamma)-L\right).
\label{magnetic-inverse-g0-1}
\end{equation}
By noting that the expression on the right-hand side should remain
positive definite, we conclude that the smallest possible value of $n$ is given by 
$n_{\rm min}=\left[ \mu L/(\pi v_F \cos{\theta})-\gamma+1\right]$, where $[\ldots]$ denotes the
integer part. This corresponds to the saturation value of the magnetic field
$B_{\rm sat} \equiv B_{n_{\rm min}}$, above which no more oscillations will be
observed.

In order to quantify the effects of interaction in the Weyl semimetal in a magnetic
field, we will study how the density of states, involving the Fermi arc modes,
oscillates as a function of the inverse magnetic field. In general, as the above
analysis shows, the period of oscillations is given by
\begin{equation}
T_{1/B}=\frac{e\pi v_F \cos{\theta}}{\mu k_0}.
\label{period-oscillations}
\end{equation}

The interaction effects on the chiral shift were analyzed in Section~\ref{secDW:perturb}
and can be summarized as follows. The full chiral shift takes the following form:
$\mathbf{b}=\mathbf{b}_{0}+\delta{b}\, \hat{\mathbf{n}}$, where $\mathbf{b}_{0}$ is the
chiral shift in absence of the magnetic field, $\delta{b}$ is the magnitude of the correction
to the chiral shift, and $\hat{\mathbf{n}}$ is the unit vector pointing
in the direction of the field. Therefore, in the {\it laboratory frame} of reference, in which
the $x$ component of the bare chiral shift vanishes, $b_{0,x} =0 $, the explicit expressions
for the components of the full chiral shift in a Weyl semimetal in a magnetic field
read
\begin{eqnarray}
b_{x} &=&  \delta{b} \sin\theta\sin\varphi,
\label{dynamical-b-x}\\
b_{y} &=& b_{0,y} + \delta{b} \cos\theta,
\label{dynamical-b-y}  \\
b_{z} &=& b_{0,z} +\delta{b} \sin\theta\cos\varphi.
\label{coordinate-connection-3}
\end{eqnarray}
Since the length of the Fermi arcs is determined by the component of the chiral shift
{\it parallel} to the surface, we
include the effects of interaction in Eq.~(\ref{period-oscillations}) by replacing the bare
arc length with its interaction modified expression; i.e.,
$k_0\to 2 b_{\parallel}=2\sqrt{b_{x}^2+b_{z}^2}$. Then, by making use of
Eqs.~(\ref{dynamical-b-x}) and (\ref{coordinate-connection-3}), we obtain our final
result for the period of oscillations in the interacting case with an arbitrary oriented
magnetic field
\begin{equation}
T_{1/B}=\frac{e\pi v_F \cos{\theta}}{2\mu\sqrt{\left(b_{0,z}\right)^2 + 2b_{0,z} \delta{b} \cos\varphi  \sin\theta
+\left(\delta{b} \sin\theta \right)^2}}.
\label{ArcLengthInteraction}
\end{equation}
Since the Fermi arc length $2b_{||}=2\sqrt{\left(b_{0,z}\right)^2 + 2b_{0,z} \delta{b} \cos\varphi  \sin\theta
+\left(\delta{b} \sin\theta \right)^2}$ depends on the magnetic field, strictly
speaking, the period will be drifting with the varying magnetic field. The numerical results for
the period as a function of angle $\theta$ ($\varphi$) are shown in the left (right) panels of
Fig.~\ref{fig:period} for several fixed values of angle $\varphi$ ($\theta$). [Note that we display only
the results for $\theta<\pi/2$ because $T_{1/B}(\pi-\theta)=T_{1/B}(\theta)$.] The thick (thin) lines
represent the results with (without) taking interaction into account. To obtain these results we
fixed the model parameters as follows:
$b_{0,z}=10^{8}\, \mbox{m}^{-1}$,
$L=1.5\times 10^{-7}\, \mbox{m}$,
$a=0.5 \,\mbox{nm}$,
$v_F=5\times 10^{5}\,  \mbox{m/s}$,
$\mu=10\, \mbox{meV}$. In order to extract the qualitative effects in the cleanest form,
let use a moderately strong coupling, $g =1/(10\Lambda^2 l^2)$. While such a set of model
parameters is representative, it does not correspond to any specific material.

\begin{figure}[t]
\includegraphics[width=0.45\textwidth]{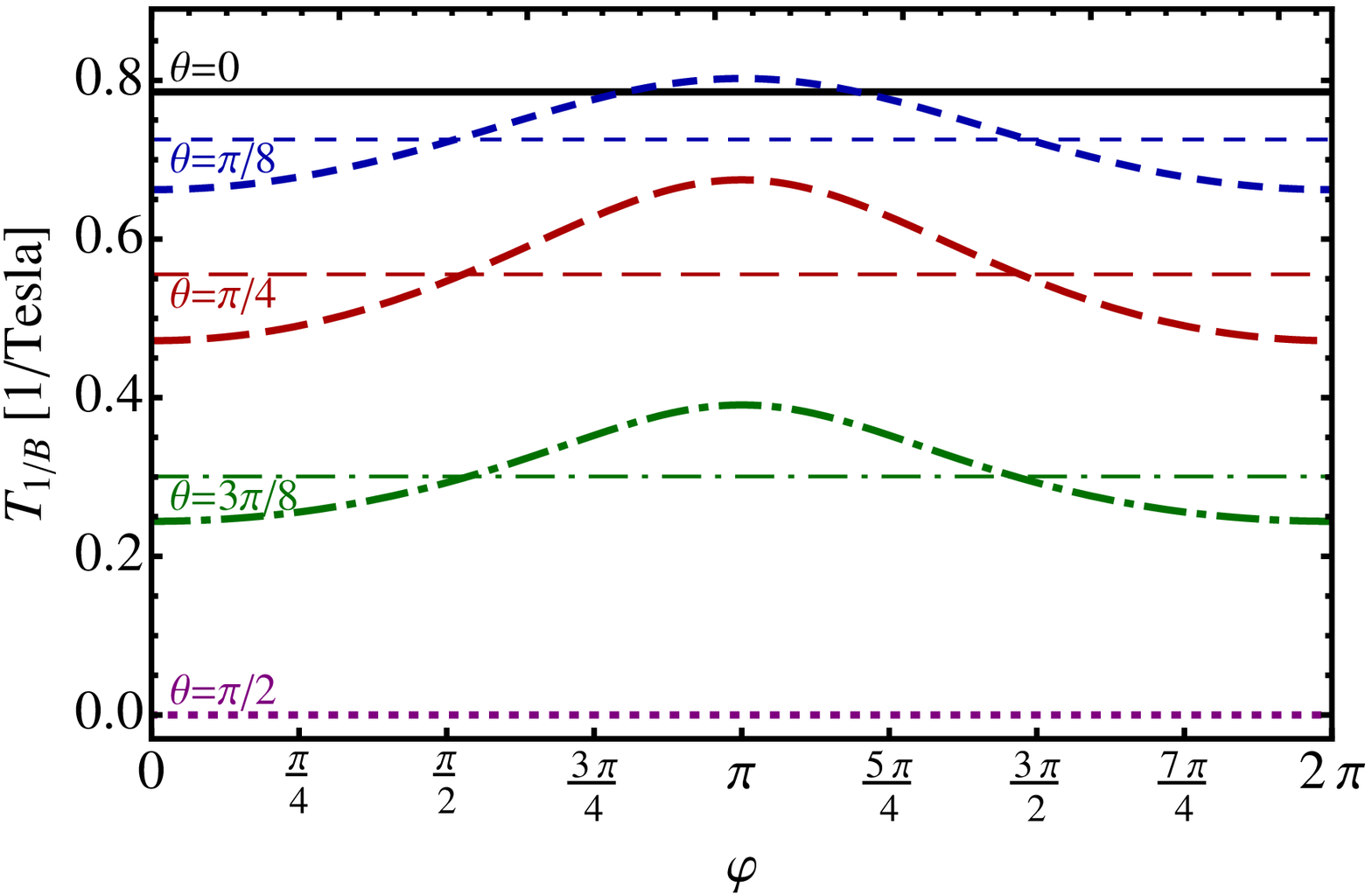}\hspace{0.05\textwidth}
\includegraphics[width=0.45\textwidth]{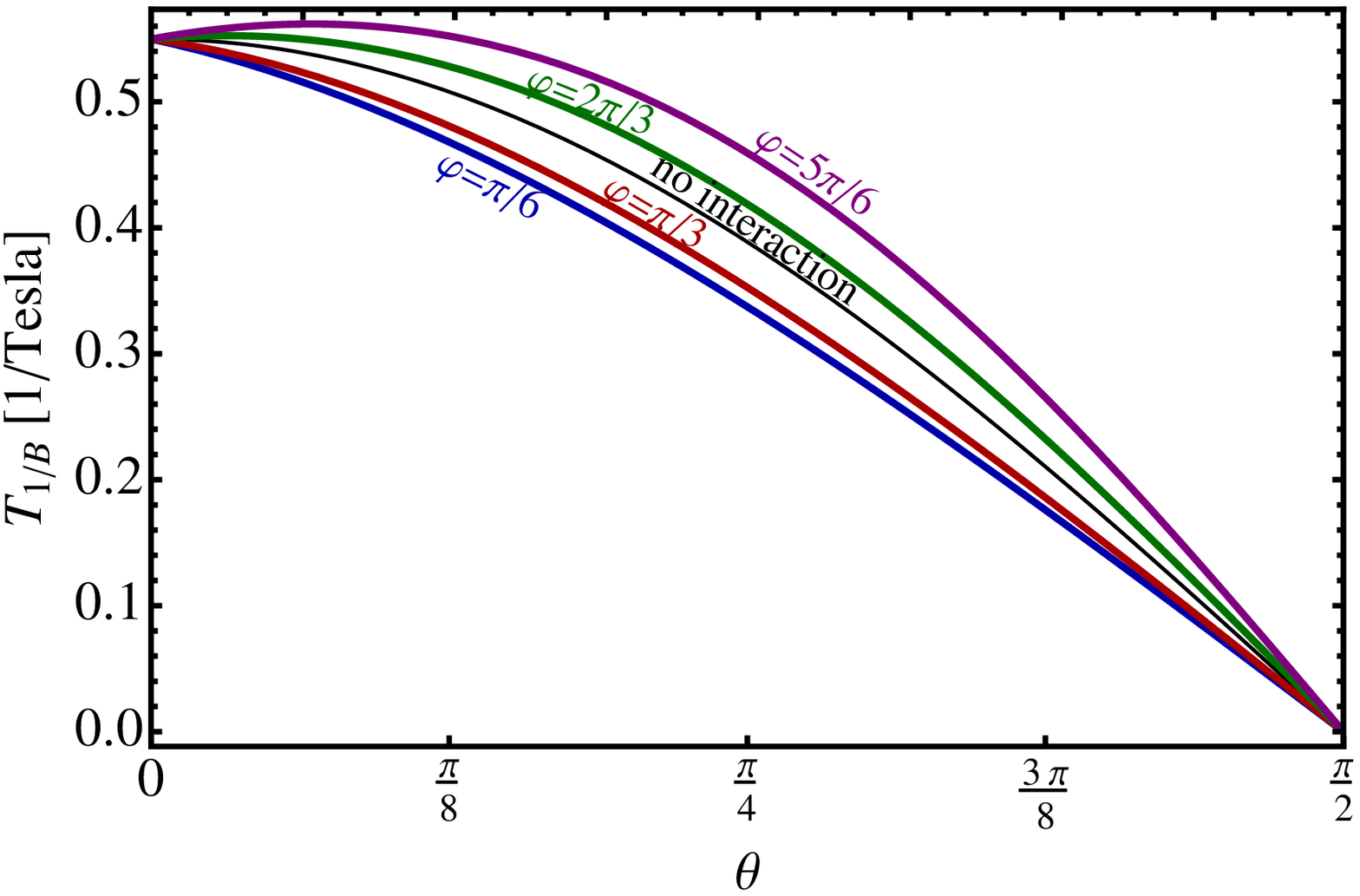}
\caption{(Color online)
The numerical results for the period of the density of state oscillations as a function of
angle $\varphi$ (left panel) and angle $\theta$ (right panel) for several fixed values of
the other angle. The results with and without interaction effects are shown by thick and
thin lines, respectively.}
\label{fig:period}
\end{figure}

As seen from the numerical results in Fig.~\ref{fig:period}, the most important qualitative
effect is the $\varphi$ dependence of the oscillation period, which appears only in
the case with interaction. The emergence of such a dependence is easy
to understand from the physics viewpoint. Indeed, in the noninteracting theory, the
chiral shift, as well as the length of the Fermi arcs $k_0$ associated with it, are
independent of the magnetic field. The situation drastically changes in the interacting
theory when a correction to the chiral shift $\delta\mathbf{b}$ parallel to $\mathbf{B}$
is generated. As follows from Eq.~(\ref{ArcLengthInteraction}), such a correction
introduces a nontrivial dependence of the Fermi arc length on $\varphi$ as soon as
the component of the magnetic field parallel to the surface is nonzero (i.e., $\theta\neq 0$).
At $\theta = 0$, the period of quantum oscillations does not depend on $\phi$ because
in this case the magnetic field is perpendicular to the surface of a semimetal and, therefore,
$\mathbf{b}_{||}=\mathbf{b}_{0,||}$. Also, as one can see on the left panel of Fig.~\ref{fig:period}
and in Eq.~(\ref{ArcLengthInteraction}), the period of quantum oscillations vanishes
at $\theta=\pi/2$. In this case the component of the magnetic field perpendicular to the
surface is absent and, therefore, the quasiclassical motion along the Fermi arcs is forbidden.
The maximum value of the peak of the period of the oscillations as a function of $\varphi$
takes place at 
\begin{equation}
\theta=\frac{1}{2}\arccos\left(\frac{(\delta{b}/b_{0,z})^2}{2 -(\delta{b}/b_{0,z})^2}\right) .
\end{equation}
Measuring this angle in experiment would make it possible to determine the value of the ratio 
$\delta{b}/b_{0,z}$ and, thus, quantify the interaction effects in Weyl semimetals. However, 
a complete fit of the angular dependence in Eq.~(\ref{ArcLengthInteraction}) to the period
obtained in experiment will allow us to extract not only $\delta{b}/b_{0,z}$, but also the value
of  $b_{0,z}$ if the chemical potential and the Fermi velocity are known.

\section{Relativistic field theories in a magnetic background as 
noncommutative field theories}
\label{sec:NoncommutativeTheories}

In this section, we will consider relativistic field theories in a magnetic background 
and their connection with the dynamics in noncommutative field theories (NCFT). 
(For reviews on NCFT, see Refs.~\cite{Douglas:2001ba,Szabo:2001kg}.) Our 
discussion is based on Refs.~\cite{Gorbar:2004ck,Gorbar:2005sx}. 

Besides being interesting in themselves, 
noncommutative theories mimic certain dynamics
in quantum mechanical models \cite{Dunne:1989hv,Bigatti:1999iz}, nonrelativistic field 
systems \cite{Iso:1992ca,Cappelli:1992yv},
nonrelativistic magnetohydrodynamical field theory \cite{Guralnik:2001ax},
and string theories \cite{Connes:1997cr,Seiberg:1999vs}. In
particular, NCFT are intimately related to the dynamics in
quantum mechanical and nonrelativistic field systems in a strong magnetic 
field \cite{Dunne:1989hv,Bigatti:1999iz,Iso:1992ca,Cappelli:1992yv,Guralnik:2001ax} and, in the case of open strings
attached to $D$-branes, to the dynamics in string theories in
magnetic backgrounds \cite{Seiberg:1999vs,SheikhJabbari:1999vm}.

In this section, we show that there is a deep
connection between the dynamics in relativistic field theories
in a strong magnetic field and that in NCFT.
Our main conclusion is that although field theories in the regime with the
lowest Landau level (LLL) dominance indeed determine a class
of NCFT, these NCFT are different from the conventional ones.
In particular, the UV/IR mixing, taking place in the
conventional NCFT \cite{Minwalla:1999px}, is absent in this case.
The reason of that is an inner structure (i.e., dynamical form-factors) of neutral
composites in these theories.

In order to be concrete, two different types of the models (theories) are considered.
In Section~\ref{NJL-NCFT}, the Nambu-Jona-Lasinio type models in a magnetic field are
discussed, and in Section~\ref{gauge-NCFT}, the relativistic gauge field theories,
QED and QCD, are considered. In Section~\ref{connection}, we summarize the main
results and further elaborate on the connection between field theories in a magnetic field
and NCFT. In Appendix~\ref{App:NCFT}, some useful formulas and relations are derived.

\subsection{The NJL model in a magnetic field as a NCFT}
\label{NJL-NCFT}

In this section, we will consider the $(d + 1)$-dimensional Nambu-Jona-Lasinio (NJL) 
models in a strong magnetic field for arbitrary spatial dimension $d \geq 2$.   
In the regime with the LLL dominance, we derive the effective action of the corresponding NCFT in the
models with a large number of fermion colors $N$ and analyze their dynamics. It will be shown
that these NCFT are consistent and quite sophisticated. An especially interesting case is that for
a magnetic field configuration with the maximal number of independent nonzero tensor components
\cite{Gavrilov:1996pz,Gorbar:2000ku}. It will be shown that these NCFT are finite for even $d$ and 
their dynamics are quasi-$(1+1)$-dimensional for odd $d$ [for even $d$, the NCFT describe a confinement dynamics 
of charged particles].

As will be shown, it is the LLL dominance that provides the exponentially damping 
(form-)factors which are responsible for finiteness of these NCFT for even $d$ and 
their quasi-$(1+1)$-dimensionality for odd $d$. Thus, besides being low-energy 
theories of the NJL models in a strong magnetic field, the NCFT based on the 
LLL dynamics are self-contained and self-consistent.

Our approach uses two different sets of composite fields for
the description of the dynamics. The first set
uses the conventional 
composite fields $\sigma(u) \sim \bar{\psi}(u) \psi (u)$
and $\pi(u) \sim i\bar{\psi}(u)\gamma_5 \psi (u)$. In this case,
besides the usual Moyal factor \cite{Douglas:2001ba,Szabo:2001kg}, 
additional exponentially damping factors
occur in the interaction vertices of the fields $\sigma(u)$
and $\pi(u)$.
These factors reflect an inner
structure of composites and play an important role in providing
consistency of these NCFT. In particular, because of them,
the UV/IR mixing is absent in these theories. In the second
approach, one considers other, ``smeared" fields $\Sigma(u)$ and $\Pi(u)$,
connected with $\sigma(u)$ and  $\pi(u)$ through a nonlocal
transformation. Then, while the additional factors are removed in the
vertices of the smeared fields, they appear in their propagators,
again resulting in the UV/IR mixing removal. By using 
the Weyl symbols \cite{Douglas:2001ba,Szabo:2001kg} of the smeared fields, we 
derive the effective action for the composites in the noncommutative coordinate space.

this section is organized as follows. In Section~\ref{NJL1}, 
in order to understand  the nature of the modified
NCFT in a clear and simple way,
we discuss the quantum mechanical model in a magnetic
field introduced in Ref.~\cite{Bigatti:1999iz}. We show that besides
the solution of Ref.~\cite{Bigatti:1999iz}, which mimics
a conventional NCFT, there is another solution, 
with an interaction vertex containing exponentially damping factors.
The existence of these two solutions reflects the possibility of
two different treatments of the case with the particle mass 
$m \to 0$ in this model. In Section~\ref{NJL2}, the effective action of
the NCFT connected with the $(3+1)$-dimensional NJL model in
a strong magnetic field is derived. In Section~\ref{NJL3},
the dynamics of this model is discussed. In Section~\ref{NJL4},
we generalize the analysis to a general case of $d+1$ dimensions 
with $d \geq 2$.

\subsubsection{Nonrelativistic model}
\label{NJL1}

In order to understand better the nature of the modified NCFT, in this
section we analyze a simple quantum mechanical 
two-dimensional system: 
a pair of unit charges of opposite sign (i.e., a dipole)
in a constant magnetic field and with a harmonic potential interaction 
between them. This model was considered in Ref.~\cite{Bigatti:1999iz}.  
It was argued there that for a strong magnetic field this simple 
system reproduces the dynamics of open strings
attached to D-branes in antisymmetric tensor backgrounds.

We will show
that important features of the modified NCFT occur already in this simple
quantum mechanical model. Its Lagrangian reads
\begin{eqnarray}
L = \frac{m}{2}\left(\dot{\mathbf{r}}_1^2 +\dot{\mathbf{r}}_2^2 \right) +
\frac{eB}{2} \left(\dot{x}_1 y_1 -\dot{y}_1 x_1 - \dot{x}_2 y_2+\dot{y}_2 x_2\right) 
- \frac{K}{2}\left(\mathbf{r}_1 - \mathbf{r}_2\right)^2 .
\label{lagrangian1}
\end{eqnarray}
It is convenient to use the center of mass and relative coordinates,
$\mathbf{X}=(\mathbf{r}_1 + \mathbf{r}_2)/2$
and $\bm{\Delta}=(\mathbf{r}_1 - \mathbf{r}_2)/2$, respectively.
In these coordinates, Lagrangian (\ref{lagrangian1}) takes the form
\begin{eqnarray}
L = m \left(\dot{\mathbf{X}}^2 + \dot{\bm{\Delta}}^2\right) +
2eB (\dot{X}_x \Delta_y - \dot{X}_y \Delta_x) -
2K\bm{\Delta}^2.
\label{lagrangian2}
\end{eqnarray}
The LLL dominance occurs when either $B \to \infty$ or $m \to 0$. 
Taking $m = 0$, the authors of \cite{Bigatti:1999iz} drop the kinetic terms in 
Lagrangian (\ref{lagrangian2})  that results in a theory of the Chern--Simons
type with only first order time derivatives. Then, they introduce 
an additional potential $V(\mathbf{r}_1)$ describing 
an interaction of the first charge with an ``impurity" centered
at the origin  
and show that the matrix element of
$V(\mathbf{r}_1)$ between dipole states contains the usual Moyal phase
that is a signature of NCFT.

Notice that this result is obtained when the limit $m \to 0$ is
taken directly in the Lagrangian. Let us now show that when one 
first solves this
problem for a nonzero $m$ and then takes the limit $m \to 0$ 
in the solution, additional exponential factors occur in the
matrix element of $V(\mathbf{r}_{1})$.

The Hamiltonian in model (\ref{lagrangian1}) is given by
\begin{eqnarray}
H = \frac{\hat{\mathbf{p}}^2 + \hat{\mathbf{d}}^2}{4m} -
\frac{eB}{m}(\Delta_x \hat{p}_y -\Delta_y \hat{p}_x) +\left(\frac{(eB)^2}{m} 
+ 2K\right)\bm{\Delta}^2,
\label{hamiltonian}
\end{eqnarray}
where $\hat{\mathbf{p}}$ and $\hat{\mathbf{d}}$ are operators of the center 
of mass and relative momenta. Since the Hamiltonian 
is independent
of the center of mass coordinates, the wave function can be 
represented in the form $\psi(\mathbf{X},\bm{\Delta}) 
= e^{i\mathbf{p}\cdot\mathbf{X}} f(\bm{\Delta})$.
Then, for $f(\bm{\Delta})$ we get the equation
\begin{eqnarray}
\left[\frac{\hat{\mathbf{d}}^2}{4m} - \frac{eB}{m}(\Delta_x p_y - \Delta_y p_x)
+ \left(\frac{(eB)^2}{m} + 2K\right)\bm{\Delta}^2\right] f(\bm{\Delta}) =
\left(E - \frac{\mathbf{p}^2}{4m}\right)f(\bm{\Delta}).
\label{dipoles-equation1}
\end{eqnarray}
Changing the variables to
\begin{eqnarray}
x &=& \Delta_x + \frac{eBp_y}{2[(eB)^2 + 2Km]},
\\
y &=& \Delta_y - \frac{eBp_x}{2[(eB)^2 + 2Km]},
\end{eqnarray}
we arrive at the following equation:
\begin{equation}
\left[\frac{K\mathbf{p}^2}{2[(eB)^2 + 2Km]} - 
\frac{\partial_x^2 + \partial_y^2}{4m} + \left(\frac{(eB)^2}{m} + 2K\right)
(x^2+y^2)\right] f(u,u^\prime) = E f(u,u^\prime).
\label{dipoles-equation2}
\end{equation}
Clearly, the first term here is the kinetic energy of the center of mass.
Note that as $m \to 0$, it coincides with the eigenvalue of the
Hamiltonian
in Ref.~\cite{Bigatti:1999iz} obtained from Lagrangian (\ref{lagrangian1}) with
$m = 0$. All other terms that are present in
Hamiltonian (\ref{dipoles-equation2}) [and reflecting the inner structure
of composite states] are absent in the Hamiltonian of Ref.~\cite{Bigatti:1999iz}.
In that case, the only information about the inner structure of
composites that is retained is given by the relations
\begin{equation}
\Delta_x = -\frac{p_y}{2eB} \qquad \mbox{and} \qquad 
\Delta_y = \frac{p_x}{2eB} 
\label{constraint}
\end{equation}
which express the relative coordinates in terms of the center of mass momentum.

Equation (\ref{dipoles-equation2}) describes a harmonic oscillator, and it can be   
solved exactly. Its spectrum contains an infinite number of composites (neutral 
bound states) with the energy eigenvalues
\begin{eqnarray}
E_{\mathbf{p},n,k} = \frac{K\mathbf{p}^2}{2r^2} + (n + k + 1) \frac{r}{m},
\label{spectrum} 
\end{eqnarray}
where $r = \sqrt{(eB)^2+2Km}$, and $n$ and $k$ are nonnegative integers. 
Note that in the limit $K \to 0$ the Lagrangian (\ref{lagrangian1}) reduces to the 
Lagrangian
of two noninteracting charged particles in a constant magnetic field 
(the Landau problem) and Eq.~(\ref{spectrum}) correctly reproduces
the Landau spectrum.

Thus, the model (\ref{lagrangian1}) 
describes an infinite number of neutral composites. 
The vector $\mathbf{p}$ is their center of 
mass momentum and the last term in (\ref{spectrum}) 
reflects their nontrivial inner structure. 
Now, in the limit $ m \to 0$,
only the LLL states with $n=k=0$ survive 
(all higher excitations decouple). The normalized LLL wave function
with the center of mass momentum $\mathbf{p}$ is given by
\begin{eqnarray}
\langle \mathbf{X},\bm{\Delta}|\mathbf{p}\rangle 
= \psi_{\mathbf{p},0,0}(\mathbf{X},\bm{\Delta}) 
= \left( \frac{r}{2\pi^3} \right)^{1/2} e^{i\mathbf{p}\cdot\mathbf{X}}
e^{-r\left(\Delta_x+\frac{eBp_y}{2r^2}\right)^2}e^{-r\left(\Delta_y-\frac{eBp_x}{2r^2}\right)^2}.
\label{wavefunctions}
\end{eqnarray}
The Gaussian exponential factors here reflect the inner 
structures of the composites. It is important that
in the limit $m \to 0$, this wave function does {\it not} 
coincide with the wave function of Ref.~\cite{Bigatti:1999iz} corresponding to
the model with the Lagrangian (\ref{lagrangian1}) at $m = 0$:
there are no Gaussian exponential factors in that case. In other
words, while in the $m \to 0$ model, there are quantum
fluctuations described by the Gaussian exponents, these
fluctuations are completely suppressed in the
model with $m \equiv 0$.

Thus, we conclude that the quantum dynamics in the limit $m \to 0$ 
in the massive model does not coincide with that in the massless one. Recall that
the same situation takes place in non-Abelian gauge theories: the
limit $m \to 0$ in a massive non-Abelian model does not yield
the dynamics of the massless one \cite{Slavnov:1970tk,vanDam:1970vg,Vainshtein:1971ip}. 
The origin of this phenomenon is the same in both cases. Because of constraints
in the massless models, the number of physical degrees of
freedom there is less than the number of degrees of freedom in
the massive ones. In the present quantum mechanical model, these
constraints are described by equation (\ref{constraint}).

Of course, there is nothing wrong with the model
(\ref{lagrangian1}) at $m = 0$. It is
mathematically consistent. However, its dynamics is very
different from that of a physical dipole in a strong
magnetic field [by a physical dipole, we understand
a dipole composed of two massive charged particle,
including the case of an infinitesimally small mass
$m \to 0$]. We also would like to point out that 
the present treatment of the dynamics in a strong magnetic field
is equivalent to the formalism
of the projection onto the LLL developed in Refs.~\cite{Girvin:1984fk,Kivelson:1987uq,Dunne:1992ew}.

If following Ref.~\cite{Bigatti:1999iz} we introduce an 
additional potential $V(\mathbf{r}_1)$ describing an interaction of the
first charge with an
impurity, the matrix element
$\langle\mathbf{k}|V(\mathbf{r}_1)|\mathbf{p}\rangle$ will describe the scattering 
of composites on the impurity in 
the Born approximation. 
In order to evaluate $\langle\mathbf{k}|V(\mathbf{r}_1)|\mathbf{p}\rangle$,
it is convenient to introduce the Fourier transform
\begin{eqnarray}
V(\mathbf{r}_1) = \int 
\frac{d^{2}q}{(2\pi)^2} \tilde{V}(\mathbf{q})e^{i{\mathbf{q}}\cdot{\mathbf{r}_1}},
\end{eqnarray}
so that
\begin{eqnarray}
\langle\mathbf{k}|V(\mathbf{r}_1)|\mathbf{p}\rangle = \int
\frac{d^{2}q}{(2\pi)^2} \tilde{V}(\mathbf{q})
\langle\mathbf{k}|e^{i\mathbf{q}\cdot(\mathbf{X}+\bm{\Delta})}|\mathbf{p}\rangle.
\end{eqnarray}
By making use of the completeness of eigenstates, i.e.,
$\int d^2X d^2\Delta\,  |\mathbf{X}, \bm{\Delta}\rangle\langle\mathbf{X}, \bm{\Delta}| = I$,
as well as Eq.~(\ref{wavefunctions}),
one can easily calculate the matrix element 
$\langle\mathbf{k}|e^{i\mathbf{q}\cdot(\mathbf{X}+\bm{\Delta})}|\mathbf{p}\rangle$.
In the limit $m \to 0$, it is
\begin{eqnarray}
\langle\mathbf{k}|e^{i\mathbf{q}\cdot(\mathbf{X}+\bm{\Delta})}|\mathbf{p}\rangle
= \delta^2\left(\mathbf{k} - \mathbf{q} - \mathbf{p}\right)
e^{-\frac{\mathbf{q}^2}{4|eB|}}e^{-\frac{i}{2}q \times k}
\label{me}
\end{eqnarray}
with the cross product $q \times k \equiv (q_x k_y-q_y k_x)/eB$.
One can see that, in addition to the standard Moyal factor $e^{-\frac{i}{2}q \times k}$, 
this vertex contains also the exponentially damping term $e^{-\frac{\mathbf{q}^2}{4|eB|}}$. 
As is easy to check, this term originates from the Gaussian factors in the wave 
function (\ref{wavefunctions}). It would be absent if we, as in Ref.~\cite{Bigatti:1999iz}, 
used the Lagrangian with $m = 0$ in Eq.~(\ref{lagrangian1}).

The general character of this phenomenon suggests that additional
exponential terms in interaction vertices should also occur in 
field theories in a strong magnetic field. This expectation 
will be confirmed in the next subsection, 
where it will be also shown that these theories
determine a class of modified NCFT.

\subsubsection{The effective action}
\label{NJL2}

In this section, we will consider the dynamics in the $(3 + 1)$-dimensional
NJL model in a strong magnetic field. Our aim is to show that this dynamics 
determines a consistent NCFT. As it will be shown in Section~\ref{NJL4}, a 
similar situation takes place in an arbitrary dimension $D = d + 1$ with the 
space dimension $d \geq 2$.

Let us consider the same NJL model with the $\mathrm{U}_L(1) \times \mathrm{U}_R(1)$ 
chiral symmetry as in Section~\ref{sec:NJL3+1General}. The Lagrangian density 
reads
\begin{eqnarray}
L = \frac{1}{2}\left[\bar{\psi}, (i\gamma^{\mu}D_{\mu})\psi \right]  +
\frac{G}{2N} \left[(\bar{\psi}\psi)^2 + (\bar{\psi}i\gamma^5\psi)^2 
\right],
\label{NJLaction}
\end{eqnarray}
cf. Eq.~(\ref{12L_NJL_U1xU1}). Once again, we assume that the fermion fields carry an additional 
``color" index $i=1,2,\ldots,N$. The covariant derivative is $D_{\mu}= \partial_{\mu} +
ieA_{\mu} $, where the external vector potential $A_{\mu} $ 
describes a constant magnetic field $B$ pointing in the $+z$ direction. We will use
interchangeably the Landau gauge (\ref{eq:Landau_gauge}) and the symmetric gauge 
(\ref{eq:symm_gauge}).

We will consider the dynamics of neutral bound states (``dipoles") in this model in  
the limit of large $N$. In this limit, as we saw in Section~\ref{sec:NJL3+1General}, 
the model becomes essentially soluble, and its nonperturbative ground state is determined 
by the magnetic catalysis phenomenon. 

At large $N$, the relevant neutral degrees of freedom  are connected with the composite 
fields $\sigma \sim \bar{\psi}\psi$ and $\pi \sim \bar{\psi}i\gamma_{5}\psi$. The formal 
definition of the effective action for these fields was given in Eqs.~(\ref{14}) and (\ref{15})
in Section~\ref{sec:NJL3+1General}. There we also derived an explicit form of the effective 
potential $V(\sigma,\pi)$ and the gap equation, see Eqs.~(\ref{16}) and (\ref{17}), respectively. 
The latter, in particular, determines the dynamical mass of fermions, $m=\langle0|\sigma|0\rangle$,
which is nonzero for any positive $G$.

Let us recall that the dynamics is dominated by the LLL in the weak coupling regime, with the 
dimensionless coupling constant $g \equiv NG\Lambda^2/4\pi^2 \ll 1$. The expression for the 
fermion mass is given by Eq.~(\ref{m_dyn_weak_3+1}), or equivalently
\begin{equation}
m^2 \simeq \frac{eB}{\pi} \exp\left(\frac{\Lambda^2}{|eB|}\right) 
\exp\left( -\frac{\Lambda^2}{g |eB|}\right), 
\label{solution}
\end{equation}
where $\Lambda$ is an ultraviolet cutoff connected with longitudinal momenta $k_{\parallel} = (k^0,k^3)$.
We assume that $\Lambda^2 \gg |eB|$. As will become clear below, there are no divergences
connected with transverse momenta  $\mathbf{k}_{\perp} = (k^1,k^2)$ in the regime with the 
LLL dominance, and therefore the ``longitudinal" cutoff removes all divergences in this model.
Notice that Eq.~(\ref{solution}) implies the following hierarchy of scales:
$\frac{|eB|}{m^2} \gg \frac{\Lambda^2}{|eB|}$. It will be shown in Section~\ref{NJL3} 
that a meaningful continuum limit $\Lambda^2 = C|eB| \to \infty$, with $C \gg 1$ and a 
finite $m$, exists in this model.

It is straightforward to calculate the interaction vertices for the quantum $\tilde{\sigma}=\sigma - m$ 
and $\pi$ fields that follow from the effective action, defined by Eqs.~(\ref{14}) and (\ref{15}). 
For example, the 3-point vertex $\Gamma_{\tilde{\sigma}\pi\pi}$ is given by
\begin{equation}
\Gamma_{\tilde{\sigma}\pi\pi} =
\int d^4 u d^4u^\prime d^4v \, \mathrm{tr}\left[S(u,u^\prime)\gamma^5\pi(u^\prime)S(u^\prime,v)
\gamma^5\pi(v)S(v,u)\tilde{\sigma}(u)\right].
\label{3point-initial}
\end{equation}
The fermion propagator $S(u,u^\prime)$ is given as a product of the Schwinger phase 
factor, $\exp\left[i \Phi_{\rm sym}(\mathbf{r}_\perp,\mathbf{r}^\prime_\perp)\right]$, and 
a translationally invariant part $\bar{S}(u-u^\prime)$, see Eq.~(\ref{eq:green}) in 
Appendix~\ref{App:A-SchwingerProp}. In the LLL approximation, the Fourier transform 
of the latter reads
\begin{equation}
\bar{S}(k) = i\, e^{-\frac{k_{\perp}^2}{|eB|}}\,
\frac{k^0\gamma^0 - k^3\gamma^3 + m}{k_0^2-k_3^2-m^2}\,
\left[1+i\gamma^1\gamma^2\mathrm{sign}(eB)\right],
\label{fourier}
\end{equation}
[for convenience, the Kronecker $\delta$-symbol with color indices is omitted in 
the propagator]. 

The explicit form of the Schwinger phase is given by
\begin{equation}
\Phi_{\rm sym}(\mathbf{r}_\perp,\mathbf{r}^\prime_\perp) = \frac{eB}{2}(x y^\prime - y x^\prime ), 
\label{phase1}
\end{equation}
in the symmetric gauge (\ref{eq:symm_gauge}), and by
\begin{equation}
\Phi_{\rm Landau}(\mathbf{r}_\perp,\mathbf{r}^\prime_\perp)  
= \Phi_{\rm sym}(\mathbf{r}_\perp,\mathbf{r}^\prime_\perp) + \frac{eB}{2}( x y - x^\prime y^\prime )
\label{phase2}
\end{equation}
in the Landau gauge (\ref{eq:Landau_gauge}). The Schwinger phase factor breaks the 
translation invariance even in the case of a constant magnetic field, although in this case 
there is a group of magnetic translations whose generators, unlike usual momenta, 
do not commute (for a more detailed discussion, see Section~\ref{sec:MagneticTranslations}).
One can easily check that the total phase along the closed fermion loop in (\ref{3point-initial}) 
is independent of a gauge, i.e., the corresponding result is gauge invariant.

We will show that, in the regime with the LLL dominance, the effective action for 
the $\sigma$ and $\pi$ fields, Eqs.~(\ref{14}) and (\ref{15}), leads to a NCFT 
with noncommutative space transverse coordinates
$\hat{x}^{a}$:
\begin{equation}
[\hat{x}^{a},\hat{x}^{b}] = i\frac{1}{eB}\epsilon^{ab} \equiv i\theta^{ab}.
\label{commrel}
\end{equation}
It is the
Schwinger phase that is responsible for this noncommutativity.
Indeed, the commutator
$[\hat{x}^{a},\hat{x}^{b}]$
is of course antisymmetric, and
the only place where an antisymmetric tensor occurs
in 3-point vertex
(\ref{3point-initial})
is the Schwinger phase (as will be shown
below, a similar situation takes place also for higher
vertices).

We begin our analysis with the observation that the LLL fermion
propagator in coordinate space factorizes into two parts,
the part depending on the transverse coordinates
$u_{\perp}=(x, y)$ and that
depending on the longitudinal coordinates
$u_{\parallel}=(t, z)$:
\begin{equation}
S(u,u^\prime) = P(u_{\perp},u^\prime_{\perp})\, S_{\parallel}(u_{\parallel}-u^\prime_{\parallel}).
\label{factorization}
\end{equation}
Indeed, taking into account expressions (\ref{fourier}) and (\ref{phase1}), we get in the 
symmetric gauge:
\begin{equation}
P(u_{\perp},u^\prime_{\perp}) =  \frac{|eB|}{2\pi}\, e^{\frac{ieB}{2}
\epsilon^{ab}u_{\perp}^{a}u_{\perp}^{\prime b}}\,e^{-\frac{|eB|}{4}(u_{\perp} - u_{\perp}^\prime)^2}
\label{projector}
\end{equation}
and
\begin{equation}
S_{\parallel}(u_{\parallel}-u^\prime_{\parallel}) = \int \frac{d^2k_{\parallel}}{(2\pi)^2}
e^{ik_{\parallel}(u_{\parallel}-u^\prime_{\parallel})} \frac{i}{k_{\parallel}\gamma^{\parallel} - m}\,
\frac{1 + i\gamma^1\gamma^2 \mathrm{sign}(eB)}{2}
\label{flatspace}
\end{equation}
[henceforth, for concreteness, we will use the symmetric gauge].
The longitudinal part $S_{\parallel}(u_{\parallel}-u^\prime_{\parallel})$ is nothing
else but a fermion propagator in 1+1 dimensions. In particular,
the matrix $[1 + i\gamma^1\gamma^2 \mathrm{sign}(eB)]/2$ is the
projector on the fermion (antifermion) states with the spin
polarized along (opposite to) the magnetic field, and therefore it
projects on two states of the four ones, as should be in 1+1
dimensions. As to the operator $P(u_{\perp},u^\prime_{\perp})$, it
is easy to check that it satisfies the relation
\begin{equation}
\int d^{2}u^{\prime}_{\perp} P(u_{\perp},u^{\prime}_{\perp})\,P(u^{\prime}_{\perp},v_{\perp}) =
P(u_{\perp},v_{\perp}),
\label{Pprojector}
\end{equation}
and therefore is a projection operator. Since $S(u,u^\prime)$ is
a LLL propagator, it is clear that $P(u_{\perp},u^\prime_{\perp})$ is a
projection operator on the LLL states.

The factorization of the LLL propagator leads to a simple structure
of interaction vertices for $\pi$ and $\tilde{\sigma}$ fields. For example, as it
is clear from expression (\ref{3point-initial}) for the 3-point vertex, a substitution 
of the Fourier transforms for the fields makes the integration over the
longitudinal and transverse coordinates completely independent. And since 
$S_{\parallel}(u_{\parallel}-u^\prime_{\parallel})$ is a
$(1 + 1)$-dimensional propagator, the integration over $u_{\parallel}$
coordinates yields a fermion loop in the $(1+1)$-dimensional Minkowski space.
It is obvious that the same is true also for higher order interaction
vertices arising from action (\ref{14}). Therefore, the dependence
of interaction vertices on longitudinal $k_{\parallel}$ momenta is standard and,
for clarity of the presentation, we will first consider the case with all external 
longitudinal momenta entering the fermion loop to be
zero. This of course corresponds to the choice of
$\pi$ and $\tilde{\sigma}$ fields independent of
longitudinal coordinates $u_{\parallel}$. The general case,
with the fields depending on both transverse and longitudinal
coordinates, will be considered in the end of this section.

Now, substituting the Fourier transforms of the fields $\pi(u_{\perp})$ and
$\tilde{\sigma}(u_{\perp})$ into Eq.~(\ref{3point-initial})
and using Eqs.~(\ref{factorization}), (\ref{projector}), and
(\ref{flatspace}), we find the following expression for the 3-point interaction
vertex $\Gamma_{\tilde{\sigma}\pi\pi}$ in the momentum space:
\begin{eqnarray}
\Gamma_{\tilde{\sigma}\pi\pi} &=& -\frac{N|eB|}{m} \int d^2u_{\parallel}
\int \frac{d^2k_1d^2k_2d^2k_3}{(2\pi)^6}
\pi(k_1)\pi(k_2)\tilde{\sigma}(k_3)\,
\delta^2(k_1+k_2+k_3)\nonumber
\\
&&\times  e^{-\frac{k_{1}^2+k_{2}^2+k_{3}^2}{4|eB|}}\,
\exp\left[-\frac{i}{2}(k_1 \times k_2 + k_1 \times k_3 +
k_2 \times k_3)\right],
\label{3point}
\end{eqnarray}
where $k_i \times k_j = k_{i}^a \theta^{ab}k_{j}^b
\equiv k_{i} \theta k_{j}$,
$\theta^{ab}=  \frac{1}{eB}\epsilon^{ab}$ (here, for convenience,
we omitted the subscript $\perp$ for the transverse coordinates).
Notice that because of the exponentially damping factors, there
are no ultraviolet divergences in this expression.

According to \cite{Douglas:2001ba,Szabo:2001kg}, an $n$-point vertex in a
noncommutative theory in momentum space has the following
structure:
\begin{equation}
\int \frac{d^Dk_1}{(2\pi)^D} \ldots  \frac{d^Dk_n}{(2\pi)^D}
\phi(k_1)\ldots  \phi(k_n) \delta^D\left(\sum_i k_i\right)
e^{-\frac{i}{2} \sum_{i<j} k_i \times k_j},
\label{vertex}
\end{equation}
where here $\phi$ denotes a generic field and the exponent
$e^{-\frac{i}{2} \sum_{i<j} k_i \times k_j}\equiv
e^{-\frac{i}{2} \sum_{i<j} k_i \theta k_j}$
is the Moyal exponent
factor. Comparing expressions (\ref{3point}) and (\ref{vertex}),
we see that apart from the factor
$e^{-\frac{k_{1}^2+ k_{2}^2+k_{3}^2}{4|eB|}}$,
the vertex $\Gamma_{\tilde{\sigma}\pi\pi}$ coincides with the standard
3-point
interaction vertex in a noncommutative theory with the commutator
$[\hat{x}^{a},\hat{x}^{b}]= i\theta^{ab} = \frac{i}{eB}\epsilon^{ab}$.

In order to take properly into account this additional factor in
the vertex, it will be convenient to introduce new, ``smeared'' fields:
\begin{eqnarray}
\Pi(u) &=& e^{\frac{\nabla_{\perp}^2}{4|eB|}}\,\pi(u),\\
\Sigma(u) &=& e^{\frac{\nabla_{\perp}^2}{4|eB|}}\,\sigma(u),
\label{smeared}
\end{eqnarray}
where $\nabla_{\perp}^2$ is the transverse Laplacian.
Then, in terms of these fields, the vertex can be rewritten in the
standard form with the Moyal exponent factor:
\begin{equation}
\Gamma_{\tilde{\Sigma}\Pi\Pi} = -\frac{N|eB|}{m} \int d^2u_{\parallel}
\int \frac{d^2k_1d^2k_2d^2k_3}{(2\pi)^6}
\Pi(k_1)\Pi(k_2)\tilde{\Sigma}(k_3)\,\delta^2\left(\sum_{i} k_i\right)\,
\exp\left(-\frac{i}{2}\sum_{i<j} k_i \times k_j\right).
\label{3point1}
\end{equation}
One can similarly analyze the 4-point interaction vertex $\Gamma_{4\Pi}$. 
We get
\begin{equation}
\Gamma_{4\Pi} = -\frac{N|eB|}{4m^2} \int d^2u_{\parallel} \int
\frac{d^2k_1 d^2k_2 d^2k_3 d^2k_4}{(2\pi)^8} \Pi(k_1)\Pi(k_2)
\Pi(k_3)\Pi(k_4) \delta^2\left(\sum_{i} k_i\right)\, 
\exp\left(-\frac{i}{2}\sum_{i<j} k_i \times k_j\right).
\label{4point}
\end{equation}
The occurrence of the smeared
fields in the vertices reflects
an inner structure (dynamical form-factors)
of $\pi$ and $\sigma$ composites on the lowest
Landau level, which is
similar to that of a dipole in the quantum mechanical problem
considered in Section~\ref{NJL1}.

As is well known, the cross product in the momentum
space corresponds to a star product in the coordinate space
\cite{Douglas:2001ba,Szabo:2001kg}:
\begin{equation}
(\Phi * \Phi)(u) =
e^{\frac{i}{2}\theta^{ab}
\frac{\partial}{\partial v^a}\frac{\partial}{\partial w^b}}
\Phi(v)\Phi(w)|_{v=w=u},
\label{starproduct}
\end{equation}
where here $\Phi$ represents the smeared fields $\Pi$ and
$\Sigma$.
By using the star product, one can rewrite the vertices
$\Gamma_{\tilde{\Sigma}\Pi\Pi}$ and $\Gamma_{4\Pi}$
in the following simple form in the coordinate space:
\begin{eqnarray}
\Gamma_{\tilde{\Sigma}\Pi\Pi} &=& -\frac{N|eB|}{4\pi^2m}
\int d^2u_{\parallel}d^2u_{\perp}\,
\tilde{\Sigma}*\Pi*\Pi\,,
\\
\Gamma_{4\Pi} &=& -\frac{N|eB|}{16\pi^2m^2} \int
d^2u_{\parallel}d^2u_{\perp}\, \Pi*\Pi*\Pi*\Pi.
\label{xspacevertices}
\end{eqnarray}

As to expressing the vertices in NCFT in
the space with noncommutative coordinates $\hat{x}^{a}$,
one should use the Weyl symbol of a field $\Phi$ there
\cite{Douglas:2001ba,Szabo:2001kg}:
\begin{eqnarray}
\hat{\Phi}(\hat{x}) &\equiv &\hat{W}[\Phi] =
\int d^Du\,\Phi(u)\,\hat{\Delta}(u),
\label{Weylsymbol}
\\
\hat{\Delta}(u) &\equiv& \int \frac{d^Dk}{(2\pi)^D} e^{ik_a\hat{x}^a}\,
e^{-ik_a u^a}.
\end{eqnarray}
The most  important property of the Weyl symbol is that the product of the Weyl symbols 
of two functions is equal to the Weyl symbol of their star product:
\begin{equation}
\hat{W}[\Phi_1]\,\hat{W}[\Phi_2] = \hat{W}[\Phi_{1}*\Phi_{2}].
\label{W1}
\end{equation}
In our case, the Weyl symbol $\hat{\Phi}$ represents $\hat{\Pi}$ and $\hat{\tilde{\Sigma}}$.
Note that the relation between the Weyl symbols of smeared and non-smeared fields is
\begin{equation}
\hat{\Phi}(\hat{x}) =
e^{\frac{\hat{\nabla}_{\perp}^2}{4|eB|}}\,\hat{\phi}(\hat{x}),
\label{Wsmeared}
\end{equation}
where the operator $\hat{\nabla}_{\perp}^2$ in the noncommutative
space acts as
\begin{equation}
\hat{\nabla}_{\perp}^2\,\hat{\phi}(\hat{x}) =
-(eB)^2\,\sum_{a = 1}^2
\left[\hat{x}^a,[\hat{x}^a, \hat{\phi}(\hat{x})]\,\right]
\end{equation}
[the latter relation follows from the definition of the derivative
in NCFT,
$ \hat{\nabla}_{\perp a}\,\hat{\phi}(\hat{x})=
-i[(\theta^{-1})_{ab}\hat{x}^b, \hat{\phi}(\hat{x})]$ \cite{Douglas:2001ba,Szabo:2001kg}\,].

In terms of $\hat{\Phi}$, the 3- and 4-point
vertices take the following form in NCFT:
\begin{eqnarray}
\Gamma_{\tilde{\Sigma}\Pi\Pi} &=& -\frac{N|eB|}{4\pi^2m}\,
\int d^2u_{\parallel}\mathbf{Tr}\,
\hat{\tilde{\Sigma}}\hat{\Pi}^2\,,
\\
\Gamma_{4\Pi} &=& -\frac{N|eB|}{16\pi^2m^2}\,
\int d^2u_{\parallel}\mathbf{Tr}\,\hat{\Pi}^4\,,
\label{ncspacevertices}
\end{eqnarray}
where the operation $\mathbf{Tr}$ is defined as in Refs.~\cite{Douglas:2001ba,Szabo:2001kg}.
As is shown in Appendix~\ref{App:NCFT1}, all interaction
vertices $\Gamma_{n\Phi}$ ($n \geq 3$) arising from action
(\ref{14}) have a similar structure.

There exists another, more convenient for practical calculations,
representation of interaction vertices in which the vertices
are expressed through the initial, non-smeared, fields
$\pi$ and $\tilde{\sigma}$.
The point is that,
due to the presence of the $\delta$-function
$\delta^2\left(\sum_i k_i\right)$,
the exponent factors
$e^{-\frac{\sum_{i=1}^n\mathbf{k}_i^2}{4|eB|}}
e^{\frac{-i}{2}\sum_{i < j} {k_i \times k_j}}$ in an
$n$-point vertex can be rewritten as
$e^{-\frac{i}{2} \sum_{i<j} k_i \times_M k_j}$, where
$k_i \times_M k_j$ is a new cross product. It is defined as
\begin{equation}
k_i \times_M k_j = k_i \Omega k_j
\label{Mcross}
\end{equation}
with the matrix $\Omega$ being
\begin{equation}
\Omega^{ab} = \frac{1}{|eB|} \left(\begin{array}{c}
i\,\mathrm{sign}(eB) \\
-\mathrm{sign}(eB)\,i \end{array}\right).
\end{equation}
We will call $k_i \times_M k_j$ an $M$ (magnetic)-cross product.
Notice that like the matrix $\theta^{ab}$, defining the cross product,
the new matrix $\Omega^{ab}$, defining the $M$-cross product, is
anti-hermitian. [Note that in the mathematical literature, the
$M$-cross product is called the Voros product \cite{Basu:2011kh,Gouba:2011rf}.]

By using the $M$-cross product, we get the following simple structure
for an $n$-point vertex in the momentum space:
\begin{equation}
\int d^2u_{\parallel}\frac{d^2k_1}{(2\pi)^2} \ldots  \frac{d^2k_n}{(2\pi)^2}
\phi(k_1)\ldots  \phi(k_n) \delta^2\left(\sum_i k_i \right)
e^{-\frac{i}{2} \sum_{i<j} k_i \times_{M} k_j}
\label{vertex1}
\end{equation}
[compare with expression (\ref{vertex})]. Here the
field $\phi$ represents initial fields $\pi$ and $\tilde{\sigma}$.

In the coordinate space, the $M$-cross product becomes
an $M$-star product:
\begin{equation}
(\phi *_M \phi)(u) =
e^{\frac{i}{2}\Omega^{ab}
\frac{\partial}{\partial v^a}\frac{\partial}{\partial w^b}}
\phi(v)\phi(w)|_{v=w=u}.
\label{Mproduct}
\end{equation}
[compare with equation (\ref{starproduct})].
By using the $M$-star product, one can express
$n$-point vertices
through the initial $\pi$ and $\tilde{\sigma}$ fields
in the coordinate space. For example, the vertices
$\Gamma_{\tilde{\sigma}\pi\pi}$ and $\Gamma_{4\pi}$
become:
\begin{eqnarray}
\Gamma_{\tilde{\sigma}\pi\pi} &=& -\frac{N|eB|}{4\pi^2m}
\int d^2u_{\parallel}d^2u_{\perp}\,
\tilde{\sigma}*_M\pi*_M\pi\,,
\\
\Gamma_{4\pi} &=& -\frac{N|eB|}{16\pi^2m^2} \int
d^2u_{\parallel}d^2u_{\perp}\, \pi*_M\pi*_M
\pi*_M\pi.
\label{xspacevertices1}
\end{eqnarray}

In fact, by using the $M$-star product, the whole
effective action (\ref{14})
can be written in a compact and explicit form
for the case of fields independent of longitudinal coordinates
$u_{\parallel}$. First, note that for constant fields, the $M$-star product
in $\Gamma_{n\phi}$ vertices
(\ref{xspacevertices1}) is reduced to
the usual product and the
vertices come from the effective potential in that case. Then,
this
implies that, up to the measure $-\int d^4u$,
the whole effective action for fields depending on transverse coordinates
coincides with the
effective potential
in which the usual product
is replaced by the $M$-star product in the part coming from the
$\mathrm{Tr} \mbox {Ln}$ term in (\ref{14}).
As to the last
term $\frac{N}{2G} \int d^4u\, (\sigma^2 + \pi^2)$ there,
it should
stay as it is. This is because
unlike the star product, the $M$-star product
and the usual one lead to different quadratic terms in the action.
Now, by using expression for the fermion propagator, we easily find
the effective potential:
\begin{equation}
V(\sigma,\pi) = \frac{N|eB|}{8\pi^2}
\left(\sigma^2 + \pi^2\right) \left[\ln\left(\frac{\sigma^2 + \pi^2}
{\Lambda^2}\right) - 1\right]
+\frac{N}{2G}\left(\sigma^2 + \pi^2\right) + O\left(\frac{\sigma^2 + \pi^2}{\Lambda^2}\right).
\label{effpotential}
\end{equation}
[Note that this agrees with the effective potential in Eq.~(\ref{16}) in the limit of weak coupling
when $\rho=\sqrt{\sigma^2+\pi^2}\to 0$.]
Then, the effective action reads:
\begin{equation}
\Gamma(\sigma,\pi) = - \frac{N|eB|}{8\pi^2} \int d^4u
\left(\left(\sigma^2 + \pi^2\right) \left[\ln\left(\frac{\sigma^2 + \pi^2}
{\Lambda^2}\right) - 1\right] \right)_{*_M}
- \frac{N}{2G} \int d^4u\, \left(\sigma^2
+ \pi^2 \right).
\label{Maction}
\end{equation}
This expression is very convenient for calculating the $n$-point
vertices $\Gamma_{n\phi}$ \cite{Sinova:2000kx}. In Appendix~\ref{App:NCFT2}, it is shown that
the $M$-star product also naturally appears in the formalism of the
projected density operators developed in Ref.~\cite{Sinova:2000kx}
for the description of the quantum Hall effect.

While the $M$-star product is useful for practical calculations,
its connection with the multiplication operation in
a noncommutative coordinate space is not direct. This is in
contrast with the star product for which relation (\ref{W1})
takes place. Therefore, it will be useful to rewrite
action (\ref{Maction}) through the star product. It can be
done by using the smeared fields $\Sigma$ and $\Pi$. The
result is
\begin{equation}
\Gamma = - \frac{N|eB|}{8\pi^2} \int d^4u
\left(\left(\Sigma^2 + \Pi^2\right)\left[\ln\left(\frac{\Sigma^2 + \Pi^2}
{\Lambda^2}\right) - 1\right] + \frac{4\pi^2}{G|eB|}(\sigma^2 +
\pi^2)\right)_{*},
\label{staraction}
\end{equation}
where we used the fact that the star product and
the usual one lead to the same quadratic terms in the action.
Notice that the fields $\sigma$ and $\pi$ are connected
with the smeared fields through the nonlocal relation
(\ref{smeared}). This relation implies that
exponentially damping form-factors are built in the propagators of the
smeared fields. As a result, their propagators decrease rapidly, as
$\exp(-k_{\perp}^2/2|eB|)$, with
$k_{\perp}^2 \to \infty$. As will be shown in the next subsection,
this property is in particular responsible for removing the UV/IR
mixing in the model.

By using relation (\ref{W1}),
it is straightforward to rewrite the action in the
NCFT through Weyl symbols:
\begin{equation}
\Gamma = - \frac{N|eB|}{8\pi^2} \int d^2u_{\parallel}\mathbf{Tr}
\left(\left(\Sigma^2 + \Pi^2\right)\left[\ln\left(\frac{\Sigma^2 + \Pi^2}
{\Lambda^2}\right) - 1\right] + \frac{4\pi^2}{G|eB|}(\hat{\sigma}^2 +
\hat{\pi}^2)\right).
\label{ncaction}
\end{equation}
The Weyl symbols $\hat{\sigma}$ and $\hat{\pi}$ are connected
with the Weyl symbols $\hat{\Sigma}$ and $\hat{\Pi}$ through
relation (\ref{Wsmeared}).

Let us now generalize expressions (\ref{Maction}), (\ref{staraction}), and (\ref{ncaction})
to the case when fields $\pi$ and $\sigma$ depend on both transverse and longitudinal 
coordinates. First, as it follows from the expansion of the $\mathrm{Tr}\mbox{Ln}$-term
in action (\ref{14}) in a series in $\pi$ and $\tilde{\sigma}$,
the general $n$-point vertex is given by
\begin{eqnarray}
\Gamma_{n\phi} &=&  -\frac{(-i)^{n+1}N}{n} \int
d^2u_{1\perp} \ldots d^2u_{n \perp}\,
d^2u_{1 \parallel}\ldots d^2u_{n \parallel} P(u_{1 \perp},u_{2 \perp})\ldots 
P(u_{n \perp},u_{1 \perp})\,\nonumber
\\
& \times & \mathrm{tr}
\left[S_{\parallel}(u_1-u_2)\left[\tilde{\sigma}(u_2) +i\gamma^5\pi(u_2)\right]\ldots 
S_{\parallel}(u_n-u_1)\left[\tilde{\sigma}(u_1)+i\gamma^5\pi(u_1)\right]\right]
\label{n-order1}
\end{eqnarray}
[notice that here
the longitudinal part of the fermion propagator $S_{\parallel}(u_{\parallel})$
does not contain color indices].
As was shown above and in Appendix~\ref{App:NCFT1}, 
the transverse part of vertex (\ref{n-order1}) can be
expressed through the $M$-star product. Therefore, the $n$-point
vertex is
\begin{eqnarray}
\Gamma_{n\phi} &=& -\frac{(-i)^{n+1}N|eB|}{2\pi n} \int d^2u_{\perp}\,
d^2u_{1 \parallel}\ldots d^2u_{n \parallel}\, \nonumber
\\
&\times &\mathrm{tr}
\left[S_{\parallel}(u_1-u_2)
\left[\tilde{\sigma}(u_{\perp},u_{2 \parallel})+i\gamma^5\pi(u_{\perp},u_{2 \parallel})\right]\ldots 
S_{\parallel}(u_n-u_1)\left[\tilde{\sigma}(u_{\perp},u_{1 \parallel})
+i\gamma^5\pi(u_{\perp},u_{1 \parallel})\right]\right]_{*M}.
\label{n-order2}
\end{eqnarray}
This relation implies that the full effective action
can be written through the $M$-star product as:
\begin{equation}
\Gamma(\sigma, \pi) = -\frac{iN|eB|}{2\pi} \int d^2u_{\perp}\,
\mathrm{Tr}_{\parallel}\left[{\cal P}_{+}\,
\mbox{Ln} \left( i\gamma^{\parallel}\partial_{\parallel} -
(\sigma + i\gamma^5\pi) \right)\right]_{*_M} -
\frac{N}{2G}\int d^4u ({\sigma}^2 + {\pi}^2),
\label{MACTION}
\end{equation}
where the projector ${\cal P}_{+}$ is
\begin{equation}
{\cal P}_{+} \equiv \frac{1 + i\gamma^1\gamma^2 \mathrm{sign}(eB)}{2}
\label{P_}
\end{equation}
(compare with action (\ref{Maction})). Here the trace $\mathrm{Tr} _{\parallel}$,
related to the longitudinal subspace, is taken in the functional sense.

As to the form of the
effective action written through the star product and its
form in the noncommutative coordinate space, they are:
\begin{equation}
\Gamma = \frac{N|eB|}{2\pi} \int d^2u_{\perp}\,
\left[-i\,\mathrm{Tr} _{\parallel}\left[{\cal P}_{+}\,
\mbox{Ln} \left( i\gamma^{\parallel}\partial_{\parallel} -
(\Sigma + i\gamma^5\Pi) \right)\right] -
\frac{\pi}{G|eB|}\int d^2u_{\parallel}({\sigma}^2+{\pi}^2)\right]_{*}
\label{starACTION}
\end{equation}
and
\begin{equation}
\Gamma = \frac{N|eB|}{2\pi} \mathbf{Tr}\,
\left[-i\,\mathrm{Tr} _{\parallel}\left[{\cal P}_{+}\,
\mbox{Ln} \left( i\gamma^{\parallel}\partial_{\parallel} -
(\hat{\Sigma} + i\gamma^5\hat{\Pi}) \right)\right] -
\frac{\pi}{G|eB|}\int d^2u_{\parallel}(\hat{\sigma}^2+\hat{\pi}^2)
\right]
\label{ncACTION}
\end{equation}
(compare with Eqs.~(\ref{staraction}) and
(\ref{ncaction}), respectively).

This concludes the derivation of the action of the
noncommutative field theory corresponding to the NJL model
in a strong magnetic field.
In the next subsection, we will consider the dynamics
in this model in more detail.

\subsubsection{The low-energy dynamics}
\label{NJL3}

In the regime with the LLL dominance, the dynamics of neutral composites
is described by quite sophisticated NCFT (\ref{ncACTION}). In this section, 
we will show that in this model (i) there exists a well defined commutative 
limit $|eB| \to \infty$ when $[\hat{x}^a, \hat{x}^b] = 0$; (ii) the universality 
class of the low-energy dynamics, with $k_{\perp} \ll \sqrt{|eB|}$, is 
intimately connected with the dynamics in the $(1+1)$-dimensional 
Gross-Neveu (GN) model \cite{Gross:1974jv}; and (iii) there is no UV/IR 
mixing.

The key point in the derivation of action (\ref{ncACTION}) was the fact that 
the LLL fermion propagator factorizes into two parts
[see Eq.~(\ref{factorization})] and that its transverse part $P(u_{\perp},u^\prime_{\perp})$
is a projection operator on the LLL states. It is quite remarkable that it 
exactly coincides with the projection operator on the LLL states in nonrelativistic 
dynamics introduced for the description of the quantum Hall effect in
Refs.~\cite{Girvin:1984fk,Bellissard:1994JMP}. This feature is, of course, 
intimately connected with the fact that the wave functions of the LLL states 
are independent of the fermion mass $m$ [see Eq.~(\ref{wavefunctions}) for 
$K = 0$]. Therefore, the transverse dynamics in this problem is universal 
and peculiarities of the relativistic dynamics reflect themselves only in the 
$(1 + 1)$-dimensional longitudinal space.

In order to study the low-energy dynamics with $k_{\perp} \ll \sqrt{|eB|}$,
it will be instructive to consider, as in Ref.~\cite{Elias:1996zu}, the following 
continuum limit: $\Lambda^2 = C|eB| \to \infty$, with $C \gg 1$ and $m$ being fixed.
Let us consider $n$-point vertex (\ref{n-order1}) in this limit. Since the projection
operator $P(u_{i \perp},u_{i+1 \perp})$ is 
\begin{equation}
P(u_{i \perp},u_{i+1 \perp}) =  \frac{|eB|}{2\pi}\,
e^{\frac{ieB}{2}
\epsilon^{ab}u^{a}_{i \perp}u^{b}_{i+1 \perp}}\,e^{-\frac{|eB|}{4}
(u_{i \perp}- u_{i+1 \perp})^2}
\label{projector1}
\end{equation}
(see Eq.~(\ref{projector})), the point with coordinates
$u_{i \perp} = u_{i+1 \perp}$, $i= 1,\ldots, n-1$, is both a saddle and
stationary
point in the multiple integral (\ref{n-order1}) in the limit
$|eB| \to \infty$. Therefore, in order to get
the leading term of the asymptotic
expansion of that integral, one can put
$u_{i \perp} = u_{n \perp} \equiv u_{\perp}$
in the arguments of all the fields $\tilde{\sigma}(u_{i \perp})
+i\gamma^5\pi(u_{i \perp})$ there.
Then, by using relation (\ref{Pprojector})
and the equality $P(u_{\perp},u_{\perp})= |eB|/2\pi$,
we easily
integrate over transverse coordinates in (\ref{n-order1})
and obtain the following asymptotic expression:
\begin{eqnarray}
\Gamma^{(as)}_{n\phi} &=& -\frac{(-i)^{n+1}}{n}\frac{N|eB|}{2\pi} \int
d^2u_{\perp}\,
d^2u_{1 \parallel}\ldots d^2u_{n \parallel} \nonumber
\\
&\times & \mathrm{tr}
\left[S_{\parallel}(u_1-u_2)[\tilde{\sigma}(u_{\perp},u_{2 \parallel})
+i\gamma^5\pi(u_{\perp},u_{2 \parallel})]\ldots 
S_{\parallel}(u_n-u_1)[\tilde{\sigma}(u_{\perp},u_{1 \parallel})
+i\gamma^5\pi(u_{\perp},u_{1 \parallel})]\right].
\label{asvertex}
\end{eqnarray}
This equation implies that as $|eB| \to \infty$, the leading asymptotic
term in the action is
\begin{equation}
\Gamma^{(as)}(\sigma, \pi) = \frac{|eB|}{2\pi} \int
d^2u_{\perp}\,
\left[-iN\mathrm{Tr} _{\parallel}\left[{\cal P}_{+}\,
\mbox{Ln} \left( i\gamma^{\parallel}\partial_{\parallel} -
(\sigma + i\gamma^5\pi) \right)\right] -
\frac{N\pi}{G|eB|}\int d^2u_{\parallel}(\sigma^2+\pi^2)\right].
\label{asympaction}
\end{equation}
This action corresponds to a {\it commutative} field theory,
as should be in the limit $|eB| \to \infty$ [indeed, the
commutator
$[\hat{x}^{a},\hat{x}^{b}] = i\frac{1}{eB}\epsilon^{ab}$ goes to
zero as $|eB| \to \infty$]. Also, since there is no hopping
term for the transverse coordinates $u_{\perp}$ in this action,
they just play the role of a label of the fields.

Let us now compare this action with the action of the $(1+1)$-dimensional GN
model \cite{Gross:1974jv}:
\begin{equation}
\Gamma_{GN}(\sigma, \pi) =
-iN\, \mathrm{Tr} \, \mbox{Ln} 
\left( i\gamma^{\mu}\partial_{\mu} - (\sigma + i\gamma^5\pi) \right) -
\frac{N}{2G_{\rm int}}\int d^2x(\sigma^2+\pi^2),\quad \mu = 0,1,
\label{GN}
\end{equation}
where $G_{\rm int}$ is a dimensionless coupling constant. One can see
that, up to the factor $|eB|/2\pi \int d^2u_{\perp}$, these
two actions coincide if the constant $G$ in (\ref{asympaction})
is identified with $2\pi G_{\rm int}/|eB|$. In particular,
with this identification, expression (\ref{solution}) for the
dynamical mass coincides with the expression for $m$
in the GN model, $m^2 = \Lambda^2 \exp\left(- 2\pi/G_{\rm int}\right)$.
Also, using Eq.~(\ref{solution}), one can
express the coupling constant $G$ in the effective potential
(\ref{effpotential}) through the dynamical mass $m$ and cutoff
$\Lambda$. Then, up to
$O[(\sigma^2 + \pi^2)/\Lambda^2]$ terms, we get
the expression independent of the cutoff:
\begin{equation}
V(\sigma,\pi) = \frac{N|eB|}{8\pi^2}
\left(\sigma^2 + \pi^2\right)\left[\ln\left(\frac{\sigma^2 + \pi^2}{m^2}\right) - 1\right].
\label{renpotential}
\end{equation}
This renormalized form of the potential coincides with the GN potential.

As to the factor $|eB|/2\pi \int d^2u_{\perp}$, its meaning is very simple. Since
density of the LLL states is equal to $|eB|/2\pi$, this factor yields the number 
of the Landau states on the transverse plane. In other words, as $|eB| \to \infty$, 
the model is reduced to a continuum set of independent $(1+1)$-dimensional 
GN models, labeled by the coordinates in the plane perpendicular to the magnetic 
field. The conjecture about such a structure of the NJL model in the limit $|eB| \to \infty$ 
was made in Ref.~\cite{Elias:1996zu} and was based on a study of the effective 
potential and the kinetic (two derivative) term in the model. The present approach
allows us to derive the whole action and thus to prove the conjecture.

The existence of the physically meaningful limit $|eB| \to \infty$ is quite noticeable. 
It confirms that the model with the LLL dominance is self-consistent. In order to 
understand its dynamics better, it is instructive to look at the dispersion relations for
$\sigma$ and $\pi$ excitations with momenta $k_{\perp} \ll \sqrt{|eB|}$ 
(see Section~\ref{sec:NJL3+1General}):
\begin{eqnarray}
E_\pi &\simeq & \biggl[\frac{m^2}{|eB|}
\ln\biggl(\frac{|eB|}{\pi m^2}\biggr)
\mathbf{k}_{\perp}^2+k_3^2\biggr]^{1/2},
\\
E_{\sigma} &\simeq & \biggl[12~m^2 +
\frac{3m^2}{|eB|}\ln\biggl(\frac{|eB|}{\pi m^2}\biggr)
\mathbf{k}^2_{\perp}+k_3^2 \biggr]^{1/2}.
\label{dispersion}
\end{eqnarray}
We find from these relations that the transverse velocity $|\mathbf{v}_{\perp}|=|\partial
E_{\pi,\sigma}/\partial \mathbf{k}_{\perp}|$ of both $\pi$ and $\sigma$ goes rapidly 
[as $O(m^2/|eB|$)] to zero as $|eB| \to \infty$. In other words, there is no hopping
between different transverse points in this limit. For a strong but finite magnetic field,
the transverse velocity is, although nonzero, very small. In this case, the $\pi$ and 
$\sigma$ composites have a cigar-like shape: while their transverse 
size is of the order of the magnetic length $l = 1/\sqrt{|eB|}$, the longitudinal size 
is of order $1/m$, and $l \ll 1/m$.

The important point is that besides being a low-energy theory of the initial NJL 
model in a magnetic field, this truncated [based on the LLL dynamics] model is 
self-contained. In particular, in this model one can consider arbitrary large values 
for transverse momenta $k_{\perp}$, although in this case its dynamics is very
different from that of the initial NJL model. In fact, by using the expression for the 
pion propagator (\ref{propagator}) written below, it is not difficult to check that 
for $k_{\perp} \gg \sqrt{|eB|}$ the dispersion relation for $\pi$ excitations takes 
the following form:
\begin{equation}
E_{\pi} \simeq \left[4m^2\left(1-\frac{\pi^2e^{- \frac{\mathbf{k}_{\perp}^2}{|eB|}}}
{\ln^2\frac{|eB|}{m^2}}\right) + k_3^2 \right]^{1/2}.
\label{dispersion1}
\end{equation}
In this regime, the transverse velocity $|\mathbf{v}_{\perp}|$ is extremely small,
$|\mathbf{v}_{\perp}| \sim \frac{m|\mathbf{k}_{\perp}|}{|eB|} e^{-\mathbf{k}_{\perp}^2/|eB|}$, 
and a $\pi$ excitation is a loose bound state moving along the $z$ direction. 
Its mass is close to the $2m$ threshold.

Thus, we conclude that the NJL model in a strong magnetic field yields an example 
of a consistent NCFT with quite nontrivial dynamics. The point that exponentially 
damping factors occur either in vertices (for the fields $\sigma$ and $\pi$) or
in propagators (for the smeared fields) plays a crucial role in its consistency. 
Let us now show that these factors are in particular responsible for removing a 
UV/IR mixing, the phenomenon that plagues conventional nonsupersymmetric 
NCFT \cite{Minwalla:1999px}.

The simplest example of the UV/IR mixing is given by a one-loop
contribution in a propagator in the noncommutative $\phi^4$ model
with the action
\begin{equation}
S = \int d^4u \left(\frac{1}{2}(\partial_{\mu}\phi)^2 -
\frac{m^2\phi^2}{2} - \frac{g^2}{4!}\phi * \phi * \phi * \phi\right).
\end{equation}
There are planar and nonplanar one-loop
contributions in the propagator of $\phi$ in this model \cite{Minwalla:1999px}:
\begin{equation}
\Gamma^{(2)}_{nc} = \Gamma^{(2)}_{pl} + \Gamma^{(2)}_{npl}
= \frac{g^2}{3(2\pi)^4} \int \frac{d^4k}{k^2+m^2} +
\frac{g^2}{6(2\pi)^4} \int \frac{d^4k}{k^2+m^2} e^{ik \times p}.
\label{one-loop}
\end{equation}
The nonplanar contribution is specific for a noncommutative theory and is 
responsible for the UV/IR mixing. Indeed, the nonplanar contribution is equal to
\begin{equation}
\Gamma^{(2)}_{npl} = \frac{g^2}{96\pi^2}\left[\Lambda_{\rm eff}^2
- m^2\ln\left(\frac{\Lambda_{\rm eff}^2}{m^2}\right) + O(1)\right],
\label{npl}
\end{equation}
where
\begin{equation}
\Lambda_{\rm eff}^2  = \frac{1}{1/\Lambda^2 - p^{i}\theta_{ij}^2p^{j}}
\end{equation}
with $\Lambda$ being cutoff. It is clear that if the external momentum $p \to 0$, 
the nonplanar contribution (\ref{npl}) diverges quadratically. On the other hand,
for a nonzero $p$, it is finite due to the Moyal phase factor $ e^{ik \times p}$ in
the second term in expression (\ref{one-loop}) (which oscillates rapidly at large $k$).
Thus, although the Moyal factor regularizes the UV divergence, it leads to an IR 
divergence of the integral, i.e., to the UV/IR mixing.

Let us now show how the exponentially damping factors in vertices (for the fields 
$\sigma$ and $\pi$) or in propagators (for the smeared fields $\Sigma$ and $\Pi$)
remove the UV/IR mixing. We will first consider the description using the fields $\sigma$
and $\pi$. As an example, we will consider the one-loop correction in the $\pi$ 
propagator generated by the four-point interaction vertex $\Gamma_{4\pi}$.
First, from action (\ref{14}), we get this propagator in tree approximation. In 
Euclidean space it is
\begin{equation}
D_{\pi}^{\rm (tree)}(p) \simeq
\frac{4\pi^2}{N|eB|\left[\left(1-e^{-\frac{p_{\perp}^2}{2|eB|}}\right)
\ln \frac{|eB|}{m^2} +
e^{-\frac{p_{\perp}^2}{2|eB|}} \int_0^1 du \frac{p_{\parallel}^2 u}{p_{\parallel}^2
u(1-u)
+ m^2}\right]}.
\label{propagator}
\end{equation}
Then, by using this $D_{\pi}^{\rm (tree)}(p)$ and Eq.~(\ref{n-order2}) for the
vertex $\Gamma_{4\pi}$, we find the following one-loop nonplanar contribution 
to the propagator:
\begin{equation}
\frac{N|eB|}{4\pi^3} \int \frac{d^4 k}{(2\pi)^4} e^{-\frac{p_{\perp}^2+
k_{\perp}^2}{2|eB|}} e^{\frac{i}{eB}(p^1k^2-p^2k^1)}
I(p_{\parallel},k_{\parallel}) D_{\pi}^{\rm (tree)}(k),
\label{nonplanar1}
\end{equation}
where
\begin{equation}
I(p_{\parallel},k_{\parallel}) = \int d^2l_{\parallel} \frac{(l_{\parallel}^2+m^2+l_{\parallel}\cdot p_{\parallel})
[(p_{\parallel}-k_{\parallel}+l_{\parallel})^2+m^2-(p_{\parallel}-k_{\parallel}+l_{\parallel})\cdot p_{\parallel}]
+p_{\parallel}^2m^2}{(l_{\parallel}^2+m^2)[(p_{\parallel}+l_{\parallel})^2+m^2]
[(p_{\parallel}-k_{\parallel}+l_{\parallel})^2+m^2][(l_{\parallel}-k_{\parallel})^2+m^2]}\,.
\nonumber
\end{equation}
Here the integral over transverse momenta $k_{\perp}$ is
\begin{equation}
\int \frac{d^2 k_{\perp}}{(2\pi)^2}
e^{-\frac{p_{\perp}^2+k_{\perp}^2}{2|eB|}}
e^{\frac{i}{eB}(p^1k^2-p^2k^1)}D_{\pi}^{\rm (tree)}(k).
\label{integr}
\end{equation}
It is clear that due to the presence of the factor $e^{-k_{\perp}^2/2|eB|}$ and 
because $D_{\pi}^{\rm (tree)}(k)$ is finite as $k_{\perp}^2 \to \infty$, this integral 
is convergent for all values of $p_{\perp}$, including $p_{\perp} = 0$, and therefore 
there is no UV/IR mixing in this case. On the other hand, if the factor 
$e^{-\frac{p_{\perp}^2+k_{\perp}^2}{2|eB|}}$ were absent in integrand (\ref{integr}), 
we would get the integral
\begin{equation}
\int \frac{d^2 k_{\perp}}{(2\pi)^2}
e^{\frac{i}{eB}(p^1k^2-p^2k^1)}D_{\pi}^{\rm (tree)}(k)\,,
\label{integr1}
\end{equation}
which diverges quadratically at $p_{\perp} = 0$, i.e., the UV/IR mixing would occur.

Let us now turn to the description using the smeared fields.
The relation (\ref{smeared}) between the fields $\pi$ and $\Pi$
implies that their propagators are related as
\begin{equation}
D_{\Pi}(p) =  e^{\frac{-p^{2}_{\perp}}{2|eB|}}  D_{\pi}(p).
\label{Ppropagator}
\end{equation}
Since $e^{\frac{-p^{2}_{\perp}}{2|eB|}}$ is an entire function,
the absence of the UV/IR mixing in the propagator $D_{\pi}$ implies
that there is no UV/IR mixing also in the propagator $D_{\Pi}$.
This conclusion
can be checked directly, by adapting the calculations of
the one-loop correction in the propagator $D_{\pi}$
to
the $D_{\Pi}$ propagator. In this case, it is
the form-factor $e^{\frac{-p^{2}_{\perp}}{2|eB|}}$, built in the
propagator $D_{\Pi}^{\rm (tree)}(p)$, that is responsible for the absence
of
the UV/IR mixing.

This concludes the analysis in $3 + 1$ dimensions.
In the next subsection, we will generalize this analysis to
arbitrary dimensions $D = d + 1$ with $d \geq 2$.

\subsubsection{Beyond 3+1 dimensions}
\label{NJL4}

In this section, we will generalize our analysis to
arbitrary dimensions $D = d + 1$ with $d \geq 2$.
We begin by considering the NJL model in a magnetic field
in $2+1$ dimensions, choosing
its Lagrangian density similar to that in 3 + 1 dimensions:
\begin{equation}
L = \frac{1}{2}[\bar{\psi}, (i\gamma^{\mu}D_{\mu})\psi]  +
\frac{G_2}{2N} \left[(\bar{\psi}\psi)^2 + (\bar{\psi}i\gamma^5\psi)^2
\right].
\label{2+1action}
\end{equation}
Here a reducible four-dimensional representation of the Dirac matrices
is used (for details, see Section~\ref{sec:MagneticCatalysis}). In a weak coupling
regime, the dynamical mass in this model is (see Section~\ref{sec:NJL2+1EffPot}) 
\begin{equation}
m=\frac{G_{2}|eB|}{2\pi}.
\label{2+1solution}
\end{equation}
The LLL propagator is obtained from the $(3+1)$-dimensional propagator
by just omitting the $z$ and $k^3$ variables there:
\begin{equation}
S(u,u^\prime)=P(u_\perp,u^\prime_\perp)S_{\parallel}(t-t^\prime),
\label{2+1propagator}
\end{equation}
where, instead (\ref{flatspace}), the expression for $S_{\parallel}(t-t^\prime)$
is:
\begin{equation}
S_{\parallel}(t-t^\prime) = \int \frac{dk_0}{2\pi}
e^{ik_0(t-t^\prime)} \frac{i}{k_0\gamma^0 - m}\,
\frac{1 + i\gamma^1\gamma^2 \mathrm{sign}(eB)}{2}.
\label{2+1flatspace}
\end{equation}
The analysis now proceeds as in the $3 + 1$ dimensional case.
The present model corresponds to a noncommutative field theory
describing neutral composites $\sigma$ and $\pi$.
Its action written
through the star product is
\begin{equation}
\Gamma_{2} = \frac{N|eB|}{2\pi} \int d^2u_\perp\,
\left[-i\,\mathrm{Tr} _{\parallel}\left[{\cal P}_{+}\,
\mbox{Ln} \left( i\gamma^0\partial_0 -
(\Sigma + i\gamma^5\Pi)\right)\right] -
\frac{\pi}{G_2|eB|}\int d t ({\sigma}^2+{\pi}^2)\right]_{*}\,,
\label{2+1starACTION}
\end{equation}
where $\Sigma$ and $\Pi$ are smeared fields (compare with
Eq.~(\ref{starACTION})).
The action can be also written directly in the noncommutative coordinate
space:
\begin{equation}
\Gamma_{2} = \frac{N|eB|}{2\pi} \mathbf{Tr}\,
\left[-i\,\mathrm{Tr} _{\parallel}\left[{\cal P}_{+}\,
\mbox{Ln} \left( i\gamma^0\partial_0 -
(\hat\Sigma + i\gamma^5\hat\Pi)\right)\right] -
\frac{\pi}{G_2|eB|}\int dt(\hat{\sigma}^2+\hat{\pi}^2)\right]
\label{2+1ncACTION}
\end{equation}
(compare with Eq.~(\ref{ncACTION})).

In the previous Sections, it was shown that in the
regime with the LLL dominance, the divergences in
(3 + 1)-dimensional model are generated only by the
$(1+1)$-dimensional longitudinal dynamics. For the
(2 + 1)-dimensional model in this regime, a stronger
statement takes place: the model is {\it finite}. It can be shown by
repeating
the analysis used in 3 + 1 dimensions. In particular,
in the continuum limit $\Lambda \to \infty$,
the effective potential in this model is finite without
any renormalizations:
\begin{equation}
V_{2}(\sigma,\pi) = \frac{N(\sigma^2 + \pi^2)}{2G_2} -
\frac{N|eB|\sqrt{\sigma^2 + \pi^2}}{2\pi}.
\label{2+1pot}
\end{equation}
Using Eq.~(\ref{2+1solution}), one can express the coupling constant
$G_2$ in the potential through $m$ and $|eB|$. Then, the potential
takes an especially simple form:
\begin{equation}
V_{2}(\sigma,\pi) = \frac{N|eB|}{2\pi}\left(\frac{\sigma^2 + \pi^2}{2m} -
\sqrt{\sigma^2 + \pi^2}\right).
\label{2+1pot1}
\end{equation}
For momenta $k \ll \sqrt{|eB|}$,
the dispersion relation for $\pi$ excitations is (see Section~\ref{sec:NJL2+1BosSpectr}):
\begin{equation}
E_\pi \simeq \frac{\sqrt{2}m}{{|eB|}^{1/2}}\sqrt{\mathbf{k}^2}.
\label{2+1dispersion}
\end{equation}
Therefore, as in $3+1$ dimensions, the velocity
$|\mathbf{v}|=|\partial E_{\pi}/\partial \mathbf{k}|$
is strongly suppressed: in the present case it is of
order $m/{|eB|}^{1/2}$.
As to the $\sigma$ excitation, its "mass", defined as the energy
at zero momentum, is very large: $M_{\sigma} \sim
(\sqrt{eB}/m)^{1/2}\,\sqrt{|eB|}$ (see Section~\ref{sec:NJL2+1BosSpectr}). Therefore, the $\sigma$-mode
decouples from the dynamics with $k \ll \sqrt{eB}$.

As in the case of $3+1$ dimensions,
this truncated
[based on the LLL dynamics] model is self-contained and
one can consider arbitrary large values for momenta
there. It is easy
to check that for $k \gg \sqrt{|eB|}$
the dispersion relation for $\pi$ excitations takes the
form
\begin{equation}
E_{\pi} \simeq m\left(2-
e^{-\frac{\mathbf{k}^2}{2|eB|}}\right).
\label{dispersion2+1}
\end{equation}
In this regime,
the velocity becomes extremely small,
$|\mathbf{v}| \sim \frac{m|\mathbf{k}|}{|eB|}
e^{-\mathbf{k}^2/2|eB|}$, and a $\pi$ excitation is a loosely bound
state. Its mass is close to the $2m$ threshold.

As was shown in Section~\ref{NJL3}, in the
limit $|eB| \to \infty$ the $(3+1)$-dimensional model is reduced
to a continuum set of independent
(1+1)--dimensional Gross--Neveu models labeled by the
coordinates in the plane perpendicular to the magnetic field.
Similarly to that, in the case of 2 + 1 dimensions, in the limit
$|eB| \to \infty$ the model is reduced to a set of (0+1)-
dimensional (i.e., quantum mechanical) models labeled by
two spatial coordinates.

A new feature of the $(2+1)$-dimensional
model is a confinement dynamics for charged
particles: they do not propagate
in a magnetic background.
On the other
hand, since neutral composites are free to propagate in a magnetic
field,
one can define asymptotic states and
$S$-matrix for them. The $S$-matrix should be unitary in the
subspace of neutral composites.

Let us now consider the case of higher dimensions
$D=d+1$ with $d>3$. First of all, recall
that for an even $d$, by using spatial rotations,
the noncommutativity tensor $\theta^{ab}$ in a noncommutative theory
with $[\hat{x}^a,\hat{x}^b]=i\theta^{ab}$
can be reduced to the following canonical skew-diagonal
form with skew-eigenvalues $\theta^a,\, a=1,\ldots,d/2$ \cite{Douglas:2001ba,Szabo:2001kg}:
\begin{equation}
\theta^{ab} = \left(%
\begin{array}{ccccccccc}
  0 & \theta^1 & & & & & & & \\
  -\theta^1 & 0 & & & & & & & \\
  & & & \cdot & & & & & \\
  & & & & \cdot & & & & \\
  & & & & & \cdot & & & \\
  & & & & & & & 0 & \theta^{d/2}  \\
  & & & & & & & -\theta^{d/2} & 0
\end{array}%
\right).
\label{canonical}
\end{equation}
If $d$ is odd, then the number of canonical skew-eigenvalues of
$\theta^{ab}$
is equal to $[d/2]$, where $[d/2]$ is the integer part
of
$d/2$, and the canonical form of $\theta^{ab}$ is similar to
(\ref{canonical})
except that there are additional one zero column and one
zero row.

On the other hand, a constant magnetic field in $d$ dimensions is
also characterized by $[d/2]$ independent parameters, and the strength tensor
$F^{ab}$ can be also reduced to the canonical skew-diagonal
form \cite{Gavrilov:1996pz,Gorbar:2000ku}:
\begin{equation}
F^{ab} = \sum_{c=1}^{[d/2]} B^{c} (\delta^a_{2c-1}\delta^b_{2c}
-\delta^b_{2c-1}\delta^a_{2c}).
\end{equation}
The corresponding nonzero components of the vector potential are equal to
\begin{equation}
\mathbf{A}  = \left(-\frac{B^1u^2}{2},\,\frac{B^1u^1}{2},\, \ldots  \,,
-\frac{B^{[d/2]}u^{2[d/2]}}{2},\,
\frac{B^{[d/2]}u^{2[d/2]-1}}{2}\right).
\end{equation}
Thus, we see that there is one-to-one mapping between the skew-eigenvalues
of the noncommutativity tensor $\theta^{ab}$ and the independent parameters 
of the spatial part of the electromagnetic strength tensor $F^{ab}$ in a space of 
any dimension $d \geq 2$.

Chiral symmetry breaking in the NJL model in a strong magnetic field in dimensions 
with $d>3$ was studied in Ref.~\cite{Gorbar:2000ku}. By using results of that paper, 
it is not difficult to extend our analysis in $3+1$ and $2+1$ dimensions to the case
of $d > 3$. The crucial point in the analysis is the structure of the Fourier transform 
of the translationally invariant part of the LLL propagator. If all $B^a$ are nonzero,
one can show that it is
\begin{equation}
\bar{S}_{[d/2]}(k) =
i\, \exp\left[-\sum_{a=1}^{[d/2]}\frac{k_{2a-1}^2 +\,
k_{2a}^2}{|eB^a|}\right]\,
\frac{k_{\parallel}\gamma^{\parallel} + m}{k_{\parallel}^2-m^2}\, \Pi_{a=1}^{[d/2]}
\left[1+ i\gamma^{2a-1}\gamma^{2a}\mathrm{sign}(eB^a)\right],
\label{d+1fourier}
\end{equation}
where $k^{\parallel} = k^0$ if $d$ is even and $k^{\parallel} = (k^0,k^d)$ if $d$ is
odd. If some $B^c=0$, then, for each $c$, the longitudinal part $k^{\parallel}$ gets 
two additional components, $k^{2c-1}$ and $k^{2c}$, and the corresponding terms 
are absent in the transverse part of expression (\ref{d+1fourier}). Thus, like in $3+1$ 
and $2+1$ dimensions, the LLL propagator factorizes into the transverse and longitudinal
parts. The projection operator $P_{n}(u_{\perp},u^\prime_{\perp})$ on the LLL is now 
equal to the direct product of the projection operators (\ref{projector}) in the 
$x^{2a-1}x^{2a}$-planes with nonzero $B^{a}$ [here the subscript $n$ is the number 
of nonzero independent components of $F^{ab}$].

Because of that, it is clear that the NJL model in a strong magnetic field in a space of 
arbitrary dimensions $d \geq 2$ corresponds to a noncommutative field theory with 
parameters $\theta^{ab}$ expressed through the magnetic part of the strength tensor 
$F^{ab}$. Its action is [compare with expressions (\ref{ncACTION}) and (\ref{2+1ncACTION})]:
\begin{equation}
\Gamma_{n} = N \mathbf{Tr}\,\left[
-\frac{i\Pi_{a=1}^{n}|eB^a|}{(2\pi)^{n}}
\, \mathrm{Tr} _{\parallel}\left[{\cal P}^{+}_n\,
\mbox{Ln} \left( i\gamma^{\parallel}\partial_{\parallel} -
(\hat{\Sigma} + i\gamma^5\hat{\Pi})\right)\right] -
\frac{1}{2G_d}\int d^{D-2n}u_{\parallel}\,(\hat{\sigma}^2+\hat{\pi}^2)
\right]\,,
\label{d+1ncACTION}
\end{equation}
where $n$ is the number of nonzero independent components of $F^{ab}$
and the projector ${\cal P}^{+}_n$ equals the direct product of projectors 
(\ref{P_}) in the $x^{2a-1}x^{2a}$-planes with nonzero $B^{a}$. In particular,
for a magnetic field configuration with the maximal number $n = [d/2]$
of independent nonzero tensor components, the dynamics is quasi-$(1+1)$-dimensional 
for odd $d$ and finite for even $d$. In the latter case the model describes a 
confinement dynamics of charged particles. Also, as all $|eB^a| \to \infty$, 
the model is reduced either to a continuum set of $(1+1)$-dimensional GN 
models labeled by $d-1$ spatial coordinates (odd $d$) or to a set of quantum 
mechanical models labeled by $d$ spatial coordinates (even $d$).

In the next section, we will describe the connection between gauge theories
in a magnetic field and NCFT. It will be shown that in that case this connection
is more sophisticated and interesting than that for the NJL model.

\subsection{Gauge theories in a magnetic field as NCFT}
\label{gauge-NCFT}

In the previous section, the connection between the dynamics in the NJL model
in a strong homogeneous magnetic field and that in NCFT has been studied.
The main conclusion was that although relativistic field theories in the regime 
with the lowest Landau level (LLL) dominance indeed determine a class
of NCFT, these NCFT are different from the conventional ones considered 
in the literature. In particular, the UV/IR mixing, taking place in the conventional 
NCFT \cite{Minwalla:1999px}, is absent in these theories. The reason of that 
is an inner structure (i.e., dynamical form factors) of electrically neutral 
composites in these theories. We emphasize that in order to establish
the connection between dynamics in a homogeneous magnetic field and
dynamics in NCFT, it is necessary to consider neutral fields. The point is that
in homogeneous magnetic backgrounds, momentum is a good quantum number 
only for neutral states and therefore one can introduce asymptotic states and 
S-matrix only for them.

While studies of the origins of the noncommutativity in relativistic quantum field 
theories in a magnetic field are interesting in themselves, it is even more important 
that they lead to new physical results. In particular, as was shown in Section~\ref{NJL2}, 
the NCFT approach allows one to derive interaction vertices for neutral composites. 
These vertices automatically include [in the form of the cross (Moyal) product, see 
Section~\ref{NJL2}] {\it all} powers of transverse derivatives [i.e., the derivatives with 
respect to coordinates orthogonal to a magnetic field]. This result is quite noticeable 
because the dimension of the transverse subspace is two and it is very seldom that 
one can with a good accuracy calculate vertices for composites in quantum field
models with spatial dimensions higher than one.

In the analysis in Section~\ref{NJL-NCFT}, it was shown that there exist two 
equivalent descriptions of the dynamics of the NJL model in a magnetic 
field as NCFT. In the first description, one uses the conventional
composite operators $\sigma(u) \sim \bar{\psi}(u) \psi (u)$
and $\pi(u) \sim i\bar{\psi}(u)\gamma_5 \psi (u)$. In this case,
besides the usual Moyal factor, the additional Gaussian-like (form-)factor
$e^{-\left(\sum_{i=1}^n\mathbf{k}_{\perp i}^2\right)/{4|eB|}}$
occurs in $n$-point interaction vertices of the fields $\sigma(u)$
and $\pi(u)$. Here $\mathbf{k}_{\perp i}$ is a momentum
of the $i$-th composite in a plane orthogonal to the magnetic field.
This form factor reflects an inner structure of composites and play an 
important role in providing consistency of the NCFT. In particular, because 
of them, the UV/IR mixing is absent in these theories. 

In the second description, one considers other, ``smeared" fields $\Sigma(u)$ and $\Pi(u)$,
connected with $\sigma(u)$ and  $\pi(u)$ through a nonlocal transformation. Then, 
while the additional factors are removed in the vertices for the smeared fields, they 
appear in their propagators, again resulting in the UV/IR mixing removal.

The Gaussian form of the exponentially damping form factors reflects the
Landau wave functions of fermions on the LLL. The form factors are intimately
connected with the holomorphic representation in the problem of quantum
oscillator (for a review of the holomorphic representation, see Ref.~\cite{Faddeev:1991bo}). 
Indeed, in the problem of a free fermion in a magnetic field, 
the dynamics in a plane orthogonal to the magnetic field
is an oscillator-like one.\footnote{In particular, the holomorphic representation
is widely used for the description of the quantum Hall effect \cite{Girvin:1984fk,Kivelson:1987uq}.}
And because weak short-range interactions between fermions in the
NJL model in a magnetic field do not change this feature of the dynamics, 
the form factors in that model have the Gaussian form. 

But what happens
in the case of more sophisticated dynamics, such as those with
long-range interactions  in gauge models? To find the answer
to this question is the primary goal of this section.

To answer this question, we will extend the analysis of Section~\ref{NJL-NCFT}
to the more complicated cases of QED and QCD in a strong magnetic field. 
It will be shown that in these gauge models, the connection of the dynamics 
with NCFT is much more sophisticated than that in the NLJ model. It is not 
just that the damping form-factors are not Gaussian in these models
but there does not exist an analogue of the smeared fields at all.
As a result, their interaction vertices cannot be
transformed into the form of vertices in conventional NCFT. On the other
hand, it is quite remarkable that, by using the Weyl symbols 
of the fields (see Section~\ref{NJL-NCFT}), their vertices
can nevertheless be represented
in the space with noncommutative spatial coordinates. The dynamics
they describe correspond to complicated nonlocal NCFT. We will call these
theories type II nonlocal NCFT. The name type I nonlocal NCFT will be
reserved for models similar to the NJL model in a magnetic field, for
which smeared fields exist. In both these cases, the term ``nonlocal"
reflects the point that, besides the Moyal factor, additional form factors
are present in the theories.

The crucial distinction between these two types of models is
in the characters of their interactions. While the interaction in the
NJL-like
models is local (short-range), it is long-range in gauge theories.
This point is reflected in a much richer structure of neutral composites
in the latter. We believe that both these types of
nonlocal NCFT can be relevant not only for relativistic field theories
but also for nonrelativistic systems in a magnetic field.
In particular, while type I NCFT can be relevant for the
description of the quantum Hall effect in condensed matter systems
with short-range interactions,
type II NCFT can be relevant in studies of this effect in
condensed matter systems with long-range interactions (such as carbon materials).

\subsubsection{Chiral symmetry breaking in QED in a magnetic field}
\label{gauge-NCFT-1}

The central dynamical phenomenon in relativistic field theories in a magnetic
field is the phenomenon of the magnetic catalysis. For QED, it was discussed in detail
in Section~\ref{sec:MagCatQED}. In this section, we will briefly describe those features
of this phenomenon in QED, which are relevant for the present purposes.

The crucial point of the analysis in Section~\ref{sec:MagCatQED}
was to recognize that there exists a special nonlocal (and non-covariant)
gauge in which the improved rainbow (ladder) approximation is reliable 
in this problem, i.e., there is a consistent truncation of the system of the 
Schwinger--Dyson equations. We recall that in the improved rainbow (ladder)
approximation the vertex $\Gamma^{\nu}(x,y,z)$ is taken to be bare and the
photon propagator is taken in the one-loop approximation. The full photon 
propagator in this gauge has the form [compare with (\ref{Dmunu-non-local})]
\begin{equation}
{D}_{\mu\nu}(k)=
i\frac{g^{\parallel}_{\mu\nu}}{k^2+k_{\parallel}^2\,
\Pi(k_{\perp}^2, k_{\parallel}^2)}
+ \,i\frac{g^{\perp}_{\mu\nu}k^2
- (k^{\perp}_{\mu}k^{\perp}_{\nu} + k^{\perp}_{\mu}
k^{\parallel}_{\nu} + k^{\parallel}_{\mu}k^{\perp}_{\nu})}
{(k^2)^2},
\label{non-local-NCFT}
\end{equation}
where $\Pi(k_{\perp}^2, k_{\parallel}^2)$ is the polarization operator and the 
symbols $\perp$ and $\parallel$ in $g_{\mu\nu}$ and $k_\mu$ are related to 
the transverse $(1,2)$ and longitudinal $(0,3)$ space-time components, 
respectively (we consider a constant magnetic field $B$ directed in the $+z$
direction). Because of the spin polarization for the positively (negatively) 
charged LLL fermion states along (opposite) to the magnetic field, the transverse
degrees of freedom decouple from the LLL dynamics, and
only the first term in photon propagator $D_{\mu\nu}$ (\ref{non-local-NCFT}),
proportional to $g^{\parallel}_{\mu\nu}$, is relevant. Therefore, as the full 
photon propagator in this special gauge, one can take the Feynman-like 
noncovariant propagator
\begin{equation}
{D}_{\mu\nu}(k)=
i\frac{g^{\parallel}_{\mu\nu}}{k^2+k_{\parallel}^2
\Pi(k_{\perp}^2, k_{\parallel}^2)}.
\label{trunc}
\end{equation}
It is important that this propagator does
not lead to infrared mass singularities in loop corrections in a
vertex (see Section~\ref{sec:MagCatQED}) that makes the improved rainbow approximation
to be reliable in this gauge (because of
mass singularities in covariant gauges, the loop corrections in the vertex
are large there.).

As was pointed in Section~\ref{sec:MagCatQED}, because the kinematic region
$m_{\rm dyn}^2 \ll |k_{\parallel}^2|\,,\, k_{\perp}^2 \ll |eB|$ is mostly
responsible for generating the fermion mass, the polarization operator can be calculated in one-loop approximation. It is
\begin{equation}
\Pi(k_{\perp}^2,k_{\parallel}^2)
\simeq -\frac{2\tilde{\alpha}_{b} |eB|}{\pi k_{\parallel}^2},
\label{polarization}
\end{equation}
where $\tilde{\alpha}_{b} \equiv N_f\alpha_b =\frac{N_{f}e_{b}^2}{4\pi}$.
Here $N_f$ is the number of fermion flavors and $\alpha_b$ is the QED 
running coupling related to the magnetic scale $\sqrt{|eB|}$.

In this approximation, photon propagator (\ref{trunc}) becomes a propagator of a free
massive boson with $M_{\gamma}^2=2\tilde{\alpha}_{b}|eB|/\pi$:
\begin{equation}
{D}_{\mu\nu}(u)=
\frac{i}{(2\pi)^4} \int d^4k\, e^{-iku}\,
\frac{g^{\parallel}_{\mu\nu}}{k^2 - M_{\gamma}^2}.
\label{trunc1}
\end{equation}
The corresponding  
approximation is reliable when the parameter $\tilde{\alpha}_b$ is small, i.e.,
$\tilde{\alpha}_b \ll 1$. The dynamically generated mass of fermions is
\begin{equation}
m_{\rm dyn} = C |eB|^{1/2} F(\tilde{\alpha}_{b}) \exp\left[-\frac{\pi N_f}
{\tilde{\alpha}_{b}\ln(C_1/\tilde{\alpha}_{b})}\right],
\label{mass}
\end{equation}
where $F(\tilde{\alpha}_{b}) \simeq  (\tilde{\alpha}_{b})^{1/3}$,
$C_1 \simeq 1.82$ and $C$ is a numerical constant of order one.
[This is the same as the expression in Eq.~(\ref{dynmass}), but with the replacement 
$N\alpha \to \tilde{\alpha}_{b}$.]

Note also that in the case of large $N_f$, when one can use the $1/N_f$ expansion, 
this approximation is reliable for arbitrary $\tilde{\alpha}_b$ \cite{Gusynin:2003dz}.
In particular, in the strong coupling limit for large $N_f$, the dynamical mass takes 
the form \cite{Gusynin:2003dz}
\begin{equation}
m_{\rm dyn} \simeq \sqrt{|eB|} \exp(-N_f).
\label{mass1}
\end{equation}

\subsubsection{The effective action of QED in a magnetic field}
\label{gauge-NCFT-2}

In this section, we analyze the dynamics in the chiral symmetric QED in a strong magnetic field.
We will consider both the weakly coupling regime, with $\tilde{\alpha}_b \ll 1$, and (for large $N_f$)
the strongly coupling regime with $\tilde{\alpha}_b \gtrsim 1$. In both these cases, one can use the
results of the analysis in Section~\ref{sec:MagCatQED}.

The chiral symmetry in this model is $\mathrm{SU}(N_f)_{L} \times \mathrm{SU}(N_f)_{R}$. 
The generation of a fermion mass breaks this symmetry down to $\mathrm{SU}(N_f)_{V}$ 
and, as a result, $N_{f}^2 - 1$ neutral Nambu-Goldstone (NG) composites 
$\pi^A$, $A = 1, 2,\ldots,N_{f}^2 - 1$, occur (we do not consider here the anomalous 
$\mathrm{U}(1)_A$). Our aim in this section is to derive the interaction vertices for $\pi^A$ 
in the regime with the LLL dominance and clarify whether their structure corresponds to a NCFT.

Integrating out the photon field $A_{\mu}$, we obtain the following
nonlocal effective action for fermions in QED in a magnetic field:
\begin{equation}
S = \int d^4u\, \bar{\psi}(u) i\gamma^{\mu}{\cal D}_{\mu}\psi (u) -
2i\pi \alpha_{b}\int d^4ud^4u^\prime\,
\bar{\psi}(u)\gamma^{\mu}\psi(u)
D_{\mu\nu}^{(0)}(u-u^\prime)\bar{\psi}(u^\prime)\gamma^{\nu}\psi(u^\prime),
\label{initialaction}
\end{equation}
where, in the lowest order, the bare propagator corresponding to
propagator (\ref{trunc}) is
\begin{equation}
D_{\mu\nu}^{(0)}(u-u^\prime) = \frac{i}{(2\pi)^4} \int d^4k\, e^{-ik(u-u^\prime)}\,
\frac{g^{\parallel}_{\mu\nu}}{k^2},
\label{bare}
\end{equation}
and the vector potential $A_{\mu} $ in the covariant derivative 
${\cal D}_{\mu} = \partial_{\mu} + ieA_{\mu} $ in (\ref{initialaction}) 
describes a constant magnetic field $B$ directed in the $+z$ direction. 
The vector potential here is assumed to be in the symmetric
gauge (\ref{eq:symm_gauge}).
Following the auxiliary field method developed for theories with
nonlocal interaction in Refs.~\cite{Kleinert:1976xz,Kugo:1978ct}, we add the term
\begin{equation}
\Delta S = - 2i\pi \alpha_{b}\int d^4ud^4u^\prime\,
\mathrm{tr}\left\{\gamma^{\mu}\,
[\varphi_{a}^{b}(u,u^\prime)-
\psi_{a}(u)\bar{\psi}^{b}(u^\prime)]\,\gamma^{\nu}\,[\varphi_{b}^{a}(u^\prime,u) -
\psi_{b}(u^\prime)\bar{\psi}^a(u)]
\right\}D_{\mu\nu}^{(0)}(u-u^\prime)
\label{auxil}
\end{equation}
in the action. Here $\varphi_{a}^{b}(u,u^\prime)$ is a bilocal auxiliary field
with
the indices $a$ and $b$ from the fundamental representation of
$\mathrm{SU}(N_f)$. Then, we obtain the action
\begin{eqnarray}
S &=& \int d^4u\, \bar{\psi}i\gamma^{\mu}{\cal D}_{\mu}\psi  -
4i\pi \alpha_{b}
\int d^4ud^4u^\prime\, \bar{\psi}(u)\gamma^{\mu}\varphi(u,u^\prime)\gamma^{\nu}\psi(u^\prime)
D_{\mu\nu}^{(0)}(u-u^\prime)
\nonumber\\
&-& 2i\pi \alpha_{b}\int d^4ud^4u^\prime\, \mathrm{tr}\,[\gamma^{\mu}\,
\varphi(u,u^\prime)\,\gamma^{\nu}\,\varphi(u^\prime,u)]\, D_{\mu\nu}^{(0)}(u-u^\prime)
\label{intermediateaction}
\end{eqnarray}
[here, for clarity of the presentation, we omitted the $\mathrm{SU}(N_f)$
indices]. Integrating over fermions, we find
\begin{eqnarray}
S(\varphi) &=& -i\mbox{TrLn}\left[\gamma^{\mu}i{\cal D}_{\mu}\delta^4(u-u^\prime) -
4i\pi \alpha_{b} \gamma^{\mu}\varphi(u,u^\prime)\gamma^{\nu}
D_{\mu\nu}^{(0)}(u-u^\prime)\right]
\nonumber\\
&-&2i\pi \alpha_{b} \int d^4ud^4y\, \mathrm{tr}[\gamma^{\mu}
\varphi(u,u^\prime)\gamma^{\nu}\varphi(u^\prime,u)]
D_{\mu\nu}^{(0)}(u-u^\prime),
\label{action}
\end{eqnarray}
where
$\mathrm{Tr}$ and $\mbox{Ln}$ are taken in the functional sense.

Following Ref.~\cite{Kugo:1978ct}, we can expand $\varphi(u,u^\prime)$ as
\begin{eqnarray}
\varphi(u,u^\prime)&=& \varphi_0(u,u^\prime) + \tilde{\varphi}(u,u^\prime), 
\label{exp}
\\
\tilde{\varphi}(u,u^\prime) &=& \sum_n \int \frac{d^4P}{(2\pi)^4} \phi_n(P)\chi^{(l)}_n(u,u^\prime;P).
\label{expansion}
\end{eqnarray}
Here $\varphi_0(u,u^\prime)$ satisfies the equation
\begin{equation}
\frac{\delta S}{\delta \varphi} = 0,
\label{extremum}
\end{equation}
which is equivalent to the Schwinger--Dyson equation
\begin{equation}
S_{(l)}^{-1}(u,u^\prime) = S^{-1}_0(u,u^\prime) -
4\pi \alpha_b \gamma^{\mu}S_{(l)}(u,u^\prime)\gamma^{\nu} D_{\mu\nu}^{(0)}(u-u^\prime),
\label{SDequation}
\end{equation}
where $S_0$ is the bare fermion propagator and $S_{(l)} \equiv \varphi_0$ is the full fermion 
propagator in the rainbow (ladder) approximation. As to Eq.~(\ref{expansion}), $\phi_n(P)$ is 
a field operator describing a neutral composite $|n,P\rangle$ and $\chi^{(l)}_n(u,u^\prime;P)$ 
are solutions of the {\it off-mass-shell} Bethe--Salpeter  equation in the ladder approximation,
\begin{equation}
\chi^{(l)}(u,u^\prime;P) = 4\pi \alpha_{b} \lambda(P) \int d^4u_1d^4u^\prime_{1}\,
S_{(l)}(u,u_{1})\,\gamma^{\mu}
\chi^{(l)}(u_1,u^\prime_{1};P)\,\gamma^{\nu}S_{(l)}(u^\prime_{1},y)\,
D_{\mu\nu}^{(0)}(u_1-u^\prime_{1}).
\label{BSequation}
\end{equation}
The insertion of factor $\lambda(P) \neq  1$ in this equation allows us to consider 
off-mass-shell states with an arbitrary mass $M^2 = P^2$. The on-mass-shell states 
correspond to $\lambda(P) = 1$.

Using Eqs.~(\ref{exp}) and (\ref{SDequation}), the action (\ref{action}) can be rewritten as
\begin{eqnarray}
S(\tilde{\varphi}) &=& -i\mbox{TrLn}\left[S_{(l)}^{-1}(u,u^\prime) -
4\pi \alpha_{b}
\gamma^{\mu}\tilde{\varphi}(u,u^\prime)\gamma^{\nu} D_{\mu\nu}^{(0)}(u-u^\prime)\right]
\nonumber\\
&-& 2i\pi \alpha_{b}\int d^4ud^4y\, \mathrm{tr}[\gamma^{\mu}
(\varphi_0(u,u^\prime) + \tilde{\varphi}(u,u^\prime))\gamma^{\nu}
(\varphi_0(u^\prime,u) + \tilde{\varphi}(u^\prime,u))] D_{\mu\nu}^{(0)}(u-u^\prime).
\label{action1}
\end{eqnarray}
Expanding now the action $S(\tilde{\varphi})$ in powers of $\tilde{\varphi}$ and ignoring 
its part that does not depend on $\tilde{\varphi}$, we obtain
\begin{eqnarray}
S(\tilde{\varphi}) &=& \sum_{n=2}^{\infty} \frac{i}{n} \int
d^4u_1d^4u^\prime_{1}\,\ldots\,d^4u_n
d^4u^\prime_n\,
\mathrm{tr}\,[S_{(l)}(u_1,u^\prime_{1})\varphi_D(u^\prime_{1},u_2)S_{(l)}(u_2,u^\prime_2)
\varphi_D(u^\prime_2,u_3)
\,\ldots \,S_{(l)}(u_{n-1},u^\prime_n)\varphi_D(u^\prime_n,u_1)\,]
\nonumber\\
&-& 2i\pi \alpha_{b}\int d^4ud^4y \mathrm{tr}[\gamma^{\mu}
\tilde{\varphi}(u,u^\prime)\gamma^{\nu}\tilde{\varphi}(u^\prime,u)] D_{\mu\nu}^{(0)}(u-u^\prime),
\label{action011}
\end{eqnarray}
where
\begin{equation}
\varphi_D(u,u^\prime) = 4\pi \alpha_{b}
\gamma^{\mu}\tilde{\varphi}(u,u^\prime)\gamma^{\nu}D_{\mu\nu}^{(0)}(u-u^\prime).
\end{equation}
Because $\varphi_0$ satisfies the Schwinger--Dyson equation (\ref{extremum}),
the term linear in $\tilde{\varphi}$ is absent
in (\ref{action011}).

As is clear from the discussion in Section~\ref{gauge-NCFT-1},
one should use the improved rainbow (ladder) approximation
in the present problem.
The Schwinger--Dyson equation for the fermion
propagator in this approximation takes the form
\begin{equation}
S^{-1}(u,u^\prime) = S^{-1}_0(u,u^\prime) - 4\pi \alpha_b \gamma^{\mu}S(u,u^\prime)\gamma^{\nu}
D_{\mu\nu}(u-u^\prime),
\label{SDequation1}
\end{equation}
where the photon propagator $D_{\mu\nu}(x)$ is given in Eq.
(\ref{trunc1}).
The off-mass-shell Bethe-Salpeter equation
in the improved ladder approximation is given by
\begin{equation}
\chi(u,u^\prime;P) = 4\pi \alpha_{b} \lambda(P) \int d^4u_1d^4u^\prime_{1}\,
S(u,u_{1})\,\gamma^{\mu}
\chi(u_1,u^\prime_{1};P)\,\gamma^{\nu}S(u^\prime_{1},u^\prime)\,D_{\mu\nu}(u_1-u^\prime_{1}).
\label{BSequation1}
\end{equation}
The comparison of Eqs.~(\ref{SDequation1}), (\ref{BSequation1})
with Eqs.~(\ref{SDequation}),
(\ref{BSequation}) suggests that in the improved
rainbow (ladder) approximation
the effective action (\ref{action011}) should be
replaced by the following one:
\begin{eqnarray}
S(\tilde{\varphi}) &=& \sum_{n=2}^{\infty} \frac{i}{n} \int
d^4u_1d^4u^\prime_{1}\,\ldots \,d^4u_n
d^4u^\prime_n\,
\mathrm{tr}\,[S(u_1,u^\prime_{1})\varphi_D(u^\prime_{1},u_2)S(u_2,u^\prime_2)
\varphi_D(u^\prime_2,u_3)
\,\ldots \,S(u_{n-1},u^\prime_n)\varphi_D(u^\prime_n,u_1)\,]
\nonumber \\ 
&-& 2i\pi \alpha_{b}\int d^4ud^4y \mathrm{tr}[\gamma^{\mu}
\tilde{\varphi}(u,u^\prime)\gamma^{\nu}\tilde{\varphi}(u^\prime,u)] D_{\mu\nu}(u-u^\prime),
\label{action11}
\end{eqnarray}
where
\begin{equation}
\varphi_D(u,u^\prime) = 4\pi \alpha_{b}
\gamma^{\mu}\tilde{\varphi}(u,u^\prime)\gamma^{\nu}D_{\mu\nu}(u-u^\prime)
\end{equation}
and
\begin{equation}
\tilde{\varphi}(u,u^\prime)=\sum_n \int \frac{d^4P}{(2\pi)^4}
\phi_n(P)\chi_n(u,u^\prime;P)
\label{expansion1}
\end{equation}
(compare with Eq.~(\ref{expansion})). Here $\chi_n(u,u^\prime;P)$ are
solutions of the off-mass-shell Bethe-Salpeter equation (\ref{BSequation1}).

As in the case of the NJL model (see Section~\ref{NJL-NCFT}), 
the LLL fermion propagator in the improved rainbow approximation in QED
factorizes into two parts: the part depending on the
transverse coordinates $u_{\perp} = (u,u^\prime)$ and that depending on the
longitudinal coordinates $u_{\parallel} = (t,z)$,
\begin{equation}
S_{LLL}(u,u^\prime) = P(u_{\perp},u^\prime_{\perp})S_{\parallel}(u_{\parallel}-u^\prime_{\parallel}).
\label{LLLpropagatorAnotherForm}
\end{equation}
Here $P(u_{\perp},u^{\prime}_{\perp})$ the projection operator on the LLL states
\ref{NJL-NCFT}, which in the symmetric gauge is
\begin{equation}
P(u_{\perp},u^\prime_{\perp}) = \frac{|eB|}{2\pi}
e^{\frac{ieB}{2}\epsilon^{ab}u_{\perp}^au_{\perp}^{\prime b}}
e^{-\frac{|eB|}{4}(u_{\perp}-u_{\perp}^\prime)^2}.
\label{projectorMore}
\end{equation}
The first exponential factor in $P(u_{\perp},u^\prime_{\perp})$ is the
Schwinger phase \cite{Schwinger:1951nm}. Its presence is dictated by the
group of magnetic translations in this problem (for more details, see 
Section~\ref{connection}). As was shown in Section~\ref{NJL-NCFT},
it is the Schwinger phase that is responsible for producing the Moyal
factor (a signature of NCFT) in interaction vertices.

As to the longitudinal part, in the improved rainbow approximation
it has the form \cite{Gusynin:1998zq,Gusynin:1999pq,Gusynin:2003dz}
\begin{equation}
S_{\parallel}(u_{\parallel}-u^\prime_{\parallel}) = \int \frac{d^2k_{\parallel}}{(2\pi)^2}
e^{ik_{\parallel}(u_{\parallel}-u^\prime_{\parallel})} \frac{i}{k_{\parallel}\gamma^{\parallel} - m(k_{\parallel}^2)}\,
\frac{1 + i\gamma^1\gamma^2 \mbox{sign}(eB)}{2},
\label{flatspace2}
\end{equation}
i.e., it has the form of a fermion propagator in 1+1 dimensions.
The dynamical mass function $m(k_{\parallel}^2)$ is
essentially constant for
$k_{\parallel}^2 \lesssim  |eB|$ and
rapidly decreases for $k_{\parallel}^2 > |eB|$ \cite{Gusynin:1998zq,Gusynin:1999pq,Gusynin:2003dz}.
Therefore, a simple and reliable approximation for $m(k_{\parallel}^2)$
is
\begin{equation}
m(k_{\parallel}^2) = \theta (|eB| - k_{\parallel}^2)\, m_{\rm dyn},
\label{mfunction}
\end{equation}
where $\theta (x)$ is the step function and $m_{\rm dyn}$ is the fermion pole
mass (\ref{mass}) or (\ref{mass1}). [This conclusion was later confirmed
in Refs.~\cite{Alexandre:2001vu,Elizalde:2002ca}.]

The operators  $\phi_n(P)$ in equation (\ref{expansion1}) describe all
possible neutral fermion-antifermion composites. The description of their 
interaction vertices in QED in a magnetic field is quite a formidable problem.
Henceforth, we limit ourselves to considering only the interaction
vertices for the NG boson states $|A;P\rangle $ and their operators
$\phi^{A}(P)$. For a brief discussion concerning other states, see
Section~\ref{connection}.\footnote{As is well known, in some approximations, 
the Bethe-Salpeter equation is plagued by the appearance of spurious solutions. 
Because we restrict ourselves to calculating the vertices for the NG bosons, which 
are manifestly physical, no such problem occurs in this study.}

In a magnetic field,
the wave function of the states $|A;P\rangle $ satisfying Bethe-Salpeter
equation (\ref{BSequation1})
has the following form \cite{Gusynin:1995nb}:
\begin{equation}
\chi^{A}(u,u^\prime;P) \equiv \langle 0|T\psi(u)\bar{\psi}(u^\prime)|A;P\rangle  = e^{-iPX}
e^{ier^{\mu}A_{\mu} (X)} \tilde{\chi}^{A}(r;P),
\label{chitilde}
\end{equation}
where $r=u-u^\prime$, $X=(u+u^\prime)/2$ and, that is very important,
the function $\tilde{\chi}^{A}(r;P)$ is independent of the center of mass 
coordinate $X$. This fact reflects the existence of the group of magnetic 
translations in the present problem. As in the case of the fermion propagator, 
the presence of the Schwinger factor $e^{ier^{\mu}A_{\mu} (X)}$ in expression 
(\ref{chitilde}) is dictated by this symmetry (see Section~\ref{connection}).
As to the $\mathrm{SU}(N_f)$ structure of $\tilde{\chi}^{A}(r;P)$, it is
\begin{equation}
\tilde{\chi}^{A}(r;P) = \frac{\lambda^{A}}{2} \tilde{\chi}(r;P),
\label{chitilde1}
\end{equation}
where $\lambda^{A}$ are $N_{f}^2 - 1$ matrices in the fundamental
representation of $\mathrm{SU}(N_f)$.

Now, transforming Bethe-Salpeter equation (\ref{BSequation1})
into momentum space, we get
\begin{eqnarray}
\tilde{\chi}^{A}(p;P) &=& \frac{16\pi \alpha_{b}
\lambda(P)}{|eB|^2} \int
\frac{d^2q_{\perp}d^2A_{\perp}d^2k_{\perp}d^2k_{\parallel}}{(2\pi)^6}
e^{i(P_{\perp}-q_{\perp}) \times (A_{\perp}-p_{\perp})}
e^{-\frac{(\mathbf{p}_{\perp}+\mathbf{A}_{\perp})^2}{2|eB|}}
e^{-\frac{\mathbf{q}_{\perp}^2}{2|eB|}}
\nonumber\\
&\times& S_{\parallel}\left(p_{\parallel}+\frac{P_{\parallel}}{2}\right)
\gamma^{\mu} \tilde{\chi}^{A}(k;P)\gamma^{\nu}
S_{\parallel}\left(p_{\parallel}-\frac{P_{\parallel}}{2}\right)
D_{\mu\nu}(k_{\parallel}-p_{\parallel},\mathbf{k}_{\perp}-\mathbf{A}_{\perp}),
\label{BSmomentum}
\end{eqnarray}
where $p_{\perp} \times q_{\perp} \equiv \frac{\epsilon^{ab}p^aq^b}{eB}$
is the Moyal cross product.
Then, introducing the variable $\mathbf{u}_{\perp}= \mathbf{A}_{\perp}-
\mathbf{k}_{\perp}$ and representing the wave function $\tilde{\chi}^{A}$ as
\begin{equation}
\tilde{\chi}^{A}(p;P) = e^{-\frac{\mathbf{p}_{\perp}^2}{|eB|}} e^{-iP_{\perp}
\times p_{\perp}} f^{A}(p;P),
\label{tildechi}
\end{equation}
we integrate over $q_{\perp}$ in (\ref{BSmomentum}) and find the following
equation for the function $f^{A}(p;P)$:
\begin{eqnarray}
f^{A}(p;P) &=& \frac{8 \alpha_{b}\lambda(P)}{|eB|} \int
\frac{d^2u_{\perp}d^2k_{\perp}d^2k_{\parallel}}{(2\pi)^4}\,
e^{iP_{\perp} \times u_{\perp}}
e^{-\frac{(\mathbf{u}_{\perp} + \mathbf{k}_{\perp})^2}{|eB|}}
e^{-\frac{\mathbf{k}_{\perp}^2}{|eB|}}\,
\nonumber \\
&\times& S_{\parallel}\left(p_{\parallel}+\frac{P_{\parallel}}{2}\right)
\gamma^{\mu} f^{A}(k;P)\gamma^{\nu} S_{\parallel}\left(p_{\parallel}-\frac{P_{\parallel}}{2}\right)
D_{\mu\nu}(k_{\parallel}-p_{\parallel},\mathbf{u}_{\perp}).
\label{f-equation1}
\end{eqnarray}

Because the right hand side of equation (\ref{f-equation1}) does not
contain
$p_{\perp}$, we conclude that $f^{A}(p;P)$ does not depend on $p_{\perp}$,
i.e., it
is a function of $p_{\parallel}$ and $P$ only.
[It is not difficult to
convince yourself that this fact is a
direct consequence of the factorization of the LLL propagator.]
Therefore, we
can explicitly integrate over $k_{\perp}$ and get
\begin{equation}
f^{A}(p_{\parallel};P) = 4\pi \alpha_{b}\lambda(P) \int
\frac{d^2u_{\perp}d^2k_{\parallel}}{(2\pi)^4}
e^{iP_{\perp} \times u_{\perp}}
e^{-\frac{\mathbf{u}_{\perp}^2}{2|eB|}}
S_{\parallel}\left(p_{\parallel}+\frac{P_{\parallel}}{2}\right)
\gamma^{\mu} f^{A}(k_{\parallel};P)\gamma^{\nu} S_{\parallel}\left(p_{\parallel}-\frac{P_{\parallel}}{2}\right)
D_{\mu\nu}(k_{\parallel}-p_{\parallel},\mathbf{u}_{\perp}).
\label{f-equation2}
\end{equation}
Henceforth, we will consider the dynamics for the case of zero longitudinal
momenta, $P_{\parallel}=0$, i.e., for $\phi^{A}(P) = (2\pi)^2\delta^2(P_{\parallel})\phi^{A}(P_{\perp})$,
when the function $\tilde{\varphi}(u,u^\prime)$ in Eq.~(\ref{expansion1}) is
\begin{equation}
\tilde{\varphi}(u,u^\prime)= \int \frac{d^2P_{\perp}}{(2\pi)^2} \phi^{A}(P_{\perp})
\chi^{A}(u,u^\prime;P_{\perp})
\label{expansion2}
\end{equation}
with the Bethe-Salpeter wave function $\chi^{A}$
depending only on $P_{\perp}$.
For $P_{\parallel}=0$,
the Fourier transform of the fields $\phi^{A}$
in coordinate space depends only on
the coordinates $u_{\perp}$.
Because in this problem the noncommutative geometry is
connected
only with the transverse coordinates, this dependence is
the most relevant for our purposes. The case with nonzero
longitudinal momenta $P_{\parallel}$ is quite cumbersome and is
considered in the Appendix~\ref{App:gauge-NCFT}.

We would like to emphasize that in the case of nonlocal interactions, such as 
those in QED, the bound state problem with nonzero $P$ is quite formidable. 
In fact, as will become clear below (see also Appendix~\ref{App:gauge-NCFT}),
it is the very special property of the factorization of the longitudinal and transverse 
dynamics in the LLL approximation that will allow us to succeed in the derivation 
of explicit expressions for interaction vertices for NG fields $\phi^{A}$ for nonzero $P$.

When $P_{\parallel} = 0$, one can check that, up to the
factors $\lambda(P_{\perp})$ and $e^{iP_{\perp} \times u_{\perp}}$,
the structure of equation (\ref{f-equation2}) is similar
to the Bethe-Salpeter equation for NG bosons with
$P_{\perp} = P_{\parallel} = 0$ considered
in Ref.~\cite{Gusynin:1995nb}. By
using the analysis performed in that work, we immediately find that
\begin{equation}
f^{A}(p_{\parallel};P_{\perp}) = S_{\parallel}(p_{\parallel})F^{A}(p_{\parallel};P_{\perp})\gamma^5
{\cal P}_{+} S_{\parallel}(p_{\parallel}),
\label{pion}
\end{equation}
where $F^{A}(p_{\parallel};P_{\perp})$ is a scalar function. It
satisfies the
following
equation in Euclidean space:
\begin{equation}
F^{A}(p_{\parallel};P_{\perp}) = 8\pi \alpha_{b} \lambda(P_{\perp}) \int
\frac{d^2u_{\perp}d^2k_{\parallel}}{(2\pi)^4}
\frac{F^{A}(k_{\parallel};P_{\perp})}{k_{\parallel}^2+m^2}
\frac{e^{iP_{\perp} \times u_{\perp}} e^{-\frac{\mathbf{u}_{\perp}^2}{2|eB|}}}
{(k_{\parallel}-p_{\parallel})^2+\mathbf{u}_{\perp}^2 + M^{2}_{\gamma}}.
\label{Aequation}
\end{equation}
In the derivation of this equation, Eq.~(\ref{trunc1}) was used.

Now, taking into account Eqs.~(\ref{chitilde}), (\ref{tildechi}),
and (\ref{pion}), and integrating explicitly over $p_{\perp}$,
we find that the Bethe-Salpeter wave function in coordinate space is
\begin{equation}
\chi^A(u,u^\prime;P_{\perp}) = P(u_{\perp},u^\prime_{\perp}) \int \frac{d^2p_{\parallel}}
{2(2\pi)^2} e^{i\mathbf{P}_{\perp}\frac{\mathbf{r}_{\perp}+\mathbf{y}_{\perp}}{2}}
e^{-ip_{\parallel}(u_{\parallel}- u^\prime_{\parallel})}
e^{-\frac{\mathbf{P}_{\perp}^2}{4|eB|}}
e^{\frac{\epsilon^{ab}P_{\perp}^a(u_{\perp}^b-u^{\prime b}_{\perp}) \mathrm{sign}(eB)}{2}}
S_{\parallel}(p_{\parallel})F^{A}(p_{\parallel};P_{\perp})\gamma^5
{\cal P}_{+} 
S_{\parallel}(p_{\parallel}).
\label{auxfield}
\end{equation}
Then, inserting the bilocal field $\tilde{\varphi}(u,u^\prime)$ from Eq.~(\ref{expansion2}) into
action (\ref{action11})
and using the Bethe-Salpeter
equation (\ref{BSequation1}), we obtain the following explicit form
for the effective action:
\begin{eqnarray}
S(\tilde{\varphi}) &=& \sum_{n=2}^{\infty} \frac{i}{n}
\int d^4u_1d^4u^\prime_{1}\ldots d^4u_nd^4u^\prime_n
\, \int \frac{d^2P_{\perp}^1\,\ldots\,d^2P_{\perp}^n}{(2\pi)^{2n}}\,
\phi^{A_1}(P^{\perp}_1)\,\ldots\,\phi^{A_n}(P^{\perp}_n)\,
\nonumber \\
&\times& \frac{\mathrm{tr}\,
[S_{LLL}^{-1}(u_1,u^\prime_{1})\chi^{A_1}(u^\prime_{1},u_2;P^{\perp}_1)\,\ldots\,
S_{LLL}^{-1}(u_{n-1},u^\prime_n)\chi^{A_n}(u^\prime_n,u_1;P^{\perp}_n)]}{\Pi_{i=1}^n
\lambda(P_i^{\perp})}\nonumber \\
&-&
\frac{i}{2} \int d^4u_1d^4u^\prime_{1}d^4u_2d^4u^\prime_2
\int \frac{d^2P_{\perp}^1d^2P_{\perp}^2}{(2\pi)^4}\,
\phi^{A_1}(P^{\perp}_1)\phi^{A_2}(P^{\perp}_2)\,
\nonumber \\
&\times& 
\frac{\mathrm{tr}[
\chi^{A_1}(u_1,u^\prime_{1};P^{\perp}_1)
S_{LLL}^{-1}(u^\prime_{1},u^\prime_2)\chi^{A_2}(u^\prime_2,u_2;P^{\perp}_2)
S_{LLL}^{-1}(u_2,u_1)]}{\lambda(P^{\perp}_1)}.
\label{action2}
\end{eqnarray}
From the effective action (\ref{action2}), one can obtain the $n$-point vertices of 
NG bosons. In fact, using Eq.~(\ref{auxfield}), the factorized form of the fermion
propagator (\ref{LLLpropagatorAnotherForm}), and the fact that
$P(u_{\perp},u^\prime_{\perp})$ is a projection operator, we can integrate over
all coordinates in the effective action (\ref{action2}), similarly as it
was done in the case of the NJL model in Section~\ref{NJL2}.
Then, we get the following expression for the effective action:
\begin{equation}
S(\phi) =  \sum_{n=2}^{\infty} \Gamma_n,
\end{equation}
where the interaction vertices $\Gamma_n$, $n > 2$, are
\begin{eqnarray}
\Gamma_n &=& \frac{\pi i|eB|}{2^{n-1} n} \int d^2u_{\parallel} \int
\frac{d^2k_{\parallel}}{(2\pi )^2}
\int \frac{d^2P_1^{\perp}}{(2\pi )^2}\,\ldots\,\frac{d^2P_n^{\perp}}
{(2\pi )^2}\,
\delta^2\left(\sum_{i=1}^n \mathbf{P}_{i}^{\perp}\right)\,
\phi^{A_1}(P_1^{\perp})\,\ldots\,
\phi^{A_n}(P_n^{\perp})\,
\nonumber \\
&\times& 
\mathrm{tr}\left[S_{\parallel}(k_{\parallel})F^{A_1}(k_{\parallel};P_1^{\perp})\gamma^5
{\cal P}_{+} \ldots\,S_{\parallel}(k_{\parallel})
F^{A_n}(k_{\parallel};P_n^{\perp})
\gamma^5 {\cal P}_{+} \right] \frac{e^{-\frac{i}{2}\sum_{i<j}
P_i^{\perp} \times P_j^{\perp}}}{\Pi_{i=1}^n \lambda(P_i^{\perp})},
\label{nvertex}
\end{eqnarray}
and the quadratic part of the action is
\begin{eqnarray}
\Gamma_2 &= & -\frac{i|eB|}{16\pi} \int d^2u_{\parallel} \int
\frac{d^2k_{\parallel}}{(2\pi)^2}
\int \frac{d^2P_{\perp}}{(2\pi )^2}\,
\frac{\lambda(P_{\perp}) - 1}{\lambda^2(P_{\perp})}\,\phi^{A_1}(P_{\perp})
\nonumber \\
&\times&  \mathrm{tr}\left[S_{\parallel}(k_{\parallel})F^{A_1}(k_{\parallel};P_{\perp})\gamma^5
{\cal P}_{+}  S_{\parallel}(k_{\parallel})F^{A_2}(k_{\parallel};-P_{\perp})\gamma^5
{\cal P}_{+} \right]\,\phi^{A_2}(-P_{\perp}) .
\label{quadratic}
\end{eqnarray}
For the expressions of the vertices in the case of nonzero longitudinal
momenta, see Appendix~\ref{App:gauge-NCFT}.

In the next subsection, we will discuss the connection of the
structure of vertices (\ref{nvertex}) with vertices in NCFT.

\subsubsection{Type I and Type II Nonlocal NCFT}
\label{gauge-NCFT-3}

The derivation of the expressions for vertices (\ref{nvertex}) and quadratic part 
of the action (\ref{quadratic}) is one of the main results of this section
(the generalization of these expressions for the case of nonzero longitudinal 
momenta are given in Eqs.~(\ref{nvertexl}) and (\ref{quadraticl}) in 
Appendix~\ref{App:gauge-NCFT}). Let us discuss the connection of the structure 
of these vertices with vertices in NCFT. According to \cite{Douglas:2001ba,Szabo:2001kg}, 
an $n$-point vertex in a noncommutative theory in momentum space has the following
canonical form:
\begin{equation}
\int \frac{d^Dk_1}{(2\pi)^D}\ldots \frac{d^Dk_n}{(2\pi)^D}
\phi(k_1)\ldots  \phi(k_n) \delta^D\left(\sum_i k_i\right)
e^{-\frac{i}{2} \sum_{i<j} k_i \times k_j},
\label{NCvertex}
\end{equation}
where here $\phi$ denotes a generic field and the exponent 
$e^{-\frac{i}{2} \sum_{i<j} k_i \times k_j}\equiv e^{-\frac{i}{2} \sum_{i<j} k_i \theta k_j}$
is the Moyal exponent factor. Here the antisymmetric matrix $\theta^{ab}$ determines
the commutator of spatial coordinates:
\begin{equation}
[\hat{x}^{a},\hat{x}^{b}] = i\theta^{ab}\,.
\end{equation}
When can the vertex (\ref{nvertex}) be transformed into the
conventional form (\ref{NCvertex})? In order to answer this question,
it will be convenient to ignore for a moment the fact that
$F^{A}(p_{\parallel};P_{\perp})$ in (\ref{nvertex})
is a solution of equation (\ref{Aequation}) and consider it as an
arbitrary function of the momenta $p_{\parallel}$ and $P_{\perp}$.
Then, comparing expressions (\ref{nvertex}) and (\ref{NCvertex}),
it is not difficult to figure out that if the function $F(p_{\parallel};P_{\perp})$,
defined as
\begin{equation}
F^{A}(p_{\parallel};P_{\perp}) = (\lambda^A/2)\,F(p_{\parallel};P_{\perp}),
\label{F1}
\end{equation}
has the factorized form
\begin{equation}
F(p_{\parallel};P_{\perp}) = F_{\parallel}(p_{\parallel})F_{\perp}(P_{\perp}),
\label{factor}
\end{equation}
then there exists a map of the fields $\phi^{A}(P_{\perp})$ into new fields in terms 
of which vertices $\Gamma_n$ (\ref{nvertex}) take the conventional form (\ref{NCvertex}). 
Indeed, let us introduce new fields
\begin{equation}
\Phi^{A}(P_{\perp}) = \frac{F_{\perp}(P_{\perp})}
{\lambda(P_{\perp})}\phi^{A}(P_{\perp}).
\label{smear}
\end{equation}
Then, after integrating over $k_{\parallel}$ and taking trace over Dirac matrices, 
we get the conventional form for $\Gamma_n$:
\begin{equation}
\Gamma_n = C_{n}|eB|
\int d^2u_{\parallel}
\int \frac{d^2P_1^{\perp}}{(2\pi )^2}\,\ldots\,\frac{d^2P_n^{\perp}}{(2\pi
)^2}\,
\delta^2\left(\sum_{i=1}^n \mathbf{P}_{i}^{\perp}\right)
\mathrm{tr}\left[\bar{\Phi}(P_1^{\perp})
\,\ldots\,
\bar{\Phi}(P_n^{\perp})\right]
e^{-\frac{i}{2}\sum_{i<j}
P_i^{\perp} \times P_j^{\perp}},
\label{convertex}
\end{equation}
where $\bar{\Phi}\equiv (\lambda^{A}/2)\Phi^{A}$ and $C_{n}$ is some constant. The propagator of these fields is determined 
from the quadratic part (\ref{quadratic}) of the action:
\begin{equation}
\Gamma_2 = C_{2}|eB| \int d^2u_{\parallel}
\int \frac{d^2P_{\perp}}{(2\pi )^2}\,
\left[\lambda(P_{\perp}) - 1\right]\,
\mathrm{tr}\left[\bar{\Phi}(P_{\perp})\,\bar{\Phi}(-P_{\perp})\right].
\label{quadratic1}
\end{equation}
Thus, in terms of
the new fields $\Phi^{A}(P_{\perp})$, vertices (\ref{nvertex}) can
be transformed into the canonical form. As will be shown in the
next subsection, there exists a special dynamical regime in QED with
a large number of fermion flavors $N_f$ and $M^{2}_{\gamma}
\gtrsim |eB|$ in which the constraint (\ref{factor}) can be fulfilled. 
In fact, this dynamical regime is essentially the same as that in the 
NJL model in a strong magnetic field (see Section~\ref{NJL-NCFT}). 
In this case, the function $F$ (\ref{F1}) is $p_{\parallel}$ independent. 
Following Ref.~\cite{Gorbar:2004ck}, the fields $\Phi^{A}(P_{\perp})$ with 
a built-in form factor will be called smeared fields.

In coordinate space, the interaction vertices (\ref{convertex}) of the 
smeared fields take the form:
\begin{equation}
\Gamma_n = \frac{C_{n}}{4\pi^2}|eB|
\int d^2u_{\parallel}\, \int d^2u_{\perp}\,
\mathrm{tr}\left[\bar{\Phi}(u_{\perp})*\ldots 
*\bar{\Phi}(u_{\perp})\right],
\label{coordinate1}
\end{equation}
where the symbol $*$ is the conventional star product
\cite{Douglas:2001ba,Szabo:2001kg} relating to the transverse coordinates.
In the space with noncommutative transverse
coordinates $\hat{X}_{\perp}^a$, $a=1,2$, these vertices can be
represented as
\begin{equation}
\Gamma_n = \frac{C_{n}}{4\pi^2}|eB|
\int d^2u_{\parallel} {\bf Tr}\,\,[\hat{\bar{\Phi}}(\hat{X}_{\perp})\ldots 
\hat{\bar{\Phi}}(\hat{X}_{\perp})],
\label{ncspace1}
\end{equation}
where $\hat{\bar{\Phi}}(\hat{X})$ is the Weyl symbol of
the field $\bar{\Phi}(X)$,
the operation ${\bf Tr}$ is defined as in Refs.~\cite{Douglas:2001ba,Szabo:2001kg}, and
\begin{equation}
[\hat{X}^{a}_{\perp},\hat{X}^{b}_{\perp}] =
i\frac{1}{eB}\epsilon^{ab} \equiv
i\theta^{ab},\,\,a,b=1,2.
\label{commutator}
\end{equation}
We will refer to theories with a factorized function
$F(p_{\parallel};P_{\perp})$ as
type I nonlocal NCFT.

In the case when the function $F(p_{\parallel};P_{\perp})$ in
Eq.~(\ref{F1}) is not factorized, one cannot
represent interaction vertices
(\ref{nvertex}) in
the form of vertices
of a conventional NCFT. This is the case for $F(p_{\parallel};P_{\perp})$
in the integral equation (\ref{Aequation}) with
$M^{2}_{\gamma} \ll |eB|$ (see Section~\ref{gauge-NCFT-5}).
However, even in this case,
we still can represent vertex
(\ref{nvertex}) as a nonlocal vertex in the noncommutative
space. Indeed, we can rewrite (\ref{nvertex}) as
\begin{equation}
\Gamma_n = \frac{i|eB|}{2^{n+1}\pi n} \int d^4u
\left[\, V_{n}^{A_1\ldots A_n}(-i\nabla^{\perp}_1,\ldots,-i\nabla^{\perp}_n)\, \
\phi^{A_1}(u_{1 \perp})*\ldots *\phi^{A_n}(u_{n \perp})\,\,
\right]\Big|_{u_{1 \perp}=u_{2 \perp}=\ldots =u_{\perp}},
\label{coordinate2}
\end{equation}
where the coincidence limit
$u_{1 \perp}=u_{2 \perp}=\ldots =u_{\perp}$ is taken after
the action of a
nonlocal operator $V_{n}^{A_1\ldots A_n}$ on the fields $\phi^{A_i}$.
In momentum space, the
operator $V_{n}^{A_1\ldots A_n}$ is
\begin{equation}
V_n^{A_1\ldots A_n}(P^{\perp}_1,\ldots,P^{\perp}_n) = \int
\frac{d^2k_{\parallel}}{(2\pi)^2}
\mathrm{tr}\left[S_{\parallel}(k_{\parallel})F^{A_1}(k_{\parallel};P_1^{\perp})\gamma^5
{\cal P}_{+} \,\ldots\,S_{\parallel}(k_{\parallel})
F^{A_n}(k_{\parallel};P_n^{\perp})
\gamma^5 {\cal P}_{+} \right]
\frac{1}{\Pi_{i=1}^n \lambda(P_i^{\perp})}.
\label{V}
\end{equation}
By using the fact that the Weyl symbol of the derivative in noncommutative space is 
given by the operator $\hat{\nabla}_{\perp a}$ acting as \cite{Douglas:2001ba,Szabo:2001kg} 
\begin{equation}
\hat{\nabla}_{\perp a} \hat{\phi}(\hat{X}_{\perp}) = -i
\left[(\theta^{-1})_{ab}\hat{X}_{\perp}^b,
\hat{\phi}(\hat{X}_{\perp})\right],
\label{derivative}
\end{equation}
we obtain the following form for the interaction vertices in the noncommutative space:
\begin{equation}
\Gamma_n = \frac{i|eB|}{2^{n+1}\pi n} \int d^2u_{\parallel} {\bf Tr} \left[\,\{
V_n^{A_1\ldots A_n}(-i\hat{\nabla}^{\perp}_1,\ldots,-i\hat{\nabla}^{\perp}_n)\, \
\hat{\phi}^{A_1}(\hat{X}^{\perp}_1)\ldots \hat{\phi}^{A_n}
(\hat{X}^{\perp}_n)\}_{\hat{X}^{\perp}_1=
\hat{X}^{\perp}_2=\ldots =\hat{X}^{\perp}}\right].
\label{ncspace2}
\end{equation}

It is clear that NCFT with such vertices are much more complicated than
type I nonlocal NCFT discussed above. We will call them type II nonlocal
NCFT. As will be shown in Sections~\ref{gauge-NCFT-5} and \ref{gauge-NCFT-6}, 
QED and QCD in a strong magnetic field yield examples of such theories.
For the case of nonzero longitudinal momentum $P_{\parallel}$, the counterparts 
of expressions (\ref{coordinate2}) and (\ref{ncspace2}) are derived in the 
Appendix~\ref{App:gauge-NCFT} (see Eqs.~(\ref{coordinate3}) and (\ref{noncommspace3}) there).

\subsubsection{QED with large $N_f$ in a strong magnetic field and NCFT:
dynamical regime with local interactions}
\label{gauge-NCFT-4}

Let us now consider the dynamical regime with such large $N_f$ that
the $1/N_f$ expansion is reliable, and the coupling
$\tilde{\alpha}_{b}$ is so strong that
$M_{\gamma}^2=2\tilde{\alpha}_{b}|eB|/\pi$
in (\ref{trunc1}) is of order $|eB|$ or larger.
In this case,
we have a NJL model with the current-current local interaction,
in which the coupling constant $G =  4\pi \alpha_{b}/M_{\gamma}^2 =
2\pi^2/{N_f|eB|}$ and
the ultraviolet cutoff $\Lambda_{\parallel}$ connected with longitudinal momenta
is $\Lambda_{\parallel}^2 = |eB|$
(see Ref.~\cite{Gusynin:2003dz}).
The dynamical fermion mass is now given by expression (\ref{mass1}) and
the mass
function $m(k^2)$ is a constant, $m(k^2)=m_{\rm dyn}$.
In this regime,
Eq.~(\ref{Aequation}) takes the form
\begin{equation}
F(p_{\parallel};P_{\perp}) = \frac{8\pi \alpha_{b} \lambda_{L}(P_{\perp})}
{M_{\gamma}^2} \int
\frac{d^2u_{\perp}d^2k_{\parallel}}{(2\pi)^4}
\frac{F(k_{\parallel};P_{\perp})}{k_{\parallel}^2+m_{\rm dyn}^2}
e^{iP_{\perp} \times u_{\perp}}\, e^{-\frac{\mathbf{u}_{\perp}^2}{2|eB|}},
\label{NJLequation}
\end{equation}
where the function $F(p_{\parallel};P_{\perp})$ is defined in Eq.~(\ref{F1}).
The subscript $L$ in $\lambda_{L}(P_{\perp})$ reflects the consideration
of the limit with local interactions here. Since the right-hand side of
equation (\ref{NJLequation}) does not depend on
$p_{\parallel}$, the function $F(p_{\parallel};P_{\perp})$ is
$p_{\parallel}$-independent, $F(p_{\parallel};P_{\perp}) = F(P_{\perp})$, i.e.,
this dynamics relates to the type I nonlocal NCFT considered in
the previous section. Then, we immediately find from (\ref{NJLequation}) that
\begin{equation}
\lambda_{L}(P_{\perp}) = \frac{\pi M_{\gamma}^2
e^{\frac{P_{\perp}^2}{2|eB|}}}
{\alpha_{b} |eB|\ln\frac{|eB|}{m_{\rm dyn}^2}} =
\frac{2N_{f}e^{\frac{P_{\perp}^2}{2|eB|}}}{\ln\frac{|eB|}{m_{\rm dyn}^2}},
\label{NJLlambda}
\end{equation}
where we used $M_{\gamma}^2 = 2\tilde{\alpha}_{b}|eB|/\pi
\gg m_{\rm dyn}^2$.
Using now the on-mass-shell
condition
$P_{\perp} \to 0$, $\lambda_{L}(P_{\perp}) \to 1$ for the NG bosons
in equation (\ref{NJLlambda}),
we arrive at the following gap equation for $m_{\rm dyn}$:
\begin{equation}
\frac{2N_f}{\ln\frac{|eB|}{m_{\rm dyn}^2}} = 1.
\label{gapequation}
\end{equation}
This gap equation yields expression (\ref{mass1}) for the mass.
It also implies that
$\lambda_{L}$ (\ref{NJLlambda}) can be rewritten in a very simple
form:
\begin{equation}
\lambda_{L}(P_{\perp}) = e^{\frac{P_{\perp}^2}{2|eB|}}.
\label{NJLlambda1}
\end{equation}

The choice of the off-mass-shell
operators $\phi^A$ in expansion
(\ref{expansion2}) is not unique. They are determined by the
choice of their propagator. If one chooses the conventional
composite NG fields $\pi^A = (G/2) \bar{\psi}\gamma_5\lambda^{A} \psi$
as $\phi^A$,
their propagator is
\begin{equation}
D^{AB}_{\pi} = \frac{8\pi^2\,\,\delta_{AB}}{|eB|\ln\frac{|eB|}
{m_{\rm dyn}^2}
(1 - e^{\frac{-P^2_{\perp}}{2|eB|}})}=
\frac{8\pi^2\,\,\delta_{AB}}{|eB|\ln\frac{|eB|}
{m_{\rm dyn}^2}\left[1 - \lambda_{L}^{-1}(P_{\perp})\right]}=
\frac{4\pi^2\,\,\delta_{AB}}
{N_f|eB|\left[1 - \lambda_{L}^{-1}(P_{\perp})\right]}.
\label{NJLpropagator}
\end{equation}
Up to a factor of $2$, this propagator coincides with that in the NJL model in a 
magnetic field, see Section~\ref{NJL-NCFT} and Ref.~\cite{Gorbar:2004ck}.\footnote{In 
Ref.~\cite{Gorbar:2004ck}, the NJL model with $N_c$  fermion colors and the 
chiral group $\mathrm{U}(1)_{L} \times \mathrm{U}(1)_{R}$ was considered. 
Therefore, when comparing with the results of Ref.~\cite{Gorbar:2004ck}, here 
one should take $N_{c}=1$ and replace the flavor matrices $\lambda^{A}/2$
by $1$. Since $\mathrm{tr}(\lambda^{A}/2)^2 = 1/2$, there is an additional factor 
$2$ in the propagator of $\pi^A$ in the present model.\label{foot6}}

From propagator (\ref{NJLpropagator}) and
Eq.~(\ref{quadratic}) we find
the function $F(P_{\perp})$:
\begin{equation}
F(P_{\perp})  =
2\lambda^{1/2}_{L}(P_{\perp}) = 2
e^{\frac{P_{\perp}^2}{4|eB|}}.
\end{equation}

Interaction vertices (\ref{nvertex}) are nonzero only for even $n$ and
they take now the form
\begin{equation}
\Gamma_n = -\frac{2|eB|(-1)^{\frac{n}{2}}} {n(n-2)m_{\rm dyn}^{n-2}}
\int d^2u_{\parallel}
\int \frac{d^2P_1^{\perp}}{(2\pi )^2}\,\ldots\,\frac{d^2P_n^{\perp}}{(2\pi
)^2}\,
\delta^2\left(\sum_{i=1}^n \mathbf{P}_{i}^{\perp}\right)
 \mathrm{tr}\left[\bar{\pi}(P_1^{\perp})
\,\ldots\,
\bar{\pi}(P_n^{\perp})\right]
e^{-\frac{\sum_{i}P_{i\perp}^2}{4|eB|}}
e^{-\frac{i}{2}\sum_{i<j}
P_i^{\perp} \times P_j^{\perp}}, 
\label{NJLnvertex1}
\end{equation}
where $\bar{\pi}\equiv (\lambda^{A}/2)\pi^{A}$.
These expressions for the vertices agree with those found in 
Section~\ref{NJL-NCFT} (see footnote~\ref{foot6}).

Apart from the exponentially damping factor $e^{-\sum_{i}P_{i\perp}^2/4|eB|}$, 
the form of these vertices coincide with that in NCFT with noncommutative space 
transverse coordinates $\hat{X}^{a}_{\perp}$ satisfying commutation relation
(\ref{commutator}). The appearance of the additional Gaussian (form-) factors 
in vertices reflects an inner structure
of composites $\pi^A$ in the LLL
dynamics. These form factors, reflecting the Landau wave functions
on the LLL,
are intimately connected with the holomorphic
representation in the problem of a free fermion in a strong
magnetic field \cite{Girvin:1984fk,Kivelson:1987uq}. The short-range interactions between
fermions in this dynamical regime
do not change their Gaussian form.
As was shown
in Section~\ref{NJL-NCFT}, because of these form factors, the UV/IR mixing is
absent in the model.
 
As we discussed in the previous section,
in order to take properly into account these
form factors,
it is convenient to introduce new, smeared, fields:
\begin{equation}
\bar{\Pi}(X) = e^{\frac{\nabla_{\perp}^2}{4|eB|}}\,\bar{\pi}(X),
\end{equation}
where $\nabla_{\perp}^2$ is the transverse Laplacian.
Then, in terms of the smeared fields, the vertices can be rewritten in the
standard form with the Moyal exponent factor only:
\begin{eqnarray}
\Gamma_n &=& = -\frac{2|eB|(-1)^{\frac{n}{2}}}{n(n-2)m_{\rm dyn}^{n-2}}
\int d^2u_{\parallel}
\int \frac{d^2P_1^{\perp}}{(2\pi )^2}\,\ldots\,\frac{d^2P_n^{\perp}}{(2\pi
)^2}\,
\delta^2\left(\sum_{i=1}^n \mathbf{P}_{i}^{\perp}\right)
\mathrm{tr}\left[\bar{\Pi}(P_1^{\perp})
\,\ldots\,
\bar{\Pi}(P_n^{\perp})\right]
e^{-\frac{i}{2}\sum_{i<j}
P_i^{\perp} \times P_j^{\perp}}.
\label{NJLnvertex}
\end{eqnarray}
But now the form factor occurs in the
propagator of the smeared fields:
\begin{equation}
D^{AB}_{\Pi}(P_{\perp}) =
e^{\frac{-P^{2}_{\perp}}{2|eB|}}  D^{AB}_{\pi}(P_{\perp}).
\label{NJLpropagator1}
\end{equation}
In this case, it is again
the form factor $e^{\frac{-P^{2}_{\perp}}{2|eB|}}$, now built in the
propagator $D^{AB}_{\Pi}(P_{\perp})$, that is responsible for the absence
of the UV/IR mixing.

The extension of the present analysis to the case with nonzero
longitudinal momenta $P_{\parallel}$ is straightforward for this
NJL-like dynamical regime (see the Appendix~\ref{App:gauge-NCFT}). 
The results coincide with those obtained in Section~\ref{NJL-NCFT}.

Therefore, the dynamics in this regime relates to type I nonlocal NCFT. As was 
discussed in Section~\ref{gauge-NCFT-3} (see also Section~\ref{NJL-NCFT}), 
the $n$-point vertex (\ref{NJLnvertex}) can be rewritten in the coordinate space
in the standard NCFT form with the star product. Moreover, as was shown in 
Section~\ref{NJL-NCFT}, in the space with noncommutative transverse
coordinates, one can derive the effective action for the composite fields in this model.
Thus, here we reproduced the results of Section~\ref{NJL-NCFT}  by using the
method of bilocal auxiliary fields.

\subsubsection{QED with weak coupling in a strong magnetic field
and type II nonlocal NCFT}
\label{gauge-NCFT-5}

In this section, we will consider the dynamics of QED
in a magnetic field in a weak coupling regime, when
the coupling $\tilde{\alpha}_{b}$ is small (the number of flavors
$N_f$ can now be arbitrary).
As will become clear in a moment, this dynamics yields an example
of a type II nonlocal NCFT.

According to the analysis in Section~\ref{gauge-NCFT-1}, the integral equation for
$F(p_{\parallel};P_{\perp})$ in this
regime is
\begin{equation}
F(p_{\parallel};P_{\perp}) = 8\pi\alpha_b\lambda_{W}(P_{\perp}) \int
\frac{d^2u_{\perp}d^2k_{\parallel}}{(2\pi)^4}
\frac{F(k_{\parallel};P_{\perp})}{k_{\parallel}^2+m_{\rm dyn}^2}
\frac{e^{iP_{\perp} \times u_{\perp}} e^{-\frac{\mathbf{u}_{\perp}^2}{2|eB|}}}
{(k_{\parallel}-p_{\parallel})^2+\mathbf{u}_{\perp}^2+M_{\gamma}^2},
\label{QEDequation}
\end{equation}
where $m_{\rm dyn}$ is given in Eq.~(\ref{mass}) and
the subscript $W$ in $\lambda_{W}$ reflects the consideration of
the weak coupling regime.
Unlike the integral equation (\ref{NJLequation}), the kernel of this
equation does not have a separable form. Therefore, the function
$F(p_{\parallel};P_{\perp})$ is not factorized in this case and the present
dynamics relates to type II nonlocal NCFT. According to
the analysis in Section~\ref{gauge-NCFT-3}, its $n$-point vertices can be
written
either through the star product in the form (\ref{coordinate2})
in coordinate space
or in the form (\ref{ncspace2})
in the noncommutative space.

In order to illustrate the difference of this dynamics from that
considered in the previous section, it will be instructive to
analyze it in a special limit with $P_{\perp}^2 \gg |eB|$.\footnote{While 
in the dynamical regime with the LLL dominance longitudinal momenta 
should satisfy the inequality $P_{\parallel} \ll \sqrt{|eB|}$, transverse momenta 
can be large.\label{foot7}} 
We will show that in this limit the approximation with 
$F(p_{\parallel};P_{\perp})$ being independent of $p_{\parallel}$ is 
quite good and, therefore, the dynamics in this limit can be considered 
as approximately relating to type I nonlocal NCFT. However, as will be 
shown below, the form factor in this dynamics is very different from
the Gaussian form factor that occurs in the NJL-like dynamics.
This point reflects a long-range character of
the QED interactions.

We start the analysis of integral equation (\ref{QEDequation})
for $P_{\perp}^2 \gg |eB|$
by considering the integral
\begin{equation}
I = \int \frac{d^2u_{\perp}}{(2\pi)^2}
\frac{e^{iP_{\perp} \times u_{\perp}}
e^{-\frac{\mathbf{u}_{\perp}^2}{2|eB|}}}
{q_{\parallel}^2+\mathbf{u}_{\perp}^2+M_{\gamma}^2} =
\int \frac{d^2u_{\perp}}{(2\pi)^2}
\frac{e^{i\bm{\Delta}_{\perp}\mathbf{u}_{\perp}}
e^{-\frac{\mathbf{u}_{\perp}^2}{2|eB|}}}
{q_{\parallel}^2+\mathbf{u}_{\perp}^2+M_{\gamma}^2},
\end{equation}
where $q_{\parallel} = k_{\parallel} - p_{\parallel}$,
$\bm{\Delta}_{\perp}=\frac{\mathbf{P}_{\perp}}{eB}$
and, for convenience,
we made the change of variable $u_{\perp}^1 \to -u_{\perp}^2$,
$u_{\perp}^2 \to u_{\perp}^1$
in the last
equality.
By representing $e^{-\mathbf{u}_{\perp}^2/2|eB|}$ and
$(q_{\parallel}^2+\mathbf{u}_{\perp}^2+M_{\gamma}^2)^{-1}$
through their Fourier transforms, we obtain
\begin{equation}
I = \int d^2\Delta_{\perp 1}\,
\frac{|eB|\,e^{-\frac{|eB|\bm{\Delta}_{\perp 1}^2}{2}}}
{2\pi}\,\,\,
\frac{K_0(|\bm{\Delta}_{\perp} -
\bm{\Delta}_{\perp 1}|\,\sqrt{q_{\parallel}^2+M_{\gamma}^2})}{2\pi},
\label{involution}
\end{equation}
where $K_0(z)$ is the Bessel function of imaginary argument \cite{1980tisp.book.....G}.
For $P_{\perp}^2 \gg |eB|$, when $|\bm{\Delta}_{\perp}| \gg 1$, one can neglect the
dependence of $K_0$ on $\bm{\Delta}_{\perp 1}$. Then, using the asymptotics 
of $K_0(z)$ at $z \to +\infty$, we find that
\begin{equation}
I \approx \frac{1}{2}\left(\frac{|eB|}
{2\pi |P_{\perp}|(q_{\parallel}^2+M_{\gamma}^2)^{1/2}}
\right)^{1/2}
e^{-\frac{|P_{\perp}|(q_{\parallel}^2+M_{\gamma}^2)^{1/2}}{|eB|}},
\end{equation}
where $|P_{\perp}| \equiv \sqrt{\mathbf{P}_{\perp}^2}$. Since this $I$ as a function 
of $q_{\parallel}$ exponentially decreases starting from $M_{\gamma}$, it is sufficient
to take into account only the region with $q_{\parallel} \lesssim M_{\gamma}$
and approximate $I$ there by
\begin{equation}
I \approx \frac{1}{2}\left(\frac{|eB|}{2\pi
|P_{\perp}|M_{\gamma}}\right)^{1/2}
e^{-\frac{|P_{\perp}|M_{\gamma}}{|eB|}}.
\label{I5}
\end{equation}

Thus, we conclude that for $P_{\perp}^2 \gg |eB|$,
a good approximation for equation (\ref{QEDequation}) is
to integrate over $k_{\parallel}^2$ up to $M_{\gamma}^2$ and
to use
expression (\ref{I5}) for $I$. This implies that for large $P_{\perp}^2$
the function $F(p_{\parallel};P_{\perp})$ can be taken independent of
$p_{\parallel}$.
Then, we find from Eq.~(\ref{QEDequation})
that
\begin{equation}
\lambda_{W}(P_{\perp}) \simeq \frac{1}{\alpha_b\ln\frac{M_{\gamma}^2}
{m_{\rm dyn}^2}}
\left(\frac{2\pi |P_{\perp}|M_{\gamma}}{|eB|}\right)^{1/2}
e^{\frac{|P_{\perp}|M_{\gamma}}{|eB|}}.
\label{QEDlambda}
\end{equation}

By choosing $F(P_{\perp}) = 2\lambda_{W}^{1/2}(P_{\perp})$,
we obtain
the following pion propagator  from Eq.~(\ref{quadratic})
(compare with Eq.~(\ref{NJLpropagator})):
\begin{equation}
D^{AB}(P_{\perp}) = \frac{8\pi^2\delta_{AB}}{|eB|\ln \frac{M^{2}_{\gamma}}
{m_{\rm dyn}^2}
\left[1-\lambda_{W}^{-1}(P_{\perp})\right]}.
\label{QEDpropagator}
\end{equation}
And, using Eq.~(\ref{nvertex}), we find
the corresponding vertices:
\begin{eqnarray}
\Gamma_n &=& \frac{2\pi i|eB|}{n}
\int d^2u_{\parallel} \int \frac{d^2k_{\parallel}}{(2\pi )^2}
\int \frac{d^2P_1^{\perp}}{(2\pi )^2}\,\ldots\,\frac{d^2P_n^{\perp}}{(2\pi )^2}\,
\delta^2\left(\sum_{i=1}^n \mathbf{P}_{i}^{\perp}\right)
\nonumber \\
& \times &
\mathrm{tr}\left[S_{\parallel}(k_{\parallel})\gamma^5 \bar{\pi}(P_{1}^{\perp})
{\cal P}_{+} \ldots\,S_{\parallel}(k_{\parallel})\bar{\pi}(P_{n}^{\perp})
\gamma^5 {\cal P}_{+} \right]\, \Pi_{i=1}^{n}\,
\lambda_{W}^{-1/2}(P_i^{\perp})\,
e^{-\frac{i}{2}\sum_{i<j}
P_i^{\perp} \times P_j^{\perp}}.
\label{QEDnvertex1}
\end{eqnarray}

Now, let us compare expressions (\ref{QEDpropagator}) and (\ref{QEDnvertex1})
with their counterparts (\ref{NJLpropagator}) and (\ref{NJLnvertex1})
in the case of local interactions. While the behaviors of
propagators (\ref{NJLpropagator}) and (\ref{QEDpropagator})
are similar
for large $P^{2}_{\perp} \gg |eB|$ (they both approach a constant as
$P^{2}_{\perp} \to \infty$), the behaviors of vertices
(\ref{NJLnvertex1}) and (\ref{QEDnvertex1})
are quite different.
While the form factor in vertices (\ref{NJLnvertex1}) has
the Gaussian form $e^{-\frac{\sum_{i}P_{i\perp}^2}{4|eB|}}$, the form
factor in vertices (\ref{QEDnvertex1}) is proportional to
$(|P_{\perp}|M_{\gamma}/|eB|)^{-1/4}\,\,
e^{\frac{-|P_{\perp}|M_{\gamma}}{2|eB|}}$
and, therefore, decreases much slower for large $P^{2}_{\perp}$.
The reason of that is a nonlocal character of the interactions
in QED in a weak coupling limit. To see this more clearly,
let us compare integral equations (\ref{NJLequation}) and
(\ref{QEDequation}). The transition to the local interactions
corresponds to
the replacement of the propagator
$[(k_{\parallel}- p_{\parallel})^2+\mathbf{u}_{\perp}^2+M_{\gamma}^2]^{-1}$
in Eq.~(\ref{QEDequation}) by $M_{\gamma}^{-2}$. This in turn leads
to the replacement of the Bessel
function
$K_0(|\bm{\Delta}_{\perp}-\bm{\Delta}_{\perp 1}|\,
\sqrt{q_{\parallel}^2+M_{\gamma}^2})$
in Eq.~(\ref{involution}) by the delta function $2\pi/M_{\gamma}^2\,\,
\delta^{2}(\bm{\Delta}_{\perp}-\bm{\Delta}_{\perp 1})$. The
substitution of the
delta function in $I$ (\ref{involution}) leads to the Gaussian form
factor,
which, therefore, is a signature of short-range interactions.

\subsubsection{Chiral dynamics in QCD in a magnetic field and type II
nonlocal NCFT}
\label{gauge-NCFT-6}

In this section, we will show that the chiral dynamics in QCD in
a strong magnetic field relates to type II nonlocal NCFT. Here under
strong magnetic fields, we understand the fields satisfying
$|eB| \gg \Lambda^{2}_{\rm QCD}$, where
$\Lambda_{\rm QCD}$ is the QCD
confinement scale.

A crucial difference between the dynamics in QED and QCD in
strong magnetic backgrounds is of course the property of asymptotic
freedom and confinement in QCD. The infrared dynamics in quantum
chromodynamics is much richer and more sophisticated. As was shown 
in Ref.~\cite{Miransky:2002rp} and Section~\ref{sec:MagCatQCDgluo},
the confinement scale $\lambda_{\rm QCD}(B)$ in QCD in a strong magnetic field
can be much less than the confinement scale $\Lambda_{\rm QCD}$ in the vacuum.
As a result, an anisotropic dynamics of confinement is realized with a rich and
unusual spectrum of very light glueballs.

On the other hand, the chiral dynamics in QED and QCD in  strong magnetic 
backgrounds have a lot in common. The point is that the region of momenta 
relevant for the chiral symmetry breaking dynamics is $m_{\rm dyn}^2 \ll |k^2| \ll |eB|$ 
(see Section~\ref{sec:MagCatQCDgap}). If the magnetic field is so strong that the dynamical
fermion mass $m_{\rm dyn}$ is much larger than the confinement scale
$\lambda_{\rm QCD}(B)$, the running coupling $\alpha_{s}$ is small for such
momenta. As a result, the dynamics in that region is essentially
Abelian. Indeed, while the contribution of (electrically neutral) gluons
and ghosts in the polarization operator is proportional to
$k^2$, the fermion contribution is proportional to $|eB|$,
similarly to the case of QED in a magnetic field
[see Eq.~(\ref{polarization}) and Eq.~(\ref{polarization1})]. As a result,
the fermion contribution dominates in the relevant region with $|k^2| \ll |eB|$.

Because of the Abelian like structure of the dynamics in this
problem, one can use the results of the analysis in
QED in a magnetic field, by introducing appropriate
modifications (see  Section~\ref{sec:MagCatQCDgap}). One of the 
modifications is that the chiral symmetry
in QCD in a magnetic field is different from that in QED. Indeed,
since the background magnetic field breaks explicitly the global chiral
symmetry that interchanges the up and down quark flavors (having
different electric charges),
the chiral
symmetry in this problem is $\mathrm{SU}(N_{u})_{L}\times \mathrm{SU}(N_{u})_{R} \times
\mathrm{SU}(N_{d})_{L}\times \mathrm{SU}(N_{d})_{R}\times \mathrm{U}^{(-)}(1)_{A}$,
where $N_{u}$ and $N_{d}$ are the numbers of up and down quarks,
respectively (the total number of quark flavors is $N_{f}
=N_{u}+N_{d}$).
The $\mathrm{U}^{(-)}(1)_{A}$ is connected with the current which is an
anomaly free linear combination
of the $\mathrm{U}^{(d)}(1)_{A}$ and $\mathrm{U}^{(u)}(1)_{A}$ currents.
[The $\mathrm{U}^{(-)}(1)_{A}$ symmetry is of course absent if either
$N_d$ or $N_u$ is equal to zero].
The generation of quark masses
breaks this symmetry spontaneously down to $\mathrm{SU}(N_{u})_{V}\times
\mathrm{SU}(N_{d})_{V}$ and, as a result, $N_{u}^{2}+N_{d}^{2}-1$ {\it neutral}
NG bosons occur.

Another modification is connected with the presence of a new quantum
number, the color. As was shown in Ref.~\cite{Miransky:2002rp} 
and Section~\ref{sec:MagCatQCDinfty}, there exists
a threshold value of the number of colors $N^{\rm thr}_c$
dividing the theories with essentially different dynamics.
For the number of colors $N_c \ll N^{\rm thr}_c$, an anisotropic dynamics of
confinement with the confinement scale $\lambda_{\rm QCD}(B)$
much less than $\Lambda_{\rm QCD}$ and
a rich spectrum of light glueballs is realized. For $N_c$ of order
$N^{\rm thr}_c$ or larger, a conventional confinement dynamics
with $\lambda_{\rm QCD}(B) \simeq \Lambda_{\rm QCD}$ takes place.
The threshold value $N^{\rm thr}_c$ grows rapidly with the
magnetic field.
For example, for $\Lambda_{\rm QCD}= 250 \mbox{ MeV}$ and $N_u =1$, $N_d =2$,
the threshold value is
$N^{\rm thr}_c \gtrsim 100$ for $|eB| \gtrsim (1\mbox{
GeV})^2$. We will consider both the case with $N_c \ll N^{\rm thr}_c$
and that with $N_c \gtrsim N^{\rm thr}_c$.

For $N_c \ll N^{\rm thr}_c$, the dynamical mass $m_{\rm dyn}^{(q)}$
of a $q$-th quark is (see Section~\ref{sec:MagCatQCDinfty}):
\begin{equation}
m_{\rm dyn}^{(q)} \simeq C \sqrt{|e_{q}B|}
\left(c_{q}\alpha_{s}\right)^{1/3}
\exp\left[-\frac{2N_{c}\pi}{\alpha_{s} (N_{c}^{2}-1)
\ln(C_{1}/c_{q}\alpha_{s})}\right]
\label{gapNLFT}
\end{equation}
(compare with expression (\ref{mass}) for the dynamical mass in QED).
Here $e_{q}$ is the electric charge of the $q$-th quark,
the numerical factors $C$ and $C_1$
are of order one and the constant $c_{q}$ is defined as
\begin{equation}
c_{q} = \frac{1}{6\pi}(2N_{u}+N_{d})\left|\frac{e}{e_{q}}\right|.
\end{equation}
The strong coupling $\alpha_{s}$ in Eq.~(\ref{gapNLFT}) is
related to the scale $\sqrt{|eB|}$, i.e.,
\begin{equation}
\frac{1}{\alpha_{s}} \simeq b\ln\frac{|eB|}{\Lambda_{\rm QCD}^2},
\quad
b=\frac{11 N_c -2 N_f}{12\pi}.
\label{coupling2}
\end{equation}

In QCD, there are two sets of NG bosons related to
the $\mathrm{SU}(N_{u})_{V}$ and $\mathrm{SU}(N_{d})_{V}$ symmetries. Their Bethe-Salpeter
wave functions are defined as in Eq.~(\ref{chitilde}) with the
superscript $A$ replaced by $A_u$
($A_d$) for the set connected
with the $\mathrm{SU}(N_{u})_{V}$ ($\mathrm{SU}(N_{d})_{V}$). The
corresponding Bethe-Salpeter equations have the form
of equation (\ref{f-equation1}) with the coupling $\alpha_b$
replaced by the strong coupling $(N_{c}^2 -1)\alpha_s/2N_c$,
where $(N_{c}^2 -1)/2N_c$ is the quadratic Casimir invariant
in the fundamental representation of $\mathrm{SU}(N_c)$.

Besides these NG bosons, there is one NG boson connected with
the anomaly free $\mathrm{U}^{(-)}(1)_A$ discussed above. Its Bethe-Salpeter wave function
is defined as in
Eqs.~(\ref{chitilde}) and (\ref{chitilde1}) but with the
matrix $\lambda^A$ replaced by the traceless matrix
$\tilde{\lambda}^{0}/2 \equiv (\sqrt{N_{d}/N_{f}}\lambda^{0}_{u} -
\sqrt{N_{u}/N_{f}}\lambda^{0}_{d})/2$  \cite{Miransky:2002rp}.
Here $\lambda^{0}_{u}$
and $\lambda^{0}_{d}$ are proportional
to the unit matrices in the up and down flavor sectors, respectively.
They are  normalized as the $\lambda^A$ matrices:
$\mathrm{tr}[(\lambda^{0}_u)^2]= \mathrm{tr}[(\lambda^{0}_d)^2]= 2$.

The polarization operator in the propagator of gluons has the form similar to
that for photons (see Section~\ref{sec:MagCatQCDgap}):
\begin{equation}
\Pi(k_{\parallel}) \simeq -\frac{\alpha_{s}}{\pi}
\sum_{q=1}^{N_{f}}\frac{|e_{q}B|}{k^{2}_{\parallel}}
\label{polarization1}
\end{equation}
[compare with Eq.~(\ref{polarization})].
This expression implies that the gluon mass is
\begin{equation}
M_{g}^2= \sum_{q=1}^{N_{f}}\frac{\alpha_{s}}{\pi}|e_{q}B|=
(2N_{u}+N_{d}) \frac{\alpha_{s}}{3\pi}|eB|.
\label{M_g_another}
\end{equation}
It is clear that the chiral dynamics in this case
is similar to that in QED in a weak coupling regime considered
in Section~\ref{gauge-NCFT-5} and it relates to the type II nonlocal NCFT.
In the noncommutative space, the expressions
for the vertices of each of the two sets of NG bosons have the form in
Eq.~(\ref{ncspace2}) (Eq.~(\ref{noncommspace3}) in Appendix~\ref{App:gauge-NCFT})
for the case when their fields are independent of
(depend on) longitudinal coordinates, respectively. Notice
that in the leading approximation, the vertices do not mix
NG fields from the different $\mathrm{SU}(N_{u})_{V}$ and $\mathrm{SU}(N_{d})_{V}$
sets: the mixing is
suppressed by powers of the small coupling $\alpha_s$
(an analogue of the Zweig-Okubo rule). On the other hand,
the NG boson related to the $\mathrm{U}^{(-)}(1)_A$
interacts
with NG bosons from both these sets without any suppression.

Let us now turn to the case with large $N_c$, in particular, to
the 't Hooft limit $N_c \to \infty$. Just a look at expression (\ref{M_g_another})
for the gluon mass is enough to recognize that the dynamics in this limit
is very different from that considered above. Indeed,
as is well known, the strong coupling constant $\alpha_s$ is proportional
to $1/N_c$ in this limit. More precisely, it rescales as  [compare 
with Eq.~(\ref{tilde})]
\begin{equation}
\alpha_s = \frac{\tilde{\alpha}_s}{N_c},
\end{equation}
where the new coupling constant $\tilde{\alpha}_s$ remains finite
as $N_c \to \infty$. Then, expression (\ref{M_g_another}) implies that
the gluon mass goes to zero in this limit. This in turn implies
that the appropriate approximation is now not
the improved rainbow (ladder) approximation but the
rainbow (ladder) approximation
itself, when the gluon propagator in the gap (Schwinger-Dyson) equation
and in the Bethe-Salpeter equation is taken to be bare with $\Pi(k_{\parallel}) = 0$
\cite{Miransky:2002rp}. In other words, gluons are massless and
genuine long range interaction take place in this regime.
The dynamical mass of quarks now is
\begin{equation}
m^{(q)}_{\rm dyn} = C \sqrt{|e_{q}B|}
\exp\left[-{\pi}
\left(\frac{\pi}{4\tilde{\alpha}_s}\right)^{1/2}
\right],
\label{m-infinity}
\end{equation}
where the constant $C$ is of order one. As was shown in Ref.~\cite{Miransky:2002rp} 
and Section~\ref{sec:MagCatQCDinfty}, this expression is a good approximation for the 
quark mass when $N_c$ is of order $N^{\rm thr}_c$ or larger. As to the Bethe-Salpeter 
equation, repeating the analysis of Section~\ref{gauge-NCFT-5}, it is easy to show that, 
unlike the QED case, the amplitude $F(p_{\parallel},P_{\perp})$, defined in 
Section~\ref{gauge-NCFT-3}, is not factorized in this dynamical regime
even for large transverse momenta $P_{\perp}^2 \gg |eB|$. Therefore,
the dynamics with large $N_c$ yields even a more striking example of
the type II nonlocal NCFT than the previous dynamical regime with
$N_c \ll N^{\rm thr}_c$.

\subsection{Concluding remarks on field theories in a magnetic field as NCFT}
\label{connection}

In Sections~\ref{NJL-NCFT} and \ref{gauge-NCFT}, we established the connection 
between field theories in a magnetic field and NCFT. 
The main result of Section~\ref{NJL-NCFT} is that in any dimension $D=d+1$ with $d \geq 2$,
the NJL model in a strong magnetic field determines a consistent NCFT. These NCFT 
are quite sophisticated that is reflected in their action (\ref{d+1ncACTION}) expressed
through the smeared fields $\Sigma$ and $\Pi$ with built-in exponentially damping 
form-factors. These form-factors occur in the propagators of the smeared fields
and are responsible for removing the UV/IR mixing that plagues conventional 
nonsupersymmetric NCFT \cite{Minwalla:1999px}. As an alternative, one can 
also use the composites fields $\sigma$ and $\pi$. In this case, the form-factors 
occur in their interaction vertices and this again leads to the removal of the
UV/IR mixing.

An especially interesting case is that for a magnetic field configuration with 
the maximal number $[d/2]$ of independent nonzero tensor components. 
In that case, the dynamics is quasi-$(1+1)$-dimensional for odd $d$ and 
finite for even $d$. How can it be, despite the fact that the initial NJL model 
is nonrenormalizable for $d \geq 2$? And, moreover, how can it happen in 
theories in which neutral composites propagate in a bulk of a space of 
{\it arbitrary} high dimensions? The answer to these questions is straightforward.
The initial NJL model in a strong magnetic field and the truncated model
based on the LLL dynamics are essentially identical only in infrared, with 
momenta $k \ll \sqrt{|eB|}$. At large momenta, $k \gg \sqrt{|eB|}$, these two 
models are very different. It is the LLL dominance that provides the exponentially 
damping (form-)factors which are responsible for finiteness of the present model 
for even $d$ and its quasi-$(1+1)$-dimensional character for odd $d$. Thus, 
besides being a low-energy theory of the NJL model in a strong magnetic field, 
the NCFT based on the LLL dynamics is self-contained and self-consistent.

As was discussed in Section~\ref{NJL1}, the exponentially damping factors
occur also in nonrelativistic quantum mechanical models. In particular, they 
are an important ingredient of the formalism of the projection onto the LLL 
developed for studies of condensed matter systems in 
Refs.~\cite{Girvin:1984fk,Kivelson:1987uq,Dunne:1992ew}.
Note that such factors occur also in a NCFT corresponding to the relativistic 
scalar $\mathrm{O}(N)$ model in a magnetic field \cite{Salim:2006ky}.\footnote{It is
important, however, that there is an essential difference between the dynamics of the spin
$1/2$ fermions and spinless scalars in a magnetic field. Because of the dispersion relations in 
Eqs.~(\ref{spectr-n=0}), (\ref{spectr-n>0}) ($2 + 1$ dimensions) and  in Eq.~(\ref{3}) ($3 + 1$
dimensions), the LLL states of the fermions are gapless for a zero mass, which implies that the LLL contribution dominates in law energy dynamics. On the other hand, because dispersion relations for scalars can be obtained from those of the fermions with the substitution $n \to n + 1/2$, the LLL states for them are gapped even for a zero mass, and therefore the LLL contribution is of the same order as that of the higher Landau levels with $n \sim O(1)$.}

It is then natural to ask why do such factors not appear also in string theories 
in a magnetic field? The answer to this question is connected with a completely 
different way that open strings respond to a strong $B$ field. It can be seen 
already on the classical level. Indeed, due to the boundary conditions at the 
ends of open strings, their length {\it grows} with $B$ until the string tension 
compensates the Lorentz forces exerted at the ends of strings \cite{SheikhJabbari:1999vm}. 
In contrast to that, in quantum field and condensed matter systems, charged particles,
which form neutral composites, move along circular orbits in a magnetic field, and their
radius {\it shrinks} with increasing $B$. This leads to the Landau type wave functions
of composites and, therefore, to the exponentially damping (form-)factors either in 
vertices (for $\sigma$ and $\pi$ fields) or in propagators (for smeared fields).

Therefore, unlike the dynamics of neutral composites in
condensed matter and quantum field systems, open strings in a
magnetic background do lead to the conventional NCFT.
Since these theories are supersymmetric, the UV/IR mixing
affects only the constants of renormalizations and does not
destroy their consistency \cite{Girotti:2000gc}.
Thus, different physical systems in a magnetic
fields lead to different classes of consistent NCFT. In fact, 
in this regard, quantum mechanical systems in
a strong magnetic field are special. As was shown in
Section~\ref{NJL1}, depending on two different treatments of
the case with the mass $m=0$, they determine either
conventional NCFT, as it is done in Ref.~\cite{Bigatti:1999iz}, or
NCFT with exponentially damping form-factors.

The NJL models in a magnetic field lead to type I nonlocal NCFT.
The main result of Section~\ref{gauge-NCFT} is that the chiral dynamics in QED and 
QCD in a strong magnetic field determine more complicated nonlocal NCFT (type II nonlocal NCFT).
These NCFT are quite different from both the NCFT considered in the literature and
the NCFT corresponding to the NJL model in a magnetic field. 
While in type I NCFT there exists a field transformation that puts interaction vertices 
in the conventional form (with a cost of introducing an exponentially damping form factor
in field propagators), no such a transformation exists for type II nonlocal NCFT.

The reason of this distinction between the two types of models is in the characters 
of their interactions, being short-range in the NJL-like models and long-range in 
gauge theories. While the influence of the short-range interactions on the LLL 
dynamics is quite minor, the long-range interactions change essentially that 
dynamics. As a result, the structure of neutral composites and manifestations 
of nonlocality in gauge theories in strong magnetic backgrounds are much richer.

What is the origin of the connection between field theories in a magnetic field and NCFT?
As was emphasized in Ref.~\cite{Gorbar:2004ck} and in Section~\ref{gauge-NCFT-1}, it is 
the Schwinger phase in the LLL fermion propagator and Bethe-Salpeter wave functions of
neutral composites that leads to the Moyal factor (a signature of NCFT) in interaction 
vertices of neutral composites. But what is the origin of the Schwinger phase itself? 
One can argue that it reflects the existence of the group of magnetic translations in 
an external magnetic field \cite{Zak:1964zz} (for a brief discussion of the corresponding
group, see Section~\ref{sec:MagneticTranslations}). The group generators and their 
commutators in the symmetric gauge Eq.~(\ref{eq:symm_gauge}) were given in 
Eqs.~(\ref{mtrans2}) and (\ref{algebra}), respectively. As one can see, all commutators 
equal zero for neutral states, implying that the momentum $\vec{P}=(P_1,P_2,P_3)$ of their 
center of mass is a good quantum number.

It is easy to check that the structure of the generators (\ref{mtrans2}) implies the presence
of the Schwinger phase in the matrix elements of the time ordered bilocal operator 
$T\left[\psi(u) \bar{\psi}(u^\prime)\right]$ taken between two arbitrary neutral states, 
$|a;P_a\rangle$ and $|b;P_b\rangle$. More precisely,
\begin{equation}
M_{ba}(u,u^\prime;P_b,P_a) \equiv
\langle P_b;b|T\psi(u)\bar{\psi}(u^\prime)|a;P_a\rangle = e^{-i(P_a - P_b)X}
e^{ier^{\mu}A_{\mu} (X)} \tilde{M}_{ba}(r;P_b,P_a),
\label{M}
\end{equation}
where $r=u-u^\prime$, $X=(u+u^\prime)/2$ (compare with Eq.~(\ref{chitilde})).
The second exponent factor on the right hand side of this equation
is the Schwinger phase. Taking $|a;P_a\rangle$ and $|b;P_b\rangle$ to be
the vacuum state $|0\rangle$, we get the fermion propagator. And taking
$|b;P_b\rangle = |0\rangle$ and $|a;P_a\rangle$ to be a state of some
neutral composite, we get the Bethe-Salpeter wave function of this
composite. Thus, the group of magnetic translations is  in the heart of the 
connection between the field dynamics in a magnetic field and NCFT.
In particular, one of the consequences of this consideration is that 
although the treatment of neutral composites other than NG bosons 
is much more involved, one can be sure that, in a strong magnetic field, 
the noncommutative structure of interaction vertices of those neutral 
composites is similar to that of the vertices for NG bosons that was
derived in Sections~\ref{gauge-NCFT-2} and \ref{gauge-NCFT-3}, 
and in Appendix~\ref{App:gauge-NCFT}.

Now, what is the form of interaction vertices which include such
{\it elementary} electrically neutral fields as photon and gluon
fields in QED and QCD, respectively. This problem in QED was
considered in Ref.~\cite{Gorbar:2005az}. As was shown there, in the
LLL approximation, fermion loops infect a noncommutative structure 
for $n$-point photon vertices and, as a result, the Moyal factor occurs 
there. The situation in QCD is more subtle. In that case, besides induced 
gluon vertices generated by quark loops, there are also triple and quartic 
gluon vertices in the initial QCD action. There is of course no Moyal factor 
in the latter. Still, since the Abelian approximation is reliable in the description
of the chiral dynamics in a strong magnetic field in QCD (see the discussion 
in Sections~\ref{sec:MagCatQCD} and \ref{gauge-NCFT-6}), the situation for 
this dynamics in QCD is similar to that in QED. However, the description of the
anisotropic confinement dynamics in QCD, relating to the deep infrared region, 
is much more involved, see Ref.~\cite{Miransky:2002rp} and Section~\ref{sec:MagCatQCD}.

Another point we want to address here is the reliability of the LLL
approximation in a strong magnetic field. As to the chiral dynamics,
there are solid arguments that it is a reliable approximation for
it (see the discussion in Section~\ref{sec:MagneticCatalysis}). 
In particular, its reliability was shown explicitly in the NJL model 
in a magnetic field in the leading order in $1/N_{c}$ expansion. 
On the other hand, as was shown in Ref.~\cite{Gorbar:2005az}, 
the cumulative effect of higher Landau levels can be important 
for $n$-point photon vertices, at least in some kinematic regions 
(a nondecoupling phenomenon). It is however noticeable that in 
the kinematic region with momenta $k_{i\perp}^2 \gg |k_{i\parallel}^2|$, 
which provides the dominant contribution in the chiral dynamics 
(see Section~\ref{sec:MagneticCatalysis}), the LLL contribution is
dominant  \cite{Gorbar:2005az}. Thus, although this question deserves
further study, the assumption about the LLL dominance in the chiral
dynamics in relativistic field theories seems to be well justified.

We believe that both these types of nonlocal NCFT can be relevant not only 
for relativistic field theories but also for nonrelativistic systems in a magnetic 
field. In particular, while type I NCFT can be relevant for the description of the 
quantum Hall effect in condensed matter systems with short-range interactions 
\cite{Iso:1992ca,Cappelli:1992kf,Polychronakos:2001mi}, type II NCFT can be 
relevant in studies of this effect in condensed matter systems with long-range 
interactions. Concrete examples of such systems are provided by carbon materials, 
in particular, graphene, where Coulomb-like interactions take place, see 
Section~\ref{sec:QHEgraphene}.

\section{Topics not covered in this review}
\label{sec:OtherTopics}

As the sizable volume of the present report demonstrates, there is a large number of interesting 
applications of relativistic and quasirelativistic quantum field theories, in which external magnetic 
fields play an essential role in underlying physics. It should be emphasized, however, that despite our 
intention to include as many topics as possible, we could not really cover everything. The actual 
range of ideas and applications is much wider and is constantly growing. In order to ensure an 
adequate rigor of presentation and provide a sufficient amount of details, required for deep 
understanding of physics phenomena, here we concentrated primarily on the topics that we had 
a personal research experience with. 

In order to do a partial justice to the topics not covered in this report, here we offer a brief guide 
to some of them. We recognize that even this extended list of physics phenomena is unavoidably
going to be incomplete. Therefore, we apologize in advance to the authors whose ideas happened 
to be omitted. This is probably the result of our unintentional bias or lack of knowledge.

\subsection{Color superconductivity in a magnetic field}

The theory of strong interactions predicts that quark matter at sufficiently high densities and 
sufficiently low temperatures is a color superconductor \cite{Barrois:1977xd,Bailin:1980bi,
Bailin:1981mt,Iwasaki:1994ij,Alford:1997zt,Rapp:1997zu,Son:1998uk,Pisarski:1999tv,
Schafer:1999jg,Hong:1999fh,Hsu:1999mp}. (For reviews, see Refs.~\cite{Shovkovy:2004me,Alford:2007xm}.)
A rigorous treatment of underlying nonperturbative dynamics of Cooper pairing is possible 
because of asymptotic freedom in QCD. Indeed, this property ensures that dense quark matter 
is weakly interacting at sufficiently high densities. As usual, the dominant Cooper pairing 
occurs in a spin-zero channel. (Note, however, that spin-zero Cooper pairing may not be 
possible in some cases because it could come in conflict with $\beta$-equilibrium and charge 
neutrality in bulk quark matter \cite{Shovkovy:2003uu,Rajagopal:2005dg}.)

The simplest superconducting states with spin-zero Cooper pairing are the two-flavor color 
superconducting (2SC) phase \cite{Alford:1997zt,Rapp:1997zu} and color-flavor-locked (CFL) 
phase \cite{Alford:1998mk,Shovkovy:1999mr}. In both cases, the Cooper pairs are made 
of quarks of different flavors. This is a simple consequence of the Pauli principle and the 
specifics of the attractive channel between colored quarks. Even though all Cooper pairs 
are electrically charged, it appears that both types of spin-zero color superconductors 
are characterized by an unbroken $\tilde{\mbox{U}}(1)_{\rm em}$ gauge symmetry 
in the ground state \cite{Shovkovy:2004me,Alford:2007xm}. The corresponding new photon 
field and the new electric charges of quarks are not the same as in vacuum, but this is of 
little importance. The key property of such superconductors is that they are not {\it electromagnetic}
superconductors. This, in turn, implies that there is no electromagnetic Meissner effect 
and a constant magnetic field is not expelled from the interior of color superconductors 
\cite{Alford:1999pb,Gorbar:2000ms,Sedrakian:2002mk}. For the discussion of the 
electromagnetic Meissner effect in several proposed spin-one color superconducting 
phases, see Ref.~\cite{Schafer:2000tw,Alford:2002rz,Schmitt:2003xq,Schmitt:2003aa}. 
(As for the physics properties of spin-one color superconductors in a magnetic field, some 
of them are discussed in Refs.~\cite{Feng:2009vt,Wu:2011qj}.)

The fact that the magnetic field can penetrate spin-zero color superconductors can have 
interesting implications. In particular, it can modify the strength of pairing in different channels 
and, thus, affect the underlying structure and the physical properties of color superconductors. 
Over the years, this topic attracted a lot of attention because it may have interesting 
applications in compact stars. Many studies are devoted to the magnetic CFL phase 
\cite{Ferrer:2005vd,Ferrer:2006vw,Manuel:2006gu,Ferrer:2007iw,Noronha:2007wg,
Fukushima:2007fc,Feng:2011fj,Feng:2011qx,Feng:2012dqa,Feng:2012ma,Wen:2013yra,Ren:2014xra}. 
Even in the absence of the magnetic field, the CFL phase has truly amazing properties. This 
type of quark matter is electrically neutral on its own and, thus, does not need any additional 
electrons \cite{Rajagopal:2000ff}. Just like the physical vacuum, it has a broken chiral symmetry 
and happens to be an electric insulator. At the same time, unlike the vacuum, it is superfluid 
and a very efficient heat conductor. The properties are further enriched by the presence of the 
magnetic field. For a short review, see Ref.~\cite{Ferrer:2012wa}.

There is also a large amount of studies of magnetic 2SC phase \cite{Fayazbakhsh:2010gc,
Fayazbakhsh:2010bh,Yu:2012jn,Fayazbakhsh:2013em}. Unlike
the CFL phase, the 2SC phase has a nonzero density of unpaired quarks that are charged 
with respect to the low-energy electromagnetism. It also has a nonzero density  of 
electrons that are required by $\beta$-equilibrium and charge neutrality. Because of this, 
the 2SC phase has metallic properties. Still, many effects of a constant magnetic field on 
the 2SC pairing are similar to those in the CFL phase. 

Among the related ideas regarding the magnetic fields and dense quark matter, we should 
also mention the proposal that superstrong magnetic fields could be generated as a result of 
an enhancement from gluon vortices in color superconductors \cite{Ferrer:2006ie,Ferrer:2007uw},
as well as the suggestion that dense quark matter should be in a ferromagnetic state that either 
coexists with color superconductivity \cite{Tatsumi:2003bk,Tatsumi:2005ys} or replaces it 
\cite{Iwazaki:2005nr}. If correct, the beauty of such arguments is that the characteristic 
magnetic fields produced by these mechanisms are determined by the QCD scale, i.e.,
 $10^{18}~\mbox{G}$. Such strong fields in the stellar cores could provide a simple 
 explanation for the magnetars with the surface magnetic fields of the order of 
 $10^{15}~\mbox{G}$ \cite{Kouveliotou:1998ze,Rea:2011fa,Olausen:2013bpa}.

\subsection{Superconductivity of vacuum in a magnetic field}

Another interesting idea is the possibility of turning the QCD 
vacuum into a superconductor by a sufficiently strong magnetic field \cite{Chernodub:2010qx,Chernodub:2011mc}.
The strength of the magnetic field needed for such a transition is estimated to be of the order of $10^{20}~\mbox{G}$. 
The resulting ground state is expected to be an anisotropic inhomogeneous superconductor that allows 
nondissipative currents along the direction of the magnetic field. This intriguing type of superconductivity
is suggested to be driven by the interaction of the magnetic moments of charged $\rho^{\pm}$ vector 
mesons with the external magnetic field. 

Conceptually, this idea is similar to a condensation of the $W$-bosons in the standard model 
of electroweak interactions in a superstrong magnetic field proposed long time ago in 
Refs.~\cite{Nielsen:1978rm,Ambjorn:1988gb,Ambjorn:1988tm}. After taking into account 
the potential energy of the magnetic moment, one finds that the {\it effective} mass squared 
of the corresponding mesonic vector states is given by $m_{\rho^{\pm},{\rm eff}}^2(B) = m_{\rho^{\pm}}^2 -|eB|$, 
where $m_{\rho^{\pm}}^2$ is the square of the actual mass of the charged $\rho$-mesons 
\cite{Chernodub:2010qx,Chernodub:2011mc}. Of course, the latter may also be affected by 
the presence of the field via the field-modified masses of constituent quarks and the change 
of the binding energy inside the composite $\rho$-meson state. 

By using simple model calculations \cite{Chernodub:2010qx,Chernodub:2011mc} and some indications
from lattice simulations \cite{Braguta:2011hq,Braguta:2013uc,Braguta:2014ksa}, it was argued that 
the $\rho$-meson condensation is indeed possible in sufficiently strong magnetic fields. This implies that,
even if the $\rho$-meson mass $m_{\rho^{\pm}}$ increases  with the magnetic field (e.g., because of the 
increase of the constituent quark masses due to magnetic catalysis), there should still exist a sufficiently 
large value of the magnetic field, at which $m_{\rho^{\pm},{\rm eff}}(B)$ vanishes and a transition to a 
superconducting state occurs. The corresponding new state has an elaborate vertex lattice structure 
\cite{Chernodub:2011gs,Chernodub:2012fi} that is built out of charged as well as neutral $\rho$-meson
condensates. For a recent review of the corresponding ideas, we refer the reader to Ref.~\cite{Chernodub:2014rya}.

Despite the apparent plausibility of the scenario presented above, the authors Ref.~\cite{Hidaka:2012mz} 
argued that the magnetic field driven superconductivity of the QCD vacuum cannot happen. By making 
use of rather general arguments, relying on the Vafa-Witten theorem, the authors of Ref.~\cite{Hidaka:2012mz} 
claimed that the key ingredient of the proposed scenario, i.e., a charged vector meson condensation, 
is forbidden in the vectorlike gauge theories such as QCD. On the other hand, the author 
of the original proposal disagrees with such a conclusion \cite{Chernodub:2012zx} by arguing that there 
is no direct disagreement with the Vafa-Witten theorem because the charged vector meson condensates 
lock relevant internal global symmetries of QCD with the electromagnetic gauge group. Later, 
a different group of authors \cite{Li:2013aa} also suggested that the strong version of the 
Vafa-Witten theorem may be inapplicable to the case of the charged $\rho$-meson condensation. 

Among the recent model studies, there are those that favor the $\rho$-meson condensation 
\cite{Frasca:2013kka,Liu:2014uwa}, while there are others that do not \cite{Andreichikov:2013zba}. 
The analogs of inhomogeneous states with a $\rho$-meson condensation were also found in 
holographic models \cite{Erdmenger:2010zz,Callebaut:2011ab,Bu:2012eca,Bergman:2012na,
Bu:2012mq,Callebaut:2013wba}. It is not clear, however, if the approximations used in such 
models could even in principle reproduce the limitations of the Vafa-Witten theorem. Therefore,
we think that the debate about the superconductivity of the QCD vacuum in a magnetic field 
is not over yet.

\section{Summary}
\label{sec:Summary}

The main theme of this report can be stated as follows: there exist a number of interesting 
properties of relativistic field theories that are driven or strongly influenced by the presence 
of an external magnetic field. In most cases, the essence of underlying physics has roots
in the characteristic Landau level spectrum of relativistic fermions in a magnetic field. Not
only Landau levels are highly degenerate, but also the low-energy dynamics is often 
dominated by the spin-polarized lowest Landau level. 

We started the report with the detailed exposition of the mechanism of magnetic catalysis of 
dynamical symmetry breaking within the framework of the NJL models in $2+1$ and $3+1$ 
dimensions. The short-range interaction in such models provides an instructive presentation 
of the core ideas in their simplest form. As explained, the physics of magnetic catalysis is
connected with the following key properties of charged massless fermions in a constant 
magnetic field: (i) a finite density of states at zero energy (i.e, the energy separating particle 
and antiparticle states) and (ii) the effective dimensional reduction $D\to D-2$ of the fermion 
dynamics at low energies. 

The combined effect of the finite density of states and the dimensional reduction is a strong
enhancement of the fermion-antifermion pairing dynamics that triggers a chiral (flavor)
symmetry breaking instability. In many ways, the mechanism is reminiscent of the famous 
Cooper instability in low-temperature superconductivity. The resulting ground state is
characterized by a nonvanishing chiral (flavor) condensate and a dynamically generated
mass of fermion excitations (i.e., an energy gap in the spectrum). 

Having established a clear understanding of the underlying physics, we argue that the 
magnetic catalysis is a model-independent phenomenon that should be expected in a 
wide range of field-theoretical models. As a large number of studies using various 
models and methods confirm, this is indeed the case. To illustrate this in the context 
of renormalized field theories, in this report we presented detailed studies of several 
types of the gauge theories in a magnetic field. While the long-range interaction makes 
the analysis much more sophisticated, the underlying physics remains the same 
as in the models with short-range interactions. When addressing the implications of
the magnetic catalysis in QCD, we also encountered interesting properties of induced
anisotropic gluodynamics as a byproduct. 

Historically, the discovery of graphene and subsequent studies of the unusual realization 
of the quantum Hall effect came after the physics of magnetic catalysis was clearly 
understood. We followed this historical order and discussed the corresponding ideas 
in the context of graphene in detail. It should be noted that this was one of the first few 
serious attempts to apply the ideas of magnetic catalysis to condensed matter systems
\cite{Khveshchenko:2001zz,Khveshchenko:2001zza,Gorbar:2002iw}. 
(A few studies that mention other possible condensed matter applications appeared in 
Refs.~\cite{Semenoff:1998bk,Ferrer:2001fz}.) It is natural, therefore, that we dedicated 
a section of this report to the discussion of quantum Hall effect in graphene. 

The profound role of the magnetic probes in graphene started from the unambiguous
experimental confirmation that its quasiparticles are indeed described by Dirac fermions.
This followed from the abnormal features of the observed quantum Hall effect in weak 
magnetic fields that matched the theoretical predictions of relativistic Dirac theory. The 
subsequent studies of graphene revealed that new quantum Hall states can form in 
sufficiently strong magnetic fields. As we argue in this report, some of them could be 
explained in principle by the same physics as the magnetic catalysis. Others require 
generalizations that also capture the physics of quantum ferromagnetism. Of course,
the low-energy physics of graphene is much more complicated than the simplest relativistic 
models. A large number of explicit symmetry breaking terms and various intrinsic solid state
effects make the theory of quantum Hall physics much more intricate. The ultimate 
judgment of the validity of theoretical predictions comes from experiment.

We also discussed the physics of chiral asymmetry in magnetized relativistic matter. 
In a way, this is a natural extension of the physics of relativistic vacuum in a magnetic 
field to the case of nonzero density of matter. The possibility that some physics properties 
of such matter in a semiclassical regime are related to the chiral anomaly in quantum field 
theory is rather unusual and intriguing. However, this is indeed the case in the chiral separation 
and chiral magnetic effects, observed even in a free theory. In essence, both effects are based 
on the unique spin-polarization nature of the lowest Landau level. It is suggested 
that these effects may have observational consequences in condensed matter 
physics, heavy-ion collisions, the early Universe, and cosmology. 

We show that the chiral separation effect has further implications in an interacting theory. 
It induces a chiral asymmetry in all Landau levels and affects the low-energy physics 
dominated by the states in the vicinity of the Fermi surface. The resulting chirally 
asymmetric Fermi surface appears to be a superposition of the Fermi surfaces 
of (mostly) left-handed and (mostly) right-handed fermions. The two Fermi surfaces 
are shifted with respect to each other along the direction of the magnetic field. The 
underlying physics of this phenomenon is connected to the fact that the chiral 
asymmetry of the lowest Landau level is promoted to all Landau levels by 
interaction effects. As argued, the resulting chiral asymmetry could have interesting
implications for the physics of compact stars. 

The popularity of graphene and other exotic materials in condensed matter physics
naturally evolved to the point of accepting that 3D analogs of graphene may also exist. 
Several new classes of 3D quasirelativistic materials, the so-called Dirac and Weyl 
(semi-)metals, were proposed. The existence of Dirac (semi-)metals were experimentally 
confirmed. While the Weyl (semi-)metals have not been discovered yet, the common 
belief is that it is just a matter of time before they are also confirmed by experiment. 
In this report, we discussed a few effects associated with a constant magnetic field 
in Dirac (semi-)metals. As we pointed out, one of the consequences of applying the 
magnetic field is a spontaneous transformation of Dirac into Weyl metals. The 
underlying mechanism is the same as in dense relativistic matter in compact stars.

In the last section of the report, we emphasized that field theories in a magnetic field 
determine a new class of noncommutative field theories. In particular, the UV/IR mixing, 
usually taking place in the conventional noncommutative field theories, is absent in this 
case. The reason for this is the exponentially damping (form-)factors both in vertices 
and in the propagators of fields. Such field theories can be relevant not only 
for relativistic systems but also for nonrelativistic systems in a magnetic field. 
The concrete examples of such systems are provided by graphene and Dirac 
semimetals.

\section*{Acknowledgments}
\addcontentsline{toc}{section}{Acknowledgments}

The authors are grateful to E.~V.~Gorbar, V.~P.~Gusynin, D.-K.~Hong,  Junji~Jia, L.~Xia, 
P.~K.~Pyatkovskiy, G.~W.~Semenoff, S.~G.~Sharapov, P.~Sukhachov, X.~Wang, and 
L.~C.~R.~Wijewardhana for fruitful collaboration and numerous discussions over the years. 
At various times, the authors also benefited from discussions with many other researchers. 
While it may be impossible to mention everybody, they would like to acknowledge useful 
discussions with Jens~Andersen, Gokce~Basar, Maxim~Chernodub, Tom~Cohen, 
Gerald~Dunne, Gergo~Endrodi, Efrain~Ferrer, Bo~Feng, Eduardo Fraga, Kenji~Fukushima, 
Elena~Gubankova, Simon~Hands, Tetsuo~Hatsuda, Igor~Herbut, Yoshimasa~Hidaka, 
Defu~Hou, Mei~Huang, Vivian~Incera, Karl~Landsteiner, Dmitri~Kharzeev, 
Jinfeng~Liao, Vadim~Loktev, Berndt~M{\"u}ller, Y.~Jack~Ng, Jorge~Noronha, 
Tony~Rebhan, Hai-Cang~Ren, Dirk~Rischke, Bittan~Roy, Andreas~Schaefer,  
Andreas~Schmitt, Dam~Son, Motoi~Tachibana, Tanmay~Vachaspati, Qun~Wang, 
Lang~Yu, Eric~Zhitnitsky, and Pengfei~Zhuang. 

A part of this review was written when V.A.M. visited the Kobayashi-Maskawa Institute at Nagoya 
University and the Institute for Theoretical Physics at J.W.~Goethe University. He expresses his 
gratitude to the Kobayashi-Maskawa Institute and the Helmholtz International Center ``HIC for FAIR"
at Goethe University for financial support, as well as his appreciation to Prof. Koichi Yamawaki 
and Prof. Dirk Rischke for their warm hospitality. Also, when working on this review, I.A.S. visited 
the Institute of High Energy Physics at the Chinese Academy of Sciences. He would like to thank 
Prof.~Mei~Huang, Dr.~Jingyi~Chao, Dr.~Lang~Yu, and other members of the Institute for their 
kind hospitality and for creating a stimulating working atmosphere. 

The work of V.A.M. was supported by the Natural Sciences and Engineering Research Council 
of Canada. The work of I.A.S. was supported in part by the Chinese Academy of Sciences Visiting
Professorship for Senior International Scientists and by U.S. National Science Foundation under 
Grant No.~PHY-1404232.

\appendix
\renewcommand*{\thesection}{\Alph{section}}
\addcontentsline{toc}{section}{Appendices}

\section{Fermion propagator in a magnetic field}
\label{App:FermionProp}

In this Appendix, we present several different forms of the fermion propagator
in an external magnetic field. To start with, we will quote the Schwinger result. Its 
derivation \cite{Schwinger:1951nm} is based on the proper-time method and 
allows one to reduce the derivation of the propagator to the calculation of a path 
integral for a simple quantum mechanical evolution operator. This method is 
well known and will be omitted here. The Schwinger representation has a 
compact form in coordinate space and is very convenient in many applications. 
On the other hand, the Landau level structure of the energy spectrum is hidden 
and far from being obvious in the Schwinger propagator. Of course, it can 
be made explicit after some transformations. Instead of following this approach, 
we will use another method for deriving the fermion propagator in a magnetic field. 
By construction, the new method has the Landau level representation built in. 
As an added benefit, it also allows an easy generalization to the case of nonzero
chemical potential, chiral shift, chiral chemical potential, and other parameters that 
can be generated dynamically.

\subsection{Schwinger propagator}
\label{App:A-SchwingerProp}

In the coordinate space, the fermion propagator is
\begin{equation}
S(u,u^\prime) = (i\hat{D}+m)_u\langle u |\frac{-i}{m^2+\hat{D}^2} | u^\prime \rangle
= (i\hat{D}+m)_u\int^\infty_0ds\langle u|\exp[-is(m^2+\hat{D}^2)]|u^\prime\rangle,
\label{eq:Sxy}
\end{equation}
where $\hat{D}\equiv\gamma^\mu D_\mu$ and $D_\mu$ is the covariant derivative
in Eq.~(\ref{eq:dmu}).

The matrix element $\langle u|e^{-is(m^2+\hat{D}^2)}| u^\prime \rangle$ can be calculated by 
using the Schwinger (proper-time) approach \cite{Schwinger:1951nm}. The result reads
\begin{equation}
\langle u |e^{-is(m^2+\hat{D}^2)}| u^\prime \rangle=
  \frac{e^{-i\frac{\pi}{4}}}{8(\pi s)^{3/2}} e^{i\left(S_{\rm cl}-sm^2\right)}
\left(eBs\cot(eBs)+  \gamma^1\gamma^2eBs\right),
\label{eq:xy}
\end{equation}
where
\begin{equation}
S_{\rm cl} = \Phi(\mathbf{r}_\perp,\mathbf{r}_\perp^\prime) - \frac{1}{4s} (u-u^\prime)^\nu C_{\mu\nu}(u-u^\prime)^\mu.
\end{equation}
In the last expression, $\Phi(\mathbf{r}_\perp,\mathbf{r}_\perp^\prime) = 
- e\int^{u}_{u^\prime} A_\lambda  dz^\lambda$ is the Schwinger phase, 
in which the integral is calculated along the straight line. We also introduced the shorthand notation 
$C_{\mu\nu}=g_{\mu\nu}+ (F^2)_{\mu\nu}\left[1-eBs\cot(eBs)\right]/B^2$. From Eqs.~(\ref{eq:Sxy}) 
and (\ref{eq:xy}), we finally obtain the following explicit expression for the fermion propagator:
\begin{equation}
S(u,u^\prime) =e^{i\Phi(\mathbf{r}_\perp,\mathbf{r}_\perp^\prime)} \bar{S} (u-u^\prime),
\label{eq:green}
\end{equation}
where the translational part of the propagator reads
\begin{equation}
\bar{S} (u) = \int^\infty_0 \frac{ds}{8(\pi s)^{3/2}}
e^{-i(\frac{\pi}{4}+sm^2)} e^{-\frac{i}{4s}(u^\nu C_{\nu\mu}u^\mu)}
 \left[\left(m+\frac{1}{2s} \gamma^\mu C_{\mu\nu} u^\nu 
 + \frac{e}{2} \gamma^\mu F_{\mu\nu}  u^\nu\right) \left( es B \cot (eBs) 
+\frac{es}{2} \gamma^\mu\gamma^\nu F_{\mu\nu} 
\right)\right].
\label{eq:green-transl-inv}
\end{equation}
The Fourier transform $\bar{S}(k) = \int d^3x e^{ikx}\bar{S} (x)$ of this  
propagator is given by 
\begin{equation}
\bar{S}(k) = \int^\infty_0 ds \exp\left[-ism^2+isk_0^2-is
\mathbf{k}^2
\frac{\tan(eBs)}{eBs} \right] \left( \hat{k}+m - (k^2\gamma^1-k^1\gamma^2)\tan(eBs)\right)
\left(1 - \gamma^1\gamma^2\tan(eBs)\right).
\label{eq:Fourier-transl-inv}
\end{equation}

\subsection{Fermion propagator in the Landau-level representation}
\label{CSE:App:Landau-level-rep}

Here we present an alternative method of derivation of the fermion propagator in
an external magnetic field. This method sufficiently general to allow the inclusion 
of nonzero chemical potential $\mu$, chiral shift $\Delta$ and several other types 
of parameters that can be generated dynamically. Also, in this method, a simple 
procedure can be used to include the dependence of all the parameters on the 
Landau level index $n$. 

By definition, the fermion propagator is given by the following matrix element:
\begin{equation}
G(u,u^{\prime}) =  i\langle u|\left[
(i\partial_t+\mu)\gamma^0 -(\bm{\pi} \cdot\bm{\gamma}) - \Sigma
\right]^{-1}|u^\prime\rangle,
\label{GuuSigma}
\end{equation}
where $u=(t,\mathbf{r})$ and $\mathbf{r}=(x,y,z)$, and the generalized self-energy 
function has the following form:
\begin{equation}
\Sigma = m -\mu_5\gamma^0\gamma^5 -\Delta\gamma^3\gamma^5 -a \gamma^0\gamma^3
-i\tilde{\mu}\gamma^1\gamma^2 - m_5 \gamma^5.
\label{generalized-self-energy}
\end{equation}
By assumption, the direction of the magnetic field coincides with the $z$-direction and the 
corresponding vector potential is taken in the Landau gauge $A^k = -By \delta^{k}_{1}$. 
The canonical momenta are defined as $\pi^{k} \equiv i (\partial^k + i e A^k)$ where $k=1,2,3$.

Let us introduce the modified self-energy
\begin{equation}
\tilde{\Sigma}= - \gamma^1\Sigma\gamma^1
= m -\mu_5\gamma^0\gamma^5 -\Delta\gamma^3\gamma^5  -a \gamma^0\gamma^3
+i\tilde{\mu}\gamma^1\gamma^2 + m_5 \gamma^5,
\end{equation} 
which differs from $\Sigma$ only by the signs of the last two terms, and rewrite the 
matrix element in Eq.~(\ref{GuuSigma}) as follows:
\begin{eqnarray}
G(u,u^{\prime})&=& i\langle u| 
  \left[ (i\partial_t+\mu)\gamma^0 -(\bm{\pi} \cdot\bm{\gamma}) + \tilde{\Sigma}\right]
\left\{
 \left[ (i\partial_t+\mu)\gamma^0 -(\bm{\pi} \cdot\bm{\gamma})  - \Sigma\right]
 \left[ (i\partial_t+\mu)\gamma^0 -(\bm{\pi} \cdot\bm{\gamma})  + \tilde{\Sigma}\right]
 \right\}^{-1}|u^\prime\rangle
\nonumber\\
&=& i\langle u| \left[(i\partial_t+\mu)\gamma^0 -(\bm{\pi} \cdot\bm{\gamma}) + \tilde{\Sigma}\right]
 \Bigg[
 (i\partial_t+\mu)^2-\bm{\pi}^2+i\gamma^1\gamma^2 eB+\tilde{\mu}^2+\mu_5^2+m_5^2-m^2-\Delta^2-a^2
 \nonumber\\
 &&
 + 2 \left(a \pi^3+m_5 \mu_5 \right)\gamma^0
 + 2 \left[m_5 \Delta - a (i\partial_t+\mu) \right]\gamma^3
 -2[\Delta \pi^3+\mu_5(i\partial_t+\mu) ]\gamma^5
 \nonumber\\
 &&
 + 2 \left(a m+m_5 \tilde{\mu} \right)\gamma^0\gamma^3 
 +2(m\mu_5-\tilde{\mu}\pi^3)\gamma^0\gamma^5
 +2[m\Delta+\tilde{\mu}(i\partial_t+\mu) ]\gamma^3\gamma^5
 \Bigg]^{-1}
|u^\prime\rangle,
\label{propagator1}
\end{eqnarray}
Because of the preserved translational invariance in the $t$- and $z$-directions, it is convenient 
to perform the Fourier transforms in the corresponding coordinates,
\begin{equation}
G(\omega,k^{3};\mathbf{r}_{\perp},\mathbf{r}_{\perp}^{\prime})
=\int dt \, dz\, e^{i\omega (t-t^\prime)-ik^{3} (z-z^\prime)}G(u,u^\prime).
\label{Gomegak3Fourier}
\end{equation}
The corresponding result reads
\begin{equation}
G(\omega,k^{3};\mathbf{r}_{\perp},\mathbf{r}_{\perp}^{\prime})
=i\left[(\omega+\mu)\gamma^0 -(\bm{\pi}_{\mathbf{r}\perp} \cdot\bm{\gamma}_{\perp})
-k^3\gamma^3 + \tilde{\Sigma}\right]
\langle\mathbf{r}_{\perp} | \left({\cal M}-\bm{\pi}_{\perp}^{2} +i\gamma^1\gamma^2 eB\right)^{-1} |
\mathbf{r}_{\perp}^{\prime}\rangle,
\label{propagator-FT}
\end{equation}
where $\bm{\pi}_{\mathbf{r}\perp}$ is the differential operator of the canonical momentum in the
coordinate space spanned by vector $\mathbf{r}_{\perp}=(u,u^\prime)$. The matrix function ${\cal M}$ is given 
by
\begin{eqnarray}
{\cal M} &=& (\omega+\mu)^2-(k^3)^2+\tilde{\mu}^2+\mu_5^2+m_5^2-m^2-\Delta^2-a^2
 + 2 \left(a k^3+m_5 \mu_5 \right)\gamma^0
 + 2 \left[m_5 \Delta -a (\omega+\mu)\right]\gamma^3
 \nonumber\\
 &&
 - 2 [\Delta k^3+\mu_5(\omega+\mu) ]\gamma^5
 +2 \left(a m+m_5 \tilde{\mu} \right)\gamma^0\gamma^3 
 +2(m\mu_5-\tilde{\mu}k^3)\gamma^0\gamma^5
 +2[m\Delta+\tilde{\mu}(\omega+\mu) ]\gamma^3\gamma^5 .
\label{matrix-M}
\end{eqnarray}
By noting that all three operators (i.e., $\bm{\pi}_{\perp}^{2}$, $ie B \gamma^1\gamma^2$, 
and ${\cal M}$) inside the matrix element on the right hand side of Eq.~(\ref{propagator-FT}) 
commute, we will utilize their common basis of eigenfunctions. 

Let us start by presenting the eigenfunctions of the differential operator $\bm{\pi}_{\perp}^{2}$. 
As is well known, it has the eigenvalues $(2n+1)|eB|$ with $n=0,1,2,\dots$ \cite{1930ZPhy...64..629L}. 
As is easy to check, in the Landau gauge used, the corresponding normalized wave functions read
\begin{eqnarray}
\psi_{np}(\mathbf{r}_{\perp})\equiv \langle \mathbf{r}_{\perp}  | n p \rangle
=\frac{1}{\sqrt{2\pi l}}\frac{e^{-\frac{\xi^2}{2}}}{\sqrt{2^nn!\sqrt{\pi}}}
H_n\left(\xi\right) e^{i p x},
\end{eqnarray}
where $l=1/\sqrt{|eB|}$ is the magnetic length, $\xi = y/l+pl\,\mathrm{sign}(eB)$, and 
$H_{n}(\xi)$ are the Hermite polynomials \cite{1980tisp.book.....G}. These wave 
functions satisfy the following normalization and completeness conditions:
\begin{eqnarray}
\int d^{2}\mathbf{r}_{\perp} \,\langle n p |\mathbf{r}_{\perp} \rangle \langle \mathbf{r}_{\perp}  | n^{\prime} p^{\prime}  \rangle = 
\int d^{2}\mathbf{r}_{\perp}\,\psi^{*}_{np}(\mathbf{r}_{\perp})\psi_{n^{\prime}p^{\prime}}(\mathbf{r}_{\perp}) &=& \delta_{nn^{\prime}}
\delta(p-p^{\prime}),
\\
\sum\limits_{n=0}^{\infty}\int\limits_{-\infty}^{\infty} dp \,
\langle \mathbf{r}_{\perp}  | n p\rangle \langle n p |\mathbf{r}_{\perp}^{\prime}  \rangle 
=\sum\limits_{n=0}^{\infty}\int\limits_{-\infty}^{\infty} dp\,
\psi_{np}(\mathbf{r}_{\perp}) \psi^{*}_{np}(\mathbf{r}_{\perp}^{\prime}) &=& \delta(\mathbf{r}_{\perp}-\mathbf{r}_{\perp}^{\prime}),
\label{completeness}
\end{eqnarray}
respectively. Then, by making use of the spectral expansion of the unit operator (\ref{completeness}),
we can rewrite the matrix element on the right hand side of Eq.~(\ref{propagator-FT}) as follows:
\begin{eqnarray}
\langle\mathbf{r}_{\perp}|\left( {\cal M}-\bm{\pi}_{\perp}^{2} + i\gamma^1\gamma^2 eB \right)^{-1}|
\mathbf{r}_{\perp}^{\prime}\rangle 
&=&\sum\limits_{n=0}^{\infty}\int\limits_{-\infty}^{\infty} dp\,
\langle \mathbf{r}_{\perp}  | n p\rangle 
\left[ {\cal M}-(2n+1)|eB|+ i\gamma^1\gamma^2 eB \right]^{-1} 
\langle n p| \mathbf{r}_{\perp}^{\prime}  \rangle  \nonumber\\
&= & \frac{e^{i\Phi(\mathbf{r}_{\perp},\mathbf{r}_{\perp}^{\prime})}}{2\pi l^2} 
\exp\left(-\frac{(\mathbf{r}_{\perp}-\mathbf{r}_{\perp}^{\prime})^2}{4l^2}\right)
\sum\limits_{n=0}^\infty\frac{L_{n}\left(z_\perp\right)}
{{\cal M}-(2n+1)|eB| + i\gamma^1\gamma^2 eB },
\label{auxillary-prop}
\end{eqnarray}
where $L_n(z)$ are Laguerre polynomials \cite{1980tisp.book.....G}, 
\begin{equation}
\Phi(\mathbf{r}_{\perp},\mathbf{r}_{\perp}^{\prime})
= e\int_{\mathbf{r}_{\perp}^{\prime}}^{\mathbf{r}_{\perp}}d\mathbf{r}_{\perp} \cdot \mathbf{A}(r_{\perp})
= - s_{\perp}\frac{(x-x^\prime)(y+y^\prime)}{2l^2}
\end{equation}
is the Schwinger phase, and
\begin{equation}
z_\perp = \frac{(\mathbf{r}_{\perp}-\mathbf{r}_{\perp}^{\prime})^2}{2l^2} 
=  \frac{(x-x^{\prime})^2+(y-y^{\prime})^2}{2l^2}. 
\end{equation}
In order to calculate the integral over the quantum number $p$ in Eq.~(\ref{auxillary-prop}), 
we made use of formula $7.377$ from Ref.~\cite{1980tisp.book.....G},
\begin{equation}
\int\limits_{-\infty}^\infty\,e^{-x^2}H_m(x+y)H_n(x+z)dx
=2^n\pi^{1/2}m!z^{n-m}L_m^{n-m}(-2yz),
\end{equation}
which assumes $m\le n$. By definition, $L^{\alpha}_n(z)$ are the generalized Laguerre
polynomials, and $L_n(z) \equiv L^{0}_n(z)$ \cite{1980tisp.book.....G}.

The matrix operator $ieB\gamma^1\gamma^2$ has eigenvalues $\pm |eB|$ in the 
eigenstates with fixed values of the spin projection on the direction of the magnetic 
field. Instead of introducing the spinor eigenstates explicitly, it is more convenient 
to make use of the projection operators on the corresponding subspaces,
\begin{equation}
{\cal P}_{\pm} =\frac{1}{2}\left[1\pm i\gamma^{1}\gamma^{2}\mathrm{sign} (eB)\right].
\end{equation}
By inserting the unit matrix $I_4 = {\cal P}_{-}+{\cal P}_{+}$ in the expression for the 
sum over $n$ in Eq.~(\ref{auxillary-prop}), we can rewrite it in a more convenient form,
\begin{equation}
\sum\limits_{n=0}^\infty \frac{{\cal P}_{+}L_{n}\left(z_\perp\right)}{{\cal M}-(2n+1)|eB|+|eB|}+ 
\sum\limits_{n=0}^\infty \frac{{\cal P}_{-}L_{n}\left(z_\perp\right)}{{\cal M}-(2n+1)|eB|-|eB|}
=\sum\limits_{n=0}^\infty\left[{\cal P}_{+}L_{n}\left(z_\perp\right)+ {\cal P}_{-}L_{n-1}\left(z_\perp\right)\right]
\frac{1}{{\cal M}-2n|eB|},
\label{A9}
\end{equation}
where we shifted the summation index by $1$ in the second sum and assumed that 
$L_{-1}(z)\equiv 0$. Since the projectors ${\cal P}_{\pm}$ commute with the matrix 
$({\cal M}- 2n|eB|)^{-1}$, it is irrelevant whether they are multiplied on the left or on 
the right. However, in the derivation below, it will be more convenient to keep the 
matrix factor $({\cal M}-2n|eB|)^{-1}$ on the far right.

By substituting the last expression into Eq.~(\ref{auxillary-prop}), the matrix element 
takes the following form:
\begin{equation}
\langle\mathbf{r}_{\perp}|
\left[{\cal M}-\bm{\pi}_{\perp}^{2}-ieB\gamma^1\gamma^2\right]^{-1}
|\mathbf{r}_{\perp}^{\prime}\rangle
=\frac{e^{i\Phi(\mathbf{r}_{\perp},\mathbf{r}_{\perp}^{\prime})}}{2\pi l^{2}}
\exp\left(-\frac{(\mathbf{r}_{\perp}-\mathbf{r}_{\perp}^{\prime})^2}{4l^2}\right)
\sum\limits_{n=0}^\infty
\left[{\cal P}_{+}L_{n}\left(z_\perp\right)+ {\cal P}_{-}L_{n-1}\left(z_\perp\right)\right]
\frac{1}{{\cal M}-2n|eB|}.
\end{equation}
When substituting this matrix element into the expression for the fermion propagator in 
Eq.~(\ref{propagator-FT}), it will be desirable to write the final expression in a form with 
the Schwinger phase as an overall factor on the left. In order to arrive at such a result, 
one should take into account, however, that the phase does not commute with the 
(differential) momentum operator $\bm{\pi}_{\mathbf{r}\perp}$. Indeed, 
\begin{eqnarray}
\pi_{x}e^{i\Phi(\mathbf{r}_{\perp},\mathbf{r}_{\perp}^{\prime})} &=& 
e^{i\Phi(\mathbf{r}_{\perp},\mathbf{r}_{\perp}^{\prime})}\left(-i\partial_{x}
+ s_\perp  \frac{y-y^{\prime}}{2l^{2}}\right),\\
\pi_{y}e^{i\Phi(\mathbf{r}_{\perp},\mathbf{r}_{\perp}^{\prime})} &=& 
e^{i\Phi(\mathbf{r}_{\perp},\mathbf{r}_{\perp}^{\prime})}\left(-i\partial_{y} 
- s_\perp \frac{x-x^{\prime}}{2l^{2}}\right).
\end{eqnarray}
By making use of these relations, we finally derive the expression for the full propagator 
(\ref{propagator-FT}) in the following form:
\begin{equation}
G(\omega,k^{3};\mathbf{r}_{\perp},\mathbf{r}_{\perp}^{\prime})
=e^{i\Phi(\mathbf{r}_{\perp},\mathbf{r}_{\perp}^{\prime})}
\bar{G}(\omega,k^{3};\mathbf{r}_{\perp}-\mathbf{r}_{\perp}^{\prime}),
\label{Gomegak3-a}
\end{equation}
where the translation invariant part of the propagator is
\begin{equation}
\bar{G}(\omega,k^{3};\mathbf{r}_{\perp}-\mathbf{r}_{\perp}^{\prime}) 
= i \frac{e^{-z_\perp/2}}{2\pi l^{2}} 
\sum\limits_{n=0}^\infty
\Bigg\{\left[(\omega+\mu)\gamma^0 -k^3\gamma^3 + \tilde{\Sigma} 
\right]\left[  L_{n}(z_\perp){\cal P}_{+}+L_{n-1}(z_\perp){\cal P}_{-}\right] 
-\frac{i}{l^2}  \bm{\gamma}\cdot\left(\mathbf{r}_{\perp} - \mathbf{r}_{\perp}^{\prime}\right)
L^{1}_{n-1}(z_\perp) 
\Bigg\}\frac{1}{{\cal M}- 2n|eB|}  .
\label{propagatorG-FT}
\end{equation}
Note that the ordering of the matrix factors in this expression is very important 
because matrix ${\cal M}$ does not commute with the expression in the curly 
brackets. 

Let us also calculate the Fourier transform of the propagator (\ref{propagatorG-FT}) 
with respect to the perpendicular spatial coordinates,
\begin{equation}
\bar{G}(\omega,\mathbf{k}) = \int d^2\mathbf{r}_{\perp}\, 
e^{-i (\mathbf{k}_{\perp} \cdot \mathbf{r}_{\perp})}
\bar{G}(\omega,k^{3};\mathbf{r}_{\perp}),
\end{equation}
where, by definition, $\mathbf{k}=(k^1,k^2,k^3)$ and $\mathbf{k}_{\perp}=(k^1,k^2)$. 
When cylindrical coordinates $r_\perp = |\mathbf{r}_{\perp}|$ and $\phi = \arctan(y/x)$ 
are used, the angular integration is reduced to the following two table integrals: 
\begin{eqnarray}
\int_{0}^{2\pi} e^{-ik_{\perp}r_{\perp}\cos \phi } d\phi &= & 2\pi J_{0}(k_{\perp}r_{\perp}),\nonumber\\
\int_{0}^{2\pi} e^{-ik_{\perp}r_{\perp}\cos \phi } \cos \phi d\phi &= & -2i \pi J_{1}(k_{\perp}r_{\perp}),
\end{eqnarray}
where $J_{n}(x)$ is the Bessel function, and the remaining integral over the radial 
coordinate $r_\perp$ can be performed by using formula $7.421.4$ from 
Ref.~\cite{1980tisp.book.....G}, 
\begin{equation}
 \int_0^\infty x^{\nu+1}e^{-\beta x^2}L^{\nu}_{n}\left(\alpha x^2\right)J_{\nu}(xy)dx
=\frac{y^{\nu}}{(2\beta)^{\nu+1}}\left(\frac{\beta-\alpha}{\beta}\right)^n 
e^{-\frac{y^2}{4\beta}}L^{\nu}_{n}\left( \frac{\alpha y^2}{4\beta(\alpha-\beta)}\right).
\end{equation}
In the end, the result for the Fourier transform reads
\begin{equation}
\bar{G}(\omega,\mathbf{k}) = ie^{-k_{\perp}^2 l^{2}}\sum_{n=0}^{\infty}
(-1)^nD_{n}(\omega,\mathbf{k})\,\frac{1}{{\cal M}-2n|eB|},
\label{GDn-new}
\end{equation}
where the numerator of the $n$th Landau level contribution is determined by
\begin{equation}
D_{n}(\omega,\mathbf{k}) = 
2\left[(\omega+\mu)\gamma^0 -k^3\gamma^3 + \tilde{\Sigma} \right]
\left[{\cal P}_{+}L_n\left(2 k_{\perp}^2 l^{2}\right)
-{\cal P}_{-}L_{n-1}\left(2 k_{\perp}^2 l^{2}\right)\right]
 + 4(\mathbf{k}_{\perp}\cdot\bm{\gamma}_{\perp}) 
 L_{n-1}^1\left(2 k_{\perp}^2 l^{2}\right).
\label{Dn}
\end{equation}
By making use of the explicit form of ${\cal M}$ in Eq.~(\ref{matrix-M}), the 
last matrix factor in Eq.~(\ref{GDn-new}) can be also inverted explicitly,
\begin{eqnarray}
\frac{1}{{\cal M}-2n|eB|} &=& 
\frac{(\omega+\mu)^2-(k^3)^2-m^2+m_5^2-a^2-\Delta^2+\tilde{\mu}^2+\mu_5^2-2n|eB|}{U_n}\nonumber\\
&&
- \frac{2 (a k^3+m_5 \mu_5 )}{U_n} \gamma^0  
- \frac{2 [m_5 \Delta -a (\omega+\mu)]}{U_n} \gamma^3
+ \frac{2 [\Delta k^3+\mu_5(\omega+\mu) ]}{U_n} \gamma^5\nonumber\\
&&
- \frac{2 (a m+m_5 \tilde{\mu} )}{U_n} \gamma^0\gamma^3
- \frac{2(m\mu_5-\tilde{\mu}k^3)}{U_n} \gamma^0\gamma^5
- \frac{2[m\Delta+\tilde{\mu}(\omega+\mu) ]}{U_n} \gamma^3\gamma^5,
\label{inverse-M-2neB}
\end{eqnarray}
where the common denominator is given by
\begin{eqnarray}
U_n &=& 
\left[(\omega+\mu)^2-(k^3)^2-m^2+m_5^2+a^2+\Delta^2-\tilde{\mu}^2-\mu_5^2-2n|eB|\right]^2
+8n|eB| \left(a^2+\Delta^2-\tilde{\mu}^2-\mu_5^2\right)
\nonumber\\
&&
-4\left[ \Delta(\omega+\mu)+a\mu_5+m\tilde{\mu}+\mu_5k^3 \right]^{2} .
\label{U_inv-simple}
\end{eqnarray}
In this general case, the Landau level energies are determined by the location of the 
poles of the propagator, i.e., $U_n=0$. Two special cases of this propagator for  
$a=m_5=\mu_5=0$ and $a=m_5=\tilde{\mu}=0$ were previously presented in 
Refs.~\cite{Gorbar:2011ya} and \cite{Xia:2014wla}, respectively. 

As is easy to check, in the simplest case when the Dirac mass $m$ is the only nonvanishing 
parameter in the self-energy (\ref{generalized-self-energy}), the matrix in Eq.~(\ref{inverse-M-2neB}) 
is proportional to the unit matrix, i.e.,
\begin{equation}
\left.\frac{1}{{\cal M}-2n|eB|}\right|_{\mu_5,\Delta,a,\tilde{\mu},m_5=0} =
\frac{1}{(\omega+\mu)^2-(k^3)^2-m^2-2n|eB|},
\end{equation}
and the propagator in Eq.~(\ref{GDn-new}) coincides with the well known result in 
Ref.~\cite{Chodos:1990vv}. Strictly speaking, we should also specify the prescription 
for handling the poles of the propagator. This is achieved by the replacement $\omega 
\to \omega +i\epsilon\, \mathrm{sign}(\omega)$ in the denominator of the propagator, 
which insures that the all states with energies less then the chemical potential are 
occupied. It can be also shown that it is  equivalent to the Schwinger one.

\subsection{Schwinger parametrization for the fermion propagator at $B\neq 0$ and $\mu\neq 0$}
\label{CSE:App:proper-time-rep-mu}

The proper-time representation for the fermion propagator in a constant external magnetic 
field was obtained long time ago by Schwinger \cite{Schwinger:1951nm}. A naive generalization 
of the corresponding representation to the case of a nonzero chemical potential (or density)
does not work however. This is due to the complications in the definition of the causal Feynman 
propagator in the complex energy plane when $\mu\neq 0$. The correct analytical 
properties of such a propagator describing particles above Fermi surface propagating forward 
in time and holes below Fermi surface propagating backward in time are implemented by 
introducing an appropriate $i\epsilon$ prescription. In particular, one replaces $k_0+\mu$ 
with $k_0+\mu+i\epsilon\, \mathrm{sign}(k_0)$, where $\epsilon$ is a vanishingly small 
positive parameter.
For example, in the Landau level representation, the Fourier transform of the translation 
invariant part of the fermion propagator is defined as follows:
\begin{equation}
\bar{S}(k) = ie^{-k_{\perp}^2  l^{2}}\sum_{n=0}^{\infty}
\frac{(-1)^n D_{n}(k)}{ [k_0+\mu+i\epsilon\, \mathrm{sign}(k_0)]^2 -m^2-k_{3}^2-2n|eB|},
\label{CSE:Fourier-tranlation-inv-S}
\end{equation}
where the residue at each individual Landau level is determined by
\begin{equation}
D_{n}(k) = 2\left[(k_0+\mu)\gamma^{0}+m-k^{3}\gamma^3\right]
\left[{\cal P}_{+}L_n\left(2 k_{\perp}^2  l^{2}\right)
-{\cal P}_{-}L_{n-1}\left(2 k_{\perp}^2  l^{2}\right)\right]
 + 4(\bm{k}_{\perp}\cdot\bm{\gamma}_{\perp}) L_{n-1}^1\left(2 k_{\perp}^2  l^{2}\right),
\end{equation}
where $L^{\alpha}_n(x)$ are associated Laguerre polynomials. 

Let us start by reminding the usual Schwinger's proper-time representation at zero fermion density, i.e., 
\begin{equation}
\frac{1}{[k_0+i\epsilon\, \mathrm{sign}(k_0)]^2-{\cal M}_n^2} 
\equiv \frac{1}{k_0^2-{\cal M}_n^2+i\epsilon} 
= -i\int_{0}^{\infty} d s e^{is(k_0^2-{\cal M}_n^2+i\epsilon)},
\label{CSE:proper-time-concept}
\end{equation}
where ${\cal M}_n^2 = m^2+k_{3}^2+2n|eB|$. It is important to emphasize that the convergence 
of the integral and, thus, the validity of the representation 
are ensured by having the positive parameter $\epsilon$ in the exponent. Unfortunately, 
such a representation fails at finite fermion density. Indeed, by taking into account that 
\begin{equation}
\frac{1}{ [k_0+\mu+i\epsilon\, \mathrm{sign}(k_0)]^2-{\cal M}_n^2} \equiv
\frac{1}{(k_0+\mu)^2-{\cal M}_n^2+i\epsilon\, \mathrm{sign}(k_0)\mathrm{sign}(k_0+\mu)},
\end{equation}
we see that the sign of the $i\epsilon$ term in the denominator is not fixed any more. The corresponding 
sign is determined by the product of $\mathrm{sign}(k_0)$ and $\mathrm{sign}(k_0+\mu)$ and 
can change, depending on the values of $k_0$ and $\mu$. For example, while it is positive
for $|k_0|>|\mu|$, it turns negative when $|k_0|<|\mu|$ and $k_0 \mu<0$. This seemingly 
innocuous property causes a serious problem for the integral representation utilized in 
Eq.~(\ref{CSE:proper-time-concept}). The sign changing $i\epsilon$ term in the exponent 
invalidates the representation at least for a range of quasiparticle energies.

In order to derive a modified proper-time representation for the fermion propagator, 
we will make use of the following identity:
\begin{eqnarray}
&&\frac{1}{ [k_0+\mu+i\epsilon\, \mathrm{sign}(k_0)]^2-{\cal M}_n^2 }
=\frac{\theta(|k_0|-|\mu|)}{(k_0+\mu)^2-{\cal M}_n^2+i\epsilon}
+\theta(|\mu|-|k_0|) 
\left(\frac{\theta(k_0 \mu)}{(k_0+\mu)^2-{\cal M}_n^2+i\epsilon}
+\frac{\theta(-k_0 \mu)}{(k_0+\mu)^2-{\cal M}_n^2-i\epsilon} \right) \nonumber \\
&&\hspace{1in}= \frac{1}{(k_0+\mu)^2-{\cal M}_n^2+i\epsilon}
-\theta(|\mu|-|k_0|) \theta(-k_0 \mu)\left(
\frac{ 1 }{(k_0+\mu)^2-{\cal M}_n^2+i\epsilon}
-\frac{1}{(k_0+\mu)^2-{\cal M}_n^2-i\epsilon}
\right)
\nonumber \\
&&\hspace{1in}=  \frac{1}{(k_0+\mu)^2-{\cal M}_n^2+i\epsilon}
+2 i \pi \, \theta(|\mu|-|k_0|) \theta(-k_0 \mu)\delta \left[(k_0+\mu)^2-{\cal M}_n^2\right].
\label{CSE:pole-identity}
\end{eqnarray}
The first term on the right-hand side of 
Eq.~(\ref{CSE:pole-identity}) has a vacuumlike $i\epsilon$ prescription and, thus, allows a 
usual proper-time representation. The second term is singular and represents the 
additional ``matter" piece, which would be lost in the naive proper-time representation.
After making use of this identity, we derive the following modified proper-time representation 
for the propagator:
\begin{eqnarray}
\bar{S}(k) &=& e^{-k_{\perp}^2  l^{2}}
\sum_{n=0}^{\infty} (-1)^n D_{n}(k)
\int_{0}^{\infty} ds\, e^{i s [(k_0+\mu)^2-m^2-k_{3}^2-2n|e B|+i\epsilon]}\nonumber \\
&-&\theta(|\mu|-|k_0|)\theta(-k_0\mu) e^{-k_{\perp}^2  l^{2}}
\sum_{n=0}^{\infty} (-1)^n D_{n}(k)
\Bigg[
\int_{0}^{\infty} ds\, e^{i s [(k_0+\mu)^2-m^2-k_{3}^2-2n|e B|+i\epsilon]}
+\int_{0}^{\infty} ds\, e^{-i s [(k_0+\mu)^2-m^2-k_{3}^2-2n|e B|-i\epsilon]}
\Bigg].\nonumber \\
\end{eqnarray}
In order to perform the sum over the Landau levels, we use the following result for the 
infinite sum of the Laguerre polynomials:
\begin{equation}
\sum\limits_{n=0}^{\infty} z^n L_n^\alpha(x) = \frac{1}{(1-z)^{1+\alpha}}\exp\left(\frac{x z}{z-1}\right).
\label{CSE:table-sum}
\end{equation}
Then, we obtain
\begin{eqnarray}
\bar{S}(k)&=& 
\int_{0}^{\infty} ds\, e^{i s [(k_0+\mu)^2-m^2-k_{3}^2+i\epsilon]-i k_{\perp}^2  l^{2} \tan(s |e B|)}
\left[(k_0+\mu)\gamma^{0}+m-\mathbf{k}\cdot\bm{\gamma}
+(k^1\gamma^2-k^2\gamma^1)\tan(s e B)\right]
\left[1 - \gamma^1\gamma^2 \tan(s e B) \right] 
\nonumber \\
&-&  \theta(|\mu|-|k_0|) \theta(-k_0\mu) \nonumber \\
&\times &
\Bigg\{
\int_{0}^{\infty} ds 
    e^{i s [(k_0+\mu)^2-m^2-k_{3}^2+i\epsilon]-i k_{\perp}^2  l^{2} \tan(s |e B|)} 
\left[(k_0+\mu)\gamma^{0}+m-\mathbf{k}\cdot\bm{\gamma}
+(k^1\gamma^2-k^2\gamma^1)\tan(s e B)\right]
\left[1 - \gamma^1\gamma^2 \tan(s e B) \right]\nonumber \\
&&+\int_{0}^{\infty} ds 
    e^{- i s [(k_0+\mu)^2-m^2-k_{3}^2-i\epsilon]+i k_{\perp}^2  l^{2} \tan(s |e B|)}
    \left[(k_0+\mu)\gamma^{0}+m-\mathbf{k}\cdot\bm{\gamma}
    - (k^1\gamma^2-k^2\gamma^1)\tan(s e B)\right]
\left[1 + \gamma^1\gamma^2 \tan(s e B) \right]\Bigg\} .
\nonumber \\
\label{CSE:S-prop-time-mu1}
\end{eqnarray}
This is a very convenient alternative representation for the fermion propagator in a constant 
external magnetic when $\mu\neq 0$. It allows, in particular, a straightforward derivation of the 
expansion in powers of the magnetic field. To zeroth order in magnetic field, we obtain
\begin{eqnarray}
\bar{S}^{(0)}(k) = \bar{S}^{(0)}_{\rm vac}(k)+\bar{S}^{(0)}_{\rm mat}(k),
\label{CSE:S0-vac-plus-matter}
\end{eqnarray}
where
\begin{eqnarray}
\bar{S}^{(0)}_{\rm vac}(k)= 
\int_{0}^{\infty} ds\, e^{i s [(k_0+\mu)^2-m^2-\mathbf{k}^2+i\epsilon]}\,\left[(k_0+\mu)\gamma^{0}+m-\mathbf{k}\cdot\bm{\gamma}\right]
\end{eqnarray}
and
\begin{eqnarray}
\bar{S}^{(0)}_{\rm mat}(k)=-2\pi \, \theta(|\mu|-|k_0|) \theta(-k_0\mu)\,\left[(k_0+\mu)\gamma^{0}+m-\mathbf{k}\cdot\bm{\gamma}\right]\,\,
\delta \left[(k_0+\mu)^2-m^2-\mathbf{k}^2 \right]
\label{CSE:S-prop-time-eB0}
\end{eqnarray}
are the vacuum and matter parts, respectively. After integration of the proper time and making use 
of the identity in Eq.~(\ref{CSE:pole-identity}), we find that this is identical to the usual free fermion 
propagator (\ref{CSE:free-term}) in the absence of the field. 

Expanding the expression in Eq.~(\ref{CSE:S-prop-time-mu1}) to linear order in magnetic field, 
we also easily obtain the following linear in $B$ correction to the fermion propagator:
\begin{eqnarray}
\bar{S}^{(1)}(k) &=& - \gamma^1\gamma^2 e B \Bigg\{
\int_{0}^{\infty} s ds\, e^{i s [(k_0+\mu)^2-m^2-\mathbf{k}^2+i\epsilon]} 
+2i\pi  \theta(|\mu|-|k_0|) \theta(-k_0\mu) 
\delta^{\prime}\left[(k_0+\mu)^2-m^2-\mathbf{k}^2 \right]
\Bigg\}\nonumber\\
&&\times \left[(k_0+\mu)\gamma^{0}+m-k^3\gamma^3\right].
\label{CSE:S-prop-time-eB1}
\end{eqnarray}
After integration over the proper time and making use of an identity obtained from Eq.~(\ref{CSE:pole-identity}) 
by differentiating with respect to ${\cal M}_n^2$, we obtain Eq.~(\ref{CSE:linear-term}).

\section{Propagator of composite fields}
\label{App:CompProp}

\subsection{Kinetic term of the low-energy effective action}
\label{App:C1}

Let us now consider the derivation of the kinetic term (\ref{eq:Lk})
in the low-energy effective action:
\begin{equation}
{\cal L}_k = N \frac{F_1^{\mu\nu}}{2} (\partial_\mu\rho_j\partial_\nu\rho_j)
+N \frac{F^{\mu\nu}_2}{\rho^2} (\rho_j\partial_\mu\rho_j)  (\rho_i\partial_\nu\rho_i),\label{eq:Lkapend}
\end{equation}
where $\bm{\rho}=(\sigma,\tau,\pi)$ and $F^{\mu\nu}_1$, $F^{\mu\nu}_2$ depend
on the $\mathrm{U}(2)$-invariant $\rho^2=\sigma^2+\tau^2+\pi^2$. The definition
$\Gamma_k=\int d^3x{\cal L}_k$ and Eq.~(\ref{eq:Lkapend}) imply that the form
of the functions $F^{\mu\nu}_1$ and $F^{\mu\nu}_2$ is determined from the equations:
\begin{eqnarray}
 \left. N^{-1}\frac{\delta^2\Gamma_k}{\delta\sigma(x)\delta\sigma(0)}
 \right|_{\tau=\pi=0,\sigma=const} &=&-
 \left.  (F^{\mu\nu}_1+2F^{\mu\nu}_2)
\right|_{\tau=\pi=0,\sigma=const}
\partial_\mu\partial_\nu\delta^3(x), \\
 \left. N^{-1}\frac{\delta^2\Gamma_k}{\delta\pi(x)\delta\pi(0)}
 \right|_{\tau=\pi=0,\sigma=const}
    &=& \left. - F^{\mu\nu}_1
\right|_{\tau=\pi=0,\sigma=const}
\partial_\mu\partial_\nu\delta^3(x).\label{eq:d2G}
\end{eqnarray}
Here $\Gamma_k$ is the part of the effective action (\ref{eq:effact})
containing terms with two derivatives. Eq.~(\ref{eq:effact}) implies that
$\Gamma_k=\tilde{\Gamma}_k$.  Therefore, we find from Eq.~(\ref{eq:d2G}) that
\begin{equation}
F^{\mu\nu}_1 = -\frac{N^{-1}}{2} \int d^3xx^\mu x^\nu
\frac{\delta^2\tilde{\Gamma}_k}{\delta\pi(x)\delta\pi(0)}
=-\frac{N^{-1}}{2} \int
d^3 x x^\mu x^\nu   \frac{\delta^2\tilde{\Gamma}}{\delta\pi(x)\delta\pi(0)}
\end{equation}
(Henceforth, we will not write explicitly the condition $\tau=\pi=0$,
$\sigma=\mbox{const}$). Taking into account the definition of the fermion
propagator,
\begin{equation}
iS^{-1}=i\hat{D}-\sigma,
\end{equation}
we find from Eq.~(\ref{eq:tildefact}) that
\begin{equation}
\frac{\delta^2\tilde{\Gamma}}{\delta\pi(x)\delta\pi(0)} =
-i\mathrm{tr} \Big(\bar{S}(x,0)i\gamma^5\bar{S}(0,x)i\gamma^5\Big) 
=-i\int\frac{d^3kd^3q}{(2\pi)^6}e^{iqx} \mathrm{tr}
\Big(\bar{S}(k)i\gamma^5\bar{S}(k+q)i\gamma^5\Big)
\end{equation}
[the functions $\bar{S}(x)$ and $\bar{S}(k)$ are given in
Eqs.~(\ref{eq:green}) -- (\ref{eq:Fourier-transl-inv})].
Therefore,
\begin{equation}
F^{\mu\nu}_1=-\frac{iN^{-1}}{2}\int \frac{d^3k}{(2\pi)^3}\mathrm{tr}
\Bigg(\bar{S}(k)i\gamma^5\frac{\partial^2\bar{S}(k)}{\partial k_\mu\partial
k_\nu}i\gamma^5\Bigg).             \label{eq:F1}
\end{equation}
In the same way, we find that
\begin{eqnarray}
F^{\mu\nu}_2 &=&-\frac{iN^{-1}}{4} \int \frac{d^3k}{(2\pi)^3}\mathrm{tr}
\Big(\bar{S}(k)\frac{\partial^2\bar{S}(k)}{\partial k_\mu\partial
k_\nu}\Big)-\frac{1}{2}F^{\mu\nu}_1=\label{eq:F2}\\
&=&-\frac{iN^{-1}}{4}\int\frac{d^3k}{(2\pi)^3}\mathrm{tr}
\Bigg(\bar{S}(k)\frac{\partial^2\bar{S}(k)}{\partial k_\mu\partial k_\nu} -
\bar{S}(k)i\gamma^5 \frac{\partial^2\bar{S}(k)}{\partial k_\mu\partial
k_\nu}i\gamma^5\Bigg).\nonumber
\end{eqnarray}
Taking into account the expression for $\bar{S}(k)$ in Eq.~(\ref{eq:Fourier-transl-inv})
(with $m=\sigma$), we get
\begin{eqnarray}
\frac{\partial^2\bar{S}(k)}{\partial k^0\partial k^0} &=& 2il^4 \int^\infty_0
dt\, t\exp \left[R(t)\right]\Big\{\sigma(1+s_{\perp}\gamma^1\gamma^2T)+ 3k^0\gamma^0(1+s_{\perp}\gamma^1\gamma^2T)
-k^i\gamma^i(1+T^2) \nonumber\\
&+& 2itl^2(k^0)^2 \sigma(1+ s_{\perp}\gamma^1\gamma^2T)+2itl^2(k^0)^3\gamma^0(1+s_{\perp}\gamma^1\gamma^2T)-
 2itl^2(k^0)^2(k^i\gamma^i)(1+T^2)\Big\}, \\
\frac{\partial^2\bar{S}(k)}{\partial k^j\partial k^j}&=&-2il^4 \int^\infty_0
dt T\exp\left[R(t)\right]\Big\{\sigma(1+s_{\perp}\gamma^1\gamma^2T)- k^i\gamma^i(1+T^2)-2k^j\gamma^j(1+T^2)
+k^0\gamma^0(1+s_{\perp}\gamma^1\gamma^2T) \nonumber\\
&-&2iTl^2(k^j)^2\sigma (1+s_{\perp}\gamma^1\gamma^2T)
-2iTl^2(k^j)^2 k^0\gamma^0(1+s_{\perp} \gamma^1\gamma^2T)+2iTl^2(k^j)^2k^i\gamma^i(1+T^2)\Big\}
\label{eq:d2S}
\end{eqnarray}
($i,j=1,2$; there is no summation over $j$), where $s_{\perp}= \mathrm{sign} (eB)$, $T=\tan t$, and
\begin{equation}
R(t) = -it(\sigma l)^2+it(k^0)^2-il^2 \mathbf{k}^2 T.
\end{equation}
Eqs.~(\ref{eq:Fourier-transl-inv}), (\ref{eq:F1}), and (\ref{eq:F2}) imply that the off-diagonal terms 
of $F^{\mu\nu}_1$ and $F^{\mu\nu}_2$ are equal to zero. The diagonal terms are determined 
from Eqs.~(\ref{eq:Fourier-transl-inv}), (\ref{eq:F1}) -- (\ref{eq:d2S}). After tedious, but 
straightforward derivation, we obtain 
\begin{eqnarray}
F^{00}_1 &=& \frac{l}{12\pi^{3/2}} \int^\infty_0 d\tau
\frac{\sqrt{\tau}}{\sinh\tau} e^{-(\sigma l)^2\tau} \Big[(\sigma l)^2\tau
\cosh\tau +\frac{3}{2}\cosh\tau+ \frac{\tau}{\sinh\tau}\Big] 
\nonumber\\
&=&\frac{l}{8\pi}
\Bigg(\frac{1}{\sqrt{2}}\zeta\Big(\frac{3}{2}, \frac{(\sigma l)^2}{2} +1\Big) +
(\sigma l)^{-3}\Bigg),\\
F^{00}_2 &=&- \frac{l(\sigma l)^2}{12\pi^{3/2}} \int^\infty_0 d\tau
\tau^{3/2}e^{-(\sigma l)^2\tau} \coth\tau 
=- \frac{l}{16\pi} \Bigg(\frac{(\sigma l)^2}{2\sqrt{2}} \zeta
\Big(\frac{5}{2}, \frac{(\sigma l)^2}{2}+1\Big) +(\sigma
l)^{-3}\Bigg),\\
F^{11}_1 &=& F^{22}_1=\frac{1}{4\pi\sigma},\\
F^{11}_2 &=& F^{22}_2 =\frac{l(\sigma l)^2}{4\pi^{3/2}} \int^\infty_0 d\tau
\tau^{-1/2} e^{-(\sigma l)^2\tau} \coth \tau (1-\tau\coth\tau)  \nonumber\\
&=& \frac{l}{8\pi} \Bigg(\frac{(\sigma l)^4}{\sqrt{2}}\zeta \Big(\frac{3}{2},
\frac{(\sigma l)^2}{2}+1\Big)+ \sqrt{2}(\sigma l)^2 \zeta \Big(\frac{1}{2},
\frac{(\sigma l)^2}{2}+1\Big) + 2\sigma l-(\sigma l)^{-1}\Bigg).
\end{eqnarray}
Here we used the table integral in Eq.~(\ref{eq:zeta}), as well as the following 
integrals \cite{1980tisp.book.....G}:
\begin{eqnarray}
\int^\infty_0 \frac{\tau^{\mu-1}e^{-\beta\tau}}{\sinh^2\tau}d\tau &=& 2^{1-\mu}
\Gamma(\mu)\left[2\zeta\left(\mu-1,\frac{\beta}{2}\right)- \beta\zeta \left(\mu, \frac{\beta}{2}\right)\right], \quad \mu>2,\\
\int^\infty_0 \tau^{\mu-1}e^{-\beta\tau}\coth^2\tau d\tau &=&
\beta^{-\mu}\Gamma(\mu) + \int^\infty_0
\frac{\tau^{\mu-1}e^{-\beta\tau}}{\sinh^2\tau} d\tau,\quad \mu>2, \\
\int^\infty_0\frac{\tau^{\mu-1}e^{-\beta\tau}\coth\tau}{\sinh^2\tau}d\tau &=&
\frac{\mu-1}{2} \int^\infty_0
 \frac{\tau^{\mu-2}e^{-\beta\tau}}{\sinh^2\tau}d\tau
- \frac{\beta}{2}\int^\infty_0
\frac{\tau^{\mu-1}e^{-\beta\tau}}{\sinh^2\tau}d\tau, \quad \mu>3.
\end{eqnarray}

\subsection{General form of the propagator for composite fields}
\label{App:C2}

In this Appendix we analyze the next-to-leading order in $1/N$ expansion
in the $(2+1)$-dimensional NJL model at zero temperature. Our main goal
is to show that the propagator of the neutral NG bosons $\pi$ and $\tau$
have a $(2+1)$-dimensional structure in this approximation and that (unlike
the $(1+1)$-dimensional Gross-Neveu model \cite{Gross:1974jv,Witten:1978qu}) the $1/N$
expansion is reliable in this model. 

A review of the $1/N$ expansion in $(2+1)$-dimensional four-fermion
interaction models can be found in Ref.~\cite{Rosenstein:1990nm}. For our
purposes, it is sufficient to know that this perturbative expansion is
given by Feynman diagrams with the vertices and the propagators of
fermions and composite particles $\sigma$, $\pi$ and $\tau$ calculated
in leading order in $1/N$. In leading order, the fermion propagator
is given in Eqs.~(\ref{eq:green}) -- (\ref{eq:Fourier-transl-inv}).
As follows from Eq.~(\ref{eq:lag2}), the Yukawa coupling of
fermions with $\sigma$, $\tau$ and $\pi$ is $g_Y=1$ in this
approximation. The inverse propagators of $\sigma$, $\tau$ and $\pi$
are \cite{Rosenstein:1990nm,Gusynin:1991wm,Miransky:1992bj}:
\begin{equation}
D^{-1}_{\bm{\rho}}(x)=N\left(\frac{\Lambda}{g\pi}\delta^3(x)
+i\mathrm{tr} \left[S(x,0)T_{\bm{\rho}}S(0,x)T_{\bm{\rho}}\right]
\right),
\end{equation}
where $\bm{\rho}=(\sigma,\tau,\pi)$ and $T_{\sigma}=1$,
$T_{\tau}=\gamma^3$,
$T_{\pi}=i\gamma^5$. Here $S(x,0)$ is the fermion propagator
(\ref{eq:green}) with the mass $m_{\rm dyn}=\bar{\sigma}$ defined
from the gap equation (\ref{eq:gap2}). For completeness, we
write down the explicit expression for the Fourier transform of
the propagators of the NG bosons:
\begin{eqnarray}
D^{-1}_{\tau}(k)&=&D^{-1}_{\pi}(k)
=\frac{N}{4\pi^{3/2}l}\int_0^\infty \frac{ds}{\sqrt{s} } \int_{-1}^1  du  
e^{-s (m l)^2} \left(1-e^{R(s,u) } \right)  \coth(s)   \nonumber\\
&-&\frac{N}{4\pi^{3/2}l}\int_0^\infty \sqrt{s} ds \int_{-1}^1  du  
e^{-s (m l)^2+R(s,u)} \left[
\frac{1-u^2}{2} \coth(s)  (k_0 l)^2 + \frac{u\sinh(su)\coth(s)-\cosh(su)}{2\sinh(s)}  (\mathbf{k} l)^2 \right],
 \nonumber\\
 \label{D-pi-corrected}
\end{eqnarray}
where
\begin{equation}
R(s,u)=\frac{s}{4}(1-u^2)(k_0 l)^2- \frac{\cosh{s}-\cosh{su}}{2\sinh{s}}(\mathbf{k} l)^2.
\end{equation}
Actually, for our purposes, we need to know the form of these
propagators at small momenta only. We find from
Eqs.~(\ref{eq:Lk}), (\ref{eq:Fij}):
\begin{equation}
D_{\tau}(k)=D_{\pi}(k)=-\frac{4\pi\bar{\sigma}}{N}f^2(\bar{\sigma}l)
\left(k_0^2-f^2(\bar{\sigma}l)\mathbf{k}^2\right)^{-1}
\label{eq:NG}
\end{equation}
where
\begin{equation}
f(\bar{\sigma}l)=\left(\frac{2}{\bar{\sigma}l}\right)^{1/2}
\left(\frac{1}{\sqrt{2}}\zeta\left(\frac{3}{2},\frac{(\bar{\sigma}l)^2}{2}
+1\right)+(\bar{\sigma}l)^{-3}\right)^{-1/2}
\end{equation}
[compare with Eq.~(\ref{v_pi_general})]. We note that the corrected expression for the propagator 
of composite fields in Eq.~(\ref{D-pi-corrected}) differs from that in Ref.~\cite{Gusynin:1994va}.

The crucial point for us is that, because of the dynamical mass $m_{\rm dyn}$,
the fermion propagator is soft in the infrared region
(see Eq.~(\ref{eq:poles})) and that the propagators of the $\tau$ and $\pi$
(\ref{eq:NG}) have a $(2+1)$-dimensional form in the infrared region (as
follows from Eqs.~(\ref{eq:Lk}), (\ref{eq:Fij}) the propagator of $\sigma$
has of course also a $(2+1)$-dimensional form).

\begin{figure}[t]
\begin{center}
\includegraphics[width=0.4\textwidth]{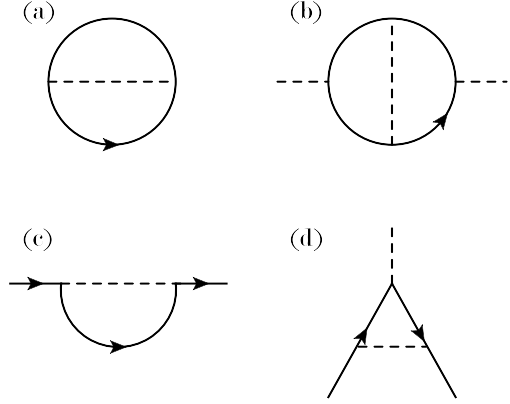}
\caption{(Color online) Next-to-leading diagrams in the $1/N$ expansion. Solid lines denote 
the fermion propagator and dashed lines denote the propagators of $\sigma$,
$\tau$, and $\pi$ in the leading order in $1/N$.}
\label{fig_diagram1}
\end{center}
\end{figure}

Let us begin by considering the next-to-leading corrections in the
effective potential. The diagram which contributes to the effective
potential in this order is shown in Fig.~\ref{fig_diagram1}a. Because of the structure
of the propagators pointed out above, there are no infrared divergences
in this contribution to the potential. (Note that this is in contrast
to the Gross-Neveu model: because of a $(1+1)$-dimensional form of
the propagators of the NG bosons, this contribution is logarithmically
divergent in the infrared region in that model, i.e. the
$1/N$ expansion is unreliable in that case).
Therefore, the diagram in Fig.~\ref{fig_diagram1}a leads to a finite, $O(1)$, correction to
the potential $V$ (we recall that the leading contribution in $V$ is of
order N). As a result, at sufficiently large values of N, the gap
equation in next-to-leading order in $1/N$ in this model admits a
nontrivial solution $\bar{\rho}\neq 0$. Since the potential depends only on
the radial variable $\rho$, the angular variables $\theta$ and $\varphi$
($\bm{\rho}=(\rho\cos\theta,
\rho\sin\theta\cos\varphi, \rho\sin\theta\sin\varphi)$),
connected with the $\tau$ and $\pi$, appear
in the effective lagrangian only through their derivatives. This in turn
implies that the $\tau$ and $\pi$ retain to be gapless NG modes in the
next-to-leading order in $1/N$.

Let us now consider the next-to-leading corrections to the propagators of
these NG modes. First of all, note that in a constant magnetic field, the
propagator of a neutral local field $\varphi(x)$, $D_{\varphi}(u,u^\prime)$, is
translation invariant, i.e., it depends on $(u-u^\prime)$. This immediately
follows from the fact that the operators of space translations (\ref{algebra})
take the canonical form for neutral fields (the operator of time
translations is $i\partial/\partial t$ for both neutral and charged
fields in a constant magnetic field).  The diagrams contributing to the
propagators of the NG modes in this order are shown in Fig.~\ref{fig_diagram1}b. Because
of the dynamical mass $m_{\rm dyn}$ in the fermion propagator, this contribution
is analytic at $k_{\mu}=0$. Since at large $N$ the gap equation has a
nontrivial solution in this approximation, there is no contribution of
$O(k^0)\sim const$ in the inverse propagators of $\tau$ and $\pi$.
Therefore, the first term in the momentum expansion of this contribution
has the form $C_1k_0^2-C_2\mathbf{k}^2$, where $C_1$ and $C_2$ are functions
of $\bar{\sigma}l$, i.e. the propagators take the following form
in this approximation:
\begin{eqnarray}
D_{\tau}(k)=D_{\pi}(k)\stackrel{k\to 0}{=}
-\frac{4\pi\bar{\sigma}}{N}f^2(\bar{\sigma}l)\left[\left(1-\frac{1}{N}
\tilde{C}_1(\bar{\sigma}l)\right)k_0^2-\left(f^2(\bar{\sigma}l)-
\frac{1}{N}\tilde{C}_2(\bar{\sigma}l)\right)\mathbf{k}^2\right]^{-1}.
\end{eqnarray}
[compare with  Eq.~(\ref{eq:NG})].

Because of the same reasons, there are also no infrared divergences either
in the fermion propagator (see Fig.~\ref{fig_diagram1}c) or in the Yukawa vertices
(see Fig.~\ref{fig_diagram1}d) in this order. Therefore, at sufficiently large values of $N$,
the results retain essentially the same as in leading order in $1/N$.

We believe that there should not be principal obstacles to extend this
analysis for all orders in $1/N$.

\section{Thermodynamic potential in NJL model}
\label{App:ThermoPotNJL}

In this Appendix we will derive the thermodynamic potential $V_{\beta,\mu}$
in the $(2+1)$- and $(3+1)$-dimensional and NJL models, defined by Eqs.~(\ref{eq:lag1})
(\ref{12L_NJL_U1xU1}), respectively. We will use the notation $\mu$ is the chemical potential,
$T$ for the temperature, and $\beta=1/T$ for the inverse temperature.

\subsection{Thermodynamic potential in $(2+1)$-dimensional NJL model}
\label{App:ThermoPot2+1}

As is well known \cite{Dolan:1973qd}, the partition function
\begin{equation}
Z_{\beta,\mu}=\mathrm{Tr}\Big[\exp(-\beta H')\Big]
\end{equation}
is expressed through a path integral over fields of a system. Here 
$H'=H-\mu\int\bar{\Psi}\gamma^0\Psi d^2x$ and $H$ is the Hamiltonian 
of the model. In the NJL model (\ref{eq:lag1}), the path integral is
\begin{eqnarray}
Z_{\beta,\mu} &=& \int [d\Psi] [d\bar{\Psi}] [d\sigma][d\tau][d\pi] \exp
\Big\{ i\int^{-i\beta}_0dt\int d^2x  \Big[\bar{\Psi}iS^{-1}\Psi-\frac{1}{2G} \rho^2\Big]\Big\}
\end{eqnarray}
where $\rho^2=\sigma^2+\tau^2+\pi^2$, $S$ is the fermion propagator
(\ref{eq:green}) with $m$ replaced by $\sigma+\gamma^3\tau+i\gamma^5\pi$,
and while the fermion fields satisfy the antiperiodic boundary conditions,
\begin{equation}
\Psi|_{t=0}=-\Psi|_{t=-i\beta},\qquad
\bar{\Psi}|_{t=0}=-\bar{\Psi}|_{t=-i\beta},\label{eq:bound}
\end{equation}
the boson fields satisfy the periodic boundary conditions.

In order to calculate the thermodynamic potential $V_{\beta,\mu}(\rho)$, it
is sufficient to consider configurations with $\tau=\pi=0$ and $\sigma=\mathrm{const}$.
Then, the potential is defined as:
\begin{eqnarray}
\exp \Big\{-\beta V_{\beta,\mu}[\int d^2x]\Big\}
&=& \int[d\Psi][d\bar{\Psi}]\exp \Big\{i\int^{-i\beta}_0dt\int
d^2x  \Big[\bar{\Psi}iS^{-1}\Psi-\frac{1}{2G}\sigma^2\Big]\Big\}.
\end{eqnarray}
At the leading order in $1/N$, this potential defines the thermodynamic
properties of the system.

As is known \cite{Dolan:1973qd}, in the formalism of the imaginary time, the
thermodynamic potential $V_{\beta,\mu}$ can be obtained from the
representation for the effective potential $V$, at $T=0$ and $\mu=0$,
by replacing
\begin{eqnarray}
\int\frac{d^3k}{(2\pi)^3} &\to & \frac{i}{\beta}\sum^{+\infty}_{n=-\infty}
\int\frac{d^2k}{(2\pi)^2}, \label{T:integral-sum}\\
k^0 &\to & i\omega_n+\mu
\end{eqnarray}
where $\omega_n=\frac{\pi}{\beta}(2n+1)$ are the fermionic Matsubara frequencies,
required by the antiperiodic boundary conditions in the imaginary time (\ref{eq:bound}). 
Then, using the representation for the effective potential in Section~\ref{sec:NJL2+1EffPot} 
and the expression for the fermion propagator in Eq.~(\ref{eq:Fourier-transl-inv}), we get
\begin{eqnarray}
V_{\beta,\mu}(\sigma) &=& \frac{\sigma^2}{2G}+\frac{N}{2\pi\beta l^2}
\int^\infty_0\frac{dt}{t} e^{-tl^2(\sigma^2-\mu^2)}\coth t \,
\Theta_2\left(2\pi t\frac{\mu l^2}{\beta}\Big| 4i\pi t\frac{l^2}{\beta^2}
\right),   \label{eq:pot}
\end{eqnarray}
where
\begin{equation}
\Theta_2(u|\tau)=2\sum^\infty_{n=0} e^{i\pi\tau\left(n+\frac{1}{2}\right)^2} \cos
\left[(2n+1)u\right]
\end{equation}
is the second Jacobian theta function \cite{1980tisp.book.....G}.

By using the identity \cite{1980tisp.book.....G}
\begin{equation}
\Theta_2(u|\tau)=\left(\frac{i}{\tau}\right)^{1/2} e^{-i\frac{u^2}{\pi\tau}} \Theta_4
\left(\frac{u}{\tau}\Big| -\frac{1}{\tau}\right),
\end{equation}
where
\begin{equation}
\Theta_4(u|\tau)=1+2\sum^\infty_{n=1}(-1)^n e^{i\pi n^2\tau}\cos(2nu)
\end{equation}
is the fourth Jacobian theta function, one can rewrite the relation
(\ref{eq:pot}) as
\begin{eqnarray}
V_{\beta,\mu}(\sigma) &=& \frac{\sigma^2}{2G} +\frac{N}{4\pi^{3/2}l^3}
\int^\infty_0\frac{dt}{t^{3/2}}  e^{-tl^2\sigma^2}\coth t \, \Theta_4\left(\frac{i}{2}\mu{\beta}\Big|
\frac{i}{4\pi t}\left(\frac{\beta}{l}\right)^2\right) 
\\
&=& V(\sigma)+\frac{N}{2\pi^{3/2}l^3}\int^\infty_0 \frac{dt}{t^{3/2}} \coth
t\sum^\infty_{n=1}(-1)^n\cosh (\mu\beta\ n)  \exp\left[-\left(t\sigma^2l^2+\frac{\beta^2n^2}{4tl^2}\right)\right],
\label{eq:Vbmu}
\end{eqnarray}
where $V(\sigma)$ is the effective potential (\ref{eq:poten}). Let us show that this representation is formally 
equivalent to that in Eq.~(\ref{eq:temperature}).

By using the series
\begin{equation}
\coth t=1+2\sum^\infty_{m=1} e^{-2tm},
\end{equation}
the expression for $\tilde{V}_{\mu,\beta}=V_{\mu,\beta}-V$ in
Eq.~(\ref{eq:Vbmu}) can be rewritten as
\begin{eqnarray}
\tilde{V}_{\mu,\beta} &=& \frac{N}{\pi l^2\beta}\sum^\infty_{n=1} (-1)^n
\frac{\cosh(\mu\beta n)}{n}  \left[e^{-\beta\sigma n}
+ 2\sum^\infty_{m=1}\exp \left(-\beta\sigma n \sqrt{1+\frac{2m}{(\sigma l)^2}}\right)
\right].
\end{eqnarray}
Here we also used the relations \cite{1980tisp.book.....G}:
\begin{eqnarray}
\int^\infty_0dx x^{\nu-1}\exp \left(-\frac{\beta}{x}-\gamma x\right) &=& 2
\left(\frac{\beta}{\gamma}\right)^{\nu/2} K_\nu \left(2\sqrt{\beta\gamma}\right), \nonumber\\
K_{-\frac{1}{2}} (z) = K_{\frac{1}{2}} (z) &=& \left(\frac{\pi}{2z}\right)^{1/2} e^{-z},
\end{eqnarray}
where $K_\nu(z)$ is a modified Bessel function.

Since
\begin{equation}
\sum^\infty_{n=1} (-1)^n\frac{e^{\alpha n}+e^{-\alpha n}}{n} e^{-\beta n} =-
\ln\left(1+e^{-2\beta}+2e^{-\beta}\cosh\alpha\right),
\end{equation}
we find that
\begin{equation}
\tilde{V}_{\mu,\beta} =- \frac{N}{2\pi\beta l^2} \left\{\ln
\left[1+e^{-2\beta\sigma}+2e^{-\beta\sigma} \cosh (\mu\beta)\right] 
+2 \sum^\infty_{m=1} \ln
\left[1+e^{-2\beta\sigma\sqrt{1+\frac{2m}{(\sigma l)^2}}} +2
 e^{-\beta\sigma\sqrt{1+\frac{2m}{(\sigma l)^2}}} \cosh(\mu\beta)\right]\right\}.
\end{equation}
This expression for the thermodynamic potential at nonzero temperature and 
chemical potential was derived in Refs.~\cite{Gusynin:1994va,Andersen:1995ia}. 
It is now easy to check that the expression for the thermodynamic potential
$V_{\beta,\mu}=V+\tilde{V}_{\beta,\mu}$ coincides with that in
Eq.~(\ref{eq:temperature}).

\subsection{Thermodynamic potential in $(3+1)$-dimensional NJL model}
\label{App:ThermoPot}

The derivation is very similar to the $(3+1)$-dimensional case. The result reads
\begin{eqnarray}
V^{(3+1)}_{\beta,\mu}(\rho) &=& V^{(3+1)}(\rho)+\frac{N}{4\pi^{2}l^4}\int^\infty_0 \frac{dt}{t^{2}} \coth
t\sum^\infty_{n=1}(-1)^n\cosh (\mu\beta n)  \exp\left[-\left(t\rho^2l^2+\frac{\beta^2n^2}{4tl^2}\right)\right]
\nonumber\\
&=&V^{(3+1)}(\rho)+\frac{N}{\pi^{2}l^2}\sum^\infty_{n=1}(-1)^n\frac{\cosh (\mu\beta n)}{\beta n}
\left[\rho K_{1}\left(n \beta \rho\right)+2\sum^\infty_{k=1}\sqrt{\rho^2+\frac{2k}{l^2}}K_{1}\left( n \beta \sqrt{\rho^2+\frac{2k}{l^2}}\right)
\right],
\end{eqnarray}
where $V^{(3+1)}(\rho)$ is the effective potential in Eq.~(\ref{16}). The above expression appears to be 
divergent $|\mu|> m$  and, thus, it is not very useful for practical calculations. A better representation of 
the thermodynamic potential is given by \cite{Elmfors:1993wj,Elmfors:1993bm,Elmfors:1994mq}
\begin{equation}
V^{(3+1)}_{\beta,\mu}(\rho) = V^{(3+1)}(\rho)
-\frac{N}{2\beta \pi^2l^2}\int_{0}^{\infty}dk_3
\left\{\ln\left[1+e^{-\beta\left(\sqrt{\rho^2+k_3^2}-\mu\right)}\right]
+2\sum_{n=1}^{\infty}
\ln\left[1+e^{-\beta\left(\sqrt{\rho^2+k_3^2+2n/l^2}-\mu\right)}\right]
+\left(\mu\to-\mu\right)
\right\}.
\end{equation}
In the special case of zero temperature, the potential takes a simpler form:
\begin{eqnarray}
V^{(3+1)}_{\mu}(\rho) &=& V^{(3+1)}(\rho)-\frac{N}{4\pi^2l^2}
\left(|\mu|\sqrt{\mu^2-\rho^2}-\rho^2\ln\frac{|\mu|+\sqrt{\mu^2-\rho^2}}{\rho}\right)\theta\left(|\mu|-|\rho|\right)
\nonumber\\
&-&\frac{N}{2\pi^2l^2}\sum^\infty_{n=1}\left(
|\mu|\sqrt{\mu^2-\rho^2-2n/l^2}-\left(\rho^2+2n/l^2\right)
\ln\frac{|\mu|+\sqrt{\mu^2-\rho^2-2n/l^2}}{\sqrt{\rho+2n/l^2}}\right)\theta\left(|\mu|-\sqrt{\rho+2n/l^2}\right).
\nonumber\\
\end{eqnarray}
Note that the above contribution due to a nonzero chemical potential coincides with the result obtained in 
Refs.~\cite{Persson:1994pz}.

\section{The analysis of Bethe-Salpeter equation in QED}
\label{GepEqSolutionQED}

In this Appendix, we present the technical details for solving the Bethe-Salpeter 
equation in QED in the  ladder and improved latter approximations.

\subsection{Solution of the Bethe-Salpeter equation in QED in the ladder approximation}

Let us start from solving the Bethe-Salpeter equation in QED in the ladder approximation. 
The corresponding problem in the Feynman gauge was reduced to integral equation 
(\ref{D143}). Here we present an approximate analytical solution to that equation. 

It is convenient to break the momentum integration in Eq.~(\ref{D143})
into two regions and expand the kernel appropriately for each region:
\begin{equation}
A(p^2)=\frac{\alpha}{2\pi}\left(\int_0^{p^2}\frac{dk^2A(k^2)}
{k^2+m^2_{\rm dyn}}\int_0^\infty\frac{dz\exp(-z l^2/2)}
{p^2+z}
+\int_{p^2}^\infty\frac{dk^2A(k^2)}{k^2+m^2_{\rm dyn}}\int_0^\infty\frac{dz\exp(-z l^2/2)}{k^2+z}\right).
\label{D145}
\end{equation}
By introducing dimensionless variables $x=p^2 l^2/2$, $y=k^2 l^2/2$, and $a=m^2_{\rm dyn} l^2/2$, 
we rewrite Eq.~(\ref{D145}) in the form
\begin{equation}
A(x)=\frac{\alpha}{2\pi}\left(g(x)\int_0^{x}\frac{dyA(y)}
{y+a^2}+\int_x^\infty\frac{dyA(y)g(y)}{y+a^2}\right),
\label{D146}
\end{equation}
where
\begin{equation}
g(x)=\int_0^\infty\frac{dze^{-z}}{z+x}=-e^x\mbox{Ei}(-x).\label{D147}
\end{equation}
The solutions of integral
equation (\ref{D146}) satisfy the second order differential equation
\begin{equation}
A^{\prime\prime}-\frac{g^{\prime\prime}}{g^\prime}A^\prime-\frac{\alpha}
{2\pi}g^\prime\frac{A}{x+a^2}=0,\label{D148}
\end{equation}
where the prime denotes the derivative with respect to $x$. The boundary
conditions to this equation follow from the integral equation
(\ref{D146}):
\begin{eqnarray}
\left. \frac{A^\prime}{g^\prime}\right|_{x=0} &=& 0,   
\label{D149}\\
\left. \left(A-\frac{gA^\prime}{g^\prime}\right)\right|_{x=\infty} &=& 0.
\label{D150}
\end{eqnarray}
Note that function $g(x)$, which is defined by Eq.~(\ref{D147}), satisfies the following relations:
\begin{equation}
g^\prime=-\frac{1}{x}+g(z),\qquad g^{\prime\prime}=\frac{1}{x^2}-\frac{1}{x}
+g(x),\label{D151}
\end{equation}
and has the following asymptotic behavior:
\begin{eqnarray}
g(x)&\sim& \ln\frac{e^{-\gamma}}{x},\quad x\to 0,\nonumber\\
g(x)&\sim&\frac{1}{x}-\frac{1}{x^2}+\frac{2}{x^3},\quad
x\to\infty.\label{D152}
\end{eqnarray}
By making use of Eqs.~(\ref{D151}) and (\ref{D152}), we find that the differential equation 
(\ref{D148}) has two independent solutions that behave as $A(x)\simeq \mbox{const}$ and 
$A(x)\propto \ln(1/x)$ near $x=0$, and as $A(x)\simeq \mbox{const}$ and $A(x)\propto 1/x$ 
near $x=\infty$, respectively. The infrared boundary condition (\ref{D149}) leaves only the 
solution with regular behavior, $A(x)\simeq \mbox{const}$, while the ultraviolet boundary 
condition gives an equation to determine $a=a(\alpha)$. To find analytically $a(\alpha)$ we 
will solve the approximate equations in regions $x\ll 1$ and $x\gg 1$ and then match the 
solutions at the point $x=1$. This will provide an insight into the critical behavior of the 
solution at $\alpha\to 0$. A numerical study of the full Eq.~(\ref{D143}) reveals the same 
approach to criticality.

In the region $x\ll 1$, Eq.~(\ref{D148}) is reduced to a hypergeometric type equation:
\begin{equation}
A^{\prime\prime}+\frac{1}{x}A^\prime+\frac{\alpha}{2\pi}\frac{A}{x(x+
a^2)}=0.\label{D153}
\end{equation}
The regular solution at $x=0$ has the form
\begin{equation}
A_1(x)=C_1F\left(i\nu,-i\nu,1;-\frac{x}{a^2}\right),\quad
\mbox{where}\quad 
\nu=\sqrt{\frac{\alpha}{2\pi}},\label{D154}
\end{equation}
and $F(a,b,c;z)$ is the hypergeometric function \cite{1980tisp.book.....G}.
In the region $x\gg 1$ Eq.~(\ref{D148}) takes the form
\begin{equation}
A^{\prime\prime}+\frac{2}{x}A^\prime+\frac{\alpha}{2\pi}\frac
{A}{x^2(x+a^2)}=0.\label{D155}
\end{equation}
The solution satisfying ultraviolet boundary condition (\ref{D150}) is
\begin{equation}
A_2(x)=C_2\frac{1}{x} F\left(\frac{1+i\mu}{2},\frac{1-i\mu}{2};2;-\frac{a^2}{x}\right),\quad
\mbox{where}\quad 
\mu=\sqrt{\frac{2\alpha}{\pi a^2}-1}.\label{D156}
\end{equation}
Equating now logarithmic derivatives of $A_1$ and $A_2$ at $x=1$ we
arrive at the equation determining the quantity $a(\alpha)$:
\begin{equation}
\frac{d}{dx}
\left.\left(
\ln\frac{xF\left(i\nu,-i\nu;1;-\frac{x}{a^2}\right)}
{F\left(\frac{1+i\mu }{2},\frac{1-i\mu}{2};2;-\frac{a^2}{x}\right)}
\right)
\right|_{x=1}=0.
\label{D157}
\end{equation}
Note that up to now we have not made any assumptions on the value of
the parameter $a$. Let us seek now for solutions of Eq.~(\ref{D157}) with $a\ll 1$
(which corresponds to the assumption of the LLL dominance).
Then, the hypergeometric function in denominator of Eq.~(\ref{D157}) can be replaced
by 1 and we are left with equation
\begin{equation}
-\frac{1}{a^2}\nu^2F\left(1+i\nu,1-i\nu;2;-\frac{1}{a^2}\right)
+F\left(i\nu,-i\nu;1;-\frac{1}{a^2}\right)=0,\label{D158}
\end{equation}
where we used the formula for differentiating the hypergeometric
function \cite{1980tisp.book.....G}
\begin{equation}
\frac{d}{dz}F(a,b,c;z)=\frac{ab}{c}F(a+1,b+1,c+1;z).\label{D159}
\end{equation}
Now, because of $a\ll 1$, we can use the formula of asymptotic
behavior of hypergeometric function at large values  of its argument
$z$ \cite{1980tisp.book.....G}:
\begin{equation}
F(a,b,c;z)\sim\frac{\Gamma(c)\Gamma(b-a)}{\Gamma(b)\Gamma(c-a)}(-z)^{-a}
+\frac{\Gamma(c)\Gamma(a-b)}{\Gamma(a)\Gamma(c-b)}(-z)^{-b}.\label{D160}
\end{equation}
Then, Eq.~(\ref{D158}) is reduced to the following one:
\begin{equation}
\cos\left(\nu\ln\frac{1}{a^2}+\arg\Sigma(\nu)\right)=0, \\
\end{equation}
where 
\begin{equation}
\Sigma(\nu)=\frac{1+i\nu}{2}\frac{\Gamma(1+2i\nu)}{\Gamma^2(1+i\nu)}.
\label{D161}
\end{equation}
Therefore, by solving for $a^2$ we derive 
\begin{equation}
m^2_{\rm dyn}=2|eB|\exp\left(-\frac{\pi(2n+1)/2-\arg\Sigma(\nu)}{\nu}\right),
\label{D162}
\end{equation}
where $n$ is a nonnegative integer. 
The $\arg\Sigma(\nu)$ can be rewritten as
\begin{equation}
\arg\Sigma(\nu)=\arctan\nu+\arg\Gamma(1+2i\nu)-2\arg\Gamma(1+i\nu)
\label{D163}
\end{equation}
and in the limit $\nu\to 0$ Eq.~(\ref{D157}) takes the form
\begin{equation}
m^2_{\rm dyn}=2|eB|\mbox{e}\exp\left(-\frac{\pi}{2\nu}(2n+1)\right)=2|eB|
\mbox{e}\exp\left(-\pi\sqrt{\frac{\pi}{2\alpha}}(2n+1)\right)
\label{D164}
\end{equation}
(the second factor $\mbox{e}$ here is $\mbox{e}\simeq 2.718$ and not the
electric charge!).
The stable vacuum corresponds to the largest value of $m^2_{\rm dyn}$
with $n=0$, i.e.,
\begin{eqnarray}
m_{\rm dyn}=C\sqrt{|eB|}\exp\left[-\frac{\pi}{2}\left(\frac{\pi}{2\alpha}\right)^{1/2}\right],
\label{App:m_dyn}
\end{eqnarray}
where the constant $C=O(1)$.

Let us now turn to considering the general covariant gauge (\ref{photon_D_mu-nu}).  As
is known, the ladder approximation is not gauge invariant.  However,
let us show that because the present effect is due to the infrared
dynamics in QED, where the coupling constant is small, the leading term
in $\ln(m_{\rm dyn}^2 l^2)$,
$\ln(m_{\rm dyn}^2 l^2) \simeq -\pi\sqrt{\pi/2\alpha}$,
is the same in all covariant gauges.

Acting in the same way as before, we find that the wave function
$\varphi(p)$ now takes the form
\begin{eqnarray}
\varphi(p)=\gamma_5\left(1 - i\gamma^1\gamma^2\right) \left(A(p^2)+\hat pC(p^2)\right)
\label{31.23}
\end{eqnarray}
where the functions $A(p^2)$ and $C(p^2)$ satisfy the equations:
\begin{eqnarray}
A(p^2)&=&\frac{\alpha}{2\pi^2} \int \frac{d^2kA(k^2)}{
k^2+m^2_{\rm dyn}} \int^\infty_0
\frac{dx(1-\lambda x l^2/4)\exp(-x l^2/2)}{
(\mathbf{k}-\mathbf{p})^2+x}~,  \label{31.24}\\
C(p^2) &=& \frac{\alpha\lambda}{4\pi^2} \int \frac{d^2kC(k^2)}{
k^2+m^2_{\rm dyn}}\left[2k^2-(\mathbf{k}\cdot\mathbf{p})-\frac{k^2(\mathbf{k}\cdot\mathbf{p})}{p^2}\right]
 \int^\infty_0 \frac{dx\exp(-x l^2/2)}{\left[(\mathbf{k}-\mathbf{p})^2+x\right]^2}~.
\label{31.25}
\end{eqnarray}
One can see that the dominant contribution on the right-hand side of
Eq.~(\ref{31.24}) (proportional to $[\ln m^2_{\rm dyn} l^2]^2$ and formed
at small $k^2$) is independent of the gauge parameter $\lambda$.
Thus, the leading contribution in $\ln(m_{\rm dyn}^2 l^2)$,
$\ln(m_{\rm dyn}^2 l^2) \simeq -\pi\sqrt{\pi/2\alpha}$,
is expected to be gauge-invariant. This is also supported by a more careful analysis. 

In an arbitrary covariant gauge, the function $g(x)$ is replaced by
\begin{equation}
\tilde g(x)=\int_0^\infty\frac{dze^{-z}(1-\lambda z/2)}{z+x}=g(x)+
\frac{1}{2}\lambda xg^\prime(x).\label{D165}
\end{equation}
As it is easy to verify, this does not change equation (\ref{D153}) in
the region $x\ll 1$. At $x\gg 1$ we have
\begin{eqnarray}
\tilde g^\prime(x)\sim
-\frac{2-\lambda}{2x^2}+2\frac{1-\lambda}{x^3},\nonumber\\
\frac{\tilde g^{\prime\prime}(x)}{\tilde g^\prime(x)}=-\frac{2}{x}\frac
{2-\lambda-6\frac{1-\lambda}{x}}{2-\lambda-4\frac{1-\lambda}{
x}}.\label{D166}
\end{eqnarray}
Therefore, in any gauge, except $\lambda=2$, the differential equation
at $x\gg 1$ takes the form
\begin{equation}
A^{\prime\prime}+\frac{2}{x}A^\prime+\frac{\alpha(2-\lambda)}{4\pi}\frac
{A}{x^2(x+a^2)}=0 \label{D167}
\end{equation}
with asymptotic solution $A(x)\propto1/x$. In the gauge $\lambda=2$,
instead of Eq.~(\ref{D167}), we have
\begin{equation}
A^{\prime\prime}+\frac{3}{x}A^\prime
+\frac{\alpha}{\pi}\frac{A}{x^3(x+a^2)}=0,\label{D168}
\end{equation}
which gives more rapidly decreasing behavior $A(x)\propto 1/ x^2$ when
$x\to\infty$. Repeating the previous analysis, we are led to
expression (\ref{m_dyn}) for $m_{\rm dyn}$.

\subsection{Solution of the Bethe-Salpeter equation in QED beyond ladder approximation}
\label{App:BetheSalpeterQEDbeyond}

Here we give the technical details for solving the Bethe-Salpeter equation 
beyond ladder approximation. The corresponding problem was reduced to 
an integral equation (\ref{masseq}). By starting from that equation and 
integrating over the angular coordinate, we arrive at
\begin{equation}
B(p^2)=\frac{\alpha}{2\pi}\int\frac{dk^2B(k^2)}
{k^2+B^2(k^2)}K(p^2,k^2)
\label{scalareq}
\end{equation}
with the kernel
\begin{equation}
K(p^2,k^2)=\int_0^\infty\frac{dz\exp(-zl^2/2)}
{\sqrt{(p^2+k^2+M_\gamma^2e^{-zl^2/2}+z)-4p^2k^2}}.
\end{equation}
To study Eq.~(\ref{scalareq}) analytically, we break up the
momentum integration into two regions and expand the kernel
appropriately for each region
\begin{equation}
B(p^2) = \frac{\alpha}{2\pi}
\left(\int_0^{p^2}\frac{dk^2B(k^2)}{k^2+B^2(k^2)}
\int_0^\infty\frac{dz\exp(-zl^2/2)}{p^2+M_\gamma^2
e^{-z l^2/2}+z}
+\int_{p^2}^\infty\frac{dk^2B(k^2)}{k^2+B^2(k^2)}
\int_0^\infty\frac{dz\exp(-zl^2/2)}
{k^2+M_\gamma^2e^{-zl^2/2}+z} \right).
\label{approxeq}
\end{equation}
Introducing dimensionless variables $x=p^2l^2/2,\,y=k^2l^2/2$ and
also the dimensionless mass function $B(p^2)/\sqrt{2|eB|} \to
B(x)$, we rewrite the last equation in the form
\begin{equation}
B(x)=\frac{\alpha}{2\pi}\left(g(x)\int_0^x
\frac{dyB(y)}{y+B^2(y)}+\int_x^\infty\frac{dyB(y)g(y)}
{y+B^2(y)}\right), \label{inteq} 
\end{equation}
where
\begin{equation}
g(x)=\int_0^\infty\frac{dze^{-z}}
{z+x+\frac{\bar\alpha}{\pi}e^{-z}}.
\end{equation}
 The solutions of
the integral equation (\ref{inteq}) satisfy the second-order
differential equation
\begin{equation}
B^{\prime\prime}-\frac{g^{\prime\prime}}{g^\prime}B^\prime
-\frac{\alpha}{2\pi}g^\prime\frac{B}{x+B^2(x)}=0,
\label{diffeq}
\end{equation}
where the prime denotes derivative with respect to $x$. The
boundary conditions are
\begin{equation}
\frac{B^\prime}{g^\prime}\Bigg|_{x=0}=0,
\label{IRBC}
\end{equation}
\begin{equation}
\left(B-\frac{gB^\prime}{g^\prime}\right)\Bigg|_{x=\infty}=0.
\label{UVBC}
\end{equation}
The function $g(x)$ has asymptotic behavior
\begin{eqnarray}
g(x)&\sim&\ln\frac{1+\frac{\bar\alpha}{\pi}}{x+\frac{\bar\alpha}{\pi}},\qquad x\ll 1, \\
g(x)&\sim&\frac{1}{x},\qquad x\gg 1.
\end{eqnarray}
We consider now the linearized version of Eq.~(\ref{diffeq}) when
the term $B^2(x)$ in denominator is replaced by a constant
$B^2(0) \equiv a^2$ [$B(p=0)\equiv m_{\rm dyn}$]: the numerical
analysis shows that it is an excellent approximation. The two
independent solutions of that equation near the point $x=\infty$
behave as $B(x)\sim {\rm const}$ and $B(x)\sim 1/x$, and the ultra-violet boundary condition (\ref{UVBC})
selects the last one.

In the region $x\ll 1$, the equation takes the form
\begin{equation}
B^{\prime\prime}+\frac{1}{x+\frac{\bar\alpha}{\pi}}
B^\prime+\frac{\alpha}
{2\pi}\frac{B}{(x+\frac{\bar\alpha}{\pi})(x+a^2)}=0.
\label{lineareq}
\end{equation}
Introducing the variable $x+a^2=-z(\bar\alpha/\pi -a^2)$,
Eq.~(\ref{lineareq}) can be rewritten in the form of an equation
for the hypergeometric function,
\begin{equation}
z(1-z)\frac{d^2B}{dz^2}-z\frac{dB}{dz}-\frac{\alpha}{2\pi}B=0.
\label{hyper}
\end{equation}
The general solution to Eq.~(\ref{hyper}) has the form
\begin{equation}
B(z)=C_1u_1+C_2u_2,
\label{genersol}
\end{equation}
where
\begin{equation}
u_1=zF(1+i\nu,1-i\nu;2;z),
\end{equation}
\begin{equation}
u_2=(-z)^{-i\nu}F\left(i\nu,1+i\nu;1+2i\nu;\frac{1}{z}\right)+(-z)^{i\nu}
F\left(-i\nu,1-i\nu;1-2i\nu;\frac{1}{z}\right),
\end{equation}
$\nu=\sqrt{\alpha/2\pi}$. From the infrared boundary condition
(\ref{IRBC}), one gets
\begin{equation}
\frac{C_2}{C_1}=-\left. \frac{u_1^\prime}{u_2^\prime}\right|_{x=0}.
\end{equation}
Equating now the logarithmic derivatives of solution
(\ref{genersol}) (at $x\ll 1$) and $1/x$ (at $x\gg 1$) at the
point $x=1$, we arrive at the equation determining the quantity
$a(\alpha)$ ({\it i.e.}, the dynamical mass $m_{\rm dyn}$):
\begin{equation}
\varphi\equiv A_1B_2-A_2B_1=0,
\label{phieq}
\end{equation}
where
\begin{equation}
A_i=\left. \left(u_i^\prime+u_i\right)\right|_{x=1},\qquad
B_i=\left.u_i^\prime\right|_{x=0}.
\end{equation}
Since the variable $z$ is
\begin{equation}
z=-\frac{x+a^2}{\frac{\bar\alpha}{\pi}-a^2}\Bigg|_{x=0}\simeq
-\frac{\pi}{\bar\alpha}a^2,\quad z
=-\frac{x+a^2}{\frac{\bar\alpha}{\pi}-a^2}
\Bigg|_{x=1}\simeq -\frac{\pi}{\bar\alpha}
\end{equation}
(we suppose $a^2\ll\bar\alpha/\pi$), in what follows we need
asymptotic behavior of $u_i(z),u_i^\prime (z)$ at small  and
large negative values of its argument $z$. Using corresponding
formulas from \cite{BatemanErdelyi_Vol1}, we find for small values of $z$
($|z|\ll 1$) 
\begin{eqnarray}
u_1&\simeq& z\left(1+\frac{1+\nu^2}{2}z\right)+ O(z^2)\\
 u_2&\simeq&2\,\mbox{Re}\left\{\frac{\Gamma(1+2i\nu)}{\Gamma^2(1+i\nu)}
 \left[ \nu^2z(\ln(-z) -h_0)+1\right]+O(z^2\ln z)\right\},\\
  u_1^\prime &\simeq& 1+(1+\nu^2) z +O(z^2), \\
u_2^\prime&\simeq&2\,\mbox{Re}\left\{\frac{\Gamma(1+2i\nu)}
{\Gamma^2(1+i\nu)}
 \nu^2\left[\ln(-z)+1 -h_0\right]+O(z\ln z)\right\},
\end{eqnarray}
where
\begin{equation}
h_0=1-2\gamma-\psi(i\nu) -\psi(1+i\nu).
\end{equation}
 At large $|z|\gg 1$ we have
\begin{eqnarray}
u_1 &\simeq& -\frac{1}{\nu}\sqrt{\frac{\tanh(\pi\nu)}{\pi\nu}}
\sin\left[\nu\ln(-z)+\Phi(\nu)\right]+O(z^{-1}),\\
u_2 &\simeq& 2 \cos\left(\nu\ln (-z)\right) +O(z^{-1}),\\
u_1^\prime &\simeq& 
-\frac{1}{z}\sqrt{\frac{\tanh(\pi\nu)}{\pi\nu}}
\cos\left(\nu\ln(-z)+\Phi(\nu)\right)+O(z^{-2}), \\
u_2^\prime &\simeq& -\frac{2\nu}{z} \sin\left(\nu\ln(-z)\right)
+O(z^{-2}),
\end{eqnarray}
where
\begin{eqnarray}
\Phi(\nu)&=&\mbox{arg}
\left(\frac{\Gamma(1+2i\nu)}{\Gamma^2(1+i\nu)}\right)
=\sum_{n=1}^{\infty}(-1)^{n+1}\frac{2(2^{2n}-1)\zeta(2n+1)}
{2n+1}\nu^{2n+1} \\ &\simeq& 2
\zeta(3)\nu^{3}-6\zeta(5)\nu^{5}+\dots.
\end{eqnarray}
By making use of these asymptotes, we obtain the following
expressions for $A_i$ and $B_i$:
\begin{eqnarray}
A_1 &=& \left.\left(\frac{du_1}{dx}+u_1\right)\right|_{x=1}\simeq
-\sqrt{\frac{\tanh(\pi\nu)}{\pi\nu}}
\left[\cos\left(\nu\ln\frac{\pi}{\bar\alpha}\right)
+\frac{1}{\nu}\sin\left(\nu\ln\frac{\pi}
{\bar\alpha}+\Phi(\nu)\right)\right],
\\ A_2 &=&\left.\left(\frac{du_2}{dx}+ u_2\right)\right|_{x=1}
\simeq2 \cos\left(\nu\ln\frac{\pi}
{\bar\alpha}\right)+2\sin\left(\nu\ln
\frac{\pi}{\bar\alpha}\right),\\
 B_1 &=& \left.\frac{du_1}{dx} \right|_{x=0} 
\simeq -\frac{\pi}{\bar\alpha}, \\
B_2&=& \left.\frac{du_2}{dx} \right|_{x=0} \simeq -\ln\frac{\pi
a^2}{\bar\alpha}.
\end{eqnarray}
And, finally, the solution to Eq.~(\ref{phieq}) reads
\begin{equation}
a^2=\frac{m_{\rm dyn}^2}{2|eB|}\simeq
\frac{N\alpha}{\pi}\exp\left[-\frac{1}{\nu}
\cot\left(\nu\ln\frac{\pi}{N\alpha}\right)\right]
\simeq\left(\frac{N\alpha }{\pi}\right)^{2/3}
\exp\left(-\frac{2\pi}{\alpha\ln (\pi/N\alpha)}\right),
\quad \mbox{as} \quad \alpha\to 0.
\label{analdynmass}
\end{equation}

\section{Effective action for composite operators and free energy density}
\label{App:EffActionThermoEnergy}

In this Appendix, we present a short discussion regarding the use of the effective 
action for composite operators \cite{Cornwall:1974vz}. We will use this method not 
only for deriving the gap equations, but also for calculating the free energy density
valid at their solutions. Such a free energy density is needed in order to identify the 
energetically most favorable solution that determines the true ground state of the 
dynamical system. We will demonstrate the method by using the example of a
$(3+1)$-dimensional NJL model, see for example Eq.~(\ref{CSE-CME:NJLmodel}) 
in Section~\ref{ChiralAsymNJL}, but allow a rather general ansatz of the full propagator, 
given in Appendix~\ref{CSE:App:Landau-level-rep}. The result for the free energy 
density will also generalized to $(2+1)$-dimensional models useful in studies of 
graphene. 

Let us note that, unlike the effective potential in the large $N$ approximation in 
Section~\ref{NJL3+1:EffPot} and its generalization for nonzero temperatures in 
Appendix~\ref{App:ThermoPotNJL}, the current approach is more suitable at 
weak coupling. 

In the mean-field approximation, the corresponding two-loop effective action 
$\Gamma$ takes the following form:
\begin{eqnarray}
\Gamma(G) = {-i}\,\mathrm{Tr}\left[\mbox{Ln} G^{-1} +S^{-1}G-1\right]
&+&\frac{ G_{\rm int}}{2}\int dt \int d^{3}\mathbf{r} \Bigg\{
\left(\mathrm{tr}\left[G(u,u)\right]\right)^{2}
-\left(\mathrm{tr}\left[\gamma^{5}G(u,u)\right]\right)^{2}
\nonumber\\
&-& \mathrm{tr}\left[G(u,u)G(u,u)\right]
+\mathrm{tr}\left[\gamma^{5}G(u,u)\gamma^{5}G(u,u)\right]\Bigg\} .
\label{potential}
\end{eqnarray}
The trace, the logarithm, and the product $S^{-1}G$ are taken in the functional sense. 
The gap equation is obtained by requiring that the full fermion propagator $G$ corresponds 
to the variational extremum of the effective action, $\delta \Gamma/\delta G=0$, the 
explicit form of which reads
\begin{equation}
G^{-1}(u,u^\prime)  = S^{-1}(u,u^\prime)
- i G_{\rm int} \left\{ G(u,u) -  \gamma^5 G(u,u) \gamma^5 
-\mathrm{tr}[G(u,u)] +  \gamma^5\, \mathrm{tr}[\gamma^5G(u,u)]\right\}
\delta^{4}(u- u^\prime). 
\label{gap-App}
\end{equation}
The structure of this equation and its solutions are discussed in Section~\ref{ChiralAsymNJL}.

\subsection{Free energy density in $3+1$ dimensions}
\label{App:FreeEnergyDensity3+1}

Let us turn to the problem of the free energy density $\Omega$. It can be obtained 
from the effective action $\Gamma$ by evaluating the result at a solution to the gap 
equation, $\Omega = -\Gamma/{\cal TV}$, where ${\cal TV}$ is a space-time volume. 
By taking into account the general form of the gap equation (\ref{gap-App}), we can rewrite
the two-loop part in Eq.~(\ref{potential}) in the same form of the one-loop expression. 
Then, at a solution to the gap equation, the free energy density takes a much simpler 
form,
\begin{equation}
\Omega = \frac{i}{{\cal TV}}\mathrm{Tr}\left[\mbox{Ln} G^{-1} +\frac{1}{2}\left(S^{-1}G-1\right)\right].
\label{omega}
\end{equation}
It may be appropriate to mention that this is a rather general result that will hold 
true in essentially any model in the mean-field approximation. Of course, the final
results for the free energy will not be the same in all models because the full 
propagators will satisfy different gap equations and, thus, differ in details. 

By making use of the following Fourier transform of the Green's function
$G(u,u^\prime)$:
\begin{equation}
G(u,u^{\prime})=\int\limits_{-\infty}^{\infty}\frac{d\omega}{2\pi}\,
e^{-i\omega(t-t^\prime)}G(\omega;\mathbf{r},\mathbf{r}^\prime),
\end{equation}
[compare with Eq.~(\ref{Gomegak3Fourier})] we rewrite the free energy density as
\begin{equation}
\Omega = \frac{i}{{\cal V}}\int\limits_{-\infty}^{\infty}\frac{d\omega}{2\pi}\mathrm{Tr}\left[\ln
G^{-1}(\omega)
+\frac{1}{2}\left(S^{-1}(\omega)G(\omega)-1\right)\right],
\label{effpot-integral-in-omega}
\end{equation}
the functional operation $\mathrm{Tr}$ includes now only the integration over the 
spatial coordinates and the trace over matrix indices.

Integrating by parts the logarithm term in Eq.~(\ref{effpot-integral-in-omega}) and 
omitting the irrelevant surface term (independent of the physical parameters), 
we arrive at the expression
\begin{equation}
\Gamma=\frac{i}{{\cal V}}\int\limits_{-\infty}^{\infty}\frac{d\omega}{2\pi}\mathrm{Tr}
\left[-\omega\frac{\partial G^{-1}(\omega)}
{\partial\omega}\,G(\omega)+\frac{1}{2}\left( S^{-1}(\omega)
\,G(\omega)-1\right)\right] .
\label{effpot2}
\end{equation}
When the wave function renormalization effects are negligible in the full propagator,
as is the case for example in Eq.~(\ref{GuuSigma}), we can use the following result:
\begin{equation}
\frac{\partial G^{-1}(\omega)}{\partial\omega} =
-i\gamma^{0}\delta(\mathbf{r}-\mathbf{r}^\prime).
\end{equation}
and finally obtain 
\begin{eqnarray}
\Omega &=& -\int\limits_{-\infty}^{\infty}\frac{d\omega}{4\pi}\int
\frac{d^{3}k}{(2\pi)^{3}}
\mathrm{tr}\left\{\left[(\omega-\mu_0)\gamma^{0}+(\mathbf{k}\cdot\bm{\gamma})+m_0\right]
\bar{G}(\omega,\mathbf{k})+i\right\}.
\label{Omega}
\end{eqnarray}
In the most general case, the propagator $\bar{G}(\omega,\mathbf{k})$ is given in
Eq.~(\ref{propagatorG-FT}) in Appendix~\ref{CSE:App:Landau-level-rep}. This result
for the free energy can be also generalized to a nonzero temperature by replacing 
the energy integration with a Matsubara sum as in Eq.~(\ref{T:integral-sum}) and 
substituting $\omega \to i\omega_k=(2k+1)\pi T$. 

In the special case when $\mu$, $m$ and $\Delta$ are the only nonvanishing dynamical 
parameters in the full propagator, the integration over the two perpendicular components 
of the momenta can be easily performed and the results is given by \cite{Gorbar:2011ya}
\begin{eqnarray}
\Omega &=& -\frac{i}{2(2\pi)^3l^2}\int d\omega dk^{3}
\left[\frac{(\omega-\mu_0)(\omega+\mu+s_{\perp} \Delta)+(k^{3})^2}{(\omega+\mu)^2 -\left(s_{\perp} \Delta+\sqrt{m^2 +(k^{3})^2}\right)^2}
     +\frac{(\omega-\mu_0)(\omega+\mu+s_{\perp} \Delta)+(k^{3})^2}{(\omega+\mu)^2 -\left(s_{\perp} \Delta-\sqrt{m^2 +(k^{3})^2}\right)^2}
\right]
\nonumber\\
&&-\frac{i}{2(2\pi)^3l^2}\int \frac{d\omega dk^{3}}{\sqrt{m^2 +(k^{3})^2}}
\left[\frac{(\omega-\mu_0)( m )^2+(k^{3})^2(\omega+\mu+s_{\perp} \Delta)}
           {(\omega+\mu)^2 -\left(s_{\perp} \Delta+\sqrt{m^2 +(k^{3})^2}\right)^2}
     -\frac{(\omega-\mu_0)( m )^2+(k^{3})^2(\omega+\mu+s_{\perp} \Delta)}
           {(\omega+\mu)^2 -\left(s_{\perp} \Delta-\sqrt{m^2 +(k^{3})^2}\right)^2}
\right]
\nonumber\\
&-& \frac{i}{(2\pi)^3l^2}\sum\limits_{n=1}^{\infty}
\int d\omega dk^{3}\left[
 \frac{(\omega-\mu_0)(\omega+\mu)+(k^{3})^2+2n|e B |}{(\omega+\mu)^2-\left(E_{k^{3},n}^{+}\right)^2}
+\frac{(\omega-\mu_0)(\omega+\mu)+(k^{3})^2+2n|e B |}{(\omega+\mu)^2-\left(E_{k^{3},n}^{-}\right)^2}
\right]
\nonumber\\
&-& \frac{i}{(2\pi)^3l^2}\sum\limits_{n=1}^{\infty}
\int  \frac{d\omega dk^{3}}{\sqrt{m^2 +(k^{3})^2}} \left[
 \frac{(k^{3})^2 s_{\perp} \Delta}{(\omega+\mu)^2-\left(E_{k^{3},n}^{+}\right)^2}
-\frac{(k^{3})^2 s_{\perp} \Delta}{(\omega+\mu)^2-\left(E_{k^{3},n}^{-}\right)^2}
\right].
\end{eqnarray}
In the special case when $m ,\mu\neq 0$ and $\Delta=\tilde\mu = 0$ 
(which correspond to the solution of type I in Section~\ref{ChiralAsymNJLMCshift}), 
the final part of the free energy becomes \cite{Gorbar:2011ya}
\begin{eqnarray}
\Omega_{m,\mu}^{\rm fin} &=& -\frac{1}{(2\pi l)^2} \left(\frac{ m ^2}{2}
+\mu_0\sqrt{\mu^2- m ^2}\mathrm{sign}(\mu)\theta(\mu^2- m ^2)
\right)
-\frac{2}{(2\pi l)^2} \sum\limits_{n=1}^{\infty}\left[
\frac{ m ^2}{2}+2n|e B |\ln\frac{\sqrt{2n|e B |}}{\sqrt{ m ^2+2n|e B |}}\right]\nonumber\\
&-& \frac{2}{(2\pi l)^2} \sum\limits_{n=1}^{\infty}\left[
\mu_0\mathrm{sign}(\mu)\sqrt{\mu^2- m ^2-2n|e B |}
+2n|e B | \ln\frac{\sqrt{ m ^2+2n|e B |}}{\sqrt{\mu^2- m ^2-2n|e B |}+|\mu|}
\right]\theta(\mu^2- m ^2-2n|e B |).\nonumber\\
\label{AppAOmega0}
\end{eqnarray}
Note that, in the calculation, we subtracted an infinite constant term:
\begin{equation}
\frac{2}{(2\pi l)^2} \int_{0}^{\infty} dk^{3}\left(k^{3}+2\sum\limits_{n=1}^{\infty} \sqrt{(k^{3})^2+2n|e B |}\right).
\end{equation}
The sum over the Landau levels in the first line of Eq.~(\ref{AppAOmega0}) still contains a logarithmic divergence, i.e.,
\begin{equation}
\Omega_{m,\mu}^{\rm div}\simeq -\frac{ m ^4}{\left(4\pi\right)^2} \sum_{n=1}^{\infty}\frac{1}{n}
\simeq -\frac{ m ^4}{\left(4\pi\right)^2} \ln (\Lambda l)^2.
\end{equation}
In numerical calculations, we use of the smooth cutoff (\ref{kappa}) to regularize this expression.
  
In the case when $\Delta,\mu\neq 0$ and $ m =\tilde\mu = 0$
(which correspond to the solution of type II in Section~\ref{ChiralAsymNJLMCshift}), 
after doing the subtraction of an infinite constant term and explicitly 
performing the integration, we arrive at \cite{Gorbar:2011ya}
\begin{eqnarray}
\Omega_{\Delta,\mu} &\simeq & -\frac{\mu_0\mu}{(2\pi l)^2}
   +\frac{s_{\perp} \Delta(\mu_0-\mu)}{(2\pi l)^2} \ln\frac{\Lambda}{|\mu|}\nonumber\\
&-&\frac{2}{(2\pi l)^2} \sum\limits_{n=1}^{\infty} \left[\mu_0\mathrm{sign}(\mu)\sqrt{\mu^2-2n|e B |}
+2n|e B |\ln\frac{\sqrt{2n|e B |}}{|\mu|+\sqrt{\mu^2-2n|e B |}}
\right]\theta\left(\mu^2-2n|e B |\right).
\label{AppAOmegaDelta0}
\end{eqnarray}

\subsection{Free energy density in $2+1$ dimensions}
\label{App:FreeEnergyDensity2+1}

For model studies of graphene in Section~\ref{sec:QHEgraphene}, we need to have a 
generalization of the free energy density valid in $2+1$ dimensions. The corresponding  
derivation in models with short-range interactions follows exactly the same steps as in 
the $(3+1)$-dimensional case above, and the general form of the result for $\Omega$ 
will be look the same as in Eq.~(\ref{Omega}), but without the integration over the 
longitudinal momentum, $\int dk_3/(2\pi)$. Note also that, in the case of graphene the 
trace over the matrix indices in Eq.~(\ref{Omega}) includes an additional sum over
two spin states $s=\pm$ and the speed of light ($c=1$) in the kinetic term is replaced by the 
Fermi velocity $v_F$.
 
In models with long-range (e.g., Coulomb type) interaction, the result for the free energy 
density should also contain the one-loop photon contribution. To leading order, however, 
the latter can be neglected because the photon propagator is independent of the dynamical 
order parameters, responsible for the quantum Hall effect. Then, the expression for the 
free energy density will be again the same as in models with short-range interaction. 
(The corresponding formal definition of the leading order approximation may be unreliable
because the photon screening effects can be large and their dependence on the order 
parameters may be important too.)

After integrating over the spatial momenta in the corresponding free energy expression 
analogous to Eq.~(\ref{Omega}), we will obtain
\begin{equation}
\Omega=-\frac{i}{(4\pi l)^2}\sum_{s=\pm}
\int\limits_{-\infty}^{\infty}d\omega\, \mathrm{tr}_D
\sum\limits_{n=0}^{\infty}
\frac{(\omega-\bar{\mu}_{s})\left[\omega+\mu_{s}+i\tilde{\mu}_{s}\gamma^0\gamma^1\gamma^2
-i\Delta_{s}\gamma^1\gamma^2+\tilde{\Delta}_{s}\gamma^0\right]{\cal P}_{n}+
4v_{F}^2|eB_{\perp}|n\theta(n-1)}
{(\omega+\mu_{s}+i\tilde{\mu}_{s}\gamma^0\gamma^1\gamma^2)^2-
(\tilde{\Delta}_{s}-i\Delta_{s}\gamma^0\gamma^1
\gamma^2)^2-2v_{F}^2|eB_{\perp}|n} ,
\end{equation}
where we introduced the shorthand notation ${\cal P}_{n} = 1+i\gamma^1\gamma^2 \mbox{sign}(eB) 
+\left[1-i\gamma^1\gamma^2 \mbox{sign}(eB)\right]\theta(n-1)$ and 
dropped an infinite divergent term independent of the physical parameters.
The trace $\mathrm{tr}_D$ in this expression is taken over the Dirac indices only.
Normalizing $\Omega$ by subtracting its value at $\tilde{\Delta}_{s}=\tilde{\mu}_{s}
=\mu_{s}=\Delta_{s}=\bar{\mu}_{s}=0$, we obtain
\begin{eqnarray}
\Omega &=&-\frac{i}{(4\pi l)^2}\sum_{s=\pm}
\sum\limits_{n=0}^{\infty}\int\limits_{-\infty}^{\infty}d\omega\,
\mathrm{tr}_D\left[
\frac{(\omega-\bar{\mu}_{s})[\omega+\mu_{s}+i\tilde{\mu}_{s}\gamma^0\gamma^1\gamma^2
-i\Delta_{s}\gamma^1\gamma^2+\tilde{\Delta}_{s}\gamma^0]{\cal P}_{n}+
4v_{F}^2|eB_{\perp}|n\theta(n-1)}
{(\omega+i\epsilon\mbox{sign}({\omega})
+\mu_{s}+i\tilde{\mu}_{s}\gamma^0\gamma^1\gamma^2)^2-
(\tilde{\Delta}_{s}-i\Delta_{s}
\gamma^0\gamma^1\gamma^2)^2-2v_{F}^2|eB_{\perp}|n}\right.\nonumber\\
&-&\left.\frac{\omega^{2}{\cal P}_{n}+4v_{F}^2|eB_{\perp}|n\theta(n-1)}
{\left(\omega+i\epsilon\mbox{sign}
({\omega})\right)^{2}-2v_{F}^2|eB_{\perp}|n}
\right]. 
\end{eqnarray}
In the case of zero temperature, integrating over $\omega$ and taking the trace, we find the
following expression for the free energy density:
\begin{eqnarray}
&&\Omega=-\frac{1}{8\pi l^2}\sum_{s=\pm} \Bigg\{\!
\left[\mu_{s}+\bar{\mu}_{s}-\tilde{\mu}_{s}-(\tilde{\Delta}_{s}+\Delta_{s})
\mbox{sign}(eB_{\perp})\right] \mbox{sign}(\mu_{s}-\tilde{\mu}_{s})
\theta(|\mu_{s}-\tilde{\mu}_{s}|-|\tilde{\Delta}_{s}+\Delta_{s}|)
\nonumber\\
&&+\left[\tilde{\Delta}_{s}+\Delta_{s}-(\mu_{s}+\bar{\mu}_{s}-
\tilde{\mu}_{s})\mbox{sign}(eB_{\perp})\right]\,\mbox{sign}(\tilde{\Delta}_{s}+\Delta_{s})
\theta(|\tilde{\Delta}_{s}+\Delta_{s}|-|\mu_{s}-\tilde{\mu}_{s}|)
\nonumber\\
&&\left. +2\sum_{n=1}^{\infty} \left[ \left[
(\mu_{s}+\bar{\mu}_{s}-\tilde{\mu}_{s})\mbox{sign}(\mu_{s}-\tilde{\mu}_{s})-
2\epsilon_{B}\sqrt{n}\right] \theta
\left(|\mu_{s}-\tilde{\mu}_{s}|-E_{ns}^{+} \right) +
\frac{(\tilde{\Delta}_{s}+\Delta_{s})^4\theta(E_{ns}^{+}-|\mu_{s}-\tilde{\mu}_{s}|)}
{E_{ns}^{+}(E_{ns}^{+}+\epsilon_{B}\sqrt{n})^2}\right]
\right.\nonumber\\
&&+\left[\tilde{\mu}_{s} \to -\tilde{\mu}_{s},\,\Delta_{s} \to -
\Delta_{s},\,\mbox{sign}(eB_{\perp}) \to
-\mbox{sign}(eB_{\perp})\right] \Bigg\}, \label{eff-pot}
\end{eqnarray}
where we used the shorthand notation $E_{ns}^{\pm}=\sqrt{n\epsilon_{B}^2+(\tilde{\Delta}_{s} \pm
\Delta_{s})^2}$ and $\epsilon_{B}=\sqrt{2v_{F}^2|eB_{\perp}|}$.

\section{Additional technical details about noncommutative field theories}
\label{App:NCFT}

\subsection{Generic form of vertices in NCFT corresponding to NJL model in a magnetic field}
\label{App:NCFT1}

In this Appendix, we will show that in the case of fields independent of the longitudinal 
coordinates $u_{\parallel}$, all their interaction vertices $\Gamma_{n\phi}$ ($n \geq 3$)
can be rewritten through the star product.

The relevant part of the $n$-point vertex $\Gamma_{n\phi}$ is the part which includes the
integration over transverse coordinates. It has the form:
\begin{equation}
\Gamma^{\perp}_{n\phi} \equiv \int d^2x_1 \ldots  d^2x_{n}
P(x_1,x_2)\phi(x_2)P(x_2,x_3)\phi(x_3) \ldots 
P(x_n,x_1)\phi(x_1),
\label{perp1}
\end{equation}
where $P(x_1,x_2)$ is the transverse part of the fermion propagator
written in Eq.~(\ref{projectorMore}) [here, for convenience, we omitted 
the subscript $\perp$ in transverse coordinates].

Expressing the fields
$\phi$ through their Fourier transforms, one can explicitly integrate
over $x_i$ coordinates in (\ref{perp1})
[the integrals are Gaussian].
It can be done step by step. First, we find
\begin{equation}
I_1(x_1,x_3) = \int d^2x_2 P(x_1,x_2) e^{i\mathbf{k}_2\cdot\mathbf{r}_2} P(x_2,x_3)
= P(x_1,x_3) e^{-\frac{\mathbf{k}_{2}^2}{2|eB|}}
e^{\frac{1}{2}\mathrm{sign}(eB)\epsilon^{ab}k_2^a(x_1-x_3)^b}
e^{\frac{i}{2}\mathbf{k}_2\cdot(\mathbf{r}_1 + \mathbf{r}_3)}.
\end{equation}

The second step leads to an expression with a similar structure:
\begin{eqnarray}
I_2(x_1,x_4) &=& \int d^2x_3 I_1(x_1,x_3) e^{i\mathbf{k}_3\cdot\mathbf{r}_3} P(x_3,x_4)
\nonumber \\
&=& P(x_1,x_4) e^{-\frac{\mathbf{k}_2^2+\mathbf{k}_3^2+\mathbf{k}_2\cdot\mathbf{k}_3}{2|eB|}}
e^{-\frac{i}{2eB}\epsilon^{ab}k_2^ak_3^b}
e^{\frac{1}{2}\mathrm{sign}(eB)\epsilon^{ab}(k_2+k_3)^a(x_1-x_4)^b}
e^{\frac{i}{2}(\mathbf{k}_2+\mathbf{k}_3)\cdot(\mathbf{r}_1+\mathbf{r}_4)}.
\end{eqnarray}
Proceeding in this way
until the integration over $x_n$, we encounter the
integral
\begin{equation}
I_{n-1}(x_1,x_1) = \int d^2x_n
I_{n-2}(x_1,x_n) e^{i\mathbf{k}_n \cdot \mathbf{r}_n}P(x_n,x_1).
\label{Ilast}
\end{equation}
It closes the fermion loop because the
last argument in $P(x_n,x_1)$ coincides with the first
argument of $I_{n-2}$. Because of that,
the result of this integration is especially simple:
\begin{equation}
I_{n-1}(x_1,x_1) = \frac{|eB|}{2\pi}
e^{-\frac{\sum_{i=2}^n\mathbf{k}_i^2+\sum_{2 \le i < j}^n
\mathbf{k}_i\cdot\mathbf{k}_j}{2|eB|}}\,
e^{-\frac{i}{2eB}(\sum_{2 \le i < j}^n\epsilon^{ab}k_i^ak_j^b)}\, 
e^{i(\sum_{i=2}^{n}\mathbf{k}_i\cdot\mathbf{r}_i)},
\label{int}
\end{equation}
where the equality $P(x_1,x_1)= |eB|/2\pi$ was used. The last integration
over $\mathbf{r}_1$ yields
\begin{equation}
I_n = \int d^2x_1 I_{n-1}(x_1,x_1) e^{i\mathbf{k}_1\cdot\mathbf{r}_1} =
2\pi|eB|\delta^2 \left(\sum_{i=1}^n \mathbf{k}_i\right)\,
e^{-\frac{\sum_{i=2}^n\mathbf{k}_i^2+\sum_{2 \le i < j}^n
\mathbf{k}_i\mathbf{k}_j}{2|eB|}}\,
e^{-\frac{i}{2eB}(\sum_{2 \le i < j}^n\epsilon^{ab}k_i^ak_j^b)}.
\label{int1}
\end{equation}
Here the delta function ensures the conservation of the
total transverse momentum.

Now, because of the identity
\begin{equation}
\sum_{i=2}^n\mathbf{k}_i^2+\sum_{2 \le i < j}^n \mathbf{k}_i\cdot\mathbf{k}_j =
-\sum_{1 \le i < j}^n\mathbf{k}_i\cdot\mathbf{k}_j + \left(\sum_{i=1}^n \mathbf{k}_i\right)^2
-\mathbf{k}_1 \cdot \left(\sum_{i=1}^n\mathbf{k}_i\right)
\end{equation}
and the conservation of the total momentum, we obtain the equalities
\begin{equation}
\sum_{i=2}^n\mathbf{k}_i^2+\sum_{2 \le i < j}^n \mathbf{k}_i\cdot \mathbf{k}_j =
-\sum_{1 \le i < j}^n\mathbf{k}_i\cdot\mathbf{k}_j =
\frac{1}{2}\sum_{i=1}^n\mathbf{k}_i^2
\end{equation}
and
$\sum_{2 \le i < j}^n\epsilon^{ab}k_i^ak_j^b =
\sum_{1 \le i < j}^n\epsilon^{ab}k_i^ak_j^b$.

Using these equalities,
we conclude that the
exponential term in expression
(\ref{int1}) can be rewritten through the cross
product as
$e^{-\frac{\sum_{i=1}^n\mathbf{k}_i^2}{4|eB|}}\,e^{-\frac{i}{2}
\sum_{i<j} k_i \times k_j}$ . Therefore, similarly to three and four point
vertices (\ref{3point1}) and (\ref{4point}),
a generic $n$-point interaction
vertex $\Gamma_{n\Phi}$ ($n \geq 3$) has the following structure:
\begin{equation}
\Gamma_{n\Phi} =  C_{n}\frac{N|eB|}{m^{n-2}}
\int d^2u_{\parallel}\frac{d^2k_1\ldots d^2k_n}{(2\pi)^{2n}}
\Phi(k_1)\ldots  \Phi(k_n) \delta^2\left(\sum_i k_i \right)
e^{-\frac{i}{2} \sum_{i<j} k_i \times k_j},
\label{vertexA}
\end{equation}
where here $\Phi$ represents the smeared fields $\Pi$ and
$\tilde{\Sigma}$ and $C_n$ is a numerical constant which can be easily
found by expanding the effective potential in the Taylor series
in constant fields $\pi$ and $\tilde{\sigma}$.
Equation (\ref{vertexA}) in turn
implies that the
vertex $\Gamma_{n\Phi}$
can be rewritten through the star product
in the coordinate space as
\begin{equation}
\Gamma_{n\Phi} =  C_{n}\frac{N|eB|}{4\pi^{2}m^{n-2}}
\int d^2u_{\parallel}d^2u_{\perp}\,
\Phi_1 * \Phi_2 *\ldots  * \Phi_n
\label{vertexA1}
\end{equation}
(compare with expressions in Eq.~(\ref{xspacevertices})).
In the noncommutative coordinate space, the vertex is
\begin{equation}
\Gamma_{n\Phi} =  C_{n}\frac{N|eB|}{4\pi^{2}m^{n-2}}
\mathbf{Tr}\,
\hat{\Phi}_1\hat{\Phi}_2\ldots \hat{\Phi}_n
\label{vertexA2}
\end{equation}
(compare with Eq.~(\ref{ncspacevertices})).

\subsection{Formalism of projected density operators on LLL}
\label{App:NCFT2}

In this Appendix, it will be shown that the exponentially damping factors and 
the $M$-star product naturally appear in the formalism of the projected density 
operators on the LLL states developed in studies of the quantum Hall effect in 
Ref.~\cite{Sinova:2000kx}. To be concrete, we will consider the $\Gamma_{4\pi}$ 
vertex in this formalism.

As follows from the effective action in Eq.~(\ref{14}), the
$\Gamma_{4\pi}$ vertex is given
by
\begin{equation}
\Gamma_{4\pi} = \frac{i}{4} \int d^4u d^4u^\prime d^4v d^4v^\prime\, \mathrm{tr}
\left[ S(u,u^\prime)\gamma^5 \pi(u^\prime) S(u^\prime,v) \gamma^5 \pi(v) S(v,v^\prime)
\gamma^5 \pi(v^\prime) S(v^\prime,u) \gamma^5 \pi(u) \right].
\label{A1}
\end{equation}
According to Eq.~(\ref{factorization}), the dependence on the transverse
$u_{\perp}$ and longitudinal $u_{\parallel}$ coordinates factorizes in the
LLL propagator $S(u,u^\prime)$. If
fields $\pi$ in (\ref{A1}) do not depend on $u_{\parallel}$, then it is
straightforward to integrate over the longitudinal coordinates in
this expression that yields the factor
\begin{equation}
\frac{i}{4} \int \frac{d^2k_{\parallel}}{(2\pi)^2}\, \mathrm{tr}\,
\left( \frac{1}{k_{\parallel}\gamma^{\parallel} - m}
\frac{1+ i\gamma^1\gamma^2\,\mathrm{sign}(eB)}{2} \gamma^5 \right)^4 =
-\frac{1}{8\pi m^2}.
\label{longitudinal}
\end{equation}

To get the $\Gamma_{4\pi}$ vertex, we now need to calculate
the transverse part
\begin{equation}
\int d^2u_{\perp}d^2u^\prime_{\perp}d^2z_{\perp}d^2v_{\perp}
P(u_{\perp},u^\prime_{\perp})
\pi(u^\prime_{\perp}) P(u^\prime_{\perp},z_{\perp})
\pi(z_{\perp}) P(z_{\perp},v_{\perp}) \pi(v_{\perp})
P(v_{\perp},u_{\perp}) \pi(u_{\perp}).
\label{transverse}
\end{equation}
We will use the formalism of projected density operators \cite{Sinova:2000kx} to calculate it.
The crucial point is the fact that the transverse part of the LLL fermion propagator $P(u,u^\prime)$
is the projection operator on the LLL states (henceforth we will omit the subscript $\perp$ for the 
transverse coordinates). Namely,
\begin{equation}
P(u,u^\prime) = \sum_n \langle x|n \rangle\langle n|y \rangle,
\label{Aprojector}
\end{equation}
where the sum is taken over all LLL states, which in the symmetric
gauge are
\begin{equation}
\psi_n(z,z^{*}) = \left(\frac{|eB|}{2}\right)^{\frac{n+1}{2}}
\frac{z^n}{\sqrt{\pi n!}}
e^{-\frac{1}{4}|eB|zz^{*}}
\label{wf}
\end{equation}
with $z=x^1- i\,\mathrm{sign}(eB)x^2$. Now, by using completeness relations
like
\begin{equation}
\int d^2y \langle n_1|y \rangle \pi(y) \langle y|n_2 \rangle = \langle n_1| \pi |n_2 \rangle,
\end{equation}
we obtain expression (\ref{transverse}) in the form
\begin{equation}
\sum_{n_1,\ldots,n_4} \langle n_1|\pi|n_2 \rangle\ldots \langle n_3|\pi|n_4 \rangle.
\label{A3}
\end{equation}
To get the $\Gamma_{4\pi}$ interaction vertex in the momentum space, we
will use the Fourier transforms of fields $\pi$. Then, we encounter
factors of the form:
\begin{equation}
\langle n_i|\rho_k|n_j \rangle,
\end{equation}
where $\rho_k=e^{i\mathbf{k}\cdot\mathbf{x}}=\exp\left[\frac{i}{2}
(kz^{*} + k^{*}z)\right]$, with $k=k^1 - i\,\mathrm{sign}(eB)k^2$,
is called the density operator.

In what follows, we will use the methods developed in
Refs.~\cite{Girvin:1984fk,Kivelson:1987uq,Dunne:1992ew} and, in
fact, follow very closely to Ref.~\cite{Gurarie:2001IJMPB}.

First of all, since the prefactor in expression (\ref{wf}) is analytic in $z$,
the factor $e^{\frac{i}{2}k^{*}z}$ in $\rho_k$
acts entirely within the LLL. On the other hand, another
factor $e^{\frac{i}{2}kz^{*}}$ in $\rho_k$ contains $z^{*}$
and therefore does not act within the LLL.
Actually, the following relation takes place:
\begin{equation}
\langle n|(z^{*})^s|m \rangle = \langle n|\left( \frac{2}{|eB|}\frac{\partial}{\partial z}
+ \frac{z^{*}}{2} \right)^s|m \rangle,
\label{projection}
\end{equation}
which expresses the matrix elements of $z^{*}$ between the LLL states
in terms of the operator $\hat{z} = \frac{2}{|eB|}\frac{\partial}{\partial z} + \frac{z^{*}}{2}$ 
that acts within the LLL. Therefore, on the LLL states, we can replace the density 
operator $\exp\left[\frac{i}{2}(kz^{*} + k^{*}z)\right]$ by the projected density operator
$\hat{\rho}_k = e^{\frac{i}{2}k\hat{z}}e^{\frac{i}{2}k^{*}z}$.

Now, using the projected density operators $\hat{\rho}_k$
for the $\Gamma_{4\pi}$ vertex, we get
\begin{equation}
\Gamma_{4\pi}=-\frac{1}{8\pi m^2} \int d^2u_{\parallel}\int
\frac{d^2k_1 d^2k_2 d^2k_3 d^2k_4}{(2\pi)^8}\pi(k_1)\pi(k_2)\pi(k_3)\pi(k_4)
\sum_{n_1,\ldots,n_4}(\hat{\rho}_{k_1})_{n_1n_2}(\hat{\rho}_{k_2})_{n_2n_3}
(\hat{\rho}_{k_3})_{n_3n_4}(\hat{\rho}_{k_4})_{n_4n_1}.
\label{A4}
\end{equation}
Since the LLL states form a complete basis for
the operators $\hat{\rho}_k$, we have
\begin{equation}
\sum_{n_2} (\hat{\rho}_{k_1})_{n_1n_2} (\hat{\rho}_{k_2})_{n_2n_3} =
(\hat{\rho}_{k_1}\hat{\rho}_{k_2})_{n_1n_3}.
\end{equation}
The product of two projected density operators is given by \cite{Gurarie:2001IJMPB}
\begin{equation}
\hat{\rho}_{k_1} \hat{\rho}_{k_2} = \exp \left[
\frac{\mathbf{k}_1\cdot\mathbf{k}_2}{2|eB|} -
\frac{i}{2}k_1 \times k_2
\right] \hat{\rho}_{k_1+k_2}.
\label{product}
\end{equation}
Notice that the exponent in this equation can be rewritten
through the $M$-cross product (\ref{Mcross}) as
$e^{-\frac{i}{2} k_1 \times_M k_2}$.

Therefore, we find the following expression for $\Gamma_{4\pi}$:
\begin{equation}
\Gamma_{4\pi} = -\frac{1}{8\pi m^2} \int d^2 u_{\parallel} \int
\frac{d^2k_1 d^2k_2 d^2k_3 d^2k_4}{(2\pi)^8} \pi(k_1)\pi(k_2)\pi(k_3)\pi(k_4)
\,e^{-\frac{i}{2}\sum_{i<j} k_i \times_M k_j}
\sum_{n} \left(\hat{\rho}_{k_1+k_2+k_3+k_4}\right)_{nn}.
\label{A5}
\end{equation}
Using further the relation (see Ref.~\cite{Sinova:2000kx})
\begin{equation}
\sum_{n} \left(\hat{\rho}_{k_1+k_2+k_3+k_4}\right)_{nn} = N \delta_{\sum_i\mathbf{k}_i,0}\,,
\end{equation}
where $N = S\, \frac{|eB|}{2\pi}$ is the number of states on
the LLL and $S$ is the area of the transverse plane, and
the identity
\begin{equation}
S\, \delta_{\sum_i\mathbf{k}_i,0} = (2\pi)^2\delta^2\left(\sum_i\mathbf{k}_i\right),
\end{equation}
we finally get the expression for the vertex
$\Gamma_{4\pi}$ that
coincides with
expression (\ref{xspacevertices1}).

Thus, we see that the mathematical reason for the appearance of exponentially damping 
factors and the $M$-star product is related to the algebra of the projected density operators 
(\ref{product}). Obviously, the generalization of the above calculations to an arbitrary 
interaction vertex for the $\pi$ and $\tilde{\sigma}$ fields is straightforward.

\subsection{General structure of Vertices in Type I and Type II NCFTs}
\label{App:gauge-NCFT}

In Section~\ref{gauge-NCFT-2}, we restricted our analysis to the case when
fields $\phi^A(X)$ depend only on the transverse coordinates $u_{\perp}$.
In this Appendix, we consider the general case of fields $\phi^A(X)$ depending 
on both transverse and longitudinal coordinates.

Instead of expression (\ref{expansion2}), now we have the following representation 
for the bilocal field $\tilde{\varphi}(u,u^\prime)$:
\begin{equation}
\tilde{\varphi}(u,u^\prime)= \int \frac{d^4P}{(2\pi)^4} \phi^{A}(P)
\chi^{A}(u,u^\prime;P),
\label{expansion3}
\end{equation}
where the structure of the Bethe-Salpeter wave function
$\chi^A(u,u^\prime;P)$ is described in Eqs.~(\ref{chitilde}), (\ref{tildechi})
and
(\ref{f-equation2}). While for the case $P_{\parallel}=0$ the effective action
was given in expression (\ref{action2}), it now takes
the form
\begin{eqnarray}
S(\tilde{\varphi}) &=& \sum_{n=2}^{\infty} \frac{i}{n} \int
d^4x_1d^4u^\prime_{1}\ldots d^4x_nd^4y_n
\, \int \frac{d^4P_1\,\ldots \,d^4P_n}{(2\pi)^{4n}}\,
\phi^{A_1}(P_1)\,\ldots \,\phi^{A_n}(P_n)\,
\nonumber\\
&\times& \frac{\mathrm{tr}\,
[S_{LLL}^{-1}(x_1,u^\prime_{1})\chi^{A_1}(u^\prime_{1},x_2;P_1)\,\ldots\,
S_{LLL}^{-1}(x_{n-1},y_n)\chi^{A_n}(y_n,x_1;P_n)]}{\Pi_{i=1}^n
\lambda(P_i)}\,-\,
\frac{i}{2} \int d^4u_1d^4u^\prime_{1}d^4x_2d^4y_2
\int \frac{d^4P_1d^4P_2}{(2\pi)^8}\,
\nonumber\\
&\times& \phi^{A_1}(P_1)\phi^{A_2}(P_2)\,
\frac{\mathrm{tr}[
\chi^{A_1}(x_1,u^\prime_{1};P_1)S_{LLL}^{-1}(u^\prime_{1},y_2)\chi^{A_2}(y_2,x_2;P_2)
S_{LLL}^{-1}(x_2,x_1)]}{\lambda(P_1)}.
\label{action3}
\end{eqnarray}

It is convenient to represent $f^A(p_{\parallel};P)$, defined in
Eq.~(\ref{f-equation2}), as
\begin{equation}
f^A(p_{\parallel};P) = S_{\parallel}\left(p_{\parallel}+\frac{P_{\parallel}}{2}\right) G^A(p_{\parallel};P)
S_{\parallel}\left(p_{\parallel}+\frac{P_{\parallel}}{2}\right).
\label{G}
\end{equation}
[Note that a $\gamma$-matrix structure of
$G^A(p_{\parallel};P)$ is determined from the corresponding Bethe-Salpeter
equation and it can be
different from that in
Eq.~(\ref{pion}).]
Then, by using (\ref{chitilde}) and (\ref{tildechi}) and integrating over
$p_{\perp}$, we get
\begin{equation}
\chi^A(u,u^\prime;P)= P(u_{\perp},u^\prime_{\perp}) \int \frac{d^2p_{\parallel}}
{2(2\pi)^2} e^{-iP\frac{x+y}{2}}
e^{-ip_{\parallel}(u_{\parallel}-y_{\parallel})}
e^{-\frac{\vec{P}_{\perp}^2}{4|eB|}}
e^{\frac{\epsilon^{ab}P_{\perp}^a(u_{\perp}^b-u^{\prime b}_{\perp})
\mathrm{sign}(eB)}{2}} 
S_{\parallel}\left(p_{\parallel}+\frac{P_{\parallel}}{2}\right)G^{A}(p_{\parallel};P)
S_{\parallel}\left(p_{\parallel}+\frac{P_{\parallel}}{2}\right)
\label{auxfield1}
\end{equation}
(compare with Eq.~(\ref{auxfield})). Substituting $\chi^A(u,u^\prime;P)$ in
Eq.~(\ref{action3}), we obtain the effective action in momentum space:
\begin{equation}
S(\phi) = \sum_{n=2}^{\infty} \Gamma_n,
\end{equation}
where the interaction vertices $\Gamma_n$, $n > 2$, are
\begin{eqnarray}
\Gamma_n &=& \frac{2^{3-n}\pi^3 i|eB|}{n} \int
\frac{d^2k_{\parallel}}{(2\pi )^2}
\int \frac{d^4P_1}{(2\pi )^4}\,\ldots \,\frac{d^4P_n}
{(2\pi )^4}\,
\delta^4\left(\sum_{i=1}^n P_{i}\right)\,
\phi^{A_1}(P_1)\,\ldots\,
\phi^{A_n}(P_n)\,
\nonumber\\
&\times& \mathrm{tr}
\Bigg[G^{A_1}\left(k_{\parallel} - \frac{P_1^{\parallel}}{2}; P_1\right)
S_{\parallel}(k_{\parallel} - P_1^{\parallel})
 G^{A_2}\left(k_{\parallel} - P_1^{\parallel} - \frac{P_2^{\parallel}}{2}; P_2\right)
 S_{\parallel}(k_{\parallel} - P_1^{\parallel} - P_2^{\parallel})
\nonumber\\
&\times&
\ldots G^{A_n}\left(k_{\parallel} - \sum_{i=1}^{n-1} P_i^{\parallel} - \frac{P_n^{\parallel}}{2}; P_n\right)
S_{\parallel}\left(k_{\parallel} - \sum_{i=1}^{n} P_i^{\parallel}\right)\Bigg]\,\,
\frac{e^{-\frac{i}{2}\sum_{i<j}
P_i^{\perp} \times P_j^{\perp}}}{\Pi_{i=1}^n \lambda(P_i)}
\label{nvertexl}
\end{eqnarray}
(compare with expression (\ref{nvertex})),
and the quadratic part of the action is
\begin{eqnarray}
\Gamma_2 &=&
-\frac{\pi i|eB|}{4} \int \frac{d^2k_{\parallel}}{(2\pi )^2}
\int \frac{d^4P}{(2\pi )^4}\,
\frac{\lambda(P) - 1}{\lambda^2(P)}\,\phi^{A_1}(P)
\nonumber\\
&\times&
\mathrm{tr}
\left[G^{A_1}\left(k_{\parallel} - \frac{P_{\parallel}}{2}; P\right) S_{\parallel}(k_{\parallel} - P_{\parallel})
 G^{A_2}\left(k_{\parallel} - \frac{P_{\parallel}}{2}; - P\right) S_{\parallel}(k_{\parallel})\right]
\,\phi^{A_2}(-P)
\label{quadraticl}
\end{eqnarray}
(compare with expression (\ref{quadratic})).

Further, in coordinate space, the
interaction vertices take the form similar to that in
Eq.~(\ref{coordinate2}):
\begin{equation}
\Gamma_n = \frac{i|eB|}{2^{n+1}\pi n} \int d^4u
\left[V_{n}^{A_1\ldots A_n}(-i\nabla_1,\ldots,-i\nabla_n)
\phi^{A_1}(X_1)*\ldots*\phi^{A_n}(X_n)\right]|_{X_1=X_2=\ldots =X}\,,
\label{coordinate3}
\end{equation}
where the nonlocal operator $V_{n}^{A_1\ldots A_n}$ now depends both on
transverse $\nabla_{\perp}$ and longitudinal $\nabla_{\parallel}$.
In momentum space, this operator is
\begin{eqnarray}
V_n^{A_1\ldots A_n}(P_1,\ldots,P_n) &=& \int
\frac{d^2k_{\parallel}}{(2\pi)^2}
\mathrm{tr}
\Bigg[G^{A_1}\left(k_{\parallel} - \frac{P_1^{\parallel}}{2}; P_1\right)
S_{\parallel}(k_{\parallel} - P_1^{\parallel})
 G^{A_2}\left(k_{\parallel} - P_1^{\parallel} - \frac{P_2^{\parallel}}{2}; P_2\right)
 S_{\parallel}(k_{\parallel} - P_1^{\parallel} - P_2^{\parallel})
\nonumber\\
&\times&
\ldots G^{A_n}\left(k_{\parallel} - \sum_{i=1}^{n-1} P_i^{\parallel} - \frac{P_n^{\parallel}}{2}; P_n\right)
S_{\parallel}\left(k_{\parallel} - \sum_{i=1}^{n} P_i^{\parallel}\right)\Bigg]
\,\frac{1}{\Pi_{i=1}^n \lambda(P_i)}
\end{eqnarray}
(compare with Eq.~(\ref{V})).
Finally, the interaction vertices are given by the following expression in
noncommutative space:
\begin{equation}
\Gamma_n = \frac{i|eB|}{2^{n+1}\pi n} \int d^2u_{\parallel}\,\mathbf{Tr}
\,\Bigg[\Big(V_{n}^{A_1\ldots A_n}(-i\nabla_1^{\parallel}, -i\hat{\nabla}_1^{\perp},\ldots,
-i\nabla_n, -i\hat{\nabla}_n^{\perp})\, \phi^{A_1}(X_1^{\parallel},\hat{X}_1^{\perp})
\ldots \phi^{A_n}(X_n^{\parallel},\hat{X}^{\perp})
\Big)_{X_i^{\parallel}=X^{\parallel},\,\hat{X}_i^{\perp}=\hat{X}^{\perp}}\Bigg],
\label{noncommspace3}
\end{equation}
where here the subscript
$i$ runs from 1 to $n$ (compare with expression (\ref{ncspace2})).

What is the connection between the forms of the function
$f^{A}(p_{\parallel};P)$ in the cases with zero
and nonzero longitudinal momentum $P_{\parallel}$? In regard to this question,
it is appropriate to recall
the Pagels-Stokar (PS) approximation \cite{Pagels:1979hd}
for a Bethe-Salpeter wave function $\chi^{A}(p;P)$, which is often used
in Lorentz invariant
field theories. It is assumed in this approximation
that the amputated Bethe-Salpeter wave function,
defined as
\begin{equation}
\hat{\chi}^{A}(p;P) \equiv S^{-1}\left(p +\frac{P}{2}\right)\chi^{A}(p;P)
S^{-1}\left(p -\frac{P}{2}\right),
\end{equation}
is approximately the same for the cases with zero and nonzero
$P$, i.e.,
$\hat{\chi}^{A}(p;P) \simeq \hat{\chi}^{A}(p)$ where
$\hat{\chi}^{A}(p) \equiv \hat{\chi}^{A}(p;P)|_{P=0}$.
Then, in this approximation,
the Bethe-Salpeter wave function $\chi^{A}(p;P)$ is
\begin{equation}
\chi^{A}(p;P) = S\left(p +\frac{P}{2}\right)\hat{\chi}^{A}(p)
S\left(p -\frac{P}{2}\right),
\end{equation}
i.e., the whole dependence of the Bethe-Salpeter wave function on
the momentum $P$ comes from the fermion propagator.
It is known (for a review, see
Ref.~\cite{Miransky:1994vk}) that the PS approximation can be justified
both for weak coupling dynamics and, in the case of the NJL model,
in the regime with large $N_c$.

Here we would like to suggest an
anisotropic version of the PS approximation for dynamics
in a magnetic field. The main assumption we make is that the function
$G^A(p_{\parallel};P)$, defined in Eq.~(\ref{G}), is approximately
$P_{\parallel}$ independent, i.e.,
\begin{equation}
G^A(p_{\parallel};P) \simeq F^{A}(p_{\parallel};P_{\perp})\gamma^5
\frac{1-i\gamma^1\gamma^2}{2},
\label{G1}
\end{equation}
where the function $F^{A}(p_{\parallel};P_{\perp})$ is defined in Eq.
(\ref{pion}).
Then, the whole dependence of the function
$f^{A}(p_{\parallel},P)$ (\ref{G})
on
the longitudinal momentum $P_{\parallel}$ comes from the fermion propagator:
\begin{equation}
f^A(p_{\parallel};P) = S_{\parallel}\left(p_{\parallel}+\frac{P_{\parallel}}{2}\right)
F^{A}(p_{\parallel};P_{\perp})\gamma^5 \frac{1-i\gamma^1\gamma^2}{2}
S_{\parallel}\left(p_{\parallel}+\frac{P_{\parallel}}{2}\right)
\label{G2}
\end{equation}
(compare with Eq.~(\ref{pion})).

We utilized expression (\ref{G1}) in QED in a magnetic field in the
dynamical regime with the local interaction considered in Section~\ref{gauge-NCFT-4}.
In that case, the function $F^{A}(p_{\parallel};P_{\perp})$ is a constant. Then,
using expression (\ref{nvertexl}) for $\Gamma_n$ with $G^A$ in Eq.~(\ref{G1}),
it is not difficult to derive the $n$-point vertices for fields $\phi^{A}(P)$ for
a general momentum $P$. The result coincides with that obtained in
Ref.~\cite{Gorbar:2004ck} (see also Section~\ref{NJL-NCFT} ) in the NJL model.

Expressions (\ref{G1}) and (\ref{G2}) can also be useful for the analysis of the dynamics 
in QED and QCD in a magnetic field in the weak coupling regime. As was shown in 
Sections~\ref{gauge-NCFT-5} and \ref{gauge-NCFT-6}, in those cases the function 
$F^{A}(p_{\parallel};P_{\perp})$ depends on both momenta $p_{\parallel}$ and $P_{\perp}$ 
(the dynamics in this regime relate to type II NCFT). The determination of this dependence 
is a nontrivial problem.

\end{document}